\documentclass[12pt,preprint]{aastex}

\usepackage{lscape}
\usepackage{graphicx}
\usepackage{amsmath}

\ifdefined\bwfigures
\graphicspath{{./figuresbw/}{./figures/}}
\else
\graphicspath{{./figures/}}
\fi

\usepackage{color}

\usepackage{cite}
\usepackage{natbib} 
\newcommand{\gammaRay}{$\gamma$ ray}
\newcommand{\gammaRays}{$\gamma$ rays}
\newcommand{\gammaRayHyph}{$\gamma$-ray}
\newcommand{\gammaRaysHyph}{$\gamma$-rays}

\newcommand{\NEWTEXT}[1]{\textbf{#1}}

\newcommand{\stools}{\emph{ScienceTools}}
\newcommand{\stool}[1]{\emph{#1}}

\newcommand{\gtmktime}{\stool{gtmktime}}

\newcommand{\selection}[1]{\texttt{#1}}

\newcommand{\roi}{ROI}
\newcommand{\rois}{ROIs}
\newcommand{\geant}{Geant4}

\newcommand{\subrm}[1]{_{\rm #1}}
\newcommand{\core}[1]{\ensuremath{#1\subrm{core}}}
\newcommand{\tail}[1]{\ensuremath{#1\subrm{tail}}}

\newcommand{\scalingfunction}[1]{\ensuremath{S_{#1}}}
\newcommand{\irf}[1]{\texttt{#1}}
\newcommand{\psix}{\irf{Pass~6}}
\newcommand{\pseven}{\irf{Pass~7}}
\newcommand{\aeff}{\ensuremath{A_{\rm eff}}}
\newcommand{\accept}{\ensuremath{\mathcal{A}}}
\newcommand{\fov}{{\rm Fo\kern-1ptV}}
\newcommand{\psf}{P} 
\newcommand{\psfscaling}{\scalingfunction{\psf}}
\newcommand{\Kingf}{K}
\newcommand{\Randof}{R}

\newcommand{\edisp}{D}
\newcommand{\edispscaling}{\scalingfunction{\edisp}}

\newcommand{\tobs}{\ensuremath{t_{\rm obs}}}
\newcommand{\probE}{\ensuremath{P_{\rm E}}}
\newcommand{\probCore}{\ensuremath{P_{\rm core}}}
\newcommand{\probCPF}{\ensuremath{P_{\rm CPF}}}
\newcommand{\probCAL}{\ensuremath{P_{\rm CAL}}}
\newcommand{\probTKR}{\ensuremath{P_{\rm TKR}}}
\newcommand{\probAll}{\ensuremath{P_{\rm all}}}
\newcommand{\loge}{\ensuremath{\log_{10}\left(E/\text{1~MeV}\right)}}
\newcommand{\bracketirf}[1]{\emph{#1}}

\newcommand{\allgamma}{\emph{allGamma}}

\newcommand{\secref}[1]{\S~\ref{#1}}
\newcommand{\appref}[1]{Appendix~\ref{#1}}
\newcommand{\Tabref}[1]{Table~\ref{tab:#1}}
\newcommand{\Tabrefs}[2]{Tables~\ref{tab:#1} and \ref{tab:#2}}
\newcommand{\tabref}[1]{Table~\ref{tab:#1}}
\newcommand{\Figref}[1]{Figure~\ref{fig:#1}}
\newcommand{\figref}[1]{Figure~\ref{fig:#1}}
\newcommand{\Figrefs}[2]{Figures~\ref{fig:#1} and \ref{fig:#2}}
\newcommand{\parenfigref}[1]{Fig.~\ref{fig:#1}}
\newcommand{\Eqref}[1]{Eqn.~\eqref{#1}}

\newcommand{\Fermi}{{\textit{Fermi}}}
\newcommand{\fermi}{\Fermi}

\newcommand{\trgprimitive}[1]{\texttt{#1}}

\newcommand{\tpperiodic}{\trgprimitive{PERIODIC}}
\newcommand{\tptkr}{\trgprimitive{TKR}}
\newcommand{\tpcallo}{\trgprimitive{CAL\_LO}}
\newcommand{\tpcalhi}{\trgprimitive{CAL\_HI}}
\newcommand{\tpcno}{\trgprimitive{CNO}}
\newcommand{\tproi}{\trgprimitive{ROI}}
\newcommand{\tpveto}{\trgprimitive{VETO}}

\newcommand{\obf}[1]{\textsc{#1}~filter}
\newcommand{\obfgam}{\obf{gamma}}
\newcommand{\obfhip}{\obf{hip}}

\newcommand{\obfdgn}{\obf{diagnostic}}

\newcommand{\calrange}[1]{\textsc{#1}}
\newcommand{\lexone}{\calrange{lex1}}
\newcommand{\lexeight}{\calrange{lex8}}
\newcommand{\hexone}{\calrange{hex1}}
\newcommand{\hexeight}{\calrange{hex8}}

\newcommand{\us}{$\mu$s}
\newcommand{\um}{$\mu$m}

\newcommand{\xy}{$x$--$y$}
\newcommand{\mcilwainl}{McIlwain~L}

\newcommand{\acronymlabel}[1]{\label{acronym:#1}}
\newcommand{\acronym}[3][]{\noindent\makebox[70pt]%
  {\ifx\\#1\\#2\else#1\fi\hfill}#3%
  \dotfill\secref{acronym:#2}\\}

\newcommand{\notation}[3]{\noindent\makebox[70pt]{#1\hfill}#2\dotfill#3\\}
\newcommand{\convention}[3]{\notation{#1}{#2}%
  {\secref{conv:#3}}
}

\newcommand{\filename}[1]{\emph{#1}}
\newcommand{\webpage}[1]{\scriptsize{#1}}

\newcommand{\syserrors}[2]{\makebox[42pt]{$#1$}\makebox[42pt]{$#2$}}

\newlength{\enumindent}
\newcounter{fermicnt}

\newenvironment{fermienumerate}{
  \begin{list}{\arabic{fermicnt}.}{\usecounter{fermicnt}%
  \setlength{\leftmargin}{0pt}%
  \setlength{\labelwidth}{0pt}}%
  \setlength{\itemindent}{\enumindent}%
  \setlength{\labelsep}{\enumindent}%
  \setlength{\topsep}{0pt}%
  \setlength{\itemsep}{0pt}%
}{
  \end{list}%
}

\newenvironment{fermiitemize}[1][$\bullet$]{%
  \begin{list}{#1}{%
  \setlength{\leftmargin}{0pt}%
  \setlength{\labelwidth}{0pt}}%
  \setlength{\itemindent}{\enumindent}%
  \setlength{\labelsep}{\enumindent}%
  \setlength{\topsep}{0pt}%
  \setlength{\itemsep}{0pt}%
}{%
  \end{list}%
}

\newlength{\onecolfigwidth}
\newlength{\twothirdscolfigwidth}
\newcommand{\twopanel}[4]{%
  \begin{figure}[#1]
    \centering
    \includegraphics[width=\onecolfigwidth]{#2}\hfill%
    \includegraphics[width=\onecolfigwidth]{#3} 

    #4
  \end{figure}
}

\newcommand{\threepanel}[5]{%
  \begin{figure}[#1]
    \centering
    \includegraphics[width=\twothirdscolfigwidth]{#2}\hfill%
    \includegraphics[width=\twothirdscolfigwidth]{#3}\hfill%
    \includegraphics[width=\twothirdscolfigwidth]{#4} 

    #5
  \end{figure}
}

\newcommand{\fourpanel}[6]{%
  \begin{figure}[#1]
    \centering
    \includegraphics[width=\onecolfigwidth]{#2}\hfill%
    \includegraphics[width=\onecolfigwidth]{#3}\\
    \includegraphics[width=\onecolfigwidth]{#4}\hfill%
    \includegraphics[width=\onecolfigwidth]{#5}

    #6
  \end{figure}
}

\begin{document}

\title{The \fermi\ Large Area Telescope On Orbit: Event Classification, Instrument Response Functions, and Calibration}

\author{
M.~Ackermann\altaffilmark{1}, 
M.~Ajello\altaffilmark{2}, 
A.~Albert\altaffilmark{3}, 
A.~Allafort\altaffilmark{2}, 
W.~B.~Atwood\altaffilmark{4}, 
M.~Axelsson\altaffilmark{5,6,7}, 
L.~Baldini\altaffilmark{8,9}, 
J.~Ballet\altaffilmark{10}, 
G.~Barbiellini\altaffilmark{11,12}, 
D.~Bastieri\altaffilmark{13,14}
% ...
}
\altaffiltext{1}{Deutsches Elektronen Synchrotron DESY, D-15738 Zeuthen, Germany}
\altaffiltext{2}{W. W. Hansen Experimental Physics Laboratory, Kavli Institute for Particle Astrophysics and Cosmology, Department of Physics and SLAC National Accelerator Laboratory, Stanford University, Stanford, CA 94305, USA}
\altaffiltext{3}{Department of Physics, Center for Cosmology and Astro-Particle Physics, The Ohio State University, Columbus, OH 43210, USA}
\altaffiltext{4}{Santa Cruz Institute for Particle Physics, Department of Physics and Department of Astronomy and Astrophysics, University of California at Santa Cruz, Santa Cruz, CA 95064, USA}
\altaffiltext{5}{Department of Astronomy, Stockholm University, SE-106 91 Stockholm, Sweden}
\altaffiltext{6}{The Oskar Klein Centre for Cosmoparticle Physics, AlbaNova, SE-106 91 Stockholm, Sweden}
\altaffiltext{7}{Department of Physics, Royal Institute of Technology (KTH), AlbaNova, SE-106 91 Stockholm, Sweden}
\altaffiltext{8}{Universit\`a  di Pisa and Istituto Nazionale di Fisica Nucleare, Sezione di Pisa I-56127 Pisa, Italy}
\altaffiltext{9}{email: luca.baldini@pi.infn.it}
\altaffiltext{10}{Laboratoire AIM, CEA-IRFU/CNRS/Universit\'e Paris Diderot, Service d'Astrophysique, CEA Saclay, 91191 Gif sur Yvette, France}
\altaffiltext{11}{Istituto Nazionale di Fisica Nucleare, Sezione di Trieste, I-34127 Trieste, Italy}
\altaffiltext{12}{Dipartimento di Fisica, Universit\`a di Trieste, I-34127 Trieste, Italy}
\altaffiltext{13}{Istituto Nazionale di Fisica Nucleare, Sezione di Padova, I-35131 Padova, Italy}
\altaffiltext{14}{Dipartimento di Fisica e Astronomia "G. Galilei", Universit\`a di Padova, I-35131 Padova, Italy}
% ...

\date{\vfil\null\newpage}

\begin{abstract}
The \fermi\ Large Area Telescope (\fermi-LAT, hereafter LAT), the primary
instrument on the \emph{Fermi Gamma-ray Space Telescope} (\fermi) mission, is
an imaging, wide field-of-view, high-energy \gammaRayHyph\ telescope, covering
the energy range from 20~MeV to more than 300~GeV.  During the first years of the 
mission the LAT team has gained considerable insight into the in-flight performance 
of the instrument. Accordingly, we have updated the analysis used to reduce LAT 
data for public release as well as the Instrument Response Functions (IRFs), the 
description of the instrument performance provided for data analysis. In this paper 
we describe the effects that motivated these updates.  Furthermore, we discuss how 
we originally derived IRFs from Monte Carlo simulations and later corrected those 
IRFs for discrepancies observed between flight and simulated data. We also give 
details of the validations performed using flight data and quantify the residual 
uncertainties in the IRFs. Finally, we describe techniques the LAT team has 
developed to propagate those uncertainties into estimates of the systematic 
errors on common measurements such as fluxes and spectra of astrophysical sources.
\end{abstract}


\maketitle
\pagebreak

%
\tableofcontents
\pagebreak


%

\section{INTRODUCTION}\label{sec:intro}

The \emph{Fermi Gamma-ray Space Telescope} (\fermi)\acronymlabel{Fermi} was
launched on 2008~June~11.
Commissioning of the \Fermi\ Large Area Telescope (LAT)\acronymlabel{LAT}
began on 2008 June~24~\citep{REF:2009.OnOrbitCalib}. On 2008~August~4, the LAT
began nominal science operations. Approximately one year later the LAT data were
publicly released via the \Fermi\ Science Support Center
(FSSC)\acronymlabel{FSSC}\footnote{\webpage{http://fermi.gsfc.nasa.gov/ssc}}.

The LAT is a pair-conversion telescope; individual \gammaRays\ convert to
$e^{+}e^{-}$ pairs, which are recorded by the instrument. By reconstructing the
$e^{+}e^{-}$ pair we can deduce the energy and direction of the incident
\gammaRay. Accordingly, LAT data analysis is entirely event-based: we record and
analyze each incident particle separately.  

In the first three years of LAT observations (from 2008~August~4 to
2011~August~4), the LAT read out over $1.8 \times 10^{11}$ individual events, of which
$\sim 3.4 \times 10^{10}$ were transmitted to the ground and subsequently analyzed in
the LAT data processing pipeline at the LAT Instrument Science Operations
Center (ISOC)\acronymlabel{ISOC}. Of those, $\sim 1.44 \times 10^{8}$ passed detailed
\gammaRayHyph\ selection criteria and entered the LAT public data set.

The LAT team and the FSSC work together to develop, maintain and publicly
distribute a suite of instrument-specific science analysis tools (hereafter
\stools%
\footnote{\webpage{http://fermi.gsfc.nasa.gov/ssc/data/analysis/software}})
that can be used to perform standard astronomical analyses.  
A critical component of these tools is the parametrized representations of
instrument performance: the instrument response functions
(IRFs)\acronymlabel{IRF}. 
In practice, the LAT team assumes that the IRFs can be factorized into three
parts (the validity of this assumption is studied
in~\secref{subsec:EDisp_psf_correl}):
\label{conv:aeff}\label{conv:psf}\label{conv:edisp}
\begin{fermienumerate}
\item \emph{Effective Area}, $\aeff(E,\hat{v},s)$, the product of the
  cross-sectional geometrical collection area, \gammaRay\ conversion
  probability, and the efficiency of a given event selection (denoted by $s$)
  for a \gammaRay\ with energy $E$ and direction $\hat{v}$\label{conv:LAT_dir} in the LAT frame;
\item \emph{Point-spread Function} (PSF)\acronymlabel{PSF},
  $\psf(\hat{v}^{\,\prime};E,\hat{v},s)$, the probability density to reconstruct
  an incident direction ${\hat{v}^{\prime}}$ for a \gammaRay\ with ($E,\hat{v}$)
  in the event selection $s$;
\item \emph{Energy Dispersion}, $\edisp(E^{\prime};E,\hat{v},s)$, the
  probability density to measure an event energy $E^{\prime}$ for a \gammaRay\
  with ($E,\hat{v}$) in the event selection $s$.
\end{fermienumerate}
The IRFs described above are designed to be used in a maximum
likelihood analysis%
\footnote{\webpage{http://fermi.gsfc.nasa.gov/ssc/data/analysis/scitools/likelihood\_tutorial.html}} 
as described in \citet{REF:1996.Mattox}.  Given a distribution of
\gammaRays\ $S(E,\hat{p})$\label{conv:sourceDistrib}, where $\hat{p}$\label{conv:skyDir} refers to the celestial
directions of the \gammaRays, we can use the IRFs to predict the
distribution of observed \gammaRays\ $M(E^{\prime},\hat{p}^{\prime},s)$:\label{conv:exCountsPred}
\begin{align}\label{eq:exCountsPred}
   M(E^{\,\prime},\hat{p}^{\,\prime},s) = & \int \int \int S(E,\hat{p}) \aeff(E,\hat{v}(t;\hat{p}),s) \times \nonumber \\
    & P(\hat{v}'(t,\hat{p}^{\,\prime}); E, \hat{v}(t;\hat{p}), s) D(E^{\,\prime}; E, \hat{v}(t;\hat{p}), s) dE d\Omega dt.
\end{align}
The integrals are over the time range of interest for the
analysis, the solid angle in the LAT reference frame and the energy range of the LAT.   

Note that the IRFs can change markedly across the LAT field-of-view
(\fov) \acronymlabel{FOV}.  Therefore, we define the exposure at a given energy for any
given energy and direction in the sky $\mathcal{E}(E,\hat{p})$ as
the integral over the time range of interest of the effective area for that particular direction;
\begin{equation}\label{eq:exposureDef}
\mathcal{E}(E,\hat{p},s) = \int \aeff(E,\hat{v}(t,\hat{p}),s) dt.
\end{equation}
Another important quantity is the distribution of observing time in
the LAT reference frame of any given direction in the sky (henceforth
referred to as the observing profile, and written \tobs\label{conv:tobs}), and which is closely related to the exposure:
\begin{equation}\label{eq:tobsDef}
\mathcal{E}(E,\hat{p},s) = \int \aeff(E,\hat{v},s) \tobs(\hat{v};\hat{p}) d\Omega.
\end{equation}

The absolute timing performance of the LAT has been described in
detail in \citet{REF:2009.OnOrbitCalib} and
\citet{REF:PulsarTiming:2008} and will not be discussed in this paper.

To allow users to perform most standard analyses with minimum effort, the LAT
team also provides, via the FSSC, a spatial and spectral model of the Galactic
diffuse \gammaRayHyph\ emission and a spectral template for isotropic
\gammaRayHyph\ emission\footnote{\webpage{http://fermi.gsfc.nasa.gov/ssc/data/access/lat/BackgroundModels.html}}.
In this prescription, contamination of the \gammaRayHyph\ sample from
residual charged cosmic rays (CR)\acronymlabel{CR} is included in the isotropic
spectral template. Although not part of the IRFs, this background contamination
is an important aspect of the instrument performance.

From the instrument design to the high level source analysis, the LAT team has
relied heavily on Monte Carlo (MC)\acronymlabel{MC} simulations of
\gammaRayHyph\ interactions with the LAT to characterize performance and
develop IRFs.
The high quality data produced since launch have largely validated this choice.
However, unsurprisingly, the real flight data exhibited unanticipated features
that required modifications to the IRFs.  After years of observations the LAT 
data set itself is by far the best source of calibration
data available to characterize these modifications.  

LAT event analysis has substantially improved since launch. We have applied the
accumulated updates in occasional re-releases of data, corresponding to
reprocessing the entirety of LAT data to make use of the latest available
analysis. In addition to being a resource-consuming task, re-releases require
that users download the newly available data and replace all
science analysis products. In addition, during the mission we have also made
minor improvements in the IRFs based on a better understanding of the
properties of the event analysis and an improved description of the LAT
performance. These incremental IRFs can be computed and released without
modifying existing data, and many of the analysis products remain valid.

We have released two major iterations of the data analysis since launch:
\begin{fermiitemize}
\item \psix\ indicates the event analysis scheme designed prior to launch. 
  As such, it was based exclusively on our informed estimates of the
  cosmic-ray environment at the orbit of \fermi\ and a MC-based evaluation of
  the LAT performance.
  After the commissioning phase, as data started accumulating, we observed
  phenomena that were not reproduced in the MC simulations
  (see \secref{subsec:LAT_simul} and \secref{subsec:Aeff_MC_corrections}).
  Without modifying the event analysis in any way, we opted to reduce
  systematic errors by adding these effects to the MC simulations, and we re-evaluated the LAT
  performance (in particular we calculated new IRFs, see
  \secref{subsec:Aeff_MC_corrections}). While this did not allow us to recover
  any of the lost LAT performance, it ensured that real and simulated data were
  subject to the same effects and the MC-estimated performance was therefore 
  adequate for science analysis. We have described the initial \psix\
  release (\irf{P6\_V1}) in \citet{REF:2009.LATPaper}, and the corrected IRFs
  (\irf{P6\_V3}) in \citet{REF:2009.LATPerf}.  We will discuss some improvements
  that were incorporated into the later \irf{P6\_V11} IRFs in
  \secref{subsec:Aeff_flightAEff} and \secref{subsec:PSF_onorbit}.
\item \pseven\ indicates an improved version of the event analysis, for which we
  updated parts of the data reduction process to account for known on-orbit 
  effects by making use of the large number of real events the LAT collected in
  2~years of operation. The event reconstruction and the overall analysis
  design were not modified, but the event classification was re-optimized on
  simulated data-sets including all known on-orbit effects. Large samples of
  real events were used to assess the efficiency of each step and the
  systematics involved. Particular attention was paid to increasing effective
  area below $\sim 300$~MeV where the impact of on-orbit effects was large,
  while maintaining tolerable rates of CR contamination at those energies. 
  Event class definitions were optimized based
  on comparisons of MC events and selected samples of real LAT data.
  See \secref{sec:event} for a description of \pseven.
\end{fermiitemize}

All data released prior to 2011 August 1 were based on \psix. On 2011 August 1
we released \pseven\ data for the entire mission to date, and since then all
data have been processed only with \pseven.

This paper has two primary purposes. The first is to describe \pseven\
(\secref{sec:event}), quantifying the differences with respect to \psix\
when necessary.  The second is to detail our understanding of the LAT,
and toward that end we describe how we have used flight data to
validate the generally excellent fidelity of our simulations of particle
interactions in the LAT, as well as the resulting IRFs and residual charged
particle contamination. In particular, we describe the methods and control
data samples we have used to study the residual charged particle contamination
(\secref{sec:bkg}), effective area (\secref{sec:Aeff}), PSF (\secref{sec:PSF}),
and energy dispersion (\secref{sec:EDisp}) of the LAT. Furthermore, we 
quantify the uncertainties in each case, and discuss how these uncertainties
affect high-level scientific analyses (\secref{sec:perf}).  

For convenience, we have included lists of the acronyms and abbreviations 
(\appref{app:acronym}) and notation conventions (\appref{app:conventions}) used in this paper.

\section{LAT INSTRUMENT, ORBITAL ENVIRONMENT, DATA PROCESSING AND SIMULATIONS}\label{sec:LAT}

In this paper, we focus primarily on those aspects of the LAT instrument, data,
and analysis algorithms that are most relevant for the understanding
and validation of the LAT performance.
Additional discussion of these subjects was provided in a dedicated paper
~\citep{REF:2009.LATPaper}. The calibrations of the LAT subsystems are
described in a second paper~\citep{REF:2009.OnOrbitCalib}.

\subsection{LAT Instrument}\label{subsec:LAT_instrument}

\begin{figure}[htbp]
  \centering
  \includegraphics[width=\onecolfigwidth]{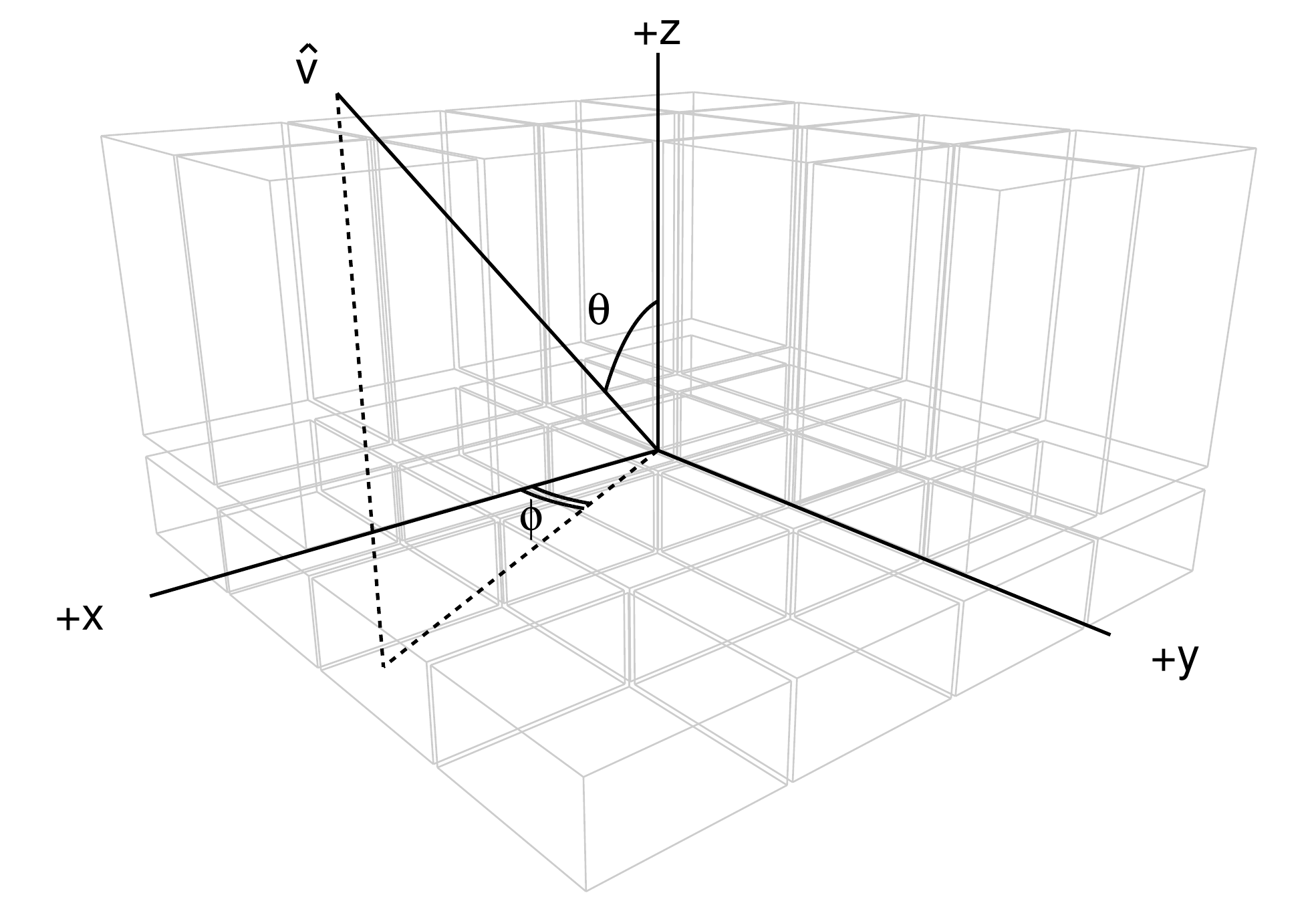}
  \caption{Schematic of the LAT, including the layout of the 16~CAL modules
    and 12 of the 16~TKR modules (for graphical clarity the ACD is not shown).
    This figure also defines the $(\theta,\phi)$ coordinate system used
    throughout the paper.}
  \label{fig:lat_layout}
\end{figure}

The LAT consists of three detector subsystems. A tracker/converter
(TKR)\acronymlabel{TKR}, comprising 18 layers of paired \xy\ Silicon Strip
Detector (SSD)\acronymlabel{SSD} planes with interleaved tungsten foils,
which promote pair conversion and measure the directions of incident
particles~\citep{REF:2007.TKRPaper}.
A calorimeter (CAL)\acronymlabel{CAL}, composed of 8.6 radiation lengths of
CsI(Tl) scintillation crystals stacked in 8 layers, provides energy 
measurements as well as some imaging capability~\citep{REF:2010.CALPaper}.
An Anticoincidence Detector (ACD)\acronymlabel{ACD}, featuring 
an array of plastic scintillator tiles and wavelength-shifting fibers, 
surrounds the TKR and rejects CR backgrounds~\citep{REF:2007.ACDPaper}.
In addition to these three subsystems a triggering and data acquisition system 
selects and records the most likely \gammaRayHyph\ candidate events
for transmission to the ground.  Both the CAL and TKR consist of 16 
modules (often referred to as \emph{towers}) arranged in a $4 \times 4$ grid.
Each tower has a footprint of $\sim 37~\text{cm} \times 37~\text{cm}$ and
is $\sim 85$~cm high (from the top of the TKR to the bottom of the CAL).  
A schematic of the LAT is shown in \figref{lat_layout}, and defines the
coordinate system used throughout this paper. Note that the z-axis corresponds
to the LAT boresight, and the incidence ($\theta$) and azimuth ($\phi$)
angles are defined with respect to the $z$- and $x$- axes respectively.\label{conv:thetaPhi}

\subsubsection{Silicon Tracker}\label{subsec:LAT_TKR}

The TKR is the section of the LAT where \gammaRays\ ideally
convert to $e^{+}e^{-}$ pairs and their trajectories are measured. A full
description of the TKR can be found in~\citet{REF:2007.TKRPaper}
and~\citet{REF:2009.LATPaper}.
Starting from the top (farthest from the CAL), the first 12 paired layers are
arranged to immediately follow converter foils, which are composed of
$\sim 3$\% of a radiation length of tungsten.  Minimizing the separation of the
converter foils from the following SSD planes, and hence the lever-arm between
the conversion point and the first position measurements, is critical to
minimize the effects of multiple scattering. This section of the TKR is
referred to as the \emph{thin} or \emph{front} section. The next 4 layers are
similar except that the tungsten converters are $\sim 6$ times thicker;
these layers are referred to as the \emph{thick} or \emph{back} section.
The last two layers have no converter; this is dictated by the TKR trigger,
which requires hits in 3 \xy\ paired adjacent layers
(see \secref{subsubsec:event_trigger}) and is therefore insensitive to
\gammaRays\ that convert in the last two layers.

\begin{figure}[htbp]
  \centering
  \includegraphics[width=\onecolfigwidth]{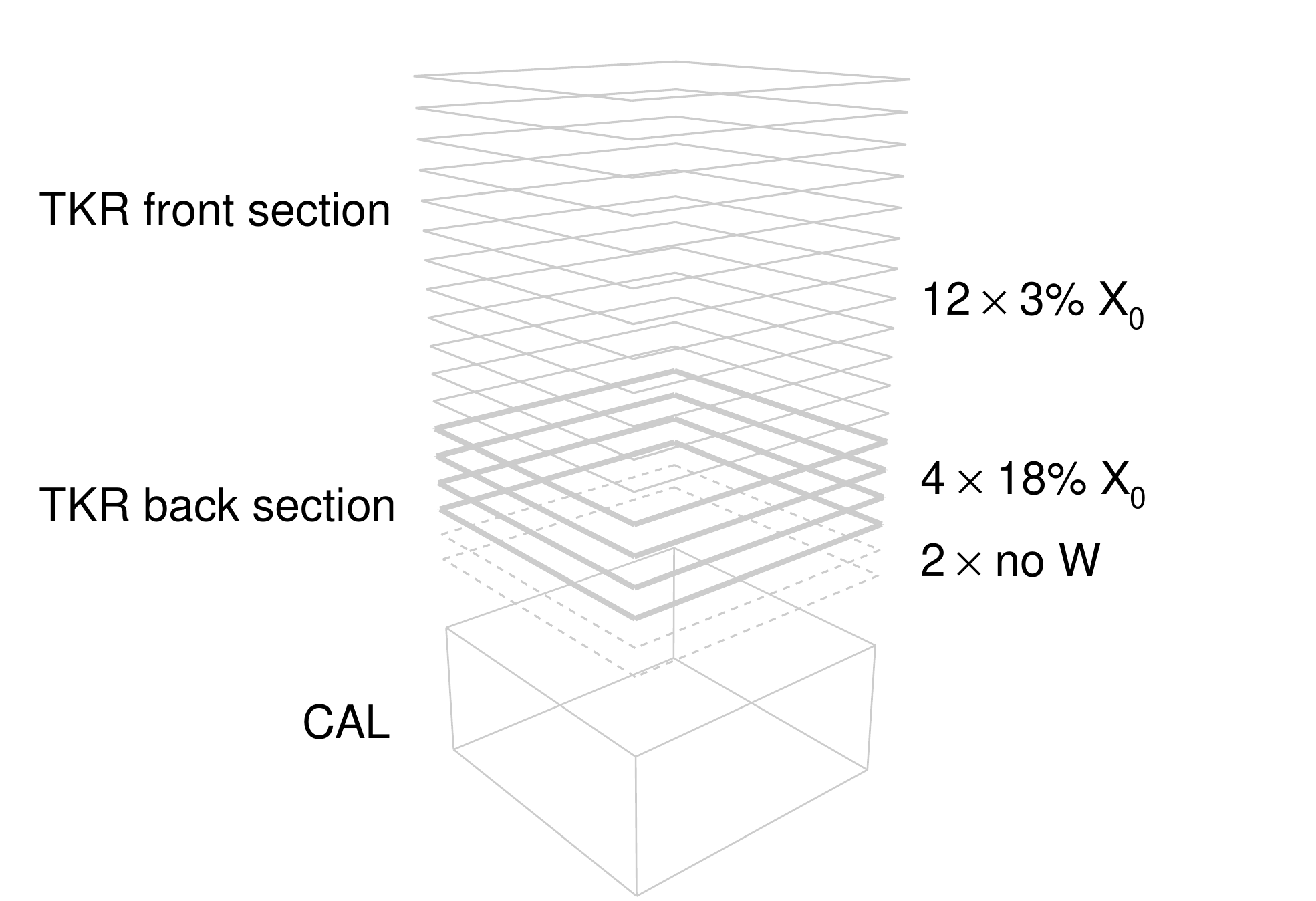}
  \caption{Schematic of a LAT tower (including a TKR and a CAL module).
    The layout of the tungsten conversion planes in the TKR is illustrated.}
  \label{fig:tower_layout}
\end{figure}

Thus the TKR effectively divides into two distinct instruments with notable
differences in performance, especially with respect to the PSF and background
contamination. This choice was suggested by the need to balance two basic
(and somewhat conflicting) requirements: simultaneously obtaining good
angular resolution and a large conversion probability.
The tungsten foils were designed such that there are approximately the same
number of \gammaRays\ (integrated over the instrument \fov) converted in the
thin and thick sections. In addition to these considerations, experience on
orbit has also revealed that the aggregate of the thick layers ($\sim 0.8$
radiation lengths) limits the amount of back-scattered particles from the CAL
returning into the TKR and ACD in high-energy events (i.e., the CAL
\emph{backsplash}) and reduces tails of showers in the TKR from events 
entering the back of the CAL. These two effects help to decrease the background
contamination in front-converting events.

After three years of on-orbit experience with the TKR we can now assess the
validity of our design decisions. The choice of the solid-state TKR
technology has resulted in negligible down time and extremely stable operation,
minimizing the necessity for calibrations.
Furthermore, the very high signal-to-noise ratio of the TKR analog readout
electronics has resulted in a single hit efficiency, averaged over the active
silicon surface, greater than 99.8\%, with a typical noise occupancy smaller
than $10^{-5}$ for a single readout channel. (We note for completeness that the
fraction of non-active area presented by the TKR is $\sim 11\%$ at
normal incidence).  As discussed below, this has yielded extremely high
efficiency for finding tracks and has been key to providing the information
necessary to reject backgrounds.

The efficiency and noise occupancy of the TKR over the first three years of operation are shown
in \figref{tkr_perf}. The variations in the average single strip noise
occupancy are dominated by one or a few noisy strips, which have been disabled at
different times during the mission. The baseline of $4 \times 10^{-6}$ is
dominated by accidental coincidences between event readouts and
charged particle tracks (see below, and \secref{subsec:LAT_DAQ}) and corresponds
to an upper limit of $\sim 3$ noise hits per event in the full LAT on average.
Since these noise hits are distributed across 16 towers and 36 layers per
tower, their effect on the event reconstruction is insignificant.

\twopanel{htb}{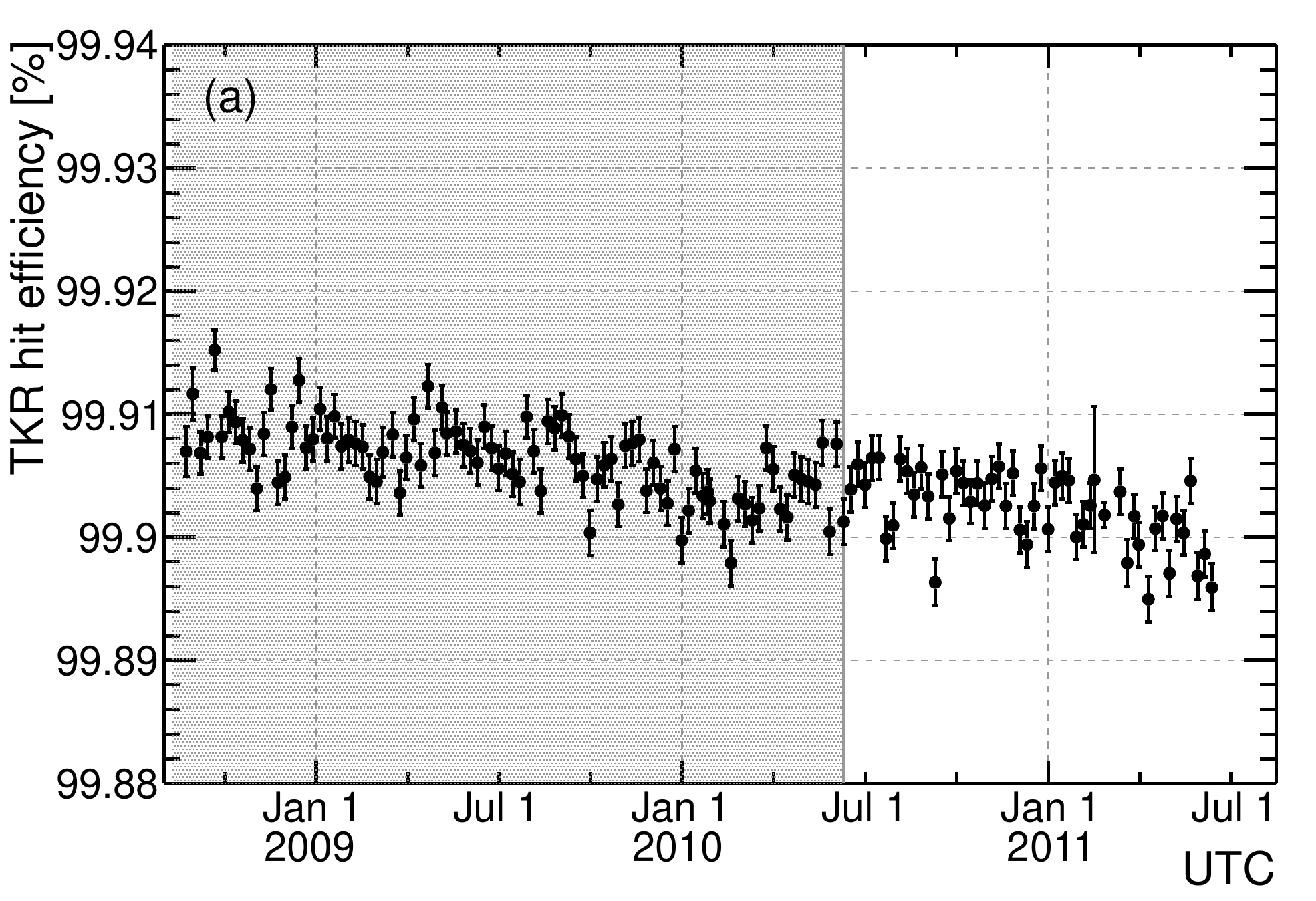}{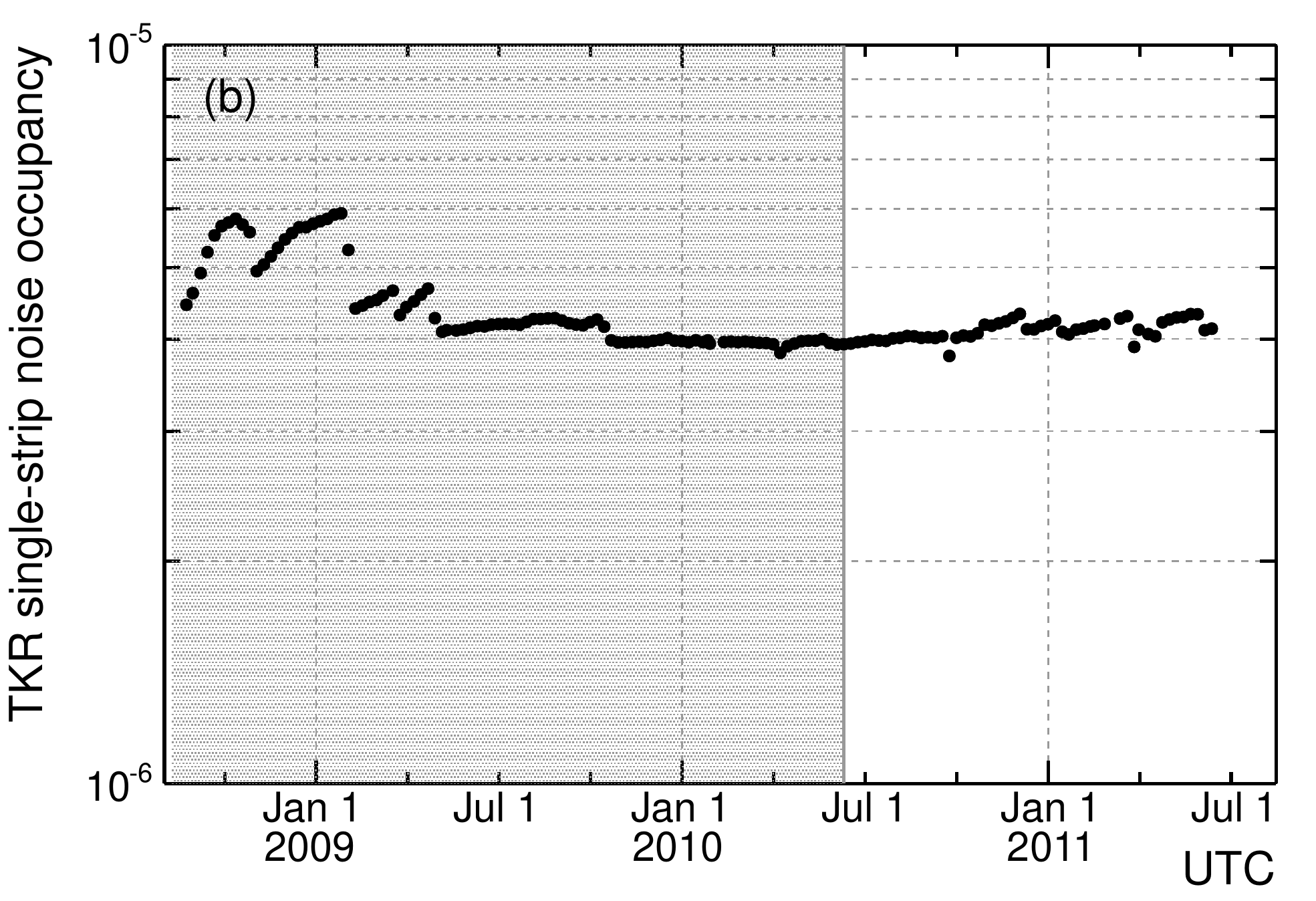}{
  \caption{(a) Average TKR hit efficiency and (b) single-strip noise occupancy
    through the first three years of the mission. Each data point is
    the average value for the full LAT over a week of data taking.
    The shaded background regions mark the first two years of operation,
    corresponding to the data selection used to calibrate the instrument
    performance.}
  \label{fig:tkr_perf}
}

The TKR readout is \emph{digital}, i.e., the readout is binary, with a single
threshold discriminator for each channel, and no pulse height information is
collected at the strip level.
The individual electronic chains connected to each SSD strip consist of a
charge-sensitive preamplifier followed by a simple CR-RC shaper with a
peaking time of $\sim 1.5$~$\mu$s. Due to the implementation details, the
baseline restoration tends to be current-limited, and the signal at the output
of the shaper is far from the exponential decay characteristic of a linear
network, with the discriminated output being high for $\sim 10$~\us\ for
Minimum Ionizing Particles (MIPs)\acronymlabel{MIP} at the nominal
$\sim 1/4$~MIP threshold setting.
As a consequence, the latched TKR strip signals are typically present
for that amount of time after the passage of a MIP. If the LAT is triggered
within this time window, these latent signals will be read out and become part
of the TKR event data. The rate of occurrence of these \emph{ghost} signals
(which may result in additional tracks when the events are reconstructed)
depends directly on the charged particle background rate.
Mitigation against this contamination in the data is discussed below
in~\secref{subsec:LAT_DAQ}, \secref{sec:event}
and~\secref{subsec:Aeff_MC_corrections}.

Each detector subsystem contributes one or more \emph{trigger primitive}
signals that the LAT trigger system uses to adjudicate whether to read out the
LAT detectors (see \secref{subsec:event_trigFilter}). The TKR trigger is a
coincidence of 3 \xy\ paired adjacent layers within a single tower (hence a
total of 6 consecutive SSD detector planes). Due primarily to power constraints,
the TKR electronic system does not feature a dedicated fast signal for
trigger purposes. Rather, the logical OR of the discriminated strip signals
from each detector plane is used to initiate a non-retriggerable one-shot pulse
of fixed length (typically 32 clock ticks or 1.6~$\mu$s), that is the basic
building block of the \emph{three-in-a-row} trigger primitive
(see \secref{subsubsec:event_trigger}). In addition, the
length (or \emph{time over threshold}) of this layer-OR signal is recorded and
included in the data stream. Since the time over threshold depends on
the magnitude of the ionization in the SSD, which in turn depends on on the
characteristics of the ionizing particle, it provides useful information for
the background rejection stage.

\begin{figure}[htbp]
  \centering
  \includegraphics[width=\onecolfigwidth]{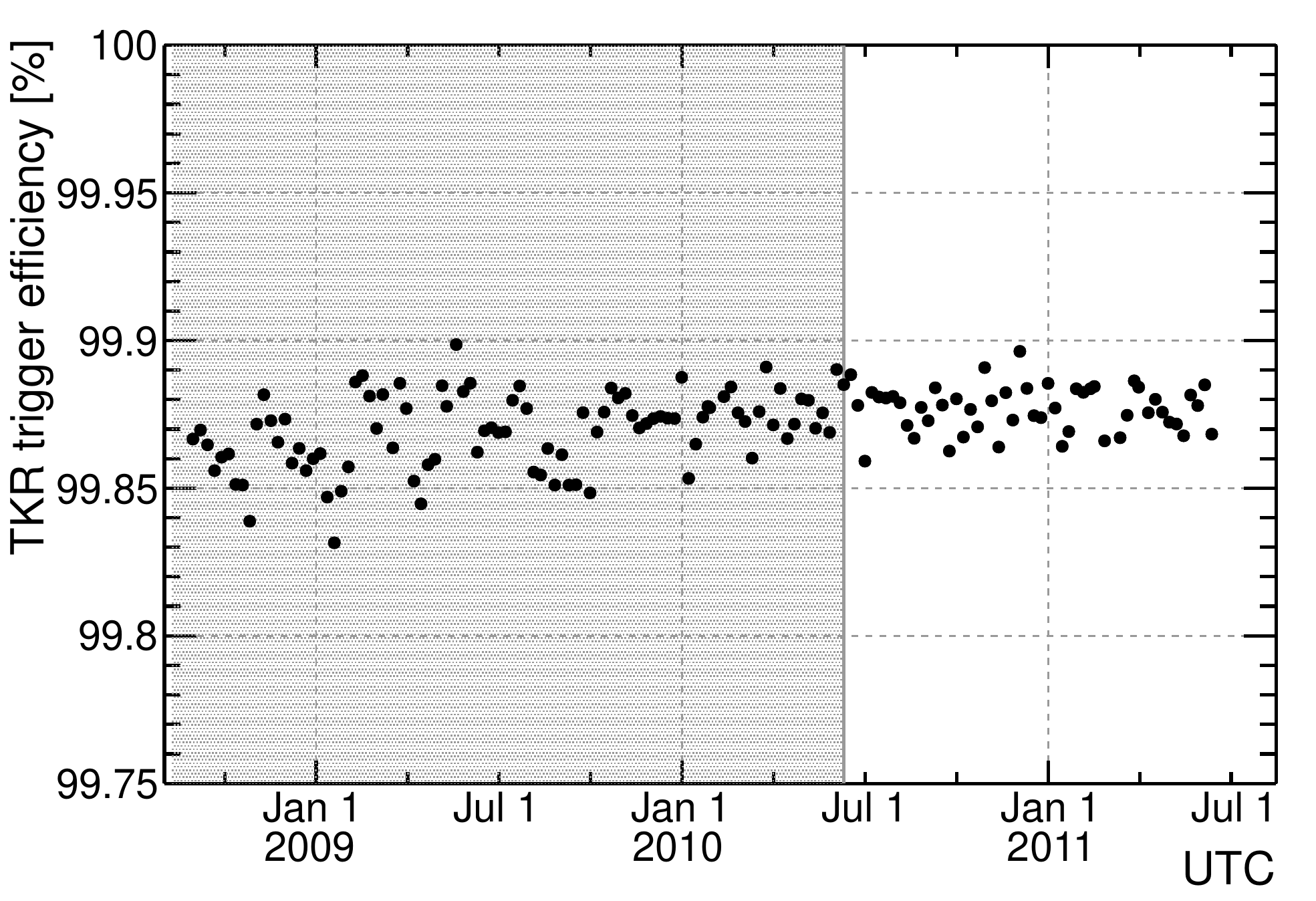}
  \caption{Average efficiency of the \tptkr\ three-in-a-row trigger through the
    first three years of the mission. Each data point is the average value for
    the full LAT over a week of data taking. The shaded background region marks
    the first two years of operation, corresponding to the data selection used
    to calibrate the instrument performance.}
\label{fig:tkr_trig_perf}
\end{figure}

The efficiency of the three-in-a-row trigger is $> 99\%$. This is due in part
to the redundancy of this trigger for the vast majority of events (i.e., by
passing through many layers of Si, most events have multiple opportunities to
form a three-in-a row). The trigger efficiency, measured in flight using MIP
tracks crossing multiple towers, is shown in \figref{tkr_trig_perf} to be
very stable.

Perhaps the most important figure of merit for the TKR is the resulting
PSF for reconstructed \gammaRayHyph\ directions. At low
energy the PSF is dominated by multiple scattering, primarily within the
tungsten conversion foils (tungsten accounts for $\sim 67$\% of the material in
the thin section and $\sim 92$\% of the material in the thick section).
At high energy the combination of the strip pitch of 228~\um, the spacing of
the planes and the overall height of the TKR result in a limiting precision for
the average conversion of $\sim 0.1^{\circ}$ at normal incidence. MC simulations 
predict that the transition to this measurement precision dominated regime
should occur between $\sim 3$~GeV and $\sim 20$~GeV. 
We will discuss the PSF in significantly more detail in \secref{sec:PSF}.

\subsubsection{Electromagnetic Calorimeter}\label{subsec:LAT_CAL}

Details of the CAL can be found in \citet{REF:2009.LATPaper} 
and \citet{REF:2010.CALPaper}.  Here, we highlight some key aspects for
understanding the LAT performance. The CAL is a 3D imaging calorimeter.
This is achieved by arranging the CsI crystals in each tower module in 8~layers,
each with 12 crystal \emph{logs} (with dimensions
$326~\text{mm}\times 26.7~\text{mm} \times 19.9~\text{mm}$) spanning the width
of the layer.
The logs of alternating layers are rotated by $90^{\circ}$ about the LAT
boresight, and aligned with the $x$ and $y$ axes of the LAT coordinate system.

Each log is read out with four photodiodes, two at each end: a large photodiode
covering low energies ($<1$~GeV per crystal), and a small photodiode covering
high energies ($<70$~GeV per crystal).
Each photodiode is connected to a charge-sensitive preamplifier whose output
drives (i) a slow ($\sim 3.5$~$\mu$s peaking time) shaping amplifier for
spectroscopy and (ii) a fast shaping amplifier ($\sim 0.5$~$\mu$s peaking time)
for trigger discrimination. In addition, the output of each slow shaper is
connected to two separate track-and-hold stages with different gains
($\times 1$ and $\times 8$).

The outputs of the four track-and-hold stages are multiplexed to a single
analog-to-digital converter. The four gain ranges (two photodiodes
$\times$ two track-and-hold gains) span an enormous dynamic range, from
$<2$~MeV to 70~GeV deposited per crystal, which is necessary to cover the full
energy range of the LAT.
A zero-suppression discriminator on each crystal sparsifies the CAL data by
eliminating the signals from all crystals with energies $<2$~MeV.
To minimize CAL data volume, each log end reports only a single range, the
lowest-gain unsaturated range (the \emph{one-range, best-range} readout)
for most events. For likely heavy ions, each log end reports all four ranges
(the \emph{four-range} readout) for calibrating the energy scale across the
different ranges (see \secref{subsubsec:event_trigger} for details of how the
readout mode is selected).

The CAL provides two inputs to the global LAT trigger. At each log end the
output of each fast shaper (for both the large and the small diode) is compared
to an adjustable threshold by a discriminator to form two separate trigger
request signals. In the standard science configuration, the discriminator
thresholds are set at 100~MeV and 1~GeV energy deposition. The 1~GeV
threshold is $>90$\% efficient for incident \gammaRays\ above 20~GeV.

For each log with deposited energy, two position coordinates are derived simply
from the geometrical location of the log within the CAL array, while the
longitudinal position is derived from the ratio of signals at opposite ends of
the log: the crystal surfaces were treated to provide monotonically decreasing
scintillation light collection with increasing distance from a photodiode.
Thus, the CAL provides a 3D image of the energy deposition for each event.

Since the CAL is only 8.6 radiation lengths thick at normal incidence, for
energies greater than a few~GeV shower leakage becomes the dominant factor
limiting the energy resolution, in particular because event-to-event variations
in the early shower development cause fluctuations in the leakage out the back
of the CAL.
Indeed, by $\sim 100$~GeV about half of the total energy in showers at normal
incidence escapes out the back of the LAT on average. The intrinsic 3D
imaging capability of the CAL is key to mitigating the degradation of the energy
resolution at high energy through an event-by-event 3D fit to
the shower profile. This was demonstrated both in beam tests and in
simulations~\citep{REF:2007.PreliminaryCUResults, REF:2010.Electrons7GeV1TeV},
achieving better than 10\% energy resolution well past 100~GeV.
The imaging capability also plays a critical role in the rejection of hadronic
showers (see \secref{subsubsec:cal_topology_analysis}). Furthermore, for events depositing more than $\sim 1$~GeV in the CAL,
imaging in the CAL can be exploited to aid in the TKR reconstruction and in
determining the event direction (see \secref{subsec:event_recon}).

We have monitored the performance of the CAL continuously over its three years
of operation. We observe a slow ($\sim1$\% per year) decrease in the
scintillation light yield of the crystals due to radiation damage in low Earth
orbit, as we anticipated prior to launch (see
also \secref{subsec:EDisp_xtalCalib}).
Although we do not yet correct for this decreased light yield, we have derived
calibration constants on a channel-by-channel basis for the mission to date.
In 2012 January we started reprocessing the full data set with these updated
calibrations, and in the future expect to maintain a quarter-yearly cadence of
updates to ensure that the calibrated values do not drift by more than 0.5\%.

\subsubsection{Anticoincidence Detector}\label{subsec:LAT_ACD}

The third LAT subsystem is the ACD, critically important for the identification
of LAT-entering charged cosmic rays. Details of its design can be found
in~\citet{REF:2007.ACDPaper} and~\citet{REF:2009.LATPaper}.
From the experience of the LAT predecessor,
the Energetic Gamma Ray Experiment Telescope (EGRET)\acronymlabel{EGRET}, came
the realization that a high degree of segmentation was required in order
to minimize \emph{self-veto} due to hard X-ray back-scattering (often referred to as
backsplash) from showers in the CAL~\citep{REF:1999.EGRET_PERF}.

The ACD consists of 25 scintillating plastic tiles covering the top
of the instrument and 16 tiles covering each of the four sides (89 in all).
The dimensions of the tiles range between 561~and~2650~cm$^2$ in geometrical surface
and between 10~and~12~mm in thickness.
By design, the segmentation of the ACD does not match that of the LAT tower
modules, to avoid lining up gaps between tiles with gaps in the TKR and CAL.
The design requirements for the ACD specified the capability to reject entering
charged particles with an efficiency $> 99.97\%$. To meet the efficiency
requirement, careful design of light collection using wavelength-shifting
fibers, and meticulous care in the fabrication to maintain the maximum light
yield were needed. The result was an average light yield of $\sim 23$
photo-electrons (p.e.)\acronymlabel{pe} for a normally-incident MIP for each
of the two redundant readouts for each tile.

The required segmentation inevitably led to less than complete hermeticity,
with construction and launch survival considerations setting lower limits
on the sizes of the gaps between tiles. Tiles overlap in one dimension,
leaving gaps between tile rows in the other.
The gaps are $\sim 2.5$~mm and coverage of these gaps is provided by bundles of
scintillating fibers (called \emph{ribbons}), read out at each end by a
photomultiplier tube (PMT)\acronymlabel{PMT}.
The light yield for these ribbons, however, is considerably less than for the
tiles: it is typically $\sim 8$ p.e./MIP and varies considerably along the
length of a bundle. Therefore, along the gaps the efficiency for detecting
the passage of charged particles is lower. However, the gaps comprise a small
fraction of the total area ($<1\%$) and accommodating them did not require
compromising the design requirements. In addition to the gaps between ACD tile
rows, the corners on the sides of the ACD have gaps that are not covered by
ribbons and must be accounted for in the reconstruction and event
classification (see \secref{subsec:event_photon_classes}).

As with the TKR and CAL, the ACD provides information used in the hardware
trigger as well as in the full reconstruction of the event. The output of each
PMT is connected to (i) a fast shaping amplifier (with $\sim 400$~ns shaping
time) for trigger purposes and (ii) two separate slow electronics chains
(with $\sim 4$~$\mu$s shaping time and different gains) to measure the signal
amplitude. The use of the fast signal in the context of the LAT global trigger
will be discussed in more detail in ~\secref{sec:event}.
Although the main purpose of the ACD is to signal the passage of charged
particles, this subsystem also provides pulse height information. 
The two independent slow signals and the accompanying circuitry for automatic
range selection accommodate a large dynamic range, from $\sim 0.1$~MIP to
hundreds of MIPs.

\begin{figure}[!ht]
  \centering
  \includegraphics[width=\onecolfigwidth]{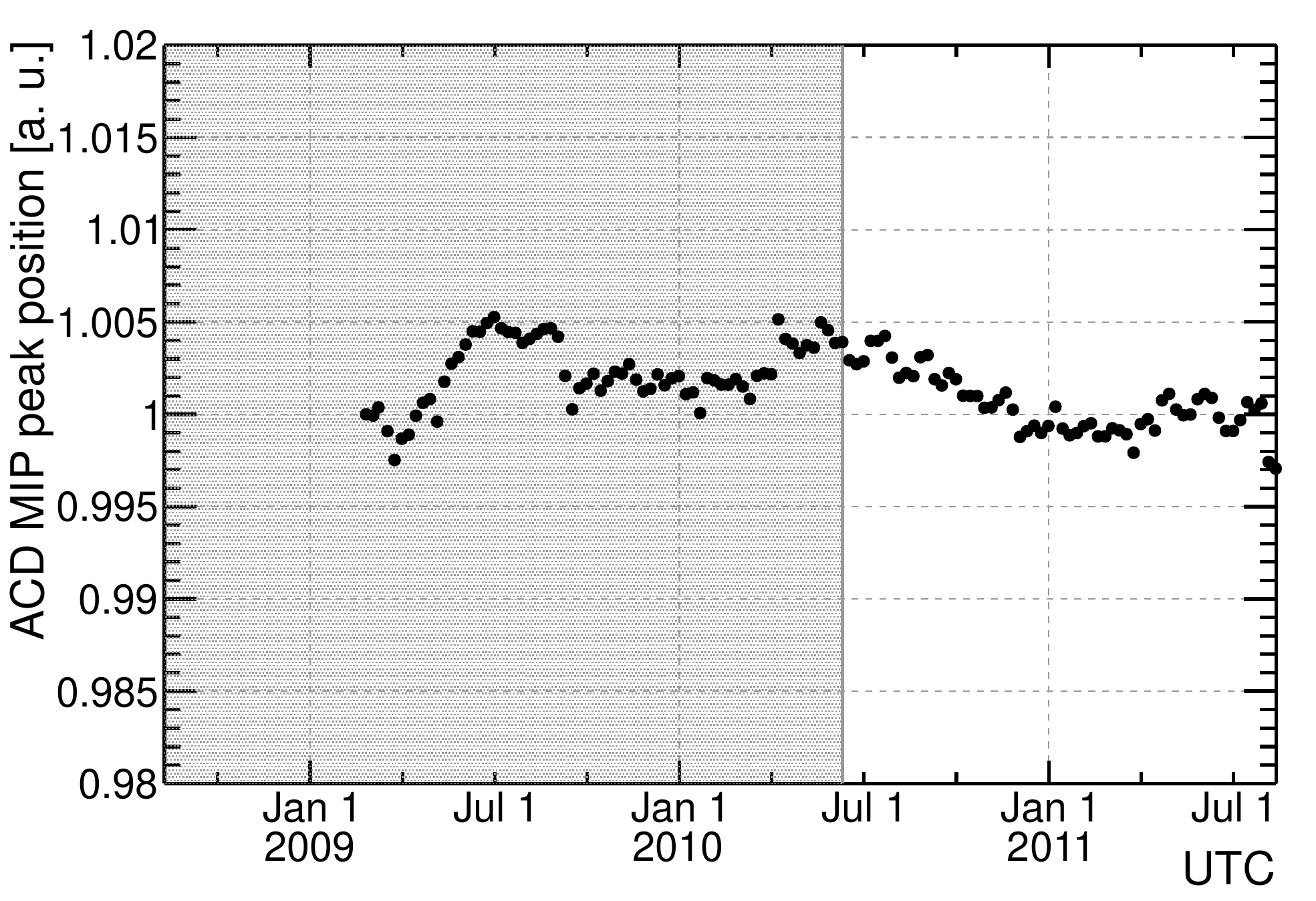}
  \caption{Relative position of the MIP peak, calibrated for the tile response
    and corrected for the incidence angle, through the first three years of the
    mission. Each data point is the average value for the 89 ACD tiles over
    a week of data taking, with the value of the first point being arbitrarily
    set to 1.  The shaded background region marks the first two years of
    operation, corresponding to the data selection used to calibrate the
    instrument performance.  Data from the first several months of the mission 
    are missing because this quantity had not yet been included as part of the
    automated data monitoring and trending.}  
\label{fig:acd_perf}
\end{figure}

As for both the TKR and the CAL, the performance of the ACD
has been continuously monitored over the past three years. The stability of the 
MIP signal is shown in \Figref{acd_perf} and in summary shows very little
degradation.  Note that slight changes in the selection criteria and
spectra of the MIP calibration event sample cause small ($< 0.5\%$) variations
in mean deposited energy per event. 

\subsubsection{Trigger and Data Acquisition}\label{subsec:LAT_DAQ}

The LAT hardware trigger collects information from the LAT subsystems and,
if certain conditions are fulfilled, initiates event readout.
Because each readout cycle produces a minimum of $26.5$~$\mu$s of dead time 
(even more if the readout buffers fill dynamically), the trigger was 
designed to be efficient for \gammaRays\ while keeping the total 
trigger rate, which is dominated by charged CRs, low enough to 
limit the dead-time fraction to less than about 10\% (which corresponds
to a readout rate of about 3.8~kHz). The triggering criteria are 
programmable to allow additional, prescaled event streams for continuous 
instrument monitoring and calibration during normal operation.  We will
defer discussion of the actual configuration used in standard science 
operations and of the corresponding performance to the more general
discussion of event processing in \secref{subsec:event_trigFilter}.

To limit the data volume to the available telemetry bandwidth, collected data
are passed to the \emph{on-board filter}. The on-board filter
(see \secref{subsubsec:event_filter}) consists of a few event selection
algorithms running in parallel, each independently able to accept a given event
for inclusion in the data stream to be downlinked.
Buffers on the input to the on-board filter can store on the order of 100
events awaiting processing.  Provided that the on-board filter
processes at least the average incoming data rate no additional dead time will be accrued because the
on-board filter is busy. The processors used for the on-board filter must also
build and compress the events for output, and the time required to make a
filter decision varies widely between events. Therefore, the event
rate that the on-board filter can handle depends on the number of events
passing the filter. In broad terms, the processors will saturate for output
rates between 1~kHz and 2.5~kHz, depending on the configuration of the on-board
filter. In practice, the average output rate is about 350~Hz, and the amount of 
dead time introduced by the on-board filter is negligible.  

\begin{figure}[!ht]
  \centering
  \includegraphics[width=\textwidth]{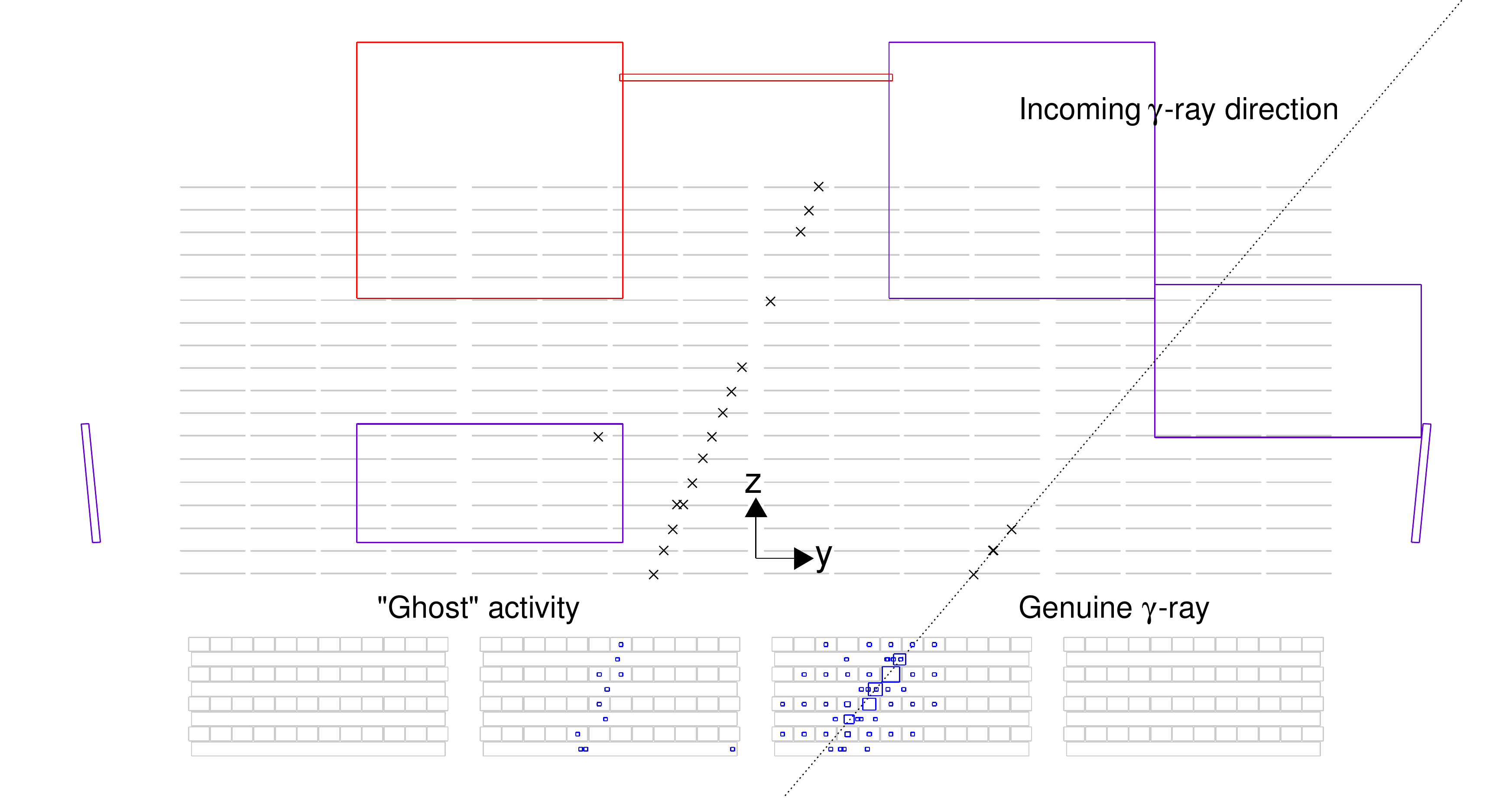}
  \caption{Example of a ghost event in the LAT ($y$--$z$ orthogonal projection).
    In addition to an 8.5~GeV back-converting \gammaRayHyph\ candidate (on the
    right) there is additional activity in all the three LAT subsystems, with
    the remnants of a charged-particle track crossing the ACD, TKR and CAL.
    The small crosses represent the clusters (i.e., groups of adjacent hit
    strips) in the TKR, while the variable-size squares indicate the
    reconstructed location of the energy deposition for every hit crystal in
    the CAL (the side of the square being proportional to the magnitude of the
    energy release). 
    The dashed line indicates the \gammaRayHyph\ direction.
    For graphical clarity, only the ACD volumes with a signal above the zero
    suppression level are displayed.}
\label{fig:ghost_event}
\end{figure}

Soon after launch, it became apparent that the LAT was recording events 
that included an unanticipated background: remnants of electronic signals from
particles that traversed the LAT a few $\mu$s before the particle that
triggered the event.
We refer to these remnants as \emph{ghosts}. An example of such an event is
shown in \figref{ghost_event}.

It is important to re-emphasize a point made in the previous sub-sections: 
many of the signals that are passed to the hardware trigger and the on-board 
filter are generated by dedicated circuits whose shaping times are
significantly shorter than for the circuits that read out the data from the same
sensors for transmission to the ground. Although the signals with longer
shaping times used for the ground event processing measure the energy deposited
in the individual channels far more precisely, they also suffer the adverse
consequence of being more susceptible to ghost particles.  
\Tabref{LAT_timeConstants} gives the characteristic times for both
the \emph{fast} signals (i.e., those used in the trigger) and the \emph{slow}
signals (i.e., those used in the event-level analysis) for each of the 
subsystems. 

\begin{table}[htb]
  \begin{center}
    \begin{tabular}{lccc}
      \hline
      Subsystem & Fast signal  & Slow signal\\
      & (trigger) & (event data)\\
      \hline
      \hline
      ACD & 0.4~\us & 4~\us\\
      CAL & 0.5~\us & 3.5~\us\\
      TKR & 1.5~\us & 10~\us\\
      \hline
    \end{tabular}
    \caption{Characteristic readout time windows for LAT subsystems. The
       TKR subsystem does not provide a dedicated fast signal for the 
       trigger: the peaking time for the shaped TKR signal is 
       $\sim 1.5$~$\mu$s (which is the relevant number for the trigger) 
       but the decay time is much longer and the average time over threshold
       for the discriminated output is of the order of 10~$\mu$s 
       for a MIP at normal incidence.}
    \label{tab:LAT_timeConstants}
  \end{center}
\end{table}

The ghost events have been the largest detrimental effect observed in flight
data. They affect all of the subsystems, significantly complicate the event
reconstruction and analysis, and can cause serious errors in event
classification:
\begin{fermiitemize}
\item they can leave spurious tracks in the TKR that do not point in the
  same direction as the incident \gammaRay\ and might not be associated
  with the correct ACD and CAL information due to the different time
  constants of the subsystems;
\item they can leave sizable signals in the CAL that can confuse the
  reconstruction of the electromagnetic shower, degrading the measurement of
  the shape and direction of the shower itself, and that can cause the energy
  reconstruction to mis-estimate the energy of the incident \gammaRay;
\item they can leave significant signals in the ACD that can be much larger
  than ordinary backsplash from the CAL and that can cause an otherwise
  reconstructable \gammaRay\ to be classified as a CR.
\end{fermiitemize}

We will discuss the effects on the LAT effective area in more detail in
\secref{sec:Aeff}. In brief: above $\sim 1$~GeV, where the average fractional
ghost signal in the CAL is small, the loss of effective area is of the order of 10\% or less.
This loss is largely due to the fact that valid \gammaRays\ can be rejected in event
reconstruction and classification if ghost energy depositions in the CAL cause
a disagreement between the apparent shower directions in CAL and TKR.
At lower \gammaRayHyph\ energies ghost signals can represent the dominant
contribution to the energy deposition in the instrument, and the
corresponding loss of the effective area can exceed $20\%$.
We also emphasize that, while these figures are averaged over orbital
conditions, the fraction of events that suffer from ghost signals, as well as
the quantity of ghost signals present in the affected events, varies by a
factor of 2--3 at different points in the orbit due to variation in the
geomagnetic cutoff and resulting CR rates.
Recovering the effective area loss due to ghost signals is one of the original
and main motivations of ongoing improvements to the reconstruction
\citep{REF:2010.Pass8}.

Finally, during extremely bright Solar Flares (SFs)\acronymlabel{SF} the
\emph{pile-up} of several $> 10$~keV X-rays within a time interval
comparable with the ACD signal shaping time can also cause \gammaRays\ to be
classified as CRs (see \appref{app:bti}).

\subsection{Orbital Environment and Event Rates}\label{subsec:LAT_ORBIT}

\fermi\ is in a 565-km altitude orbit with an inclination of $25.6^{\circ}$. The orbit 
has a period of $\sim 96$~minutes, and its pole precesses about the celestial
pole with a period of $\sim 53.4$~days.  At this inclination \fermi\ spends 
about 15\% of the time inside the South Atlantic Anomaly
(SAA)\acronymlabel{SAA}.  Science data taking is suspended while \fermi\ is
within the SAA because of the high flux of trapped
particles~\citep[for details, see][]{REF:2009.OnOrbitCalib}.

When \Fermi\ is in standard sky-survey mode the spacecraft rocks north and
south about the orbital plane on alternate orbits. Specifically, the LAT
boresight is offset from the zenith toward either the north or south orbital
poles by a characteristic rocking angle. On 2009 September 3 this rocking
angle was increased from $35^\circ$ to $50^\circ$ in order to lower
the temperature of the spacecraft batteries and thus extend their
lifetime.  As a result of this change, the amount of the Earth
limb that is subtended by the \fov\ of the LAT during survey-mode
observations increased substantially.
The most noticeable consequence is a much larger contribution to the LAT data
volume from atmospheric \gammaRays. This will be discussed more in
\secref{sec:bkg}, and further details about atmospheric \gammaRays\ can be
found in \citet{REF:2009.FermiEarthPhotons}.

The flux of charged particles passing through the LAT is usually several
thousand times larger than the \gammaRayHyph\ flux. Accordingly, several stages
of event selection are needed to purify the \gammaRayHyph\ content
(see \secref{subsec:event_trigFilter}, \secref{subsec:LAT_recon} and
\secref{sec:event}) and some care must be taken to account for contamination of
the \gammaRayHyph\ sample by charged particles that are incorrectly classified
as \gammaRays\ (see \secref{sec:bkg}).
Here we distinguish 4 stages of the event classification process:
\begin{fermienumerate}
\item \emph{hardware trigger request}: the LAT detects some traces of 
   particle interaction and starts the triggering process
  (\secref{subsec:event_trigFilter});
\item \emph{hardware trigger accept}: the event in question generates an
  acceptable trigger pattern and is read out and passed to the on-board filter
  (\secref{subsec:event_trigFilter});
\item \emph{on-board filter}: the event passes the on-board \gammaRayHyph\
  filter criteria (\secref{subsec:event_trigFilter});
\item \emph{standard \gammaRayHyph\ selection}: the event passes more stringent
  selection criteria (\secref{subsec:event_photon_classes}), such as
  \irf{P7SOURCE}, which is the selection currently being recommended for
  analysis of individual point sources, or \irf{P6\_DIFFUSE}, the selection
  recommended for analysis of point sources in the \psix\ iteration of the
  event selections.
\end{fermienumerate}

\begin{figure}[!ht]
  \centering
  \includegraphics[width=\onecolfigwidth]{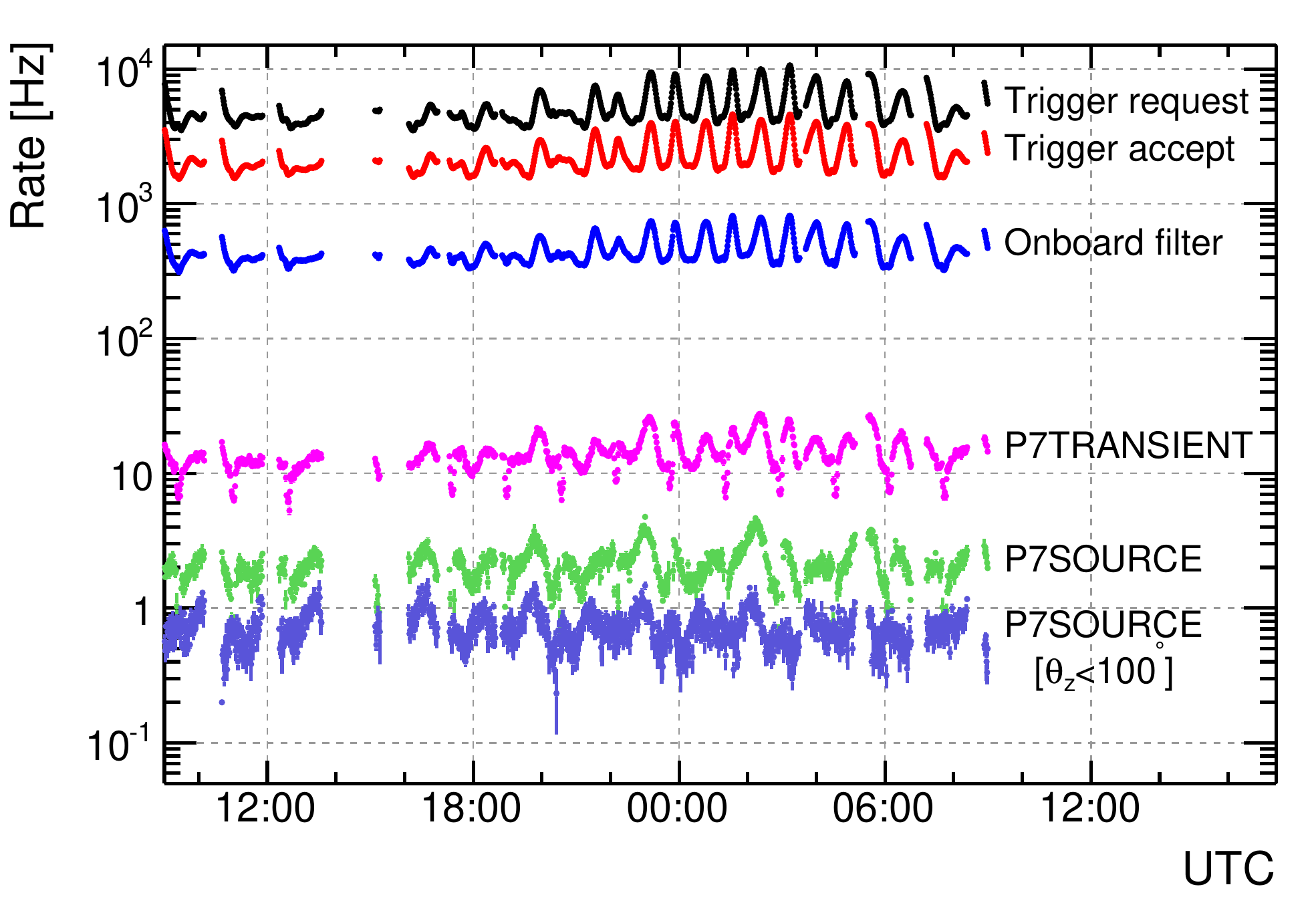}
  \caption{Rates at several stages of the data acquisition and reduction
    process on a typical day (2011 August 17). Starting from the highest, the
    curves shown are for the rates: (i) at the input of the hardware trigger
    process (trigger request), (ii) at output of the hardware trigger
    (trigger accept), (iii) at the output of the
    on-board filter, (iv) after the loose \irf{P7TRANSIENT} \gammaRayHyph\
    selection, (v) after the tighter \irf{P7SOURCE} \gammaRayHyph\ selection,
    and (vi) the \irf{P7SOURCE} \gammaRayHyph\ selection with an additional cut
    on the zenith angle ($\theta_{z} < 100^{\circ}$).  
    See \secref{sec:event} for more details about the event selection stages.}
  \label{fig:RateTimeSeries}
\end{figure}

\Figref{RateTimeSeries} shows an example of both the orbital variations of the
charged particle flux, and how the initially overwhelming CR contamination
is made tractable while maintaining high efficiency for
\gammaRays\ by several stages of data reduction and analysis.
\figref{RateLatLon} shows how the charged particle rate varies with spacecraft 
position. In particular, data taken during the parts of the orbit with the
highest background rates are more difficult to analyze for two reasons: (i)
there are simply more non-\gammaRayHyph\ events that must be rejected, and
(ii) the increased ghost signals complicate the analysis of the
\gammaRayHyph\ events (see \secref{subsubsec:Overlays}).  

A model of the particle backgrounds for the region outside the SAA was compiled
before launch as documented in \citet{REF:2009.LATPaper}: above the
geomagnetic cutoff rigidity ($\sim 10$~GV) the most abundant species is primary
galactic CR protons, at lower energies particle
fluxes are dominated by trapped protons and electrons, in addition to the
\gammaRays\ due to interaction of CRs with the atmosphere of the Earth. 
Since launch, the particle model has been updated to include new
measurements,
\citep[see, e.g.,][for the electron population]{REF:2010.Electrons7GeV1TeV}.\label{conv:earthCoords}

\threepanel{htbp}{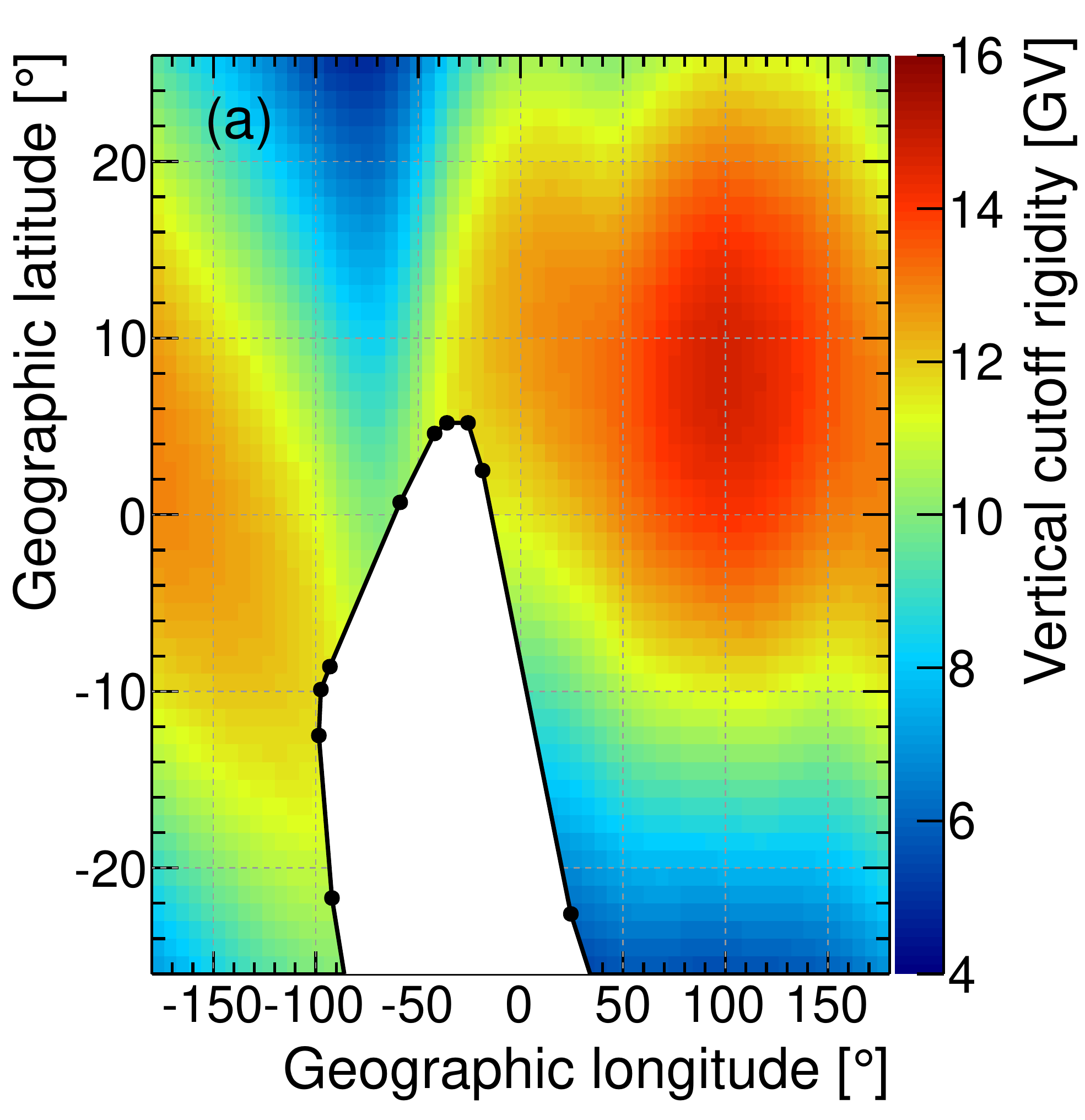}{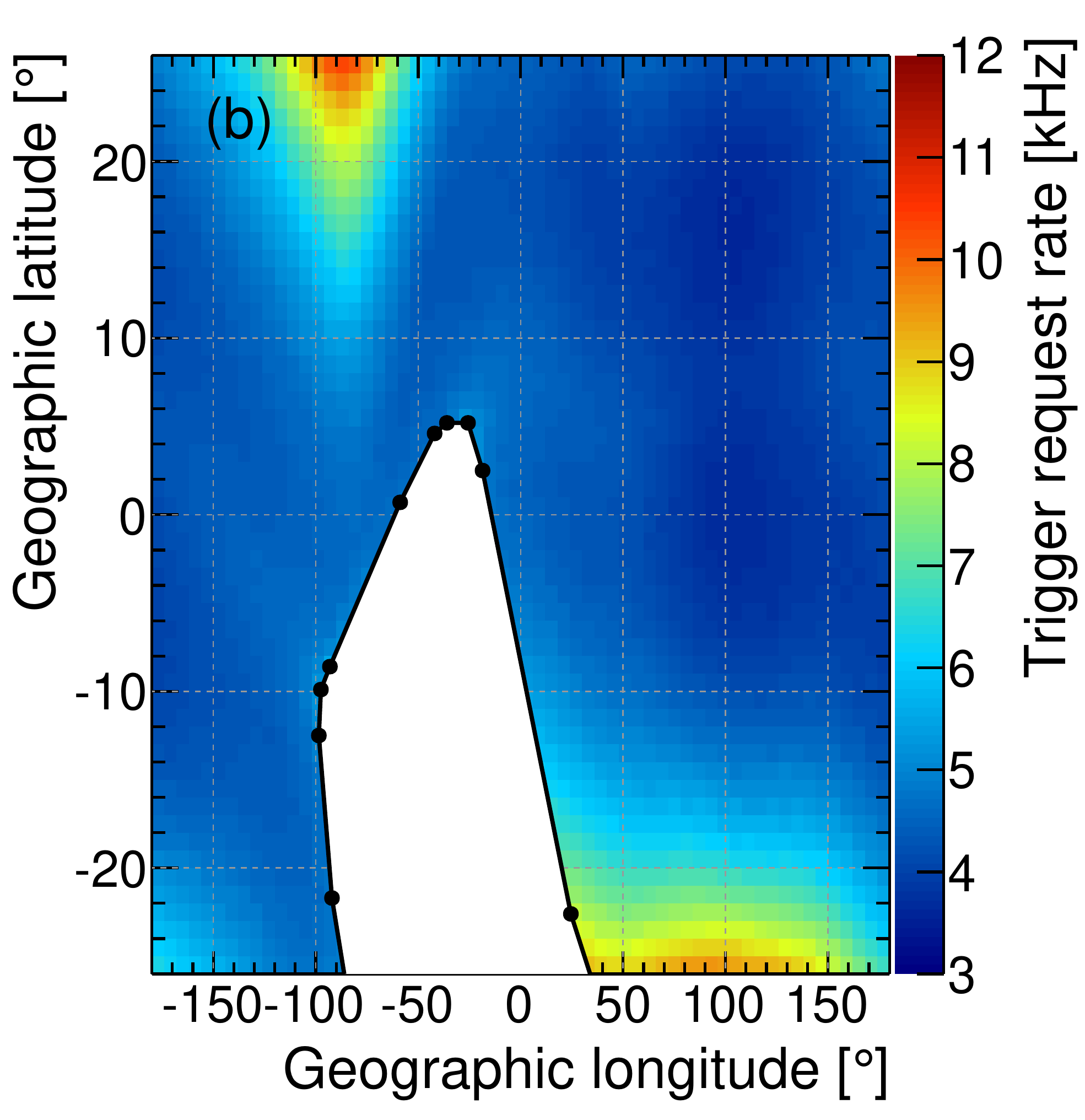}{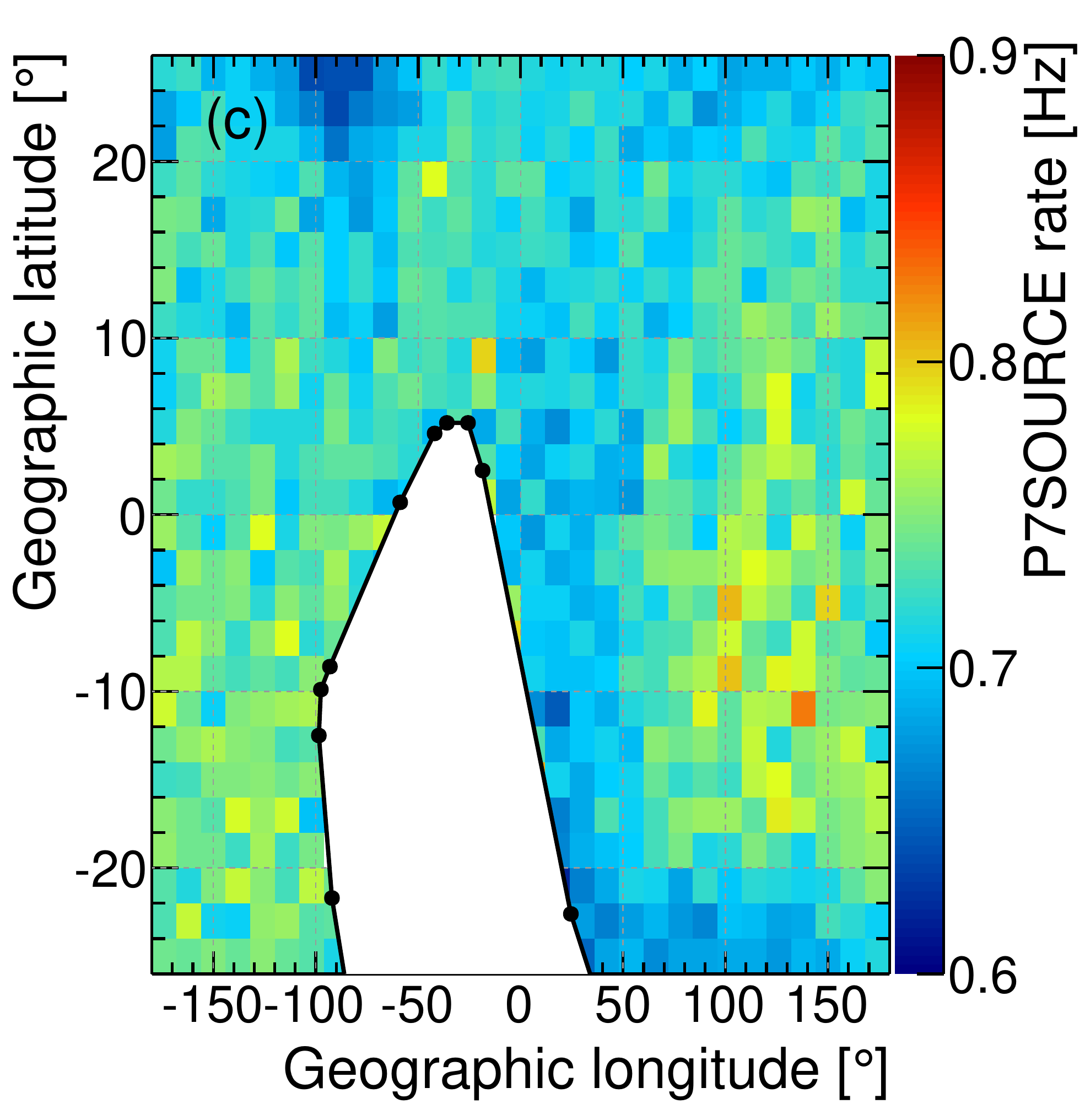}{
  \caption{Orbital background environment as a function of geographic location:
    (a) vertical geomagnetic cutoff rigidity;
    (b) mean rate of events at the input of the hardware trigger process
    (trigger requests, see~\secref{subsubsec:event_trigger}), which decreases 
    roughly as the square root of the geomagnetic cutoff rigidity;
    (c) mean rate of events in \irf{P7SOURCE} event class
    (see~\secref{subsubsec:p7source_selection}) with the additional cut on the
    zenith angle ($\theta_{z} < 100^{\circ}$). The anti-correlation
    between the \irf{P7SOURCE} event rate and the trigger request rate
    is the result of the loss of efficiency due to ghost signals
    (see \secref{subsec:LAT_DAQ} and \secref{subsubsec:Aeff_Livetime}).
    The black lines and points surrounding the white area represent the
    sides and vertices of the SAA polygon as defined for LAT operations.
    The LAT does not acquire science data while inside the SAA polygon.}
 \label{fig:RateLatLon}
}

\subsection{Observing Strategy and Paths of Sources Across the LAT
  Field-of-View}\label{subsec:LAT_obsProf}

\Fermi\ has spent over $95\%$ of the mission in \emph{sky survey} mode,
with most of the remainder split between pointed observations and calibrations.
Furthermore, in sky survey mode, the azimuthal orientation of the LAT is
constrained by the need to keep the spacecraft solar panels pointed toward the
Sun and the radiators away from the Sun.  Specifically, \NEWTEXT{in sky survey mode} \fermi\ completes one
full rotation in $\phi$ every orbit.

\begin{figure}[ht!]
  \centering
  \includegraphics[width=\onecolfigwidth]{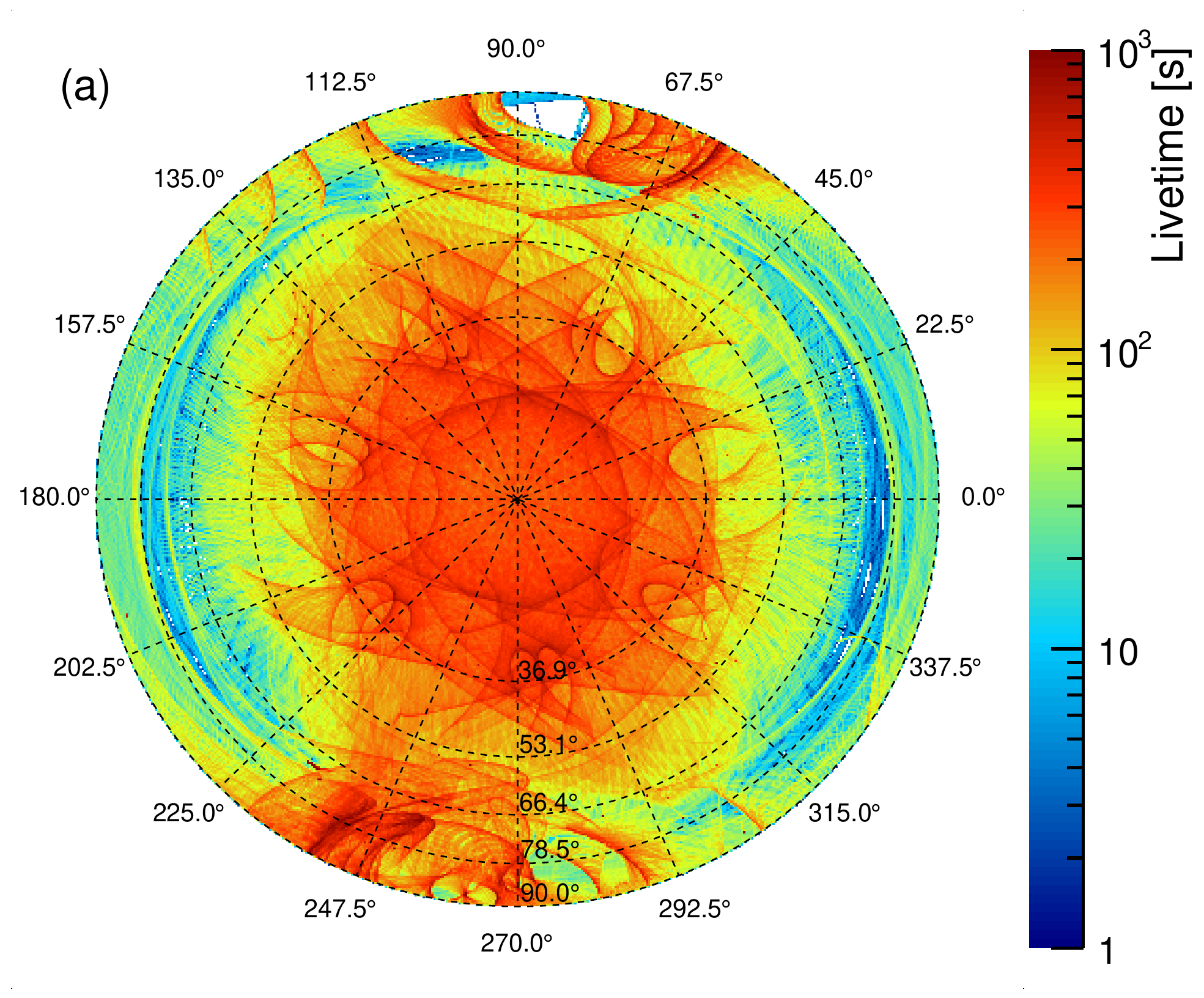}
  \includegraphics[width=\onecolfigwidth]{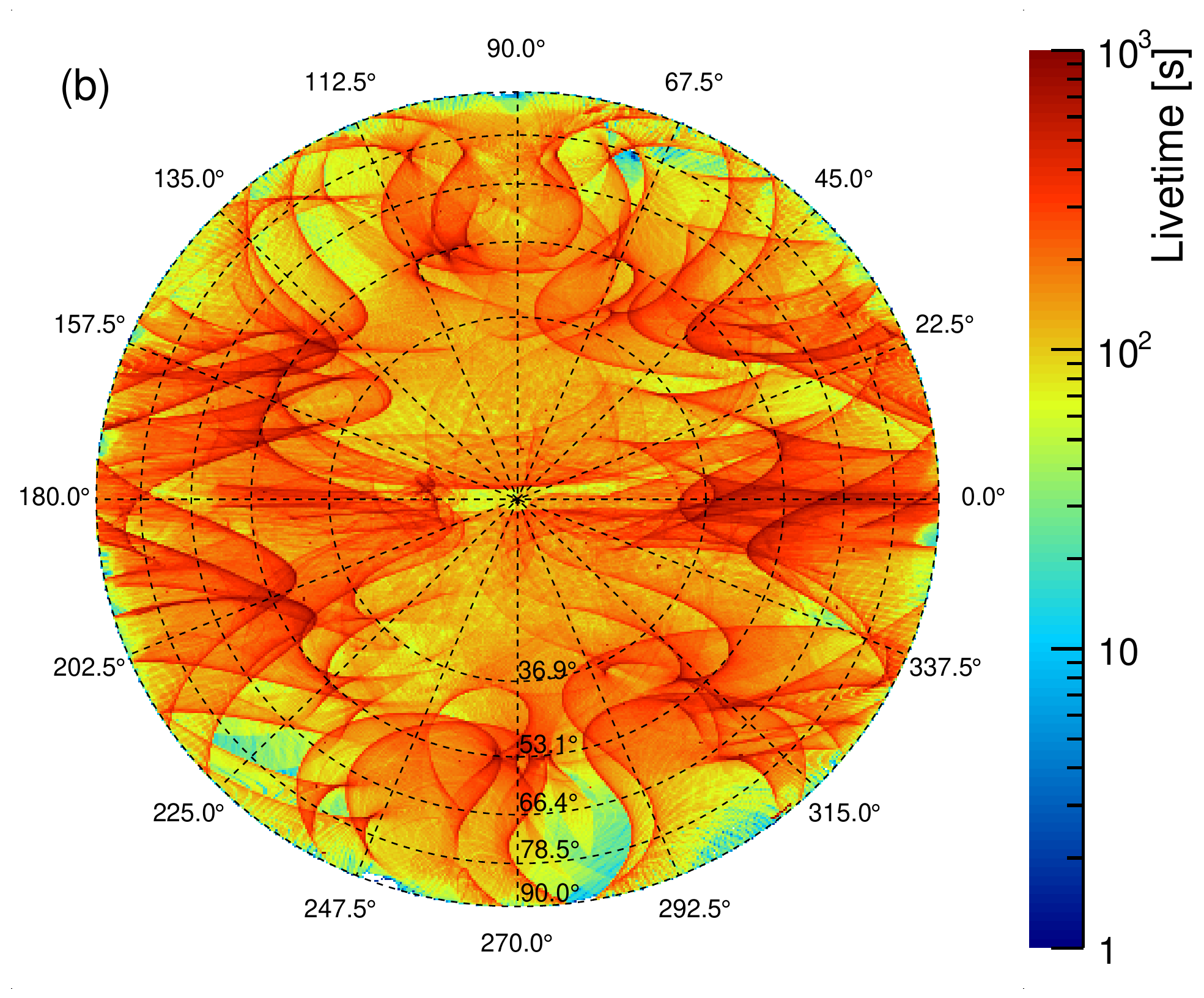}
  \caption{Live time maps in instrument coordinates for the Vela pulsar (a) and
    the Crab (b).
    These plots are made in a zenith equal area projection with
    the LAT boresight at the center of the image, and the color scale shows the
    total live time each source was at $\theta_{z} < 85^{\circ}$.  Recall that
    $\phi=0$ corresponds to the $+x$ axis of the LAT (\parenfigref{lat_layout}).
    These plots cover the same two-year time range that defines the standard
    calibration samples (see \secref{subsec:LAT_methods}).}
  \label{fig:LatPolar}
\end{figure}

Therefore, during an orbital precession period the LAT
boresight will cross a range of declinations $\pm 25.6^{\circ}$
relative to the rocking angle, \NEWTEXT{while the LAT $x$-axis will track the direction toward the Sun}.
As a result, during the sky survey mode, 
each point in the sky traces a complicated path across the LAT
\fov, the details of which depend on the declination and 
ecliptic latitude. \Figref{LatPolar} shows examples of these paths
for two of the brightest LAT sources. \NEWTEXT{Furthermore, in sky survey mode 
operations the path that the LAT boresight traces across the sky during any two orbit 
period is only slightly different than during the two previous or subsequent orbits.} 

\subsection{Ground-Based Data Processing}\label{subsec:LAT_recon}

Reconstructing the signals in the individual detector channels into a 
coherent picture of a particle interaction with the LAT for 
each of the several hundred events collected every second is a formidable task.  
We will defer detailed discussion of the event reconstruction and
classification to \secref{sec:event}; here we describe just the steps
to give a sense of the constraints.
\begin{fermienumerate}
\item \emph{Digitization}: we decompress the data and convert the information
  about signals in individual channels from the schema used in the
  electronics readout to more physically motivated schema---such as grouping
  signals in the ACD by tile, rather than by readout module.
\item \emph{Reconstruction}: we apply pattern recognition and fitting
  algorithms commonly used in high-energy particle physics
  experiments to reconstruct the event in
  terms of individual TKR tracks and energy clusters in the CAL and to associate
  those objects with signals in the ACD (see \secref{subsec:event_recon}).
\item \emph{Event analysis}: we evaluate quantities that can be used as figures
  of merit for the event from the collections of tracks, clusters and
  associated ACD information. Once we have extracted this information we apply
  multi-variate analysis techniques to extract measurements of the energy and
  direction of the event and to construct estimators that the event is in fact
  a \gammaRayHyph\ interaction (see \secref{subsec:event_CT_analysis}).
\item \emph{Event selection}: we apply the selection criteria for the
  various \gammaRayHyph\ event classes (\secref{subsec:event_photon_classes}).
\end{fermienumerate}
In addition to these procedures, the processing pipeline is
responsible for verifying the data integrity at each step and for producing and
making available all the ancillary data products related to calibration and
performance monitoring of the LAT.

On the whole, the LAT data processing pipeline
utilizes approximately 130 CPU years and 300 TB of storage each year.

\begin{table}[ht]
  \begin{center}
    \begin{tabular}{lll}
      \hline
      Stage & CPU Time & Disk Usage \\
       & [years/year] & [TB/year]\\
      \hline
      \hline
      Digitization & 1.2 & 26\\
      Reconstruction & 85 & 167\\
      Event analysis & 5 & 10\\
      Event selection & 1.3 & 60\\
      \hline
      Data monitoring & 25 & 12\\
      Calibration & 5 & 72\\
      \hline
    \end{tabular}
    \caption{Various stages of the data processing pipeline. The numbers in the
      CPU time and disk usage columns refer to a typical year of data
      taking.}
    \label{tab:data_pipeline}
  \end{center}
\end{table}

The ISOC can reprocess data with updated algorithms from any stage in the
process. However, given the quantities of data involved, reprocessing presents
challenges both to the available pool of computing power and storage space.
\Tabref{data_pipeline} gives a rough idea of the resources used by each stage
in the process. Given the available resources, reprocessing from the
\emph{Reconstruction} step is practical only once a year or less often, and
reprocessing from the \emph{Digitization} step is feasible only every few years.

\subsection{Simulated Data}\label{subsec:LAT_simul}

In order to develop the filtering, reconstruction and event selection
algorithms described in \secref{sec:event} we used a detailed  simulation of
particle interactions with \Fermi\ written in the \geant~\citep{REF:GEANT}
framework. This simulation includes particle interactions with a detailed
material model of \Fermi, as well as simulations of the uncalibrated signals
produced in the various sensors in the three subsystems.

The fact that the simulation is detailed enough to produce uncalibrated signals
for each channel allows the simulations to be truly end-to-end:
\begin{fermienumerate}
\item we maintain fidelity to the analysis of the flight data by
  processing simulated data with the same reconstruction and analysis
  algorithms as the flight data;
\item on a related but slightly different note, we simulate the data as
  seen by the trigger and the on-board software, and process the data with
  a simulation of the hardware trigger and exactly the same on-board filter 
  algorithms as used on the LAT;
\item we can merge the signals from two events into a single event. As
  described in \secref{subsubsec:Overlays}, we rely on this feature to add an
  unbiased sample of ghost signals to simulated events.
\end{fermienumerate}

\subsubsection{Ghosts and Overlaid Events}\label{subsubsec:Overlays}

The presence of ghost signals in the LAT data reduced the efficiency 
of the event selections tuned on simulated data lacking this effect. So we
developed a strategy to account for the ghosts.
One of the triggers implemented in the LAT data-acquisition system, the
\tpperiodic\ trigger (see \secref{subsubsec:event_trigger}), provides us with a
sample of ghost events: twice per second the LAT subsystems are read out
independent of any other trigger condition. For the CAL and the ACD, all the
channels, even those below threshold, are recorded (this is not possible for 
the TKR). Since these triggers occur asynchronously with the particle triggers, 
the ghost signals collected are a representative sample of the background
present in the LAT.

We merge the signals channel-by-channel from a randomly chosen periodic trigger
event into each simulated event.  In more detail, we analyze the signals in
each periodic trigger, converting the instrument signals to physical units;
analog-to-digital converter channels in the ACD and CAL, and time over
threshold signals in the TKR, are converted into deposited energy. 
Since the intensity and spectrum of cosmic rays seen by the LAT depend on the
position of the \Fermi\ satellite in its orbit, the ghost events are sorted by
the value of \mcilwainl\ \citep{REF:TrappedRadiation}\label{conv:mcilwainl} at
the point in the orbit where the event was recorded. Then, during the
simulation process, for each simulated event we randomly select a periodic trigger 
event with similar \mcilwainl, and add the energy deposits in this ``overlay event''
to those of the original simulated event, after which the combined event
is digitized and reconstructed in the usual way.

We have used the improved simulations made with this overlay technique
to more accurately evaluate the effective area of the LAT, and
to better optimize the event selection criteria when developing \pseven.
Furthermore, we are studying ways to identify and tag the ghost signals and 
remove them from consideration in the event analysis~\citep{REF:2010.Pass8}.

\subsubsection{Specific Source \gammaRayHyph\ Signal Simulation}\label{subsec:simul_sources}

We have developed interfaces between the software libraries that implement the
flux generation and coordinate transformation used by the \stools\ and our
detailed \geant-based detector simulation. This allows us to re-use the same source
models that we use with the \stools\ within our detailed \geant-based detector
simulation, insuring consistent treatment of \Fermi\ pointing history and the
source modeling.  One application for the present work was simulating
atmospheric \gammaRays\ above 10~GeV for comparison with our Earth limb
calibration sample (see \secref{subsubsec:LAT_earthLimb}).

\subsubsection{Generic \gammaRayHyph\ Signal Simulation}\label{subsec:simul_allGamma}

In addition to simulating individual sources, we also simulate uniform
\gammaRayHyph\ fields that we can use to explore the instrument response
across the entire \fov\ and energy range of the LAT.
These will henceforth be referred to as \allgamma\ simulations, and
they have these features:
\begin{fermienumerate}
\item the \gammaRays\ are generated with an $E^{-1}$ number spectrum
  (where $E$ is the \gammaRayHyph\ energy), i.e., the same number
  of \gammaRays\ are generated in each logarithmic energy bin;
\item the individual \gammaRays\ are randomly generated on a sphere with
  $6$~m$^2$ cross sectional area (i.e., large enough to contain the entire LAT
  and most of the spacecraft) centered at the origin of the LAT reference
  frame, i.e., the center of the TKR/CAL interface plane;
\item the directions of the \gammaRaysHyph\ are sampled randomly across
  $2\pi$ of downward-going (in the LAT reference frame) directions, so as to
  represent a semi-isotropic incident flux.
\end{fermienumerate}

For defining the IRFs we simulate 200 million \gammaRays\ over the range
$\loge \in{[1.25,5.75]}$ and a further 200 million \gammaRays\ over the range
$\loge \in{[1.25,2.75]}$.  The net result is to produce a data set that
adequately samples the LAT energy band and \fov\ and that far exceeds
the statistics of any single point source.

\subsubsection{Simulation of Particle Backgrounds}\label{subsec:simul_bkgnd}

In order to develop our event classification algorithms, and to quantify the
amount of residual CR background that remains in each \gammaRayHyph\ event
class, we require accurate models of both the CR background intensities and
the interactions of those particles with the LAT.  Accordingly, we use the same
\geant-based detector simulation described in the previous section to
simulate fluxes of CR backgrounds.

There are three components to the CR-induced background in the LAT energy range.
\begin{fermiitemize}
\item \emph{Primary CRs}: Protons, electrons, and heavier nuclei form the bulk
  of charged CRs. Due to the shielding by the magnetic field of the Earth, only
  particles with rigidities above a few GV (spacecraft position dependent) can
  reach the LAT orbit. Therefore, the background from primary CRs is relevant
  above a few GeV.
  The LAT event classification provides a very effective rejection of charged
  particles, up to an overall suppression factor of $10^{6}$ for CR protons
  (with a prominent contribution from the ACD). However, due to the intense
  flux of primary protons, a significant number of misclassified CRs enter even
  the \irf{P7SOURCE}, \irf{P7CLEAN}, and \irf{P7ULTRACLEAN} \gammaRayHyph\
  classes.
  Minimum-ionizing primary protons above a few GeV produce a background around
  100--300~MeV in the LAT. However, protons that inelastically scatter in the
  passive material surrounding the ACD (e.g., the micro-meteoroid shield)
  without entering the ACD can produce secondary \gammaRays\ at lower energies
  which are then indistinguishable from cosmic \gammaRays, and which we refer
  to as \emph{irreducible} background (see also \secref{subsec:bkg_irreducible}).
\item \emph{Charged secondaries from CR interactions}: CRs entering the
  atmosphere produce a number of secondaries in interactions with the
  atmosphere itself. Charged particles produced in these interactions can
  spiral back out, constrained by the magnetic field of the Earth, and
  eventually re-enter the atmosphere. These particles are predominantly protons,
  electrons, and positrons and are an important background below the
  geomagnetic cutoff rigidity. Due to the high efficiency of the ACD in
  rejecting charged particles, the dominant contribution to this background in
  the \irf{P7CLEAN} and \irf{P7ULTRACLEAN} event classes is positron
  annihilations in the passive material around the ACD. The resulting
  \gammaRays\ are again indistinguishable from cosmic \gammaRays.
\item \emph{Neutral secondaries from CR interactions}: \gammaRays\ and neutrons
  are also produced in local CR interactions. Unaffected by the magnetic field,
  they approach the LAT from the direction of the Earth with intensities
  peaking at the Earth limb. Neutrons are easily suppressed by the event
  classification scheme and do not contribute significantly to the background.
  The \gammaRayHyph\ background is suppressed by pointing away from the Earth
  and excluding events that originate from near or below the Earth limb.
  Due to the long tails of the LAT PSF (see~\secref{sec:PSF}), however, a small
  fraction of the events originating from the Earth limb are reconstructed
  with directions outside the exclusion regions. 
  Since the Earth limb is extremely bright~\citep{REF:2009.EarthAlbedo}, even
  this small fraction is an important residual background for celestial
  \gammaRays.  Since the PSF tails broaden with decreasing energy the main
  contribution is at energies below a few hundred MeV (see
  \secref{subsec:bkg_treatment}).  Similarly, a small fraction of \gammaRays\ 
  entering the LAT with incidence angles greater than $90^\circ$ are mischaracterized 
  as downward-going ($\sim 5\%$ to $\sim 0.01\%$, depending on the energy, incidence angle, 
  and the event sample). Because the Earth limb is extremely bright and some part 
  of it is almost always behind the LAT, \gammaRays\ from the limb are the dominant 
  component of ``back-entering'' background contamination. Furthermore, 
  since we mis-estimate the directions of these back-entering \gammaRays\ by $> 90^\circ$, 
  they are often reconstructed outside the Earth exclusion region.
\end{fermiitemize}
 
\begin{figure}[htb]
  \centering\includegraphics[width=\textwidth]{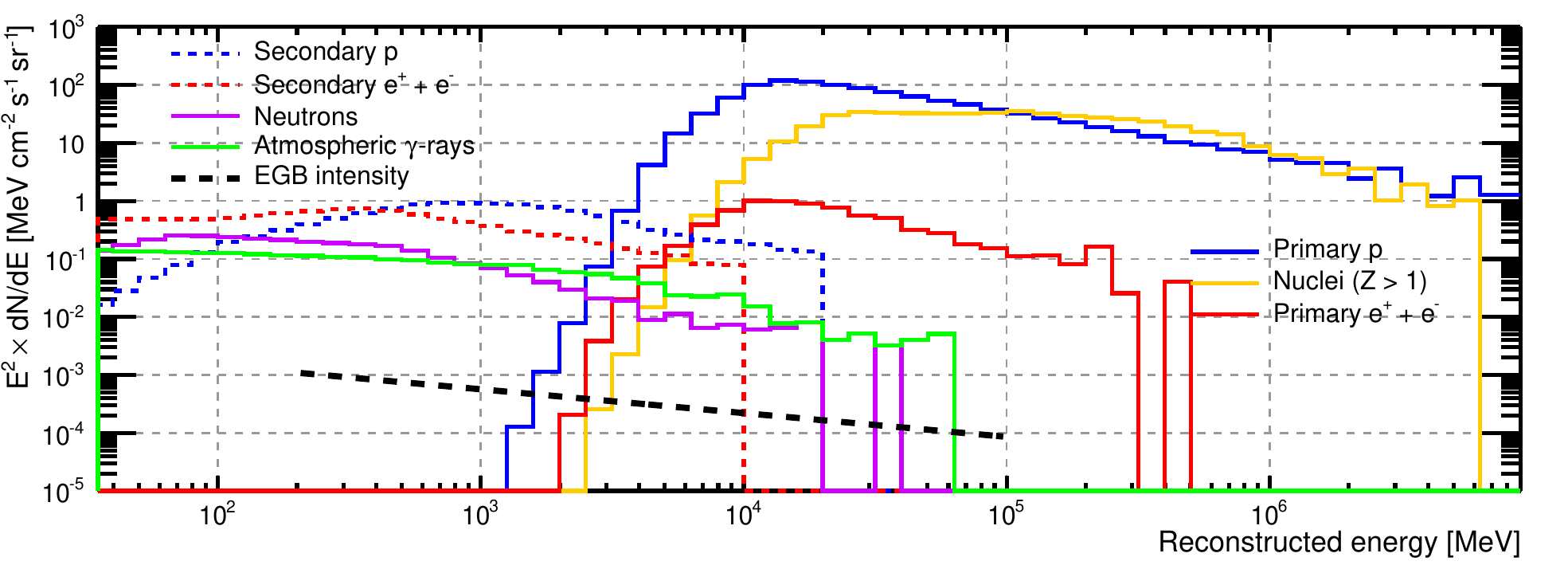}
  \caption{Model of the LAT orbital position and particle direction-averaged
    CR-induced particle intensities \citep{REF:2004.CR_MODEL} sampled from a
    64~s live time background simulation run. The intensity of the
    extragalactic diffuse background emission measured by the LAT
    \citep{REF:ExtraGalBkg} is shown for comparison. Note that the event energy
    is reconstructed under the hypothesis of a downward-going \gammaRay\
    and in general does not represent the actual energy for hadrons.}
  \label{fig:lat_bg_model}
\end{figure}

\Figref{lat_bg_model} shows the average CR-induced particle intensities
at the orbit of the LAT in the model that we use. For comparison the intensity
of the extragalactic diffuse \gammaRayHyph\ emission measured by the LAT
\citep{REF:ExtraGalBkg} is overlaid to demonstrate the many orders of
background suppression necessary to distinguish it from particle background. 
The model was developed prior to launch based on data from satellites in
similar orbits and balloon experiments \citep{REF:2004.CR_MODEL}.

As the particle rates are strongly dependent on location in geomagnetic
coordinates, the details of the orbit model are also important. For tuning the
event analysis, or for estimating the background rates for typical integration
times of months or years, the simulated time interval must be at least equal
to the precession period of the \Fermi\ orbit ($53.4$~days).
Simulating these high particle rates for such a long time interval is quite
impractical, in terms of both CPU capacity and disk storage requirements.
For studies of background rejection we usually simulate an entire precession
period to ensure a proper sampling of the geomagnetic history, but to limit the
particle counts we generate events for only a few seconds of simulated time
every several minutes, e.g., a typical configuration requires event generation
for 4 seconds every 4 minutes of time in orbit.
This partial sampling is a compromise between the limited CPU and disk usage,
and the requirement of having good statistics. Considering the LAT
background rejection power, in order to have sizable statistics after even the
first stages of the event analysis are performed, we must start with a simulated
background data set of over $10^{9}$ CRs.

\clearpage

\section{EVENT TRIGGERING, FILTERING, ANALYSIS AND CLASSIFICATION}\label{sec:event}

\begin{figure}[htb!]
  \includegraphics[width=\textwidth]{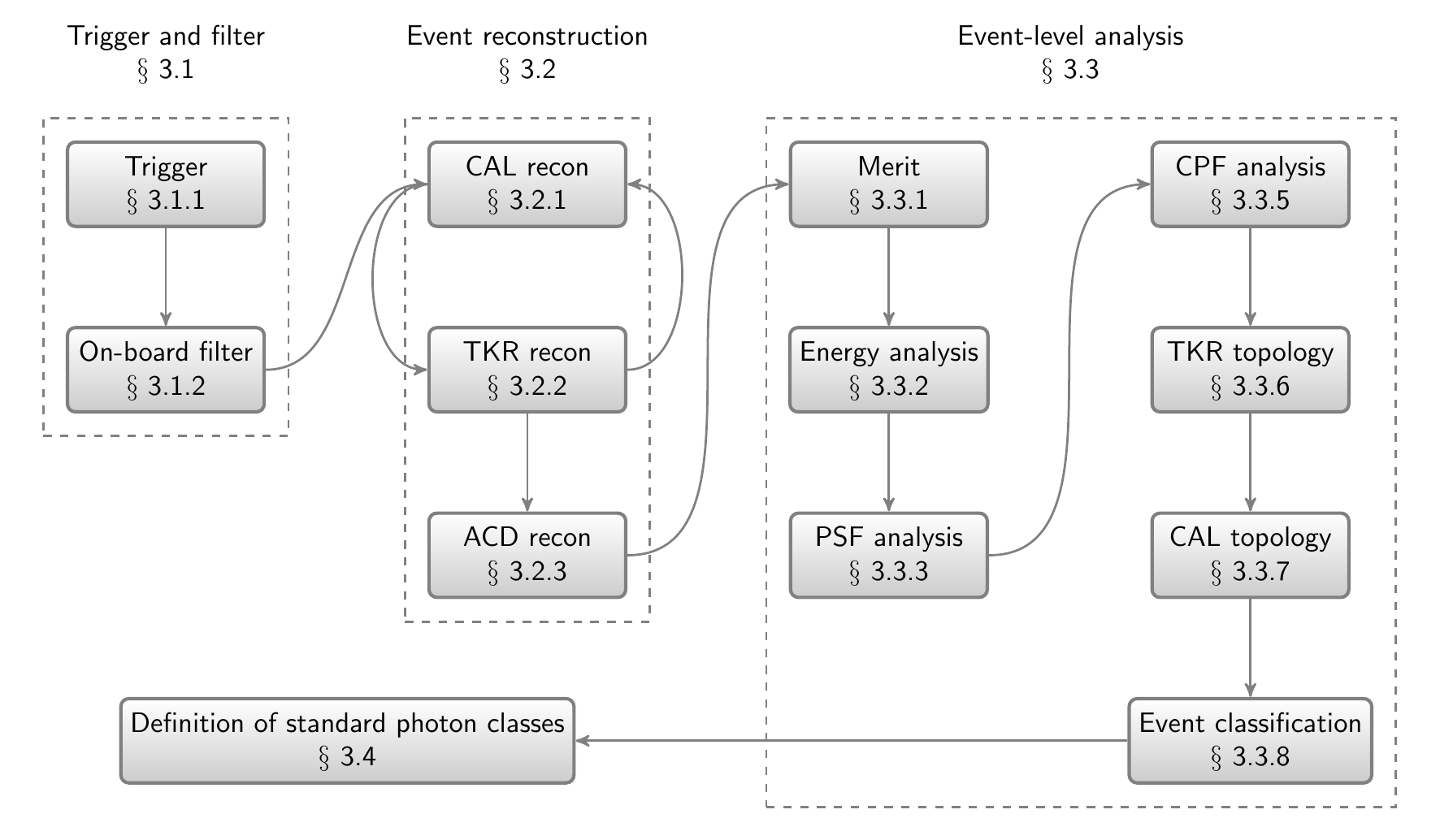}
  \caption{Logical structure of the analysis steps determining which events are
    selected for a given class. The references to section numbers are meant to
    help the reader navigate the content of~\secref{sec:event}.}
  \label{fig:Figure_11_overview}
\end{figure}

In this section we describe the analysis steps that determine which events make
it into our final \gammaRayHyph\ data sample, starting with the triggering and
filtering performed by the on-board data acquisition system
(\secref{subsec:event_trigFilter}), moving on to the reconstruction of particle
interactions in the event (\secref{subsec:event_recon}), the analysis of the
event as a whole (\secref{subsec:event_CT_analysis}) and finally the definition
of the \gammaRayHyph\ classes (\secref{subsec:event_photon_classes}).
The overall logical structure of this process is illustrated in
\figref{Figure_11_overview}.

The event analysis requires knowledge of the LAT, the physics of particle
interactions within its volumes, and of the particle backgrounds in the 
\fermi\ orbit. As described in \secref{subsec:LAT_simul}, we use large MC 
samples of \gammaRays\ and of CRs to devise the best procedures to extract
estimates of energies and incident directions, and to classify events as either
\gammaRays\ or charged particle backgrounds.

Finally, in \secref{subsec:LAT_public_data} we describe the publicly available
LAT event samples, while in \secref{subsec:LAT_methods} we describe the calibration sources,
event samples and methods we use to validate and characterize the performance of the LAT
using flight data.

\subsection{Trigger and On-Board Filter}\label{subsec:event_trigFilter}

In this section we review the event triggering, the readout of the LAT and
the filtering performed on-board in order to reduce the data volume downlinked
to ground.

\subsubsection{Triggering the LAT Readout}\label{subsubsec:event_trigger}

Each subsystem provides one or more \emph{trigger primitives}
(or \emph{trigger requests}) as detailed in the following list.
\begin{fermiitemize}
\item \tptkr\ (also known as ``three-in-a-row''): issued when three consecutive
  \xy\ silicon layer pairs---corresponding to six consecutive silicon
  planes---have a signal above threshold (nominally 0.25~MIPs).
  This signals the potential presence of a track in a tower.
  Since many tracks cross between towers and/or have more than 3 \xy\ layers
  within a tower, the \tptkr\ trigger request is very efficient. 
\item \tpcallo: issued when the signal in any of the CAL crystal ends
  crosses the low-energy trigger threshold (nominally 100~MeV). 
\item \tpcalhi: issued when the signal in any of the CAL crystal ends
  crosses the high-energy trigger threshold (nominally 1~GeV).
\item \tpveto: issued when the signal in any of the ACD tiles is above the
  veto threshold (nominally 0.45 MIP). It signals a charged
  particle crossing the tile. The trigger system has a programmable list of
  ACD tiles associated with each TKR tower; if ACD shadowing is enabled
  (which is the case for the nominal configuration) an additional \tproi\
  primitive is assembled when a \tptkr\ primitive in a tower happens 
  in coincidence with a \tpveto\ primitive in the ACD tiles
  associated to the same tower.
\item \tpcno: issued when the signal in any of the ACD tiles is above the CNO
  threshold (nominally 25~MIPs). It indicates the passage of a heavily ionizing
  nucleus (CNO stands for ``Carbon, Nitrogen, Oxygen'')\acronymlabel{CNO} and
  it is primarily used for the calibration of the CAL
  (see \secref{subsec:EDisp_xtalCalib}).
\end{fermiitemize}

The LAT has the ability to generate three other trigger primitives. We will 
refer to these as special primitives. Two of these are unused in flight and 
will not be discussed here. The third, the \tpperiodic\ trigger, runs at a nominal 
frequency of 2 Hz during all science data taking and is used for diagnostic 
and calibration purposes. 

We emphasize that although the trigger primitives provided by the
TKR and the CAL are tower-based, a trigger initiates a readout of
the whole LAT.

Trigger primitives are collected by the Central Trigger Unit.
All 256 possible combinations of the eight trigger primitives are
mapped into so-called \emph{trigger engines}.
In brief, some trigger requests are allowed to open a trigger window of fixed
duration (nominally 700~ns) and the set of primitives collected in this time
interval is compared to a table of allowed trigger conditions.
In case a trigger condition is satisfied, a global trigger
(or \emph{trigger accept}) is issued and event acquisition commences.
The LAT readout mode (i.e., enabling the CAL and ACD zero suppression and
selecting the CAL one-range or four-range readout) can be individually set
for each trigger engine.
In addition, trigger engines are scalable: for each trigger condition
a \emph{prescale} is specified, corresponding to the number of valid trigger
requests necessary to issue a single global trigger (obviously no prescale is
applied to engines intended for \gammaRayHyph\ collection).

\begin{table}[ht]
  \begin{center}
    \begin{tabular}{cccccccrr}
      \hline
      Engine & \tpperiodic & \tpcalhi & \tpcallo & \tptkr &
      \tproi & \tpcno &
      Prescale & Average rate [Hz] \\
      \hline
      \hline
      3 & 1 & $\times$ & $\times$ & $\times$ & $\times$ & $\times$ & 0 & 2 \\
      4 & 0 & $\times$ & 1 & 1 & 1 & 1 & 0 &  200 \\
      5 & 0 & $\times$ & $\times$ & $\times$ & $\times$ & 1 & 250 &  5 \\
      6 & 0 & 1 & $\times$ & $\times$ & $\times$ & 0 & 0 &  100 \\
      7 & 0 & 0 & $\times$ & 1 & 0 & 0 & 0 & 1500 \\
      8 & 0 & 0 & 1 & 0 & 0 & 0 & 0 & 400 \tablenotemark{a}\\
      9 & 0 & 0 & 1 & 1 & 1 & 0 & 0 & 700 \\
      10 & 0 & 0 & 0 & 1 & 1 & 0 & 50 & 100 \\
      \hline
    \end{tabular}
    \caption{Definition of the standard trigger engines: primitives used
      (1: required, 0: excluded, $\times$: either), prescale factors and
      typical rates.
      In this short-hand representation, engines are defined with highest precedence 
      first: each combination of trigger primitives is mapped to the engine corresponding 
      to the first condition it matches. Trigger engines 0, 1 and 2 are used by the LAT software for
      bookkeeping and to catch conditions that should not happen on
      orbit.}
    \tablenotetext{a}{In the nominal configuration for science data taking the \tpcallo\
      condition is inhibited from opening a trigger window, and therefore engine~8
      is effectively disabled.}
    \label{tab:triggers}
  \end{center}
\end{table}

The standard trigger engine definitions are listed in \tabref{triggers}.
Trigger engine~7 is particularly important for \gammaRayHyph\ events: 
it requires no special primitives, a three-in-a-row signal in the TKR
(\tptkr), absence of \tproi\ shadowing and \tpcno\ from the ACD, and no
\tpcalhi. Engine~6 handles \tpcalhi\ events and engine~9 handles 
events with enough energy deposition in the CAL to potentially cause 
\tproi\ shadowing.

Trigger engine~4 is typical for calibration data (e.g., heavy ion collection):
it requires no special primitives, three-in-a-row in the TKR (\tptkr),
a highly ionizing passage in the ACD (\tpcno) in close correspondence (\tproi)
and at least a moderate energy deposition in the CAL (\tpcallo).
Engines 5 and 10 process very few \gammaRays\ and are used primarily for calibration and monitoring,
so they are prescaled to limit the readout rate and/or to match the downlink
bandwidth allocation.

\subsubsection{Event Filtering}\label{subsubsec:event_filter}

As mentioned before, since limited telemetry bandwidth is available for data
downlink, some event filtering is required.
The on-board filter allows the coexistence of different filtering algorithms
and in fact, in the nominal science data taking configuration, all the events
are presented to three different filters:
\begin{fermiitemize}
\item the \obfgam, designed to accept \gammaRays\ (with an output average rate
  of approximately 350~Hz);
\item the \obfhip, designed to select  heavy ion event candidates, primarily
  for CAL calibration (with an output average rate of approximately
  10~Hz);
\item the \obfdgn, designed to enrich the downlinked data sample in events
  useful to monitor sensor performance and selection biases: specifically
  $\sim 2$~Hz of periodic triggers and an unbiased sample of all trigger types
  prescaled by a factor 250 (with an output average rate of approximately
  20~Hz).
\end{fermiitemize}

The \obfgam\ consists of a hierarchical sequence of veto tests, with the 
least CPU-intensive tests performed first.  If an event fails a test, it is 
marked for rejection and not passed on for further processing (however, a small
subset of events that are marked for rejection are downlinked through the
\obfdgn).
Some of the tests utilize rudimentary, two-dimensional tracks found by the
on-board processing.  The \obfgam\ has several steps (listed in order of
processing):
\begin{fermienumerate}
\item reject events that have patterns of ACD tile hits that are consistent 
  with CRs and do not have the \tpcallo\ trigger primitive asserted, making it
  unlikely that the ACD hits were caused by backsplash;
\item accept all events for which the total energy deposited in the CAL 
  is greater than a programmable threshold, currently set to 20~GeV;
\item reject events that have ACD hit tile patterns that are
  spatially associated with the TKR towers that caused the trigger,
  provided that the energy deposited in the CAL is less that a
  programmable threshold (currently set to 350~MeV);
\item reject events for which a significant energy deposition in the CAL
  (typically $> 100$~MeV) is present but the pattern of hits in the TKR is
  unlikely to produce any track;
\item reject events for which rudimentary tracks match with individual ACD
  tiles that were hit, provided the energy deposited in the CAL is less than
  some programmable amount (typically 5~GeV);
\item reject events that do not have at least one rudimentary track.
\end{fermienumerate}
Although it may seem strange to apply the requirement that there be
any tracks after cutting on matches between tracks and the ACD, recall
that the on-board filter software is highly optimized for speed, and
terminates processing of each event as soon as it is possible to reach
a decision.  Thus the testing for track matches is performed during
the track-finding stage, at the time the tracks are constructed.

\subsection{Reconstruction Algorithms}\label{subsec:event_recon}

\begin{figure}[htb!]
  \centering
  \includegraphics[width=\textwidth]{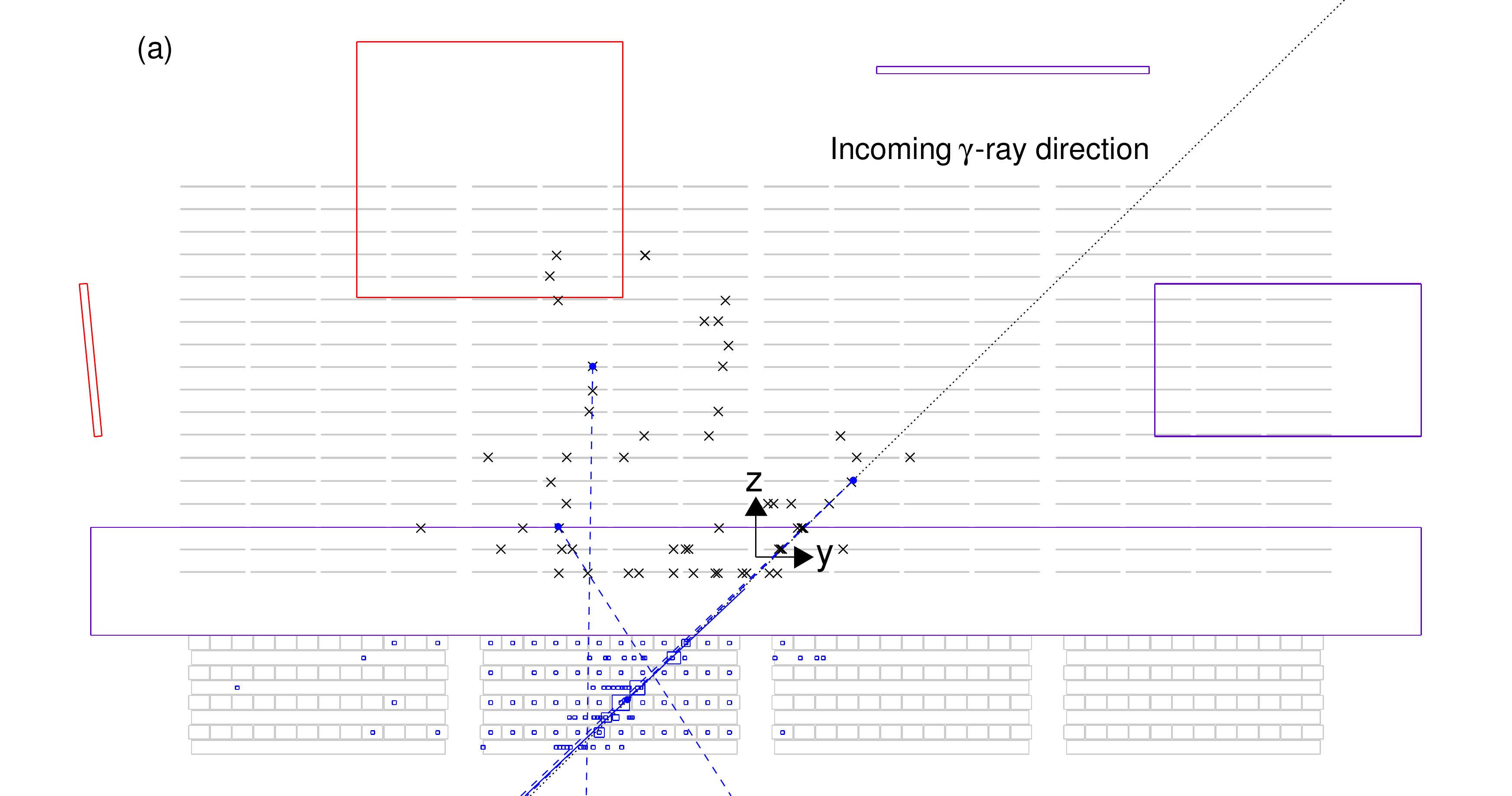}\\[10pt]
  \includegraphics[width=0.48\textwidth]{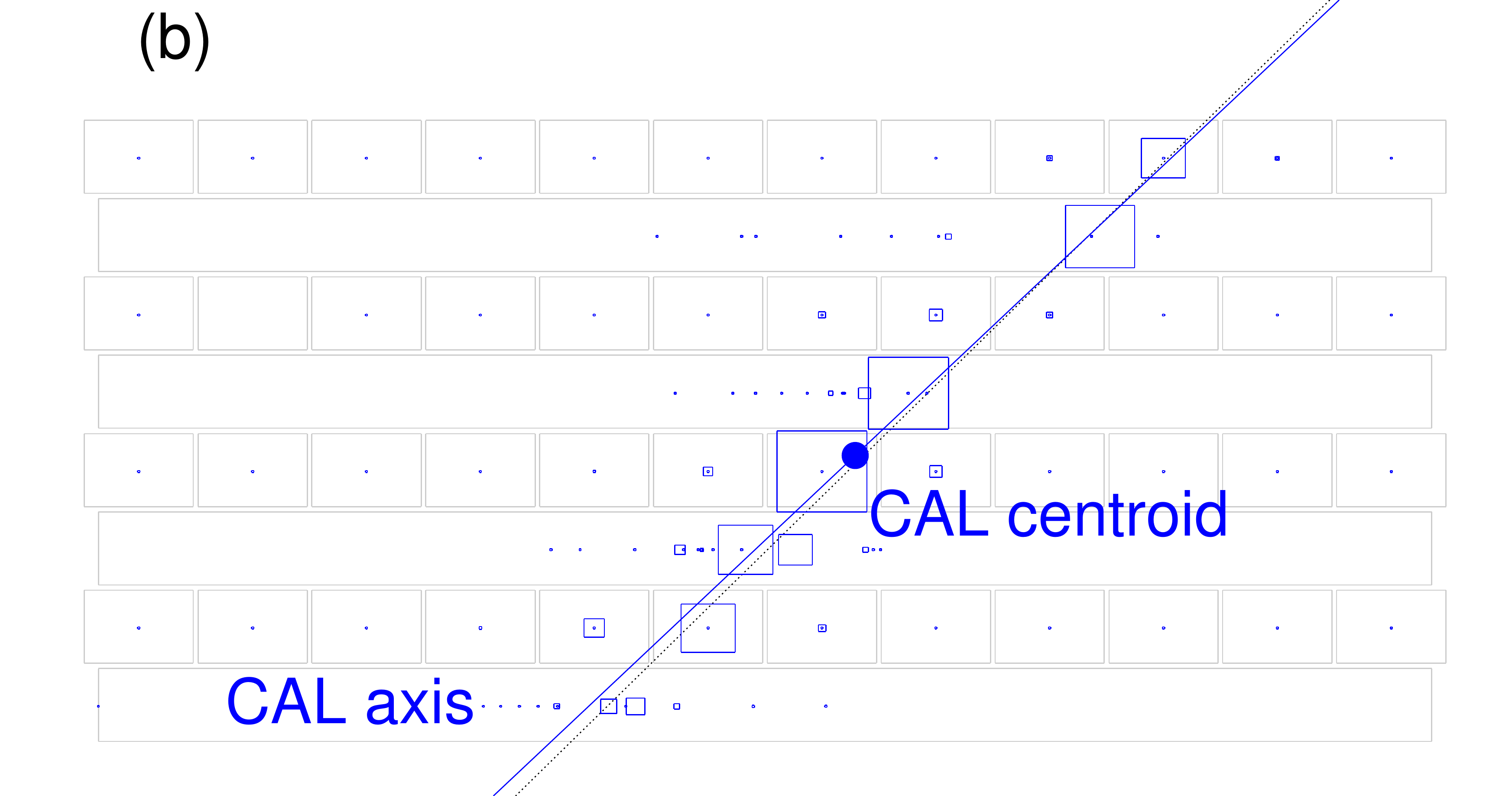}\hfill%
  \includegraphics[width=0.48\textwidth]{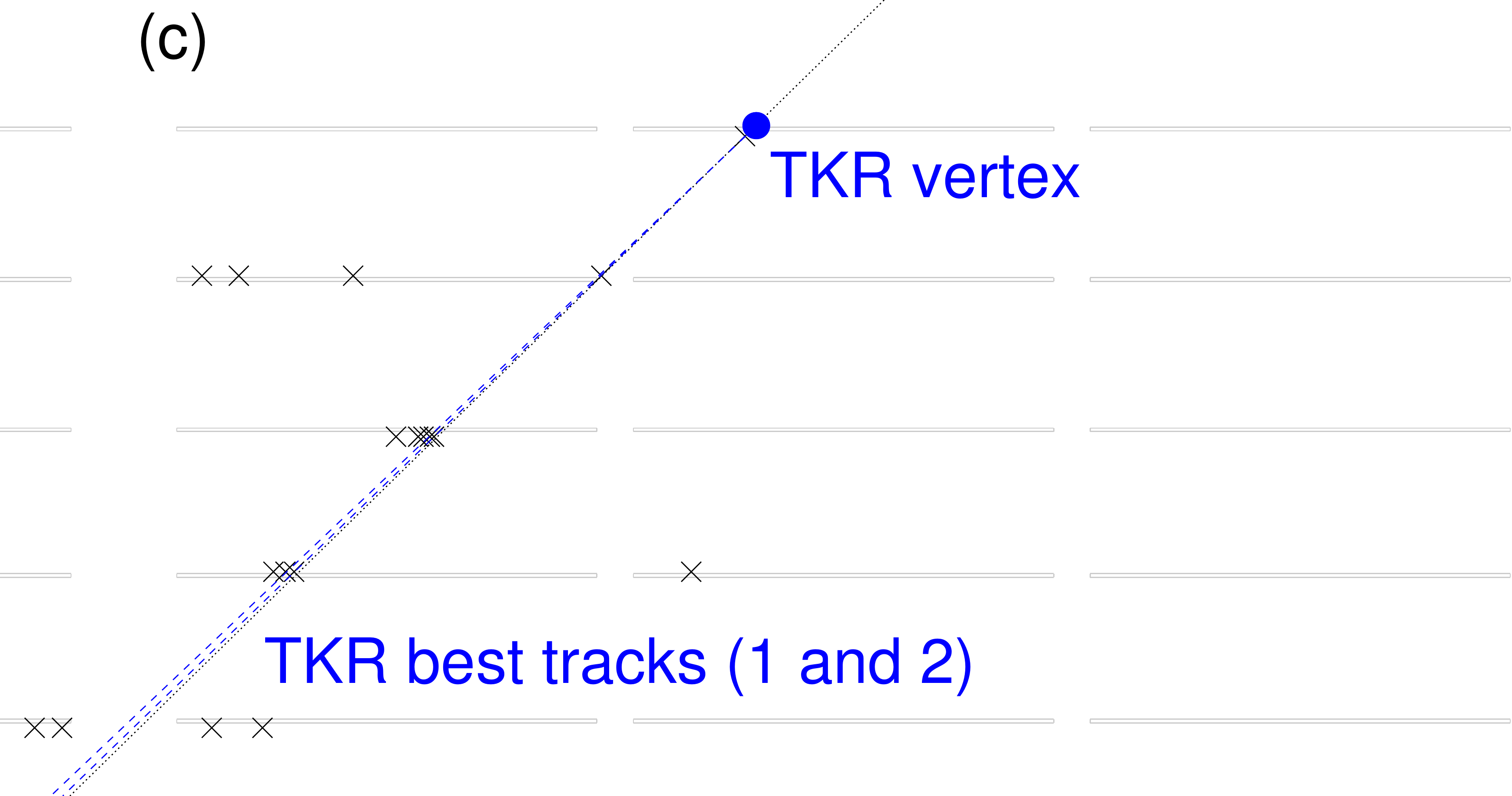}
  \caption{Event display of a simulated 27~GeV \gammaRay\ (a) and zoom
    over the CAL (b) and TKR (c) portions of the event. 
    The small crosses represent the clusters in the TKR, while the
    variable-size squares indicate the reconstructed location and magnitude of the
    energy deposition for every hit crystal in the CAL. 
    The dotted line represents the true \gammaRayHyph\ direction, the solid line
    is the CAL axis (\secref{subsubsec:cal_recon}) and the dashed lines are
    the reconstructed TKR tracks (\secref{subsubsec:cal_recon}).
    The backsplash from the CAL generates tens of hits in the TKR, with two
    spurious tracks reconstructed in addition to the two associated with
    the \gammaRay\ (note that they extrapolate away from the CAL centroid and
    do not match the CAL direction). It also generates a few hits in the ACD,
    which, however, are away from the vertex direction extrapolation and
    therefore do not compromise our ability to correctly classify the event
    as a \gammaRay.}
  \label{fig:recon_event_display}
\end{figure}

Event reconstruction translates the raw event information from the LAT
subsystems into a high-level event description under the assumption 
of a \gammaRay\ impinging on the LAT volumes within $90^\circ$ of the
boresight (see \parenfigref{recon_event_display} for an illustrative event display).

Here we will briefly summarize the event reconstruction algorithms underlying
both \psix\ and \pseven\ flavors of the event analysis; a more detailed
description is given in \citet{REF:2009.LATPaper}. We want to stress 
that \psix\ and \pseven\ use exactly the same reconstruction algorithms 
with the exception of the energy un-biasing (described in
\secref{subsubsec:cal_recon}), which was only applied to \pseven\ data.

\subsubsection{Calorimeter Reconstruction}\label{subsubsec:cal_recon}

The starting  point for the energy evaluation is the measured energy depositions
in the crystals. The centroid of the energy deposition is determined and the
principal axes  of the shower are evaluated by means of a principal moment
analysis. In the \psix\ and \pseven\ event reconstruction procedure, the energy
deposition is treated as a single quantity, with no attempt to identify
contamination from ghost signals. Work to develop an algorithm to separate the
CAL energy deposition into multiple clusters and to disentangle ghost signals is
ongoing \citep{REF:2010.Pass8}.
The amount of energy deposited in the TKR is evaluated by treating the
tungsten-silicon detector as a sampling calorimeter; this contribution is an
important correction at low energies.

We apply three algorithms to estimate the actual energy of an event: a
Parametric Correction (PC)\acronymlabel{PC}, a fit of the Shower Profile
(SP)\acronymlabel{SP} and a Maximum Likelihood (LH)\acronymlabel{LH} approach.
The energy assigned to any given event is the energy from one or the other of
these algorithms. The three methods were designed to provide the best
performance in different parts of the energy and incidence angle phase space
(in fact, the LH algorithm was only tuned up to $300$~GeV, while the SP
algorithm does not work well for events below $\sim 1$~GeV).
Accordingly, they provide different energy resolutions and their distributions
have slightly different biases (i.e., the most probable values are slightly
above or below the true energy) for different energies and incidence 
angles; more details can be found in \citet{REF:2009.LATPaper}.  

In fact, the only significant change in the \pseven\ event reconstruction
relative to \psix\ is to apply separate corrections for the biases of 
each energy estimation algorithms.  We used MC simulations to characterize 
the deviations of the most probable value of the energy dispersion from the
true energy across the entire LAT phase space for the three methods separately.
Such deviations were found to be typically of the order of few percent 
(with a maximum value of $\sim 10$\%) and always significantly smaller than the
energy resolution---with LH displaying a negative bias and PC and SP displaying
a positive bias in most of the phase space. 

We generated correction maps (as functions of \gammaRayHyph\ energy and zenith
angle) and in \pseven\ the residual average bias for all the inputs of the
final energy assignment (discussed in \secref{subsubsec:energy_analysis}) is
less than 1\% in the entire LAT phase space.

\subsubsection{Tracker Reconstruction}\label{subsubsec:tkr_recon}

For the TKR we merge and assemble clusters of adjacent hit strips into track
candidates by combinatorial analysis. We have developed two methods: for
CAL-Seeded Pattern Recognition (CSPR)\acronymlabel{CSPR} the trajectory
of the original \gammaRay\ is assumed to point at the centroid of the energy
released in the CAL; the Blind Search Pattern Recognition
(BSPR)\acronymlabel{BSPR} can be used when there is little or no energy deposit
in the CAL sensitive volumes. 

Both the CSPR and BSPR algorithms start by considering nearby pairs
of TKR clusters in adjacent layers as candidate tracks (the CSPR
algorithm limits the candidate tracks to those pointing toward the
CAL energy centroid).  Both algorithms then proceed by using a Kalman 
filtering technique \citep[see][for details about Kalman filtering and its use
in particle tracking]{REF:Kalman,REF:Fruhwirth:1987fm} which tests the
hypotheses that additional TKR
clusters were generated by the same incident particle and should
be associated to each of the candidate tracks, and adds any such clusters
to the appropriate candidate tracks. Furthermore, as each cluster is
added to candidate tracks the Kalman filter updates the estimated direction and
associated covariance matrix of those tracks. Both the CSPR and BSPR
algorithms are weighted to consider best candidate track to be the one that is
both pointing toward the CAL centroid and is the longest and straightest.
In the CSPR, the main axis of the CAL energy deposition is also considered, 
candidate tracks for which the TKR and CAL estimated directions differ significantly 
are disfavored starting at $\sim 1$~GeV, and increasingly so at higher energies.
At the completion of the CSPR algorithm the best candidate track is
selected and confirmed as a track, and the clusters in it are flagged
as used.  We iterate the CSPR algorithm until no further tracks can be
assembled from the unused TKR clusters, then proceed with the BSPR.

If more than one track is found in a given event, we apply a vertexing
algorithm that attempts to compute the most likely common origination point
of the two highest quality (i.e., longest and straightest) tracks, and,
more importantly, to use that point as a constraint in combining the momenta
of the two tracks to obtain a better estimate of the direction of the
incoming \gammaRay.

\subsubsection{ACD Reconstruction}\label{subsubsec:acd_recon}

The ACD phase of the event reconstruction starts by estimating the energy
deposited in each of the tiles and ribbons.  Subsequently, these energy
depositions are associated to each of the tracks found in the TKR.
More specifically, each track is projected to the ACD, and we calculate whether
that track intersects each ACD tile or ribbon with non-zero energy deposition.
Furthermore, if the track projection does not actually cross an ACD element
with non-zero energy deposition, we calculate the distance of closest approach 
between the track projection and the nearest such ACD element. Finally, we use
the distance calculation and the amount of energy deposited to sort all the
possible ACD-TKR associations by how likely they are to represent a charged
particle passing through the ACD and into the TKR. This ranking is used when
considering whether the event should be rejected in later analysis stages.

\subsection{Event Analysis}\label{subsec:event_CT_analysis}

The first step of the event-level analysis procedure is to extract simpler
representations of the event characteristics from the complex structures
assembled during the reconstruction phase. These high-level quantities are then
used for the final assignment of the event energy and direction (among the
outputs of the reconstruction algorithms described in
\secref{subsec:event_recon}) and for the background rejection.
The final product is an event-by-event array of simple quantities relevant
for scientific analysis: energy, direction and estimates of the probability 
a given event is a \gammaRay\ (\probAll).

\subsubsection{Extraction of ``Figure-of-Merit'' Quantities}\label{subsubsec:event_meritVars}

For each event we reduce the output of the TKR, CAL and ACD reconstruction to
a set of a few hundred figure-of-merit quantities whose analyzing power has
been studied and optimized with MC simulations.   

It is important to note that the best track (and to a lesser extent the
second best track) plays a particularly important role in later analyses.
Specifically, although we calculate figure-of-merit variables---such as the
agreement between the track direction and the axis of the CAL shower, or the
distance between the track extrapolation and the nearest ACD energy
deposition---for all the tracks in the event, many of the multi-variate analyses
described in the rest of this section consider only those figure-of-merit
variables associated with the two best tracks in the event. Furthermore, 
the figure-of-merit variables associated with the best track tend to carry significantly more weight
in the multi-variate analysis---which is of primary importance, as most ghost
tracks come from protons and heavy ions and tend to be longer and straighter
than tracks from $e^+$ and $e^-$ from a \gammaRayHyph\ conversion.

Early iterations of the event analysis split the events by energy and then
applied different selections in the various energy bands. We found that this
approach led to large spectral features at the bin boundaries. Therefore, we
chose instead to scale many of our figure-of-merit variables with energy so as
to remove most of the energy dependence. This allowed us to have a single set
of event selection criteria spanning the entire LAT energy range.

\subsubsection{Event Energy Analysis}\label{subsubsec:energy_analysis}

The second step in the event analysis is selecting one energy estimate
among those available for the event (see \secref{subsubsec:cal_recon}).
We apply a classification tree (CT)\acronymlabel{CT}
analysis~\citep{REF:ClassificationTrees,REF:BaggingTrees} to select which of
the energy reconstruction methods is most likely to provide the best estimate
of the event energy.

Because of the numerous edges and gaps between the logs in the CAL, and because
of the huge energy range and large \fov\ of the LAT, the quality of
the energy reconstruction can vary significantly from event to event. Therefore,
for each event we also apply a second CT analysis to estimate the probability
that the estimated energy is within the nominal core of the energy dispersion.
Specifically we define a \emph{scaled} energy deviation, as 
described in more detail in \secref{subsubsec:edisp_scaling}, from which most
of the energy and angular dependence is factored (i.e., the energy dispersion
in the scaled deviation is largely energy- and angle-independent).
We then train CTs that provide probability estimates this event is less than 
$2\sigma$ ($P_{2\sigma}$) or $3\sigma$ ($P_{3\sigma}$) away from the most 
probable value of the energy dispersion. Finally, we define a reconstruction
quality estimator \probE\ by combining $P_{2\sigma}$ and $P_{3\sigma}$:
\label{conv:probE}
\begin{equation}
  \probE = \sqrt{P_{2\sigma} P_{3\sigma}}.
\end{equation}
Large values of \probE\ indicate that the event is likely to be in the core
of the energy dispersion, and so have an accurate energy estimate.

In the \pseven\ analysis we did not use the energy estimates from the LH
algorithm: by construction it is a binned estimator, and energy assignments
tended to pile up at the bin boundaries (see \secref{subsec:eDispCorrections}
for a more detailed discussion of the effects of removing the LH algorithm
from consideration in the energy assignment).
The removal of the LH energies causes a somewhat degraded energy resolution
for those events where it would have been selected; we compensate for the loss
of this energy estimator by requiring a slightly more stringent cut on the
energy reconstruction quality when defining the standard event classes
(e.g., see \secref{subsubsec:p7source_selection}).

\subsubsection{Analysis of the Event Direction}\label{subsubsec:psf_analysis}

The third step selects the measured direction of the incoming \gammaRay\ from
the available options. Those options are the directions as estimated from (i)
the best track (ii) the best vertex and (iii and iv) the same two options using
the centroid of the energy deposition in the CAL as an additional constraint
on the direction of the incident \gammaRay.
Again, we use a CT analysis to combine the information about the event
and determine which of the methods is most likely to provide the best direction
measurement. 

As with the energy analysis, the quality of the direction reconstruction can
vary significantly from one event to the next. In this case, we have the
additional complication that \gammaRays\ can convert at different heights in
the TKR, giving us anywhere between 6 and 36 track position measurements.
Therefore we use an additional CT analysis to estimate the probability
$\probCore$\label{conv:probCore} that the reconstructed direction falls within the nominal 68\%
containment angle as defined by the simplified analytical
model:\label{conv:psf_68_cont}
\begin{equation}\label{eq:psf_ctbcore}
  C_{68}(E) = \sqrt{
    \left[c_0\cdot\left(\frac{E}{100\ {\rm MeV}} \right)^{-\beta}\right]^2+c_1^2}
\end{equation}
(where the values of the coefficients for front- and back-converting events
are listed in \tabref{psf_ctbcore}).
The reader will notice the similarity with the functional expression used for
the PSF prescaling described in~\secref{subsec:PSF_MonteCarlo}.

\begin{table}[htb]
  \centering
  \begin{tabular}{cccc}
    \hline
    Conversion type & $c_0$ [$^\circ$] & $c_1$ [$^\circ$] & $\beta$\\
    \hline
    \hline
    Front & 3.3 & 0.1 & 0.78\\
    Back  & 6.6 & 0.2 & 0.78\\
    \hline
  \end{tabular}
  \caption{Coefficients of the analytical model for the 68\% containment angle
    for the PSF from \Eqref{eq:psf_ctbcore} used for the multi-variate analysis
    used to evaluate the quality $p_{\rm CORE}$ of the direction reconstruction.}
  \label{tab:psf_ctbcore}
\end{table}

\subsubsection{Differences in Event Energy and Direction Analyses 
  Between \psix\ and \pseven}\label{subsubsec:reconVersions}

There is a particular subtlety to the event reconstruction that merits a brief
discussion. Specifically, the different event reconstruction algorithms we use
provide a set of choices for the best energy and direction of each event.
The stages of the event-level analysis described in the previous two subsections
select the algorithms that are more likely to provide the best estimates.
Therefore, although the event reconstruction was unchanged except for the
energy un-biasing, changes in the event-level analysis can result in
individual events being assigned slightly different directions and/or energy
estimates in \pseven\ with respect to \psix.
\figref{p7p6EnergyDirDiff} illustrates such differences for the
\irf{P7SOURCE} events above 100~MeV that were also included in 
the \irf{P6\_DIFFUSE} \gammaRayHyph\ class. 

\twopanel{htb!}{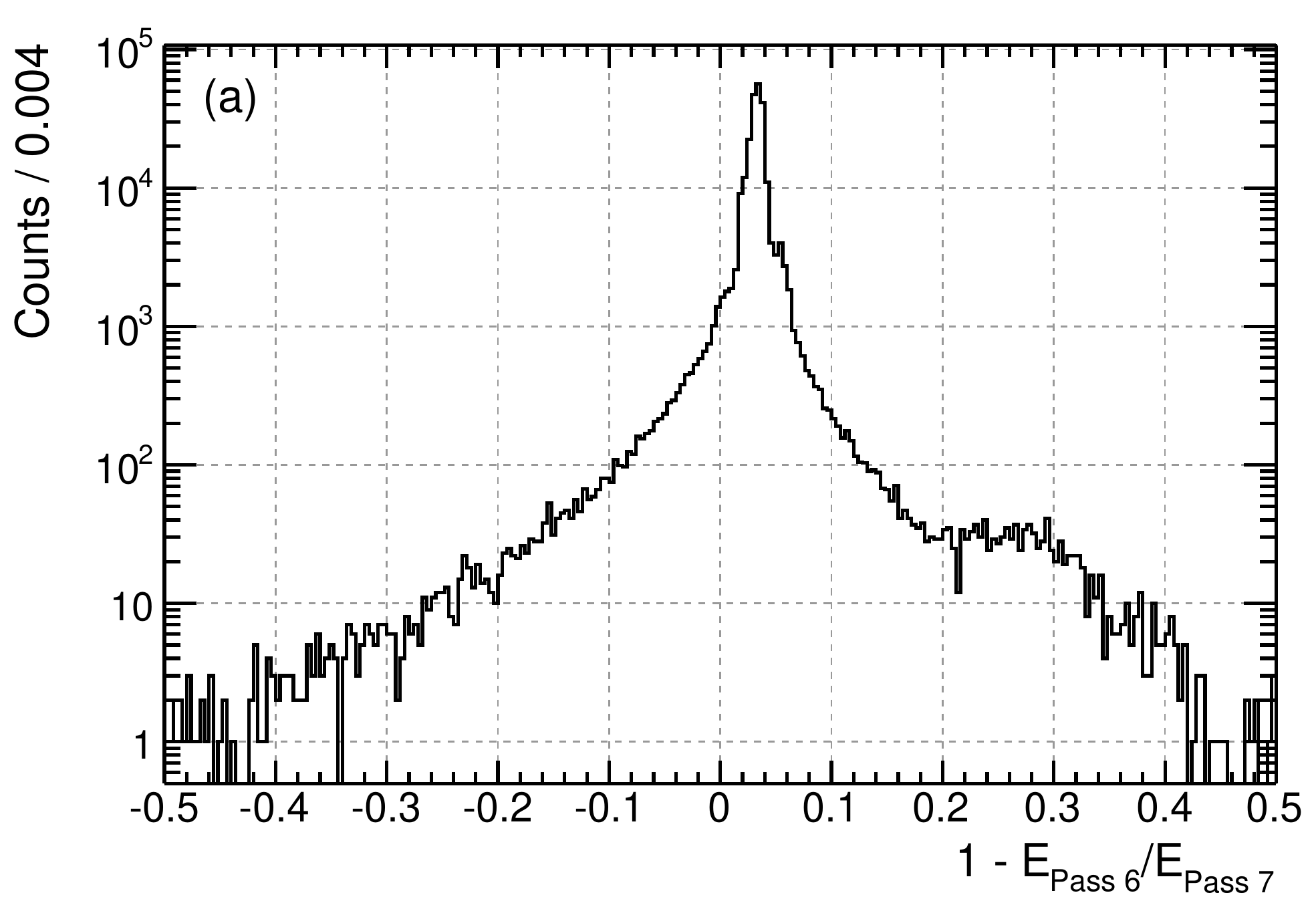}{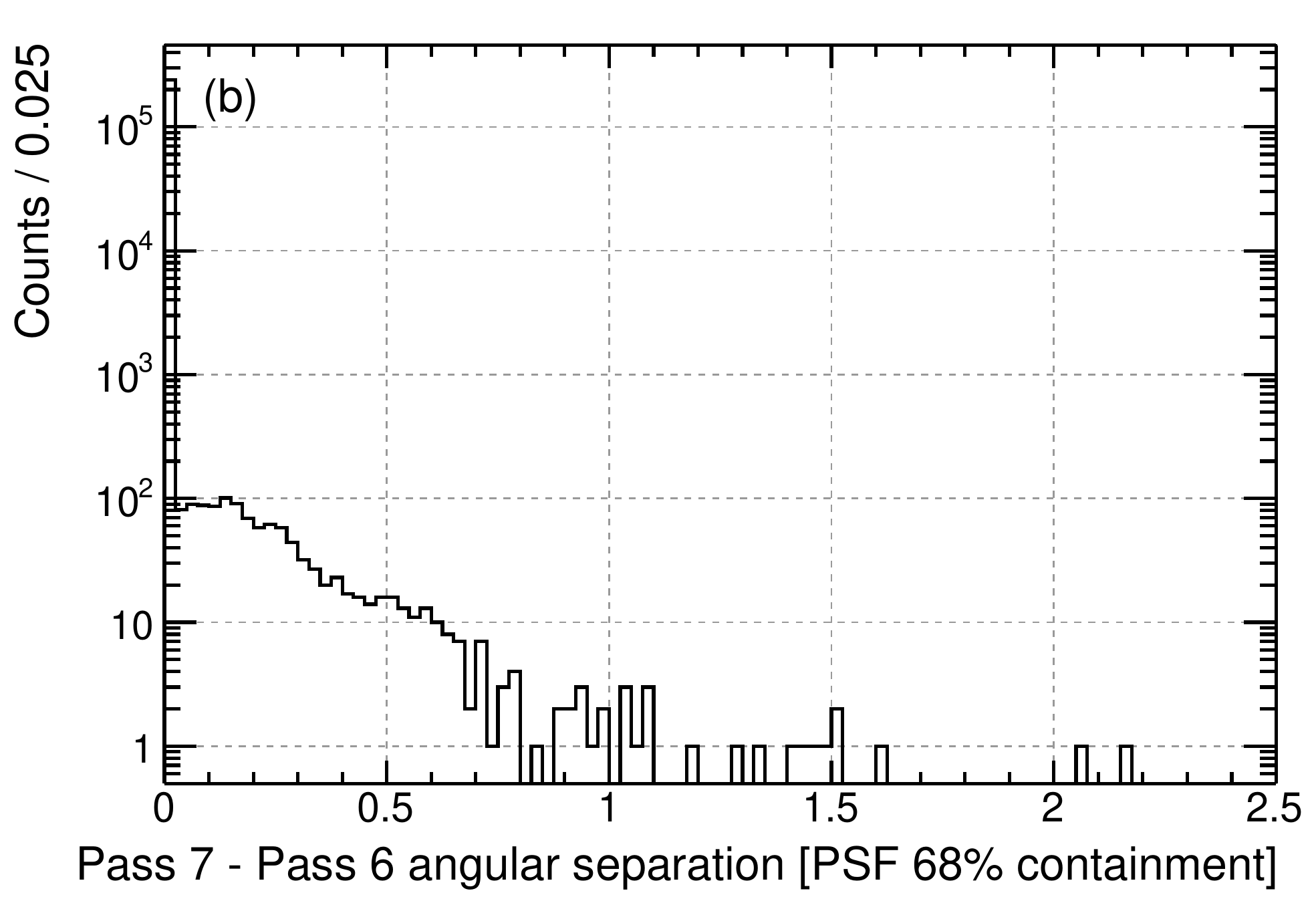}{
  \caption{Event by event differences in the reconstructed energy (a) and
    direction (b) between \psix\ and \pseven\ for a sample of \irf{P7SOURCE}
    events above 100~MeV. The two histograms are based on a week of data, with
    a zenith angle cut at $100^\circ$ to remove the bright Earth limb emission,
    typically viewed at large angle in LAT coordinates.
    In order to factor out the energy dependence of the PSF, the angular
    separation is measured in units of nominal 68\% containment at the energy
    of the event. Note that in the vast majority of the cases the same
    direction is assigned in \psix\ and \pseven. On the other hand, the
    systematic offset in energy is due to the energy un-biasing introduced in
    \pseven\ (\secref{subsubsec:cal_recon}).}
  \label{fig:p7p6EnergyDirDiff}
}

\subsubsection{Rejection of Charged Particles in the Field of View}\label{subsubsec:cpf_analysis}

The next step of the event analysis starts the process of classifying events as
\gammaRays\ or particle backgrounds, specifically identifying events for which
evidence clearly indicates that a charged particle entered the LAT from inside
the \fov.  We refer to this stage as the Charged Particle in the Field-of-view
(CPF)\acronymlabel{CPF} analysis. To accomplish this we first apply a series of
filters that classify as background those events for which the best track has
hits all the way to the edge of the TKR and (i) points to an active region of
the ACD that has significant deposited energy, or (ii) points directly to less
sensitive areas in the ACD, such as the corners or gaps between ACD tiles.  

After applying these filters we attempt to account for cases where the best
reconstructed track does not represent the incoming particle well. This can
happen for a variety of reasons, for example when the bias of the track finding
algorithm toward the longest, straightest track causes us to incorrectly assign
a backsplash particle from the CAL as the best track, or when the incoming
particle passes largely in the gaps between the TKR towers and we incorrectly
select some secondary particle as the best track. We classify as background
the events for which the total energy deposited in the ACD is too large to be
accounted for as backsplash from the CAL. It is important to note that this
particular requirement loses the benefit of the ACD segmentation, so we apply a
conservative version to all events, and a tighter version to events that have
a track that passes within an energy dependent distance of significant energy
deposition in the ACD. During extremely bright SFs this total ACD energy
requirement can cause us to misclassify many \gammaRays\ as CRs because of
energy deposition in the ACD from SF X-rays (see \appref{app:bti}).

All the events that are classified as charged particles at this stage are
removed  from further consideration as \gammaRayHyph\ candidates and are passed
to separate analyses to identify various species of charged particles;
see, e.g., \citet{REF:2009.Electrons20GeV1TeV}, \citet{REF:2010.CREPRD},
\citet{REF:2011.AbsoluteEnergyScale} and \citet{REF:2011.PositronFraction}.

Finally, we perform another CT analysis to combine all the available
information into an estimate of the probability that the event is from charged
particle backgrounds (\probCPF\label{conv:probCPF}). Although the performance
of this selection depends on energy and angle, roughly speaking more than 95\%
of the background in the telemetered data is removed by means of these cuts,
and about 10\% of the \gammaRayHyph\ sample is lost.   

We note in passing that electrons and positrons cause electromagnetic
showers that look extremely similar to \gammaRayHyph\ interactions,
and the remaining stages of the event analysis are based on the
topology (e.g., the shape, density and smoothness)  of the energy
deposition in the TKR (\secref{subsubsec:tkr_topology_analysis}) and
CAL (\secref{subsubsec:cal_topology_analysis}), and have very little
additional discriminating power against such backgrounds.

\subsubsection{TKR Topology Analysis}\label{subsubsec:tkr_topology_analysis}

In the next stage of the event analysis we use information from the TKR to
identify CR backgrounds that were not identified as such by the CPF analysis.
These events are not immediately removed from the \gammaRayHyph\ analysis but
only flagged to allow for removal from the higher purity event classes. In this part of the analysis we flag
events saturating the energy deposition in the TKR planes with a pattern
expected for heavily ionizing particles as well as events with very high energy
deposition in the first hit layers of the TKR, which is a signature of a
particle coming from the CAL that ranges out in the middle of the TKR.

After these two flags are applied, we divide the events among 5 branches for
evaluating their topologies in the TKR:  the first one collects events with a
vertex solution in the TKR, the remaining ones separate events with
one track and events with many tracks, treating separately events converting in
the thin and thick section of the TKR. This is done because each of these
different topologies has significantly different ratios of \gammaRayHyph\
signal-to-background contamination, and presents accordingly different
difficulty for selecting \gammaRays. For each branch we apply a different
initial selection to remove many CR background events and then pass the
remaining events to a CT analysis, which estimates the probability
(\probTKR\label{conv:probTKR}) the event comes from \gammaRayHyph\ interactions as opposed to charged particle
backgrounds (and more specifically hadronic charged particle backgrounds).

The variables used in the CT analysis were designed to emphasize differences 
between the characteristics of \gammaRayHyph\ conversions and hadronic
CRs. This includes distinguishing between electromagnetic and hadronic showers by
counting extra hits near the track, quantifying the complexity of the event in
the TKR and how collimated the overall track structure is, looking how deep 
into the fiducial volume the event starts, and requiring that the
ionization along the track is consistent with $e^{\pm} $ pairs.

\subsubsection{CAL Topology Analysis}\label{subsubsec:cal_topology_analysis}

Next we apply a CAL topology analysis that is similar in design to the
TKR topology analysis. The first part of this analysis consists of few
general cuts flagging events coming from the bottom and sides of the LAT.
As with the TKR topology analysis, we then split the events into five branches 
depending on the topology; for each branch we apply a cut and a CT stage.
Again, CTs are trained to select \gammaRays\ versus hadronic charged particle
backgrounds, and provide an estimate of the probability that the event
is a \gammaRay (\probCAL\label{conv:probCAL}).

The variables used in the CAL topology analysis are not only CAL derived
quantities, but also involve comparisons between what is observed in the CAL
and what is seen in the TKR. Among the important discriminants are how well
the track solution points toward the CAL energy centroid, how well the
direction derived from the CAL (via the moments analysis, see \secref{subsubsec:cal_recon}) agrees with the
track direction, the root mean square (RMS) width of the CAL shower, and the ratio
of the highest energy in a single crystal to the total energy in all crystals.

\subsubsection{Event Classification}\label{subsec:event_CT_eventLevel}

After the analyses related to the three LAT subsystems, the last stage of our
event analysis applies a final CT analysis utilizing all available
information, notably including the outputs of the CT analyses from previous
phases of the event analysis.  This second-order CT analysis is particularly
important for defining event classes with high \gammaRayHyph\ purity, as
discussed in \secref{subsec:event_photon_classes}.

At this point we have a number of specifiers of event reconstruction
and classification quality on an event-by-event basis:
\begin{fermiitemize}
\item energy reconstruction quality \probE\ (see
  \secref{subsubsec:energy_analysis});
\item direction reconstruction quality \probCore\ (see
  \secref{subsubsec:psf_analysis});
\item \gammaRayHyph\ probabilities from the CPF analysis
  (\probCPF, see \secref{subsubsec:cpf_analysis}) and the
  TKR (\probTKR, see \secref{subsubsec:tkr_topology_analysis}) and
  CAL (\probCAL, see \secref{subsubsec:cal_topology_analysis}) topology
  analysis;
\item overall \gammaRayHyph\ probability \probAll\label{conv:probAll} from the final
  classification step of the event analysis.
\end{fermiitemize}
These are the basic ingredients for defining the standard
\gammaRayHyph\ classes described in \secref{subsec:event_photon_classes}.

\subsection{Standard Event Classes for \gammaRays}\label{subsec:event_photon_classes}

\begin{figure}[htbp]
  \centering
  \includegraphics[width=0.5\textwidth]{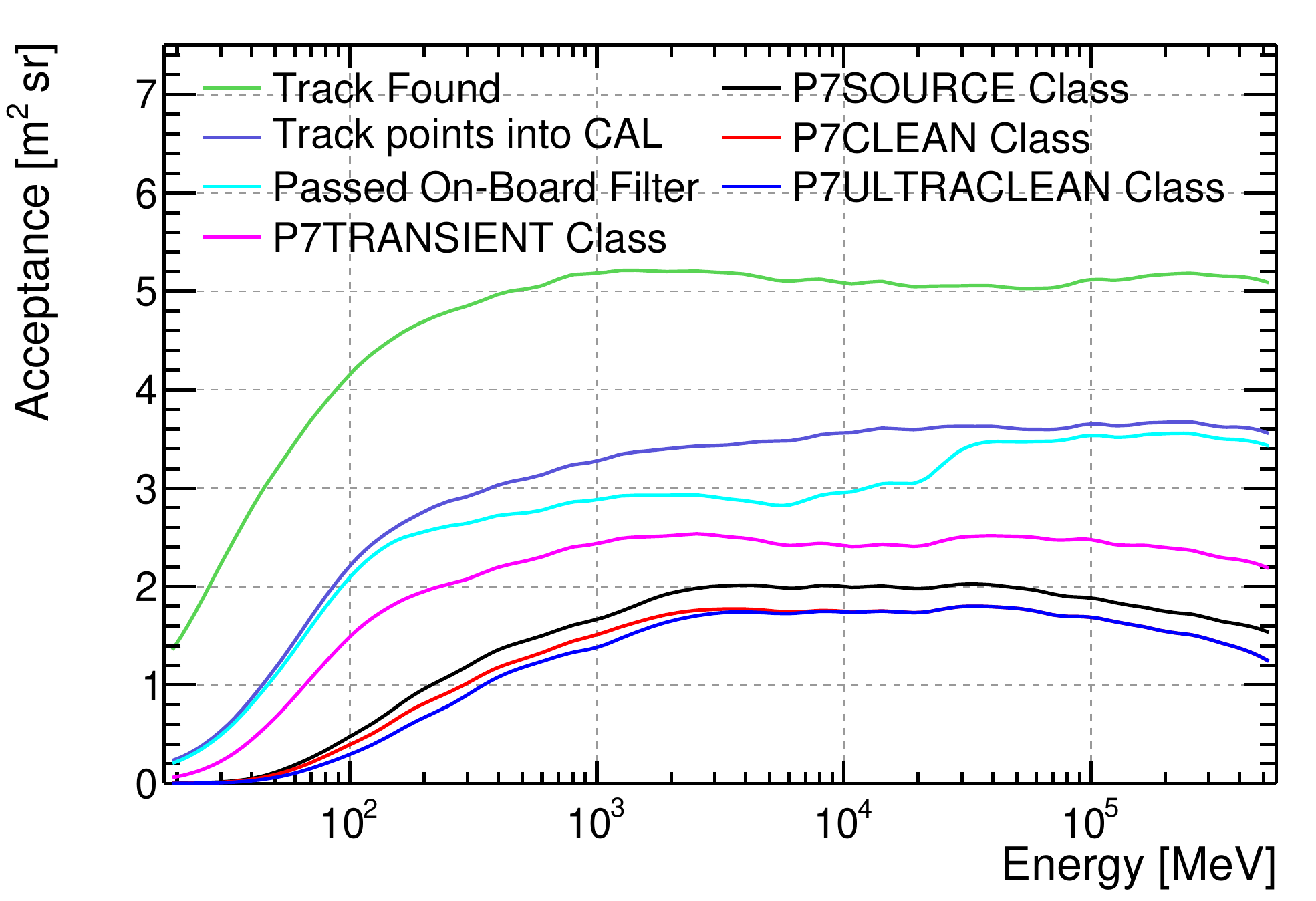}
  \caption{LAT \aeff\ integrated over the \fov\ as a function of energy
    at successive stages of the event filtering as estimated with simulated
    data. Since we require a direction and energy to use a \gammaRay\ for
    science analysis we consider only events with at least one track found and
    that pass the fiducial cuts
    (see \secref{subsubsec:track_find_fiducial_cuts}).}
  \label{fig:AcceptanceStages}
\end{figure}

Making use of the classification quantities described in the preceding
subsections we define event classes optimized for a range of astrophysical
source analyses. We note that these event classes are nested:
each succeeding selection is a strict subset of the previous ones.
The relative selection efficiencies for the event classes as well as
previous stages of the event selection process are shown in \figref{AcceptanceStages}.

\subsubsection{Track-Finding and Fiducial Cuts}\label{subsubsec:track_find_fiducial_cuts}

For the definition of the standard \gammaRayHyph\ classes we confine ourselves
to events that have some chance to be useful for most standard science analysis,
namely events for which we have enough information to reconstruct an energy and
a direction. Therefore, we require that the event has a reconstructed track to
allow for a direction estimation. Furthermore, we require that the track points
into the CAL and crosses at least 4 radiation lengths of material in the CAL
and that at least 5~MeV of energy is deposited in the CAL. This second
requirement reduces the fiducial volume of the detector by rejecting off-axis
\gammaRays\ that pass near the top of the TKR and miss the CAL. Events with no
tracks in the TKR or with less energy deposited in the CAL are not considered
further as candidate \gammaRays. The remaining data set (i.e., the events that
are passed along as potential candidates for the standard \gammaRayHyph\
classes) is still composed almost entirely of background events.

We note that these cuts remove from consideration two classes of events
that might be useful for specific, non-standard, analyses. The first class
consists of events that deposit all their energy in the TKR, either because
they range out before they reach the CAL or simply because they miss the
CAL entirely. In general the energy resolution is much poorer for these events,
though at low energies ($< 100$~MeV) the CAL does not improve the energy
resolution significantly.
These events have been used effectively in the analysis of both Gamma Ray
Bursts (GRBs)\acronymlabel{GRB}~\citep{REF:GRBLLE:2010} and SFs~\citep{REF:SolarLLE:2012}.
The second class consists of events that do not have reconstructed tracks, but
have enough information in the CAL to derive an estimate of the event
direction (though without the TKR information the angular resolution for
these events is highly degraded). These events can occur because the \gammaRay\
entered the LAT at a large incidence angle and missed most of the TKR, or simply
because the \gammaRay\ passed through the TKR without converting.

\subsubsection{\irf{P7TRANSIENT} Class Selection}\label{subsubsec:p7trans_selection}

For the analysis of brief transient sources (e.g., GRBs) a high purity is
not required as the time selection itself limits the amount of background counts
in the \roi. Accordingly, we define a \irf{P7TRANSIENT} event class with only
a few cuts, with the aim of achieving a residual background rate of a few~Hz
while maintaining a large efficiency for \gammaRays.
The list of cuts is short:
\begin{fermienumerate}
\item the event must pass minimal cuts on the quality estimators
  \probE\ and \probCore ($\probE > 0.$ and $\probCore > 0.$);
\item the event energy after all corrections must be greater than 10~MeV;
\item the event energy after all corrections must not amount to more than
  5~times the energy deposited in the CAL;
\item the CPF analysis must not flag the event as a charged
  particle $(\probCPF > 0.1)$;
\item the event must pass a relatively loose cut on \probAll ($\probAll > 0.2$).
\end{fermienumerate}

It is worth noting that although the cuts are optimized somewhat differently
in \psix: a similar series of cuts are applied to define the \irf{P6\_TRANSIENT}
event class, which is roughly equivalent to the \irf{P7TRANSIENT}
event class.

\subsubsection{\irf{P7SOURCE} Class Selection}\label{subsubsec:p7source_selection}

We have several additional considerations when defining the \irf{P7SOURCE}
event class, which is intended for the analysis of point sources.
While the \gammaRays\ from a point source are clustered on the sky near the
true source position, residual CR background can be modeled as an isotropic
component and be accounted for in the analysis of the spectrum or position of
the source. Although the exact numbers depend on the details of the PSF and the
spectrum of the source, in general we require a background rate of less than
$\sim 1$~Hz in the LAT \fov\ to ensure that we maintain a high enough
signal-to-background ratio so that this has little impact on source detection
and characterization.
Furthermore, in contrast to studies of transient sources, for sources
with an integral energy flux above 100~MeV of
$\sim 10^{-11}\text{ erg  cm}^{-2}\text{ s}^{-1}$ the precision of spectral
studies is often limited by systematic uncertainties below 1~GeV
(see \secref{subsec:perf_fluxAndIndex}), increasing the importance of the 
precision and accuracy of the event energy and direction reconstruction. 

Additionally, we note that for the \irf{P7SOURCE} event class and both
higher-purity event classes (\irf{P7CLEAN} and \irf{P7ULTRACLEAN}) we developed
and optimized the cuts using on-orbit data samples (particularly the clean
and dirty samples described in \secref{subsubsec:LAT_galRidge}) as well as MC
simulations. In particular, we performed a comparison of many event quality
parameters and reconstruction variables between the clean and dirty samples to
identify characteristics of the CR background, which is greatly enhanced 
in the dirty sample, and devised selection criteria to remove it.
Some examples of this procedure are described in more detail in
\secref{subsec:bkg_distrib}.

An important limitation of this technique is obviously that one needs a very
low ratio of residual CR background to \gammaRays\ in the clean sample to obtain
large contrast between the clean and dirty samples. It can only be applied to
data samples which already have a \gammaRayHyph-to-residual-background ratio
that is comparable to 1.  Therefore, when optimizing these selections we only
used events that are part of the \irf{P7TRANSIENT} class
(see\secref{subsubsec:track_find_fiducial_cuts}) and additionally have a high
value of \probAll\ estimator introduced in \secref{subsec:event_CT_eventLevel}.

The \irf{P7SOURCE} event class is defined starting from the \irf{P7TRANSIENT}
event class and includes tighter cuts on many of the same quantities.
Specifically, the \irf{P7SOURCE}  event class selection requires that:
\begin{fermienumerate}
\item the event must not have been flagged as background based on topological
  analysis of the CAL (\secref{subsubsec:tkr_topology_analysis}) and
  TKR (\secref{subsubsec:tkr_topology_analysis}) reconstruction, and
  must pass a tighter cut on \probE; the cut on \probE\ does reject a noticeable
  number of events, though by definition these events have less accurate energy
  estimates on average ($\probE > 0.1$ or  $\probE > 0.3$ for events
  which originally used the LH energy estimate);
\item the event must pass cuts to reject MIPs based on the event topology in the CAL and the
  energy deposited in the TKR; 
\item the event must pass a set of cuts on the agreement between the TKR and
  CAL direction reconstruction;
\item the event must pass a tighter, energy dependent, cut on \probCore\:
  $$
  \probCore > \max(0.025,0.025+0.175(3.-\log_{10}(E/1\rm{MeV}));
  $$
  \label{item:p7trans_ctbcore_cut} 
\item the event must pass a tighter cut on \probAll\:
  $$
  \probAll > \max(0.7,0.996-1.4\times10^{-4}(
  \max(5.4-\log_{10}(E/1\rm{MeV}),0)^{5.3})).
  $$
\end{fermienumerate}

As with the \irf{TRANSIENT} event classes, it is worth noting that although the
cuts are optimized somewhat differently in \psix\ a similar series of cuts are
applied to define the \irf{P6\_DIFFUSE} event class, which is roughly
equivalent to the \irf{P7SOURCE} selection.

\subsubsection{\irf{P7CLEAN} Class Selection}\label{subsubsec:p7clean_selection}

For the analysis of diffuse \gammaRayHyph\ emission, we need to reduce
the background contamination to a level of about $\sim 0.1$~Hz across the LAT
\fov, to keep it below the extragalactic \gammaRayHyph\ background
at all energies. For comparison, the total Galactic diffuse contribution is
$\sim 1$~Hz, depending on where the LAT is pointing, though most of that is
localized along the Galactic plane.

The selection of \irf{P7CLEAN} class events starts from the
\irf{P7SOURCE} event class and includes the following additional cuts:
\begin{fermienumerate}
\item the event must pass a series of cuts designed to reject specific
  backgrounds such as CRs that passed through the mounting holes in the ACD,
  or the gaps along the corners of the ACD, with minimal costs to the
  \gammaRayHyph\ efficiency (see \secref{subsec:bkg_samples});
\item the event must pass cuts on the topology of the event in the CAL
  designed to remove hadronic CRs.
\end{fermienumerate}

As with the two previous event classes, a similar series of cuts was applied
to define the \irf{P6\_DATACLEAN} event class, which is roughly
equivalent to the \irf{P7CLEAN} event class.

\subsubsection{\irf{P7ULTRACLEAN} Class Selection}\label{subsubsec:p7ultra_selection}

For the analysis of extragalactic diffuse \gammaRayHyph\ emission we need to
reduce the background contamination even further below the extragalactic
\gammaRayHyph\ background rate to avoid introducing artificial spectral
features. As illustrated in \figref{backgroundcont} the residual contamination
for the \irf{P7ULTRACLEAN} class is $\sim 40\%$ lower than that of the
\irf{P7CLEAN} class around 100~MeV (the residual levels becoming more similar
to each other as the energy increases and becoming the same above 10~GeV).

The selection of \irf{P7ULTRACLEAN} class events is relatively simple,
consisting of a tighter, energy-dependent cut on \probAll:

\begin{align*}
\probAll & > 0.996-0.0394(\max(3.26-log_{10}(E/1~\rm{MeV}),0)^{1.78} \quad
{\rm(Front)}\\
\probAll & > 0.996-0.006(\max(4.0-log_{10}(E/1~\rm{MeV}),0)^{3.0} \quad
{\rm(Back)}.
\end{align*}

\subsection{Publicly Released Data}\label{subsec:LAT_public_data}

At the time of this writing the LAT team has published results and released
data for both the \psix\ and \pseven\ event analyses, as well as several event
classes and the associated IRFs for each iteration of the event analysis.
Furthermore, as our understanding of the LAT has improved, we have updated the
IRFs for particular event classes. \Tabref{public_data} summarizes these data
sets and associated IRFs.
This paper will focus in particular on the performance of the \irf{P7SOURCE}
event class and the validation of the associated \irf{P7SOURCE\_V6} IRFs since 
this is the data set the LAT team currently recommends for most analyses.
Of the other \pseven\ event classes, \irf{P7TRANSIENT}  is recommended for the
analysis of short ($< 1000$~s) transient events such as GRBs, \irf{P7CLEAN} is
recommended for analyses requiring low CR background contamination, such as the
study of large-scale diffuse emission, and \irf{P7ULTRACLEAN} is recommended
when CR background contamination must be minimized, even at the cost of some
loss of effective area, such as when studying the extragalactic background.
Accordingly, all plots, figures and tables will be made with the \irf{P7SOURCE}
event sample and the \irf{P7SOURCE\_V6} set of IRFs unless stated otherwise.   

One other very important point is that the excellent stability of the LAT
subsystems (see \secref{subsec:LAT_instrument}) means that changes in the
instrument performance over time are not a consideration in defining the event
analyses or IRFs. The small changes in performance at the subsystem level are
easily addressed by calibrations applied during the event reconstruction
procedure~\citep{REF:2009.OnOrbitCalib}. Accordingly, to date the LAT team is
able to produce IRFs that are valid for the entire mission. This in turn
simplifies the data analysis task by removing the need to split the LAT data
by time range.

\begin{table}[ht]
  \begin{center}
    \begin{tabular}{llll}
      \hline
      Event Class & \pseven\ IRF Set & \psix\ Counterpart & \psix\ IRF Set\\
      \hline
      \hline
      \irf{P7TRANSIENT} & \irf{P7TRANSIENT\_V6} & \irf{P6\_TRANSIENT} &
      \irf{P6\_V1\_TRANSIENT}\\
      & & & \irf{P6\_V3\_TRANSIENT}\\
      \hline
      \irf{P7SOURCE} & \irf{P7SOURCE\_V6}\tablenotemark{a} & \irf{P6\_DIFFUSE} &
      \irf{P6\_V1\_DIFFUSE}\\
      & & & \irf{P6\_V3\_DIFFUSE}\\
      & & & \irf{P6\_V11\_DIFFUSE}\\
     \hline
      \irf{P7CLEAN} & \irf{P7CLEAN\_V6}\tablenotemark{a} & \irf{P6\_DATACLEAN} & 
      \irf{P6\_V3\_DATACLEAN}\\
     \hline
      \irf{P7ULTRACLEAN} & \irf{P7ULTRACLEAN\_V6} & - & - \\
      \hline
    \end{tabular}
    \caption{Publicly released event selections and IRFs. Note the slight
      change in naming conventions between \psix\ and \pseven. The \pseven\
      naming convention emphasize the point that we may release multiple IRFs
      for the same event class as we improve the IRFs and background
      rejection.}
    \tablenotetext{a}{We have also released \irf{P7SOURCE\_V6MC} and
      \irf{P7CLEAN\_V6MC} IRF sets, which feature a MC-based PSF that
      includes $\theta$~dependence that we have used when we need to 
      minimize the potential of instrument-induced variability (see \secref{sec:PSF}).}      
    \label{tab:public_data}
  \end{center}
\end{table}

\subsection{Calibration Sources and Background Subtraction Methods}\label{subsec:LAT_methods}

Because of the complexity of the LAT and of the physics simulations of
particle interactions we cannot expect the MC to perfectly reproduce the flight data.
For this reason we have developed validation data sets for the IRFs. Although
no astrophysical source has perfectly known properties, in practice there are
several sources for which accurate background subtraction allows to
extract a clean \gammaRayHyph\ sample that we can use to validate the MC
predictions. \Tabref{calibSamples} summarizes these calibration sources and
associated background subtraction techniques and MC samples. The remainder of
the section discusses the particulars of these samples.

\begin{deluxetable}{p{0.15\textheight}p{0.19\textheight}p{0.19\textheight}
    p{0.19\textheight}p{0.19\textheight}}
  \rotate
  \tablewidth{0pt}
  \tablecaption{Summary table of calibration data samples.}
  \tablehead{
    \colhead{} &
    \colhead{Vela Pulsar (\secref{subsubsec:LAT_Vela})} &
    \colhead{AGN (\secref{subsubsec:LAT_AGN})} &
    \colhead{Earth limb (\secref{subsubsec:LAT_earthLimb})} &
    \colhead{Galactic ridge (\secref{subsubsec:LAT_galRidge})}
  }
  \startdata
  Mission Elapsed Time (MET)\acronymlabel{MET} &
  239414402--302486402 &
  239414402--302486402 &
  237783738--239102693, 244395837--244401823 &
  239414402--302486402\\
  Energy range &
  30~MeV--10~GeV &
  1--100~GeV &
  10--100~GeV &
  17~MeV--300~GeV\\
  Selection &
  15$^\circ$ region of interest (\roi\acronymlabel{ROI}) around Vela & 
  6$^\circ$ \roi\ around each of the AGN\tablenotemark{a} &
  Zenith cut & 
  Clean and dirty regions (\secref{subsubsec:LAT_galRidge})\\
  Zenith cut &
  $\theta_z < 100^\circ$\tablenotemark{b} &
  $\theta_z < 100^\circ$ &
  $105^\circ < \theta_z < 120^\circ$ &
  $\theta_z < 100^\circ$\\
  Rocking angle cut\tablenotemark{c} &
  Yes & 
  Yes & 
  No & 
  Yes \\
  Data quality cut\tablenotemark{d} &
  Yes & 
  Yes & 
  Yes & 
  Yes \\
  Bkg. subtraction &
  Phase-gated & 
  Angular separation $\alpha$ to the nearest AGN & 
  Zenith angle $\theta_z$ &
  Galactic latitude\tablenotemark{e} \\
  Signal region &
  \makebox{$\phi \in [0.125,0.175]$},  \makebox{$\phi \in [0.5125,0.6125]$} & 
  \makebox{$\alpha < 0.5^{\circ}$} &
  \makebox{$111.10^{\circ} < \theta_{z} < 112.95^{\circ}$} & 
  Clean region\\
  Background region &
  \makebox{$\phi \in [0.7, 1]$} & 
  \makebox{$3.87288^{\circ} < \alpha < 4^{\circ}$} & 
  \makebox{$108.66^{\circ} < \theta_{z} < 109.57^{\circ}$},
  \makebox{$114.52^{\circ} < \theta_{z} < 115.47^{\circ}$} & 
  Dirty region\\
  MC sample &
  $\theta$-weighted \allgamma &
  $\theta$-weighted \allgamma &
  Limb flux model &
  None\\
  \enddata
  \tablenotetext{a}{The criteria used to select the AGN sample are described
    in~\secref{subsubsec:LAT_AGN}.}
  \tablenotetext{b}{The zenith angle cut for the Vela sample is
    applied over the entire $15^\circ$ \roi.}
  \tablenotetext{c}{Standard rocking angle selection
    \selection{ABS(ROCK\_ANGLE) < 52} in \gtmktime.}
  \tablenotetext{d}{Standard data quality selection
    \selection{DATA\_QUAL == 1 \&\& LAT\_CONFIG == 1} in \gtmktime.}
  \tablenotetext{e}{In the case of the Galactic ridge we cannot subtract
    background accurately enough for detailed quantitative validations;
    however, we can distinguish between regions of higher and lower
    \gammaRayHyph-to-CR ratios.}
  \label{tab:calibSamples}
\end{deluxetable}

\subsubsection{The Vela Pulsar}\label{subsubsec:LAT_Vela}

The Vela pulsar (PSR J0835$-$4510) has the largest integral flux $>100$~MeV 
of any \gammaRayHyph\ source, and has been well studied in the LAT energy range
\citep{REF:2009.VelaI,REF:2010.VelaII}.  
Furthermore, the pulsed nature of high-energy \gammaRayHyph\ emission 
gives us an independent control on the background. 
In fact, off-pulse \gammaRayHyph\ emission is almost entirely absent. These factors combine to
make the Vela pulsar an almost ideal calibration source. Unfortunately, the
spectrum of Vela cuts off at $\sim 3$~GeV and Vela is nearly undetectable above
30~GeV.  

The selection criteria we use to define our Vela calibration samples are listed
in \tabref{calibSamples}. Specific calibration samples used for particular
studies may include additional requirements. For example,
the ``\irf{P7TRANSIENT} Vela calibration sample'' includes all events in the
\irf{P7TRANSIENT} event class that pass the Vela calibration sample criteria.
We used the \filename{TEMPO2} package\footnote{\webpage{http://www.atnf.csiro.au/research/pulsar/tempo2/}} \citep{REF:2006:TEMPO2}
and a pulsar timing model\footnote{\webpage{http://fermi.gsfc.nasa.gov/ssc/data/access/lat/ephems/}} 
derived from data taken with the Parkes radio telescope \citep{REF:2009.VelaI,REF:2010:VelaParkes} 
to assign a phase to each \gammaRay.

\begin{figure}[htbp]
  \centering
  \includegraphics[width=\onecolfigwidth]{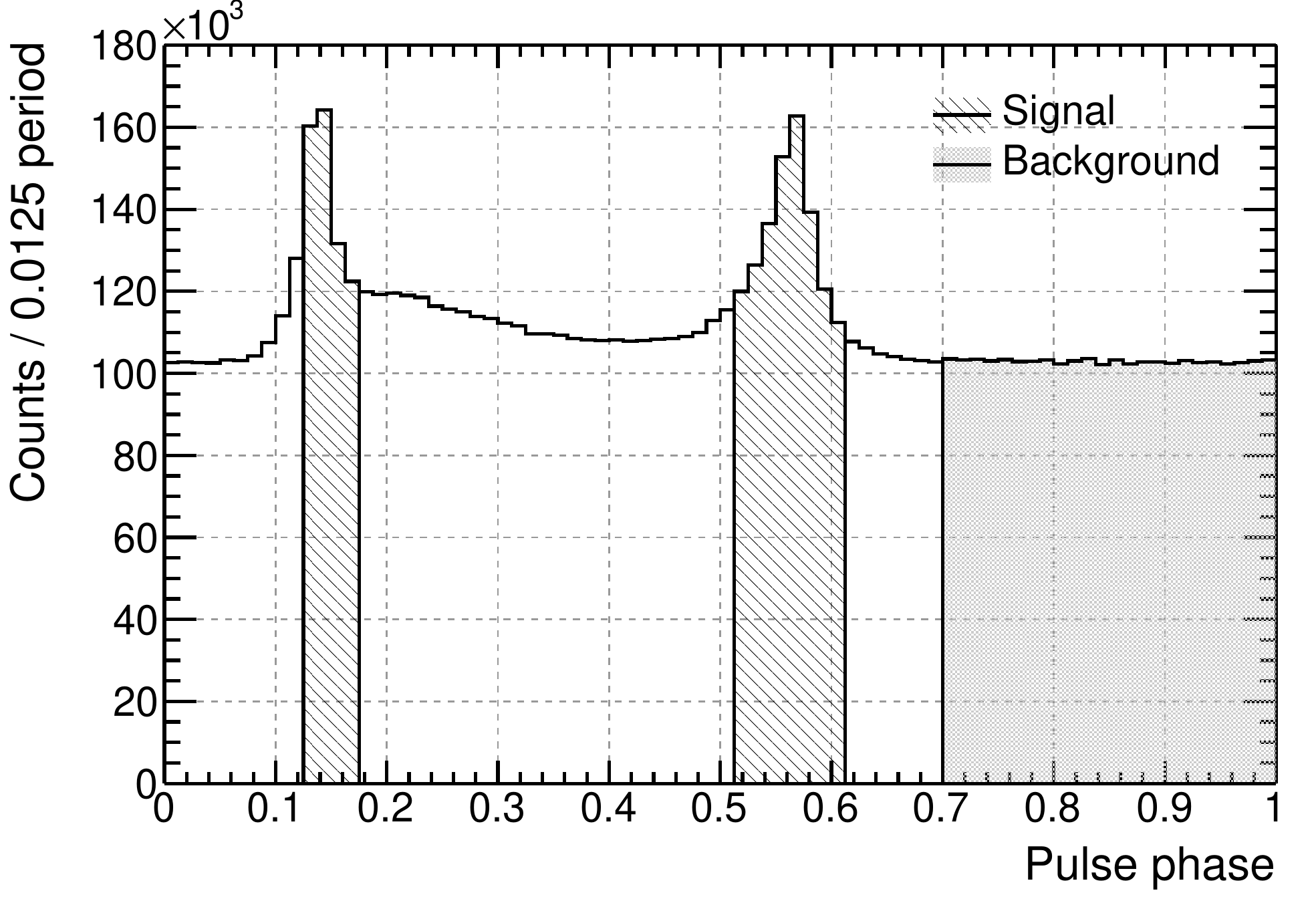}
  \caption{Phase-folded times for events in the \irf{P7TRANSIENT} Vela
    calibration sample, which includes events between 30~MeV and 10~GeV. 
    The signal and background regions, as defined in the
    background subtraction analysis, are highlighted.}
  \label{fig:vela_phase}
\end{figure}

We can achieve excellent statistical background subtraction for any
distribution (i.e., spectrum, spatial distribution, any discriminating
variable used in the event classification) by subtracting the distribution of
off-pulse events (defined as phases in the range  $[0.7,1.0]$) from the
distribution of on-peak events (defined as phases in
$[0.125,0.175]\cup [0.5125,0.6125]$). \Figref{vela_phase} shows the
distribution of phases of \gammaRays\ in the \irf{P7TRANSIENT} Vela calibration
sample, including our standard on-peak and off-pulse regions.
(Note that the PSF analysis in \secref{subsubsec:PSF_oo_pulsars} uses slightly
different definitions of on-peak ($[0.12,0.17]\cup[0.52,0.57]$) and off-pulse
$[0.8,1.0]$ regions; the slight difference in definition causes no significant
change in the results).

Admittedly, all bright pulsars are potential candidates; nonetheless for this
analysis the consequent increase in event statistics does not warrant the
procedural complication required to deal with a stacked sample of pulsars
(see also \secref{subsec:PSF_onorbit}).

\subsubsection{Bright Active Galactic Nuclei}\label{subsubsec:LAT_AGN}

At 1~GeV the 95\% containment radius for the \irf{P7SOURCE} event class is
$\sim 1.4^{\circ}$ for front-converting events and $\sim 2.5^{\circ}$ for
back-converting events (see~\secref{sec:PSF}).
Given the density of bright \gammaRayHyph\ sources in the sky, above this
energy the instrument PSF becomes narrow enough that we can use the angular
distance $\alpha$ between a \gammaRay\ and the nearest celestial
\gammaRayHyph\ source as a good discriminator for background subtraction,
particularly at high Galactic latitudes where there are fewer sources and
the interstellar diffuse emission is less pronounced. Unfortunately, no single
source is bright enough to provide adequate statistics to serve as a good
calibrator. However, by considering \gammaRays\ from a sample of bright and/or
hard spectrum Active Galactic Nuclei (AGN)\acronymlabel{AGN} that are isolated
from other hard sources we can create a good calibration sample for energies
from 1~GeV to 100~GeV.

\begin{figure}[htb!]
  \centering
  \includegraphics[width=\onecolfigwidth]{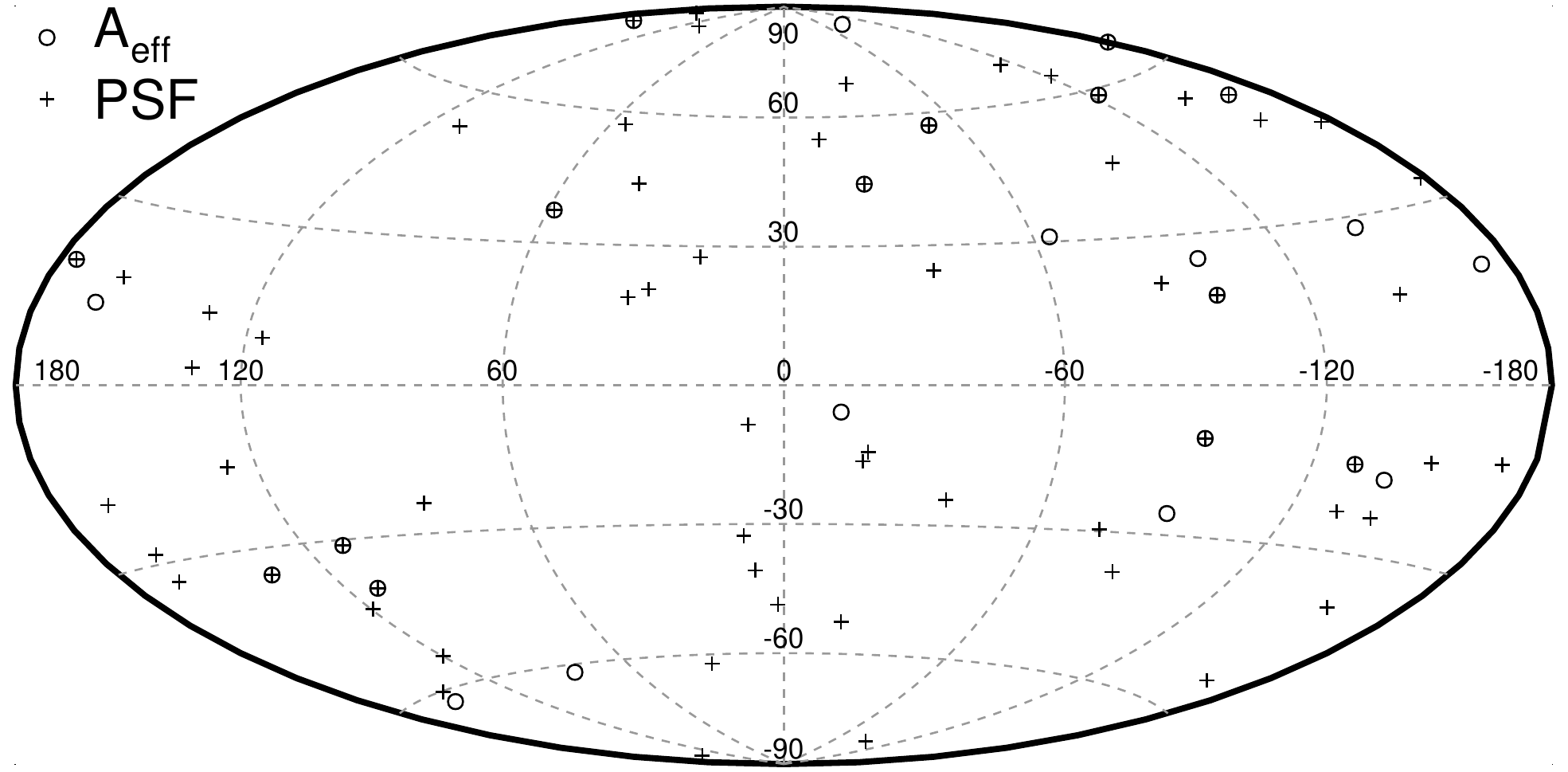}
  \caption{Positions of the AGN in the calibration samples, shown in
    a Hammer-Aitoff projection in Galactic coordinates. Circles mark the AGN
    used for \aeff\ studies and crosses those used in PSF studies.}
  \label{fig:AGN_2FGL_aitoff}
\end{figure}

\begin{figure}[htb!]
  \centering
  \includegraphics[width=\onecolfigwidth]{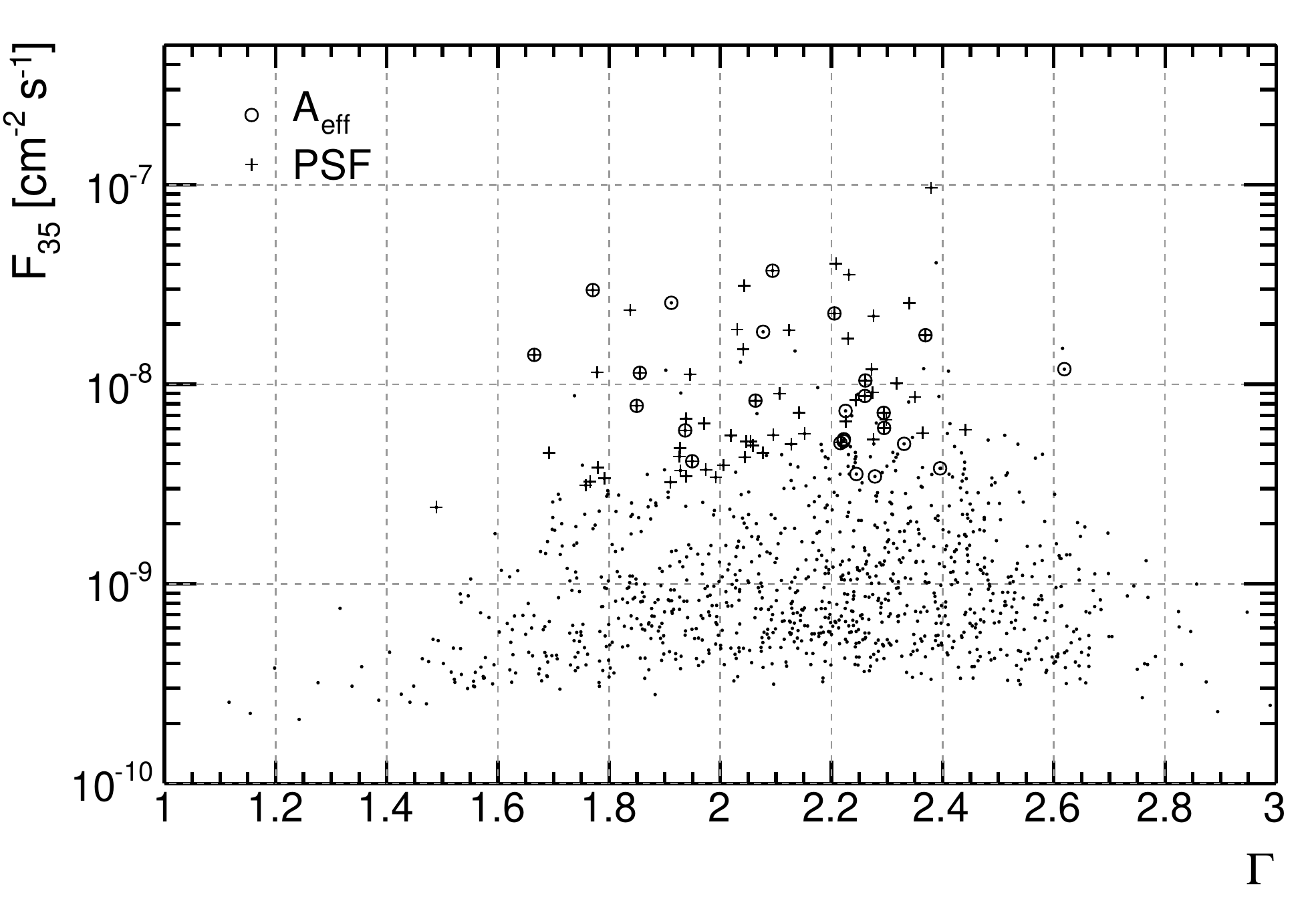}
  \caption{Integral \gammaRayHyph\ fluxes between 1~GeV and
    100~GeV ($F_{35}$) and spectral indices ($\Gamma$) of the AGN in the
    calibration samples. Circles mark the AGN used for \aeff\ studies
    and crosses those used in PSF studies. The dots mark all the other sources
    associated with AGN in the 2FGL catalog.}
  \label{fig:AGN_2FGL_params}
\end{figure}

\tabref{AGN_list} lists the AGN that we use here, \figref{AGN_2FGL_aitoff}
shows their positions, and \figref{AGN_2FGL_params} shows a comparison of their
spectral index ($\Gamma$) and integral \gammaRayHyph\ flux between 1~GeV and
100~GeV ($F_{35}$) to other AGN in the second LAT source
catalog~\citep[2FGL;\acronymlabel{2FGL}][]{REF:2011.2FGL}.
Note that different but overlapping sets of AGN are used for \aeff-related
studies and PSF-related studies. More information about the source properties
can be found in~\citet{REF:2010.1FGL} and~\citet{REF:2011.2FGL}.  

\begin{table}[hp!]
  \begin{center}
    \tiny
    \begin{tabular}{lrrll}
      \hline
      Source & l~[$^\circ$] & b~[$^\circ$] & 2FGL & Used\\
\hline
\hline
KUV 00311$-$1938 & 94.17 & $-$81.21 & J0033.5$-$1921 &PSF\\
PKS 0118$-$272 & 213.58 & $-$83.53 & J0120.4$-$2700 &PSF\\
B3 0133+388 & 132.43 & $-$22.95 & J0136.5+3905 &PSF\\
OC 457 & 130.79 & $-$14.32 & J0136.9+4751 &\aeff\ PSF\\
PKS 0208$-$512 & 276.12 & $-$61.76 & J0210.7$-$5102 &\aeff\\
PKS 0215+015 & 162.20 & $-$54.41 & J0217.9+0143 &PSF\\
S4 0218+35 & 142.60 & $-$23.49 & J0221.0+3555 &PSF\\
3C 66A & 140.14 & $-$16.77 & J0222.6+4302 &\aeff\\
AO 0235+164 & 156.78 & $-$39.10 & J0238.7+1637 &PSF\\
PKS 0250$-$225 & 209.72 & $-$62.10 & J0252.7$-$2218 &\aeff\\
PKS 0301$-$243 & 214.64 & $-$60.17 & J0303.4$-$2407 &PSF\\
NGC 1275 & 150.59 & $-$13.25 & J0319.8+4130 &PSF\\
PMN J0334$-$3725 & 240.22 & $-$54.36 & J0334.3$-$3728 &PSF\\
PKS 0420$-$01 & 195.28 & $-$33.15 & J0423.2$-$0120 &PSF\\
PKS 0426$-$380 & 240.70 & $-$43.62 & J0428.6$-$3756 &PSF\\
MG2 J043337+2905 & 170.52 & $-$12.62 & J0433.5+2905 &PSF\\
PKS 0440$-$00 & 197.21 & $-$28.44 & J0442.7$-$0017 &PSF\\
PKS 0447$-$439 & 248.81 & $-$39.91 & J0449.4$-$4350 &\aeff\ PSF\\
PKS 0454$-$234 & 223.73 & $-$34.90 & J0457.0$-$2325 &\aeff\ PSF\\
1ES 0502+675 & 143.80 & 15.90 & J0508.0+6737 &PSF\\
TXS 0506+056 & 195.40 & $-$19.62 & J0509.4+0542 &PSF\\
PKS 0537$-$441 & 250.08 & $-$31.09 & J0538.8$-$4405 &\aeff\ PSF\\
TXS 0628$-$240 & 232.68 & $-$15.00 & J0630.9$-$2406 &PSF\\
B3 0650+453 & 171.20 & 19.36 & J0654.2+4514 &\aeff\\
PKS 0700$-$661 & 276.77 & $-$23.78 & J0700.3$-$6611 &PSF\\
MG2 J071354+1934 & 197.68 & 13.61 & J0714.0+1933 &\aeff\\
B2 0716+33 & 185.06 & 19.85 & J0719.3+3306 &\aeff\ PSF\\
S5 0716+71 & 143.97 & 28.02 & J0721.9+7120 &\aeff\\
PKS 0727$-$11 & 227.77 & 3.13 & J0730.2$-$1141 &PSF\\
PKS 0735+17 & 201.85 & 18.06 & J0738.0+1742 &PSF\\
PKS 0805$-$07 & 229.04 & 13.16 & J0808.2$-$0750 &PSF\\
S4 0814+42 & 178.21 & 33.41 & J0818.2+4223 &PSF\\
PKS 0823$-$223 & 243.97 & 8.92 & J0825.9$-$2229 &PSF\\
S4 0917+44 & 175.70 & 44.81 & J0920.9+4441 &PSF\\
4C +55.17 & 158.59 & 47.94 & J0957.7+5522 &PSF\\
1H 1013+498 & 165.53 & 52.73 & J1015.1+4925 &\aeff\ PSF\\
4C +01.28 & 251.50 & 52.77 & J1058.4+0133 &PSF\\
TXS 1055+567 & 149.57 & 54.42 & J1058.6+5628 &PSF\\
Mkn 421 & 179.82 & 65.03 & J1104.4+3812 &\aeff\ PSF\\
Ton 599 & 199.41 & 78.37 & J1159.5+2914 &\aeff\ PSF\\
W Comae & 201.69 & 83.28 & J1221.4+2814 &PSF\\
4C +21.35 & 255.07 & 81.66 & J1224.9+2122 &PSF\\
PKS 1244$-$255 & 301.60 & 37.08 & J1246.7$-$2546 &\aeff\ PSF\\
PG 1246+586 & 123.74 & 58.77 & J1248.2+5820 &\aeff\ PSF\\
S4 1250+53 & 122.36 & 64.08 & J1253.1+5302 &PSF\\
3C 279 & 305.10 & 57.06 & J1256.1$-$0547 &PSF\\
OP 313 & 85.59 & 83.29 & J1310.6+3222 &\aeff\\
GB 1310+487 & 113.32 & 68.25 & J1312.8+4828 &PSF\\
PMN J1344$-$1723 & 320.48 & 43.67 & J1344.2$-$1723 &PSF\\
PKS 1424+240 & 29.48 & 68.20 & J1427.0+2347 &PSF\\
PKS 1440$-$389 & 325.64 & 18.72 & J1443.9$-$3908 &PSF\\
PKS 1454$-$354 & 329.89 & 20.52 & J1457.4$-$3540 &PSF\\
PKS 1502+106 & 11.37 & 54.58 & J1504.3+1029 &PSF\\
AP Librae & 340.70 & 27.58 & J1517.7$-$2421 &PSF\\
B2 1520+31 & 50.18 & 57.02 & J1522.1+3144 &\aeff\ PSF\\
GB6 J1542+6129 & 95.38 & 45.40 & J1542.9+6129 &PSF\\
PG 1553+113 & 21.92 & 43.95 & J1555.7+1111 &\aeff\ PSF\\
1H 1720+117 & 34.11 & 24.47 & J1725.0+1151 &PSF\\
B2 1732+38A & 64.04 & 31.02 & J1734.3+3858 &\aeff\\
PMN J1802$-$3940 & 352.44 & $-$8.42 & J1802.6$-$3940 &PSF\\
PKS 1830$-$211 & 12.15 & $-$5.72 & J1833.6$-$2104 &\aeff\\
S4 1849+67 & 97.50 & 25.03 & J1849.4+6706 &\aeff\\
TXS 1902+556 & 85.96 & 20.51 & J1903.3+5539 &PSF\\
1H 1914$-$194 & 18.24 & $-$14.30 & J1917.6$-$1921 &PSF\\
TXS 1920$-$211 & 17.17 & $-$16.26 & J1923.5$-$2105 &PSF\\
1ES 1959+650 & 98.02 & 17.67 & J2000.0+6509 &\aeff\ PSF\\
PKS 2005$-$489 & 350.37 & $-$32.61 & J2009.5$-$4850 &PSF\\
PKS 2023$-$07 & 36.89 & $-$24.39 & J2025.6$-$0736 &PSF\\
PKS 2052$-$47 & 352.58 & $-$40.38 & J2056.2$-$4715 &PSF\\
MH 2136$-$428 & 358.29 & $-$48.32 & J2139.3$-$4236 &PSF\\
PKS 2155$-$304 & 17.74 & $-$52.24 & J2158.8$-$3013 &PSF\\
BL Lacertae & 92.60 & $-$10.46 & J2202.8+4216 &\aeff\ PSF\\
PKS 2201+171 & 75.68 & $-$29.63 & J2203.4+1726 &PSF\\
B2 2234+28A & 90.12 & $-$25.66 & J2236.4+2828 &\aeff\\
3C 454.3 & 86.12 & $-$38.18 & J2253.9+1609 &PSF\\
PKS 2326$-$502 & 332.00 & $-$62.30 & J2329.2$-$4956 &PSF\\

      \hline
    \end{tabular}
    \caption{List of the Active Galactic Nuclei used in the AGN calibration
      samples for \aeff\ and PSF studies.}
    \label{tab:AGN_list}
  \end{center}
\end{table}

The selection criteria we use to define AGN calibration samples are
listed in \tabref{calibSamples}. As with the Vela calibration samples, specific
calibration samples used for particular studies may include additional
requirements (e.g., the \irf{P7TRANSIENT} AGN calibration sample).

\begin{figure}[htb!]
  \centering
  \includegraphics[width=\onecolfigwidth]{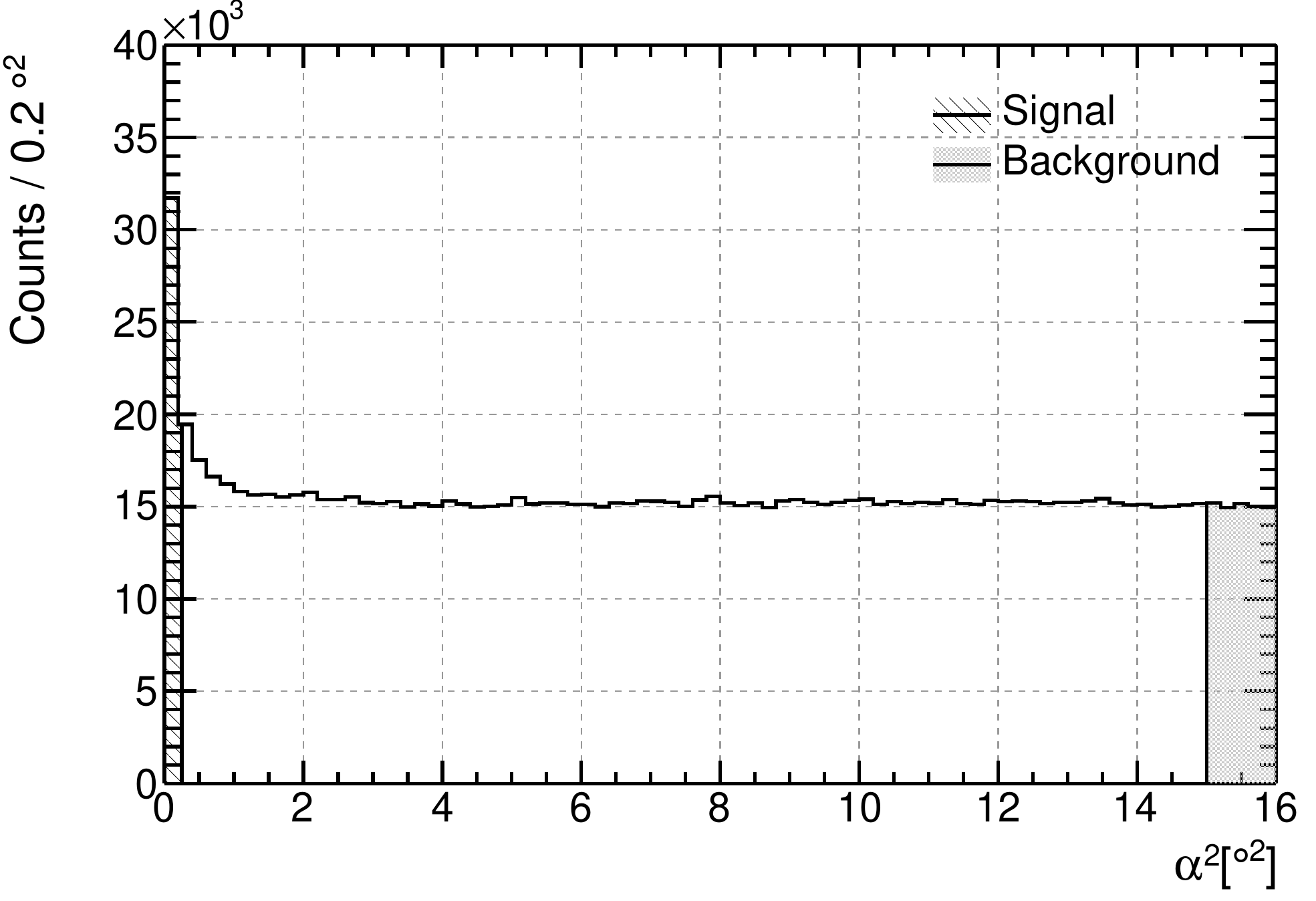}
  \caption{Square of the angular separation between reconstructed
    \gammaRayHyph\ directions and the AGN for events in the \irf{P7TRANSIENT}
    AGN calibration sample, which includes \gammaRays\ in the energy range
    1~GeV to 100~GeV. The signal and background regions, as defined in the
    background subtraction analysis, are highlighted.} 
  \label{fig:AGN_angDist}
\end{figure}

To use this calibration sample to perform background subtraction we define 
signal and background regions in terms of the angular distance $\alpha$ between
the \gammaRay\ and the closest AGN.
Specifically, we use $\alpha < 0.5^{\circ}$ for the signal region and the
annulus $3.87288^{\circ} < \alpha < 4^{\circ}$ for the background region. These
ranges are chosen so that the background region contains four times the
solid angle of the signal region. We adopted the separation between the
signal and background regions in order to minimize signal \gammaRays\
from the tails of the PSF leaking into the  background region. 
\Figref{AGN_angDist} shows the squares of the angular separations between 
\gammaRays\ and the nearest bright AGN for all \gammaRays\ in the
\irf{P7TRANSIENT} event class in $6^\circ$ regions around the 25 AGN listed as
the \aeff\ calibration sample in \tabref{AGN_list}, including
our definitions of source and background regions.

\subsubsection{The Earth Limb}\label{subsubsec:LAT_earthLimb}

The Earth's atmosphere is a very bright \gammaRayHyph\ source. Furthermore,
at energies above a few GeV the \gammaRayHyph\ flux seen by the LAT is
dominated by \gammaRays\ from the interactions of primary CR protons with the
upper atmosphere. This consideration, together with the narrowness of the PSF at
energies $> 10$~GeV, causes the Earth limb to appear as a very bright and
sharp feature, which provides an excellent calibration source.  Furthermore, we 
have selected data from 200 orbits during which the LAT was pointed near the
Earth limb as the basis of the Earth limb calibration sample. The selection
criteria we use to define the Earth limb calibration samples are listed
in~\tabref{calibSamples}. When using the Earth limb as a calibration 
source we generally limit the energy range to energies $> 10$~GeV, primarily because 
at lower energies orbital variations in the geomagnetic field significantly
affect the \gammaRayHyph\ fluxes
(however see \secref{subsec:EDisp_spectralFeatures} for a counter-example). 

For this calibration source, we define our signal region as
$111.1002^{\circ} < \theta_{z} < 112.9545^{\circ}$ and background regions just
above and below the limb: $108.6629^{\circ} < \theta_{z} < 109.5725^{\circ}$ and
$114.5193^{\circ} < \theta_{z} < 115.4675^{\circ}$. Note that these ranges are
defined to give the same solid angle ($0.06 \pi$~sr) in the signal and
background regions. 

\begin{figure}[htbp]
  \centering
  \includegraphics[width=\onecolfigwidth]{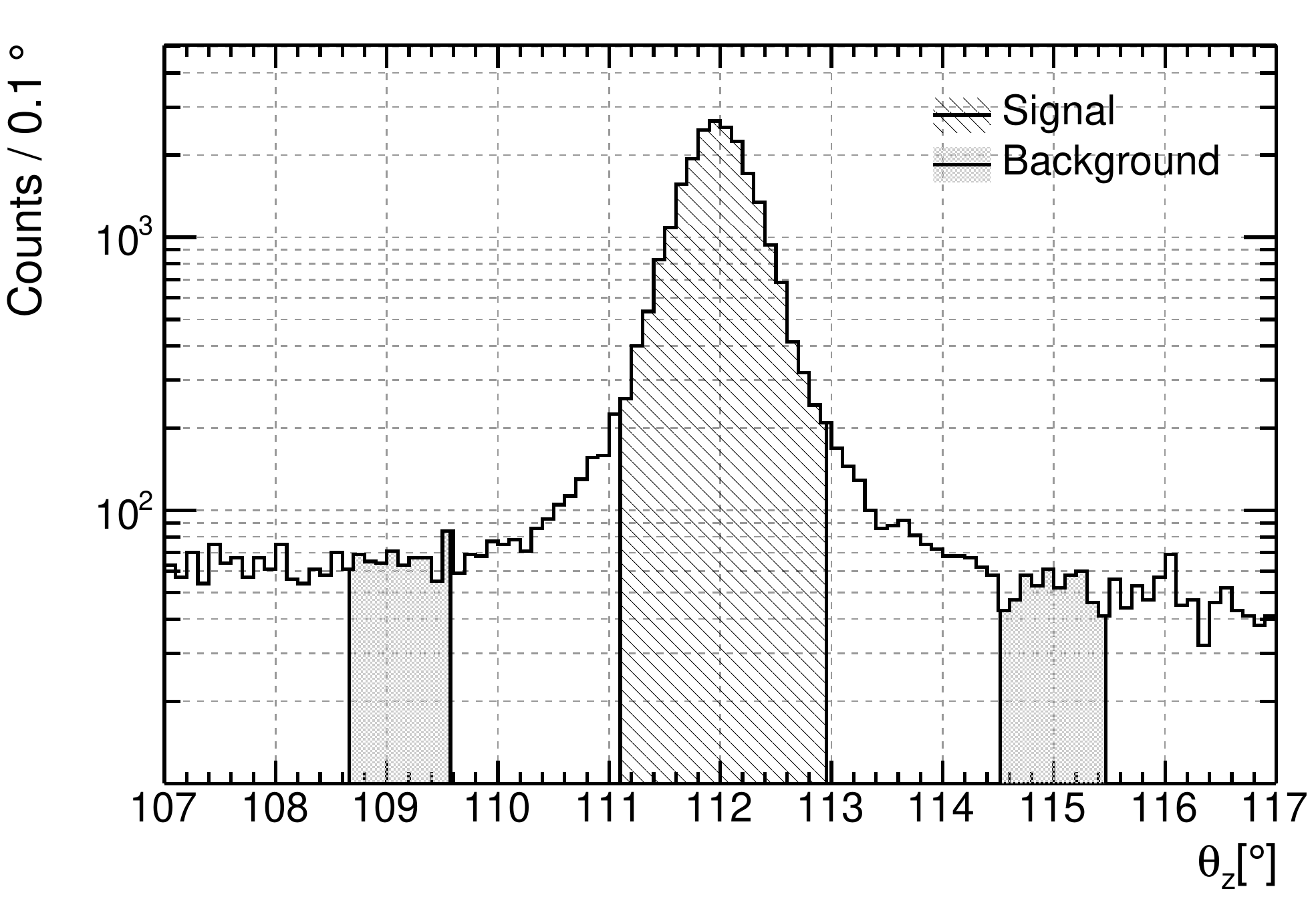}
  \caption{Angle with respect to the zenith for events in the
    \irf{P7TRANSIENT} Earth limb calibration sample, which includes 
    \gammaRays\ with energies above 10~GeV. The signal and background
    regions, as defined in the background subtraction analysis, are
    highlighted.} 
  \label{fig:zenithTheta}
\end{figure}

\Figref{zenithTheta} shows the zenith angle distribution 
for all \gammaRays\ in the \irf{P7TRANSIENT} event class for the Earth limb 
calibration sample.

\subsubsection{The Galactic Ridge}\label{subsubsec:LAT_galRidge}

At energies above $\sim 30$~GeV no single source provides enough \gammaRays\ for
a good comparison between flight data and MC. However, the combination
of bright Galactic sources and Galactic diffuse backgrounds means that there
is a very large excess of \gammaRays\ coming from the Galactic plane relative
to high Galactic latitudes.  

The intensity of the \gammaRayHyph\ emission at low latitudes in the inner
Galaxy is  more than an order of magnitude greater than at high latitudes in
the outer Galaxy.
In contrast, the intensity of the CR background is approximately isotropic for observation periods 
longer than the $53.4$-day orbital precession period. 

Unfortunately, since the Galaxy extends over much of the sky, and since the data 
set consists of several thousand orbits it is not practical to disentangle 
the variations of exposure from the spatial variations in Galactic diffuse emission without
relying on detailed modeling of the Galactic diffuse emission. Accordingly,
we use the Galactic ridge primarily when we are developing our event
selections, rather than for precise calibration of the LAT performance.

Specifically, we developed the event classes that require a high \gammaRayHyph\ 
purity, i.e., the event classes used in the analysis of celestial 
point sources and diffuse emission (see \secref{subsec:event_photon_classes}), 
in part by tuning our selection criteria to maximize the contrast between regions in 
the bright Galactic plane and at high Galactic latitudes.  This helped to mitigate the 
risk that insufficient statistics of the MC training sample or limited 
accuracy of the MC description of the geometry of the detector and the particle 
interactions in the LAT limited the discriminating power and accuracy of the 
event classification analysis.

\begin{figure}[htb!]
  \centering
  \includegraphics[width=\onecolfigwidth]{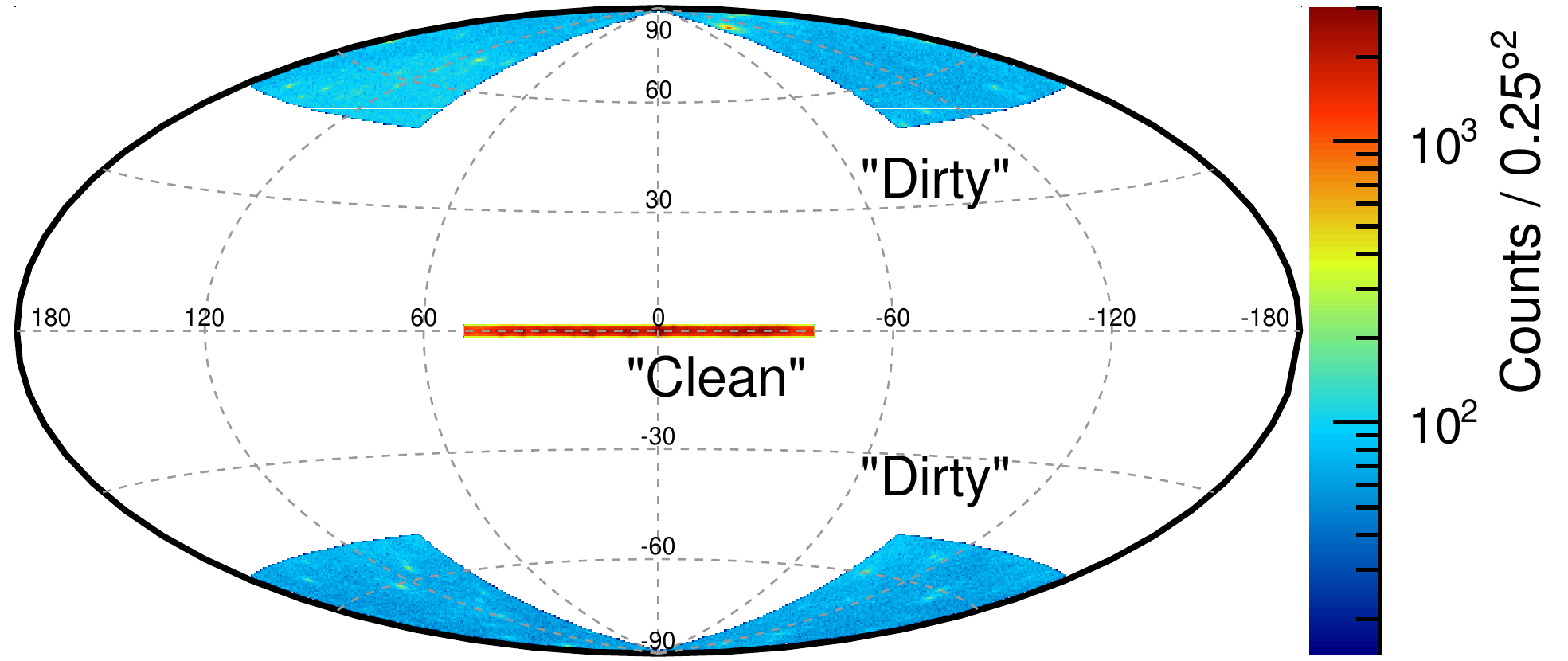}
  \caption{Definitions of the clean and dirty regions, showing the counts in
    both regions in a Hammer-Aitoff projection. This figure uses the data in
    the \irf{P7SOURCE} Galactic ridge calibration sample.}
  \label{fig:cleanDirtyAitoff}
\end{figure}

The selection criteria we use to define the Galactic ridge calibrations
samples are listed in \tabref{calibSamples}.  In particular, we use the 
region ($|b|<1.5^{\circ}$ , $-40^{\circ} < l < 50^{\circ}$, which was
selected to maximize the total \gammaRayHyph\ flux) to define a \emph{clean} 
data sample and the region ($|b|>50^{\circ}$,  $90^{\circ} < l < 270^{\circ}$) to 
define a \emph{dirty} data sample. The ratio of \gammaRays\ to CR background in
the clean region is more than an order of magnitude higher than the same ratio
in the dirty region. \Figref{cleanDirtyAitoff} shows the count maps for the
\irf{P7SOURCE} samples for both regions.  

\begin{figure}[htb!]
  \centering
  \includegraphics[width=\onecolfigwidth]{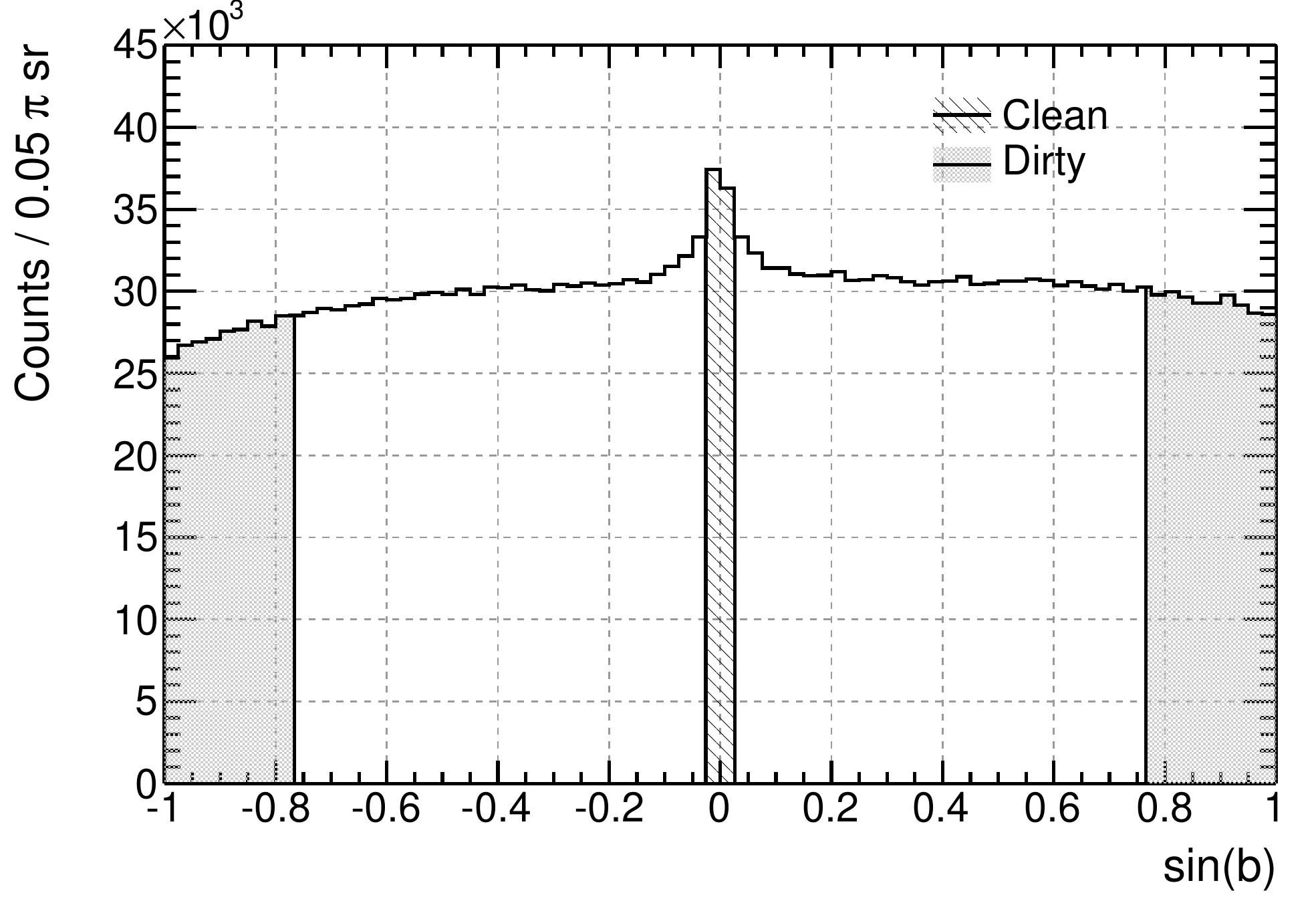}
  \caption{Sine of Galactic latitude for events above 17783~MeV in the
    \irf{P7TRANSIENT} Galactic ridge calibration sample. The Galactic
    latitude selections for the clean and dirty regions are highlighted.
    Note that the definition of the clean and dirty regions also include
    selections on Galactic longitude.}
  \label{fig:galLat}
\end{figure}

Furthermore, to give a sense of the statistics of these samples at high
energies, \Figref{galLat} shows the Galactic latitude distribution for all
\gammaRays\ in the \irf{P7TRANSIENT} event class with energies above 17783~MeV
(i.e., $\loge > 4.25$).

\subsubsection{Summary of Astrophysical Calibration Sources}

As we will see in the next sections, the IRFs depend heavily on $\theta$ (and,
to a lesser extent, on $\phi$). Therefore, any detailed comparison between
flight data and MC simulations must account for the distribution of observing
profile, particularly $\tobs(\theta)$. How best to account for the observing profile
depends on the particulars of the calibration samples.

\begin{figure}[htbp]
  \centering
  \includegraphics[width=\onecolfigwidth]{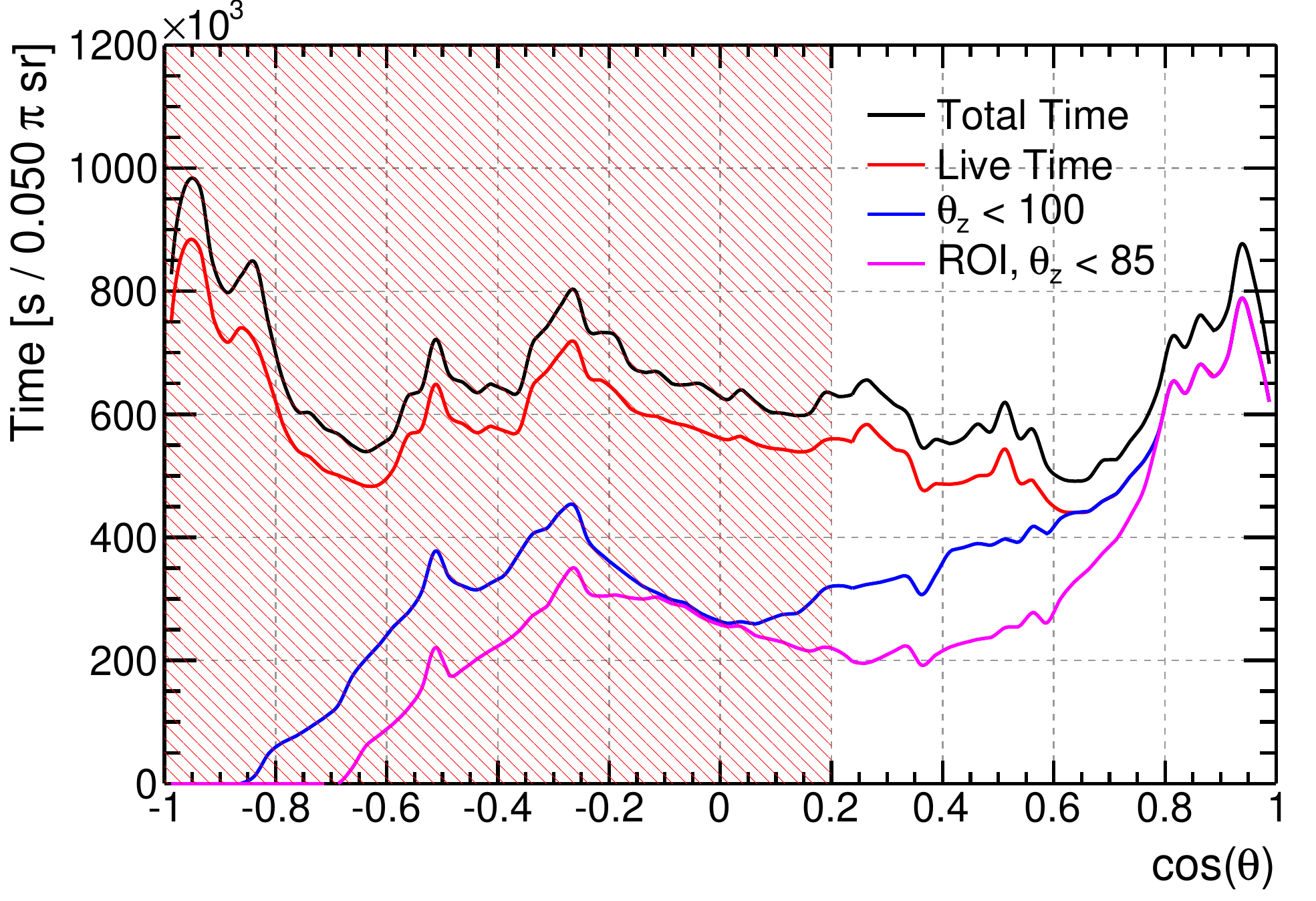}
  \caption{Vela observing profile: starting from the top, the curves show the
    accumulated time as a function of $\cos{\theta}$ for the first two years of
    the mission, the accumulated observing time (accounting for the dead time
    when the LAT triggers), the time during which Vela was less than $100^\circ$
    from the zenith, and the time during which the entire $15^\circ$
    \roi\ around Vela was less than $100^\circ$ from
    the zenith (or, equivalently, that Vela was less than $85^\circ$ from the
    zenith).
    The shaded region corresponds to the area outside the LAT \fov.}
  \label{fig:vela_obs}
\end{figure}

For any point source, the observing profile is determined by the position of
the source, the rocking angle of the LAT and the amount of time spent in survey
mode relative to pointed observations. \Figref{vela_obs} shows the observing
profile for Vela for the first two years of the mission.  Rather than produce a dedicated 
large statistics MC sample for Vela, we re-use our \allgamma\ MC sample, re-weighting the
events in that sample so as to match the Vela observing profile.   

\begin{figure}[htb!]
  \centering
  \includegraphics[width=\onecolfigwidth]{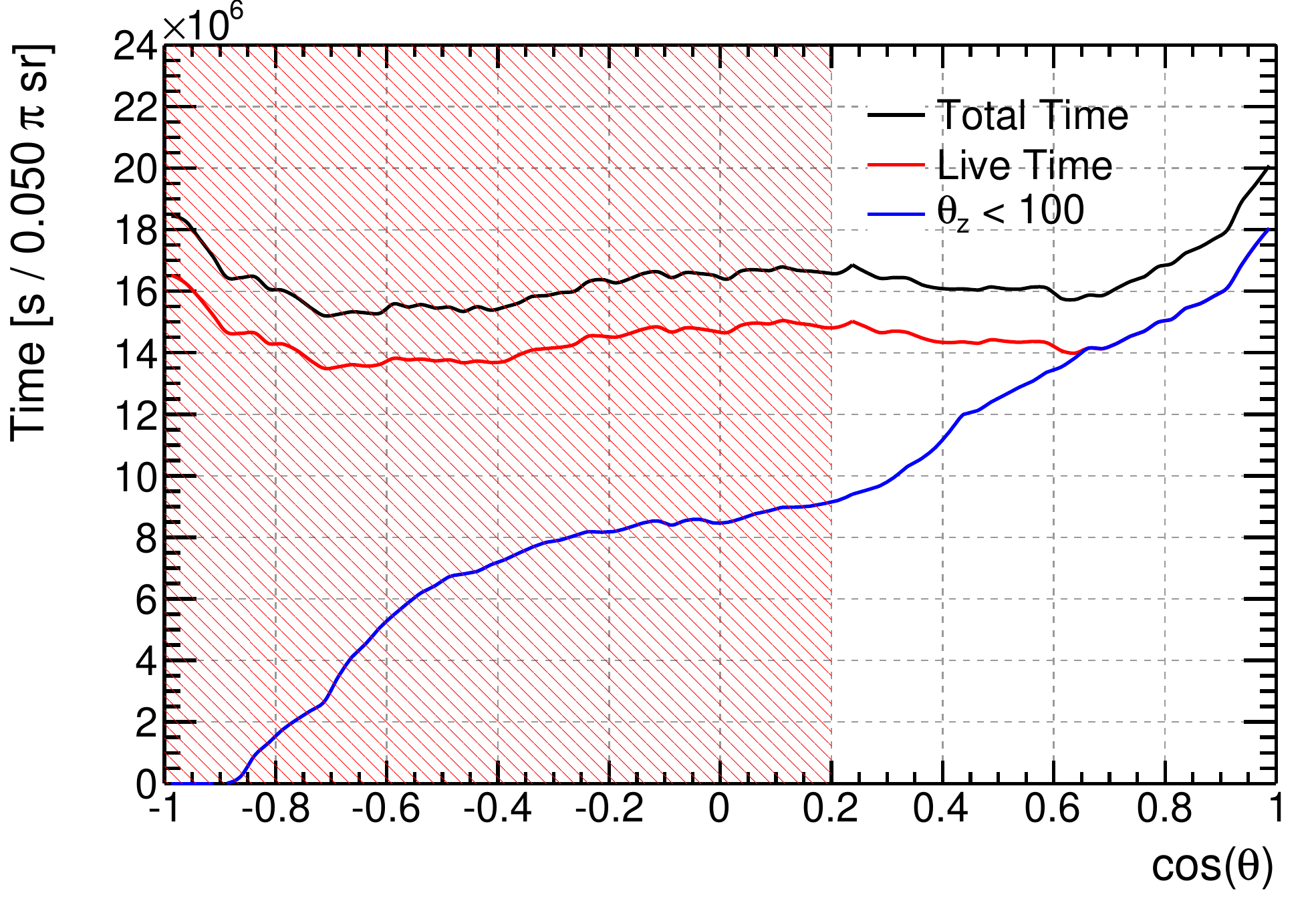}
  \caption{Sum of the observing profiles for the AGN sample: starting from the
    top, the curves show the sum of the accumulated time as a function of
    $\cos{\theta}$ for the first two years of the mission, the sum of the
    accumulated live time, and the sum of the accumulated live times during
    which each AGN was less than $100^{\circ}$ from the zenith. The shaded
    region corresponds to the area outside the LAT \fov.} 
  \label{fig:AGN_obs}
\end{figure}

Similarly, we re-weight the \allgamma\ MC 
to match the summed observing profiles of all of the AGN our sample, which is
shown in \figref{AGN_obs}. Unfortunately, since AGN are intrinsically variable,
and since the AGN in this sample span a range of fluxes, this re-weighting
technique will not work as well with this sample. On the other hand, by taking
a large set of AGN, we reduce the bias due to the variability of any one
particular source. In broad terms, our re-weighted MC sample reproduced the
$\theta$ distribution of the AGN sample to better than 2\%
(see \secref{subsec:Aeff_consistency}).  Finally, we note that since the PSF 
is narrower above $1$~GeV, and the \roi\ around each AGN is only $6^{\circ}$ 
we do not apply the \roi-based $\theta_{z}$ cut when building the AGN calibration 
samples.

Since the Earth limb is a spatially extended source, we cannot apply the
re-weighting technique we used for the Vela and AGN samples to account for
the observing profile. On the other hand, since the data set consists of only
200 orbits, and the Earth limb emission is well understood above 10~GeV we can
produce a MC simulation of the Earth limb emission for those orbits and
compare it with the flight data (see \secref{subsec:simul_sources}).

\begin{figure}[htb!]
  \centering
  \includegraphics[width=\onecolfigwidth]{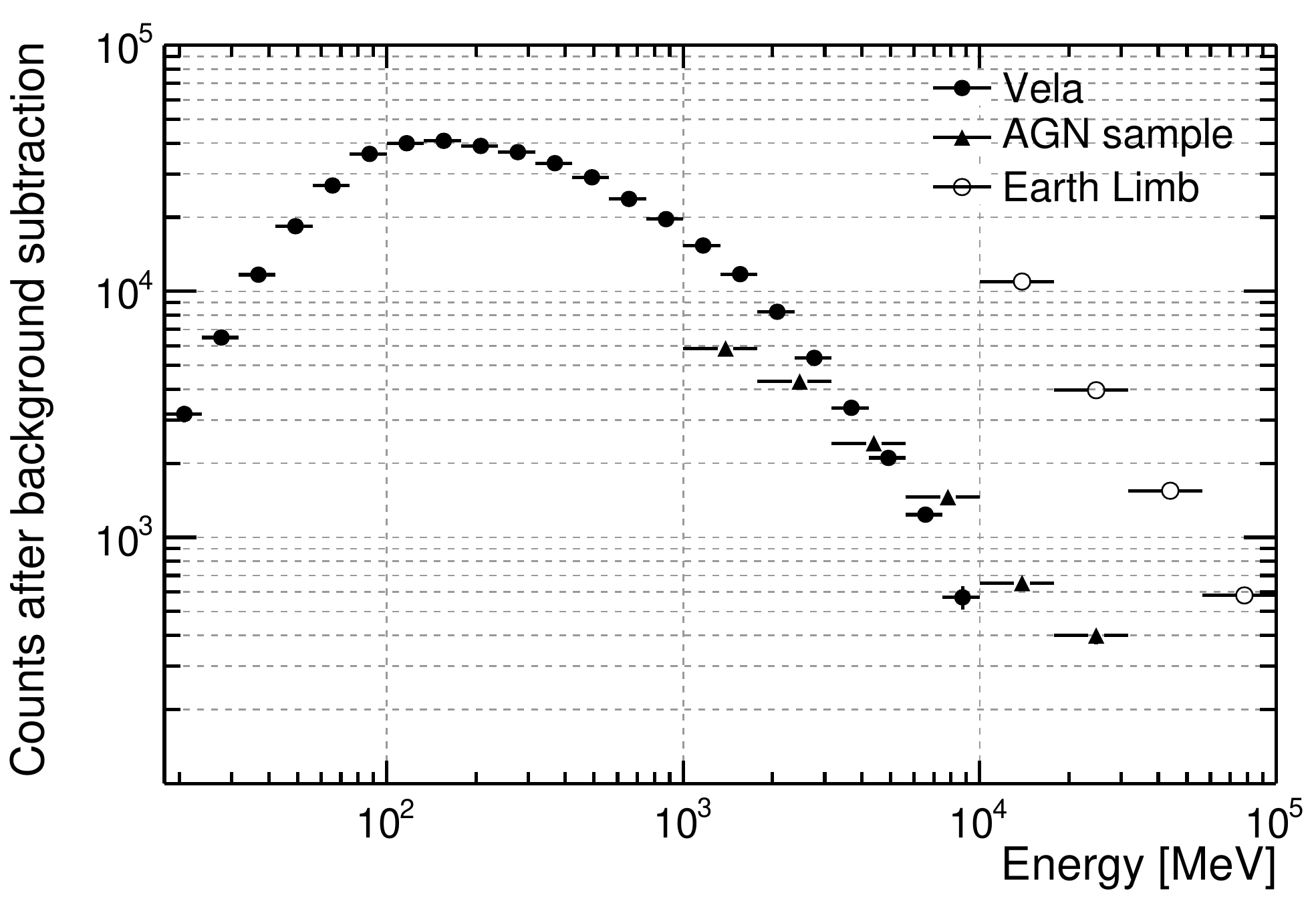}
  \caption{Excess counts in the signal regions as a function of energy
    for Vela, stacked AGN, and Earth limb calibration samples for the
    \irf{P7TRANSIENT} event class. The horizontal error bars indicate
    the energy binning used with the different samples.}
  \label{fig:method_stats}
\end{figure}

Finally, \figref{method_stats} shows the statistics available for each of the
samples. This shows that the calibration sources span most of the LAT energy
range, certainly from 30~MeV up to at least 100~GeV.  

\clearpage

\section{BACKGROUND CONTAMINATION}\label{sec:bkg}

In this section we discuss the residual particle backgrounds, methods to
estimate the contamination of LAT \gammaRayHyph\ samples by these backgrounds
and how to treat such contamination in high-level analysis.
In this context we define \emph{particle backgrounds} as all events that are
classified as \gammaRays\ in a given LAT event class but originate from
CRs or the interactions of CRs in the Earth's atmosphere. Therefore, the
particle backgrounds include both charged and neutral particles---including
secondary \gammaRays.

The LAT background model is described in detail in \secref{subsec:simul_bkgnd}. 
We focus here on describing the particle background contamination in the
high-purity event classes, i.e., the ones used for single source, source
population and diffuse emission analysis (\irf{P7SOURCE} and above). We also
focus on average background contamination for long (few months or longer)
observation periods. The CR-induced particle background is extremely variable
throughout the orbit of the LAT; therefore estimates of particle backgrounds
for brief transient sources must be derived from a dedicated analysis of the
periods of interest \citep[e.g., as done in][]{REF:GRB080825C:2009}.

\subsection{Residual Background Contamination in Monte Carlo Simulations}\label{subsec:bkg_MonteCarlo}

We can estimate the residual background in the various event classes by
propagating the LAT background model (see \secref{subsec:simul_bkgnd}) through
the full \geant-based LAT detector simulation chain and applying the event
classification analysis on the simulated data (see \secref{subsec:LAT_ORBIT},
\secref{subsec:LAT_recon} and \secref{subsec:LAT_simul}).

In comparison to the pre-launch particle background model shown in
\figref{lat_bg_model} we have implemented substantial improvements in our model
of the primary CR protons and electrons.
The effects of the geomagnetic cutoff on the directional and energy dependence
of the primary CR flux in the pre-launch were based on a dipole approximation
of the geomagnetic field. Currently, we simulate an isotropic flux of CRs, and
trace the trajectory of each particle backward through the geomagnetic
magnetic field, eliminating any particles that intersect the Earth or lower
atmosphere. We are using the current version of the International Geomagnetic
Reference Field (IGRF-11), a high order multipole model of the geomagnetic
field~\citep{REF:IGRF11}, and the publicly available trajectory tracing code
described in~\citet{REF:SmartShea}.
We oversample the primary proton and electron spectrum at high energies
to obtain sufficient statistics up to $\sim 600$~GeV in reconstructed energy, 
and obtain rate predictions by appropriately deweighting events at the oversampled 
energies.

A total of $2.2 \times 10^{12}$ primary protons and $1.6 \times 10^{8}$ primary
electrons were generated. In addition, the equivalent of 80~ks ($\sim
1$~day) of instrument live time of background events from CR secondaries produced in the Earth
atmosphere were simulated according to the spatial and spectral distributions
in the pre-launch LAT particle background model (i.e., with no trajectory
tracing). This intensive simulation effort was used to determine the CR-induced
background between 100~MeV and $\sim 600$~GeV in the \irf{P7SOURCE},
\irf{P7CLEAN}, and \irf{P7ULTRACLEAN} event classes. 

However, these simulations still have important shortcomings. For example, the modeling of
inelastic interactions of alpha particles and heavier ions does not match our
observations (see further discussion in \secref{subsec:bkg_samples}, especially 
\parenfigref{cleandirty}~b). The particle 
background model for secondaries produced in CR
interactions in the atmosphere is derived from measurements from satellites
with different orbits and during different parts of the Solar cycle, and has
been projected to the LAT orbit. Furthermore, due to our required background
suppression factors of up to $10^{6}$, even small inaccuracies in our
simulation of particle interactions with the LAT can potentially lead to large
discrepancies between predicted and actual charged particle background rates.

As a consequence we take several measures to account for any such shortcomings
and minimize possible discrepancies. First, we compare a region on the sky with
a high \gammaRay\ to residual background ratio to a region with a low
\gammaRay\ to residual background ratio to isolate those contributions caused
by the accumulation of background in the latter data set; and we compare those
contributions to the predictions from the background simulations
(see \secref{subsec:bkg_samples}). 
Second, we compare several key event parameters between data and simulation for
high-purity \gammaRayHyph\ samples to crosscheck the agreement between data and
simulation, and we slightly adjust the intensities of our CR background model
based on a fit of the shapes of two key event classification variables with
signal and background component (see \secref{subsec:bkg_distrib}).

\subsection{Estimating and Reducing Residual Background Using Test Samples}\label{subsec:bkg_samples}

\threepanel{htb}{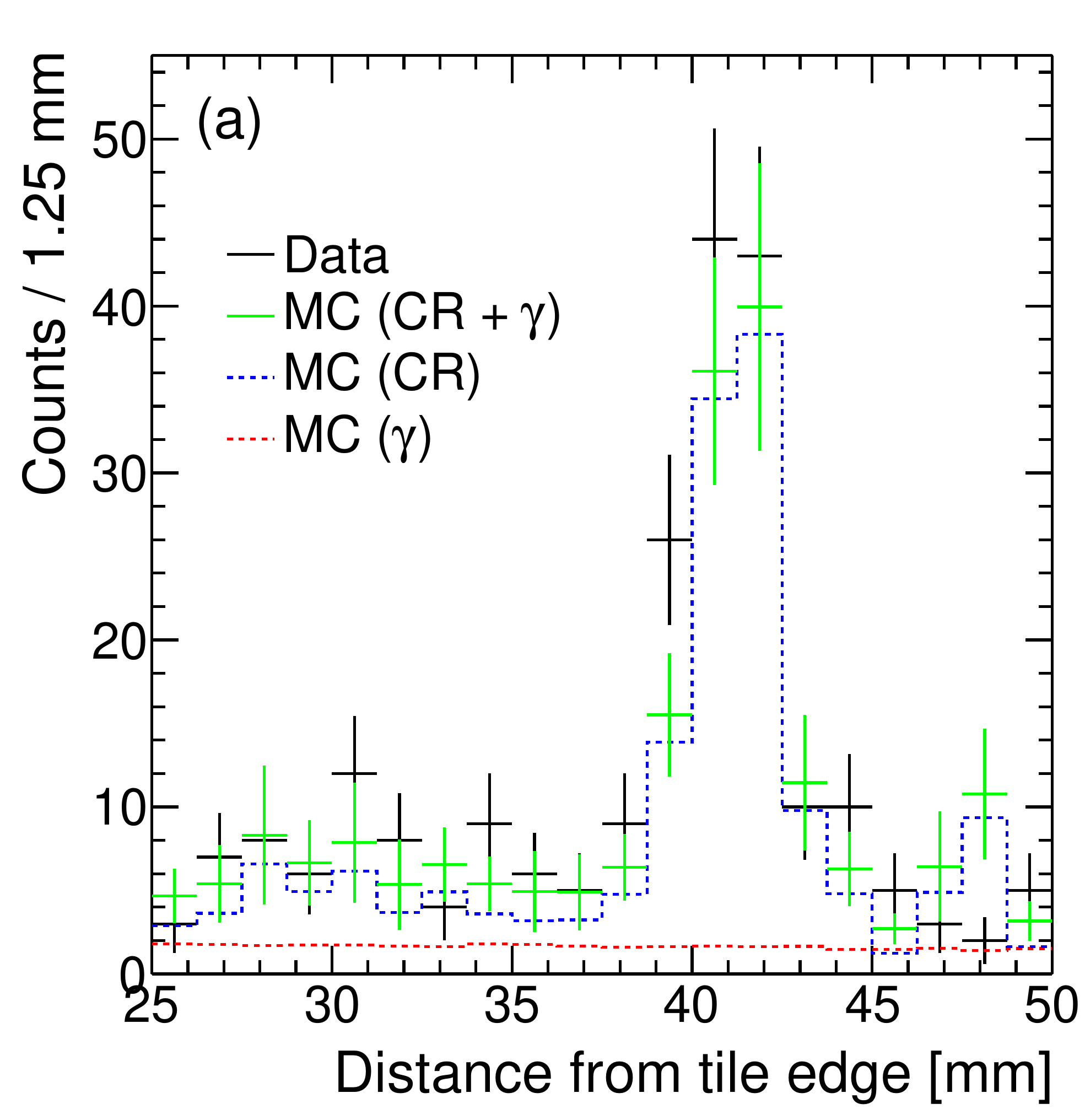}{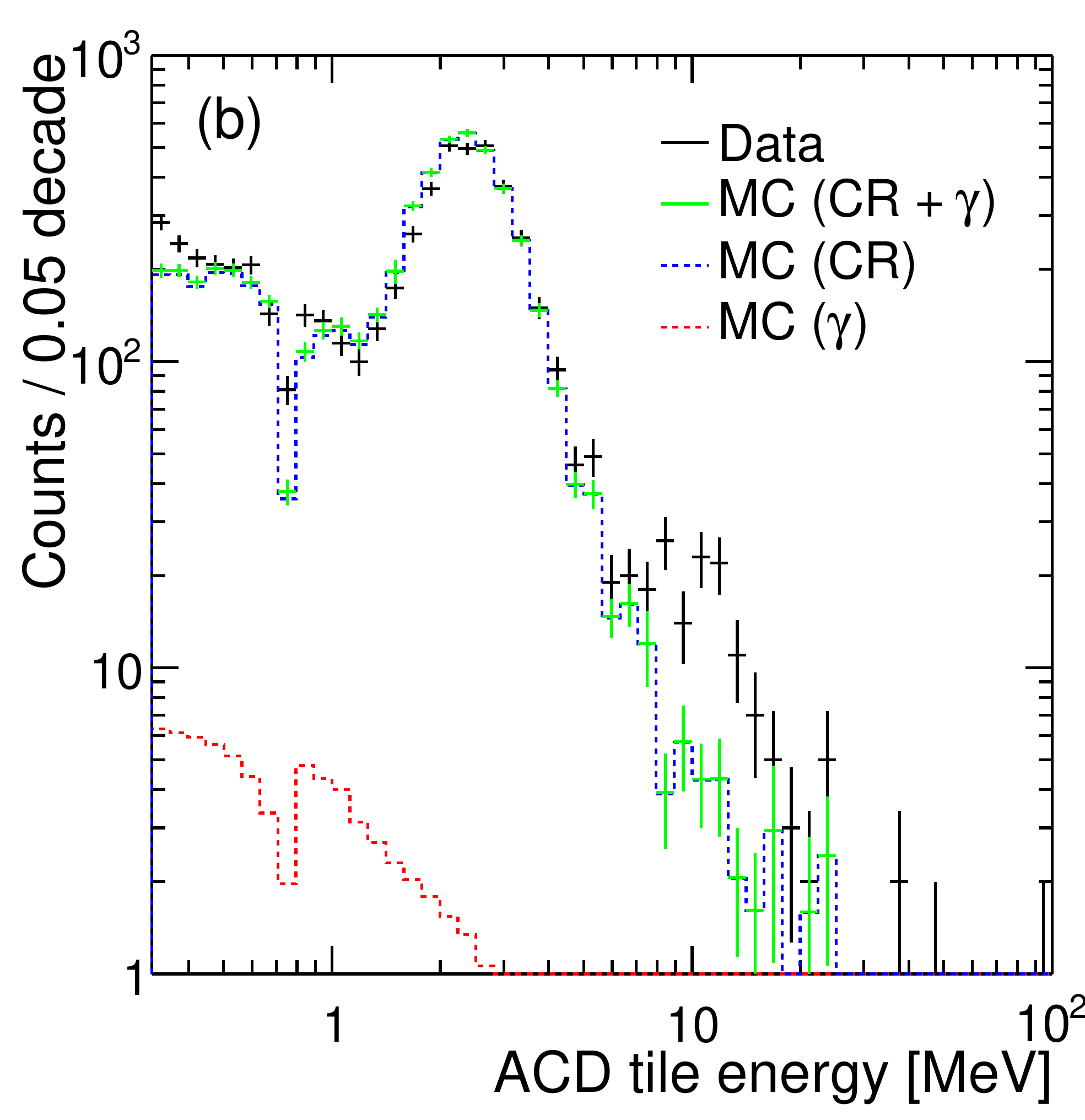}{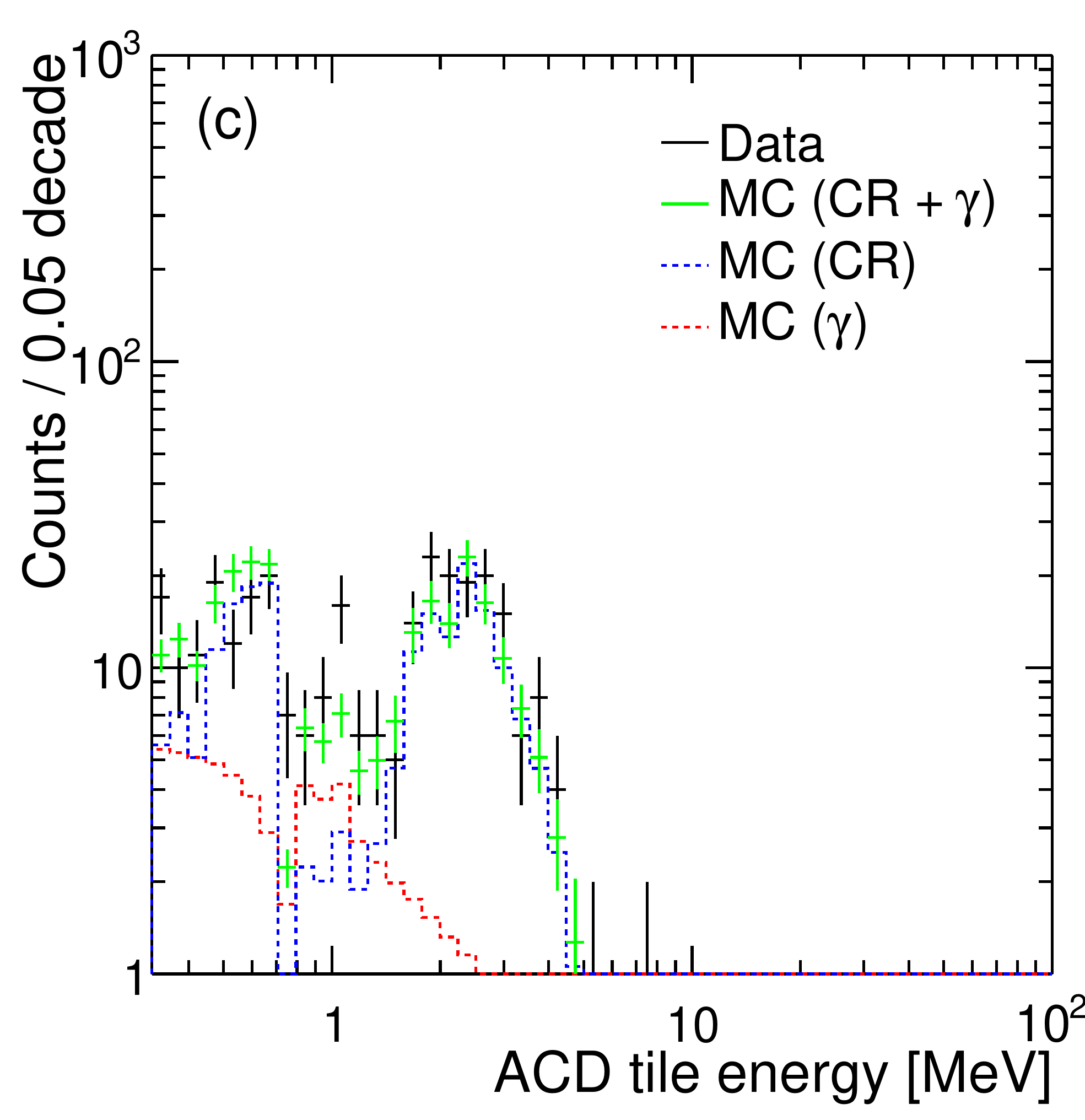}{
  \caption{(a) Distribution of minimal distance of the intersection point of the
    extrapolation of the best track in the event from the edge of the
    corresponding ACD tile; (b) distribution of energy deposited in the ACD
    tile that intersects the extrapolation of the best track in the event for a
    subset of \irf{P7TRANSIENT} events---with some \irf{P7SOURCE} cuts applied;
    (c) same as (b) but for the \irf{P7SOURCE} event class.
    The energy ranges are 30--100~GeV for (a) and 100--300~GeV for (b) and (c). 
    Only data from high Galactic latitudes $|b|>50^{\circ}$ and longitudes
    $90^{\circ}<l<270^{\circ}$, which features a low \gammaRayHyph-to-background
    ratio, are shown (all the background studies described in this section
    use $\sim 24$~months of data, from August~2008 to end of July~2010).
    The relative normalizations of the two Monte Carlo components
    (CR and $\gamma$) are adjusted to fit the data.}
  \label{fig:cleandirty}
}

We already described the usage of data samples with different
\gammaRayHyph-to-background ratios to define high-purity event classes in
\secref{subsubsec:LAT_galRidge}; and we detailed the resulting event
classification cuts in \secref{subsubsec:p7source_selection},
\secref{subsubsec:p7clean_selection}, and \secref{subsubsec:p7ultra_selection}.
A key part of the development of the event classes was using the same clean
and dirty samples to search for and eliminate residual background that is
either not simulated with sufficient accuracy or has passed the multi-variate
event selection that was trained on a limited-statistics sample of simulated
events.

Two specific examples of such improvements to the event classification are
described in this section. The first is designed to mitigate ACD inefficiencies
around the mounting holes of the ACD tiles. \Figref{cleandirty} (a) shows the
distribution of the closest distance of the extrapolation of the best track in
the event from the edge of an ACD tile. Only events that are classified as
\gammaRays\ in the \irf{P7SOURCE} class are shown. A peak starting 39~mm
from the edge is clearly visible, corresponding to the closest distance of many
mounting holes in the ACD from the edge of a tile, where charged particles
entering the LAT often leave very small or undetectable signals due to
inefficiencies in the ACD response near these mounting holes. A matching peak
is visible in the simulation, showing a good example of the detailed
description of the LAT detector model entering the MC simulation of the LAT.
This particular source of residual background has not been removed by the event
classification scheme but can be easily eliminated.
For \irf{P7CLEAN} and \irf{P7UTLTRACLEAN} classes we remove all events where
the best track extrapolates to a range between 35 and 45~mm from the closest
edge of an ACD tile that additionally produces a signal in the first
TKR layer (see \secref{subsubsec:p7clean_selection}).

The second example demonstrates the removal of effects from poorly simulated
interactions like the inelastic interactions of alpha particles and heavier
nuclei. \Figref{cleandirty} (b) shows the distribution of the energy deposited
in the ACD tile closest to the extrapolation of the best track in the event 
onto the ACD plane. The events included pass the \irf{P7TRANSIENT} selection
and additionally some of the cuts used to define the \irf{P7SOURCE} class
(omitting cuts that are effective for reducing heavy nuclei contamination).
A peak at a few MeV is clearly visible in both data and simulation corresponding
to residual protons traversing the ACD. The second peak above 10~MeV,
corresponding to residual helium, is almost completely missing in the
simulation. \Figref{cleandirty} (c) shows the same distribution after all
selection criteria for the \irf{P7SOURCE} class have been applied. The residual
helium peak has been removed and data and simulation show good agreement.

\subsection{Estimating Residual Background from Distributions of Control Quantities}\label{subsec:bkg_distrib}

There are limitations to the agreement achievable between simulated and
experimental data with methods such as those shown in the previous section.
In particular, the primary and secondary CR fluxes, which are important inputs
for our simulation, are uncertainly known. Furthermore, efficiencies of the
trigger and on-board filter might be under- or over-estimated in the
simulation. Therefore we adjust the normalization of the total residual CR
background independently in 15 energy bins between 100~MeV and $\sim 600$~GeV
to better describe the counts observed in the calibration data sample with
a low \gammaRayHyph-to-background ratio defined
in~\secref{subsubsec:LAT_galRidge}.
The scaling is based on the events in the \irf{P7SOURCE} class.
For these events, the distributions of two event properties, the \probAll\
estimator (see \secref{subsec:event_CT_eventLevel}) and the transverse size of
the particle shower in the CAL, have different shapes for
\gammaRays\ and CRs. The shapes of the distributions are sufficiently distinct
for extracting the contribution of both components by fitting a superposition of
simulated \gammaRays\ and CRs to the on-orbit data.
Although the difference between the shapes of the distributions for
\gammaRays\ and CR background decreases with increasing energy for the
\probAll\ estimator, it does increase for the transverse shower size.
Therefore the fit is performed on \probAll\ for energies E~$\leq$~3~GeV, and on
the transverse shower size for E~$>$~3~GeV.  The CR background correction
factors obtained by these fits are then used to adjust the residual background
predicted by the simulations.

\begin{figure}[htb]
  \centering
  \includegraphics[width=\onecolfigwidth]{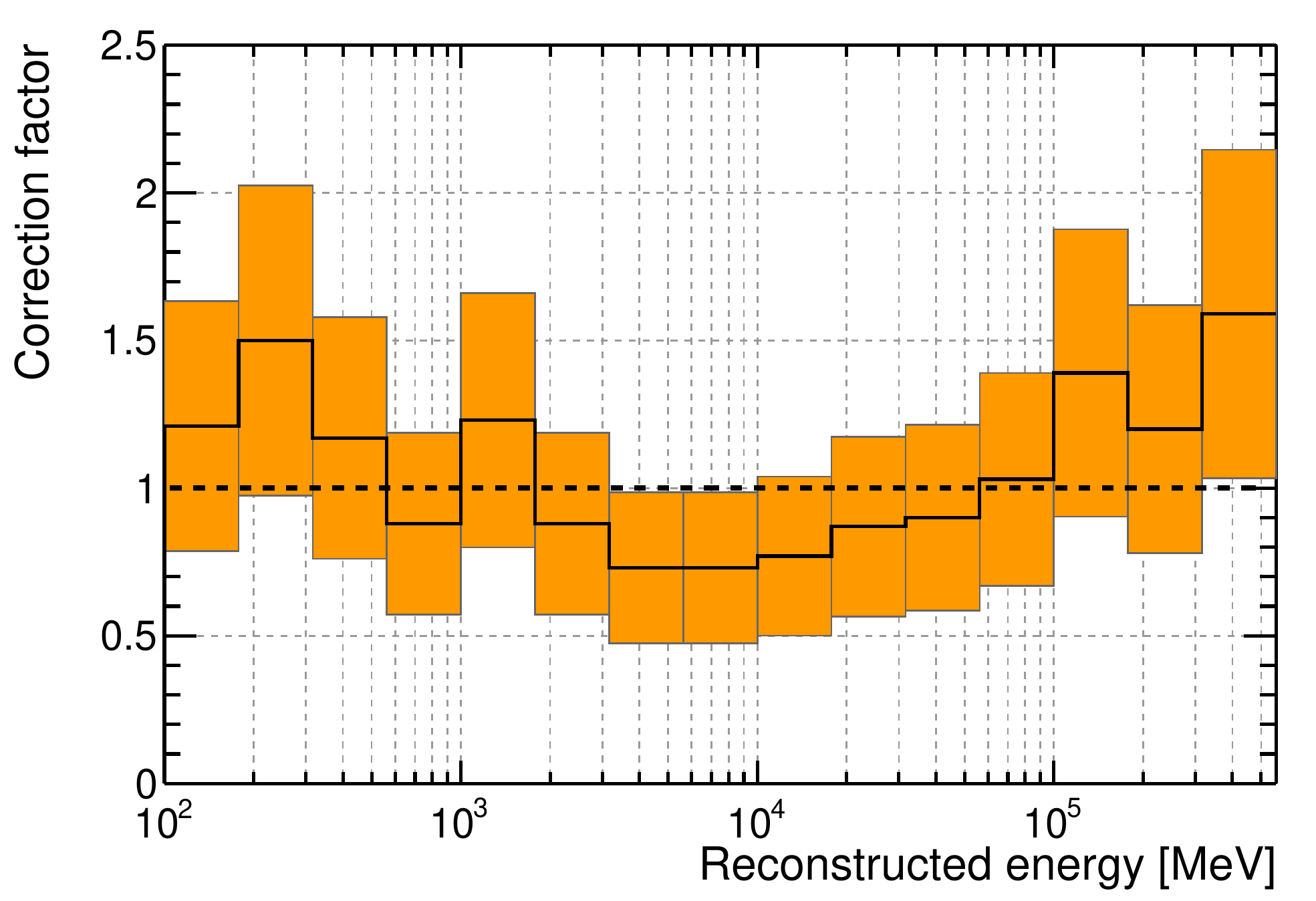}
  \caption{Correction factors for the CR background simulation
    as a function of energy, determined from a fit to the experimental data.
    The filled band shows the 35\% systematic uncertainty.  Note that the energies 
    shown here are based on reconstruction under the hypothesis that the event
    is a \gammaRay\ and most high-energy protons deposit only a small fraction
    of their total energy in the LAT.}
  \label{fig:bgadjustmentfactors}
\end{figure}

\Figref{bgnormadjust} compares the distribution of the \probAll\ estimator and
the transverse shower size between simulated and experimentally observed events
after the normalizations of the predicted CR background and of the
\gammaRayHyph\ simulation have been adjusted.
Each plot refers to a representative energy band in which the
fit was performed for the respective variable. \Figref{bgadjustmentfactors} 
shows the adjustment factors obtained in the fit as a function of energy. An
adjustment factor of 1 corresponds to the CR background intensity predicted in
the simulation. The adjustment factors vary between 0.7 and 1.6, depending on
energy.

We use the predictions of the residual CR background from the MC simulation for
the \irf{P7SOURCE}, \irf{P7CLEAN}, and \irf{P7ULTRACLEAN} event classes
multiplied by the adjustment factors in \figref{bgadjustmentfactors} as our
best estimate of the residual background.  We use the largest 
adjustment factor (1.59) an indicator of the relative uncertainty of our
determination of the residual background.  This uncertainty is found to be $\sim 35$\%,
i.e., $(1.59 - 1.00)/(1.59) = 0.37$. 

\twopanel{htb}{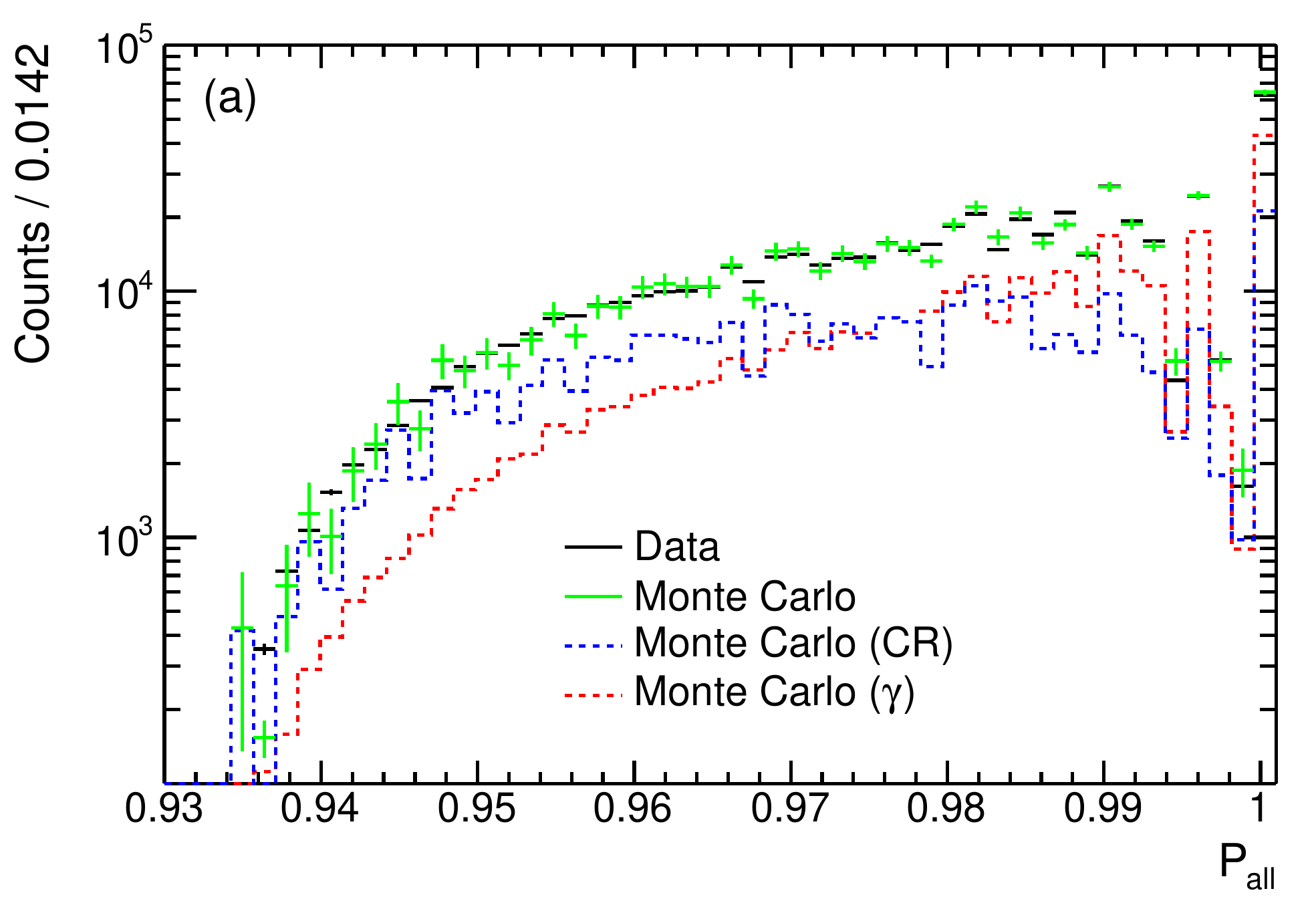}{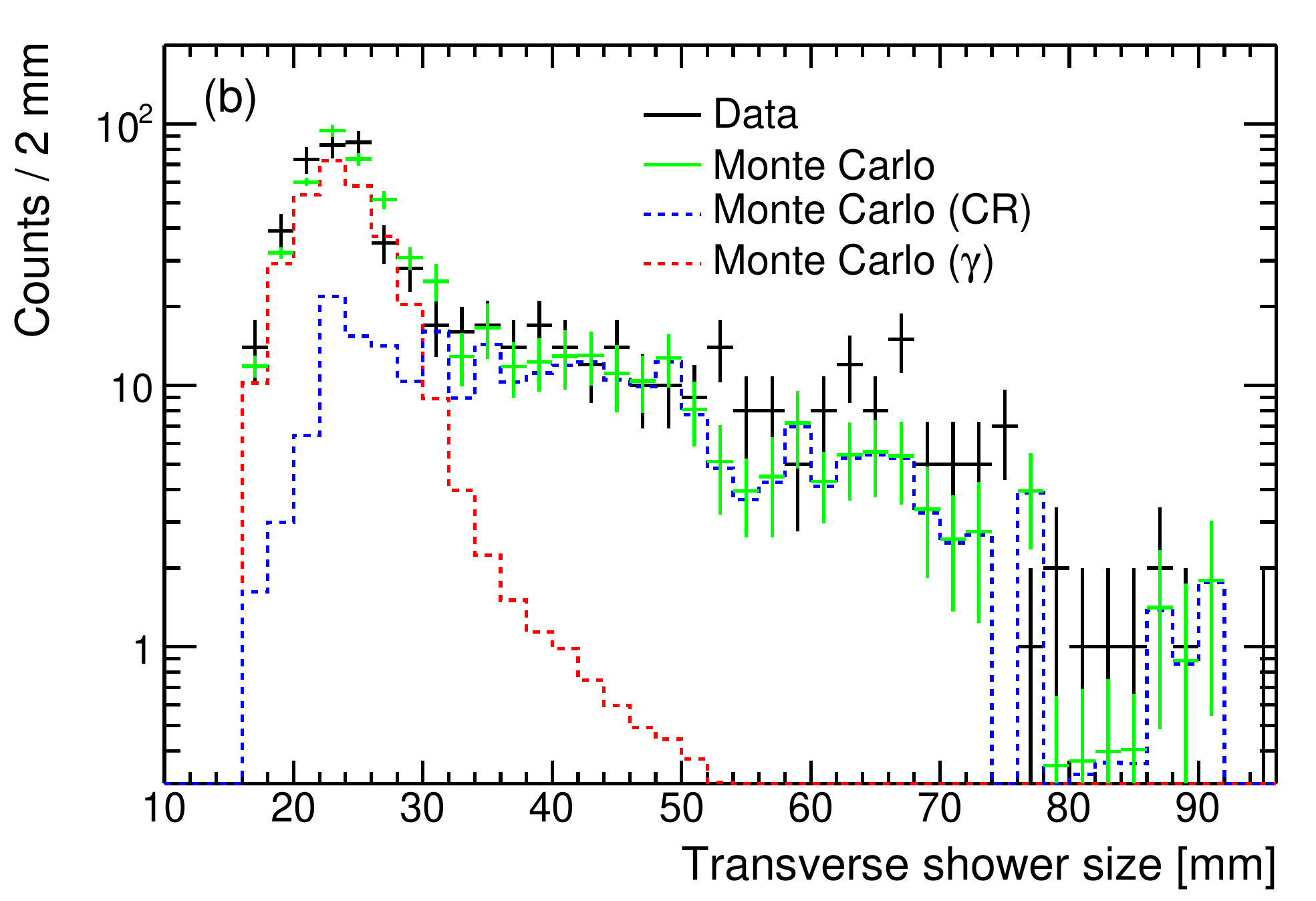}{
  \caption{(a) Distribution of the \probAll\ estimator in the energy range
    200--400~MeV for \gammaRays\ and residual CRs; (b) distribution of
    the transverse shower size in the energy range 32--64~GeV.
    The ranges are representative of the energy intervals in which the fit was
    performed for each of the variables.
    Only \irf{P7SOURCE} events from high Galactic latitudes $|b|>50^{\circ}$
    and longitudes $90^{\circ}<|l|<270^{\circ}$, which feature a low
    \gammaRayHyph-to-background ratio (i.e., the ``dirty'' calibration
    sample), are shown.
    The normalizations of the CR background simulation and the \gammaRayHyph\
    simulation are adjusted to fit the data.}
  \label{fig:bgnormadjust}
}

\Figref{backgroundcont} summarizes our best estimate of the differential
particle background rates in the three high-purity event classes for the
energy range between 100~MeV and 600~GeV. Our background model is likely
inaccurate below 100~MeV, and therefore the background contamination cannot be
reliably determined by means of Monte Carlo simulations in that energy range.

\threepanel{!htb}{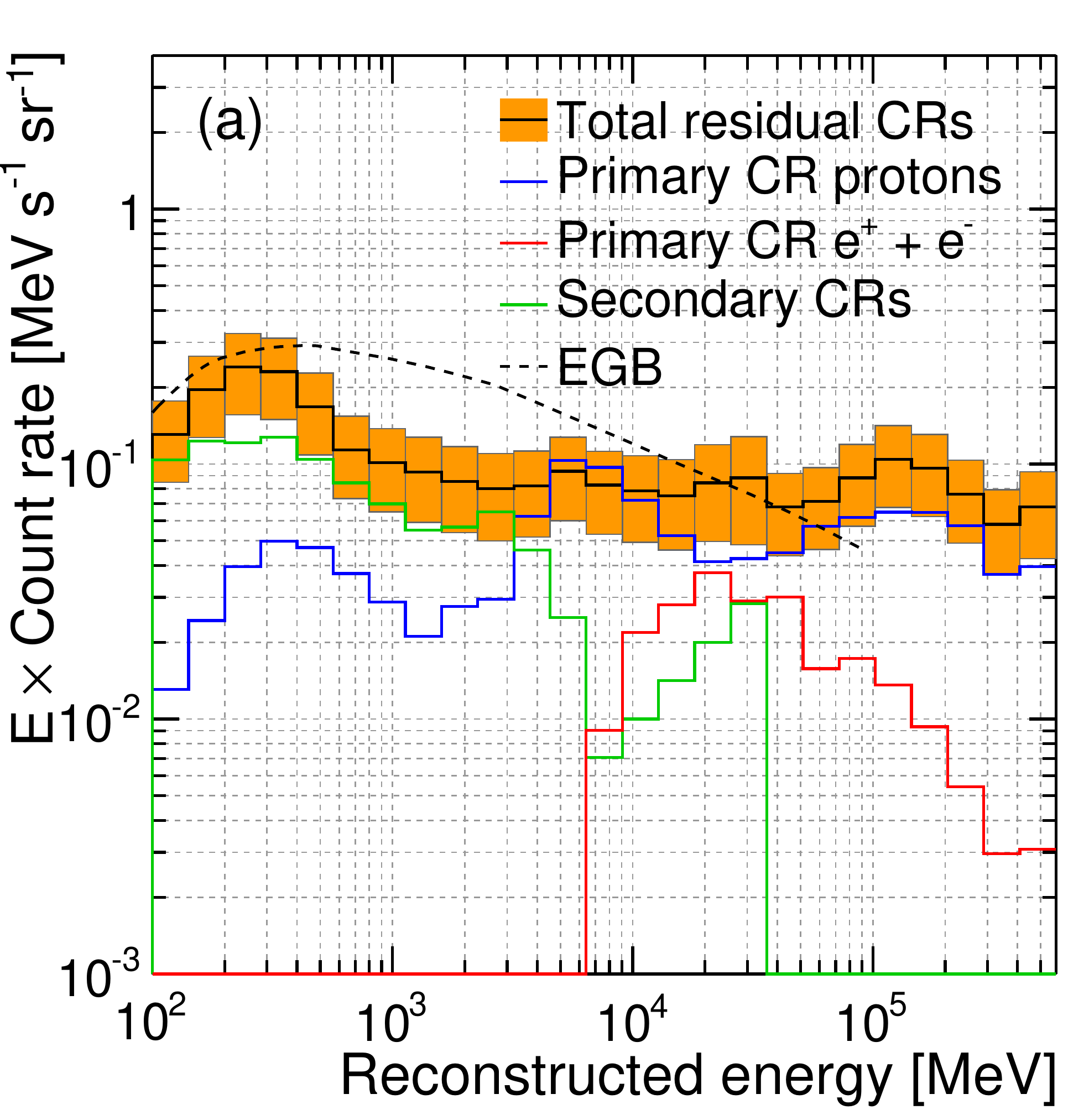}{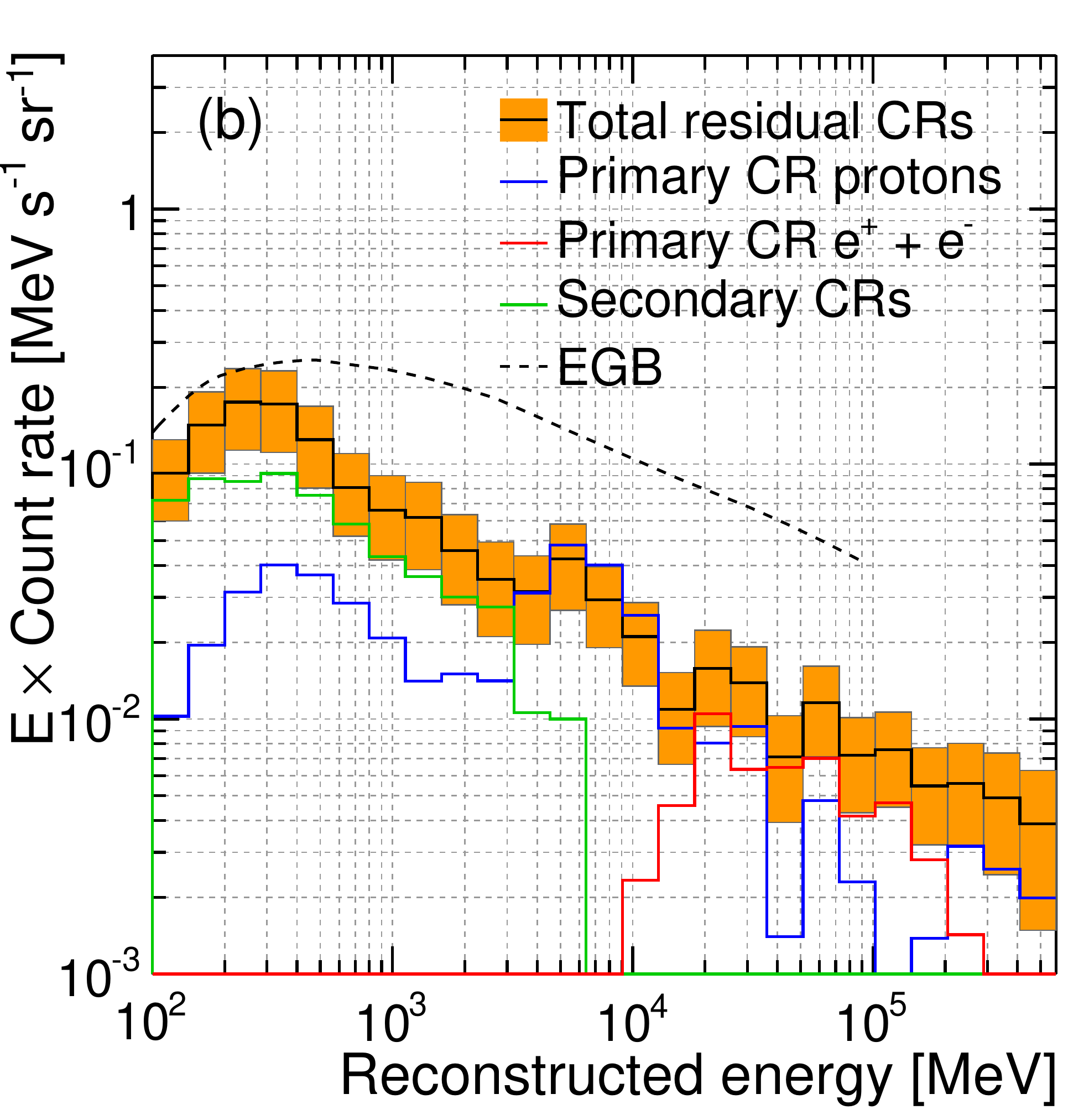}{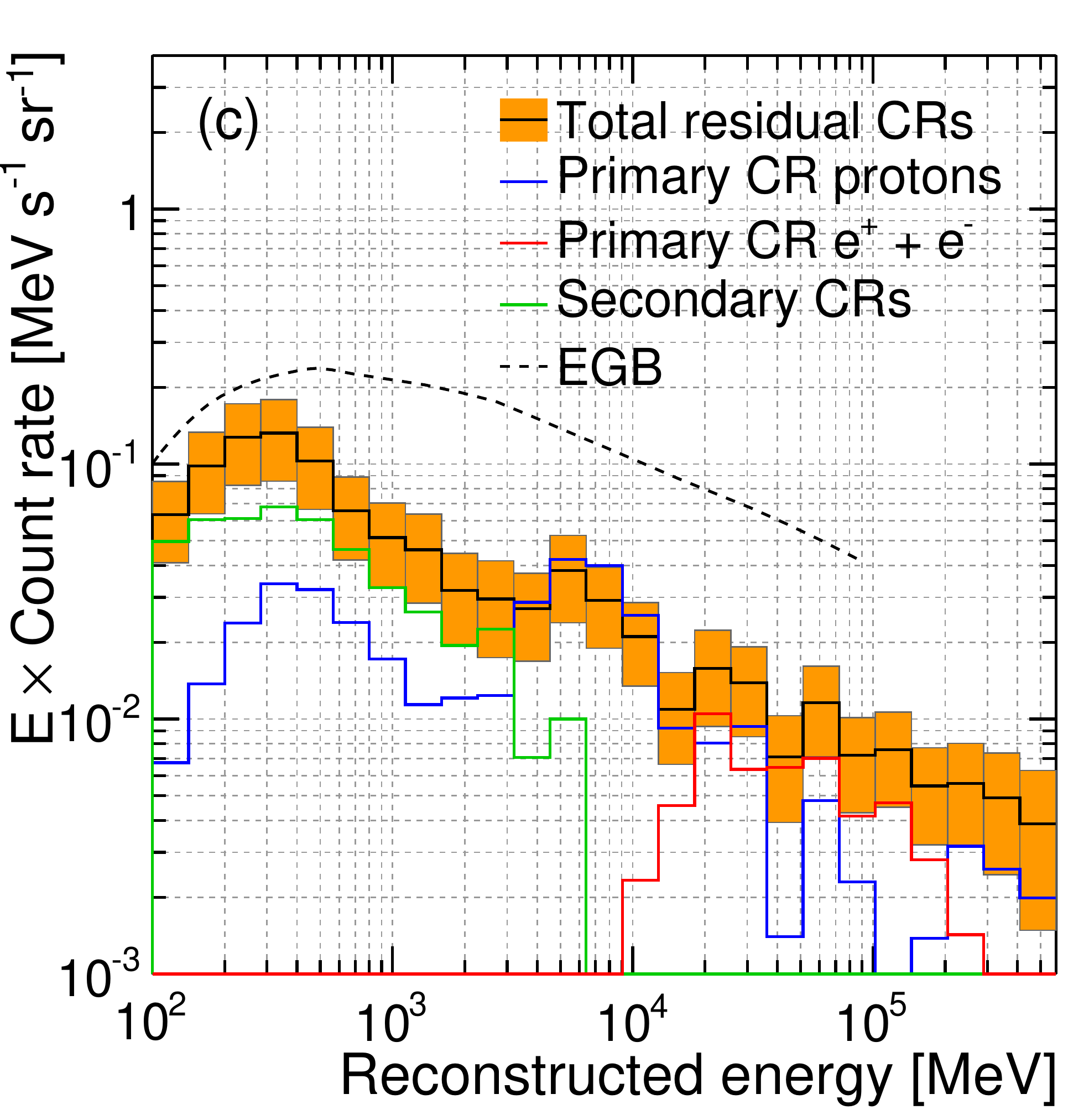}{
  \caption{Best estimates of differential rates of residual particle backgrounds
    for the \irf{P7SOURCE} (a), \irf{P7CLEAN} (b), and \irf{P7ULTRACLEAN} (c)
    event classes. Individual contributions from primary CR protons, primary
    CR electrons and the secondaries from CR interactions are shown; the
    corresponding count rates for the extragalactic \gammaRayHyph\ background
    measured by \Fermi~\citep{REF:ExtraGalBkg} are also overlaid for comparison.
    }
  \label{fig:backgroundcont}
}

\subsection{Estimating Irreducible Backgrounds}\label{subsec:bkg_irreducible}

The term ``irreducible'' was introduced in \secref{subsec:simul_bkgnd} for
CR-induced background with a well-reconstructed \gammaRay\ inside the
LAT as its only signature and is therefore indistinguishable from \gammaRays\
of cosmic origin. As the two main classes of irreducible backgrounds we listed
the CR positrons that annihilate outside the detector, and the CR protons that
inelastically scatter in the passive material surrounding the ACD.
This irreducible background is mostly limited to energies below a few GeV due
to its production mechanisms. We do not consider the \gammaRayHyph\ emission
from the Earth's atmosphere as irreducible, as only \gammaRays\ with large
errors in reconstructed directions enter the samples usually chosen for
high-level analysis. The contamination can be reduced by both a stricter
selection on event quality to reject badly reconstructed \gammaRays\ and by
larger exclusion regions around the Earth. 

An estimate of the fraction of irreducible background in the cleanest event
class (i.e., \irf{P7ULTRACLEAN}) is informative as it represents a lower limit
to the achievable background rejection. Since the irreducible background cannot
be separated from the data we can deduce its amount only based on available MC
information.

To determine the irreducible background from positron annihilations, one can
compare the relative fractions of electrons and positrons surviving at
different stages of the event selection. The secondary CR leptons between
100~MeV and 3~GeV passing the on-board filter are composed of
\makebox{$f_{\rm obf}^{-} \approx 0.28$} of electrons
and \makebox{$f_{\rm obf}^{+} \approx 0.72$} of positrons.
Technically, the positron component is the sum of a reducible and an
irreducible part; however, at this stage, the data set is still overwhelmingly
dominated by reducible charged CRs, so that the irreducible contribution
is effectively negligible:
\begin{equation}
f_{\rm obf}^{+} = f_{\rm obf,red}^{+} + f_{\rm obf,irr}^{+} \approx
f_{\rm obf,red}^{+}.
\end{equation}
The secondary CR leptons passing the \irf{P7ULTRACLEAN} selection are composed
of \makebox{$f_{\rm uc}^{-} \approx 0.10$} of electrons and
\makebox{$f_{\rm uc}^{+} = f_{\rm uc,red}^{+} + f_{\rm uc,irr}^{+} \approx 0.90$}
of positrons. Since the reducible electron and positron components are
indistinguishable in the LAT, they scale identically
\begin{equation}
f_{\rm uc,red}^{+} =  \frac{f_{\rm obf,red}^{+} f_{\rm uc}^{-}}{f_{\rm obf}^{-}}
\approx \frac{f_{\rm obf}^{+} f_{\rm uc}^{-}}{f_{\rm obf}^{-}};
\end{equation}
therefore we have:
\begin{equation}
f_{\rm uc,irr}^{+} = 1 - f_{\rm uc}^{-} - f_{\rm uc,red}^{+} \approx
\frac{f_{\rm obf}^{-} - f_{\rm uc}^{-}}{f_{\rm obf}^{-}} \approx 0.64.
\end{equation}
(i.e., $\approx 64\%$ of the secondary leptons in the \irf{P7ULTRACLEAN} event
class are irreducible background events from positron annihilations).


The amount of irreducible background below 1~GeV from inelastic scatters of
protons can be estimated by evaluating the fraction of the residual simulated
CR protons that does not enter the volume surrounded by the ACD. This is the case
for about 95\% of the simulated CR protons passing the \irf{P7ULTRACLEAN}
selection. 

\begin{figure}[htb]
  \centering
  \includegraphics[width=\onecolfigwidth]{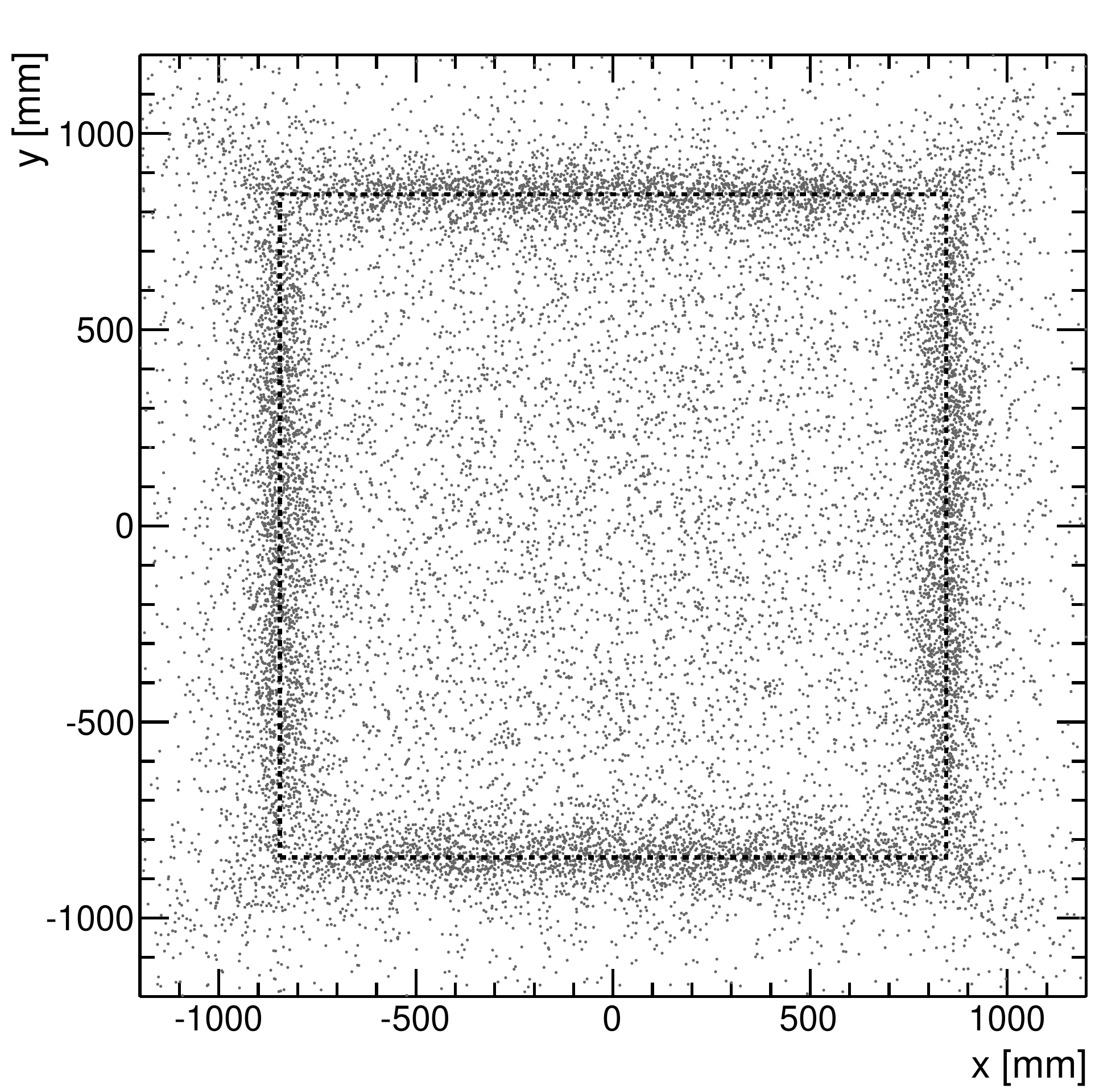}
  \caption{Positions of the projected intersections with
    the top ACD plane for simulated CR proton events surviving the
    \irf{P7ULTRACLEAN} selection. The dashed line marks the edges of the area
    covered by the ACD.}
  \label{fig:acdedges}
\end{figure}

For geometric reasons these scatters predominantly occur at the edges of the
LAT (about 75\% of the residual CR protons, while the remaining 25\% scatter
in the spacecraft body). \Figref{acdedges} shows the positions of the 
projected intersections with the top ACD plane for simulated
CR proton events surviving the \irf{P7ULTRACLEAN} selection with reconstructed
energies below 1 GeV. An enhancement of tracks from the edges of the LAT
is clearly visible, but smeared out due to the finite accuracy of the direction
reconstruction. This feature in fact suggests the possibility to suppress this
type of irreducible background by rejecting events from the edges of the LAT.
However, due to the size of the PSF at low energies tracks intersecting large
regions of the LAT surface would have to be vetoed, resulting in an
unacceptable loss of effective area.
Therefore, such a veto has not been implemented in any of the event classes.

\subsection{Treatment of Particle Backgrounds}\label{subsec:bkg_treatment}

After the event selection, even the purest class will contain residual
background from misclassified CRs. This contribution needs to be accounted
for in the modeling used for spectral and spatial analysis of celestial
\gammaRayHyph\ sources. In particular, in the maximum likelihood model fitting
framework of the \stools, the misclassified CRs must either be modeled separately or
subsumed into one of the diffuse \gammaRayHyph\ components.

The simplest approach is to add a new source to the model to account
for the residual background due to particle leakage. In most cases, since 
CR rates are related to geomagnetic (i.e., Earth) coordinates, for 
time intervals greater than a few months residual background events 
become approximately isotropically distributed in sky coordinates.

The possibility of deriving an effective background template rests mainly on
the assumption that the incidence angle dependence of \aeff\ is the same
for CRs and \gammaRays. 
In many cases we also use a simplifying assumption that the CR 
contamination rates in front- and back-converting event scale with the relative 
acceptances. However, as we will discuss below and in
\secref{subsec:bkg_highLevelAnalysis}, that is not always the case.

The heuristic nature of such a source is evident if one considers
how it changes for different event selections. When we analyze a real
\gammaRayHyph\ source using different event class selections we expect to find
the same spectral distribution within the known systematics; on the other hand,
this template depends on the amount of residual background and therefore on the
CR rejection efficiency of the event class, so the templates we derive
for the various event selections are dramatically different.
\emph{Each event class requires a dedicated background template}.

Under the above assumptions any isotropic \gammaRayHyph\ component (e.g., the
contribution of unresolved extragalactic sources) is not separable from the
background leakage by means of a standard high-level source analysis
(likelihood analysis) without additional knowledge (e.g., of spectral differences), so 
the two components are collected into a single term, simply called the \emph{isotropic template}%
\footnote{E.g., \filename{iso\_p7v6source.txt}, available at
\webpage{http://fermi.gsfc.nasa.gov/ssc/data/access/lat/BackgroundModels.html}}.
To derive an isotropic template for a given event selection we perform a
likelihood analysis of the high-latitude ($|b| > 30$) sky, including all resolved individual sources and a
model of the Galactic interstellar emission%
\footnote{E.g., \filename{gal\_2yearp7v6\_v0.fits}, available at
\webpage{http://fermi.gsfc.nasa.gov/ssc/data/access/lat/BackgroundModels.html}}, 
fitting the spectrum of the isotropic component. It follows that the derived
isotropic template depends on the assumed model for the Galactic interstellar
emission, notably on the inverse-Compton component, which is smooth and far
from negligible even at high Galactic latitudes, since the interstellar
radiation field and CR electrons are broadly distributed about the Galactic
plane. Therefore \emph{each Galactic interstellar emission model requires a
different isotropic template}. 

Between $\sim 400$~MeV and 10~GeV the assumptions mentioned above are rather
good. In the 2FGL catalog analysis \citep{REF:2011.2FGL} a single isotropic
template was used and no significant systematics were observed above 400~MeV.
Outside this energy range the rate of residual background events in the back
section is appreciably greater than for the front section: the use of a single
isotropic template (describing a weighted average of front and back background
contamination) leads to a small hardening of measured spectra of point sources
\citep{REF:2011.2FGL}. The effect is maximum for low-significance, soft
sources: on average the spectral indices of power-law spectra are hardened by
less than half of the typical uncertainty in the measured spectral index.
It is preferable to derive separate isotropic templates for front and back and
use them in a combined likelihood approach\footnote{For example,
\filename{isotropic\_iem\_front\_v02.txt} and
\filename{isotropic\_iem\_back\_v02.txt}, 
available at 
\webpage{http://fermi.gsfc.nasa.gov/ssc/data/access/lat/BackgroundModels.html}.
}, 
if front and back events are kept separate in the analysis, but the
magnitude of this effect does not warrant such a complication for many analyses.

To derive the true isotropic \gammaRayHyph\ component from the measured
isotropic component it is necessary to separately estimate and subtract the
amount of residual background contamination (see~\secref{subsec:bkg_distrib}).

The strategy to account for the CR background by means of an isotropic template,
however, fails in the case of the \gammaRayHyph\ emission from the Earth limb.
Residual background events in the \fov\ due to limb emission
reconstructed in the tails of the PSF will produce a distinct 
pattern on the sky; its shape will depend on the pointing history and the
time and energy ranges under consideration, and will be different for front-
and back-converting events. 

A more stringent cut on $\theta_z$ will reduce the
contamination at the expense of exposure in certain regions of the sky.
In particular, below 100~MeV exclusion regions to effectively eliminate the
residual background become prohibitively large and significant Earth limb
emission remains in the data sample for the commonly-used zenith angle limit
of $100^\circ$.

For the analysis leading to the 2FGL catalog, \gammaRays\ with
$\theta_z > 100^{\circ}$ were rejected. The remaining Earth
limb emission was characterized by a template derived from the residual
emission visible in the 50~MeV to 68~MeV energy band and which extended up to
about 400~MeV\footnote{Available at
\webpage{http://fermi.gsfc.nasa.gov/ssc/data/access/lat/2yr\_catalog}.}
\citep[for details see][]{REF:2011.2FGL}. However, this template 
should not be used for periods much shorter than two years.

Finally, residual background associated with mischaracterized ``back-entering'' 
\gammaRays\ (see \secref{subsec:simul_allGamma}), are another specific background that 
does not follow the \gammaRay\ acceptance. The probability to accept ``back-entering'' 
\gammaRays\ into the \irf{P7SOURCE} event selection is $\sim 1000$ times smaller than 
for ``front-entering'' \gammaRays\ and they are assigned directions roughly $180^\circ$ 
away from the true directions. We consider the effect of this background in \secref{subsec:bkg_highLevelAnalysis} 
and find that treating it as part of the isotropic background does not introduce 
significant errors into analyses of point sources.

\subsection{Propagating Systematic Uncertainties to High Level Science Analysis}\label{subsec:bkg_highLevelAnalysis}

As discussed in the previous section, residual CR background is treated as an
isotropic fictitious \gammaRayHyph\ source in high-level science analysis of
astrophysical \gammaRayHyph\ sources. As this approximation becomes less than
perfect, significant systematic uncertainties can arise.

In addition, a slight inconsistency between front and back \aeff\
(see~\secref{subsec:Aeff_consistency}) complicates the issue further, causing
additional uncertainties when deriving separately the isotropic emission for
the two selections. In general, we can quantify the resulting systematic
uncertainties by comparing estimates obtained from the front-only selection
with the full data set.

The isotropic templates derived for the 2FGL catalog analysis for
\irf{P7SOURCE\_V6} events (and released via the FSSC) can be used for analyses
spanning timescales of many months.
On short timescales, especially less than the $\sim 53.4$ day precession period of
the \Fermi\ orbit, changes in the distribution of geomagnetic latitudes through which the LAT
passes cause the residual background rates to be strongly dependent
on the exact orbital history of the spacecraft and on the CR spectra at
different geomagnetic locations. Analyses based on short time selections could
do better either by using a dedicated estimate of the CR and isotropic
backgrounds, e.g., by using a nearby control region, or take particular care
to assess the impact of a possibly incorrect spectrum and spatial distribution
of the background counts. Often this is done by allowing the isotropic
component some freedom in the fitting procedure;
see, for example, \secref{subsec:perf_variability}.

As discussed in \secref{subsec:bkg_MonteCarlo},
\secref{subsec:bkg_distrib} and \secref{subsec:bkg_irreducible} \irf{P7CLEAN}
and \irf{P7ULTRACLEAN} event classes have much lower levels of background
contamination than \irf{P7SOURCE}. Accordingly, these samples can be used to
study the dependence of any particular analysis on the level of particle
background contamination in \irf{P7SOURCE} analysis.

Finally, we have studied the distribution of residual CR backgrounds in the \irf{P7SOURCE}
event sample $b_s(E,\hat{p})$\label{conv:bkgDist} by comparing the observed counts distribution $n_s(E,\hat{p})$ in that sample
with the predicted distribution $\tilde{n}_s(E,\hat{p})$, which we obtain by scaling the distribution of the
\irf{P7ULTRACLEAN} $n_u(E,\hat{p})$ sample by the ratio of the exposure calculated with
the \irf{P7SOURCE\_V6} IRFs $\mathcal{E}_s(E,\hat{p})$ to the exposure calculated with the
\irf{P7ULTRACLEAN\_V6} IRFs $\mathcal{E}_u(E,\hat{p})$. Specifically,
\begin{align}\label{eq:nPredSource}
\tilde{n}_s(E,\hat{p}) = &
n_u(E,\hat{p})\frac{\mathcal{E}_s(E,\hat{p})}{\mathcal{E}_u(E,\hat{p})}
\nonumber \\
b_s(E,\hat{p}) = & n_s(E,\hat{p}) - \tilde{n}_s(E,\hat{p})
\end{align}

We studied the correlation between residual background and exposure as a function of energy.
The detailed results are beyond the scope of this paper, but in general $b_s$ is not strictly proportional to $\mathcal{E}_s$. This implies that the effective acceptance for residual CR backgrounds in the \irf{P7SOURCE} event sample is not the same as for \gammaRays.

Although the spatial distribution of the residual CR-background could impact studies of large-scale diffuse emission,
the variation across a typical $\sim 20^\circ$~\roi\ used when analyzing point sources is less that the variation in the exposure (2 to 5\%, 
depending on the energy).  Furthermore, for bright sources with sufficient statistics to make high-precision measurements,
the correlation factor between the source parameters and the
normalization of the isotropic component typically has a very small
magnitude ($<0.03$).  Accordingly we neglect the spatial variations in the CR-background contamination when performing point-source analyses.

\clearpage

\section{EFFECTIVE AREA}\label{sec:Aeff}

In order to correctly evaluate the spectra of astrophysical \gammaRayHyph\ 
sources we need to know the effective collecting area of the LAT.  In fact, 
\aeff\ depends on the geometrical cross section of the LAT as well as the
efficiency for converting and correctly identifying incident \gammaRays.  
Because of the complexity of determining these we use high statistics MC 
simulations to evaluate \aeff.  We then quantify any discrepancies
between simulations and flight data, and if needed, correct the MC-based \aeff\
accordingly.

As mentioned in \secref{sec:intro}, we express the effective area as a
function of the incident \gammaRayHyph\ energy and direction in the LAT
instrument frame. Therefore, the exposure ($\mathcal{E}$) at a given energy
for any point in the sky depends on the effective area and the observing
profile (see \Eqref{eq:tobsDef}).

In practice, the observing profile depends on the direction in the sky, and is 
accurately known. Therefore, the uncertainties on $\aeff(E,\theta,\phi)$
are the dominant source of instrument-related systematic error.
Of course, we must also consider the uncertainties on our measurements of the
\gammaRay\ direction (i.e., the PSF) and the \gammaRay\ energy (i.e., the
energy dispersion).
However, as we will show in the next three sections, in many cases
the uncertainty of \aeff\ is more important to the analyses than
those of the PSF and the energy dispersion.

In \secref{subsec:Aeff_MonteCarlo} and \secref{subsec:Aeff_MC_corrections} we 
will describe how we generate tables of \aeff\ as a function of energy and
incidence angle (for the front and back sections of the LAT separately),
and how we apply small corrections to those tables to account for variations
of \aeff\ with orbital position and azimuthal direction of the
incoming \gammaRay.
Then in \secref{subsec:Aeff_stepByStep} we will describe how we have validated
the MC predictions of the \gammaRayHyph\ selection efficiency for all the
stages of the analysis using calibration samples
within the flight data set, while in \secref{subsec:Aeff_flightAEff} we will
describe corrections to the \aeff\ tables motivated by disagreement between
the measured and predicted efficiency in one step of the selection process.
Finally, in \secref{subsec:Aeff_consistency}, \secref{subsec:Aeff_errors}, and
\secref{subsec:Aeff_highLevel} we will evaluate the systematic uncertainties
on \aeff\ and show how we propagate these uncertainties into estimated
systematic errors on measured astrophysical quantities such as fluxes and
spectral indices.

\subsection{Effective Area Studies with Monte Carlo Simulations}\label{subsec:Aeff_MonteCarlo}

The starting point of the \aeff\ evaluation is a dedicated \allgamma\ 
sample (\secref{subsec:simul_allGamma}). Since the \gammaRays\ are generated uniformly in $\log(E)$ and
solid angle, the effective area in any of the bins in which the parameter space
is partitioned can be expressed in terms of the total number
of generated events $N_{\rm gen}$\label{conv:ngen} and the number of events
$n_{i,j,k}$\label{conv:nobs} passing the \gammaRayHyph\ selection
criteria within the specific bin centered at $E = E_i$, $\theta = \theta_j$
and $\phi = \phi_k$:
\begin{equation}\label{eq:aeff_general}
  \aeff(E_i, \theta_j, \phi_k) = (6~{\rm m}^2) 
  \left(\frac{n_{i,j,k}}{N_{\rm gen}}\right)
  \left(\frac{2\pi}{\Delta\Omega_{j,k}}\right)
  \left(\frac{\log_{10}{E_{\rm max}} - \log_{10}{E_{\rm min}}}%
       {\log_{10}{E_{\rm max,i}} - \log_{10}{E_{\rm min,i}}}\right)
\end{equation}
where $\Delta\Omega_{j,k}$ is the solid angle subtended by the bin $j,k$ in $\theta$ and $\phi$,
$E_{\rm min}$ and $E_{\rm max}$ give the energy range of the \allgamma\ sample and
$E_{\rm min,i}$ and $E_{\rm max,i}$ are the boundaries of the $i^{\rm th}$ energy bin.
(See~\secref{subsec:simul_allGamma} for more details about the numerical
factors and about the \allgamma\ simulations in general).
In practice, since the effective area is routinely averaged over $\phi$ 
in scientific analysis, we factor out the $\phi$ dependence and rewrite
\Eqref{eq:aeff_general} as
\begin{equation}
  \aeff(E_i,\theta_j,\phi_k) = (6~{\rm m}^2) 
  \left(\frac{n_{i,j}}{N_{\rm gen}}\right)
  \left(\frac{2\pi}{\Delta\Omega_{j}}\right)
  \left(\frac{\log_{10}{E_{\rm max}} - \log_{10}{E_{\rm min}}}%
       {\log_{10}{E_{\rm max,i}} - \log_{10}{E_{\rm min,i}}}\right) \times
  R(E_i,\theta_j,\phi_k) 
\end{equation}
where $R(E,\theta,\phi)$\label{conv:RphiDep} is a small (of the order of 10\%)
correction factor whose average over $\phi$ is~$1$ by construction for any $E$
and $\theta$ (see~\secref{subsubsec:aeff_phi_dep} for more details).
Examples of effective area tables averaged over $\phi$ are shown in \figref{aeff_tables}.
\twopanel{htb!}{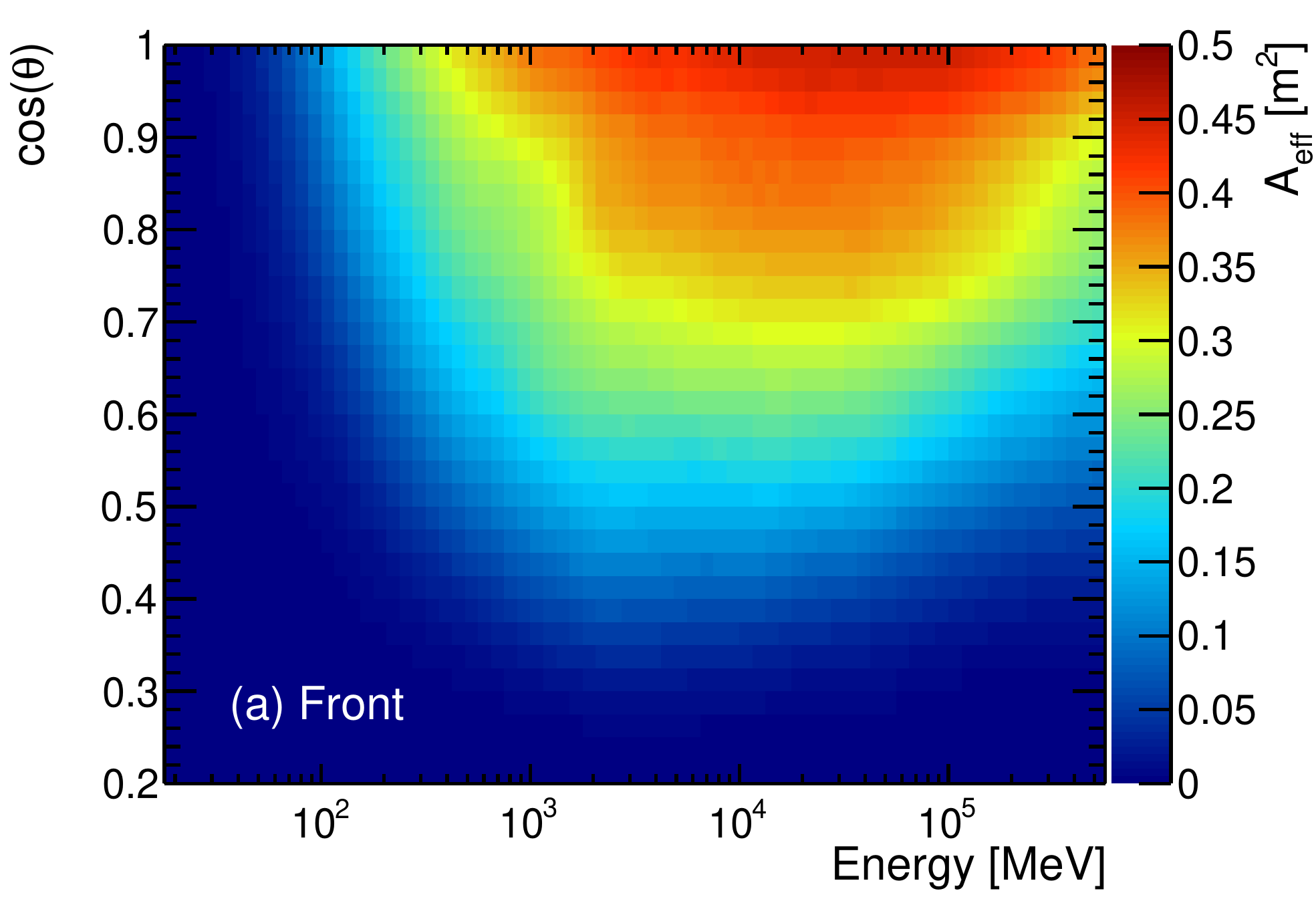}{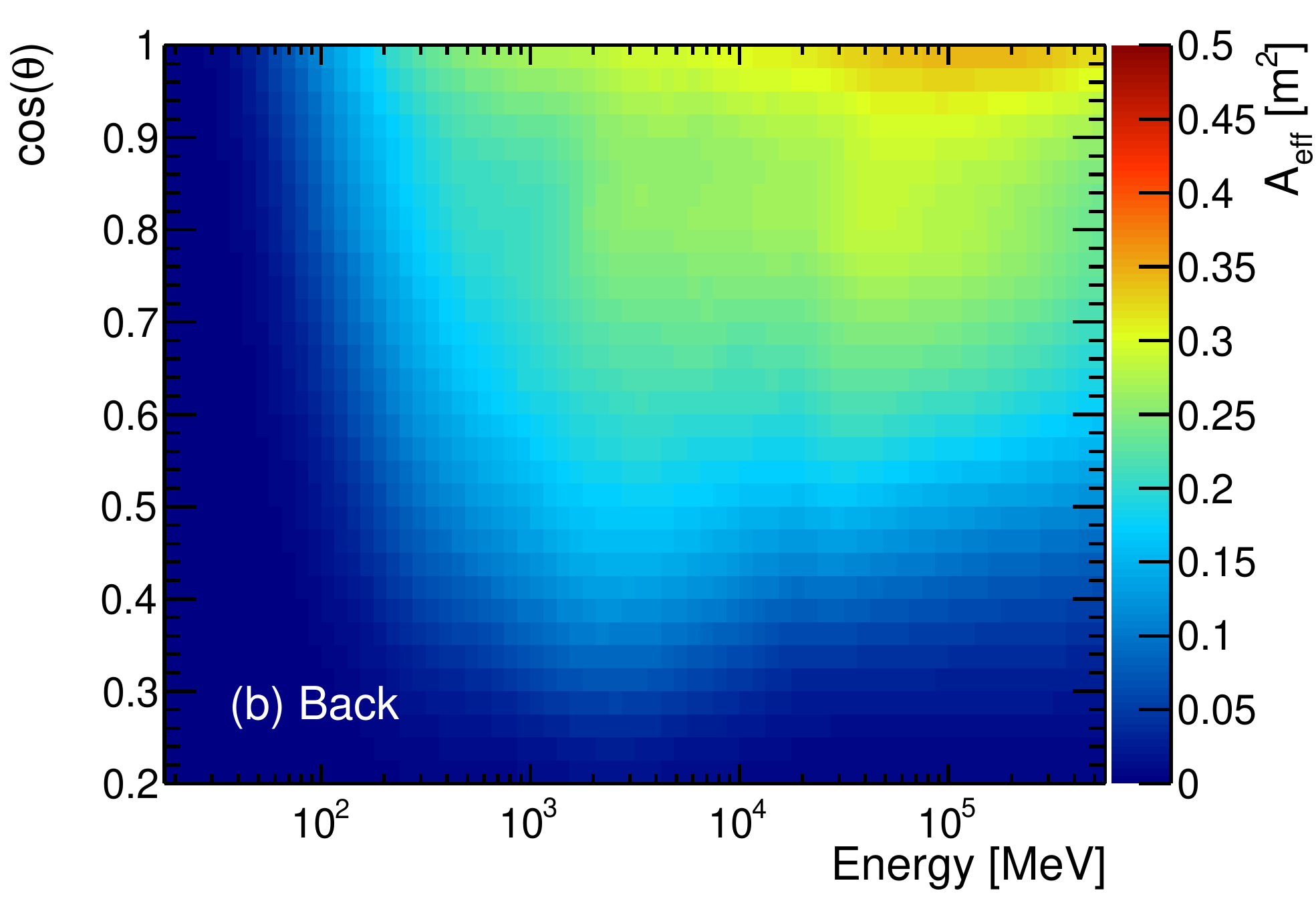}{
  \caption{Graphical representation of the effective area tables for the
    \irf{P7SOURCE\_V6} class, front (a) and back (b) sections of the LAT.}
  \label{fig:aeff_tables}
}

When describing the instrument performance, we more commonly show the effective
area at normal incidence as a function of the energy or the angular dependence
of the effective area for a given energy (usually 10~GeV), as shown
in~\figref{aeff_ene_theta_dep}.

\twopanel{htb!}{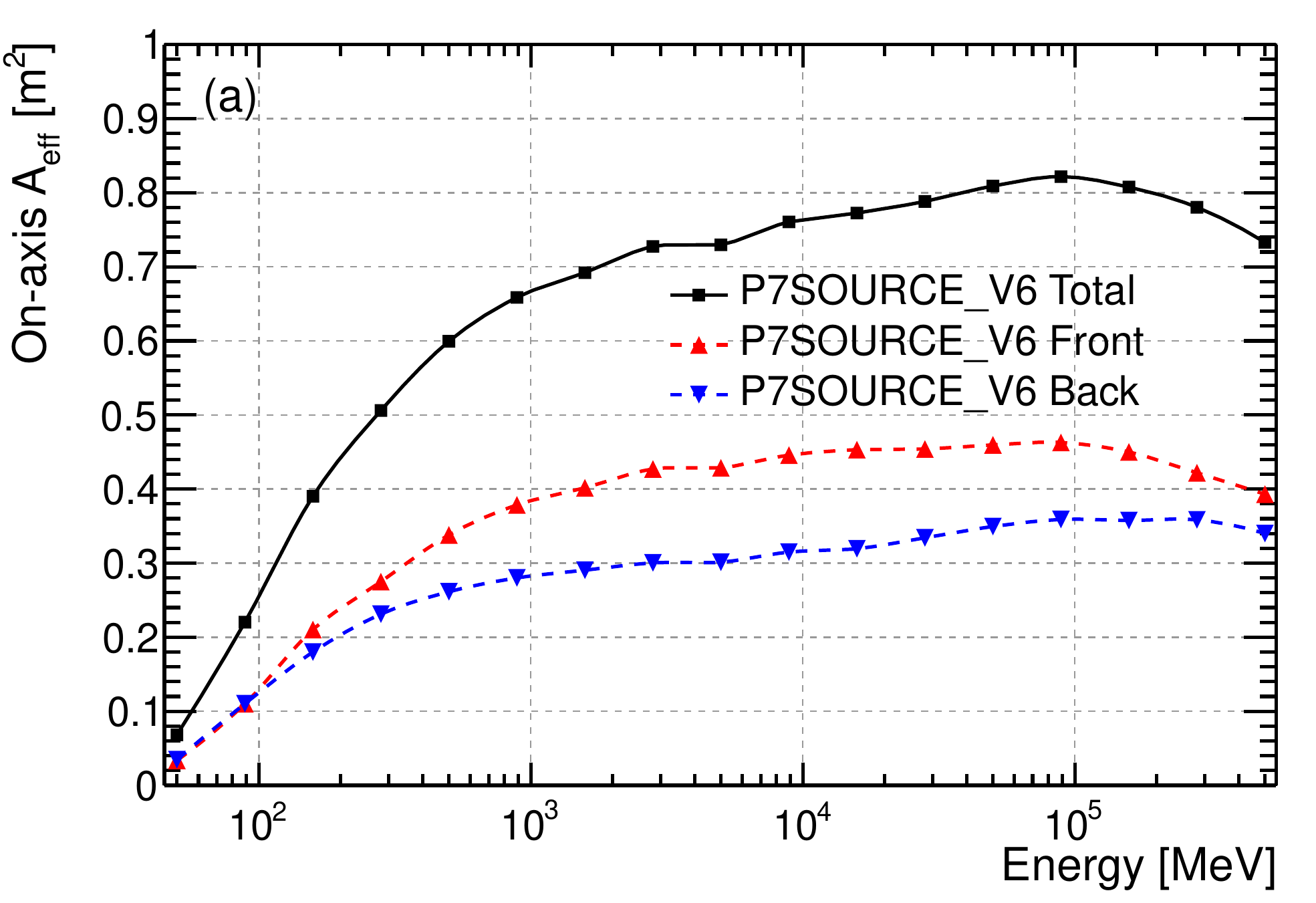}{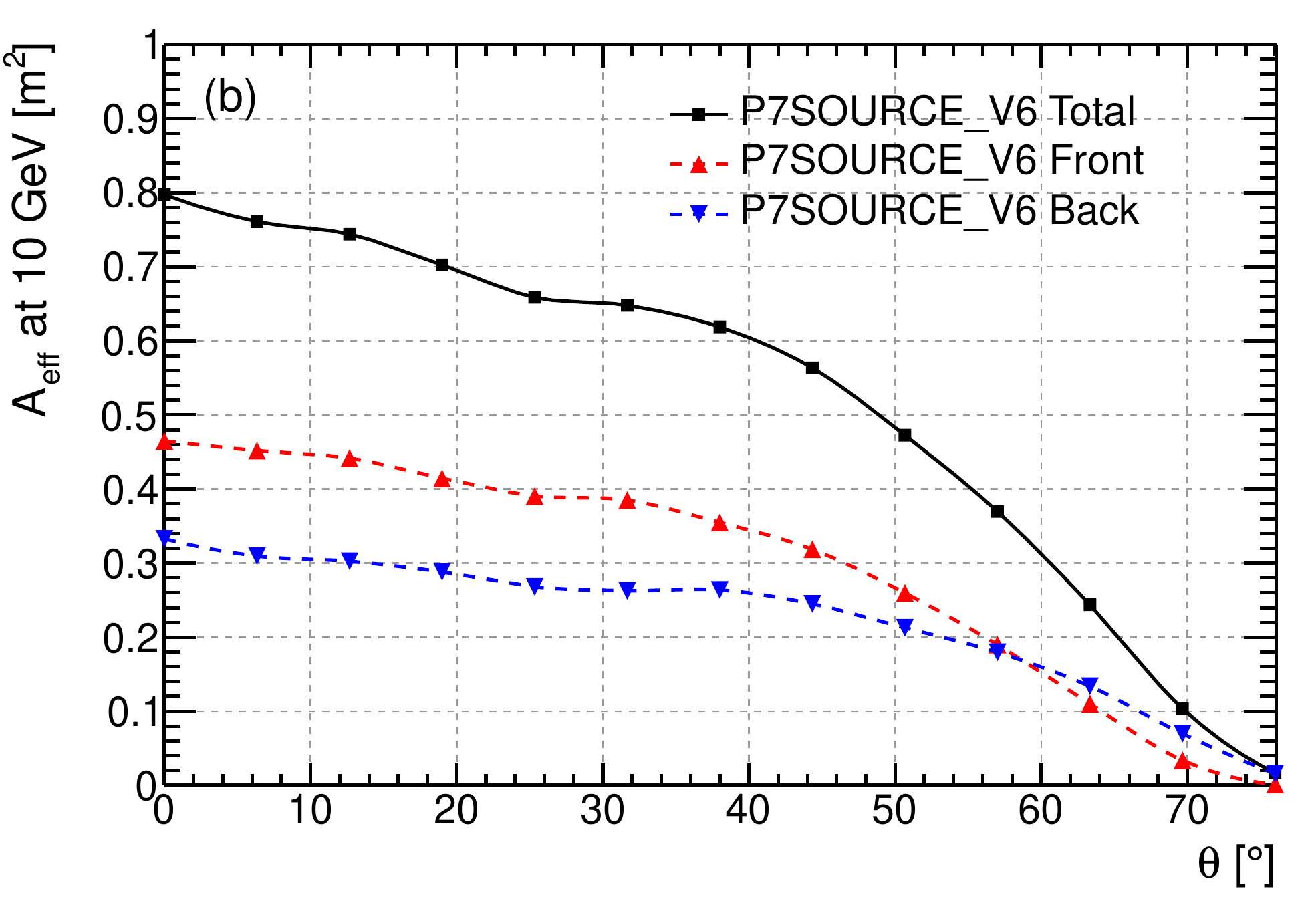}{
  \caption{On-axis effective area as a function of the energy (a) and
    angular dependence (b) of the effective area at 10~GeV for the \irf{P7SOURCE}
    class.}
  \label{fig:aeff_ene_theta_dep}
}

The integral of the effective area over the solid angle, called
the \emph{acceptance}
\begin{equation}\label{eq:acceptance}
  \accept(E) = \int \aeff(E, \theta, \phi) \,d\Omega = 
  \int_{0}^{\frac{\pi}{2}}\int_{0}^{2\pi} \aeff(E, \theta, \phi)
  \,\sin\theta \, d\theta \, d\phi,
\end{equation}
is another widely used performance measure and is shown as a function of 
energy in \figref{acceptance}.

\twopanel{htb!}{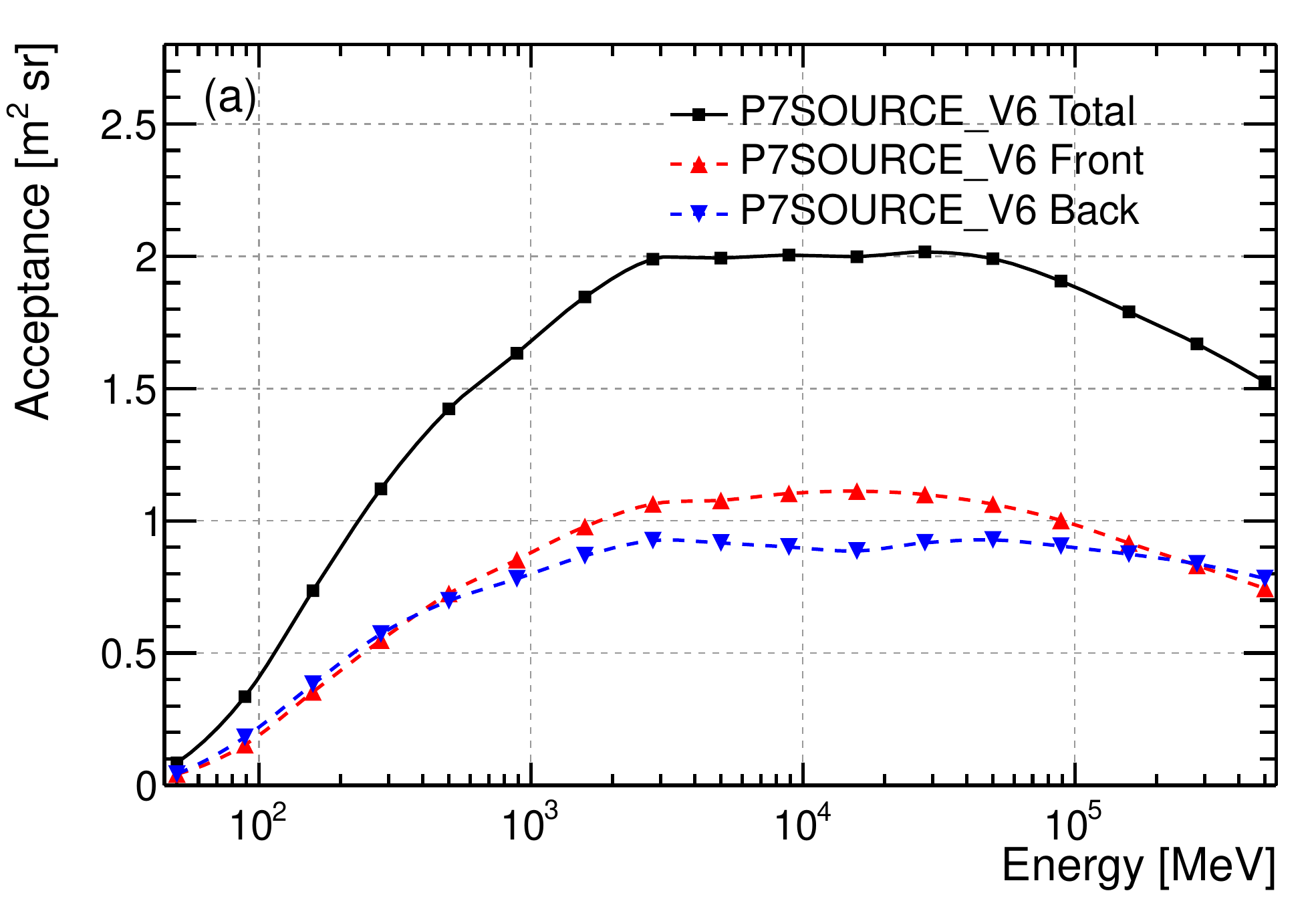}{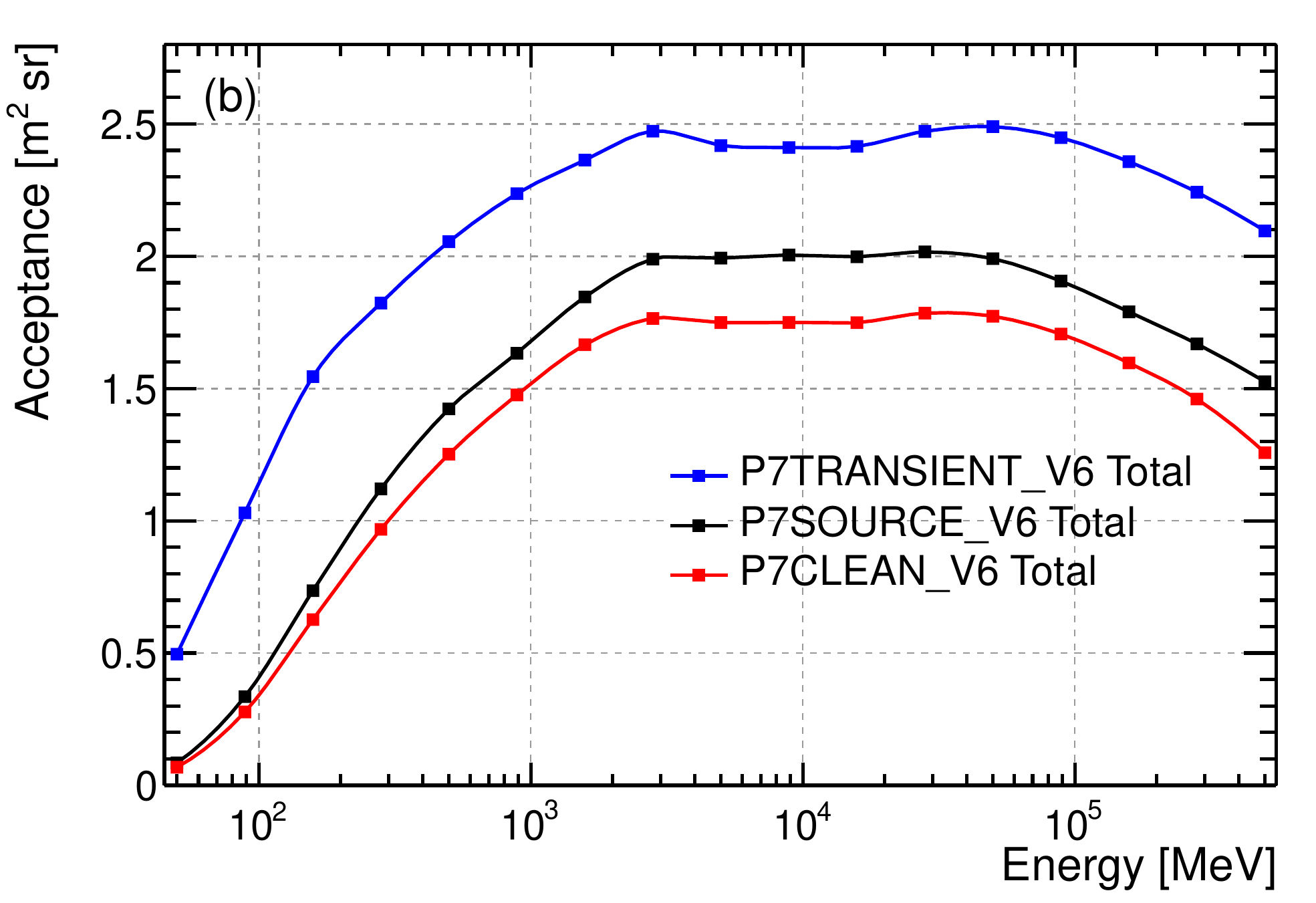}{
  \caption{Acceptance as a function of energy for the \irf{P7SOURCE} class
    (a) and for the other standard \gammaRayHyph\ classes (b).}
  \label{fig:acceptance}
}

\begin{figure}[htb!]
  \begin{center}
    \includegraphics[width=\onecolfigwidth]{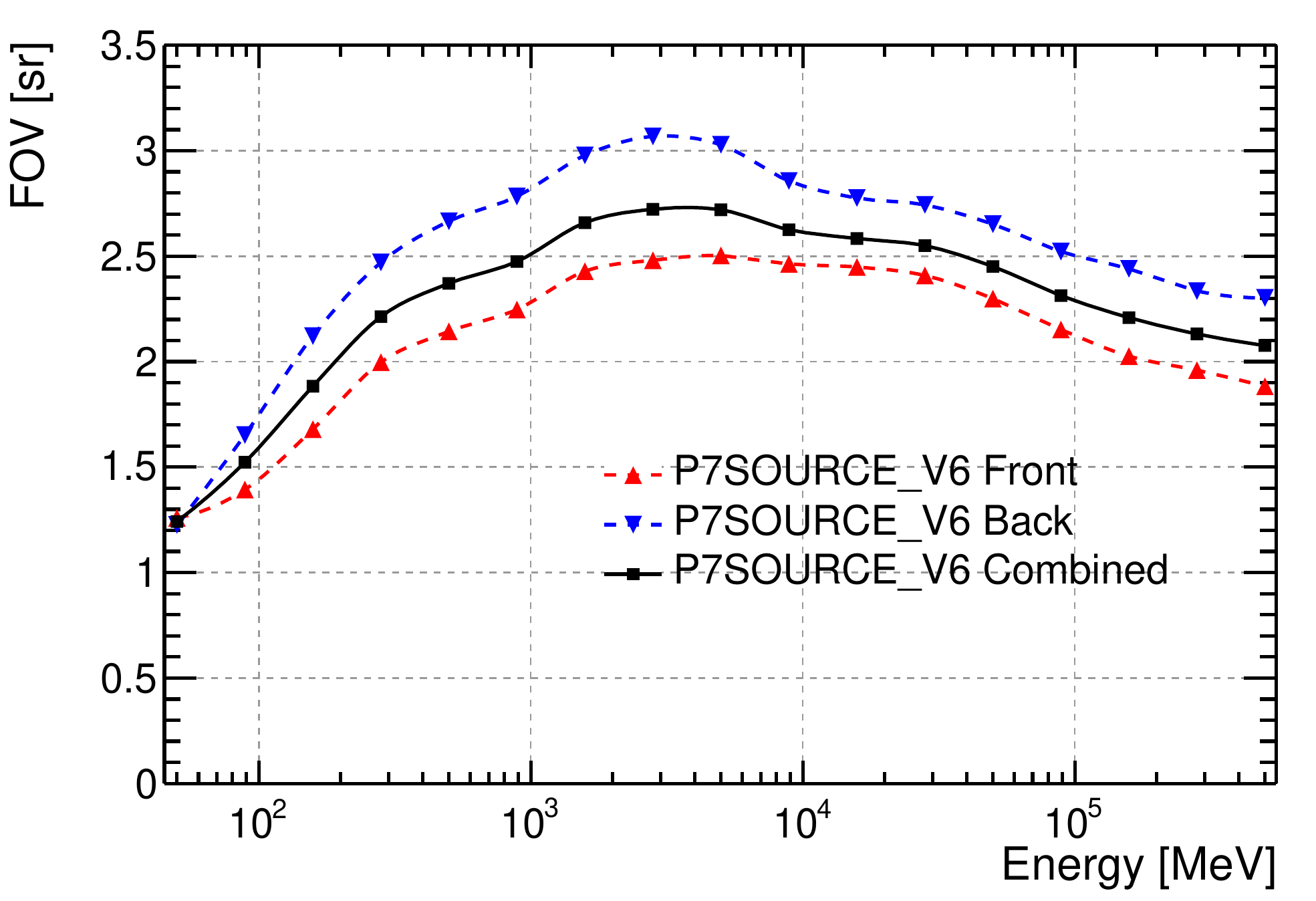}
    \caption{LAT \fov\ as a function of the energy for the \irf{P7SOURCE}
      class. Front- and back-converting events are shown separately.}
    \label{fig:fov}
  \end{center}
\end{figure}

Formally, the \fov\ is defined, at any given energy, as the
ratio between the acceptance and the on-axis effective area:
\begin{equation}
  \fov(E) = \frac{\accept(E)}{\aeff(E, \theta = 0)}
\end{equation}
\figref{fov} shows that the peak \fov\ of the LAT for the \irf{P7SOURCE} event
class is of the order of 2.7~sr between 1 and 10~GeV. At lower energies
the \fov\ decreases with energy, as \gammaRays\ converting in the TKR at large
angles pass through comparatively more material and therefore are less likely
to trigger the instrument. A similar (smaller) effect is observed at very
high energy, where, due to backsplash from the CAL, it becomes more difficult
to reconstruct events at large angles. Finally we note that, for geometrical
reasons (we require events in the standard classes to intersect the
CAL) the \fov\ for the back section is typically larger than that for
the front section.

\subsection{Corrections to the Effective Area Derived from Simulations}%
\label{subsec:Aeff_MC_corrections}

In this section we describe three refinements we made to the effective area
characterization based on experience with flight data. In all three cases we
have simulated the effects that we had previously ignored or averaged out,
but that we discovered could significantly impact particular scientific
analyses. 

\subsubsection{Correction for Ghost Events}\label{subsubsec:Aeff_ghosts}

\begin{figure}[htbp]
  \centering
  \includegraphics[width=\onecolfigwidth]{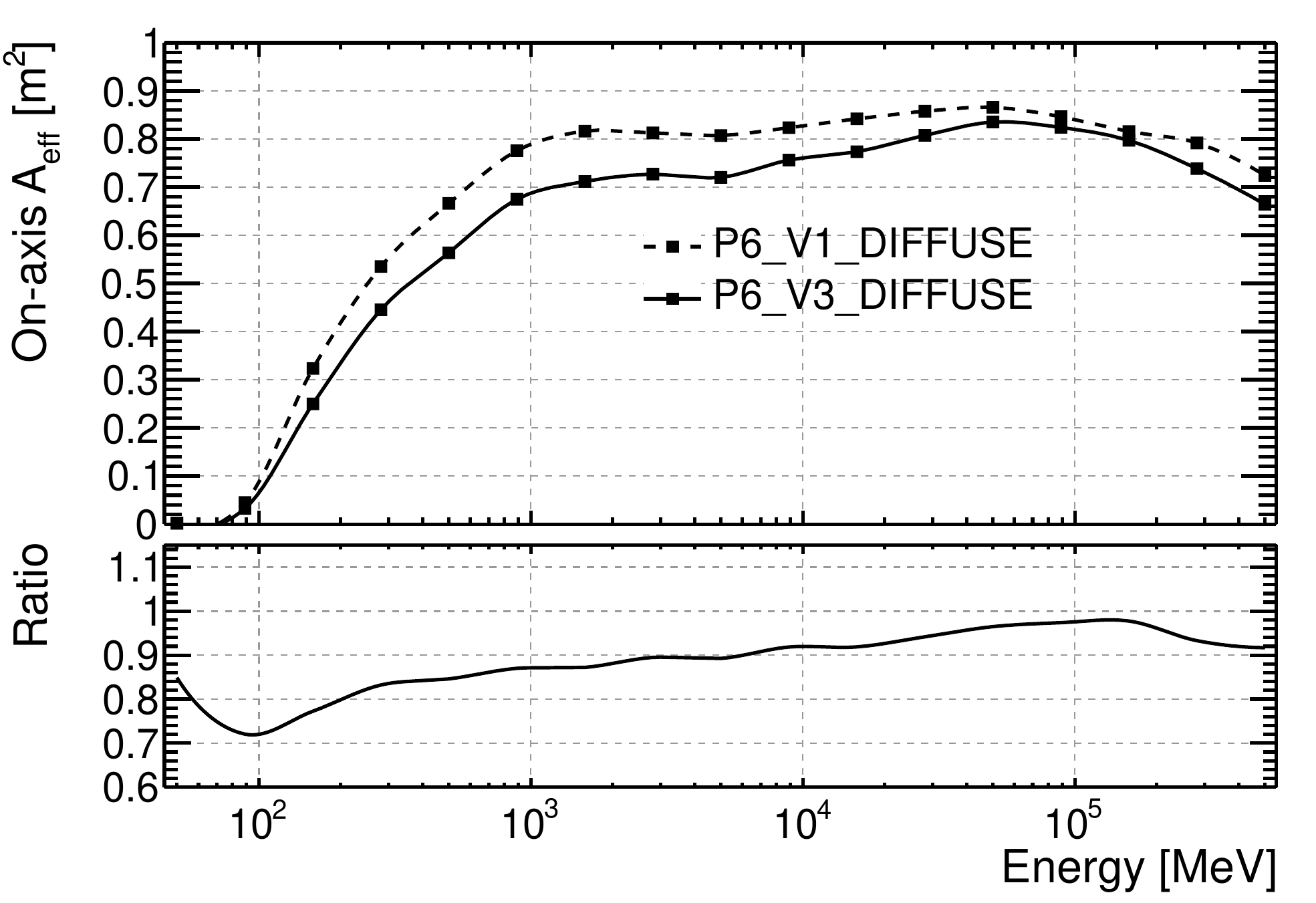}
  \caption{Effective area at normal incidence for the \irf{P6\_DIFFUSE} class,
    in the pre-launch version (\irf{P6\_V1\_DIFFUSE}, not corrected for
    ghost effects) and in the updated version including a modeling of
    ghost effects (\irf{P6\_V3\_DIFFUSE}).
    Adapted from~\citet{REF:2009.LATPerf}.}
  \label{fig:aeff_overlay}
\end{figure}

As explained in more detail in \secref{subsec:LAT_DAQ}, after the start of LAT
operations, it became apparent that ghost signals led to a significant
decrease in effective area with respect to the pre-launch estimates, for which
this effect was not considered.
The \emph{overlay} procedure used to account for this effect, first introduced
in the \irf{P6\_V3} set of IRFs, is described in detail
in~\secref{subsubsec:Overlays} and its impact on the effective
area is shown in \figref{aeff_overlay}.

\subsubsection{Live Time Dependence}\label{subsubsec:Aeff_Livetime}

The effect of ghost signals is corrected on average as described in the
previous section. A smaller correction is necessary to account for the detailed
dependence of \aeff\ on the CR rates. To account for this we need an estimator
for the rate of CRs entering the LAT; the obvious one is the trigger rate, but
technical issues make this choice impractical.
A variable that can be easily obtained from the pointing
history files and which is linearly correlated with trigger rate is the
\emph{live time fraction} $F_l$\label{conv:livetime_frac}, the fraction of the
total observing time in which the LAT is triggerable and not busy reading out a
previous event. The average value of $F_l$ is $\sim 90\%$ and varies between
82\% and 92\% over the \Fermi\ orbit.

We bin events from a sample of periodic triggers according to the corresponding
live time fraction and for each bin we produce a dedicated \allgamma\
simulation (\secref{subsec:simul_allGamma}); for each of the resulting overlay data sets we derive \aeff.
We have found the dependence on the incidence angle to be small and so we
choose to neglect it when studying this effect. Furthermore, we have found that
at a given energy \aeff\ varies linearly with the live time fraction.
We perform a linear interpolation in each energy bin in accord with
\begin{equation}\label{eq:aeff_livetime_dep}
  \aeff(E, F_l) = \aeff(E) \cdot (c_0(E) F_l + c_1(E))
\end{equation}
separately for front and back events.

\twopanel{htb!}{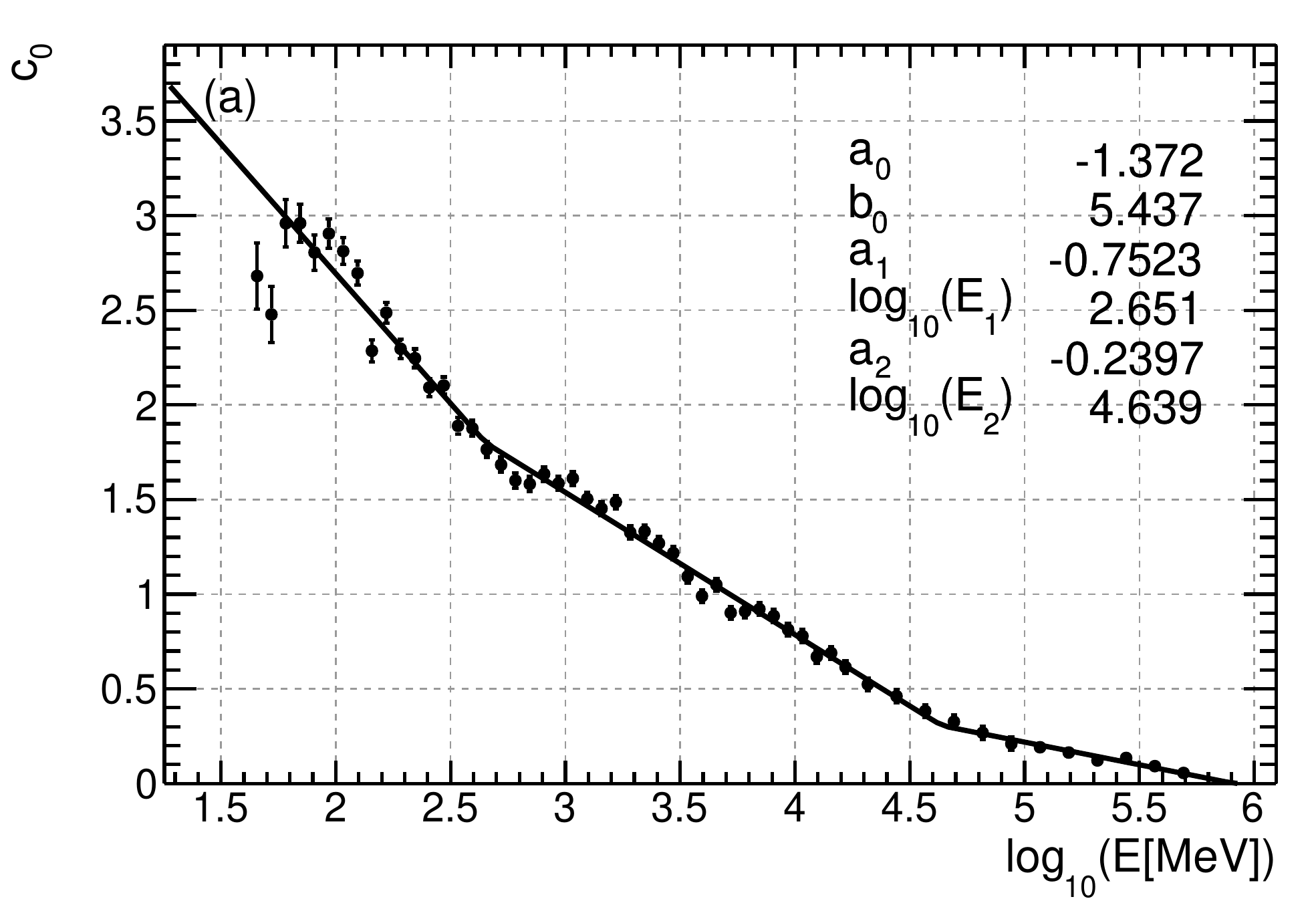}{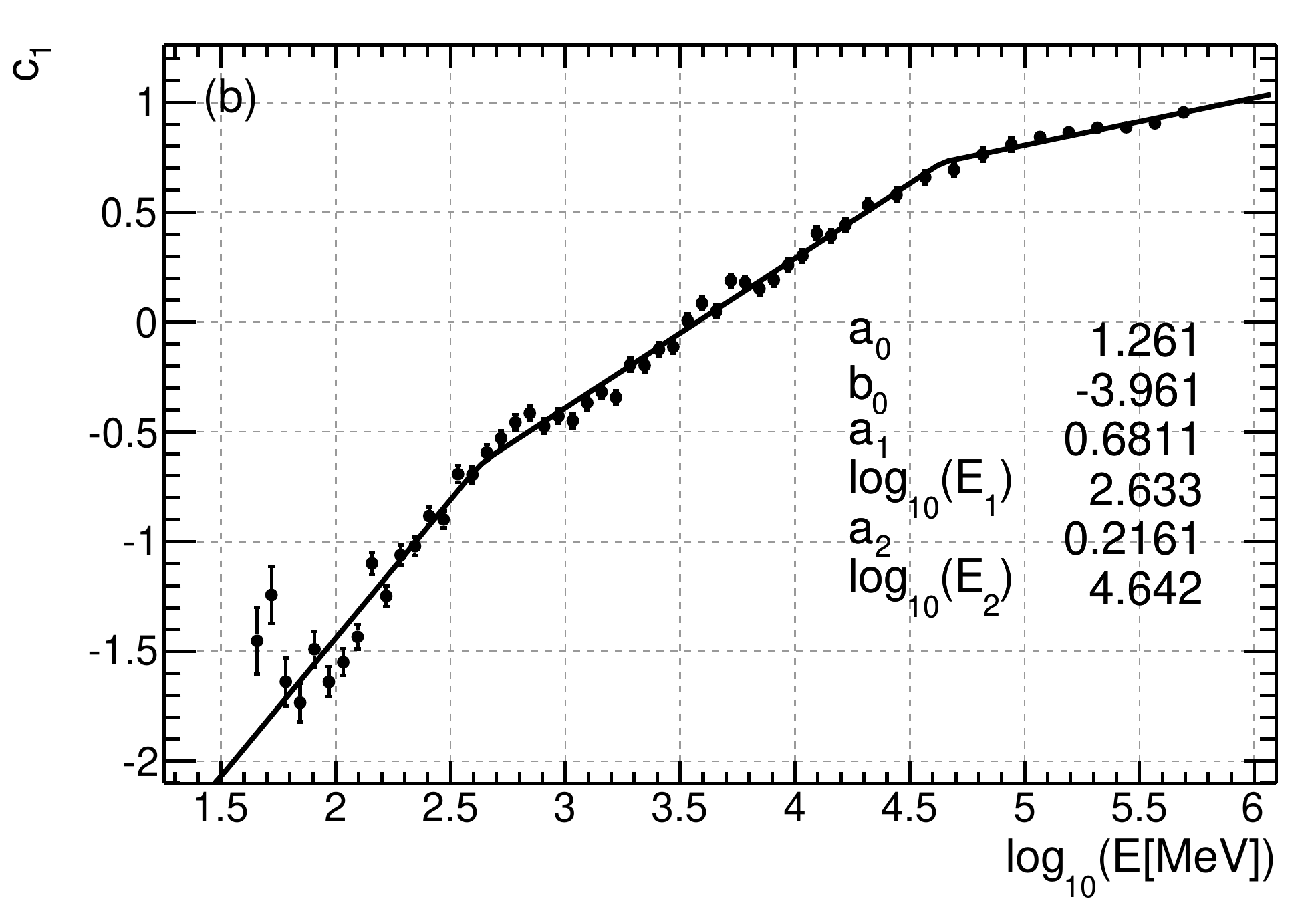}{
  \caption{Energy dependence of the linear fit parameters $c_0$ (a) and
    $c_1$ (b) for the \aeff\ live time dependence given in
    \Eqref{eq:aeff_livetime_dep} for front-converting events in the
    \irf{P7SOURCE::FRONT} event class.}
  \label{fig:aeff_lt_dep}
}

In \Figref{aeff_lt_dep} we plot $c_0$ and $c_1$ as a function of energy for
front-converting events in the \irf{P7SOURCE} event class. As shown by the solid
lines we use a simple piecewise linear fit to describe the energy dependence
and the fit parameters are stored in the \aeff\ tables.
As we mentioned, the effective area derived from \allgamma\ simulations that
have overlaid periodic trigger events is effectively corrected for the average
effect of this live time dependence, so we treat this additional correction
(which can be positive or negative) as a modulation of the tabulated \aeff.
Correction parameters are read in from the \aeff\ files of each set of IRFs,
and used to correct the calculated exposures.

The resulting corrections to the average \aeff\ can reach $-$30\% at 100~MeV,
decreasing at higher energies to $< 5\%$ above $\sim 10$~GeV.
The uncertainties in the corrections are much smaller; studies using flight
data confirm the Monte Carlo-based predictions to better than $2\%$. Note that
over a $53.4$~day orbital precession period this effect will tend to the
overall average correction described in \secref{subsubsec:Aeff_ghosts}, with
less than 1\% variation across the sky.

\NEWTEXT{Since the correction to the effective area is based on live time fraction, which 
is very strongly correlated with the CR intensity, it avoids any direct biases
from long term changes in the CR intensity associated with the influence of solar activity on the geomagnetic field.  
However, the correction does not address the possibility that 
the CR population changes during the solar cycle in such a way as to change the 
effective area dependence on the live time fraction.  Given the small change
in the daily averaged LAT trigger accept rate observed in the mission to date ($<5$\%), we neglect this effect.}\\

\subsubsection{$\phi$ Dependence}\label{subsubsec:aeff_phi_dep}

\begin{figure}[htb!]
  \centering\includegraphics[width=\onecolfigwidth]{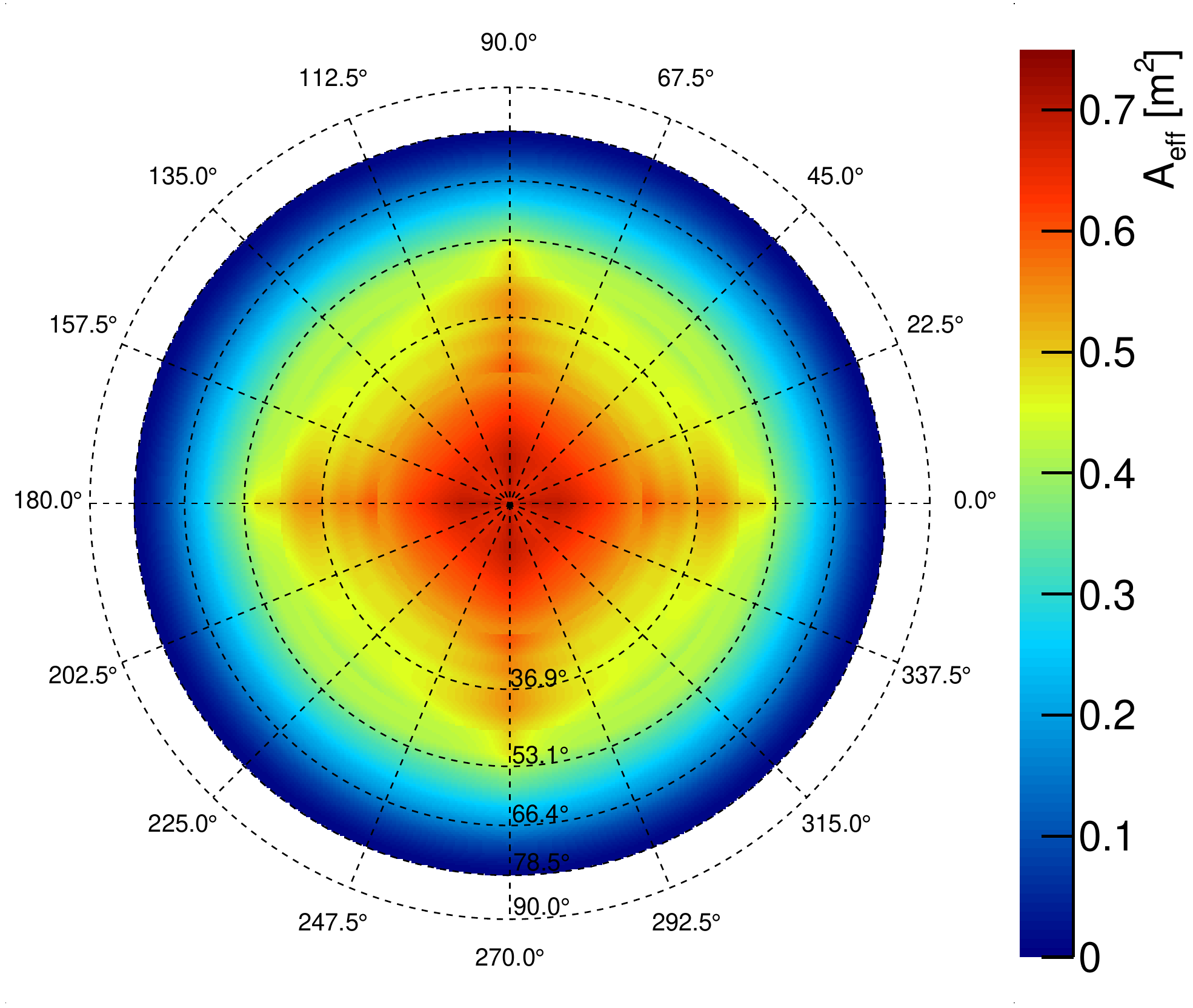}
  \caption{Total effective area at 10~GeV as a function
    of the incidence angle $\theta$ and the azimuthal angle $\phi$ for the
    \irf{P7SOURCE} event class.  The plot is shown in a zenith equal
    area projection with the LAT boresight at the center of the image;
    the concentric rings correspond to $0.2$ increments in $\cos(\theta)$.}
  \label{fig:aeff_phi_dep}
\end{figure}

The tabulated values of \aeff\ are averaged over the azimuthal
angle of incidence and shown in \figref{aeff_phi_dep}. Much of the azimuthal 
dependence of the effective area is geometrical, due to the square
shape of the LAT and the alignment of the gaps along 
the $x$ and $y$ axes.  \NEWTEXT{The RMS variation of the effective area as a 
function of $\phi$ is typically of the order of 5\% and exceeds 10\% only at low 
energies ($< 100$~MeV) or far off-axis ($\theta > 60^\circ)$ where the effective
area is small, and at very high energies ($> 100$~GeV) where the event rate is small.}

\begin{figure}[htb!]
  \centering\includegraphics[width=\onecolfigwidth]{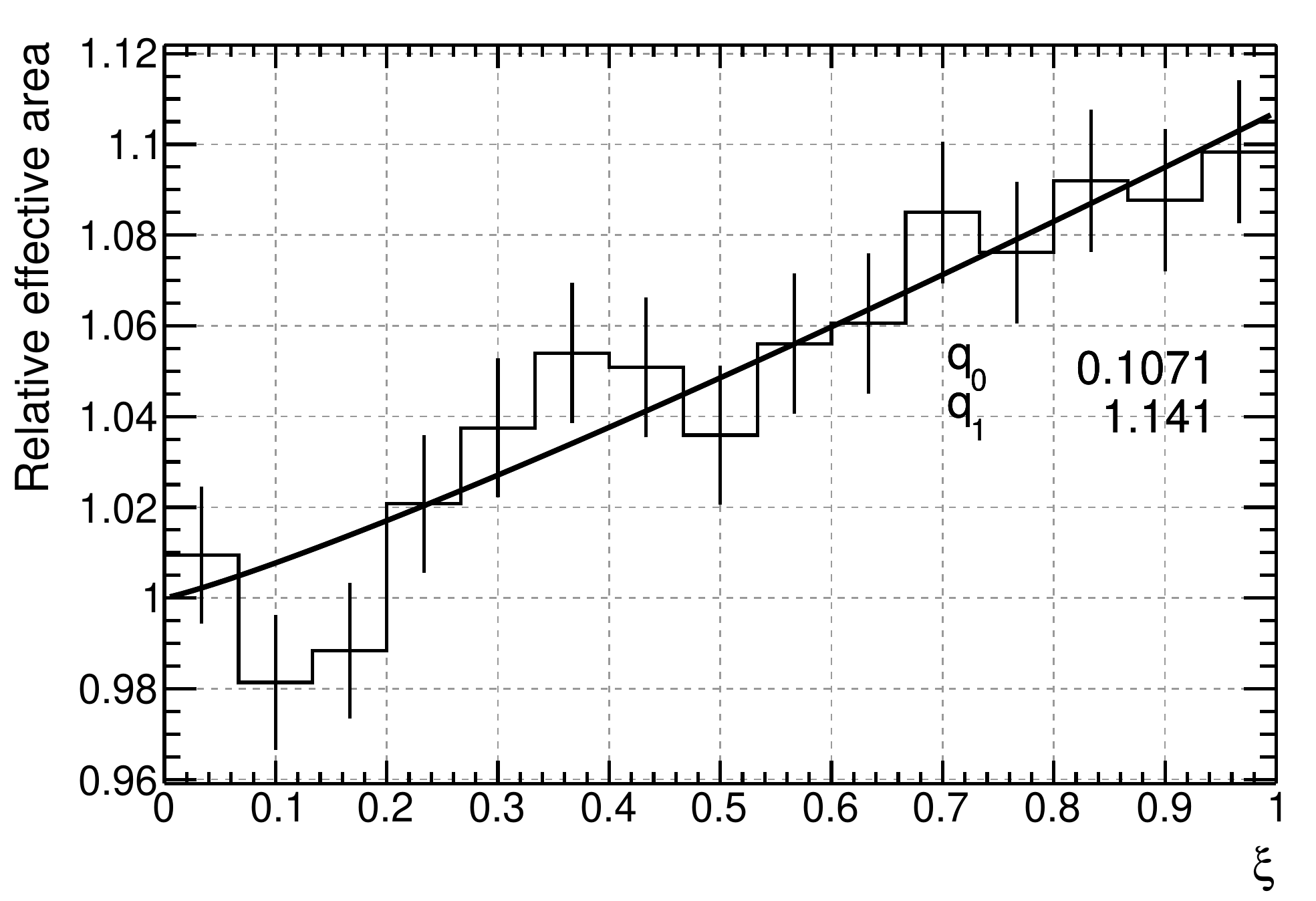}
  \caption{Example of \aeff\ azimuthal dependence \and fit.  The plot refers
  to the bin centered at 7.5~GeV and $30^\circ$ for the \irf{P7SOURCE} class,
  front section---a similar fit is performed in each $(E, \theta)$ bin. 
  The folded azimuthal angle $\xi$ is defined in \Eqref{eq:phi_xi}
  and the fit function in \Eqref{eq:xi_fit}.  Note that this plots shows 
  \aeff\ relative to \aeff\ for $\xi = 0$.}
  \label{fig:aeff_phi_dep_fit}
\end{figure}

In order to parametrize the azimuthal dependence of \aeff\ we fold the azimuthal
angle into the $[0^\circ,45^\circ]$ interval and remap it into $[0,1]$ using the
following transformation:\label{conv:foldedPhi}
\begin{equation}\label{eq:phi_xi}
  \xi = \frac{4}{\pi}\left|
  \left( \phi\!\!\!\!\mod\frac{\pi}{2} \right)- \frac{\pi}{4}\right|
\end{equation}
(the transformation maps $0^\circ$ to 1, $45^\circ$ to 0, $90^\circ$ to 1 and
so on).
The \allgamma\ events are binned in energy and $\theta$ and in each bin a
histogram of $\xi$ is fitted (see \parenfigref{aeff_phi_dep_fit} for an example).
Front- and back-converting events are treated separately.
The fitting function is:
\begin{equation}\label{eq:xi_fit}
  f(\xi) = 1 + q_0 \xi^{q_1}.
\end{equation}
The absolute scale is not important: we normalize the correction to result in
an average multiplicative factor of 1, so that the average \aeff\ is
tabulated. The fitted parameters $q_0$ and $q_1$ are stored in the \aeff\
tables.

We note in passing that the default for high-level analyses is to disregard
the azimuthal variations when calculating exposures because they average out when intervals of 
months or longer are considered\footnote{\NEWTEXT{Information on how
    to include the $\phi$ dependence 
of the effective area in exposure calculations can be found at 
{\webpage{http://fermi.gsfc.nasa.gov/ssc/data/analysis/scitools/binned\_likelihood\_tutorial.html}} and
{\webpage{http://fermi.gsfc.nasa.gov/ssc/data/analysis/scitools/help/gtltcube.txt}}.}}.
\NEWTEXT{Although the combined $\theta$ and $\phi$ dependence of the
  observing profile averages out only on year-long
timescales, the 8-fold symmetry of the LAT combined with the rotation
of the $x$-axis to track the Sun results in effective averaging over $\phi$ on short time scales.
In fact, we have found that ignoring the $\phi$ dependence of the effective area results in 
only a small variation of the exposure on 12-hour time scales ($<3\%$~RMS
at all energies).} \\

\subsection{Step by Step Performance of Cuts and Filters}\label{subsec:Aeff_stepByStep}

Before describing the studies we performed to validate our event selections
it is worth recalling that we require background rejection of $\sim 10^6$
while retaining high efficiency for \gammaRays. To achieve this we must select
events based on many different criteria that are applied in several stages
(see \secref{sec:LAT} and \secref{sec:event}), which complicates the task of
measuring the overall efficiency.
For validation purposes we specifically examine the agreement of the
selection efficiencies ($\eta$) between flight data and MC simulation.
Therefore, the most relevant quantity is
\begin{equation}\label{eq:aeff_cut_eff_ratio}
  R = \frac{\eta_{\rm data}}{\eta_{\rm mc}},
\end{equation}
\label{conv:ratioEffic}
which is in general a function of direction, energy and conversion point of
the incident \gammaRay.
The approach described in this section can be used both to quantify the
systematic uncertainties in \aeff\ and possibly to correct the IRFs derived
from the MC simulations, in cases of severe discrepancies with the flight data
(e.g., see~\secref{subsec:Aeff_flightAEff}).

\subsubsection{Background Subtraction}\label{subsec:Aeff_bkg_subtract}

To measure the efficiency of a cut on flight data, we can perform background
subtraction before and after the cut and compare the excess in the signal
region in the two cases. To the extent that the background subtraction is
correct, the cut efficiency is simply the ratio of the number of
background subtracted events before and after the cut we are testing:\label{conv:aeffEffic}
\begin{equation}\label{eq:aeff_cut_eff}
  \eta = \frac{n_{s,1} - r n_{b,1}}{n_{s,0} - r n_{b,0}},
\end{equation}
where the subscripts $s$ and $b$ indicate the signal and background regions,
the subscripts $0$ and $1$ indicate the samples before and after the cut under
test, and $r$\label{conv:ratioSize} is the ratio of the size of the background region to the size of
the signal region.
Note that the sample after the cut is a subset of the sample before the cut.
Therefore, for a reasonable sample size ($n_{s} > 10$), the statistical
uncertainty of the efficiency is:
\begin{equation}\label{eq:aeff_cut_eff_err}
  \delta\eta = \left[
    \frac{[( n_{s,0} - n_{s,1} ) + r^{2}(n_{b,0} - n_{b,1})]\eta^{2} +
      (n_{s,1} + r^2 n_{b,1})(1-\eta)^{2}}{( n_{s,0} - r n_{b,0} )^{2}}
  \right]^{1/2}.
\end{equation}
An example application of this technique using data from the Vela calibration
data set is shown in \figref{effic_ratio_example}.

\fourpanel{htb!}{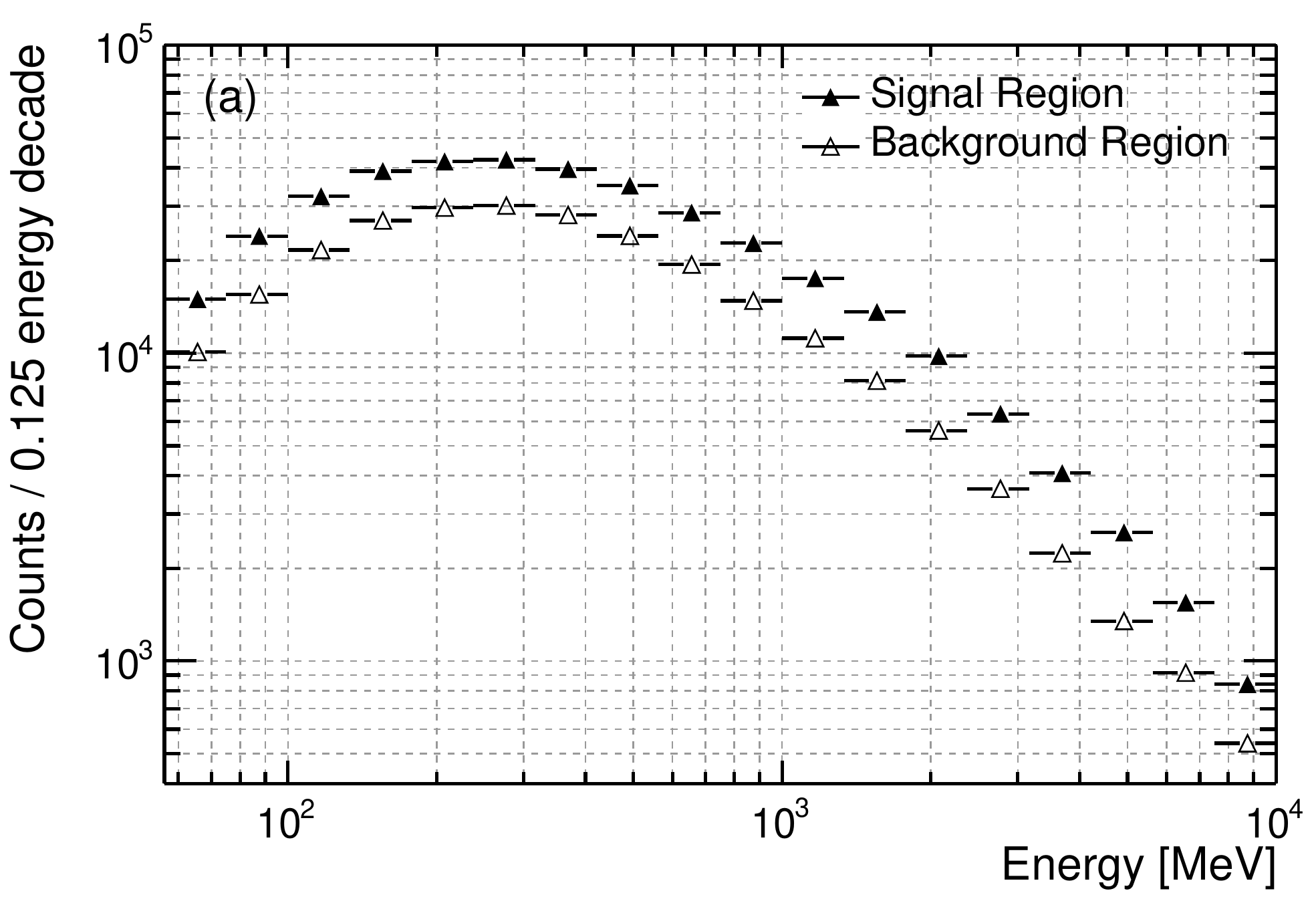}{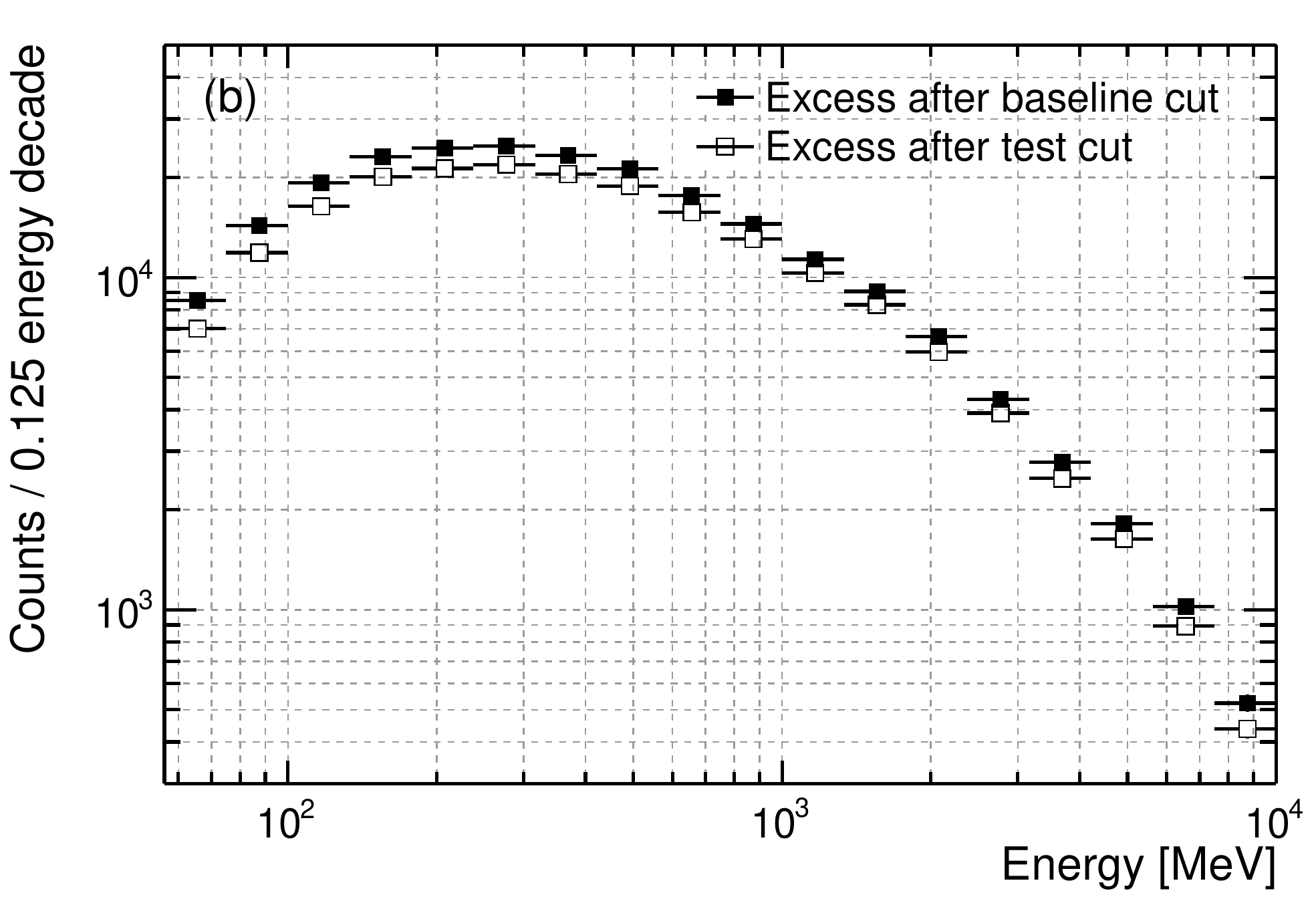}{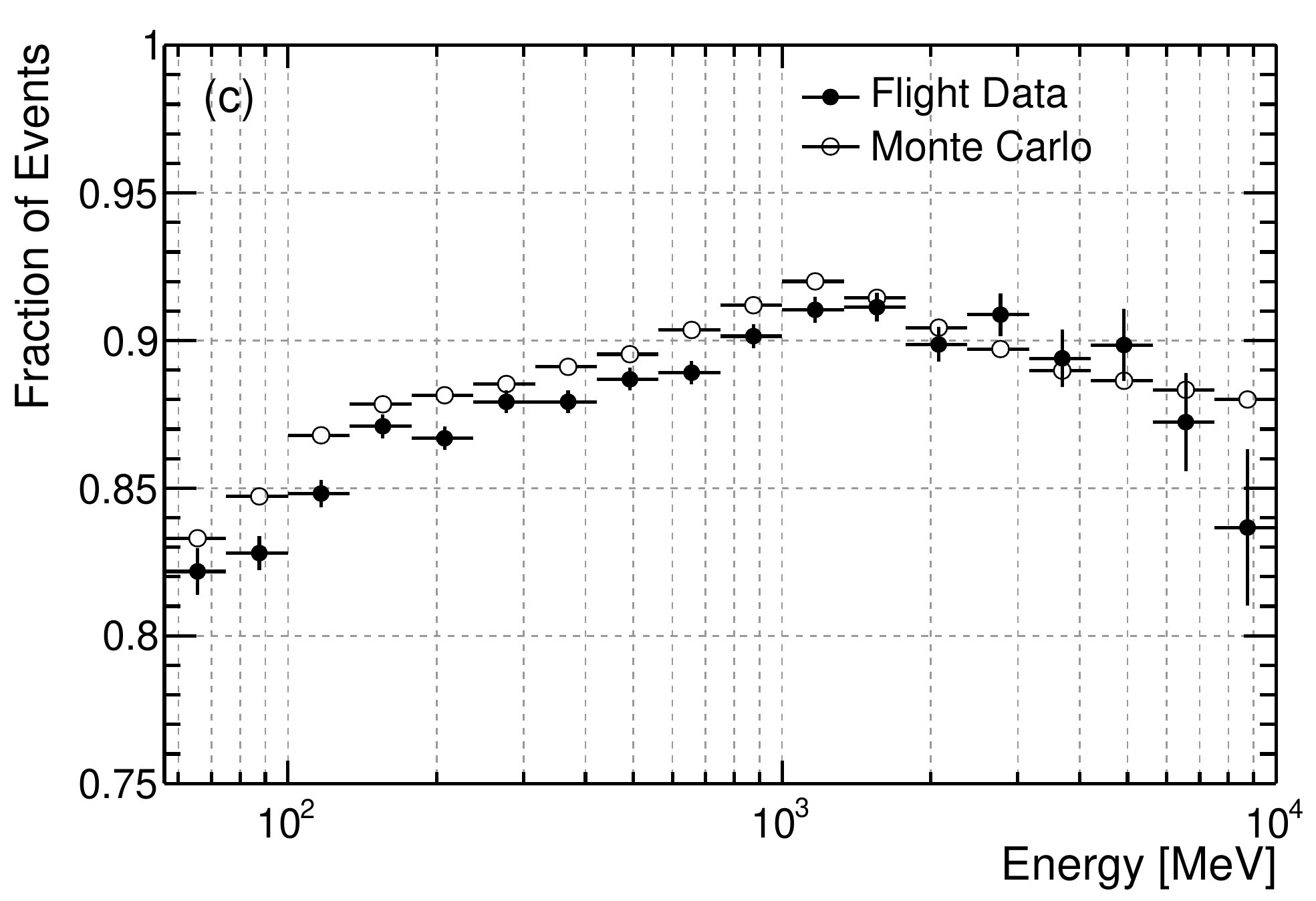}{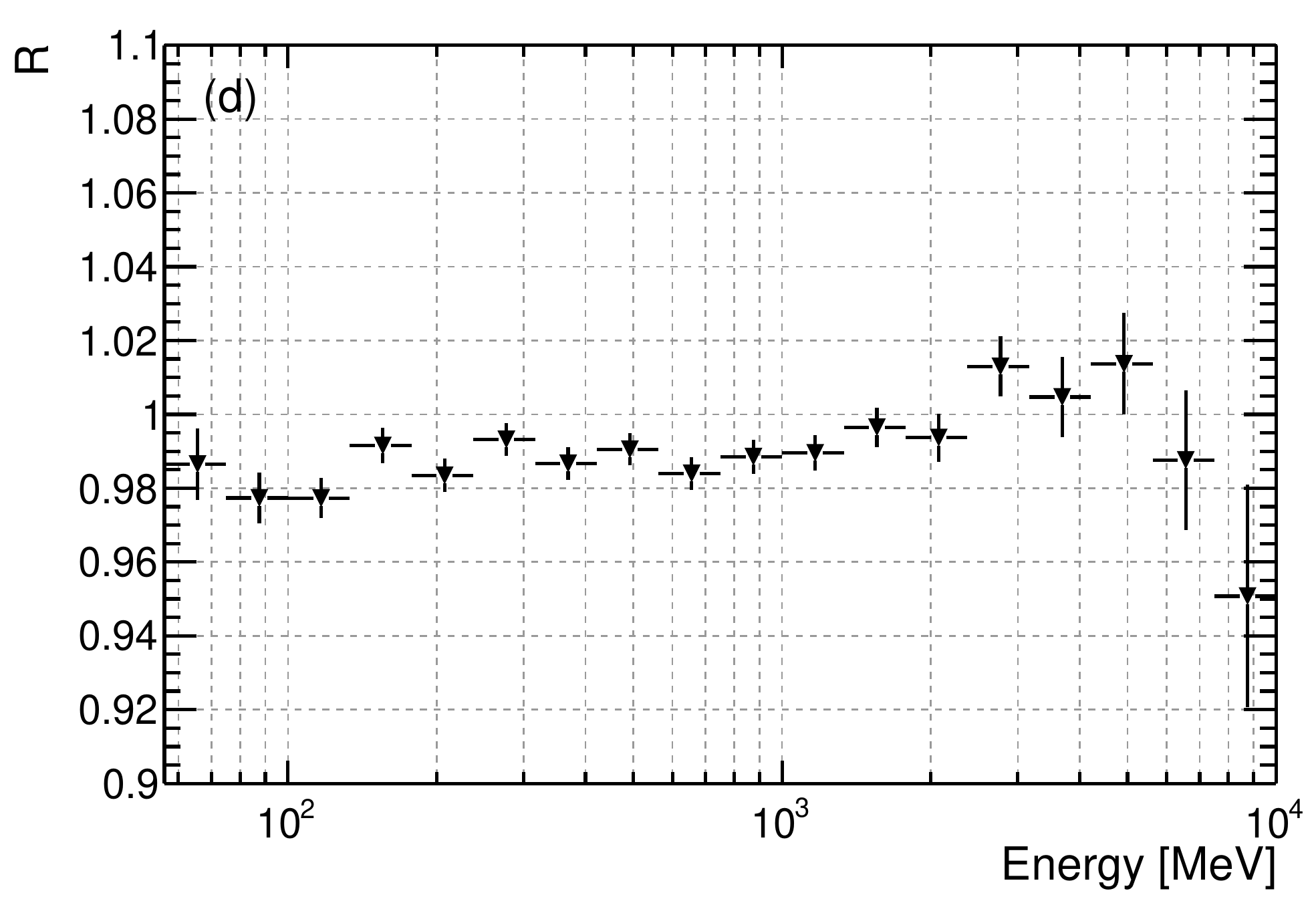}{
  \caption{Measurement of the efficiency of a cut using the background
    subtraction technique:
    (a) count spectra in the signal and background regions for the flight
    data set---note that the background counts have been scaled by 1/2 to
    account for the different phase ranges of the two regions;
    (b) excess in the signal region before and after the application of the
    test cut for the flight data sets---note that, for both (a) and (b), the
    corresponding plots for the MC data sets are not shown;
    (c) efficiency of the test cut $\eta$ for the flight and MC
    data sets;
    (d) ratio $R$ of the efficiencies as measured with flight data to the MC
    predictions.
    In this case the baseline sample is the \irf{P7SOURCE} event selection on
    the Vela calibration sample and the cut being tested is the \irf{P7CLEAN}
    event selection.}
  \label{fig:effic_ratio_example}
}

In many cases our cuts have significant overlap---two cuts may reject
many of the same events.  In such cases, measuring the efficiency of
the second cut after the first cut has been
applied would give a quite different result than measuring the efficiency of the
second cut without the first cut.  Therefore, whenever possible,
we use the background subtraction technique described above to evaluate 
the efficiency of each step of our event selection both independently of the
other steps as well after all the other cuts have been applied.

\subsubsection{Track-Finding and Fiducial Cut Efficiency}{\label{subsubsec:steps_trk_fid}}

Although the track-finding and fiducial cuts are applied midway through our
event selection process we choose to discuss them first for two reasons:
(i) we require an event direction to be able to perform a background
subtraction on flight data and (ii) we express \aeff\ as a function of $E$
and $\theta$.
Therefore, for performance studies we need to apply some minimum event
quality and fiducial cuts before considering events for analysis. The standard
cuts require at least one track found, with at least 5~MeV of energy in the
CAL and that the track extrapolates through at least 4 radiation
lengths in the CAL (see \secref{subsubsec:track_find_fiducial_cuts}).

\begin{figure}[htbp]
  \centering
  \includegraphics[width=\onecolfigwidth]{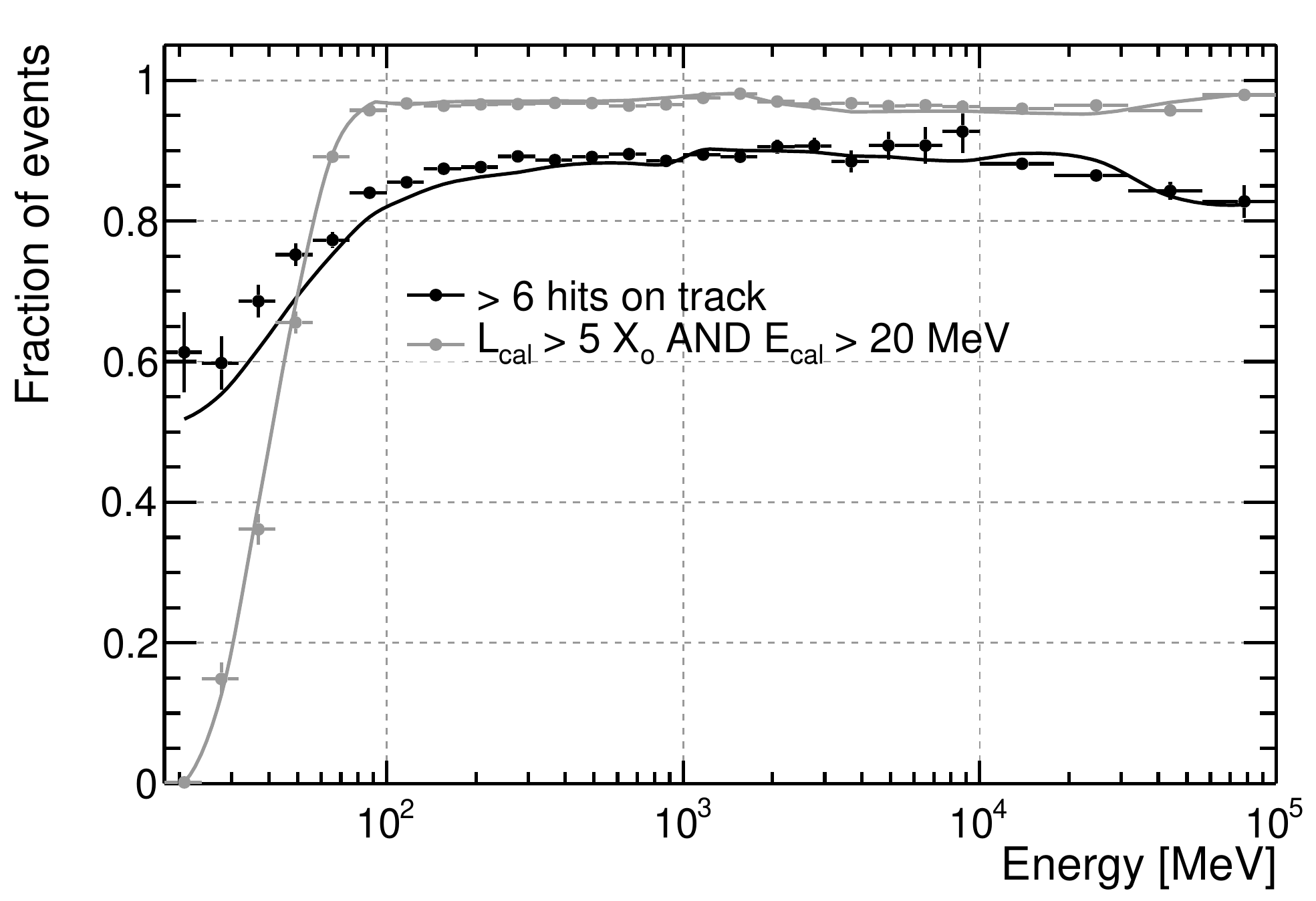}
  \caption{Validation of the track-finding and fiducial cuts. The data points 
    show the fraction of \irf{P7SOURCE} events in the Vela ($<10$~GeV) and Earth
    limb ($>10$~GeV) calibration samples that pass the fiducial cuts
    with some margin:  having more than the absolute minimum number
    of hits on a track (black points), or having more than 20~MeV of energy deposition
    and crossing more than 5 radiation lengths in the CAL (gray
    points).  The curves show the MC predictions for comparison.}
  \label{fig:effic_fid}
\end{figure}

Most events that fail these cuts have either poorly reconstructed directions,
poorly reconstructed energies, or both---which makes it difficult to study the
performance as a function of energy and direction.
Therefore, we choose to study the efficiency of the track-finding and fiducial
cuts by selecting events that \emph{almost} fail these cuts in the \irf{P7SOURCE} calibration samples.  
Specifically, we study the fraction of events that are very close to the cut thresholds
and verify that the flight data agree with our MC simulations,
see \figref{effic_fid}.

\subsubsection{Trigger Conditions and Trigger Request Efficiency}

The LAT hardware trigger, trigger configuration and on-board filter are
described in \secref{subsec:event_trigFilter}. For our purpose here we first
consider the fractions of \gammaRayHyph\ events that have one of the five
physics trigger conditions (\tptkr, \tproi, \tpcallo, \tpcalhi, \tpcno)
asserted.
Furthermore, only two of them (\tptkr\ and \tpcalhi) effectively serve to
initiate a trigger request. Of the others, \tproi\ does not exist without
\tptkr, \tpcallo\ alone does not open the trigger window and from the point of
view of selecting \gammaRayHyph\ events \tpcno\ is primarily a veto rather than
a trigger.
Since the \tpcalhi\ requires at least 1~GeV in a CAL channel, and our fiducial
cuts and quality cuts require at least one track, we are effectively using the
\tptkr\ condition as the primary trigger for \gammaRays\ up to such energies
that the \tpcalhi\ is very likely to be asserted. 

Using the \obfdgn\ events described in \secref{subsec:event_trigFilter}
we can measure the fractions of all trigger requests that have individual
primitives asserted. However, because of the high particle background rates
this does not really probe the trigger stage of the \gammaRayHyph\ selection
process.
On the other hand, for each of the five relevant trigger primitives we measure
this efficiency as a function of energy for the \allgamma\ sample as well as
the \irf{P7TRANSIENT} selection. For the \irf{P7TRANSIENT} selection we also
measure the fractions of events with each trigger primitive asserted and compare
these to the MC predictions. These are shown in \figref{effic_trig}.
The only notable discrepancy is that the MC over-predicts the fraction of
events having \tproi\ asserted at very high energies; this is likely related
to imperfect simulation of the backsplash from the CAL, and since the
\tpcallo\ and \tpcalhi\ are typically asserted for these events, it does not
affect the trigger readout decision (see \tabref{triggers}).

\twopanel{htb}{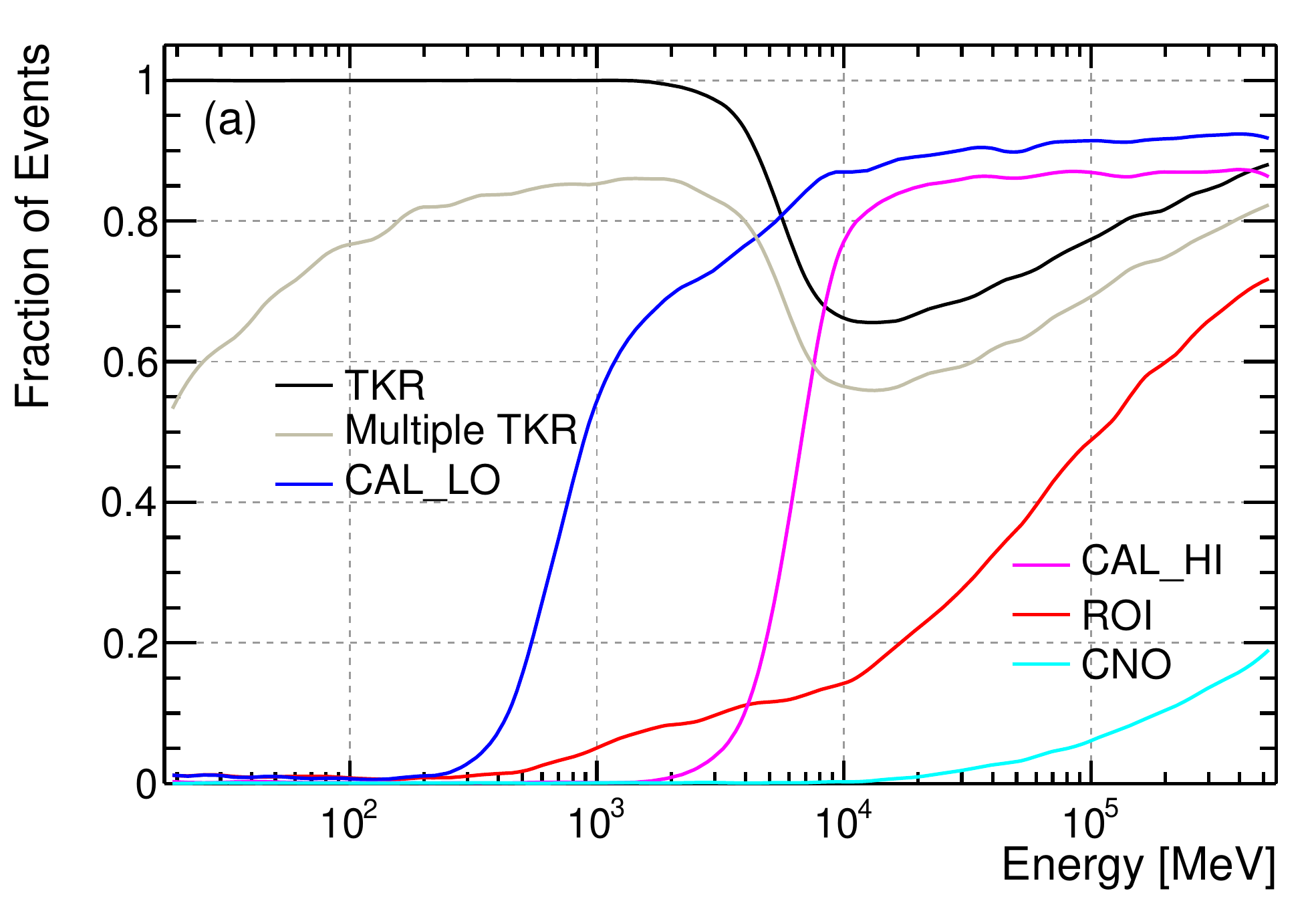}{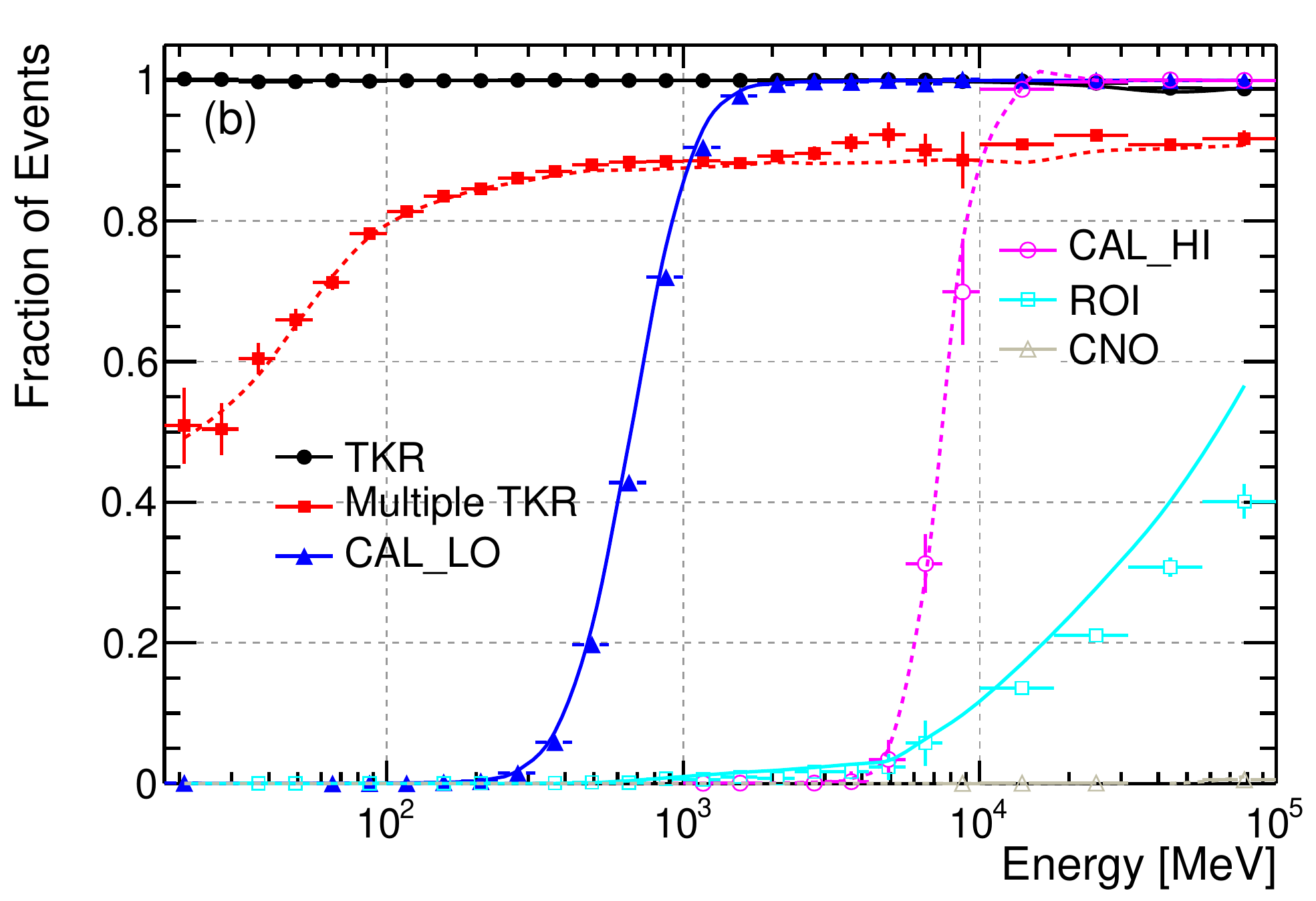}{
  \caption{(a) Fractions of all triggered events with various trigger conditions
    asserted as a function of the energy of the simulated event and (b)
    fraction of all \irf{P7TRANSIENT} events that have the same conditions
    asserted as a function of the reconstructed energy of the event, for
    MC simulations (lines) and flight data (points).
    ``Multiple TKR'' refers to the events with more than one three-in-a-row
    combinations asserted.}
  \label{fig:effic_trig}
}

\begin{fermiitemize}
\item The \tpcallo\ trigger condition requires 100~MeV of energy deposited in
  any CAL channel. This condition starts to be asserted for \gammaRays\ 
  at $\sim 300$~MeV and reaches full efficiency for events contained 
  within the CAL at $\sim 1$~GeV.
\item The \tpcalhi\ trigger condition requires 1~GeV of energy deposited in
  any CAL channel. This condition starts to be asserted for \gammaRays\
  at $\sim 3$~GeV and reaches full efficiency at $\sim 15$~GeV.
\item The \tpcno\ condition requires any ACD tile to have very large signal,
  consistent with the passage of a heavy ion. The \tpcno\ condition
  actually serves more to veto than to select an event for readout, but only becomes active
  at very high energies ($> 100$~GeV), where \tpcallo\ and \tpcalhi\ are already
  active.
\item The \tproi\ condition requires any ACD tile in a predefined region of
  interest associated with a TKR tower to have a signal above 0.45~MIP.
  The \tproi\ condition actually serves more to veto than to select 
  an event for readout and becomes active at a few GeV.  
\item Since the \tptkr\ condition serves as the primary trigger it is very
  difficult to measure the efficiency of the condition. However, we can
  estimate how well the MC simulates this efficiency by studying how well
  it models cases where the events almost fail the trigger request
  conditions. This is possible by calculating the fraction of events with
  exactly one combination of three hit layers in a row in the TKR.
  Most high-energy \gammaRays\ have many such combinations and would have
  triggered the LAT even if one hit had been lost. However, for back-converting
  low-energy events ($< 100$~MeV) the fraction of single combination events
  becomes significant.
\end{fermiitemize}

Since the fiducial cuts for our standard event selection require that
a track is found, and that the track extrapolates into the CAL, the
interplay between the trigger and the standard event selections is
actually quite simple.

\begin{fermiitemize}
\item At very low energies, the LAT trigger starts to become efficient at
  $\sim 10$~MeV and follows the efficiency of the \tptkr\ condition, which
  becomes fully efficient by $\sim 100$~MeV.
\item At around 1 GeV the \tpcallo\ condition becomes active. By design, this
  is considerably lower than the 10 GeV where the \tproi\ starts to be asserted
  because of backsplash. 
\item Above $\sim 10$~GeV the \tpcalhi\ condition becomes active and we
  no longer rely on the \tptkr\ condition as the primary driver of the
  trigger.
\end{fermiitemize}

Furthermore, events that are rejected because \tproi\ is asserted are
extremely unlikely to pass standard event class selections. Taken together,
this means that the only part of the LAT energy band where the trigger
has strong influence on \aeff\ is below $\sim 100$~MeV.

\subsubsection{On-Board Filter Efficiency}

Although the \obfgam\ has many different cuts
(see \secref{subsubsec:event_filter}), most events that are rejected by the
\obfgam\ would be either rejected by the fiducial cuts
(see \secref{subsubsec:track_find_fiducial_cuts}) or by the
\irf{P7TRANSIENT} event selection.  Accordingly, we choose to study the 
efficiency of the \obfgam\ as a whole, and only for those events which 
pass all the other selection criteria for the \irf{P7TRANSIENT} class event sample.

\begin{figure}[htbp]
  \centering
  \includegraphics[width=\onecolfigwidth]{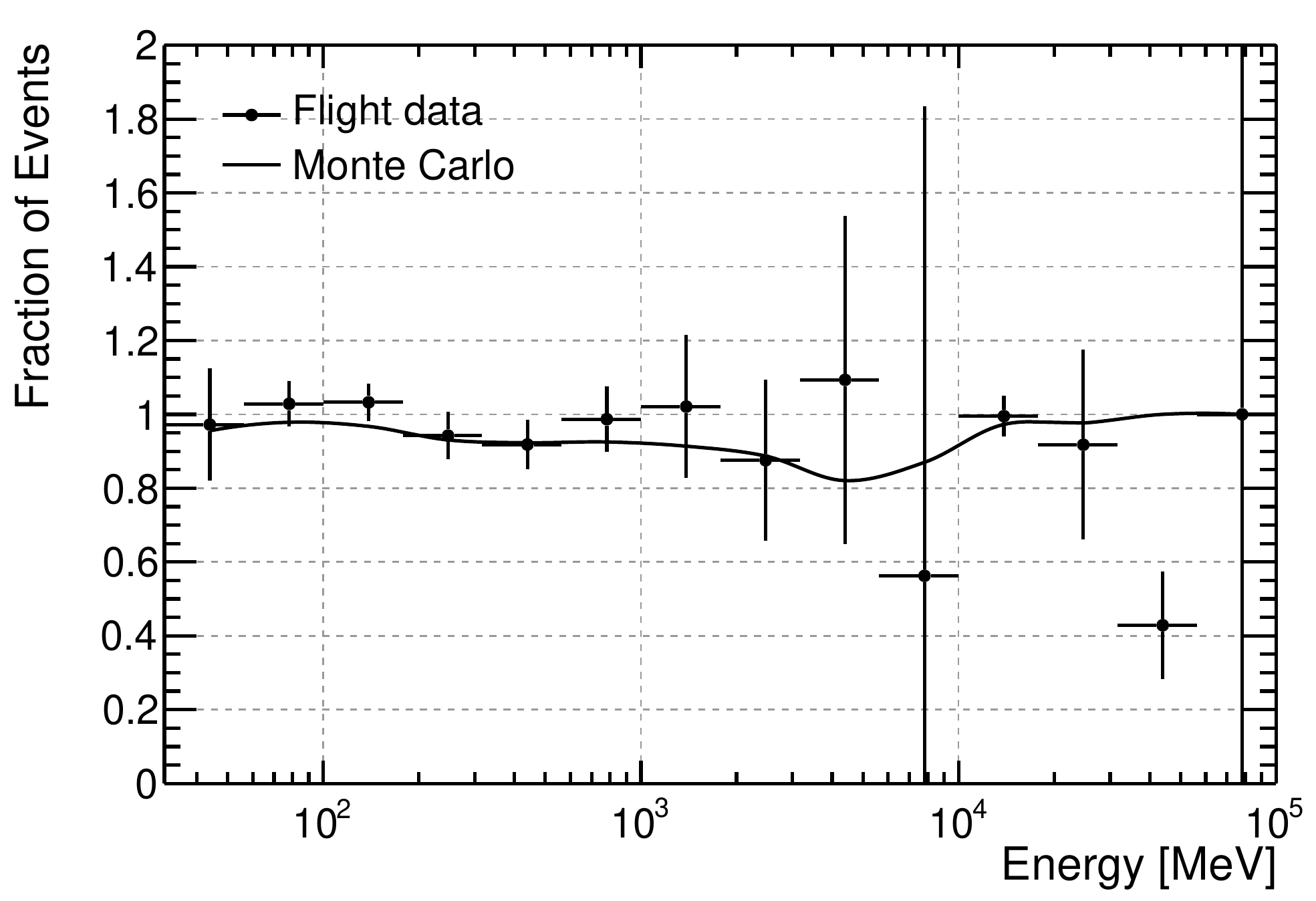}
  \caption{Fraction of events passing the \irf{P7TRANSIENT} selection
    that also pass the \obfgam\ for the \obfdgn\ flight data, compared to the
    MC predictions.  Note that point at $42$~GeV is well above the $20$~GeV
    high-energy pass criteria where the \obfgam\ becomes fully efficient,
    so the low measured value of the efficiency is likely due to a statistical 
    fluctuation in the subtracted background.}
  \label{fig:effic_obf}
\end{figure}

Since we downlink a small fraction of the events that fail the \obfgam\
(see \secref{subsec:event_trigFilter}), we can check that this is indeed the
case for this diagnostic sample.
However, the large prescale factor (250) for the \obfdgn\ and high level of
background rejection in the \irf{P7TRANSIENT} selection severely limit our
statistics for this study, and we can do little more than confirm that the
\obfgam\ is highly efficient for events in the \irf{P7TRANSIENT}
sample (see \parenfigref{effic_obf}).

\subsubsection{\irf{P7TRANSIENT} Class Efficiency}

Measuring the efficiency of the selection for the \irf{P7TRANSIENT} event
class is the most technically challenging part of the \aeff\ validation
for two reasons: (i) at the output of the \obfgam\ the background rates
are still great enough to overwhelm almost all traces of \gammaRayHyph\ signals
and (ii) the event analysis directs events that are tagged as likely due to CRs 
away from the remainder of the \gammaRayHyph\ selection
criteria, which means that many of the CT analyses are never applied, and
cannot be used to construct a cleaner sample on which we can measure
the efficiency of any of these cuts independently of the other cuts.

\Figref{effic_trans_sel}(a) shows the efficiency of each part of the
\irf{P7TRANSIENT} selection on events from the \allgamma\ sample that
have passed the \obfgam.  As stated above, the high levels of CR
background make it infeasible to use flight data to obtain stringent constraints on the efficiency of
these selection criteria.  Therefore, similarly to what we did for the track-finding and fiducial cuts, 
we also study the events that \emph{almost} fail these cuts in the \irf{P7TRANSIENT} calibration 
samples: these comparisons are shown in \Figref{effic_trans_sel}(b).

Because of the difficulty in validating the efficiency of the \irf{P7TRANSIENT}
event selection criteria, we have chosen to use the consistency checks described
in \secref{subsec:Aeff_consistency} to estimate the systematic uncertainty 
in our \aeff\ representation.

\twopanel{htb}{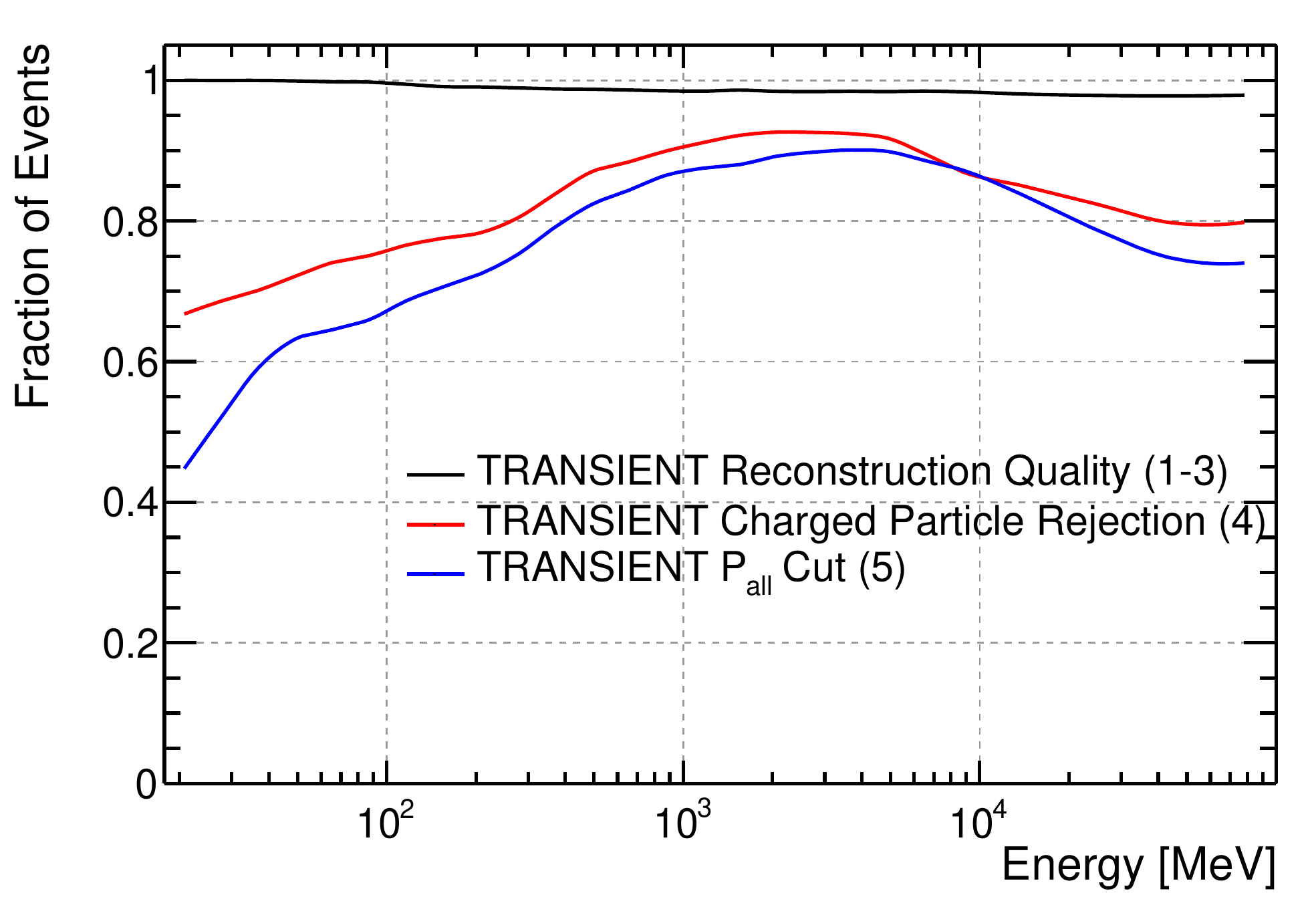}{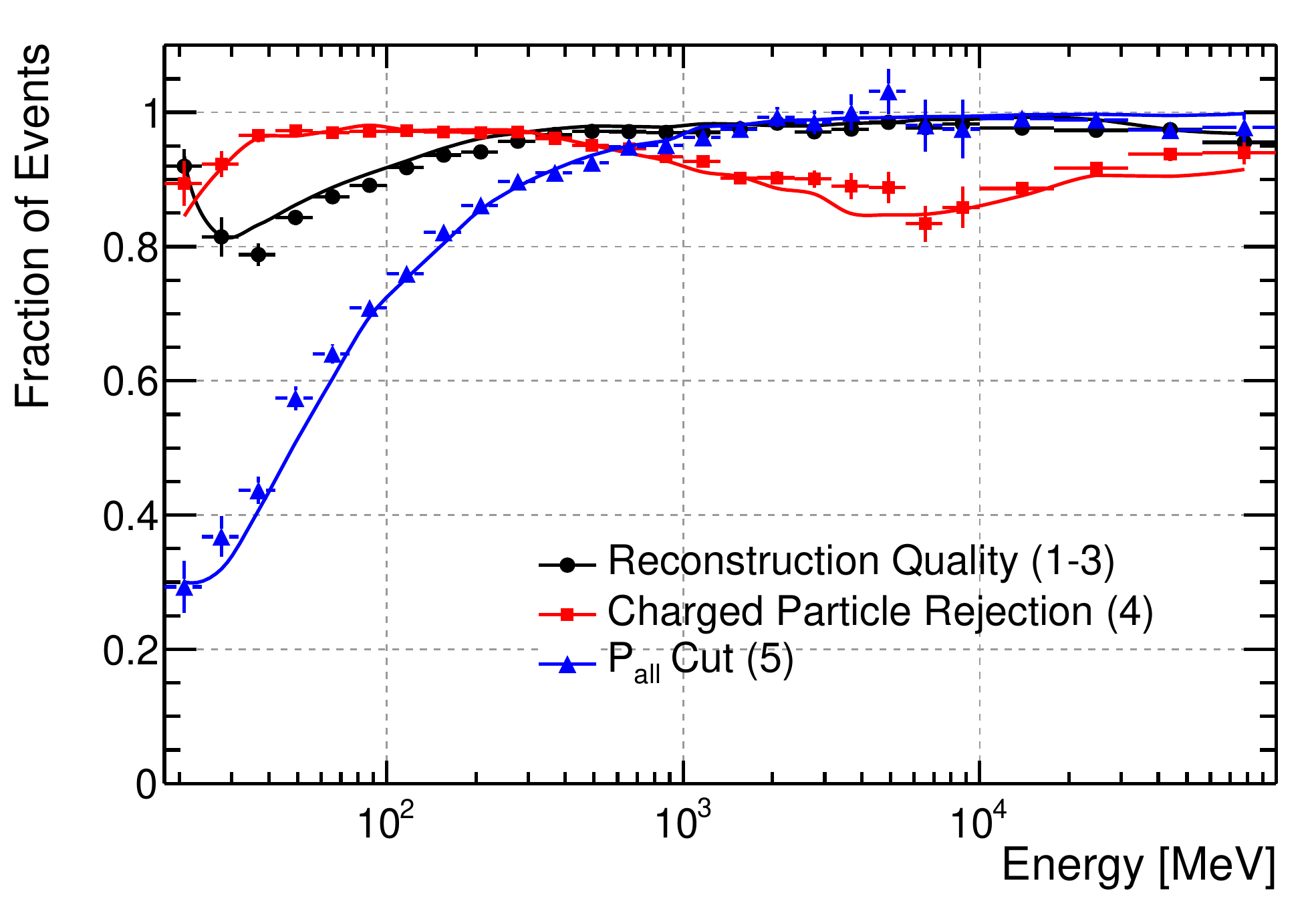}{
  \caption{(a) Fractions of events in the \allgamma\ sample passing the 
  \obfgam\ and track-finding and fiducial cuts that also pass each part of 
  the \irf{P7TRANSIENT} (\secref{subsubsec:p7trans_selection}) event selection. 
  (b) Fractions of events in the \irf{P7TRANSIENT} calibration samples that pass 
  similar cuts with some margin for MC simulations (lines) and flight data (points).
  The numbers in the legends refer to the list of cuts in \secref{subsubsec:p7trans_selection}.}
  \label{fig:effic_trans_sel}
}

\subsubsection{\irf{P7SOURCE},\irf{P7CLEAN} and \irf{P7ULTRACLEAN} Class Efficiencies}\label{subsubsec:steps_source_and_up}

Although the \irf{P7TRANSIENT} event class is dominated by residual background
across the entire LAT energy range, the background levels are at least reduced
to the point where the \gammaRayHyph\ signals in the calibration samples
described in \secref{subsec:LAT_methods} are clearly detectable. This makes
the validation of the effective area from this point on much easier. We can
compare the efficiency of each cut, as measured on flight data, to the MC
prediction with the method described in \secref{subsec:Aeff_bkg_subtract}.
These comparisons are shown in \figref{effic_source_and_up_sel}.

\twopanel{htb}{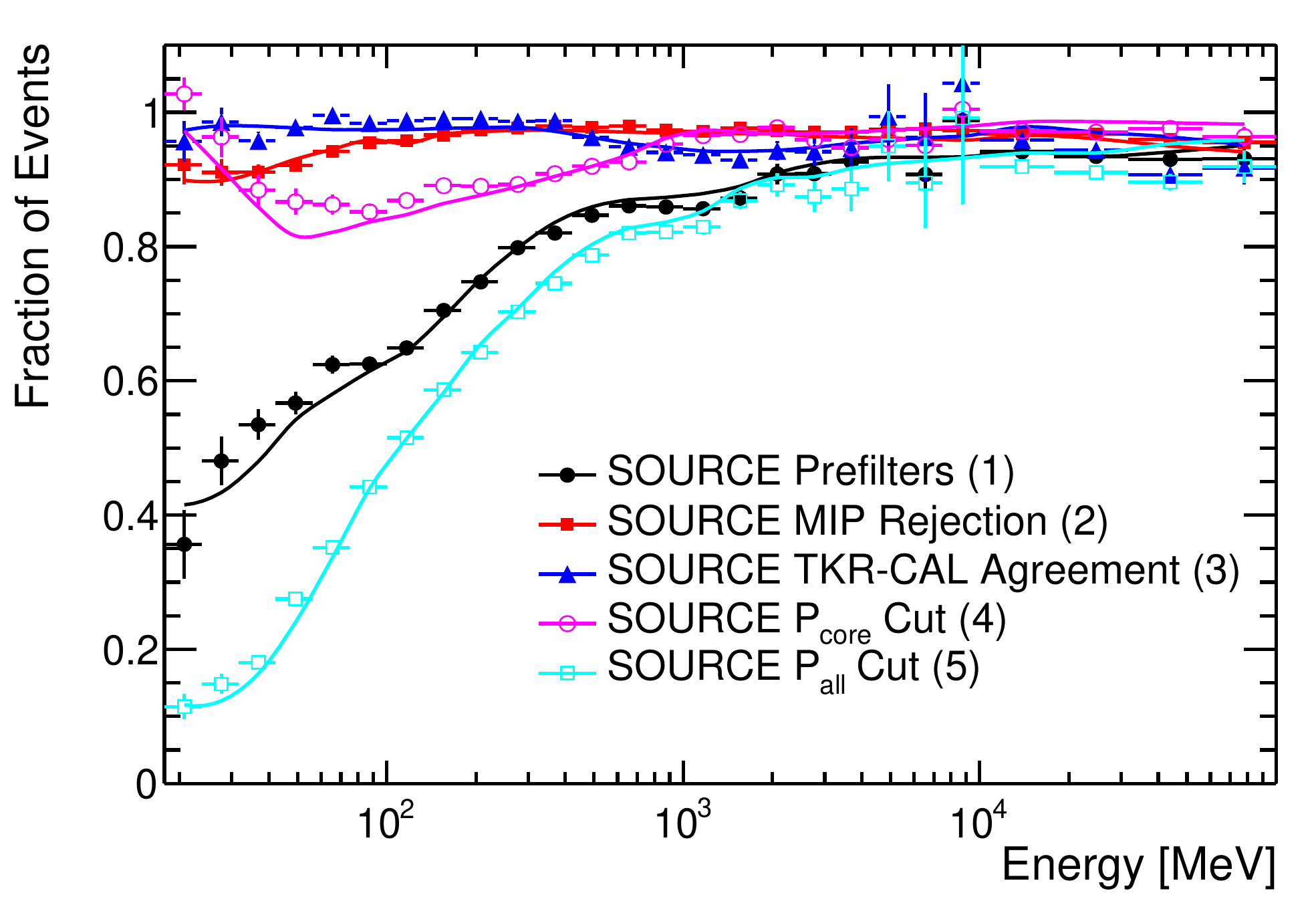}{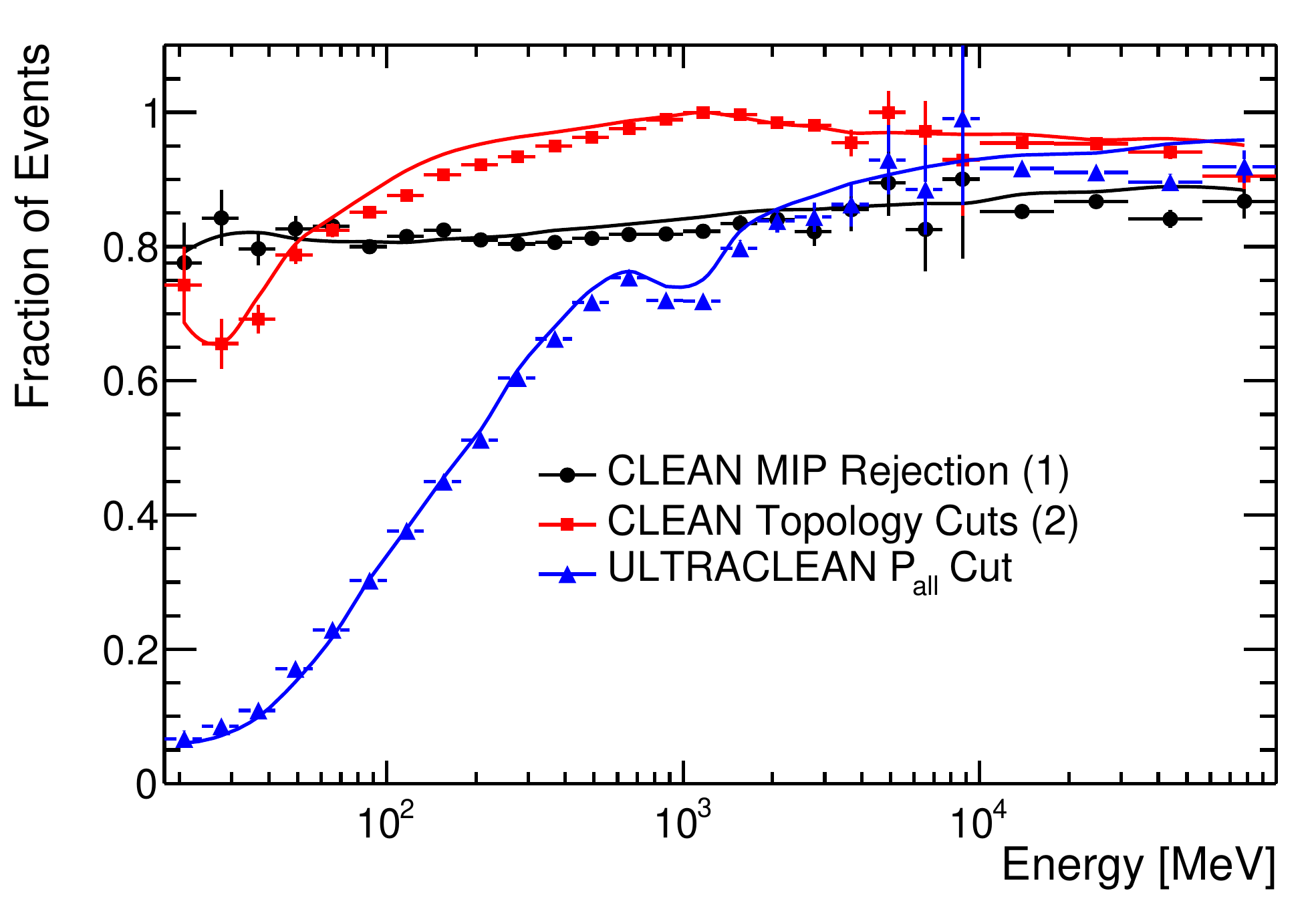}{
  \caption{Fractions of events in the \irf{P7TRANSIENT} calibration samples 
  that pass each part of the (a) \irf{P7SOURCE} event selection (\secref{subsubsec:p7source_selection}) and 
  (b) \irf{P7CLEAN} (\secref{subsubsec:p7clean_selection}) and 
  \irf{P7ULTRACLEAN} (\secref{subsubsec:p7ultra_selection}) event selections for MC 
  simulations (lines) and flight data (points). The numbers in the legends refer to 
  the list of cuts in \secref{subsubsec:p7source_selection} and \secref{subsubsec:p7clean_selection}.}
  \label{fig:effic_source_and_up_sel}
}

\subsection{In-Flight Effective Area}\label{subsec:Aeff_flightAEff}

We observed that the efficiency for one part of our \psix\ event selection was
systematically lower near 10~GeV for flight data than for the \allgamma\ we used to
evaluate the effective area, and we attempted to correct the effective area
tables to provide more accurate flux measurements for \gammaRayHyph\ sources.
To be more specific, the offending cut is the \psix-equivalent
of the \irf{P7SOURCE} cut on the quality of direction reconstruction described
in \secref{subsubsec:p7source_selection} (item \ref{item:p7trans_ctbcore_cut}
in the numbered list).  It is important to note that for \pseven\ we chose
instead to make this particular cut less stringent to avoid the need
to make such a correction, so that the in-flight corrections discussed here do
not apply to the effective area tables for the \pseven\ standard \gammaRayHyph\
classes.   Furthermore, we traced the discrepancy to limitations in
pre-launch calibration algorithm of CAL light asymmetry that resulted
in degraded position and direction resolution in the CAL:  above
$\sim 1$~GeV consistency between the TKR and CAL position and
direction measurements is a strong indicator of accurate direction reconstruction.

We measure the ratio of cut efficiency between flight data and MC as a function
of energy and incidence angle (see \figref{effRatio_2d})---as described in
\secref{subsec:Aeff_bkg_subtract}---and we use it to correct the MC
based \aeff.
Because most of our calibration sources have limited statistics relative to
the \allgamma\ samples we are forced to use fewer bins when calculating the
efficiency ratios.  Furthermore, to avoid inducing sharp spectral features in
measurements of \gammaRayHyph\ sources we smooth the energy and angle
dependence of the efficiency ratio.
Specifically, we split the data into two $\cos(\theta)$ bins: $[0.2, 0.7]$ and
$[0.7, 1.0]$.
For values between the bin centers (i.e., $\cos(\theta) \in [0.45, 0.85]$) we
perform a linear interpolation in $\cos(\theta)$.
For values outside that range we use the correction factor from the appropriate $\cos(\theta)$ bin.

\twopanel{htb}{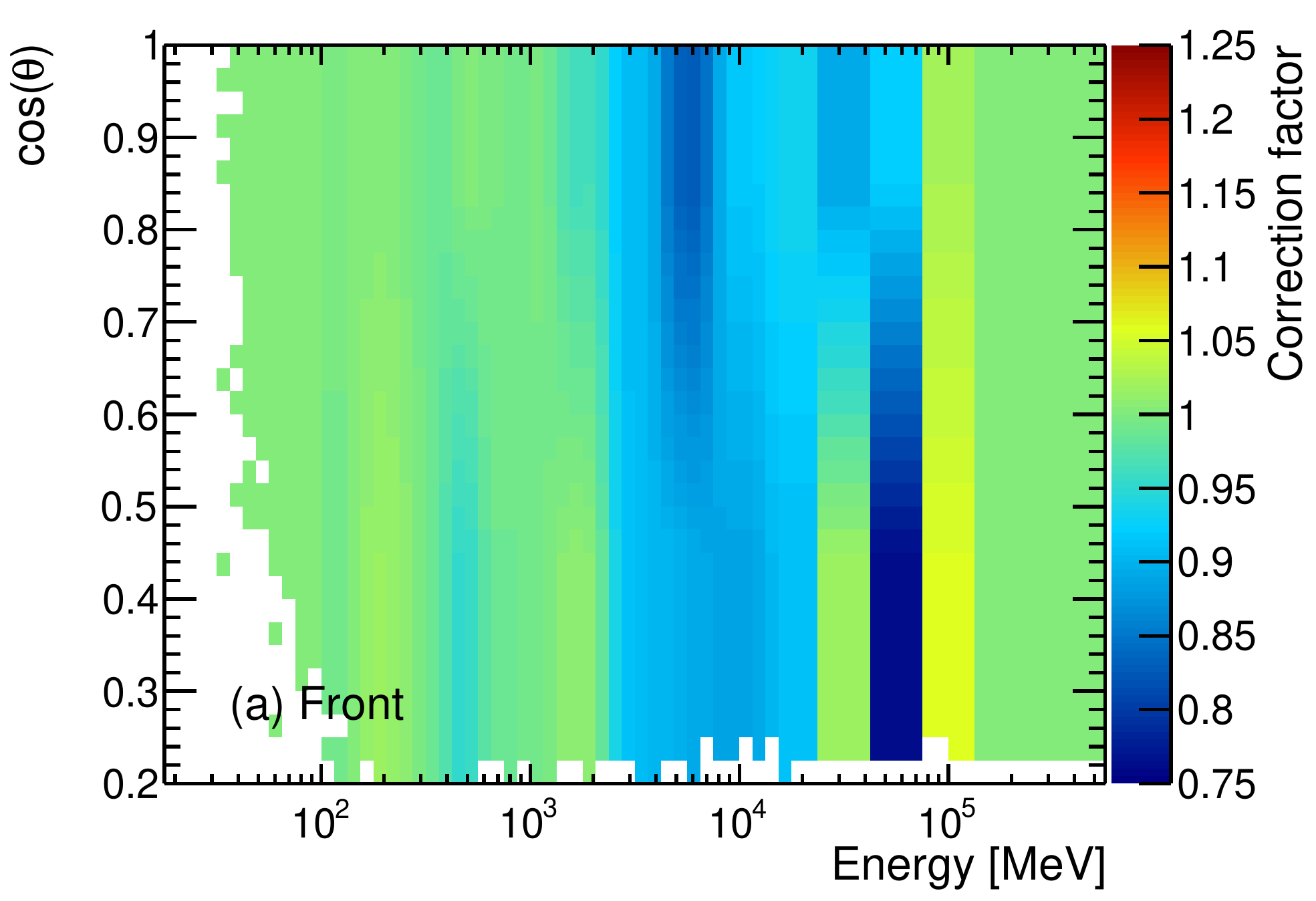}{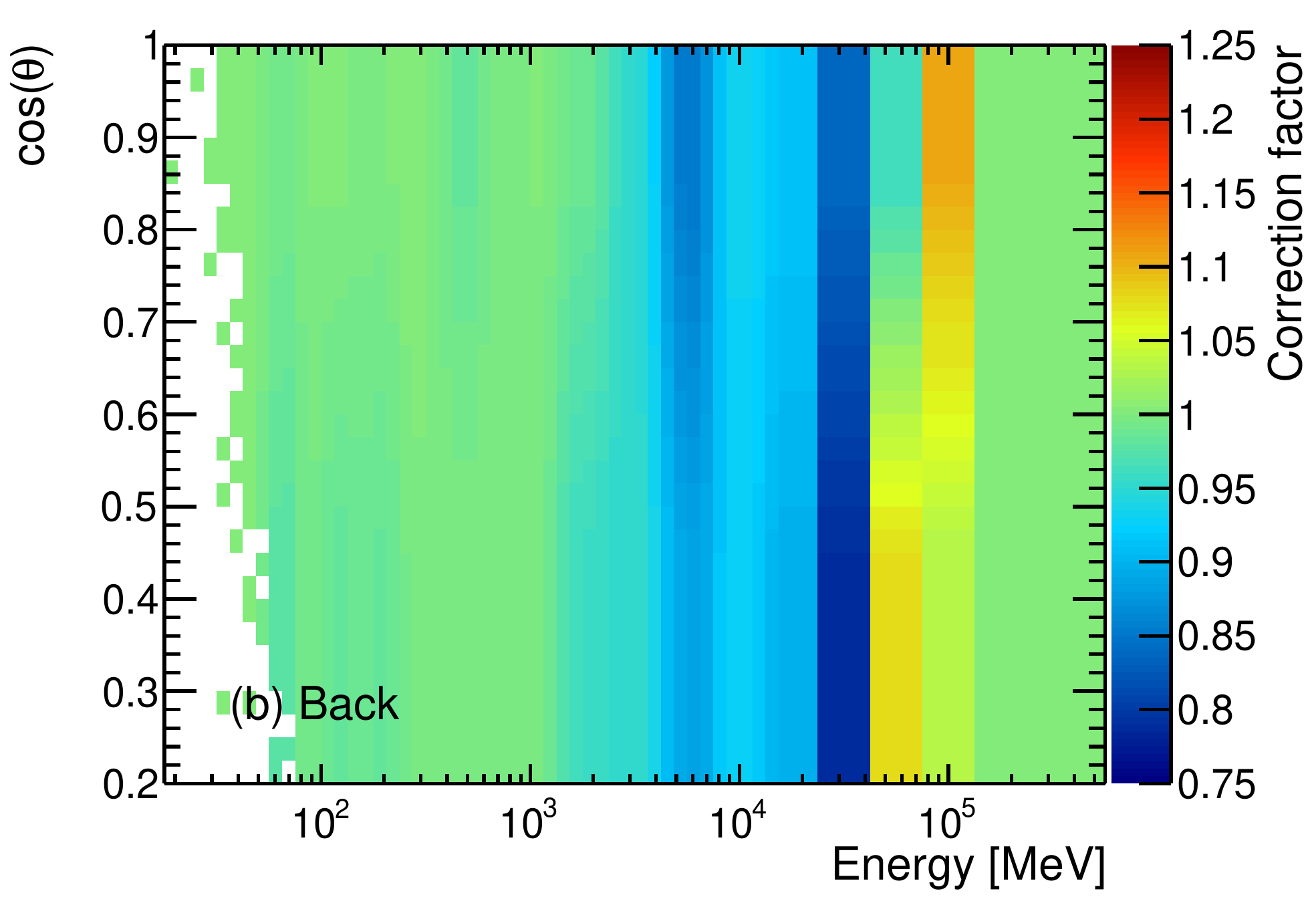}{
  \caption{Ratio of the \irf{P6\_V11\_DIFFUSE} to \irf{P6\_V3\_DIFFUSE} \aeff\
    for front-converting (a) and back-converting (b) events.
    The \irf{P6\_V11\_DIFFUSE} \aeff\ tables have a correction factor relative 
    to \irf{P6\_V3\_DIFFUSE} that is based on the ratio of the
    efficiencies between flight and simulated data for the selection cut on
    the direction reconstruction quality (see
    item~\ref{item:p7trans_ctbcore_cut} in
    \secref{subsubsec:p7source_selection}).
    The underflow bins (white areas) had no \aeff\ in \irf{P6\_V3\_DIFFUSE}.
    As we did not interpolate the correction factor along the energy axis, fluctuations
    in the correction factor lead to the vertical bands visible in this figure.}
\label{fig:effRatio_2d}
}

\begin{figure}[htb]
  \centering
  \includegraphics[width=\onecolfigwidth]{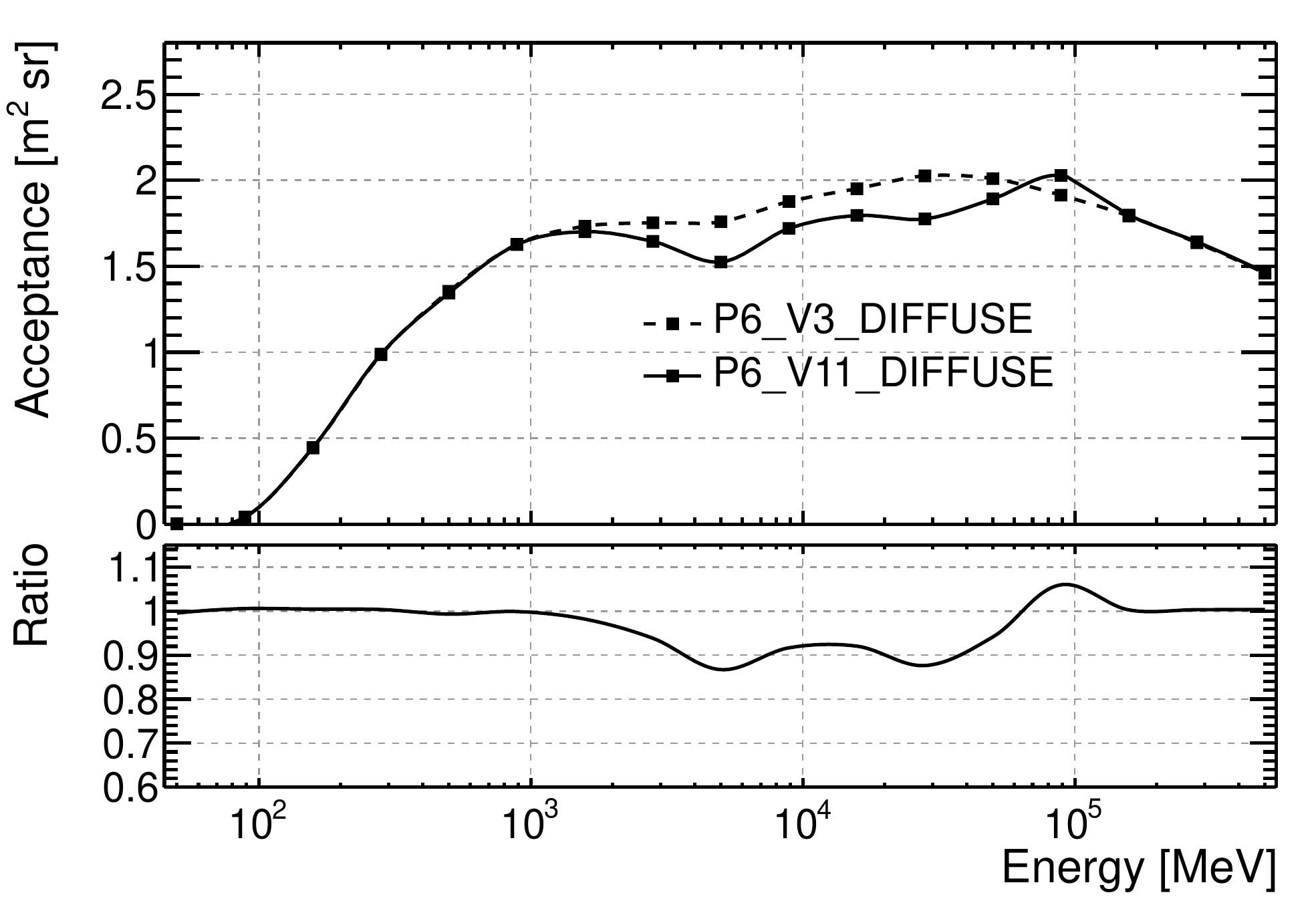}
  \caption{Comparison of acceptance between \irf{P6\_V3\_DIFFUSE} and
    \irf{P6\_V11\_DIFFUSE} IRFs. The only difference between the IRFs is the
    application of the corrections to \aeff\ for \irf{P6\_V11\_DIFFUSE}
    described in \secref{subsec:Aeff_flightAEff}.}
  \label{fig:aeff_corrected}
\end{figure}

This procedure yields a correction map that we use to convert from the MC
\aeff\ to our best estimate of the true in-flight \aeff\ for the
\irf{P6\_V11\_DIFFUSE} set of IRFs, as shown in \figref{aeff_corrected}.
Note that these IRFs should only be used together with the corresponding
rescaled models for Galactic and isotropic diffuse emission provided by the
FSSC\footnote{Specifically, \filename{gll\_iem\_v02\_P6\_V11\_DIFFUSE.fit} and
\filename{isotropic\_iem\_v02\_P6\_V11\_DIFFUSE.txt}, available at
\webpage{http://fermi.gsfc.nasa.gov/ssc/data/access/lat/BackgroundModels.html}.}.

\subsection{Consistency Checks}\label{subsec:Aeff_consistency}

In this section we will describe several consistency checks we performed to
estimate how well we understand \aeff. Each of these tests consists of
splitting a specific \gammaRayHyph\ event selection into two subsets and
comparing the fraction in each of the subsets from the flight data to the
predictions from MC simulations.

\fourpanel{h!tb}{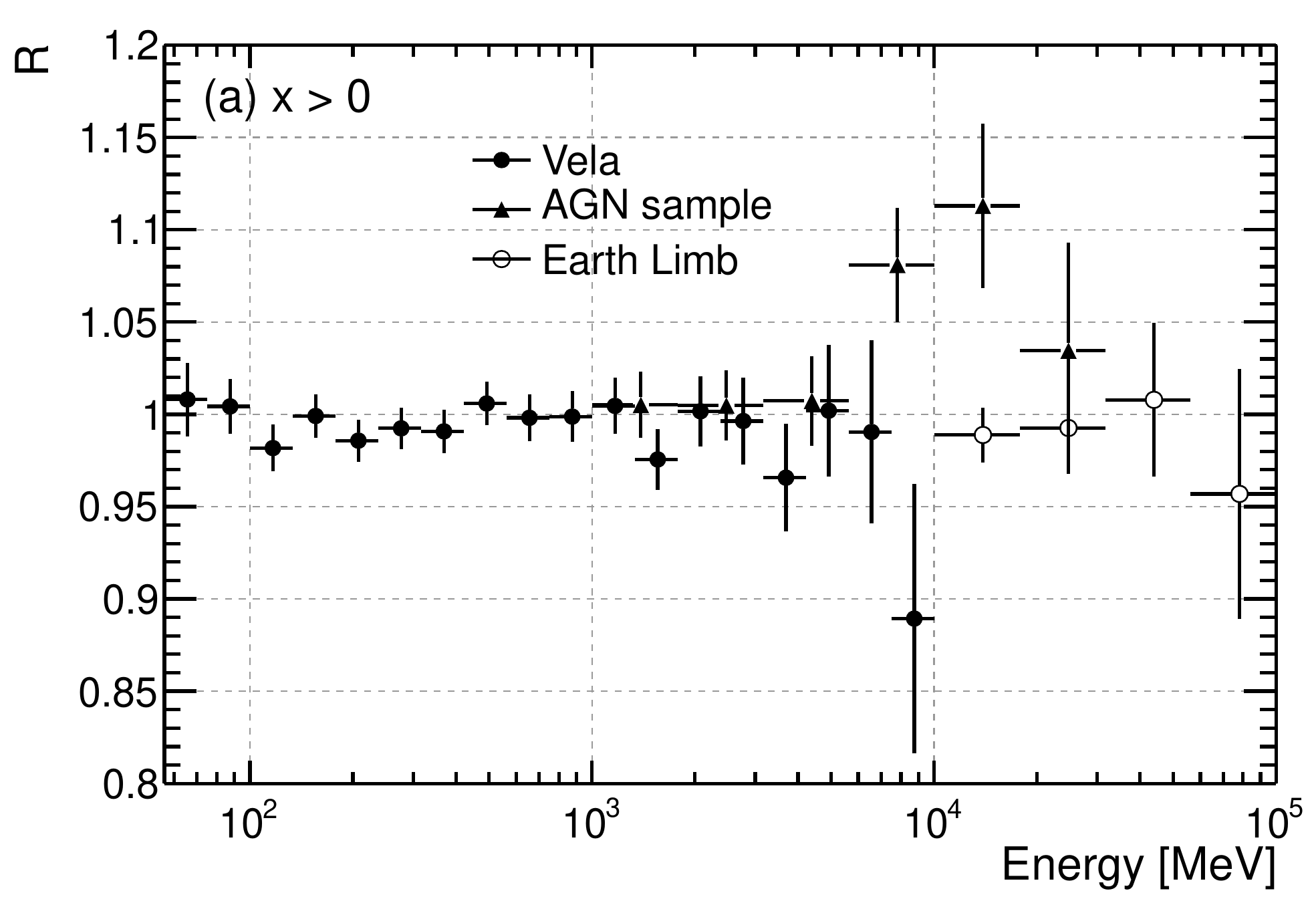}{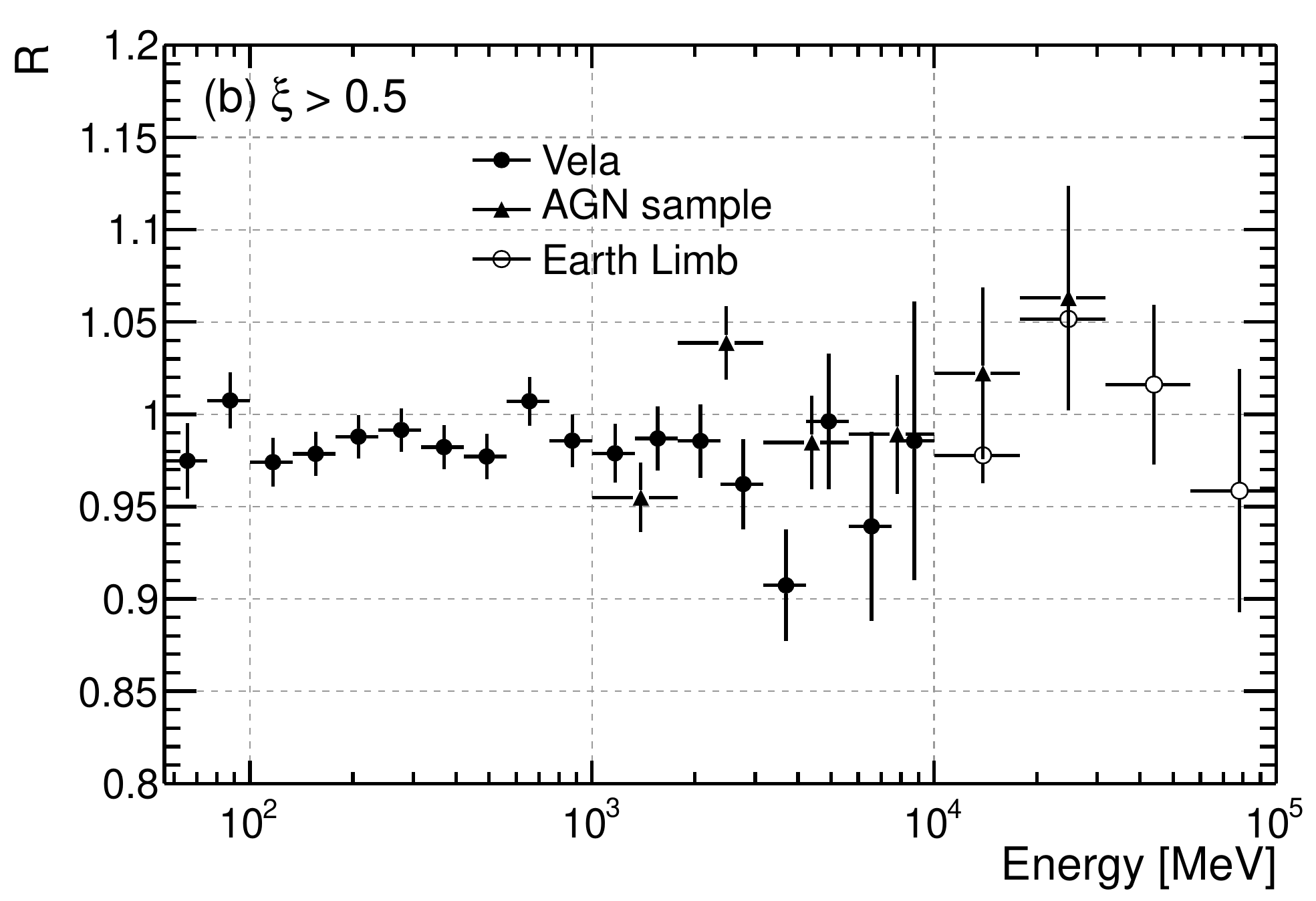}{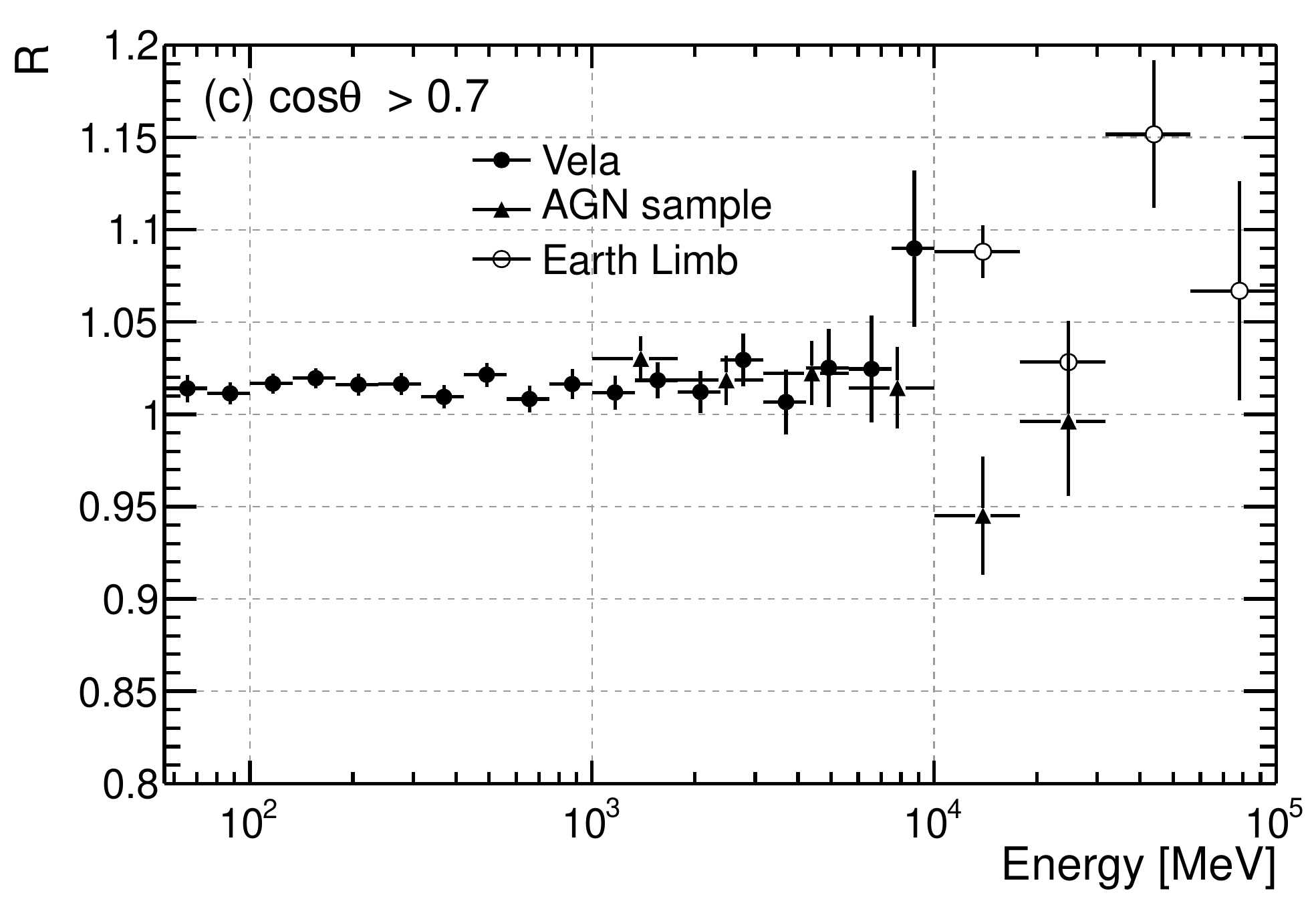}{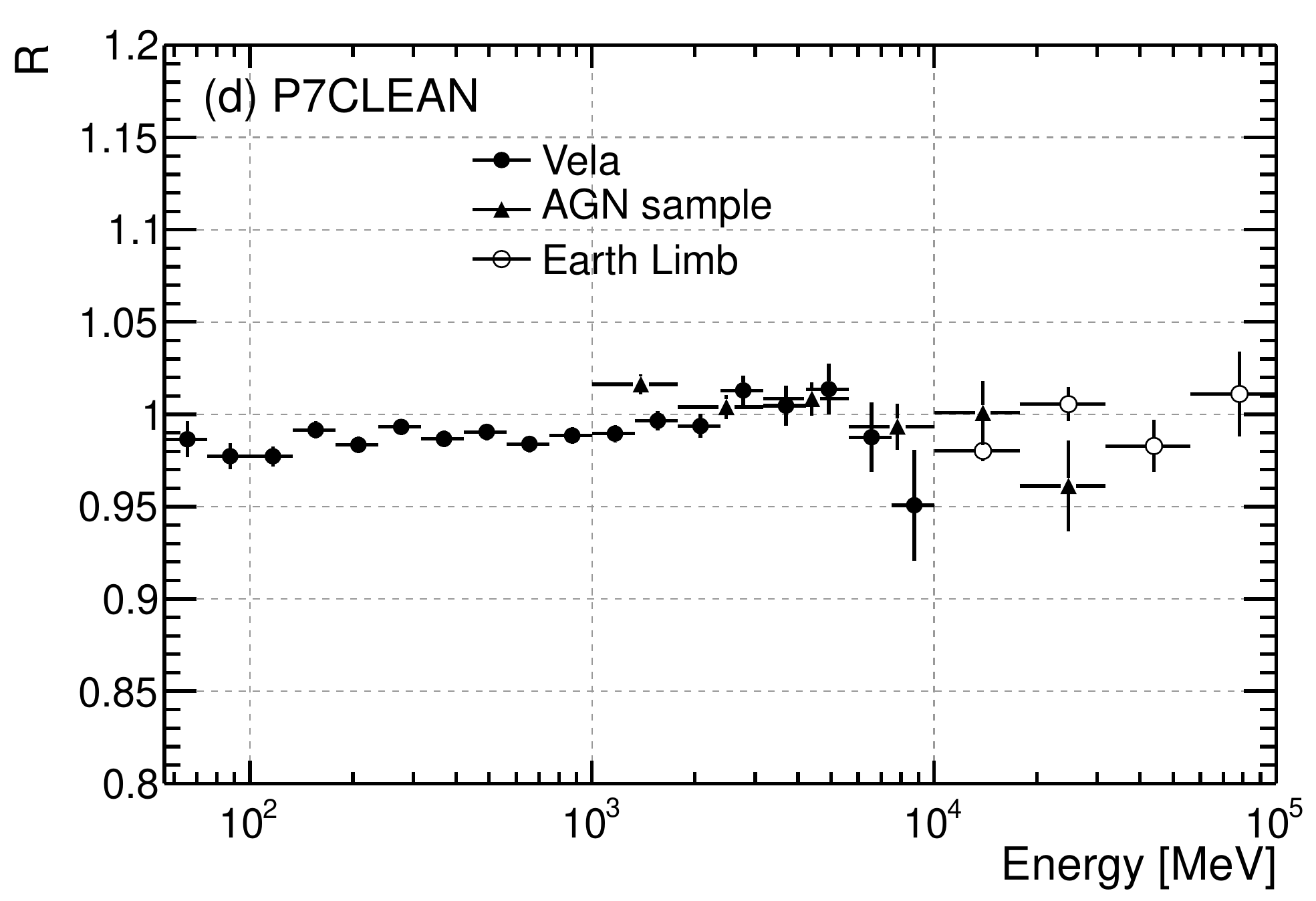}{
   \caption{Ratio of the fraction of events as seen in flight data relative to
     the prediction from MC simulations: 
     (a) \gammaRays\ converting on the $+x$ side of the LAT;
     (b) \gammaRays\ pointing in $45^{\circ}$ ranges of azimuth toward the
     corners of the LAT  (i.e., $\xi > 0.5$) ;
     (c) \gammaRays\ from near the LAT boresight, i.e., for which
     $\cos{\theta} > 0.7$;
     (d) \gammaRays\ passing the \irf{P7CLEAN} selection.}
   \label{fig:aeff_check}
}

As a simple check of our method, we split the event sample into subsets of
events converting on the $+x$ and $-x$ sides of the LAT.
As noted in \secref{subsec:Aeff_MC_corrections} the $\phi$ dependence of
\aeff\ is strongest between directions toward the corners of the LAT relative
to directions toward the sides of the LAT; accordingly, we split the data into
events coming from the sides ($\xi < 0.5$) or corners ($\xi > 0.5$), based on
\Eqref{eq:phi_xi}. Finally, we tested the $\theta$ dependence of the
\aeff\ by splitting the data into on-axis ($\cos\theta > 0.7$) and off-axis
($\cos\theta < 0.7$) subsets.
The results of these tests are shown in \figref{aeff_check}. In each of these
examples, we used the \irf{P7SOURCE} event sample as the starting point.

We also compare the fluxes we measure with 
different event classes.  By doing so we can check the accuracy of our measured
efficiency loss for each of the selection cuts to go from one event 
class to the next.   Technically we do this by asking what 
fraction of events in one event class also remain into an event class 
with tighter selection.  \Figref{aeff_check} also shows the results of comparing
the \irf{P7CLEAN} selection to the \irf{P7SOURCE} selection.

\begin{figure}[!htb]
  \centering%
  \includegraphics[width=\onecolfigwidth]{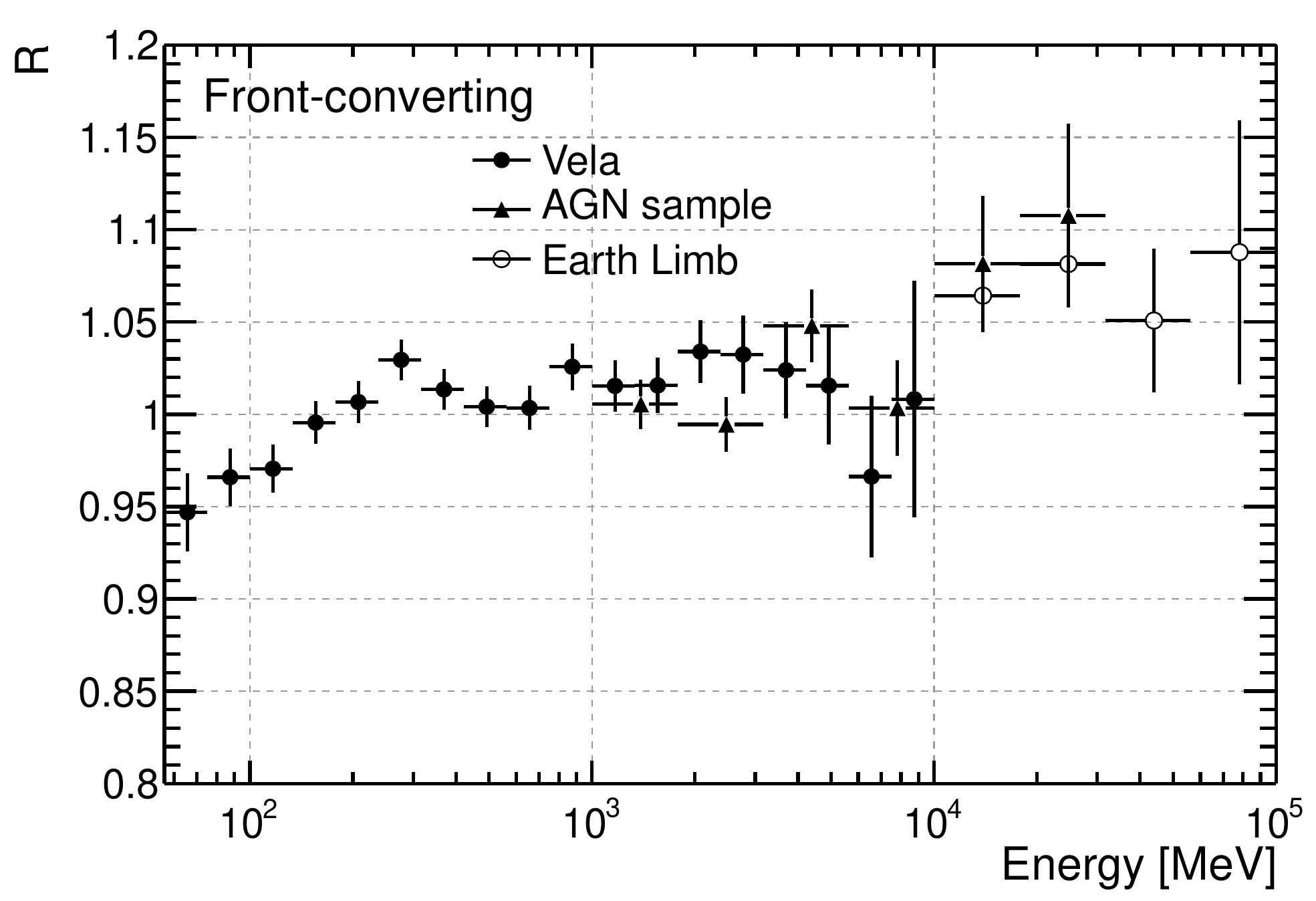}
  \caption{Ratio of the fraction of front-converting events as seen in flight
    data relative to the prediction from MC simulations:}
  \label{fig:aeff_check_fb}
\end{figure}

We also compared the fluxes we measure with front-converting events relative
to back-converting events (\parenfigref{aeff_check_fb}). Since the primary
difference between these two parts of the LAT is in the distribution
of conversion material, this test is especially sensitive to issues with our
\geant\ simulation and probes our modeling of the trigger and track-finding
efficiency. 

In each case we find that the fraction of events in each subset for the 
flight data are consistent with MC predictions to better than 15\%.  
In fact, for most of the cases the agreement is far better than that, 
closer to the 2--3\% level.  The most significant discrepancies we see are 
between front-converting and back-converting events (\parenfigref{aeff_check_fb}).

\subsection{Uncertainties on the Effective Area}\label{subsec:Aeff_errors}

\subsubsection{Overall Uncertainty of the Effective Area}\label{subsubsec:overall_aeff_errors}

\begin{figure}[htb]
  \centering\includegraphics[width=\onecolfigwidth]{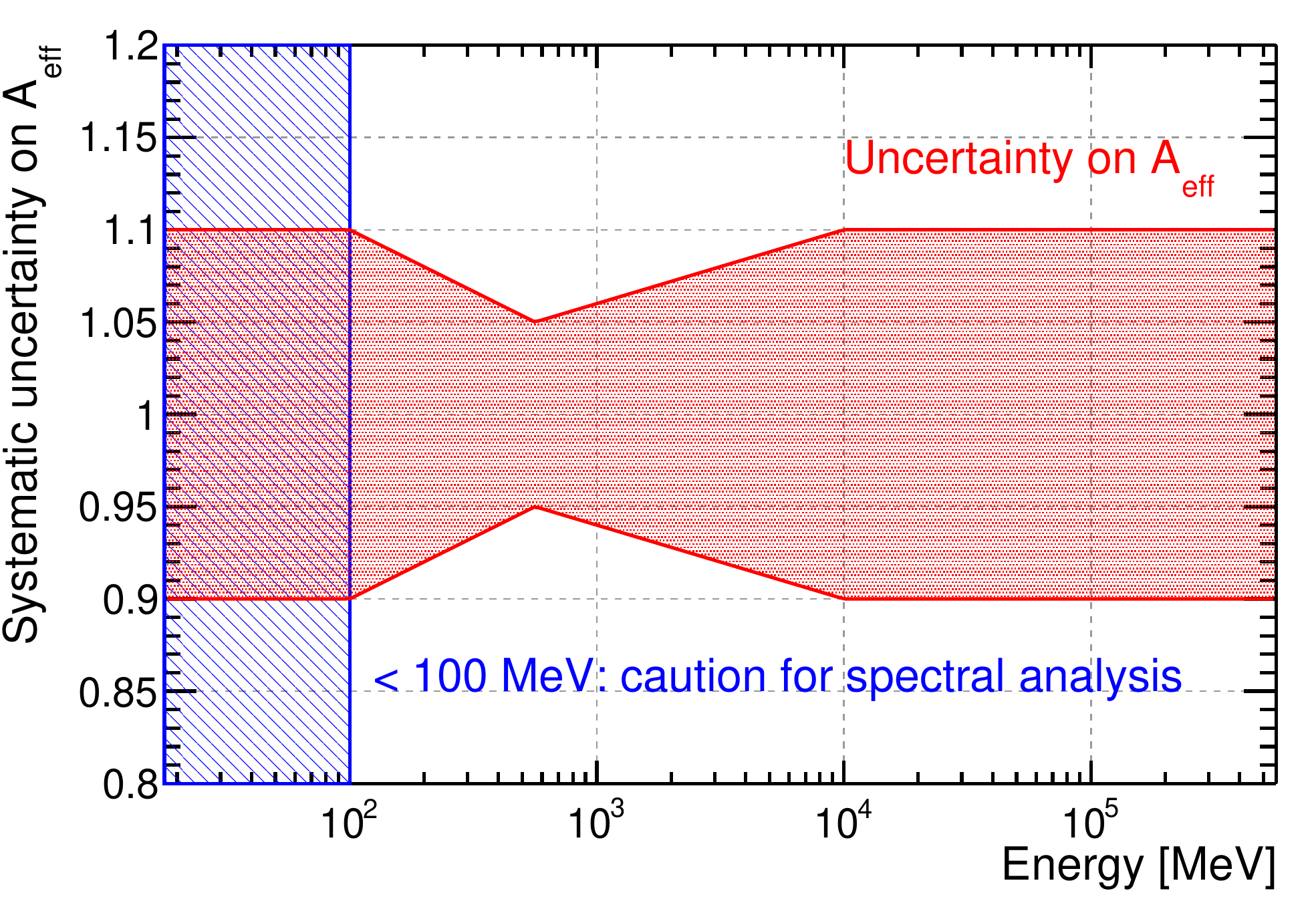}
  \caption{Systematic uncertainty band on \aeff\ as a function of energy.
    The interplay between the steeply decreasing effective area and the
    degrading energy resolution below 100~MeV and the resulting impact on
    spectral analysis will be thoroughly discussed in~\secref{sec:EDisp}.
  }
  \label{fig:Aeff_overallError}
\end{figure}

From the consistency checks described in the previous section we arrive at a
rough overall estimate of the uncertainty of \aeff, which is shown in
\figref{Aeff_overallError}. Note that this estimate is assigned simply to
account for the largest observed inconsistency, namely the mismatch between the
front-converting and back-converting events. 
Roughly speaking, for the \irf{P7SOURCE\_V6} and \irf{P7CLEAN\_V6} event 
classes these uncertainties may be be quoted as 10\% at 100~MeV, decreasing 
to 5\% at 560~MeV and increasing to 10\% at 10~GeV and above.
It is important to note that these uncertainties are statements about 
overall uncertainty of \aeff\ at various energies, and do not 
include any statement about what types of deviations we might expect 
within the stated uncertainty bands, nor about the point-to-point correlations 
in any systematic biases of \aeff. Those questions are addressed
in the next sections.

\subsubsection{Point-to-Point Correlations of the Effective Area}\label{subsubsec:aeff_point_to_point}

Since our selection criteria are generally scaled with energy (see
\secref{subsubsec:event_meritVars}), we expect any biases of \aeff\ to be
highly correlated from one energy band to the next.
This point is very important when estimating the size of potential instrumental
spectral artifacts.

We have studied the point-to-point correlation of the effective area
through the consistency checks described in \secref{subsec:Aeff_consistency},
where it is evident that the deviations from unity are not independent for
neighboring energy bins.
In order to quantify this correlation we first scale the values $r_i$ of the
data-to-Monte Carlo ratio $R$ in each of the $N$ energy bins (indexed by $i$), 
turning them into \emph{normalized deviations}\label{conv:norm_dev}
\begin{equation}
  d_i = \frac{(r_i - m)}{\sqrt{\frac{1}{(N - 1)}\sum_{i = 1}^{N}(r_i - m)^2}},
\end{equation}
and $m$ is the arithmetic mean of the $r_i$) and then we construct the
metrics\label{conv:tauCorrel}
\begin{equation}\label{eq:aeff_correlation}
  \tau_n = \frac{\sum_{i = 1 + n}^{N}(d_i - d_{i - n})^2}{(N - n)}.
\end{equation}
The quantity in \Eqref{eq:aeff_correlation} is related to a reduced
$\chi^2_{N - n}(2x)$; it is small for highly positively correlated deviations
(in which case the differences between neighboring bins are generally small),
while the expectation value is 2, for all values of $n$, for normal
uncorrelated errors.

\figref{Aeff_pointToPoint} shows this metric, for different values of $n$,
calculated on the front vs. back consistency check shown
in~\figref{aeff_check_fb}. (Only the Vela calibration data set is considered
here).
The extremely small value of $\tau_1$ indicates that neighboring logarithmic
energy bins are highly positively correlated (as can be naively inferred from
the plot in \parenfigref{aeff_check_fb}), while on the scale of half a decade in
energy ($n = 4$) there is little evidence of a correlation. This implies
that the systematic uncertainties on the effective area are not likely to
introduce significant spectral features over scales much smaller than half a
decade in energy (which is much larger than the LAT energy resolution).
The results for all the consistency checks are summarized
in~\tabref{aeff_correlation}.

\twopanel{htb!}{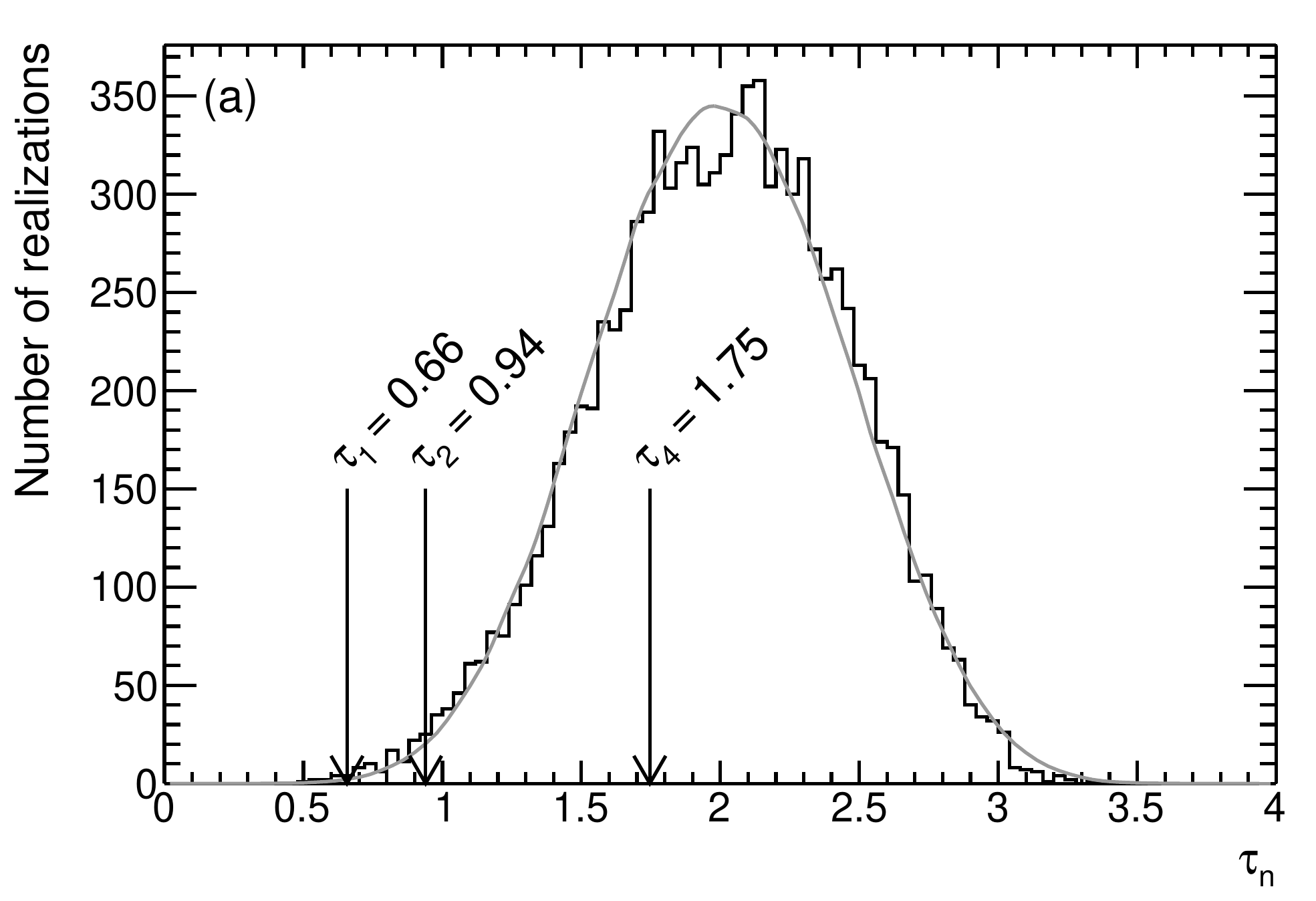}{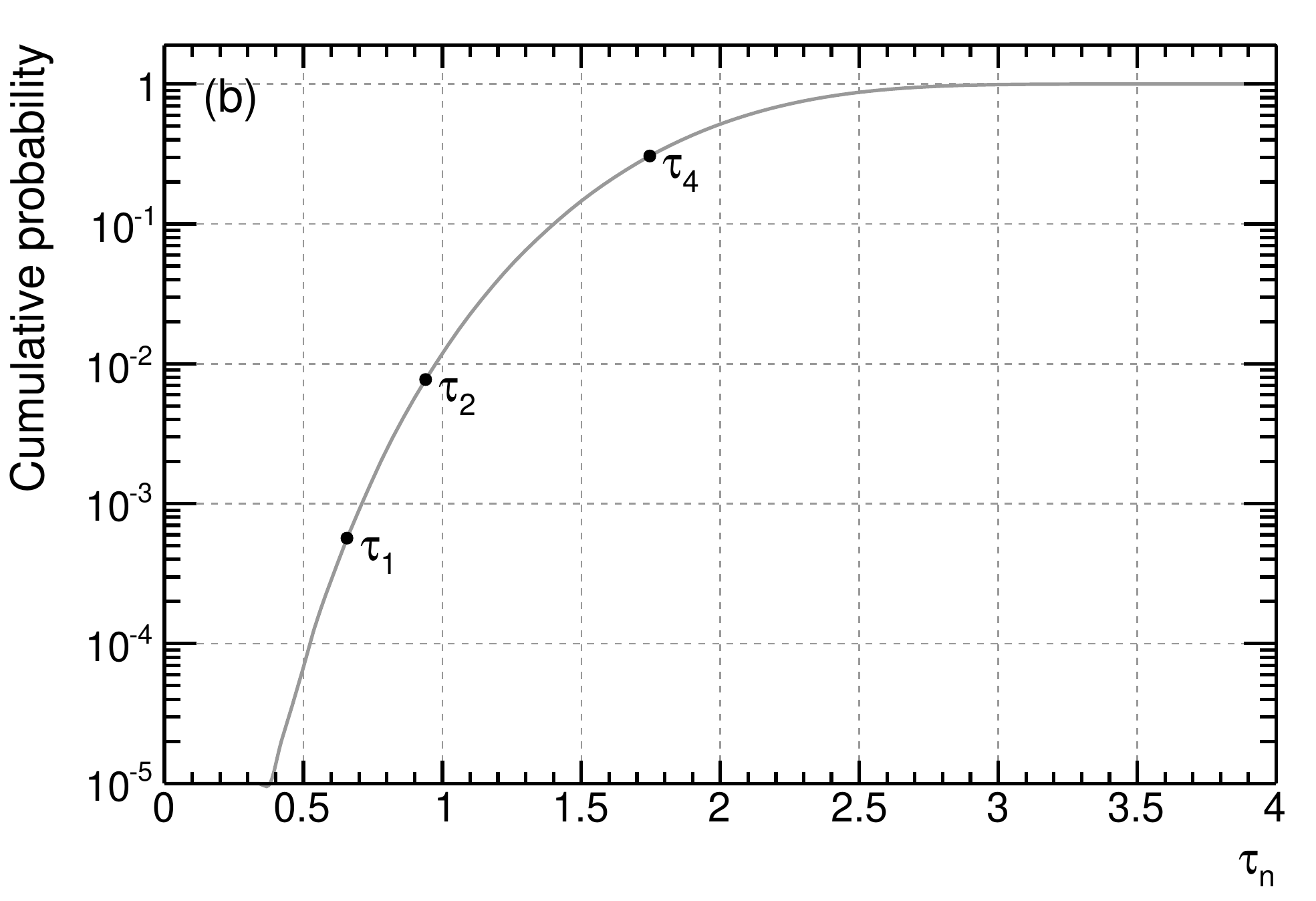}{
  \caption{(a) Values of $\tau_n$ ($n = 1$, 2 and 4) defined in
    \Eqref{eq:aeff_correlation} for the front/back consistency check with the
    Vela data set shown in~\figref{aeff_check_fb}.
    The parent distribution of $\tau_n$ for random normal uncorrelated
    deviations is shown in gray, while the black histogram represents the
    values of $\tau_n$ for 10,000 random permutations of the original data
    points. (Note that in both cases the distributions are independent of $n$.)
    (b) Cumulative probability distribution for random normal uncorrelated
    deviations.}
  \label{fig:Aeff_pointToPoint}
}

\begin{table}[htb]
  \centering
  \begin{tabular}{lcc}
    \hline
    Consistency check & $\tau_1$ & $P(\tau < \tau_1)$\\
    \hline
    \hline
    $+x$ vs. $-x$ & 1.24 & $4.9 \times 10^{-2}$\\
    Azimuthal dependence & 2.21 & 0.69\\
    $\theta$~dependence & 1.05 & $1.6 \times 10^{-2}$\\
    \irf{P7CLEAN} selection & 0.95 & $8.4 \times 10^{-3}$\\
    Front vs. back & 0.66 & $5.0 \times 10^{-4}$\\
    \hline
  \end{tabular}
  \caption{Summary of the point-to-point correlations for the consistency checks
    described in~\secref{subsec:Aeff_consistency}. Most of them indicate a
    strong positive correlation of the systematic biases between adjacent bins 
    on the effective area (the expectation value for random normal uncorrelated
    deviations is $\tau_n = 2$ for all values of n).}
  \label{tab:aeff_correlation}
\end{table}

For many analyses, especially those that are limited by statistics, it is 
enough to consider the overall uncertainty and allow for worst case deviations 
within the stated uncertainty bands. However, doing so will result in very
conservative systematic error estimates. We will discuss this in more detail in
\secref{subsec:Aeff_highLevel} when we describe techniques to propagate the
estimates of the uncertainty of \aeff\ to uncertainties on quantities such as
fluxes and spectral indices. Furthermore, we will come back to the issues of
point-to-point correlations and the potential induced spectral features when we
discuss the uncertainties associated with the energy reconstruction
in~\secref{sec:EDisp}.

\subsubsection{Variability Induced by Errors in the Effective Area}\label{subsubsec:Aeff_variability}

As a source moves across the LAT \fov\ and \aeff\ changes
with the viewing angle, any errors in the \aeff\ parametrization as a
function of $\theta$ potentially could induce artificial variability.   
We have searched for such induced variability with Vela.  We split the
data set into 12 hour periods (indexed by $i$) and compared the
number of \gammaRays\ observed ($n_{i}$) during each period with 
the number of \gammaRays\ we predict ($\tilde{n}_{i}$)\label{conv:npred} 
based on the fraction of the total exposure for Vela that we integrated 
during that 12 hour period. 

On average, our Vela calibration sample contains 230 (176) on-peak (off-pulse)
\irf{P7SOURCE} class events in the 100~MeV--10~GeV energy band every 12
hours.  Since the off-pulse region is twice the size of the on-peak region, 
background subtraction yields an average on-peak excess of $n_{i}=142$
\gammaRays\ with an average statistical uncertainty of $\sigma_{i}=16$
\gammaRays\ in each time interval. 

The exposure calculation requires several inputs:
\begin{fermienumerate}
\item the spacecraft pointing and live time history, which are binned
  in $30$~s intervals;
\item the \irf{P7SOURCE\_V6} \aeff\ parametrization, which we use when deriving
  $\aeff(E,t)$, the effective area for \gammaRays\ for Vela for each
  $30$~s interval;
\item the \irf{P7SOURCE\_V6MC} PSF (see \secref{subsec:PSF_MonteCarlo}) parametrization, which allows us to
  calculate the energy dependent containment $C(E,t,15^{\circ})$ within the $15^{\circ}$ \roi\ for
  each $30$~s interval;
\item a parametrization of the spectrum of Vela $F(E)$ so that we may correctly
  integrate $\aeff(E,t)$ and $C(E,t,15^{\circ})$ over the energy range.
\end{fermienumerate}

In order to minimize dependence on the modeled flux, we calculate the
exposure independently for 24 energy bins (which we index by $j$).
The exposure in a single time and energy bin is
\begin{equation}
  \mathcal{E}_{i,j} = \int^{E_i} \int^{t_j} \aeff(E,t)
  C(E,t,15^{\circ}) \frac{F(E)}{\int^{E_i}  F(E) dE}  dt dE.
  \label{eq:aeff_binned_exposure}
\end{equation}
We can then express the expected number of \gammaRays\ in each time and
energy bin ($\tilde{n}_{i,j}$) as a fraction of the total number of \gammaRays\
in that energy bin ($n_{j}$),
\begin{equation}
  \tilde{n}_{i,j} = \frac{\mathcal{E}_{i,j}} {\Sigma_{i} \mathcal{E}_{i,j}} n_{j}.
  \label{eq:aeff_binned_npred}
\end{equation}
Then we sum $n_{i,j}$ and $\tilde{n}_{i,j}$ across energy bins to find
$n_{i}$  and $\tilde{n}_{i}$. 

\begin{figure}[htbp]
  \centering
  \includegraphics[width=\onecolfigwidth]{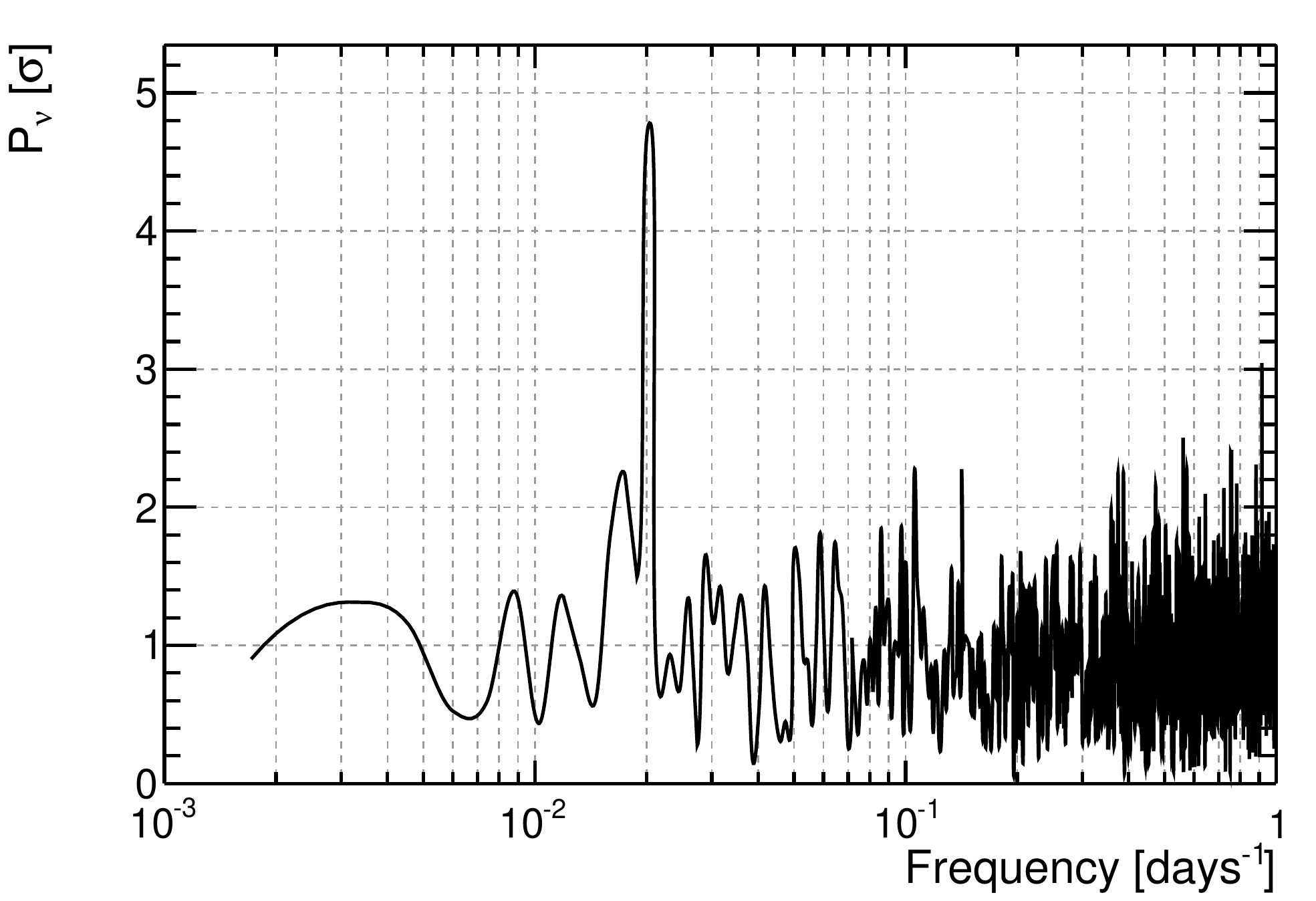}
  \caption{The discrete Fourier transform of the
    $\frac{n - \tilde{n}}{\sigma_n}$ time series. The only peak visible above
    the noise floor corresponds to the 53-day orbital precession
    period.   Note that the figure is normalized and the vertical
    scale is expressed in units of the statistical uncertainty.}    
  \label{fig:aeff_variability_fft}
\end{figure}

We have performed this analysis, dividing the first 700 days of the Vela data
sample into 1400 12-hour time intervals and using the phase-averaged flux model
\begin{equation}
F(E) \propto E^{-\Gamma}e^{-\frac{E}{E_0}}
\end{equation}
with $\Gamma = 1.38$ and $E_0 = 1.360$~GeV~\citep{REF:2009.VelaI}.
\Figref{aeff_variability} shows the fractional variation,
\begin{equation}
\frac{n_{i} - \tilde{n}_{i}}{\tilde{n}_{i}}
\end{equation}
the normalized residuals, 
\begin{equation}
\frac{n_{i} - \tilde{n}_{i}}{\sigma_{i}}
\end{equation}
and the distribution of the normalized residuals for each of the
time-intervals. The normalized residuals are very nearly normally
distributed. Furthermore, the Fourier transform of the
time-series (\parenfigref{aeff_variability_fft}) shows only a small peak
corresponding to the orbital precession period and is otherwise consistent with
Poisson noise.  Note that unlike more complicated analyses that involve fitting
the flux of a point source, this analysis is testing only the accuracy of
the  \aeff\ representation (and to a much lesser extent, the
$15^{\circ}$ containment of the PSF).  \NEWTEXT{We attribute the peak in the Fourier 
spectrum to small, incidence angle-dependent errors in the effective area. 
As the orbit precesses, the range of incidence angles sampled, and hence the
bias in the calculated exposure, varies slightly.}

\NEWTEXT{Although we performed this analysis with 12-hour time intervals, as
noted in \secref{subsec:LAT_obsProf}, the LAT boresight follows very similar paths 
across the sky during successive 2-orbit periods.  Therefore, the level of instrument-induced 
variability observed with 12-hour time intervals is likely to be indicative of the
systematic uncertainties for variability analyses down to 3-hour time scales.}

\twopanel{htb!}{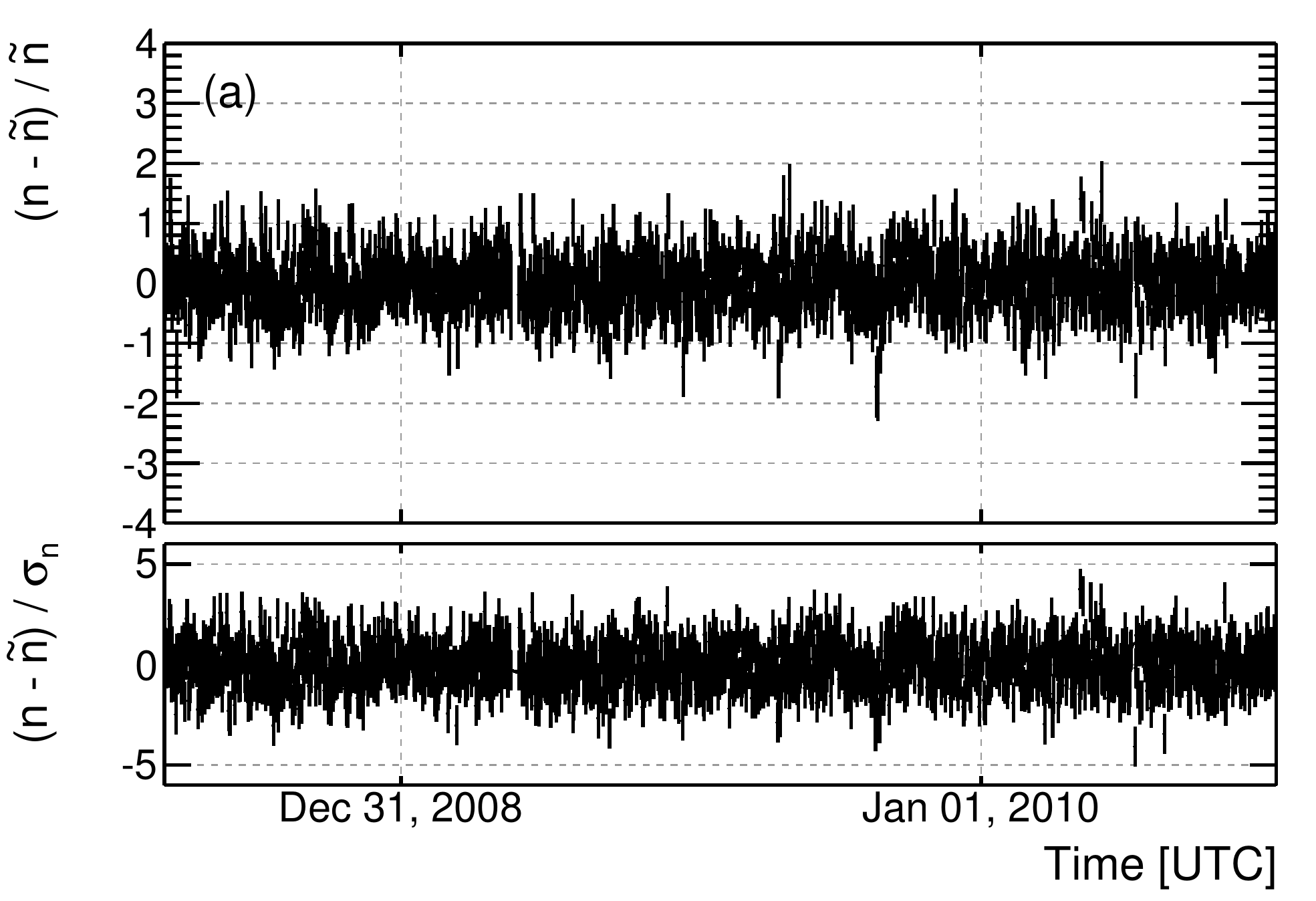}{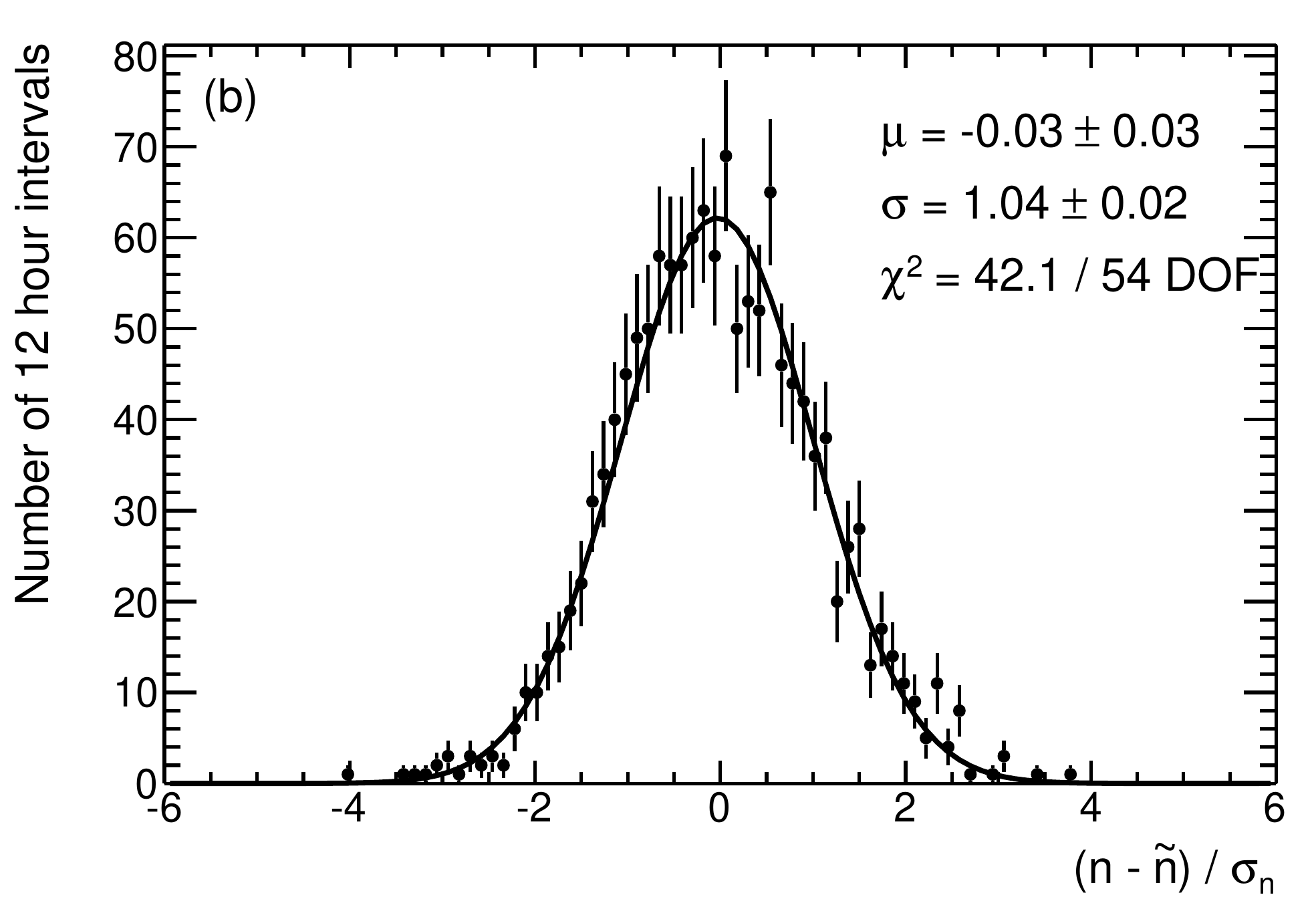}{
  \caption{Check for variability-induced by errors in the \aeff\ tables
    \irf{P7SOURCE} Vela calibration sample and the \irf{P7SOURCE\_V6MC}
    IRFs.  Panel (a) shows the fractional difference and normalized
    residuals between the observed counts in the 100~MeV to~10 GeV
    band and the prediction based on the fraction of the total
    exposure accumulated during each of 1400 12-hour
    time intervals. Panel (b) shows the normalized residuals,
    which are very well fit with a Gaussian with unit width and zero mean.}
  \label{fig:aeff_variability}
}

Although the estimate used in \secref{subsubsec:overall_aeff_errors}
that the systematic uncertainty of \aeff\ is less than the disagreement 
between the extreme cases is quite conservative for long time
scale observations, it is somewhat less conservative for shorter
observations.  For example, in observations less than the $Fermi$
orbital precession period of $\sim 53.4$~days, a particular region of
the sky might be preferentially observed at incidence
angles where the bias of \aeff\ is particularly large, 
or during parts of the orbit in which $Fermi$ is exposed to
particularly high CR background rates and the correction described in
\secref{subsubsec:Aeff_Livetime} leaves some residual bias in the calculated exposure.
Finally, we have observed that ignoring the $\phi$~dependence of the effective area 
(\secref{subsubsec:aeff_phi_dep}) can induce artificial quarter-yearly 
periodicity in the fluxes from directions near extremely bright sources,
in particular the Vela pulsar.

\subsection{Propagating Uncertainties on the Effective Area to High Level Science Analysis}\label{subsec:Aeff_highLevel}

As we hinted in the previous section, translating uncertainties on
\aeff\ into systematic errors on quantities such as fluxes and spectral
indices depends on the particular analysis and requires assumptions
about the variation of \aeff\ within the uncertainty bands.

\subsubsection{Using Custom Made IRFs to Generate an Error Envelope}

A somewhat brute force approach to this problem is to generate IRFs
that represent worst case scenarios for measuring specific quantities like
fluxes or spectral parameters and use these \emph{bracketing IRFs} to repeat the
analysis and extract the variation in the measured quantities. Of course,
the nature of the variations between the IRFs depends on the quantity in
question. 

We address this in a generic way by scaling \aeff\ by the product of the
relative systematic uncertainty
\begin{equation}\label{conv:epsilonBracket}
\epsilon(E) = \frac{\delta\aeff(E)}{\aeff(E)}
\end{equation}
and arbitrary \emph{bracketing} functions $B(E)$\label{conv:aeff_bracket}
taking values in the $[-1,1]$ interval. Specifically we define modified \aeff\
as
\begin{equation}\label{eq:aeff_bracketing}
  \aeff^{\prime}(E, \theta) = \aeff(E, \theta) \cdot
  \left( 1 + \epsilon(E)B(E) \right).
\end{equation}
      
The simplest bracketing functions, $B(E) = \pm 1$ clearly minimize and maximize
\aeff\ within the uncertainty band. On the other hand, to maximize the effect
on the spectral index in a power-law fit, we choose a functional form that
changes sign at the \emph{pivot} or decorrelation energy $E_{0}$ (i.e., the
energy at which the fitted differential flux and spectral index are
uncorrelated):
\begin{equation}
  B(E) = \pm\tanh\left( \frac{1}{k}\log(E/E_{0}) \right).
\end{equation}
The parameter $k$\label{conv:aeff_bracket_smooth_const} controls the slope of
the transition near $E_{0}$; in practice we use $k=0.13$, which corresponds to
smoothing over twice the LAT energy resolution of $\Delta E/E \sim 0.15$.  
The bracketing IRFs used for effective area studies are listed
in~\tabref{bracket_aeffFuncs}.
\begin{table}[hp!]
  \begin{center}
   \begin{tabular}{ll}
     \hline
     Name & $B(E)$\\
     \hline
     \hline
     \bracketirf{c\_flux\_lo} & $+1$\\
     \bracketirf{c\_flux\_hi} & $-1$\\
     \bracketirf{c\_index\_soft} & $+\tanh(\frac{1}{k}\log(E/E_{0}))$\\
     \bracketirf{c\_index\_hard} & $-\tanh(\frac{1}{k}\log(E/E_{0}))$\\
     \hline      
   \end{tabular}
   \caption{Bracketing \aeff\ and corresponding energy-dependent scaling
     functions used to create them.}
   \label{tab:bracket_aeffFuncs}
  \end{center}
\end{table}
\figref{bracketing_funcs} shows the bracketing functions for
\bracketirf{c\_index\_soft} and \bracketirf{c\_index\_hard} and their effects
on the on-axis \aeff.
\twopanel{htbp}{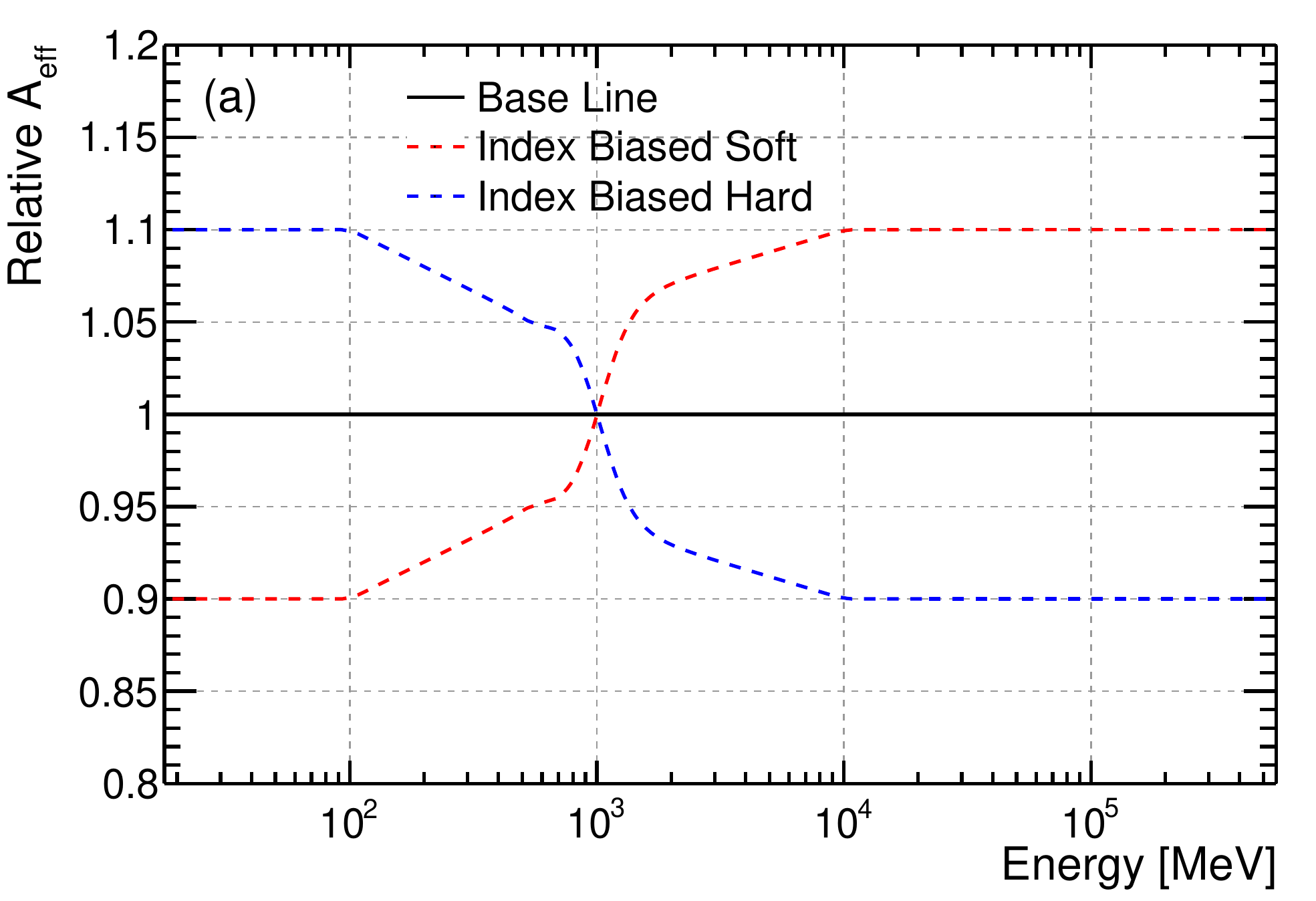}{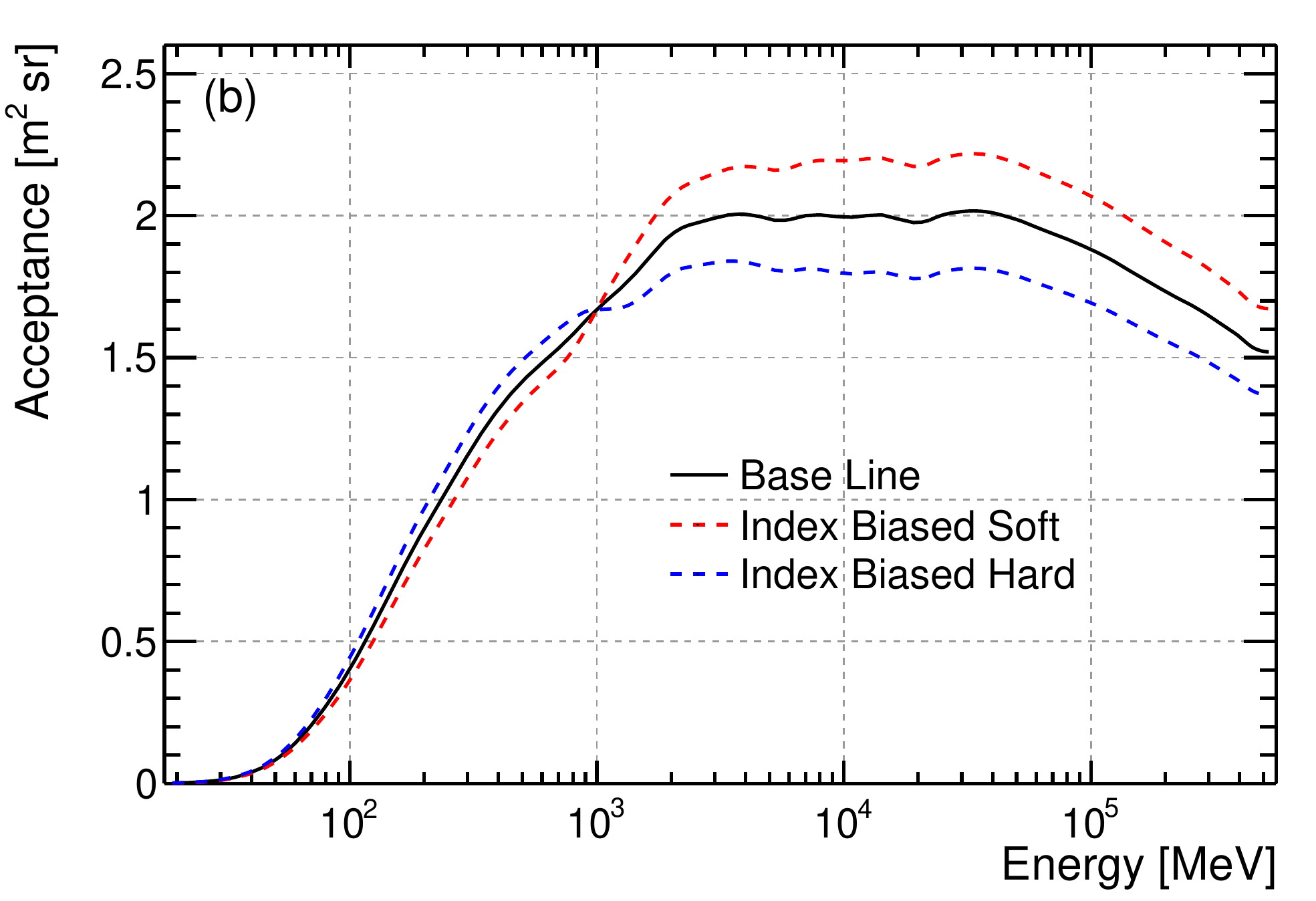}{
  \caption{Bracketing IRFs for \aeff\ designed to estimate possible
    systematic biases of the measurement of the spectral index. (a) energy
    dependence of the scaling parameter $\epsilon(E)B(E)$;
    (b) acceptance of the resulting scaled \aeff\ for \irf{P7SOURCE}.
    For this example we used a pivot energy $E_{0} = 1$~GeV.}
  \label{fig:bracketing_funcs}
}

We have studied two sources from the AGN sample to obtain estimates of the
effects of instrumental uncertainties on the measured fluxes and spectral
parameters, namely:
\begin{fermiitemize}
\item PG~1553+113---associated with 2FGL~J1555.7+1111, which has spectral index
  of \makebox{$1.66 \pm 0.02$}, making it one of the hardest bright AGN;
\item  B2~1520+31---associated with 2FGL~J1522.1+3144, which has a spectral
  index of \makebox{$2.37 \pm 0.02$} (when fit with a power-law), making it one of the softest bright AGN.
\end{fermiitemize}
In each case we used the \stools\ (version v9r25p2) to perform a series of
binned maximum likelihood fits to a $20^{\circ} \times 20^{\circ}$ region centered
at the source position over the energy range 100~MeV--100~GeV.
For each individual fit we followed the same procedure:
\begin{fermienumerate}
\item used the \gammaRayHyph\ and time interval selection criteria as for the
  Vela calibration sample (see \secref{subsec:LAT_methods});
\item included all 2FGL sources within $20^{\circ}$ in our likelihood model, with
  the same spectral parametrizations as were used in the 2FGL catalog;  
\item included models of the Galactic diffuse emission
  (\filename{ring\_2year\_P76\_v0.fits}) and the isotropic diffuse emission
  (\filename{isotrop\_2year\_P76\_source\_v1.txt}) rescaled by the inverse of
  the function used to rescale the effective area (so as to ensure the same
  distribution of expected counts from these diffuse sources);
\item freed the spectral parameters for all point sources within $8^{\circ}$
  from the center of the region, as well as the overall normalizations of both
  diffuse components.
\end{fermienumerate}
Following what was done for the 2FGL catalog, we used a simple power-law\label{conv:powerLawGamma}
\begin{equation}
  \frac{dN}{dE} = N_{0} \left(\frac{E}{E_{0}}\right)^{-\Gamma}
\end{equation}
to model the differential flux from PG~1553+113 and a ``log-parabola''\label{conv:logParAlpha}\label{conv:logParBeta}
\begin{equation}
  \frac{dN}{dE} = N_{0}\left(\frac{E}{E_{0}}\right)^{-(\alpha + \beta \ln(E/E_{0}))}
\end{equation}
to model that of B2~1520+31. (All of this will be also relevant for the tests
with bracketing PSFs described in~\secref{subsec:PSF_bracketing} and
with energy dispersion included in the likelihood fit described in~\secref{subsec:EDisp_highLevelAnalysis}).

\Tabref{bracket_PG1553p113} and~\ref{tab:bracket_B21520p31} shows the fits
results for PG~1553+113 and~B2~1520+31, respectively, using these \aeff\
bracketing functions, as well as the integral counts ($F_{25}$\label{conv:f25})
and energy ($S_{25}$\label{conv:S25}) fluxes between 100~MeV and 100~GeV.

\begin{table}[hp!]
  \begin{center}
    \begin{tabular}{lclcc}
      \hline
      Bracketing \aeff & $N_{0}$ & $\Gamma$ & $F_{25}$ & $S_{25}$ \\
 & [MeV$^{-1}$ cm$^{-2}$ s$^{-1}$] & & [cm$^{-2}$ s$^{-1}$] & [MeV cm$^{-2}$ s$^{-1}$] \\
\hline
\hline
Nominal & $2.54 \times 10^{-12}$ & 1.68 & $6.91 \times 10^{-8}$ & $1.19 \times 10^{-4}$\\
\bracketirf{c\_flux\_hi} & $2.75 \times 10^{-12}$ & 1.67 & $7.32 \times 10^{-8}$ & $1.31 \times 10^{-4}$\\
\bracketirf{c\_flux\_lo} & $2.37 \times 10^{-12}$ & 1.69 & $6.54 \times 10^{-8}$ & $1.09 \times 10^{-4}$\\
\bracketirf{c\_index\_hard} & $2.57 \times 10^{-12}$ & 1.64 & $6.45 \times 10^{-8}$ & $1.30 \times 10^{-4}$\\
\bracketirf{c\_index\_soft} & $2.53 \times 10^{-12}$ & 1.73 & $7.44 \times 10^{-8}$ & $1.11 \times 10^{-4}$\\

      \hline
    \end{tabular}
    \caption{Fit parameters and integral fluxes obtained using the \aeff\
      bracketing IRFs for PG~1553+113.  Note that the quoted precision
      is roughly equivalent to the fit uncertainties and the pivot
      energy is $E_{0} = 2240$~MeV for this source.}
    \label{tab:bracket_PG1553p113}
  \end{center}
\end{table}

\begin{table}[hp!]
  \begin{center}
    \begin{tabular}{lcllcc}
      \hline
      Bracketing \aeff & $N_{0}$ & $\alpha$ & $\beta$ & $F_{25}$ & $S_{25}$ \\
 & [MeV$^{-1}$ cm$^{-2}$ s$^{-1}$] & & & [cm$^{-2}$ s$^{-1}$] & [MeV cm$^{-2}$ s$^{-1}$] \\
\hline
\hline
Nominal & $5.23 \times 10^{-10}$ & 2.24 & 0.08 & $4.09 \times 10^{-7}$ & $1.33 \times 10^{-4}$ \\
\bracketirf{c\_flux\_hi} & $5.60 \times 10^{-10}$ & 2.26 & 0.07 & $4.44 \times 10^{-7}$ & $1.44 \times 10^{-4}$ \\
\bracketirf{c\_flux\_lo} & $4.90 \times 10^{-10}$ & 2.22 & 0.08 & $3.79 \times 10^{-7}$ & $1.24 \times 10^{-4}$ \\
\bracketirf{c\_index\_hard} & $5.20 \times 10^{-10}$ & 2.15 & 0.10 & $3.93 \times 10^{-7}$ & $1.35 \times 10^{-4}$ \\
\bracketirf{c\_index\_soft} & $5.27 \times 10^{-10}$ & 2.33 & 0.05 & $4.29 \times 10^{-7}$ & $1.32 \times 10^{-4}$ \\

      \hline
    \end{tabular}
    \caption{Fit parameters and integral fluxes obtained using the \aeff\
      bracketing IRFs for B2~1520+31.  Note that the quoted precision
      is roughly equivalent to the fit uncertainties and the pivot
      energy is $E_{0} = 281$~MeV for this source.}
    \label{tab:bracket_B21520p31}
  \end{center}
\end{table}

The ranges of the fit values indicate propagated uncertainties from
the uncertainty in \aeff\ (\tabref{bracketErrors})\label{conv:flux_prefactor}.
It is important to note that the systematic error estimates resulting from
this technique represent conservative estimates within the instrumental
uncertainties, rather than random variations.  Furthermore, many of the 
bracketing IRFs are mutually exclusive, so when considering {\em
  relative} variations between \gammaRayHyph\ sources it is more
appropriate to compare how the relative values change for each  
set of bracketing IRFs.

\begin{table}[hp!]
  \begin{center}
    \begin{tabular}{lcc}
      \hline
      Parameter & B2 1520+31 & PG 1553+113 \\
 & (soft) & (hard) \\
\hline
\hline
$\delta N_{0}/N_{0}$ & \syserrors{+7.2\%}{-6.3\%} & \syserrors{+8.0\%}{-6.9\%} \\
$\delta\Gamma$ ($\delta\alpha$) & \syserrors{+0.09}{-0.09} & \syserrors{+0.05}{-0.05} \\
$\delta\beta$ & \syserrors{+0.02}{-0.02} & {\hfill-\hfill} \\
$\delta F_{25}/F_{25}$ & \syserrors{+8.5\%}{-7.2\%} & \syserrors{+7.7\%}{-6.6\%} \\
$\delta S_{25}/S_{25}$ & \syserrors{+8.1\%}{-6.9\%} & \syserrors{+10.0\%}{-8.3\%} \\

      \hline
    \end{tabular}
    \caption{Systematic variations arising from uncertainties in the effective
      area. For the spectral index ($\Gamma$ or $\alpha$) and spectral
      curvature ($\beta$) we give the absolute variation with respect to the
      nominal value (e.g., $\delta \Gamma$). For the flux prefactor and the
      integral fluxes we give the relative variation with respect to the
      nominal value (e.g., $\delta N_{0} / N_{0}$).}    
    \label{tab:bracketErrors}
  \end{center}
\end{table}

\subsubsection{Using a Bootstrap Method to Generate an Error Envelope}

Alternatively, given a family of plausible \aeff\ curves, we can use a {\it weighted
  bootstrap} approach \citep[see][for more details]{REF:BootstrapBook:1993} for propagating the systematic uncertainties on
\aeff.  The weighed bootstrap approach is closely related to the bracketing IRFs method
described in the previous section and to the methods discussed in \citet{REF:2001.ChandraBootstrap} in
the context of the analysis of {\it Chandra} X-ray data.

\begin{figure}[htb!]
  \centering\includegraphics[width=\onecolfigwidth]{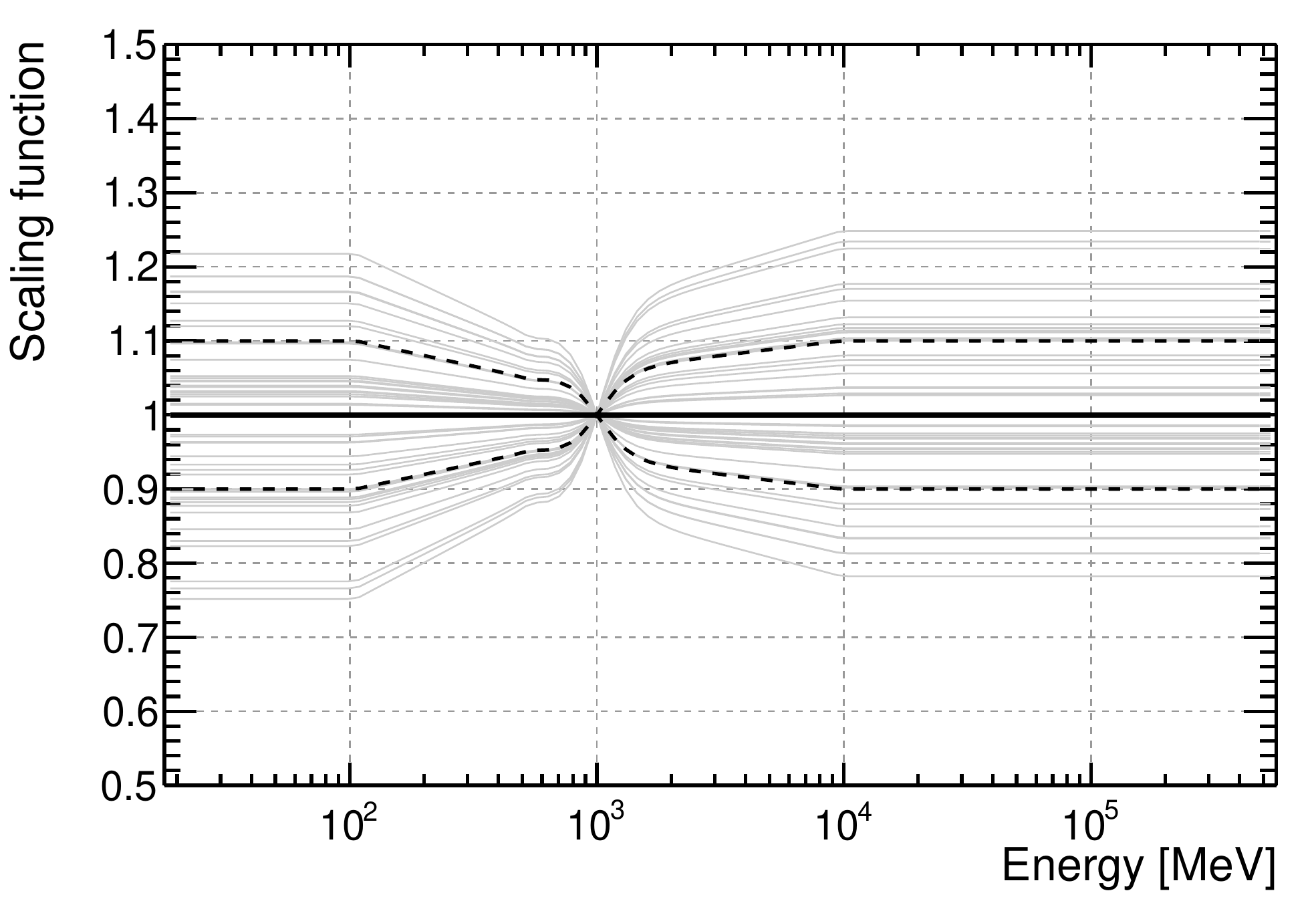}
  \caption{A family of effective area scaling functions obtained for
    different values of the normal random variable $\xi$ in
    \Eqref{eq:bootstrap_aeff_family}.}
  \label{fig:bootstrap_aeff_family}
\end{figure}

The basic idea is that, for each trial, the event data are bootstrap resampled
using a weighting based on an effective area scaling function that is drawn
from a family of plausible curves. The simplest of such families of \aeff\
curves (\parenfigref{bootstrap_aeff_family}) can be constructed starting from
\Eqref{eq:aeff_bracketing} and multiplying the scaling function
$\epsilon(E)B(E)$ by a normally distributed random number $\xi$ with
zero mean---which
effectively becomes the parameter controlling the family itself:
\begin{equation}\label{eq:bootstrap_aeff_family}
  \aeff^{\prime}(E, \theta, \xi) = \aeff(E, \theta)\cdot
  \left( 1 + \xi\epsilon(E)B(E) \right)
\end{equation}

We have found the results of the weighted bootstrap with the family of scaling functions
defined in \Eqref{eq:bootstrap_aeff_family} to be in good agreement with those
of the bracketing IRFs approach described in the previous section.
This should not be surprising as the bracketing IRFs use the $\pm 1\sigma$
excursions of the bracketing function and the weighted bootstrap draws from a
Gaussian distribution of function scalings that are also based on that same
bracketing function. 
The real benefits of the weighted bootstrap arise when one has families of
plausible effective area functions that have more complicated dependencies
(e.g., on incidence angle as well as energy) such that exposures would need to
be recalculated to apply the bracketing method.

\subsection{Comparison of \psix\ and \pseven}\label{subsec:Aeff_comparison_p6}

\begin{figure}[htb!]
  \centering
  \includegraphics[width=\onecolfigwidth]{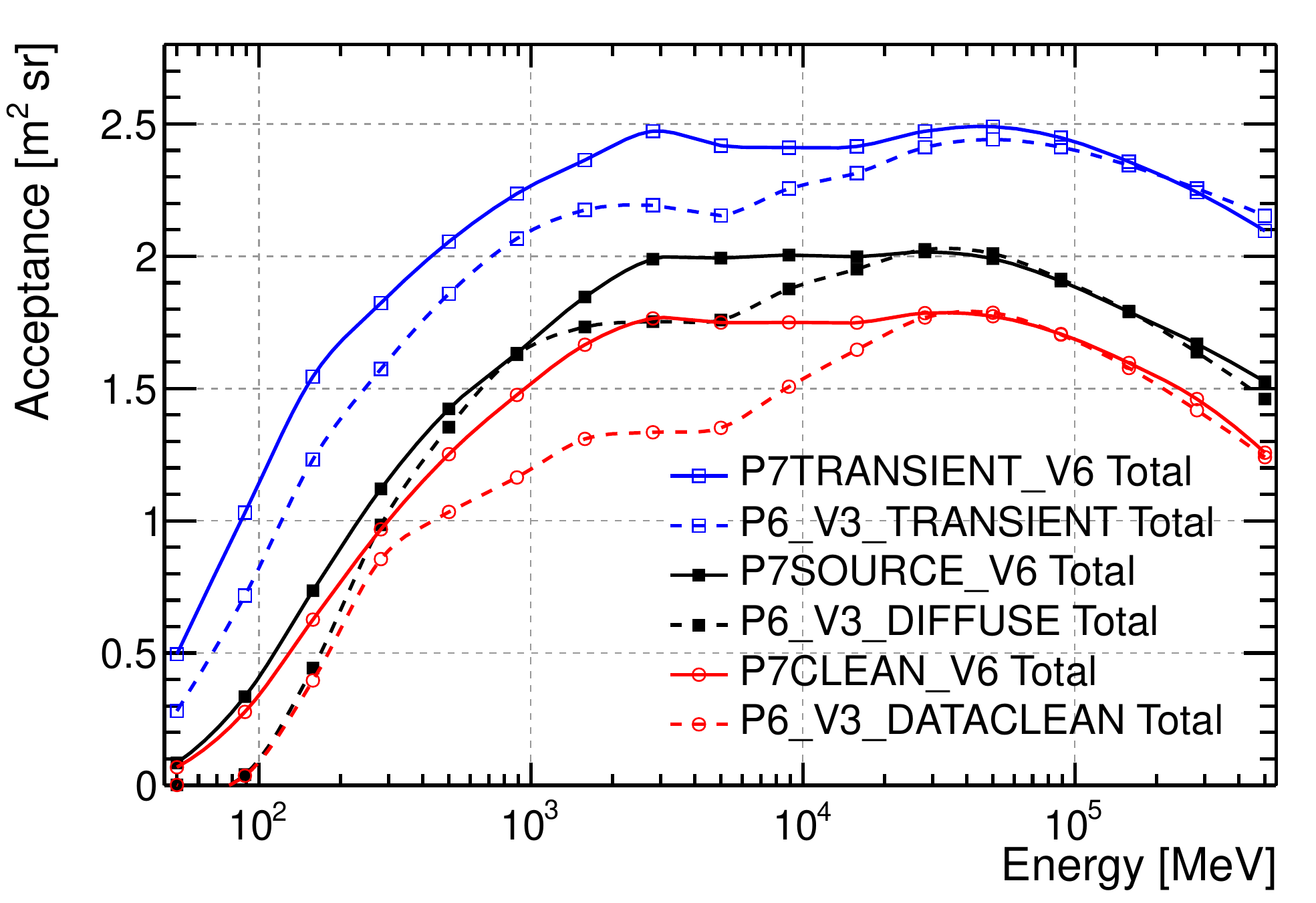}
  \caption{Comparison of the acceptances for the \pseven\ and the \psix\
    standard \gammaRayHyph\ classes.}
  \label{fig:compare_accept_p6_p7}
\end{figure}

The \pseven\ event classes were designed to meet the same high-level
analysis needs as the \psix\ classes. \Tabref{public_data} summarizes the
correspondence between the \pseven\ standard \gammaRayHyph\ classes and their
closest \psix\ equivalents.
The corresponding acceptances are shown in \figref{compare_accept_p6_p7}.

As discussed in \secref{sec:event}, the main technical improvement in
\pseven\ was optimizing the event selections on
simulations that included the ghost signals, and on flight data. The
principal outcome for \pseven\ is a substantial increase of \aeff\
below $\sim 300$~MeV, especially for the cleaner event classes.
At 100~MeV the \irf{P7SOURCE\_V6} and \irf{P7CLEAN\_V6} \gammaRayHyph\ classes
feature an acceptance of $\sim 0.3$~m$^2$~sr, to be compared with the
$\sim 0.1$~m$^2$~sr of the \psix\ equivalent
(\parenfigref{compare_accept_p6_p7}). \pseven\ has opened a window on
astronomy with the LAT data below 100~MeV, though the reader should bear in
mind the specific caveats in \secref{subsec:PSF_highLevel} and
\secref{subsec:EDisp_highLevelAnalysis} when performing spectral
analyses at these energies.

Additionally, in \pseven, the energy dependence of \aeff\ is smoother. 
This was accomplished in conjunction with obtaining a better
overall understanding of the effective area itself, owing to the
extensive use of on-orbit data to verify the fidelity of our MC
simulation at each step of the \gammaRayHyph\ selection process.

\clearpage

\section{POINT-SPREAD FUNCTION}\label{sec:PSF}

As discussed in \secref{subsec:LAT_TKR}, at low energies the PSF is determined
by multiple scattering. For example, the calculated multiple scattering for a
normally incident 100~MeV \gammaRay\ converting in the middle of a thin (front)
section tungsten foil is $\sim 3.1^{\circ}$ (space angle).
The 68\% containment angle as measured from flight data for 100~MeV 
\gammaRays\ near the LAT boresight averaged over all towers is
$\sim 3.3^{\circ}$ and is in agreement with the MC simulations
(see \secref{subsec:PSF_onorbit}). 
The small difference is due to missing measurements when the
trajectories happen to pass through regions without SSD coverage and the fact
that the electron and positron from a conversion can undergo hard scattering
processes such as bremsstrahlung.

If multiple scattering were the only consideration, the PSF should
become narrower as $E^{-1}$. The measured PSF however improves more slowly with
energy, instead falling as $\sim E^{-0.78}$
(see \secref{subsubsec:psf_analysis}). This slower improvement
relative to that expected for pure multiple scattering is also due to missed 
measurements and hard scattering processes and is 
predicted by the MC calculations.

Above a few GeV the narrowing of the PSF with energy is limited by the finite hit resolution
of the SSDs. The strip pitch of 228~\um\ and the lever arm for the
direction measurement result in a limiting precision for the average
conversion of $\sim 0.1^{\circ}$ at normal incidence. The transition
to this measurement precision-dominated regime as predicted by the 
MC should occur between $\sim 3$~GeV and $\sim 20$~GeV.
Estimates of the limiting PSF from flight data however indicate a worse
performance above $\sim 3$~GeV; the PSF 68\% containment levels off at almost
double the calculated value (i.e., $\sim 0.16^{\circ}$). This departure is one of
the few instances where the MC results significantly differ from 
real data (see \secref{subsec:PSF_onorbit}).   The LAT collaboration has
identified limitations in the pre-launch calibration algorithms of the CAL light
asymmetry \citep[for more details see][]{REF:2009.OnOrbitCalib} as the
primary cause of these discrepancies, and is assessing
the improvement of the flight data-derived PSF at high energies for data which were
reprocessed with improved calibration constants\footnote{\webpage{http://fermi.gsfc.nasa.gov/ssc/data/analysis/documentation/Cicerone/Cicerone\_LAT\_IRFs/IRF\_PSF.html}}
(see \secref{subsubsec:psf_analysis} for details of how the CAL energy
centroid is used in the event direction analysis).

\subsection{Point-Spread Function from Monte Carlo Simulations}%
\label{subsec:PSF_MonteCarlo}

Equivalently to what we stated in \secref{sec:intro}, the PSF is the
likelihood to reconstruct a \gammaRay\ with a given angular deviation
$\delta v = |\hat{v}^{\,\prime} - \hat{v}\, |$. We write it as \psf.

As for the effective area, events from a dedicated \allgamma\ MC simulation that
pass the selections for the event class in question are binned in true energy
and incidence angle.  Note again that we ignore any $\phi$ dependence of
the PSF.  \NEWTEXT{As discussed below, in almost all cases the
  $\phi$ dependence of the PSF is much weaker
than the $\theta$ dependence, and in \secref{subsec:perf_variability}
we show that ignoring the $\theta$ dependence induces 
at most a $\sim 4\%$ RMS variation of the flux.}  To allow for some additional 
smoothing of the variations of the parameters of the PSF fitting function with energy
and angle a procedure based on running averages is implemented: in each bin events
belonging to nearby bins (by default $\pm 1$~bin, where the bin sizes
are 0.1 in $\cos\theta$ and 0.25 decades in energy) along the energy and angle axes are included.
For every bin we build a histogram with the angular deviations of detected
\gammaRays. This distribution is fitted and the fit parameters are saved.
Note that, although the PSF is parametrized in the LAT reference frame, the
angular deviation is the same whether expressed in celestial or LAT reference
frames.

Since the PSF varies with $\theta$, it is often useful to consider the PSF
averaged over the observing profile (\secref{subsec:LAT_obsProf}) for a source
of interest:
\label{conv:psf_bar}
\begin{equation}
  \label{eq:PSF_ave}
  \bar{\psf}(\delta v; E) =
  \frac{\int \psf(\delta v; E, \theta) \aeff(E,\theta)
    \tobs(\theta) d\Omega}{\int \aeff(E,\theta) \tobs(\hat{\theta}) d\Omega}.
\end{equation}

\subsubsection{Point-Spread Function: Scaling and Fitting}\label{subsubsec:psf_prescaling}

In our parametrized description of the PSF most of the energy dependence is
factored into a scaling term:\label{conv:psfEScale}
\begin{equation}
  \label{eq:PSF_prescale}
  \psfscaling(E) = \sqrt{\left[c_0\cdot
      \left(\frac{E}{100\ {\rm MeV}} \right)^{-\beta}\right]^2+c_1^2}.
\end{equation}
Despite a careful investigation, we did not find a simple satisfactory
description of the $\theta$ dependence to be incorporated in the scaling
function.

When building our MC-based PSF we use a set of scaling function parameters
based on pre-launch simulations and confirmed with analysis from beam tests
with the Calibration Unit (see \secref{subsec:EDisp_beamTest} for more
details). The values of these parameters are shown in \Tabref{scalef}. Note that 
\psfscaling(E) has same functional form as for $C_{68}$ in \secref{subsubsec:psf_analysis},  
however we have updated the parameters slightly based on the scaling observed in
our \allgamma\ sample.

\begin{table}[htb]
  \centering
  \begin{tabular}{cccc}
    \hline
    Conversion type & $c_0$ [$^\circ$] & $c_1$ [$^\circ$] & $\beta$ \\
    \hline
    \hline
    Front & 3.32 & 0.022 & 0.80 \\
    Back  & 5.50 & 0.074 & 0.80 \\
    \hline
  \end{tabular}
  \caption{Parameters of the angular deviation scaling function \psfscaling\
    for the PSF parametrization.}
  \label{tab:scalef}
\end{table}

We define the \emph{scaled angular deviation} $x$\label{conv:scaledPSFvar} as:
\begin{equation}
  x = \frac{\delta v}{\psfscaling(E)}.
  \label{eq:PSF_angulardev}
\end{equation}
An example of scaled deviation is shown in \figref{psfdiff}. The effect
of the scaling is to make the profile almost independent of energy, in that
the maximum is always close to $x = 1$ for all energy bins while the PSF 68\%
containment varies by almost two orders of magnitude from 100~MeV to 100~GeV.

Before the fit is performed, each scaled deviation histogram is converted into
a probability density with respect to solid angle.
The result is illustrated in \figref{psfdiff}.

\twopanel{htb!}{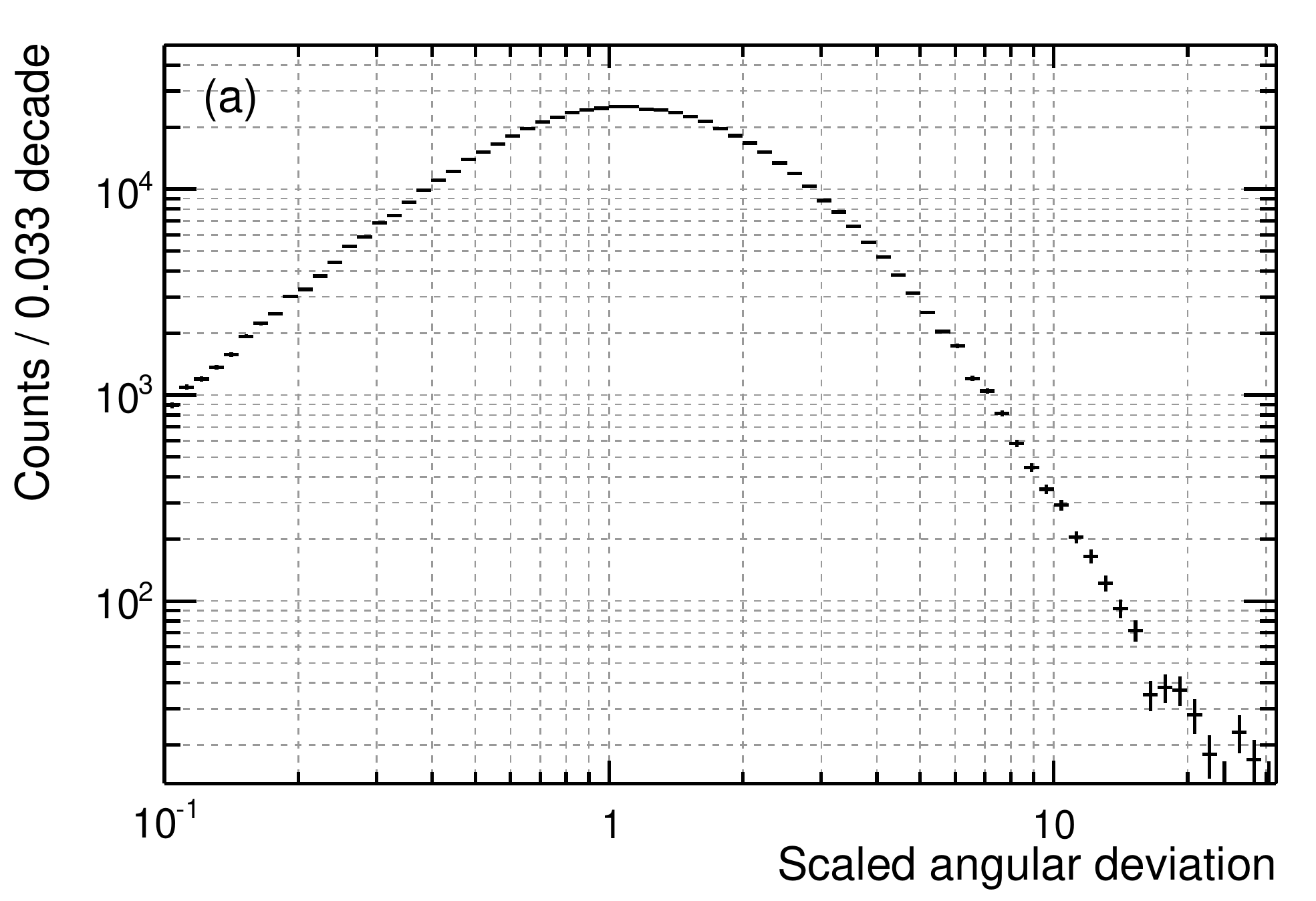}{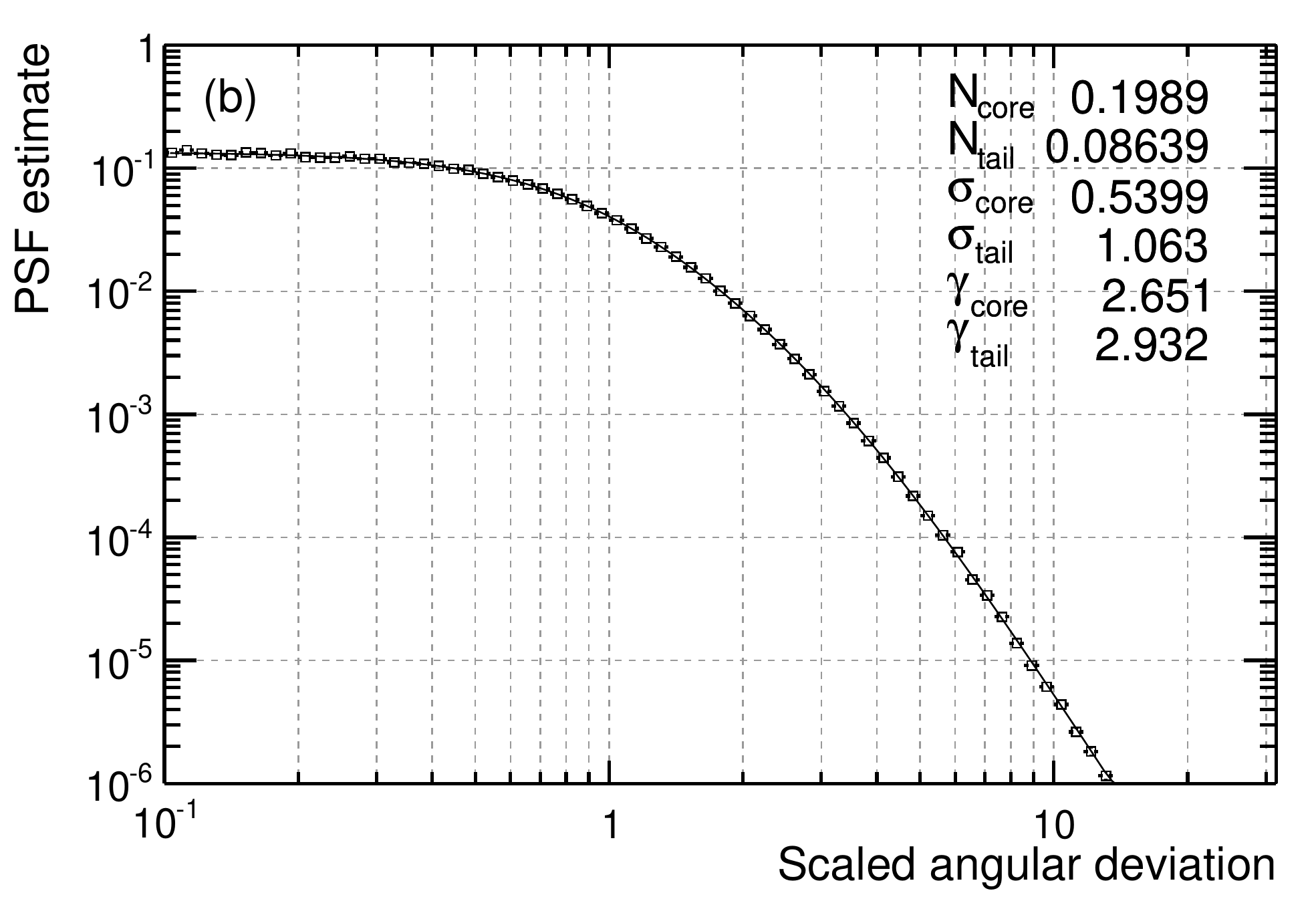}{
  \caption{(a) Scaled angular deviation histogram and (b) PSF estimate in the
    range $E=[5.6,10]$~GeV and $\theta = [26,37]^\circ$ for the \irf{P7SOURCE\_V6}
    event class, front section.}
  \label{fig:psfdiff}
}

Note that, although the scaling removes most of the energy dependence, the simulation
indicates significant variation of the PSF with $\theta$. At larger
incidence angle the tracks must cross more material in each TKR plane. At energies below
$\sim 1$~GeV this degrades the PSF owing to the increased multiple scattering,
while at higher energies (above $\sim 1$~GeV) the additional complication
of hard scattering processes in the TKR and additional hits in the TKR from the
nascent electromagnetic shower complicate the track finding and degrade the
PSF. \Figref{PSF_contain2d} shows how the scaled containment radii evolve with
energy and incidence angle.  \NEWTEXT{To test the $\phi$ dependence of the PSF we have also 
measured the containment radii independently for events with $\xi >
0.33$ and $\xi < 0.33$, where $\xi$ is the folded azimuthal angle defined by \Eqref{eq:phi_xi}. The 68\% and 95\%
containment radii for the two $\xi$ ranges differ by $< 5\%$ for all energies and angles except 
at high energies ($>10$~GeV) and low incidence angles ($\cos\theta >0.7$) for back-converting events, 
and at even higher energies ($>100$~GeV) and large incidence angles ($\cos\theta < 0.7$) for front-converting events.
Even then the maximum difference in the 68\% containment radius for
events is only $10\%$ at $10$~GeV and $25\%$ at $100$~GeV.   In summary, the
variations of the containment radii with $\phi$ are many times smaller
than the corresponding variations with $\theta$ for all but the highest
energies ($>\sim 100$~GeV).}

\fourpanel{htb!}{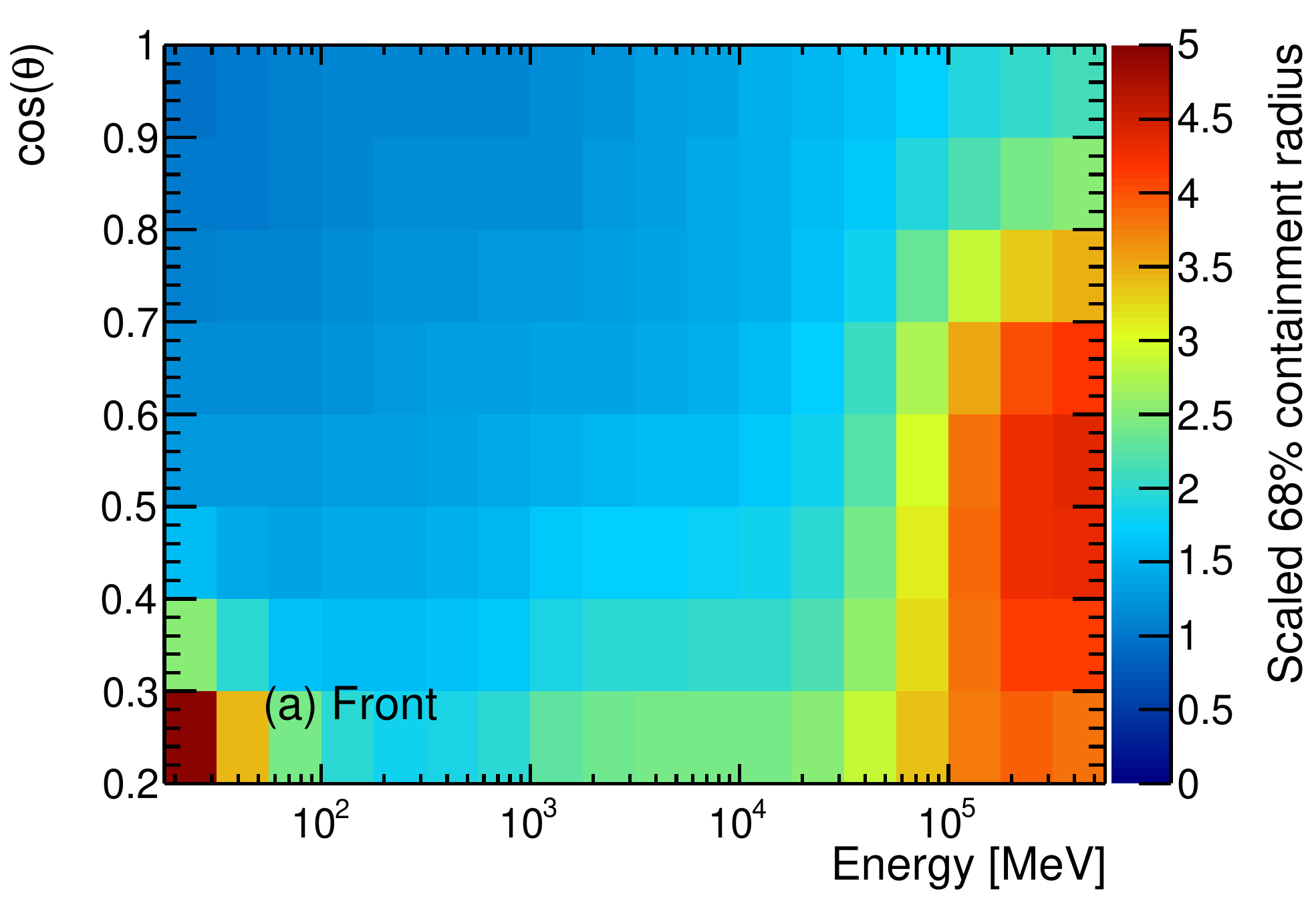}{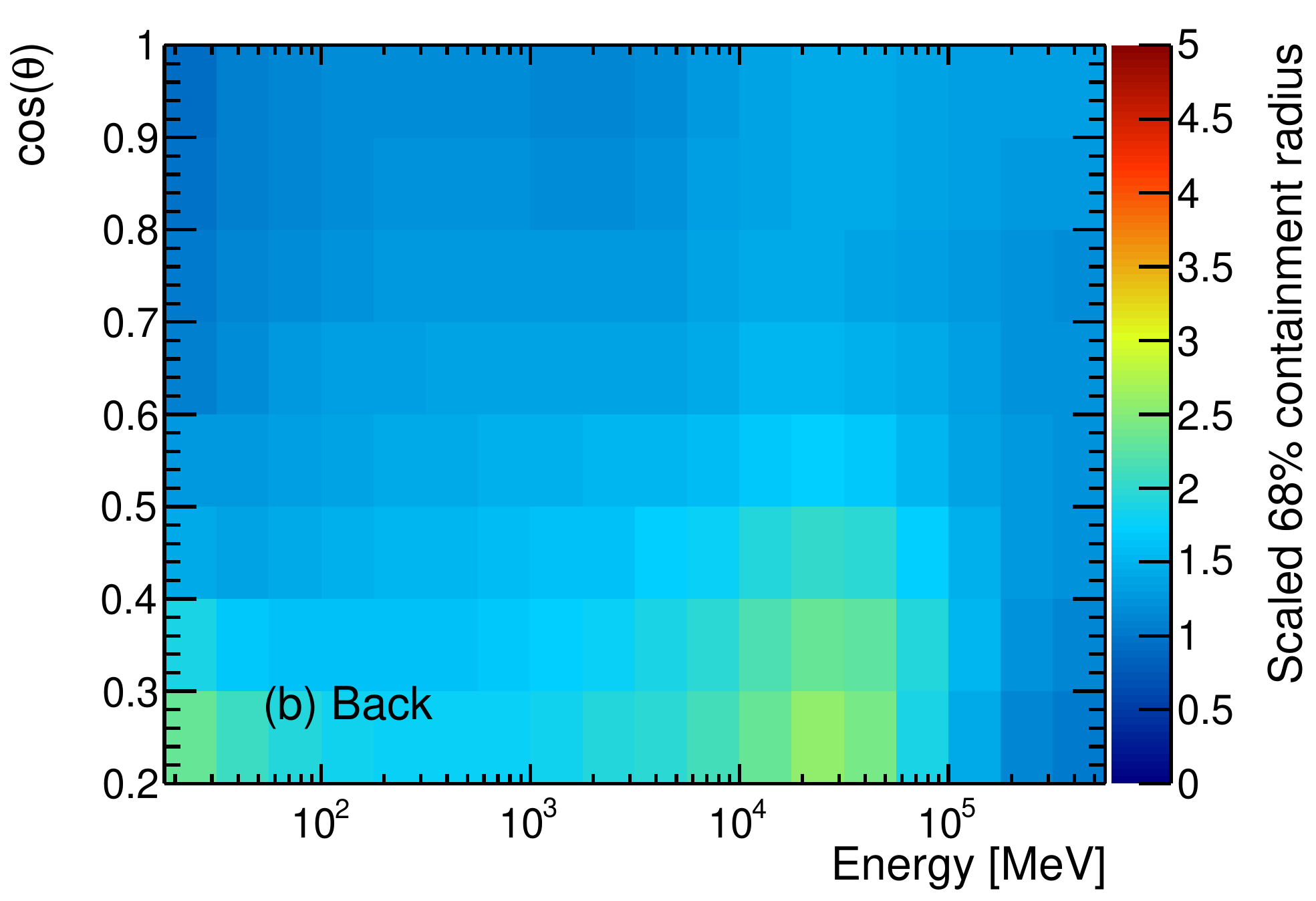}{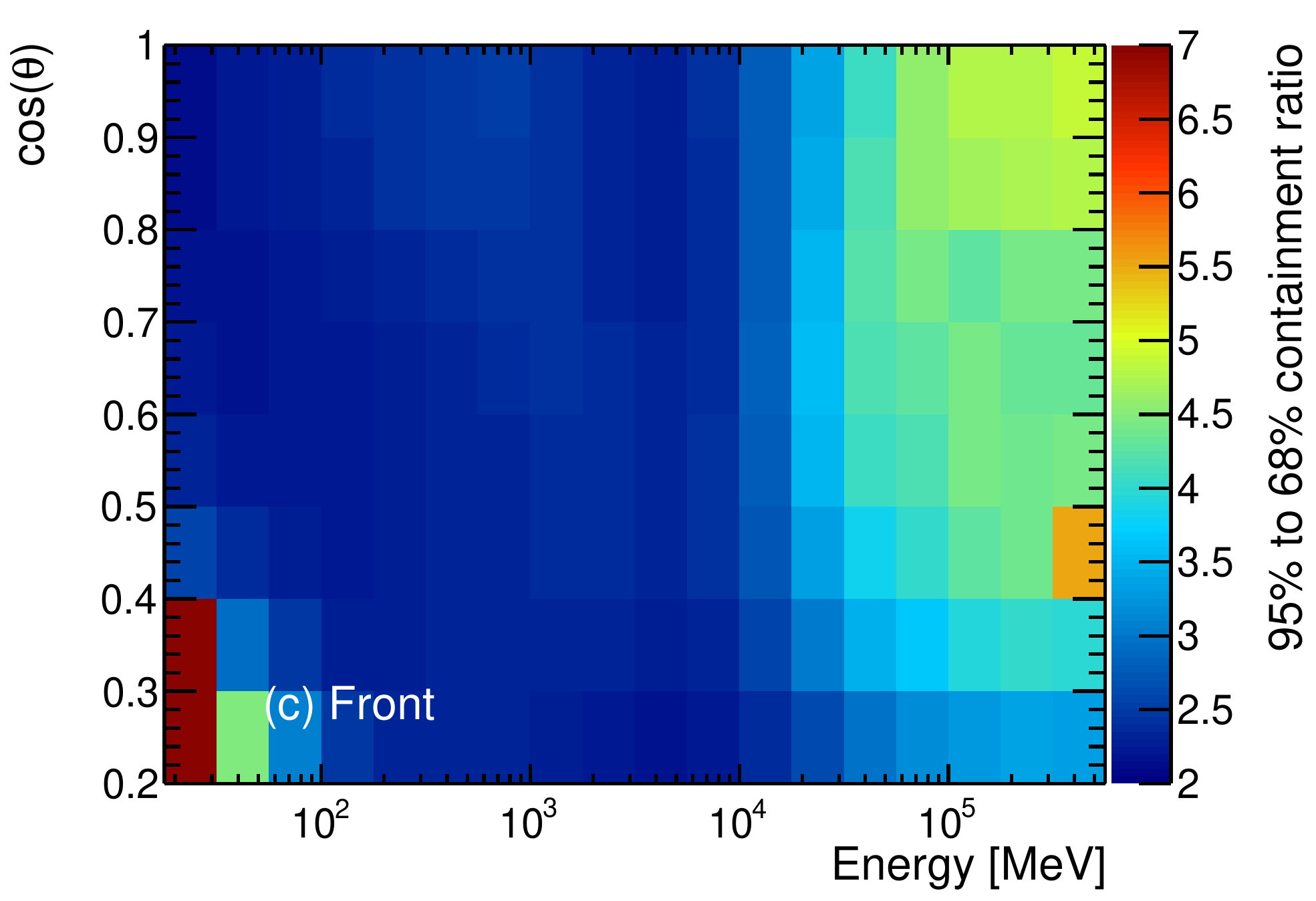}{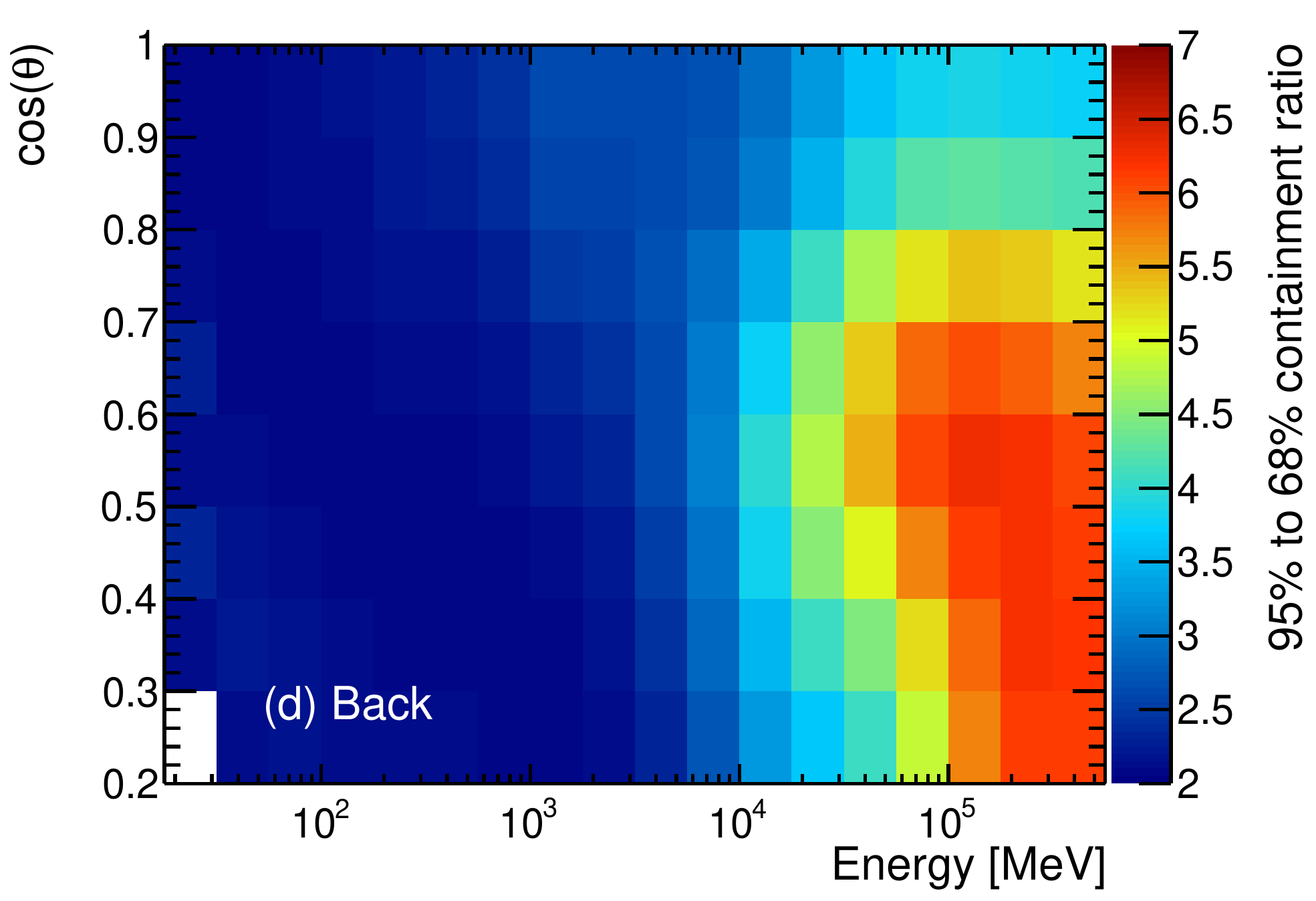}{
  \caption{Scaled 68\% containment radius (top) and ratio of 95\% to 68\%
    containment radii (bottom) as a function of $(E,\theta)$, for front-
    and back-converting events in the \irf{P7SOURCE\_V6} IRFs.}
  \label{fig:PSF_contain2d}
}

The base function for the PSF is the same as used by the \emph{XMM-Newton}
mission
\citep{REF:XMM.Psf,REF:XMM.Psf:2011}\label{conv:KingGamma}\label{conv:KingSigma}:
\begin{equation}
  \Kingf(x,\sigma,\gamma) = \frac{1}{2\pi\sigma^{2}}
  \left( 1-{\frac{1}{\gamma}}\right) \cdot
  \left[ 1+{\frac{1}{2\gamma}}\cdot {\frac{x^2}{\sigma^2}}\right]^{-\gamma};
  \label{eq:PSF_psfbase}
\end{equation}
which \citet{REF:XMM.Psf} refer to as a \emph{King function}\label{conv:king_func}\citep{REF:KING:1962}, and is isomorphic
to the well-studied Student's t-distribution \citep{REF:Students:1908}.
Note that this function is defined so as to satisfy the normalization condition:
\begin{equation}
  \int^{\infty}_{0} \Kingf(x,\sigma,\gamma) \, 2\pi x \,dx = 1;
\end{equation}
the extra $2\pi x$ comes from the integral over the solid angle
$d\Omega = \sin(x) \,dx d\phi \sim 2 \pi x \,dx$.  
However, at very low energies the PSF widens to the point that the small angle
approximation fails by more than a few percent and $\Kingf(x)\sin(x)$ must be normalized numerically.

To allow for more accurate descriptions of the tails of the distributions,
we use the sum of two King functions to represent the dependence of
the PSF on scaled deviation for a given incidence angle and energy:
\label{conv:king_f_core}
\begin{equation}
  \psf(x) = \core{f}\Kingf(x, \core{\sigma}, \core{\gamma}) +
  (1 - \core{f}) \Kingf(x, \tail{\sigma}, \tail{\gamma}).
  \label{eq:PSF_psf2king}
\end{equation}
The $\sigma$ and $\gamma$ values are stored in tables of PSF parameters as
\texttt{SCORE}, \texttt{STAIL}, \texttt{GCORE} and \texttt{GTAIL} respectively.
Because of the arbitrary normalization used in fitting the PSF function,
$\core{f}$ must be extracted from the \texttt{NTAIL} table parameter,
in conjunction with \texttt{SCORE} and \texttt{STAIL}:
\begin{equation}
  \core{f} = \frac{1}{1 + \texttt{NTAIL} \cdot
    \texttt{STAIL}^2/\texttt{SCORE}^2}.
\label{eq:PSF_f_tail}
\end{equation}

The fitting function has been revised several times since the
development of the first preliminary response functions. The version described
here is the one currently being used and is different from e.g., that used for
\irf{P6\_V3} IRFs. A description of the fit functions used in the past is
given in \secref{subsubsec:legacy_psf}.

\subsubsection{Legacy Point-Spread Function Parametrization}\label{subsubsec:legacy_psf}

The first set of publicly released IRFs, \irf{P6\_V3}, used a slightly
different PSF parametrization. Specifically, it allowed for only one $\sigma$
parameter and fixed the relative normalization of the two King functions by
constraining the two to contribute equally at $x_b = 2\sqrt{5}\sigma$.
So for \irf{P6\_V3} we used
\begin{equation}
  \psf(x)= \core{f} \Kingf(x, \sigma, \core{\gamma}) +
  (1 - \core{f}) \Kingf(x, \sigma, \tail{\gamma}),
  \label{eq:PSF_psflegacy}
\end{equation}
with
\begin{equation}
  \core{f} = \frac{1}{1 + \Kingf(x_{b},\sigma,\gamma_{\rm core}) /
    \Kingf(x_{b},\sigma,\gamma_{\rm tail} )}.
  \label{eq:PSF_ncorelegacy}
\end{equation}

\subsection{Point-Spread Function from On-Orbit Data}\label{subsec:PSF_onorbit}

During the first year of the mission we observed that for energies greater than
a few GeV the distributions of \gammaRays\ around isolated point sources were
systematically wider than the expectations based on the PSF estimated from the
MC simulations. We observed the same discrepancies for pulsars and blazars. In 
order to obtain a more accurate description of the core of the PSF for
sources that are observed at a typical range of incidence angles, starting from the 
then-current \psix\ IRFs, we derived the PSF directly from
flight data, by means of a stacking analysis of selected point sources. The
details of this analysis are described in~\citet{REF:PSFHALO}. Here we
summarize the procedure, the associated uncertainties, and the impact on
high-level source analysis.

\subsubsection{Angular Containment from Pulsars}\label{subsubsec:PSF_oo_pulsars}

In the 1--10~GeV energy range bright pulsars are excellent sources for
evaluating the in-flight PSF: not only they are among the brightest \gammaRay\
sources, providing abundant statistics, the pulsed emission is very stable and
the angular distributions of \gammaRays\ around the true positions can be
estimated readily by phase selecting \gammaRays. The Vela pulsar is the
brightest pulsar and an analysis with adequate statistics can be based on Vela
alone.
In the remainder of this section we describe the procedure for Vela only; the
extension to an arbitrary number of pulsars is straightforward
\citep[see][]{REF:PSFHALO}.

We use the \irf{P7SOURCE} Vela calibration sample, which we divided into
front- and back-converting subsamples, and bin the data in 4
energy bins per decade.  We then calculate the pulsar phase for each \gammaRay.  As mentioned in
\secref{subsubsec:LAT_Vela} we define $[0.12,0.17]\cup[0.52,0.57]$ as the
``on'' interval and $[0.8,1.0]$ as ``off'' (see \parenfigref{vela_phase} for the
phase histogram). Next we calculate the containment angles: the
position of Vela is known with a precision that greatly exceeds the LAT angular
resolution and which can be assumed as the \emph{true} source
position.  We make
histograms of the angular deviations from Vela for the on-peak and
off-pulse intervals and normalize them for the relative phase ranges.
To estimate the PSF from flight data we measure the 
containment radii from the difference between the histograms.

\subsubsection{Angular Containment from Active Galactic Nuclei}\label{subsubsec:PSF_oo_agn}

Above $\sim 10$~GeV spectral cutoffs of pulsars leave AGN as the only
attractive sources for studying the PSF.
In \citet{REF:PSFHALO} we address the potential
contribution from \emph{pair halos} around AGN and conclude that we see no
indication that this phenomenon is the explanation for the PSF being broader
than predicted. Thus, we treat AGN as point sources.  Many \gammaRayHyph\
sources are considered to be only ``associated'' with AGN, as opposed to ``firmly
identified'' \citep[see][for a discussion of the distinction]{REF:2011.2FGL},
because of the limited angular resolution of the LAT. In the present analysis we
consider only sources with high-confidence associations.  As for
pulsars, the positions of AGN are known with high precision from
other wavelengths, and the angular distances from the true directions can be
calculated. To accumulate enough statistics we stacked several sources
and performed a joint analysis. We selected AGN
from among the LAT sources with the highest significances above 10~GeV outside
the Galactic plane.
This energy limit is set by the source density: below a few GeV the LAT
PSF is broad enough that nearby sources frequently overlap.

A significant difference with respect to the pulsar analysis is the necessity
of modeling the background in evaluating the distribution of angular
deviations.  We assume that after the stacking of the sky regions far from the
Galactic plane the background count distribution can be assumed to be
isotropic. At each energy the background is modeled as a flat distribution 
normalized by the amplitude in an annulus centered on the stacked data set. 
The inner radius of this annulus was chosen to be significantly larger than the 
region containing \gammaRays\ from the stacked AGN sample.  The uncertainty of 
the containment radius in each energy bin was set to the RMS of a large sample of MC 
realizations for the signal and background distributions.


\subsubsection{Point-Spread Function Fitting}\label{subsubsec:PSF_oo_irfs}

We have developed a procedure \citep[described in detail in][]{REF:PSFHALO},
to fit our PSF model to the measured containment radii for different energy
ranges. Given the statistical limitations we use a single King function,
\Eqref{eq:PSF_psfbase}. For the same reason we do not measure the dependence of
the PSF on the incidence angle, i.e., we calculate an acceptance-weighted
average over the incidence angle.  We first fit the experimental
68\% and 95\% containment radii ($R_{68}$ and $R_{95}$\label{conv:R68}\label{conv:R95}) with \Eqref{eq:PSF_psfbase}. Then we
extract a new scaling relation.  And finally, we use the {\it fitted}
(rather that the measured) 68\% and 95\% containment radii to obtain a 
new set of PSF parameters for each energy bin.  By using the fitted 
containment radii, this procedure smooths out the statistical
fluctuations across the energy bins.

\begin{figure}[htb!]
  \centering
  \includegraphics[width=\onecolfigwidth]{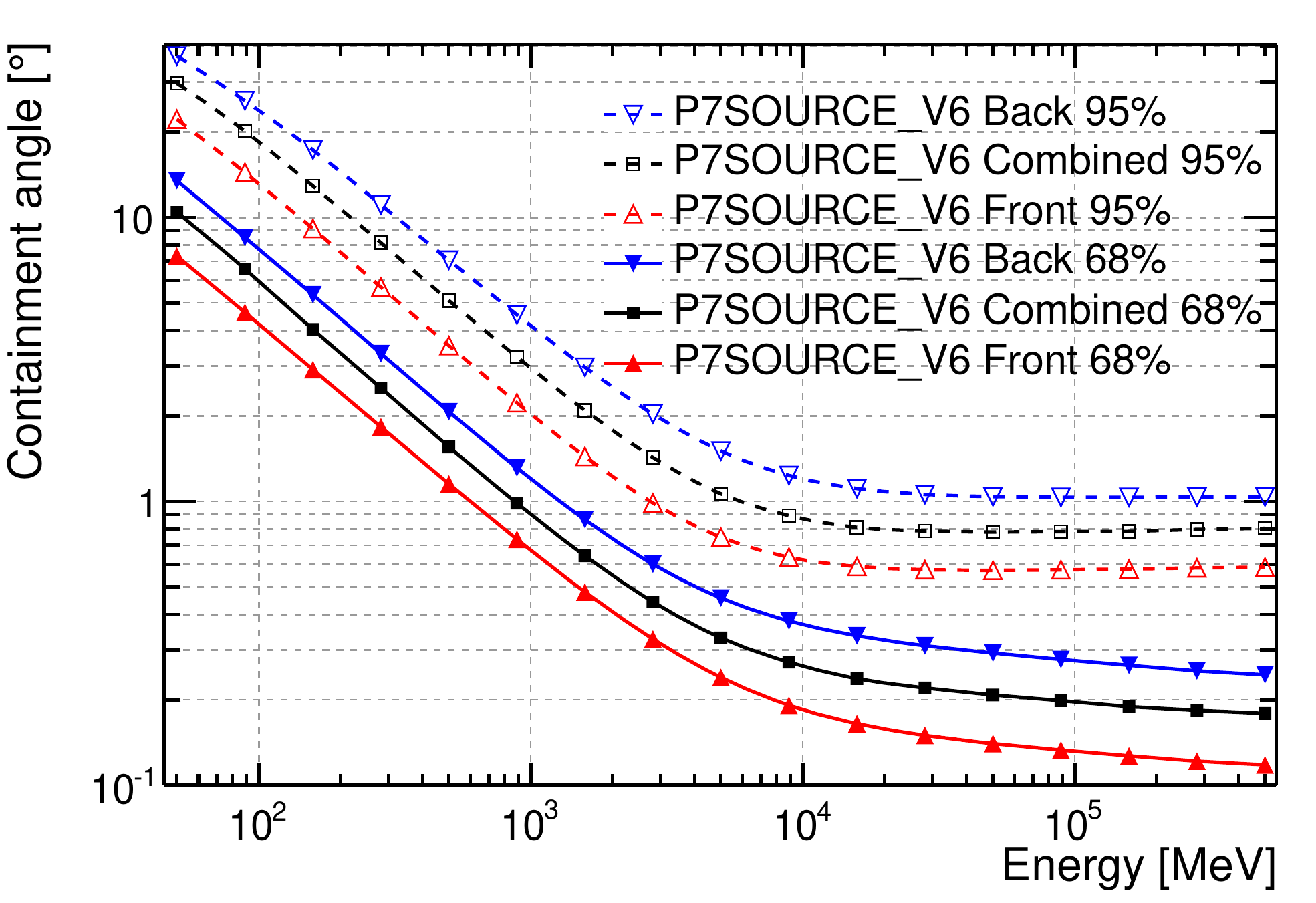}
  \caption{68\% and 95\% containment angles as a function of energy
    for the \irf{P7SOURCE\_V6} event class.}
  \label{fig:P7SOURCE_psf_energy}
\end{figure}

The 68\% and 95\% angular containment radii for the flight-based
\irf{P7SOURCE\_V6}  PSF are shown in \figref{P7SOURCE_psf_energy}.

\subsection{Uncertainties of the Point-Spread Function}\label{subsec:PSF_errors}

The uncertainty of the derived PSF was estimated by comparing the 68\%
and 95\% PSF containment radii from a set of calibration point sources
with the corresponding containment radii derived from the \irf{P7SOURCE\_V6}
PSF \citep{REF:PSFHALO}.
The 68\% and 95\% containment radii measure of the accuracy
representation of the PSF in the core and tail, respectively.  

The analysis was performed as a function of energy with
4 energy bins per decade.  To determine the
accuracy of the PSF fit as a function of incidence angle,
subsamples were also studied in which \gammaRays\ were additionally split
into three bins of $\cos\theta$ ($[0.2,0.5)$, $[0.5,0.75)$, and $[0.75,1.0])$.

\fourpanel{htb!}{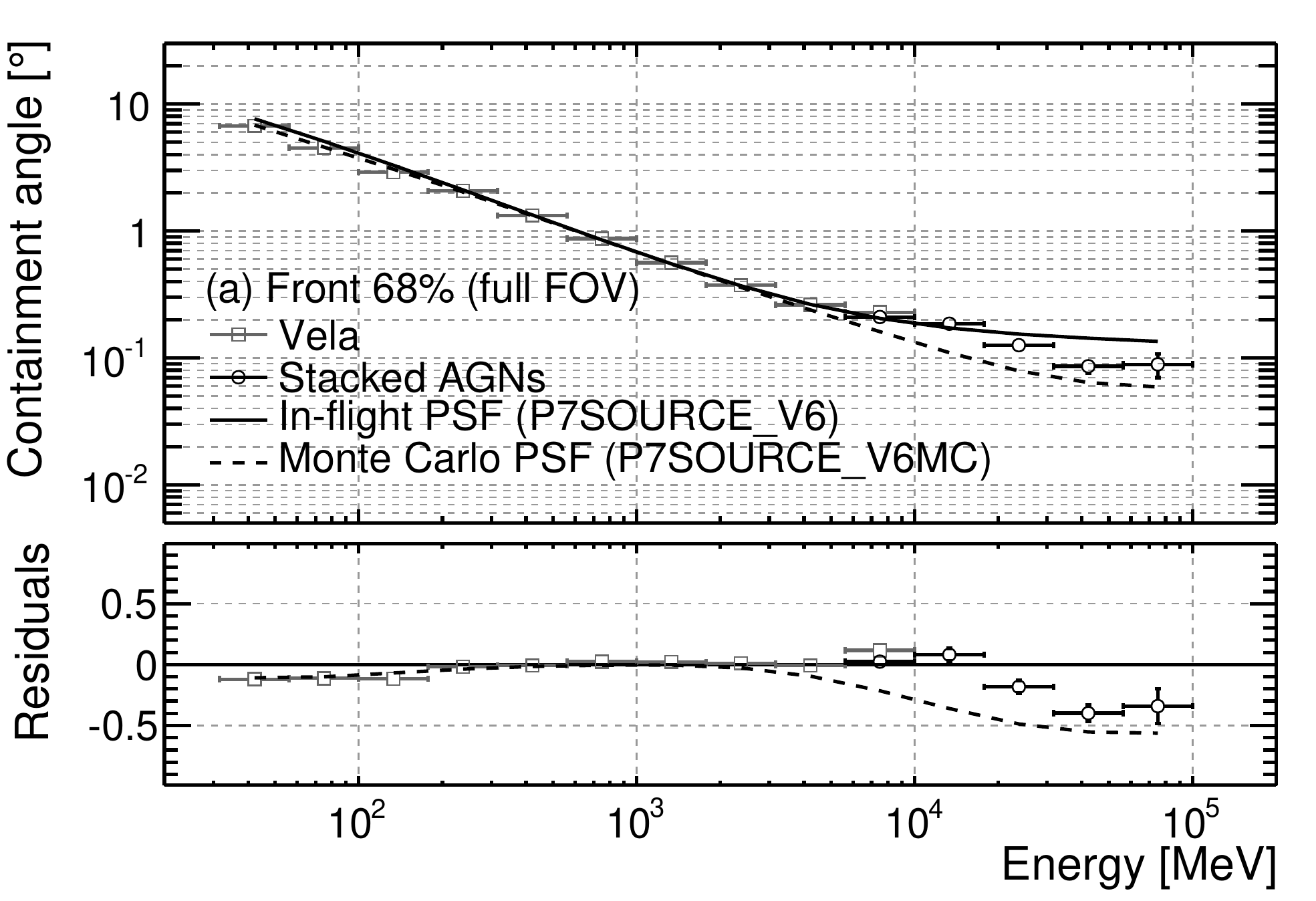}{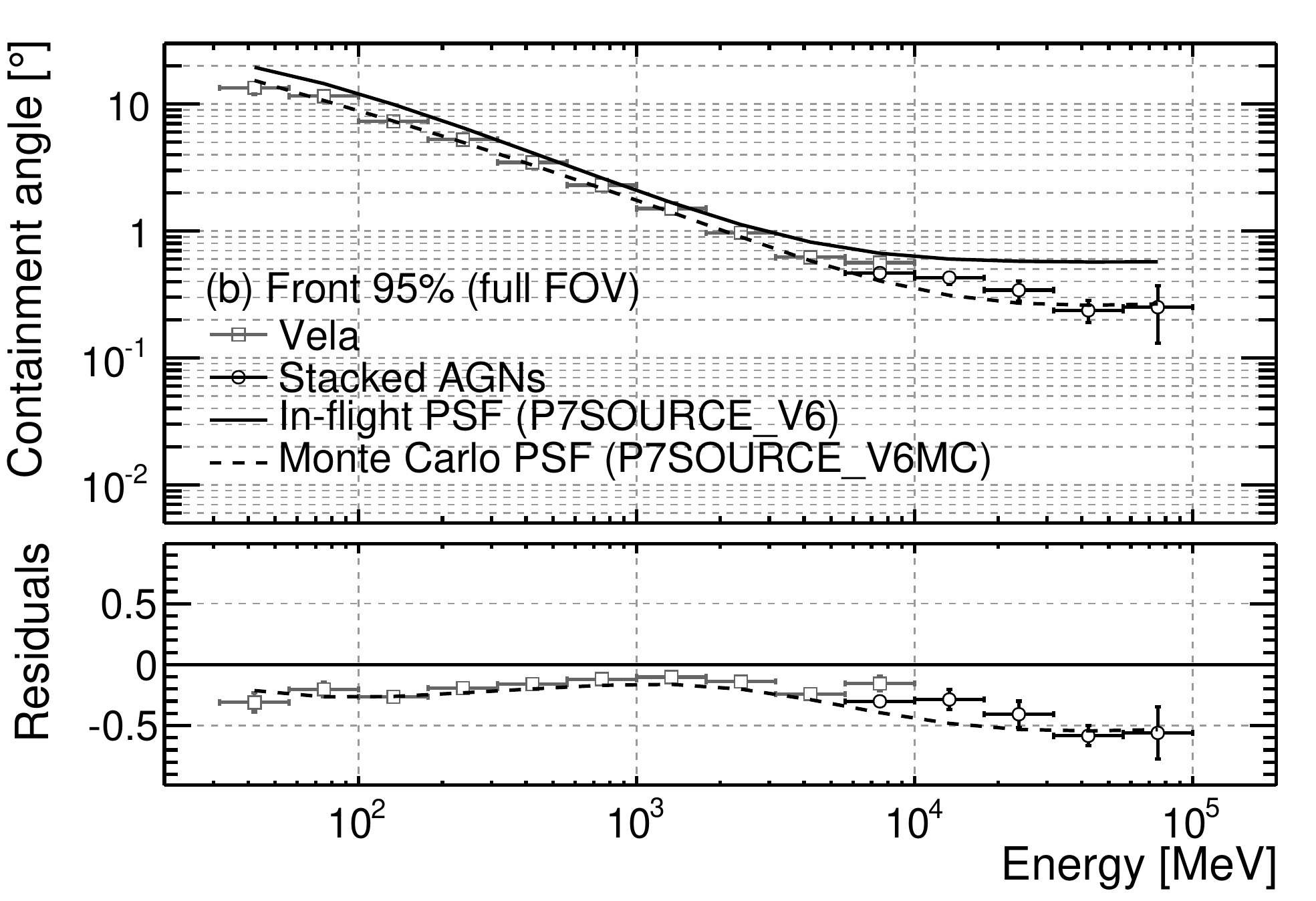}{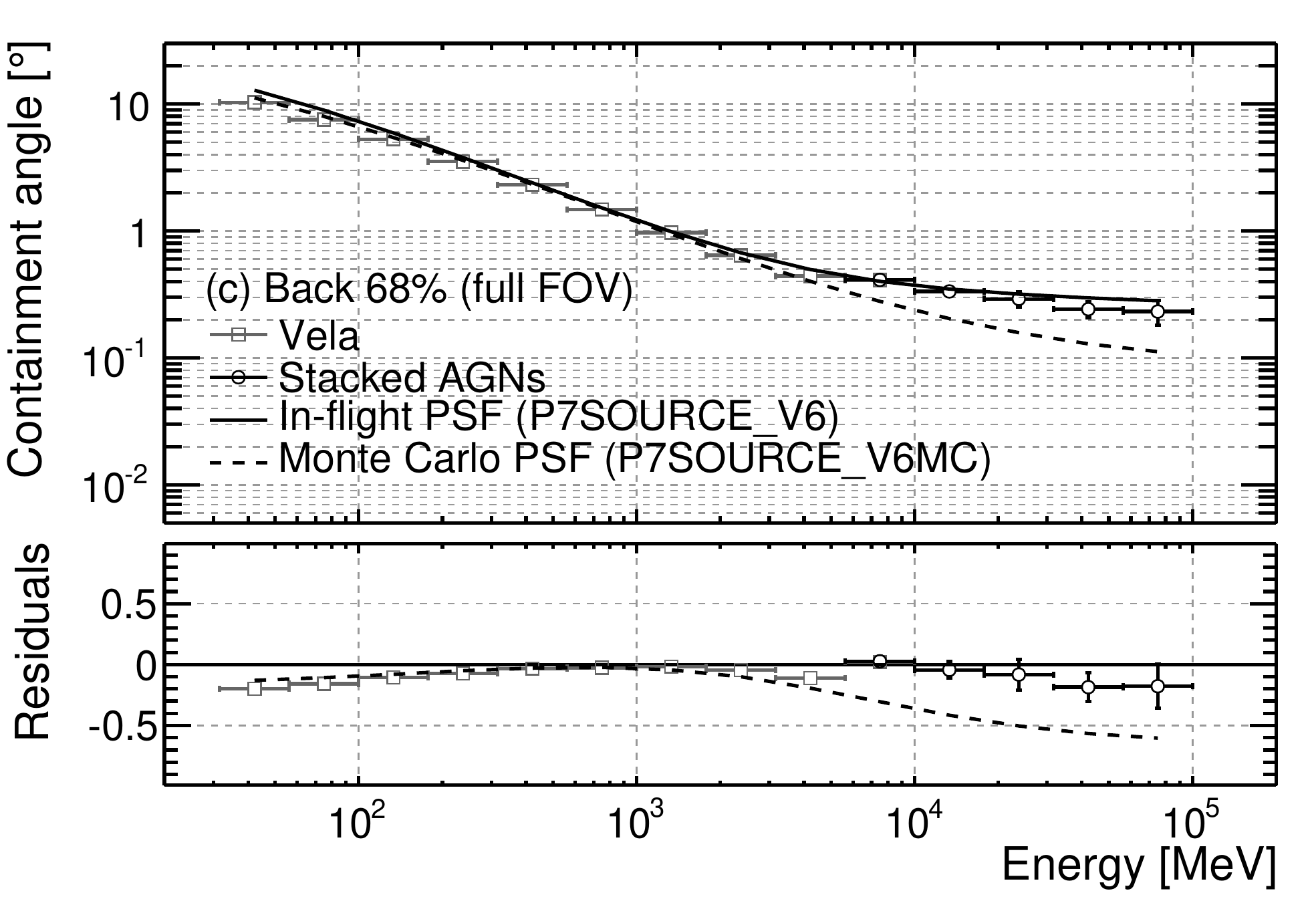}{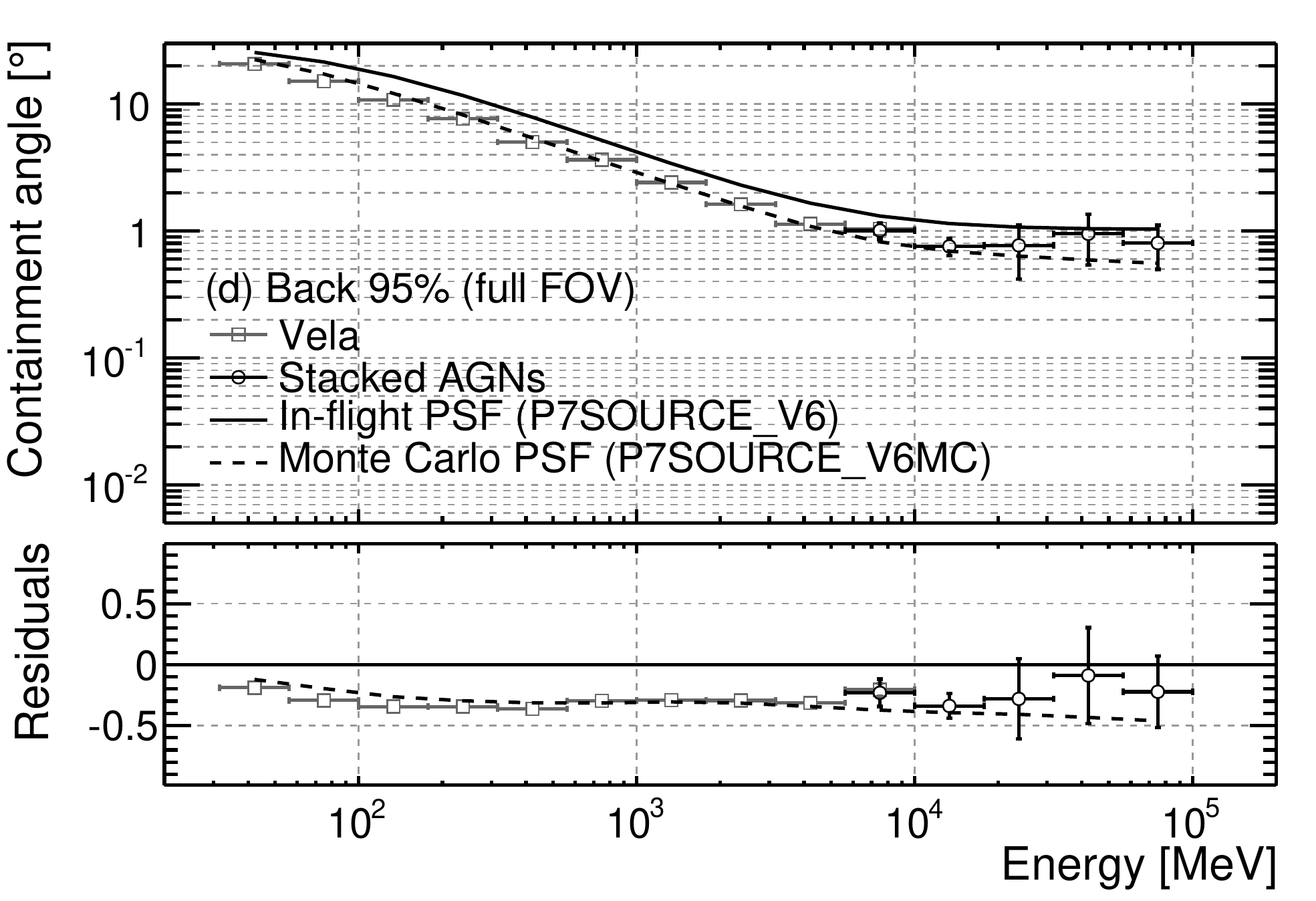}{
  \caption{68\% and 95\% containment radii for \gammaRays\ averaged over all
    incidence angles ($\cos\theta \in [0.2,1.0]$) as a function of energy for
    front (top) and back (bottom). Data points with error bars show the measured
    containment radii derived from the Vela and AGN PSF calibration data sets.
    Solid lines show the \irf{P7SOURCE\_V6} model predictions in each energy
    bin; dashed lines show the predictions from the MC simulations
    (\irf{P7SOURCE\_V6MC}). Residual plots indicate the fractional deviation
    with respect to \irf{P7SOURCE\_V6}. }
  \label{fig:psfvalidation}
}

\Figref{psfvalidation} shows the 68\% and 95\% containment
radii as a function of energy for front- and back-converting \gammaRays\ averaged over
incidence angle.  The smooth lines show the model predictions for the
MC (\irf{P7SOURCE\_V6MC}) and in-flight (\irf{P7SOURCE\_V6}) IRFs.  At
energies below 3~GeV the containment radii match the 
MC PSF with fractional residuals no larger than 10\%.  Above 3~GeV the
MC PSF begins to systematically underestimate the 68\% containment radius
 by as much as 50\% for both front- and back-converting \gammaRays.  As shown in \figref{psfvalidation}
the \irf{P7SOURCE\_V6} PSF reproduces the flattening of the energy dependence of the PSF
containment at high energies. However owing to the limitations of
using a single King function to parametrize the PSF, this model
over-predicts the PSF tails as represented by the 95\% containment radii.

\fourpanel{htb!}{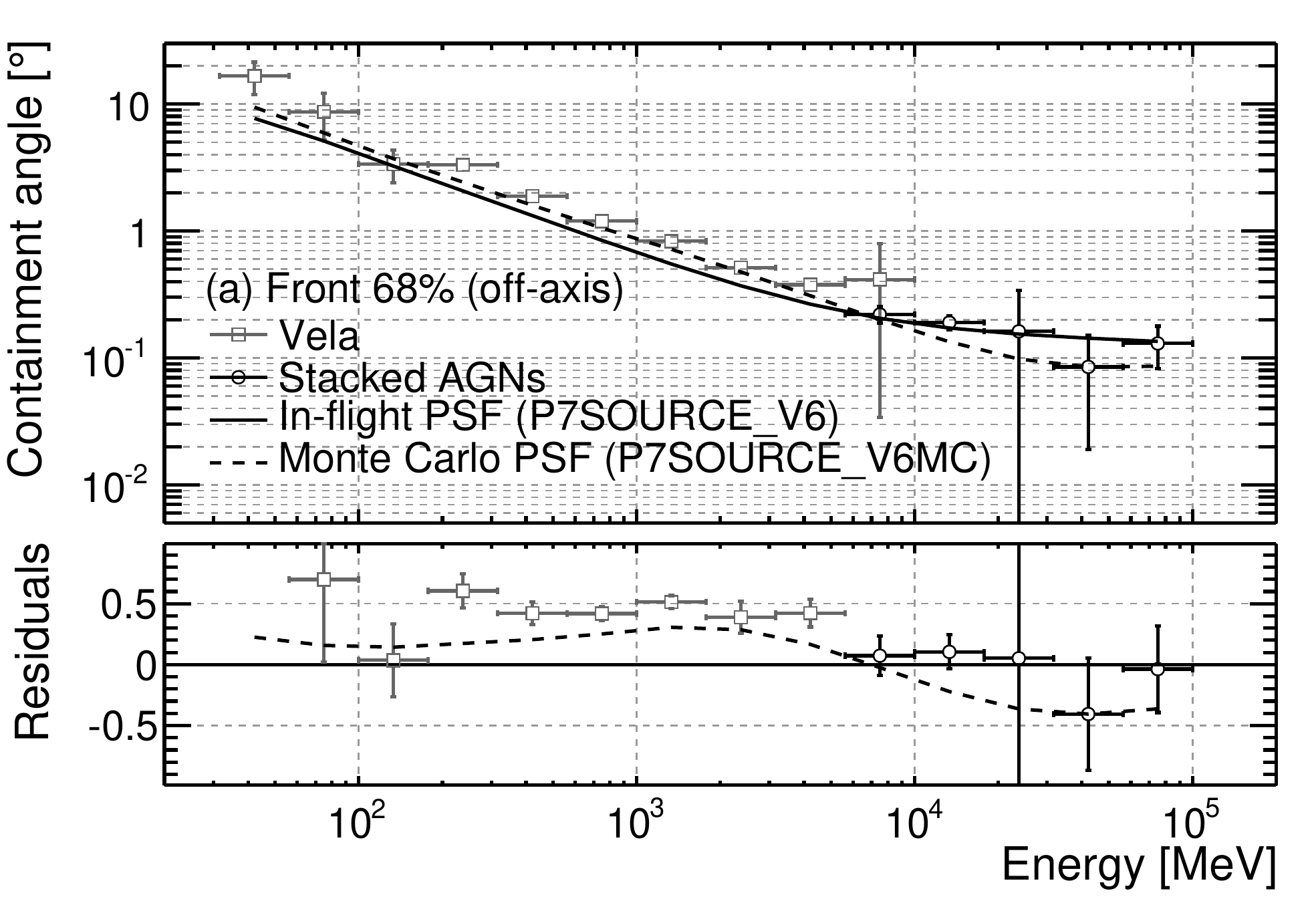}{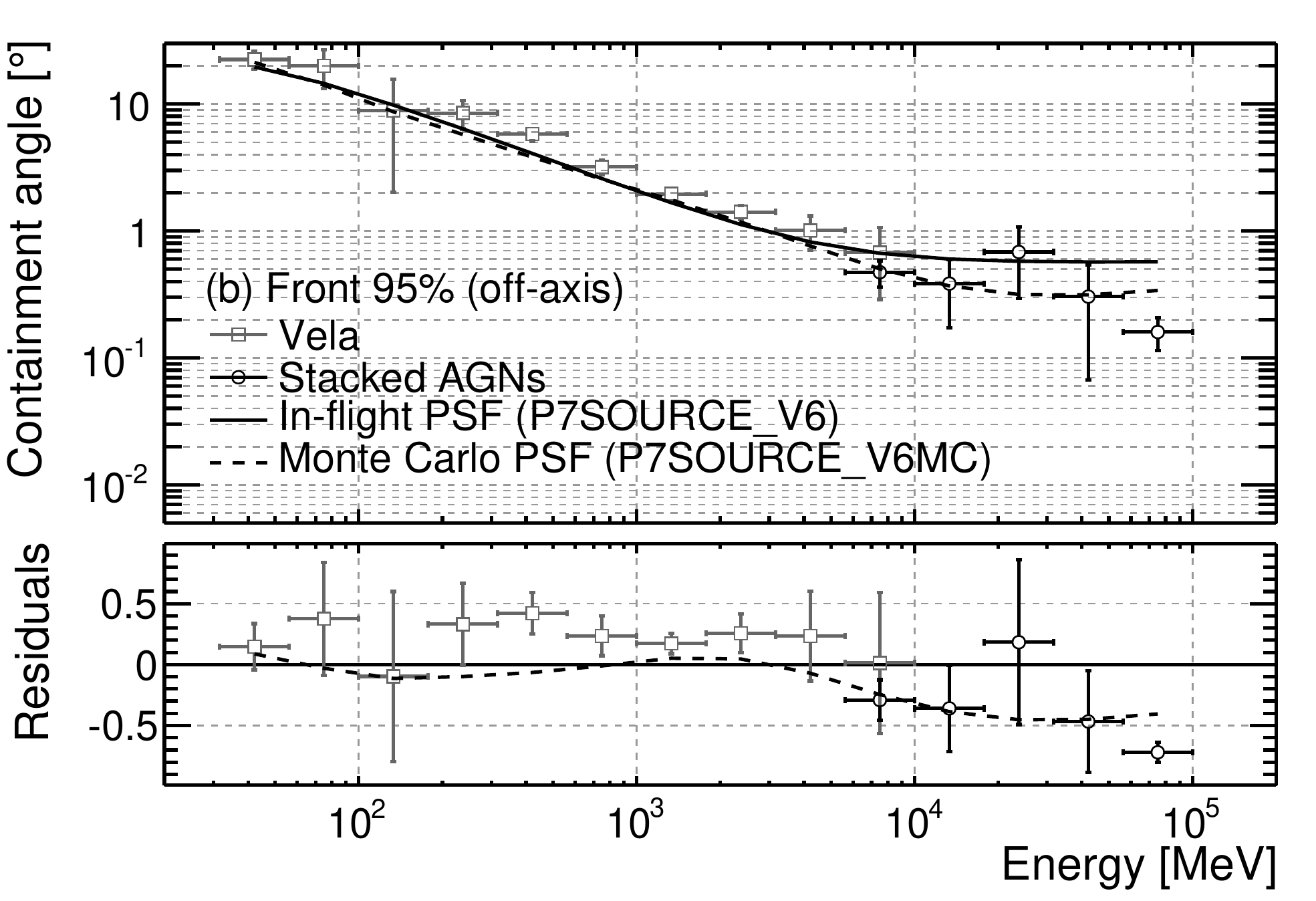}{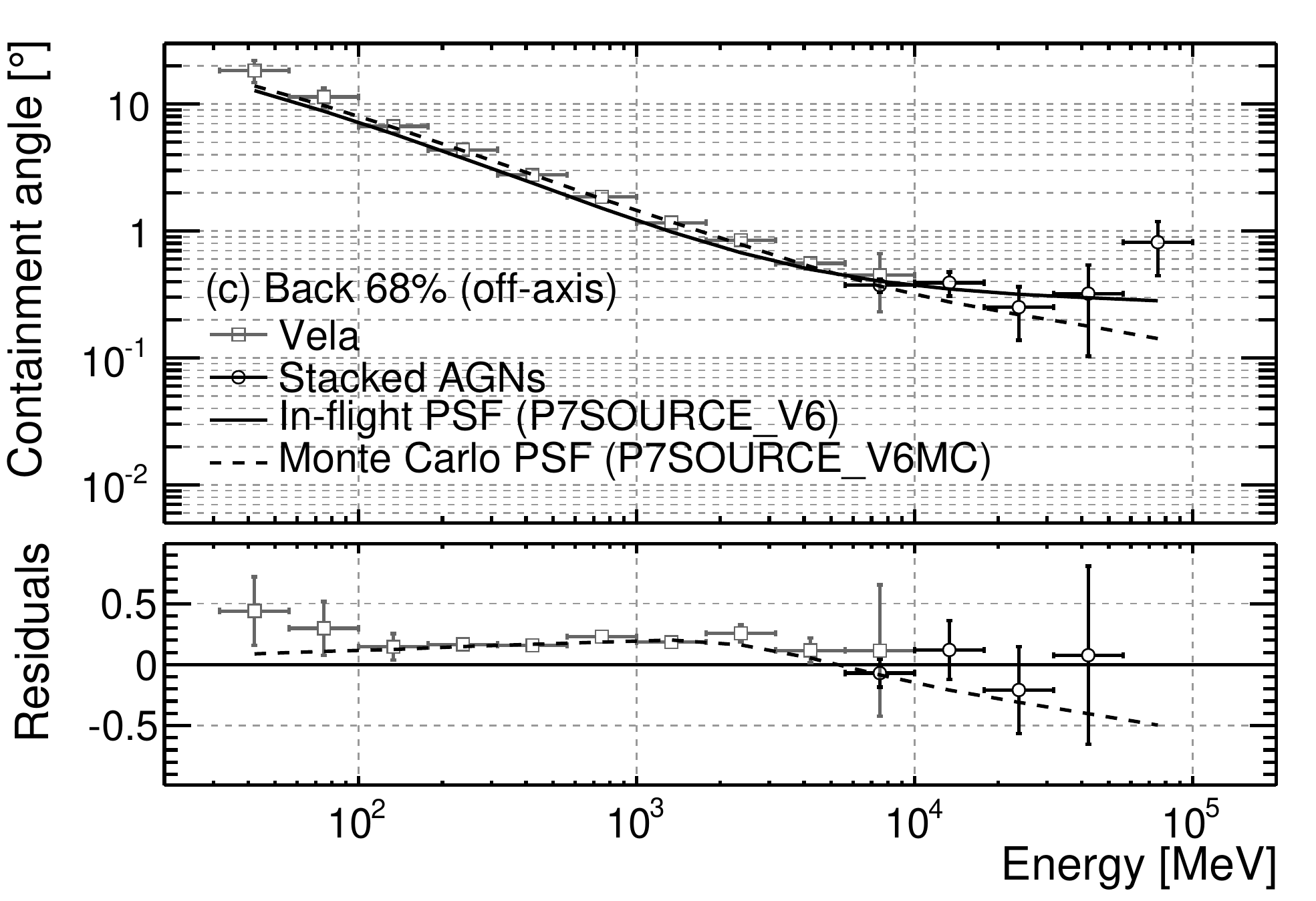}{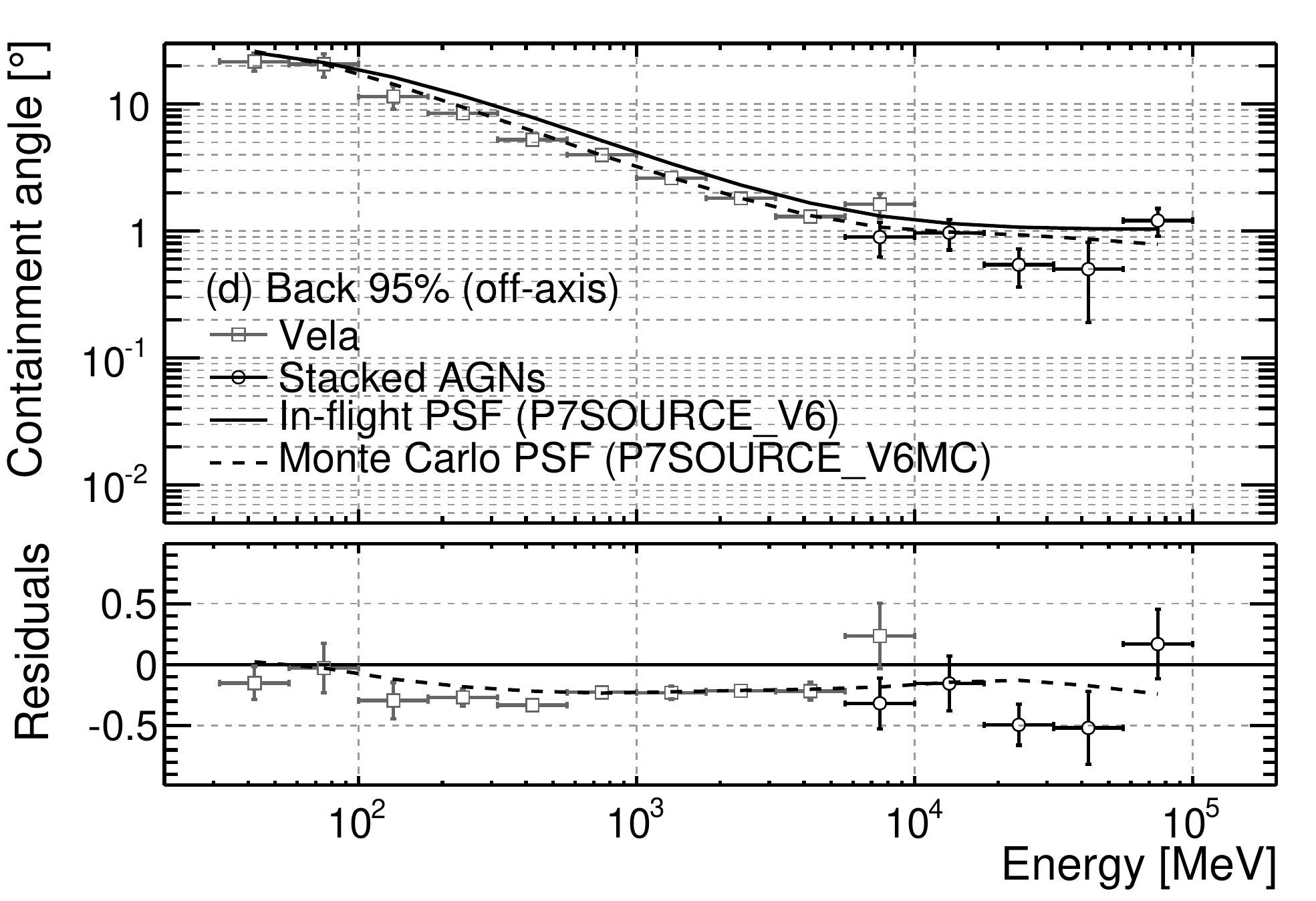}{
  \caption{68\% and 95\%  containment radii for off-axis \gammaRays\
    ($\cos\theta \in [0.2,0.5]$) as a function of energy for front (top) and
    back (bottom). Data points with error bars show the measured
    containment radii derived from the Vela and AGN PSF calibration
    data sets. Solid lines show the \irf{P7SOURCE\_V6} model predictions in
    each energy bin; dashed lines show the predictions from the MC 
    simulations (\irf{P7SOURCE\_V6MC}). Residual plots indicate the fractional
    deviation with respect to \irf{P7SOURCE\_V6}. }
  \label{fig:psfvalidation_offaxis}
}

At large incidence angles ($\cos\theta \in [0.2,0.5]$) the LAT PSF broadens by
approximately a factor of 1.5. Due to the sky survey observing mode of \fermi\ and the
decreased effective area at large incidence angles the fit of the in-flight PSF is
dominated by \gammaRays\ at smaller incidence angles ($\cos\theta \in
[0.5,1.0]$).  The variation of the PSF with incidence angle is most
relevant for the analysis of
transient phenomena in which the time-scale of interest is comparable
to or shorter than the orbital period of \fermi, such as GRBs
and short-period time-series analyses. As shown in
\figref{psfvalidation_offaxis}, the agreement of in-flight and
MC PSF models with the data appears worse at large
incidence angles ($\cos\theta \in [0.2,0.5]$) although the limited
statistics limit a rigorous comparison.  The in-flight PSF model, which
does not incorporate $\theta$~dependence, significantly under-predicts the
width of the 68\% containment radius for both front- and back-converting \gammaRays.
The effect of these discrepancies on high-level analysis will be considered in
\secref{subsec:PSF_highLevel}.

\subsection{The ``Fisheye'' Effect}\label{subsec:PSF_fisheye}

For sources observed only in a narrow range of incidence angle,
particularly near the edges of the LAT \fov, we must consider an additional
complication: particles that scatter toward the LAT boresight are more likely
to trigger the LAT and be reconstructed than particles that scatter away from
the boresight. Furthermore, since the event selections do not require the
separate reconstruction of both tracks from the $\gamma \to e^{+} e^{-}$
conversion and the reconstruction code will estimate an event direction using a
single track if no vertex is found, for some events we base our direction
estimate only on the particle that scattered more toward the LAT boresight.
This effect increases as the energy decreases, since multiple scattering causes
larger deviations. 

Over the course of a year the position of the LAT boresight relative to any
given direction in the sky is fairly uniformly distributed in azimuth; however,
for shorter periods several factors can lead to large non-uniformities
(see \secref{subsec:LAT_obsProf}). Furthermore, by construction the PSF is
averaged over azimuth. Therefore, for long ($>1$~year) integration periods the
treatment of the PSF described in the previous sections is perfectly adequate
and the fisheye effect simply results in a broadening of the PSF which is
correctly described by the IRFs. However, for shorter periods the fisheye
effect can result in systematic biases in localization, which should be
accounted for. To do so we consider the polar and azimuthal components of the
angular separation between the true and reconstructed \gammaRayHyph\ directions
in the \allgamma\  sample. Taking $\hat{v}$ as the true \gammaRayHyph\
direction (in the LAT frame), $\hat{v}^{\,\prime}$ as the reconstructed direction,
and $\hat{z}$ as the LAT boresight we can define the local polar and azimuthal
directions:
\begin{equation}\label{conv:localThetaPhi}
  \hat{\phi} = \frac{\hat{z} \times \hat{v}}{|\hat{z} \times \hat{v}|}
  \quad
  \hat{\theta} = \frac{\hat{\phi} \times \hat{v}}{|\hat{\phi} \times \hat{v}|}.
\end{equation}
Then we calculate the component of the misreconstruction along each:
\begin{equation}
  \delta\phi =
  \sin^{-1}\left( \hat{\phi} \cdot (\hat{v}^{\,\prime} - \hat{v}\,) \right)
  \quad
  \delta\theta =
  -\sin^{-1}\left( \hat{\theta} \cdot (\hat{v}^{\,\prime} - \hat{v}\,) \right)
\end{equation}
(the extra negative sign in the above equation is applied so that the fisheye
effect represents a bias toward positive values of $\delta\theta$). It is also
worth noting that simply considering the distributions of differences between
the true and reconstructed $\theta$ is complicated by the fact that the amount
of solid angle varies with $\theta$.

\Figref{psf_fisheye_offset_source} shows how the mean of $\delta\theta$ varies
as a function of energy and incidence angle for the \irf{P7SOURCE} event
selection. Although the bias can be very large at high incidence angles and
low energies, it is important to recall that (i) the PSF is quite wide in those
cases, and (ii) there is relatively little acceptance in that region.
\Figref{psf_fisheye_ratio_source} shows the ratio of the mean to the RMS of
$\delta\theta$ for the same energies and angles, and we see that except for
the furthest off-axis events and the lowest energies, the fisheye effect is a
small contributor to the overall width of the PSF when considering persistent 
\gammaRayHyph\ sources.

\twopanel{htb!}{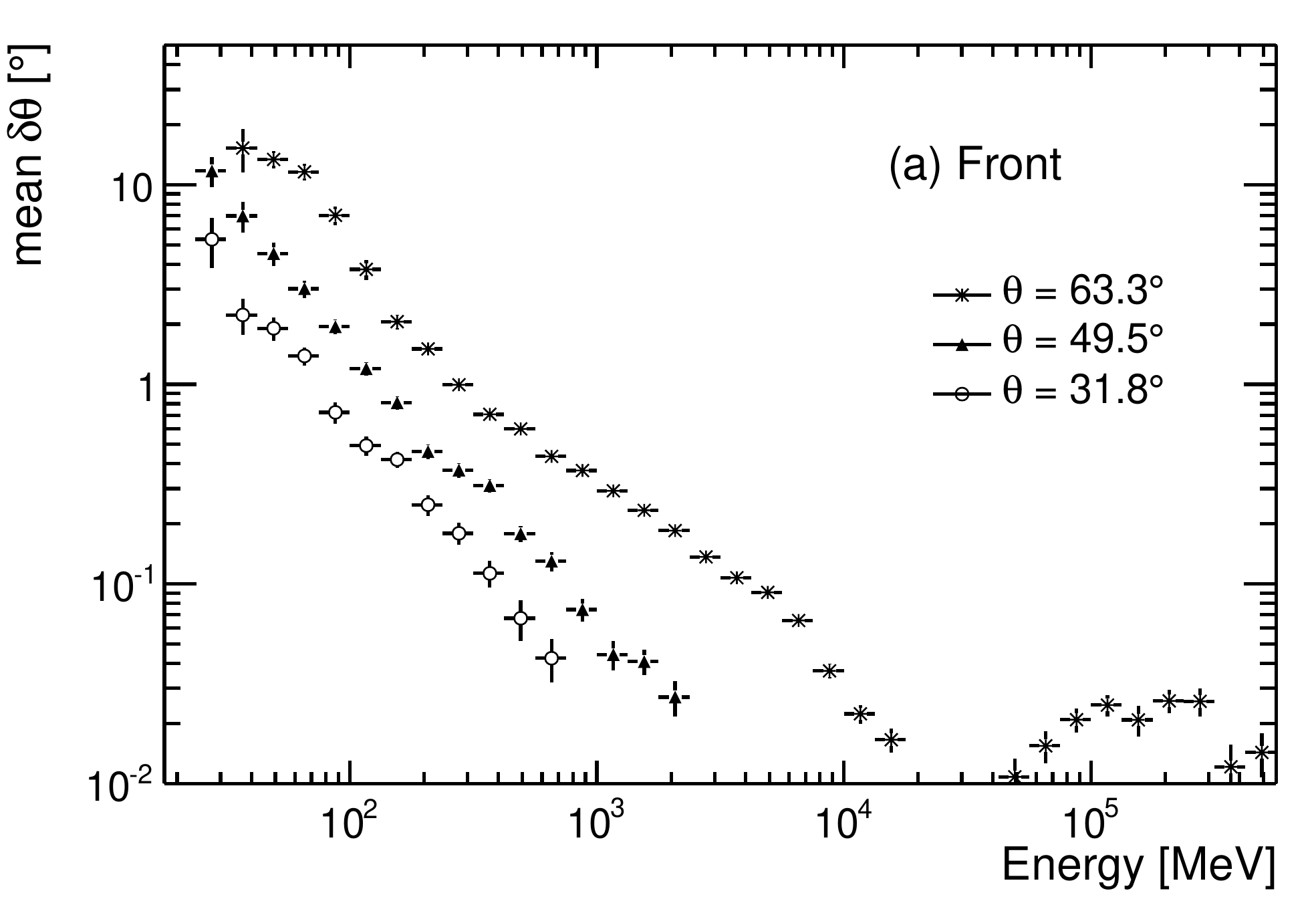}{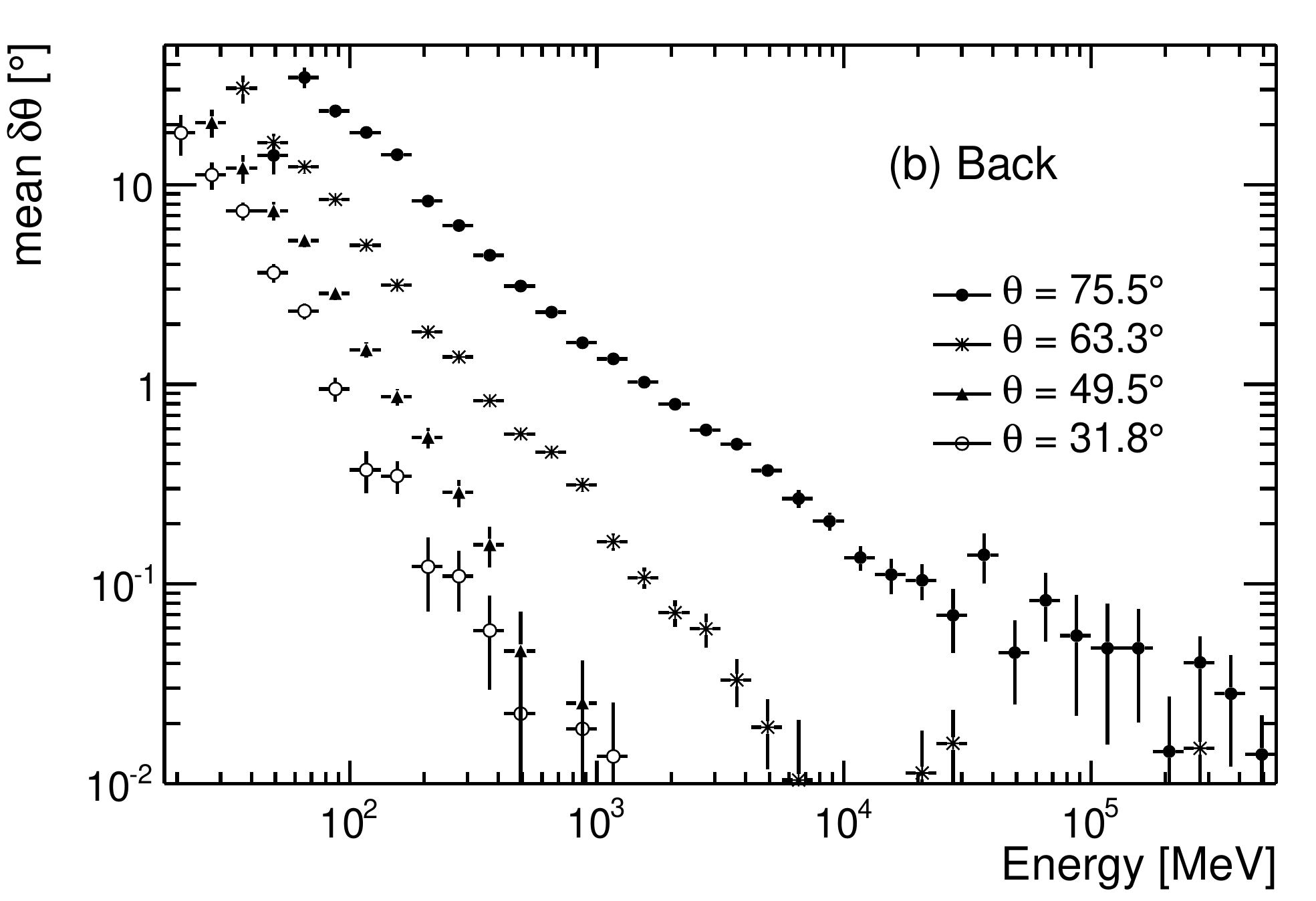}{
  \caption{Mean of $\delta\theta$ for MC simulated data as a function of energy 
    for several incidence angles for the \irf{P7SOURCE} event selection for 
    front-converting (a) and back-converting (b) events.  There are not 
    enough statistics for front-converting events at $\theta = 75.5^{\circ}$ to 
    extract reliable values.}
  \label{fig:psf_fisheye_offset_source}
}

\twopanel{htb!}{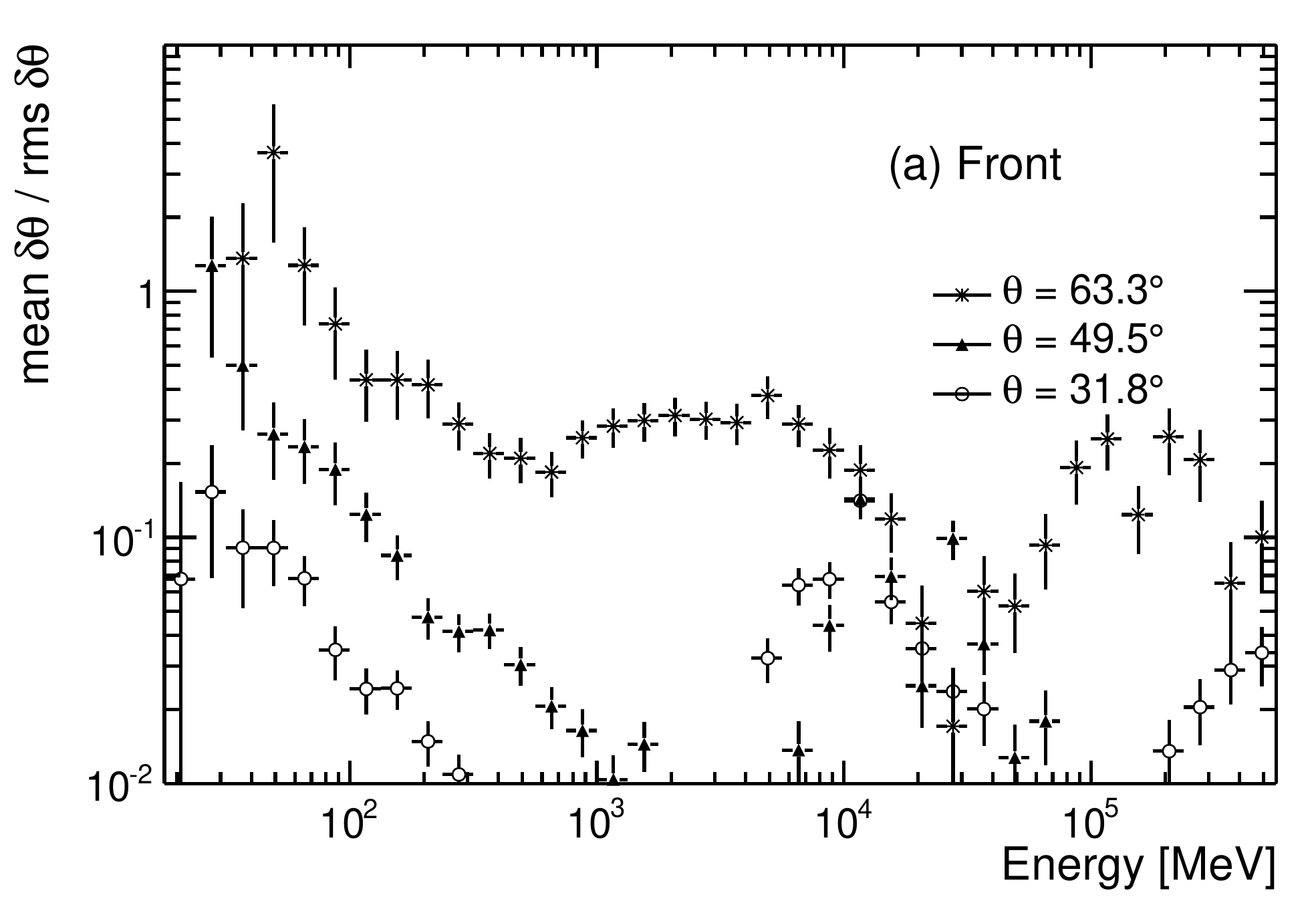}{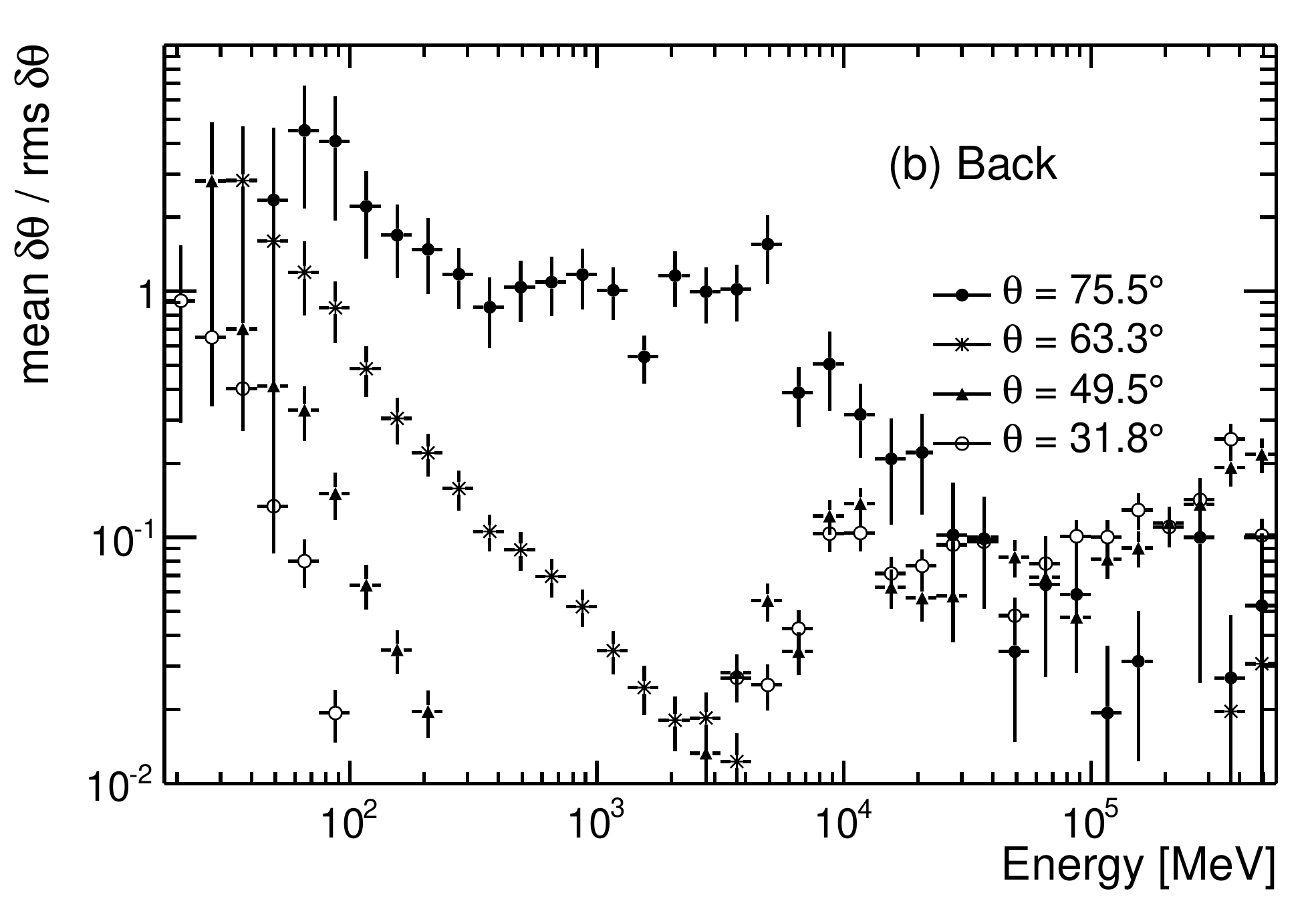}{
  \caption{Ratio of the mean to the RMS of $\delta\theta$ for MC simulated data 
    as a function of energy for several incidence angles for the \irf{P7SOURCE} 
    event selection for front-converting (a) and back-converting (b) events.
    There are not enough statistics for front-converting events at
    $\theta = 75.5^{\circ}$ to extract reliable values.}
  \label{fig:psf_fisheye_ratio_source}
}

We should recall that the \irf{P7TRANSIENT} event selection does not include as
tight constraints on the quality of the event reconstruction as the
\irf{P7SOURCE} selection. Accordingly, the fisheye effect is more pronounced 
for this event selection, as can be seen in
\Figrefs{psf_fisheye_offset_trans}{psf_fisheye_ratio_trans}.  
This is of particular importance for GRBs, for which almost all of the exposure might 
be at a single incidence angle. 
For soft GRBs in the LAT, the bias for the \irf{P7TRANSIENT} \gammaRays\ near 100~MeV that 
contribute the most to the localization can be up to $6^\circ$ at $50^\circ$ 
off axis.  \NEWTEXT{Furthermore, since the orientation of the LAT boresight relative to a 
GRB might not change significantly during the GRB outburst, these \gammaRays\ will tend to 
be biased in the same direction, causing an overall bias in the localization of the GRB.
Therefore, when statistics permit, more robust localizations can be obtained by using
events with energies $> 200$~MeV, or by using \irf{P7SOURCE} rather than
\irf{P7TRANSIENT} class events.}

\twopanel{htb!}{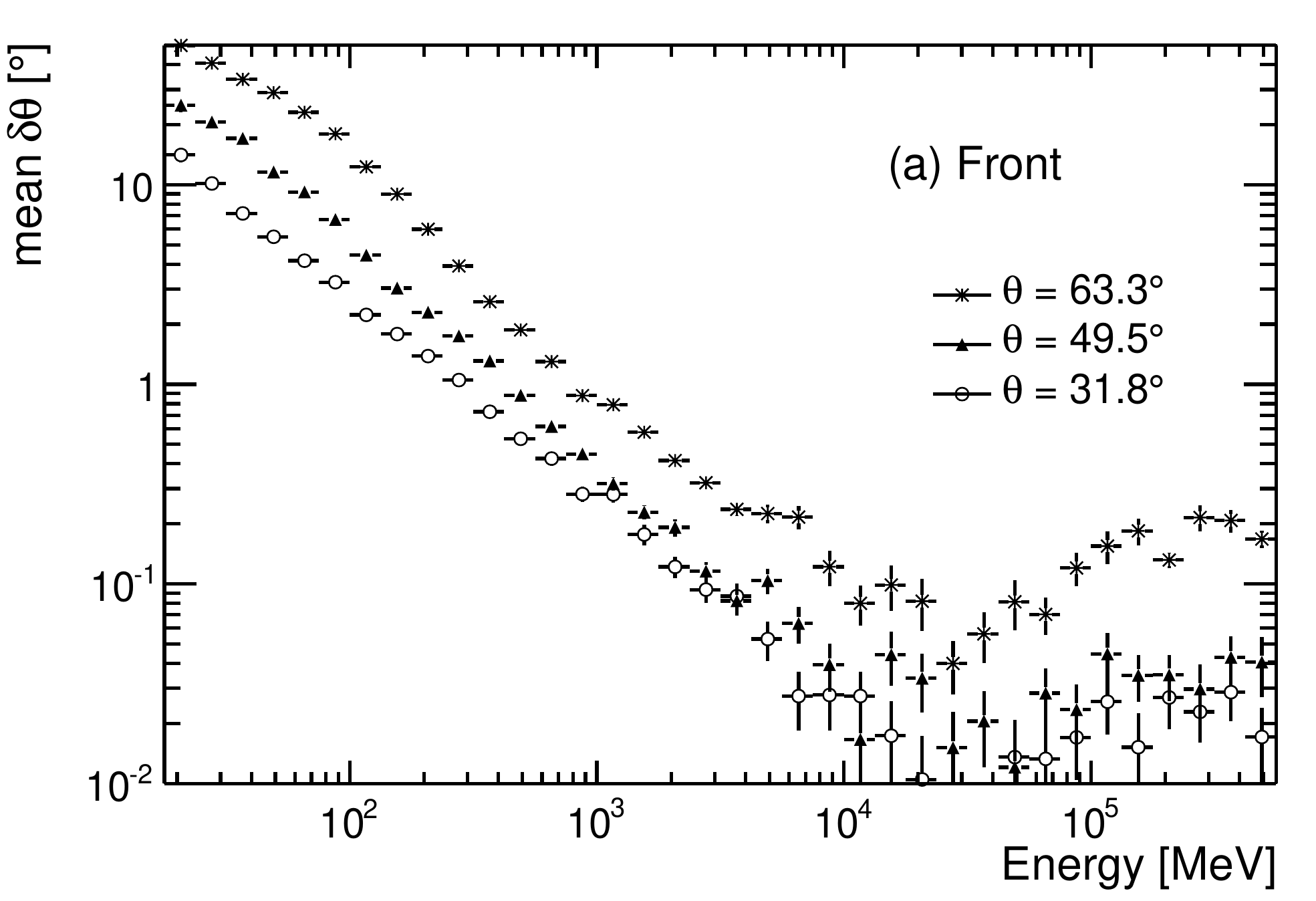}{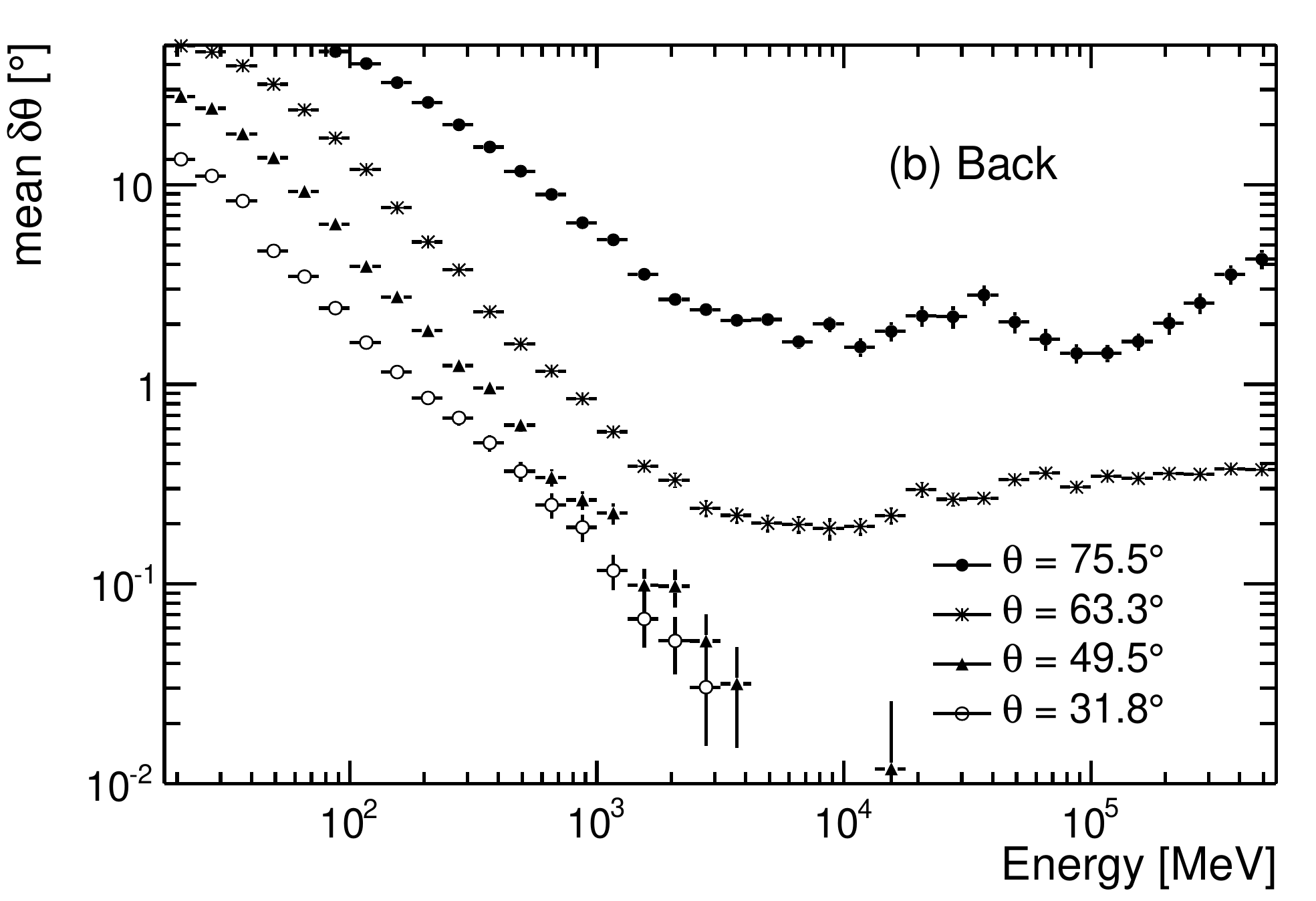}{
  \caption{Mean of $\delta\theta$ for MC simulated
    data as a function of energy for several incidence angles for the
    \irf{P7TRANSIENT} event selection for front-converting (a) and back-converting (b)
    events.  There are not enough statistics for front-converting events at 
    $\theta = 75.5^{\circ}$ to extract reliable values.}
  \label{fig:psf_fisheye_offset_trans}
}

\twopanel{htb!}{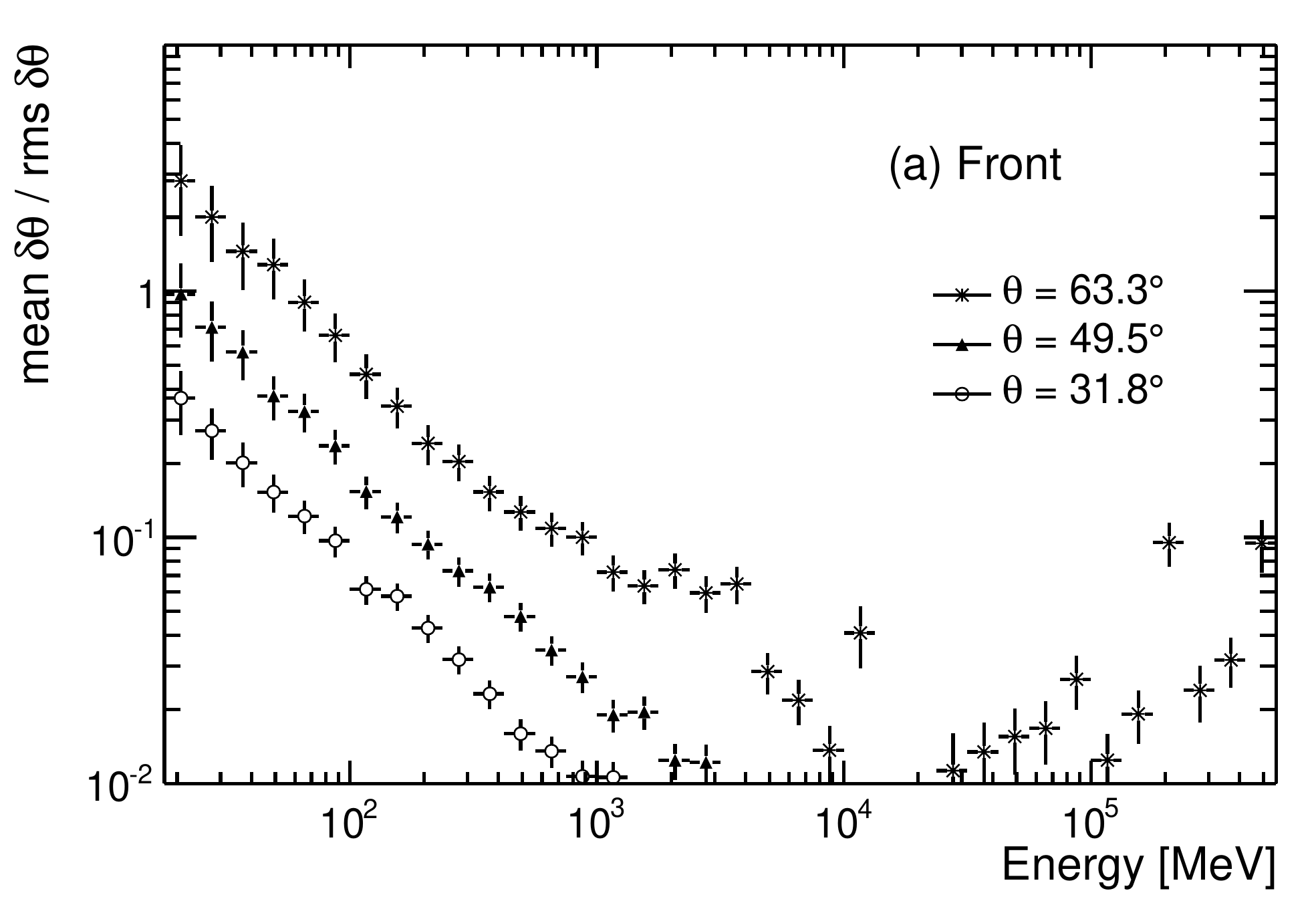}{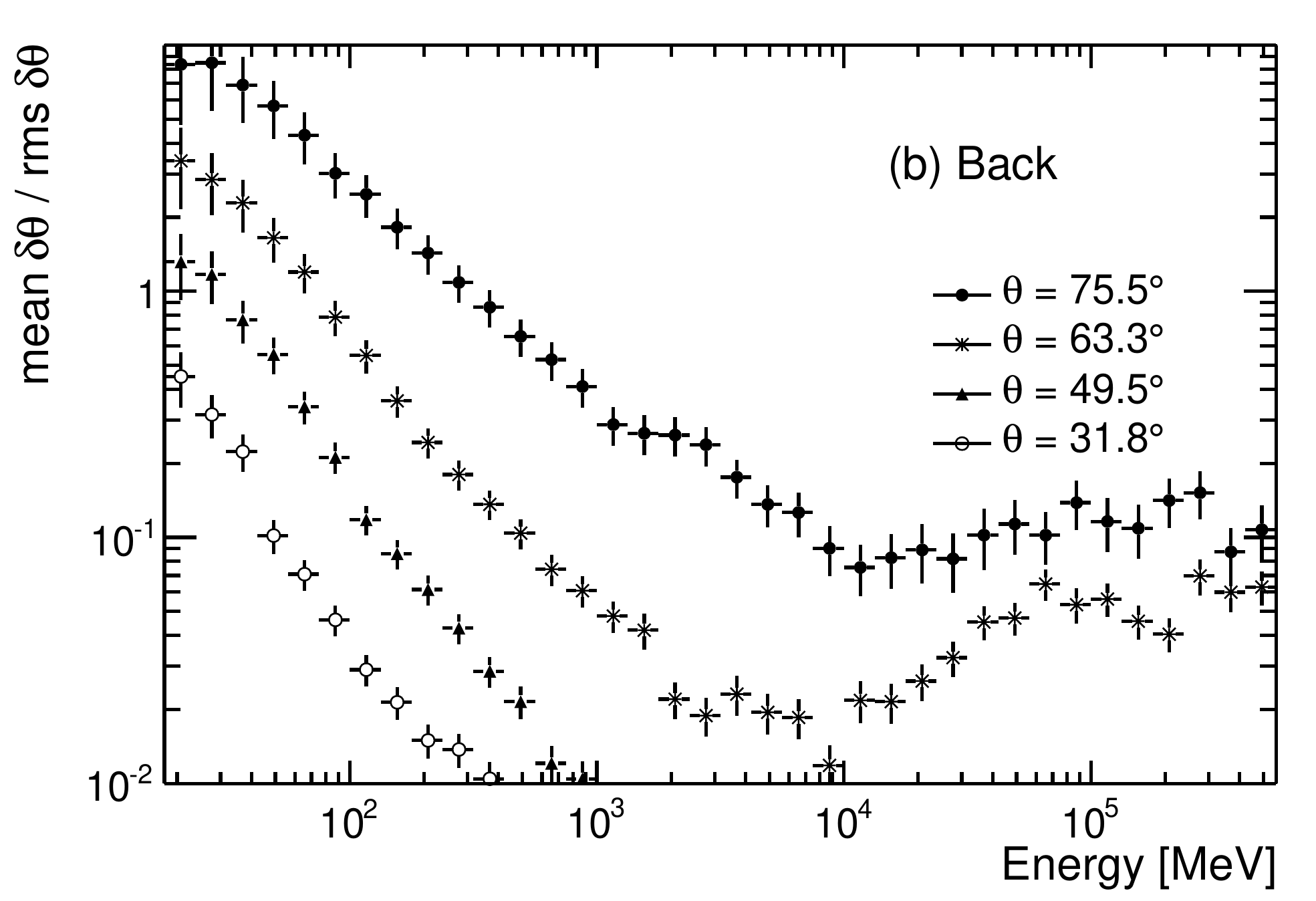}{
  \caption{Ratio of the mean to the RMS of $\delta\theta$ for MC simulated
    data as a function of energy for several incidence angles for the
    \irf{P7TRANSIENT} event selection for front-converting (a) and back-converting (b)
    events.   There are not enough statistics for front-converting events at 
    $\theta = 75.5^{\circ}$ to extract reliable values.}
  \label{fig:psf_fisheye_ratio_trans}
}

\subsection{Propagating Uncertainties of the Point-Spread Function to High Level Science Analysis}\label{subsec:PSF_highLevel}

Uncertainties in the PSF parametrization lead to an imperfect source model in
the high-level source analysis (likelihood fitting), and therefore to
systematic uncertainties in source spectra (\secref{subsec:PSF_bracketing}), 
localizations and measurements of source extensions (\secref{subsec:PSF_onextension}).

An additional source of systematic uncertainty comes from neglecting the $\theta$~dependence
of the PSF in the derivation of the parametrization from flight data.  In particular, 
this can lead to artificial variability as the PSF changes with the varying observing profile.

\subsubsection{Using Custom IRFs to Generate an Error Envelope}\label{subsec:PSF_bracketing}

We can create custom PSFs in a similar manner to that used for the effective
area (see ~\secref{subsec:Aeff_highLevel}), though with the additional
complication that we have to explore variations in the shape of the PSF as a
function of energy and incidence angle. Because the dependence of the PSF on $\gamma$ is not intuitive
we choose to express the bracketing functions in terms of the observable quantities $R_{68}$ and $r = R_{95}/R_{68}$ \label{conv:psfRatio}
rather than in terms of $\sigma$ and $\gamma$.
Specifically, in each bin of energy and incidence angle, we can define the bracketing values $R'_{68}$ and $r'$
in terms of $R_{68}$ and $r$:\label{conv:psf_bracket}
\begin{align}
  R'_{68} &= R_{68} (1 + \epsilon_{68}(E)B_{68}(E))\nonumber\\
  r' &= r  (1 + \epsilon_{r}(E)B_{r}(E)).
\end{align}
We can then solve for the King function parameters $\sigma'$ and $\gamma'$
which would correspond to these values.
  
We re-analyzed both the B2~1520+31 and the PG~1553+113 \rois\ with each of the
PSF bracketing functions listed in \Tabref{bracket_psfFuncs} using the
procedure described in \secref{subsec:Aeff_highLevel}. Based on the quality of
the fits described in \secref{subsec:PSF_errors}, in particular the residuals
on the 68\% and 95\% containment radii, we have assigned:
\begin{align}
  \epsilon_{68}(E) &= 10\% \nonumber\\
  \epsilon_{r}(E) &= 50\%.
\end{align}

\begin{table}[htb!]
  \begin{center}
    \begin{tabular}{lll}
      \hline
      Name & $B_{68}(E)$ & $B_{r}(E)$ \\
      \hline
      \hline
      \bracketirf{c\_scalehi\_t\_nom} & $+1$ & 0 \\
      \bracketirf{c\_scalelo\_t\_nom} & $-1$ & 0 \\
      \bracketirf{c\_pivothi\_t\_nom} & $\tanh(\frac{1}{k}\log(E/E_{0}))$ & 0 \\
      \bracketirf{c\_pivotlo\_t\_nom} &
      $-\tanh(\frac{1}{k}\log(E/E_{0}))$  & 0 \\
      \bracketirf{c\_nom\_t\_scalehi} & 0 & $+1$ \\
      \bracketirf{c\_nom\_t\_scalelo} & 0 & $-1$ \\
      \bracketirf{c\_nom\_t\_pivothi} & 0 & $\tanh(\frac{1}{k}\log(E/E_{0}))$ \\
      \bracketirf{c\_nom\_t\_pivotlo} & 0 &
      $-\tanh(\frac{1}{k}\log(E/E_{0}))$ \\
      \hline      
    \end{tabular}
    \caption{Bracketing PSFs and the energy-dependent scaling functions used
      to create them.   As in \secref{subsec:Aeff_highLevel} we use
      $k=0.13$, which corresponds to smoothing over $\Delta E/E \sim 0.30$.}
    \label{tab:bracket_psfFuncs}
  \end{center}
\end{table}

\Tabrefs{bracketPSF_PG1553p113}{bracketPSF_B21520p31} shows the fits
results for PG~1553+113 and B2~1520+31 as well as the integral counts and energy fluxes between
100~MeV and 100~GeV obtained using these PSF bracketing functions.  The ranges of the fit values indicate propagated uncertainties from
the uncertainty in the PSF (\tabref{bracketErrorsPSF}).

\begin{table}[htb!]
  \begin{center}
    \begin{tabular}{lclcc}
      \hline
      Bracketing PSF & $N_{0}$ & $\Gamma$ & $F_{25}$ & $S_{25}$ \\
 & [MeV$^{-1}$ cm$^{-2}$ s$^{-1}$] & & [cm$^{-2}$ s$^{-1}$] & [MeV cm$^{-2}$ s$^{-1}$] \\
\hline
\hline
Nominal & $2.54 \times 10^{-12}$ & 1.68 & $6.91 \times 10^{-8}$ & $1.19 \times 10^{-4}$\\
\bracketirf{c\_nom\_t\_scalelo} & $2.48 \times 10^{-12}$ & 1.70 & $6.94 \times 10^{-8}$ & $1.13 \times 10^{-4}$\\
\bracketirf{c\_nom\_t\_scalehi} & $2.54 \times 10^{-12}$ & 1.66 & $6.68 \times 10^{-8}$ & $1.22 \times 10^{-4}$\\
\bracketirf{c\_nom\_t\_pivotlo} & $2.46 \times 10^{-12}$ & 1.68 & $6.67 \times 10^{-8}$ & $1.15 \times 10^{-4}$\\
\bracketirf{c\_nom\_t\_pivothi} & $2.58 \times 10^{-12}$ & 1.68 & $6.99 \times 10^{-8}$ & $1.21 \times 10^{-4}$\\
\bracketirf{c\_scalelo\_t\_nom} & $2.46 \times 10^{-12}$ & 1.67 & $6.53 \times 10^{-8}$ & $1.18 \times 10^{-4}$\\
\bracketirf{c\_scalehi\_t\_nom} & $2.63 \times 10^{-12}$ & 1.70 & $7.32 \times 10^{-8}$ & $1.21 \times 10^{-4}$\\
\bracketirf{c\_pivotlo\_t\_nom} & $2.60 \times 10^{-12}$ & 1.70 & $7.30 \times 10^{-8}$ & $1.18 \times 10^{-4}$\\
\bracketirf{c\_pivothi\_t\_nom} & $2.49 \times 10^{-12}$ & 1.66 & $6.50 \times 10^{-8}$ & $1.21 \times 10^{-4}$\\

      \hline
    \end{tabular}
    \caption{Fit parameters and integral fluxes obtained using the PSF
      bracketing IRFs for PG~1553+113.  Note that the quoted precision
      is roughly equivalent to the fit uncertainties and the pivot
      energy is $E_{0} = 2240$~MeV for this source.}
    \label{tab:bracketPSF_PG1553p113}
  \end{center}
\end{table}

\begin{table}[htb!]
  \begin{center}
    \begin{tabular}{lcllcc}
      \hline
      Bracketing PSF & $N_{0}$ & $\alpha$ & $\beta$ & $F_{25}$ & $S_{25}$ \\
 & [MeV$^{-1}$ cm$^{-2}$ s$^{-1}$] & & & [cm$^{-2}$ s$^{-1}$] & [MeV cm$^{-2}$ s$^{-1}$] \\
\hline
\hline
Nominal & $5.23 \times 10^{-10}$ & 2.24 & 0.08 & $4.09 \times 10^{-7}$ & $1.33 \times 10^{-4}$ \\
\bracketirf{c\_nom\_t\_scalelo} & $5.34 \times 10^{-10}$ & 2.31 & 0.06 & $4.28 \times 10^{-7}$ & $1.33 \times 10^{-4}$ \\
\bracketirf{c\_nom\_t\_scalehi} & $5.02 \times 10^{-10}$ & 2.19 & 0.09 & $3.85 \times 10^{-7}$ & $1.30 \times 10^{-4}$ \\
\bracketirf{c\_nom\_t\_pivotlo} & $5.00 \times 10^{-10}$ & 2.22 & 0.09 & $3.86 \times 10^{-7}$ & $1.26 \times 10^{-4}$ \\
\bracketirf{c\_nom\_t\_pivothi} & $5.43 \times 10^{-10}$ & 2.29 & 0.06 & $4.34 \times 10^{-7}$ & $1.38 \times 10^{-4}$ \\
\bracketirf{c\_scalelo\_t\_nom} & $4.87 \times 10^{-10}$ & 2.21 & 0.08 & $3.77 \times 10^{-7}$ & $1.26 \times 10^{-4}$ \\
\bracketirf{c\_scalehi\_t\_nom} & $5.57 \times 10^{-10}$ & 2.27 & 0.07 & $4.40 \times 10^{-7}$ & $1.40 \times 10^{-4}$ \\
\bracketirf{c\_pivotlo\_t\_nom} & $5.27 \times 10^{-10}$ & 2.31 & 0.06 & $4.26 \times 10^{-7}$ & $1.34 \times 10^{-4}$ \\
\bracketirf{c\_pivothi\_t\_nom} & $5.16 \times 10^{-10}$ & 2.17 & 0.10 & $3.91 \times 10^{-7}$ & $1.32 \times 10^{-4}$ \\

      \hline
    \end{tabular}
    \caption{Fit parameters and integral fluxes obtained using the PSF
      bracketing IRFs for B2~1520+31. Note that the quoted precision
      is roughly equivalent to the fit uncertainties and the pivot
      energy is $E_{0} = 281$~MeV for this source.}
    \label{tab:bracketPSF_B21520p31}
  \end{center}
\end{table}

\begin{table}[htb!]
  \begin{center}
    \begin{tabular}{lcc}
      \hline
      Parameter & B2 1520+31 & PG 1553+113 \\
 & (soft) & (hard) \\
\hline
\hline
$\delta N_{0}/N_{0}$ & \syserrors{+6.6\%}{-6.8\%} & \syserrors{+3.5\%}{-3.4\%} \\
$\delta\Gamma$ ($\delta\alpha$) & \syserrors{+0.07}{-0.07} & \syserrors{+0.02}{-0.02} \\
$\delta\beta$ & \syserrors{+0.02}{-0.02} & {\hfill-\hfill} \\
$\delta F_{25}/F_{25}$ & \syserrors{+7.7\%}{-7.7\%} & \syserrors{+6.0\%}{-5.9\%} \\
$\delta S_{25}/S_{25}$ & \syserrors{+5.4\%}{-5.5\%} & \syserrors{+2.4\%}{-4.9\%} \\

      \hline
    \end{tabular}
    \caption{Systematic variations arising from uncertainties in the PSF. For
      the spectral index ($\Gamma$ or $\alpha$) and spectral curvature
      ($\beta$) we give the absolute variation with respect to the nominal
      value (e.g., $\delta \Gamma$). For the flux prefactor and the integral
      fluxes we give the relative variations with respect to the nominal value
      (e.g., $\delta N_{0} / N_{0}$).}    
    \label{tab:bracketErrorsPSF}
  \end{center}
\end{table}

The greater influence of the uncertainty of the PSF on the flux and spectral
measurements for the softer source (B2 1520+31) comes about because at lower
energies the wider PSF makes resolving sources more difficult and results
in greater correlation with the Galactic and isotropic diffuse components in the
likelihood fit, which varied by up to $\pm 4\%$ and $\pm 5\%$ respectively.

\subsubsection{Effects on Source Extension}\label{subsec:PSF_onextension}

For sufficiently long exposures, the LAT can spatially resolve a
number of \gammaRayHyph\ sources.  In the 2FGL catalog, 12 spatially
extended LAT sources were identified \citep{REF:2011.2FGL} using 24
months of LAT data; and several additional extended sources have been
recently resolved from these data by \citet{REF:SourceExt} using
special techniques for modeling the spatial
extension.  Understanding the possible spatial extension of LAT
sources is important for identifying multiwavelength counterparts, and
using a source model with the correct spatial extent produces more
accurate spectral fits and avoids biases in the model parameters.

In addition to using a correct spatial model, the accuracy of the PSF
can also affect the analyses of extended sources; and using an
incorrect PSF will result in biases both in the fitted model
parameters and in the significance of any spatial extension that is
found.  For example, flight data indicate that the MC-based
PSF in the \irf{P7SOURCE\_V6MC} IRFs is too narrow at energies
$>3$~GeV (\secref{subsec:PSF_onorbit}).  Fitting the extension of SNR IC~443 \citep{REF:2010.IC443_LAT} with
a uniform disk for data from the first two years of observations and
using the flight-determined PSF in \irf{P7SOURCE\_V6}, we find a best
fit disk radius of IC~443 of $\sigma = 0.35^\circ \pm 0.01^\circ$
\citep{REF:SourceExt}.  By contrast, fitting these same data, but
using the MC-based PSF in \irf{P7SOURCE\_V6MC}, we find a
best fit radius of $\sigma = 0.39^\circ$.  This corresponds to a $\sim
10$\% systematic bias in the measurement of the extension of IC~443.
Since IC~443 is a fairly hard source, with photon index $\Gamma =
2.2$, the bias found for softer sources would be somewhat smaller.

Use of an imperfect model of the PSF can also affect the assessment of
the statistical significance of the extension of the source.  We
determine the significance of a measured source extension using the
test statistic
\label{conv:L_ext}\label{conv:L_pt}
\begin{equation}\label{conv:ts_ext}
  \mathrm{TS}_\mathrm{ext} = 2\log(L_\mathrm{ext}/L_\mathrm{pt}),
\end{equation}
This is twice the difference in log-likelihood found when fitting the
source assuming it is spatially extended versus assuming it is a point
source.  The use and validity of this formula is described in
\citep{REF:SourceExt}.  For IC~443, using the \irf{P7SOURCE\_V6}
IRFs, we find ${\rm TS}_{\rm ext} = 640$, corresponding to a formal
statistical significance of $\sim 25\sigma$.  Performing the same
analysis using \irf{P7SOURCE\_V6MC}, we obtain ${\rm TS}_{\rm ext} =
1300$, corresponding to $\sim 37\sigma$.

\subsubsection{Effects on Variability}\label{subsec:psf_variability}

Neglecting the dependence of the PSF on incidence angle (\secref{subsec:PSF_errors}) may
introduce time-dependent biases of the estimated flux of the source, due to the
different distribution of the source position with respect to the LAT boresight
in each time interval. To estimate the potential size of such biases we
consider how great an effect on source fluxes we might see if we were to
naively use the nominal containment radius for aperture photometry
analysis (i.e., an analysis where we simply count the \gammaRays\ within a given
aperture radius).   Specifically, we use the MC based PSF to extract the
containment radii averaged over the first two years of the mission
$\bar{R}(E)$ in the direction of the Vela pulsar.  Then we split
the data into much shorter time intervals (indexed by $i$) and compute the
fraction of \gammaRays\ $C_{i}(E;{\rm vela})$ that fall within $\bar{R}(E;{\rm vela})$ for each of those shorter
intervals:
\label{conv:psf_actual_cont}
\begin{equation}
  C_i(E,{\rm vela}) =
  \frac{\int \int^{\bar{R}(E;{\rm vela})}_{0} \aeff(E,\theta; {\rm vela}) t_{\rm
      obs,i}(\theta;{\rm vela}) \sin\theta d\theta dr}
       {\int \aeff(E,\theta) t_{\rm obs,i}(\theta;{\rm vela}) \sin\theta d\theta}
       \label{eq:psf_actualContainment}
\end{equation}

Here we consider 68\% and 95\% containment radii. \Figref{psfContainError}
shows how $C_{68,i}(E;{\rm vela})$ and $C_{95,i}(E;{\rm vela})$ vary for each 12-hour time interval over
the first 700 days of routine science operations. If there  were no
$\theta$~dependence to the PSF this figure would show lines at $0.68$ and
$0.95$. 
However we clearly see that ignoring the $\theta$~dependence will cause slight
mis-estimates in the fraction of \gammaRays\ falling within the mission-averaged
$\bar{R}_{68}$ and $\bar{R}_{95}$. In fact, the width of the bands (i.e., the
spread of that mis-estimation) indicate the errors in flux we would expect
if we were making a flux estimate based on aperture photometry with aperture
cuts at $\bar{R}$.  
It is worth pointing out explicitly that the $C_{95,i}(E)$ band is significantly
narrower, simply because most of the \gammaRays\ are already contained within the
$\bar{R}_{95}(E)$ and the derivative of the PSF is smaller.

Understanding the effect of the PSF dependence on incidence angle will have
on a full likelihood analysis is much more complicated, as it depends on the
other sources in the likelihood model. However, it is reasonable to take the
width of the $C_{68}(E)$ bands as indicative of the magnitude of the effect when
analyzing sources in complex regions where source proximity is a issue and the
width of the $C_{95}(E)$ bands as indicative when analyzing isolated sources. 

\twopanel{htb!}{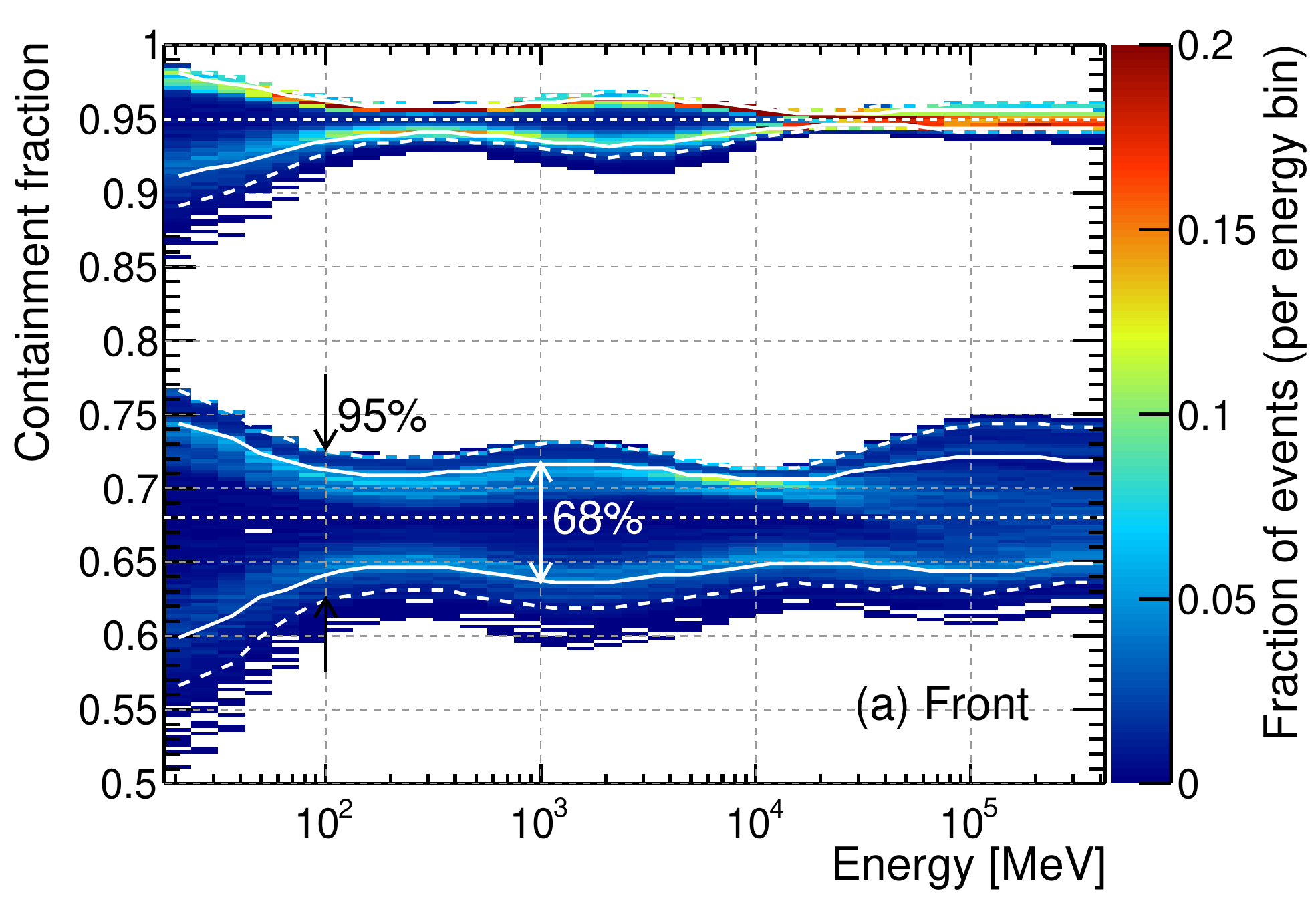}{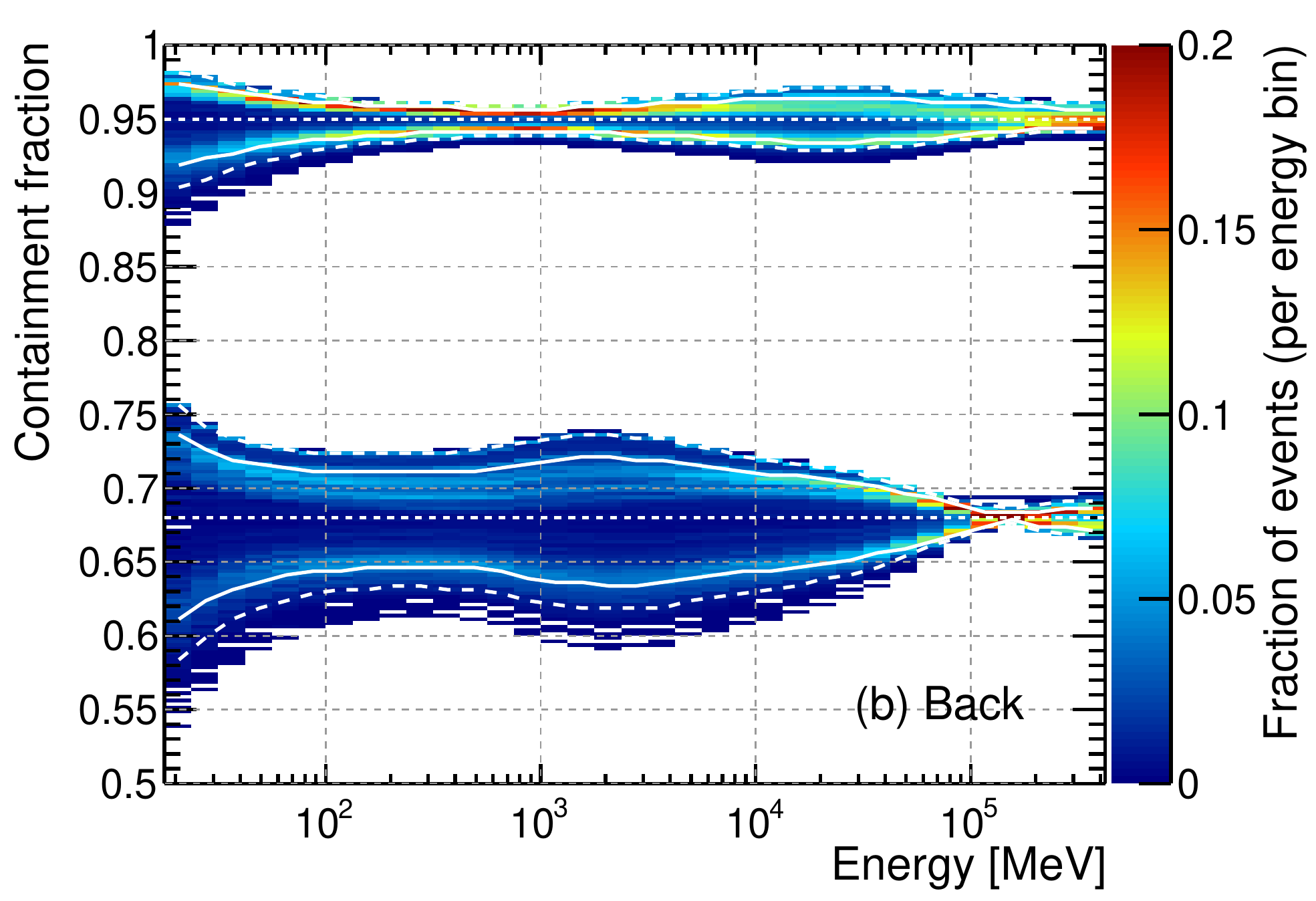}{
  \caption{Fraction of the PSF contained at the mission averaged 68\% and 95\%
    containment radii in the direction of the Vela pulsar for the \irf{P7SOURCE\_V6MC} IRFs as a function
    of energy for front-converting (a) and
    back-converting (b) \gammaRays\ for each of 1400 12-hour time intervals.}
  \label{fig:psfContainError}
}

\subsection{Comparison of \psix\ and \pseven}\label{subsec:PSF_comparison_p6}

As a consequence of the issues discussed in \secref{subsec:Aeff_flightAEff},
\pseven\ event classes have less systematic uncertainty for \aeff\ at the
expense of a slightly broader PSF across the entire parameter space. 

In addition \pseven\ standard classes (\irf{P7SOURCE} and cleaner) feature a
PSF derived from data above 1~GeV; see \secref{subsec:PSF_onorbit}.
Below 1~GeV, where the PSF is derived from MC simulations, the PSF parameters
are recalculated in order to ensure a smooth variation of the containment
levels as a function of energy (\secref{subsubsec:PSF_oo_irfs}). 

\figref{compare_psf_p6} compares the 68\% containment angles for \psix\ and
\pseven\ classes recommended for routine analyses of \gammaRayHyph\ point
sources.  While the difference between \irf{P6\_V3\_DIFFUSE} and \irf{P7SOURCE\_V6} is
due mostly to the data-derived PSF in the latter, there is also a clear difference
across all energies between the \irf{P7SOURCE\_V6} PSF and in-flight \irf{P6\_V11\_DIFFUSE} PSF.

\twopanel{htb}{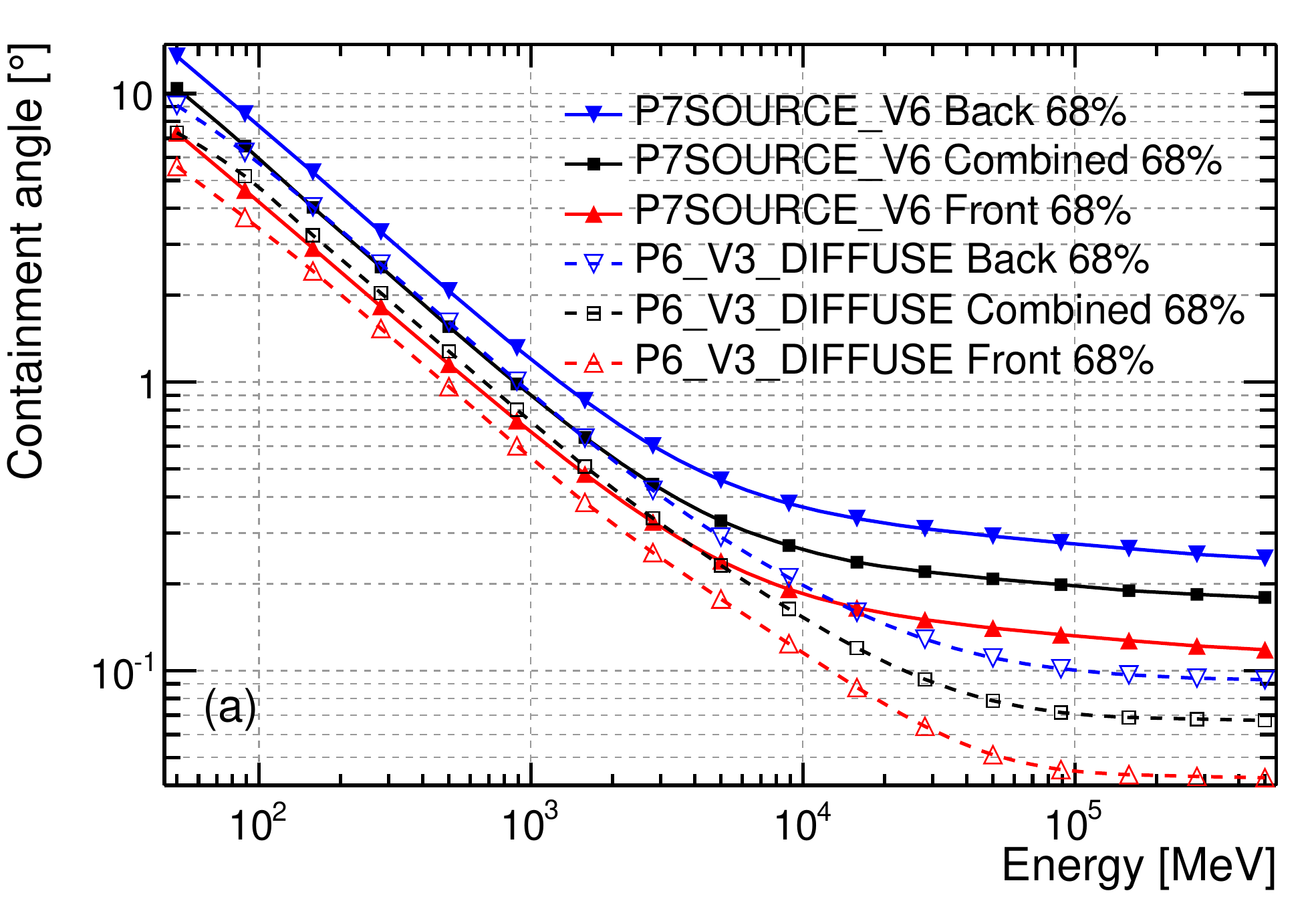}{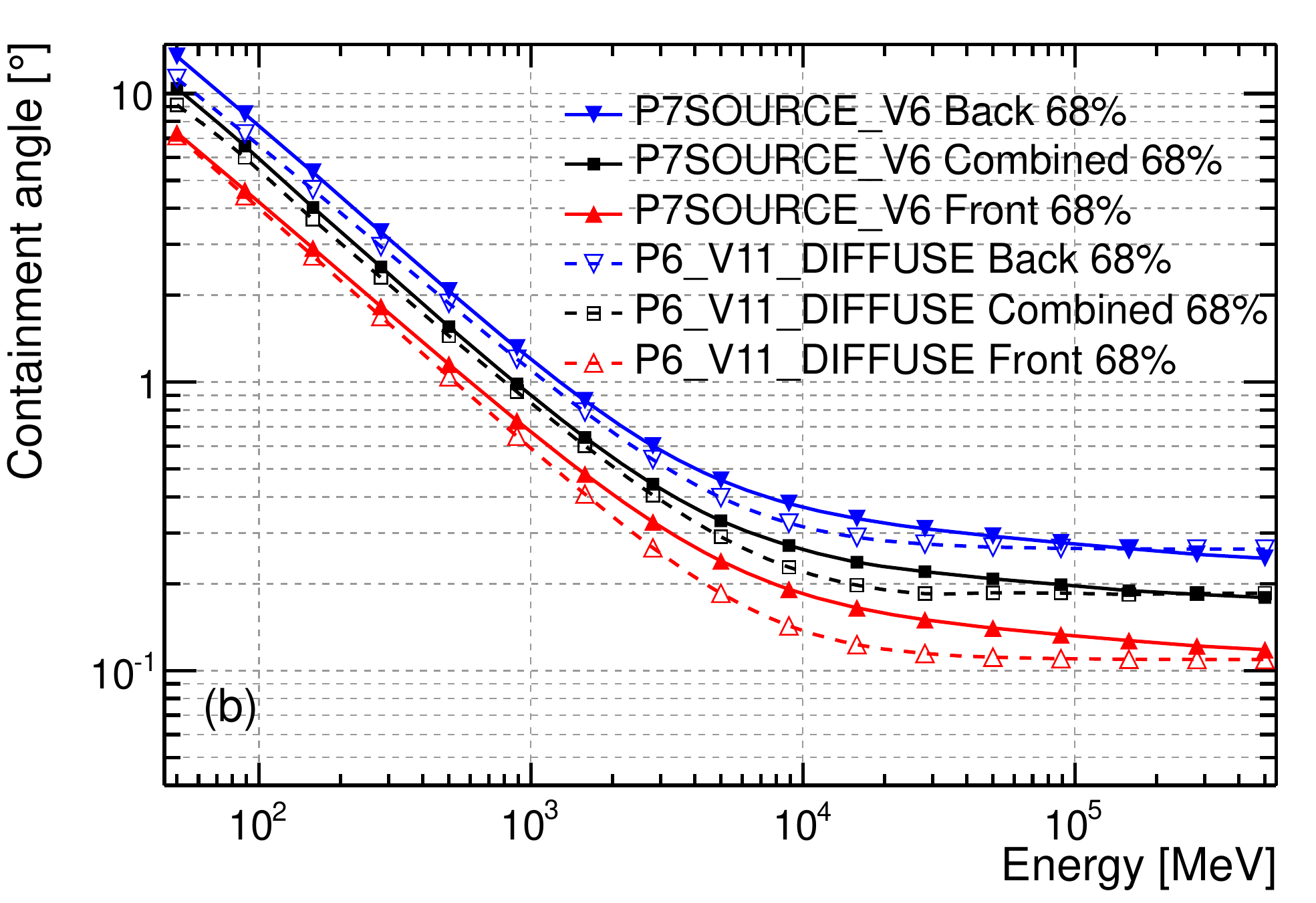}{
  \caption{Comparison of the 68\% containment radius of the PSF for
    \irf{P7SOURCE\_V6} with respect to \irf{P6\_V3\_DIFFUSE} (a) and
    \irf{P6\_V11\_DIFFUSE} (b). The \irf{P6\_V3\_DIFFUSE} is plotted
    for normal incidence; \irf{P6\_V11\_DIFFUSE} and
    \irf{P7SOURCE\_V6} do not include dependence on the incidence angle.}
  \label{fig:compare_psf_p6}
}

While the difference in the containment radii we described is typically within
the PSF uncertainties this comparison was performed with the incidence
angle dependence ignored, so the tabulated PSF is an average over the \fov\
weighted by the average exposure over long timescales. This may be an issue
for the analysis of bright sources over short timescales, when the observation
angle distribution in the LAT reference system is significantly different from
the average.
Future \pseven\ IRFs may address this by reintroducing the $\theta$ dependence.

Current \pseven\ IRFs parametrize the PSF with a
simplified version of \Eqref{eq:PSF_psf2king} for which the tail term is omitted
as described in \secref{subsubsec:PSF_oo_irfs}. 
The less accurate model of the PSF tails in the \pseven\ classes with on-orbit
PSF (to date \irf{P7SOURCE\_V6}, \irf{P7CLEAN\_V6} and \irf{P7ULTRACLEAN\_V6})
is probably not an issue for most analyses, but caution is recommended when
comparing the dispersion of \gammaRays\ around the true position of bright sources
with the expected distribution derived from the IRFs.

\clearpage

\section{ENERGY DISPERSION}\label{sec:EDisp}

The dispersion of measured energies around the true values (i.e., the energy
dispersion, or redistribution function) is generally asymmetric, with the most
prominent tail being toward lower energies. This feature, characteristic of thin
electromagnetic calorimeters (such as the LAT CAL, as opposed
to full-containment calorimeters) makes the energy redistribution difficult to
parametrize.
As a matter of fact, since most source spectra are steeply falling
with energy, low-energy tails in the energy dispersion are relatively harmless,
while overestimating the event energy can potentially lead to overestimating the
hardness of the spectrum.
In the event selections we specifically make an effort to suppress the
high-energy tails, rejecting events for which we might overcompensate with the
energy corrections (see \secref{subsubsec:cal_recon} and
\secref{subsubsec:energy_analysis}).

While the width and shape of the energy redistribution are well understood, by
default the energy dispersion is \emph{not} taken into account in the standard
likelihood analysis, primarily due to computational limitations.
Furthermore, as we will see in~\secref{subsec:EDisp_highLevelAnalysis}, the
effect of neglecting the energy redistribution is usually small enough that it
can be ignored. When that is not the case, it is easy to check the magnitude
of the bias induced with dedicated simulations. 
If needed, the energy dispersion can be accounted for in the spectral analysis
either by specifically enabling the functionality in the binned likelihood
(\stools\ version 09-26-00 or later) or by means of unfolding techniques, as we
briefly describe in \secref{subsec:EDisp_highLevelAnalysis}.

It is worth stressing that an in-flight validation of the energy response
is much more difficult than the corresponding validations of the effective area
and the PSF described in the previous section, as there is no known
astrophysical calibration source that provides a spectral line at a well defined
energy that would play the role that point sources have for the PSF
(see, however, the discussion in \secref{subsec:EDisp_absoluteScale}).
The energy reconstruction validation will be described
in \secref{subsec:EDisp_errors}.

\subsection{Energy Dispersion and Parametrization from Monte Carlo Simulations}\label{subsec:EDisp_MonteCarlo}

As for \aeff\ and PSF, in order to derive the energy dispersion events from an \allgamma\ simulation belonging
to a specified event class are binned in true energy $E$ and incidence
angle $\theta$ and the distribution of measured energy $E'$ is fitted
(and the parameters tabulated).
A complex functional form is necessary and attention is required to
avoid fit instability and overly rapid variation of the fit parameters.

\subsubsection{Scaling}\label{subsubsec:edisp_scaling}

The success of scaling the PSF distributions
(see~\secref{subsubsec:psf_prescaling}) prompted a similar approach to 
the histogramming and fitting of the energy deviations. The scaling
function \edispscaling\label{conv:edispscaling} currently employed, derived by
fitting the 68\% containment of the measured fractional energy
distribution $(E' - E)/E$ across the entire $E$--$\theta$ plane, is
\begin{equation}\label{eq:edisp_prescale_func}
  \edispscaling(E,\theta) = c_0(\log_{10} E)^2 + c_1(\cos\theta)^2 +
  c_2\log_{10} E + c_3\cos\theta + c_4\log_{10} E\cos\theta + c_5.
\end{equation}
Front- and back-converting events are treated separately.
The parameters used since the \psix\ analysis are listed in
\Tabref{scalef_edisp}. These are included in the IRF files for \psix\
and \pseven\ event classes.
\begin{table}[htbp]
  \centering
  \begin{tabular}{ccccccc}
    \hline
    & $c_0$ & $c_1$ & $c_2$ & $c_3$ & $c_4$ & $c_5$ \\
    \hline
    \hline
    Front & 0.0210 & 0.0580 & -0.207 & -0.213 & 0.042 & 0.564  \\
    Back  &  0.0215 & 0.0507 & -0.220 & -0.243 & 0.065 & 0.584 \\
    \hline
  \end{tabular}
  \caption{Numerical values of the coefficients defining the energy resolution
    scaling function \edispscaling\ in \Eqref{eq:edisp_prescale_func} for the
    \psix\ and \pseven\ event classes.}
  \label{tab:scalef_edisp}
\end{table}

The scaling function \edispscaling\ allows us to define a
\emph{scaled energy deviation} $x$\label{conv:scaledEDispvar} for which much of the energy and angular
dependence is already accounted for:
\begin{equation}\label{eq:edisp_scale_dev}
  x = \frac{(E' - E)}{\edispscaling(E, \theta) E}
\end{equation}
This scaled value is calculated for each simulated event and histogrammed,
as shown in \Figref{scalededisp}. Scaling the distribution achieves
two main effects: the width of the core of the distribution is almost constant
for all energies and incidence angles, and the distribution thus expressed
is significantly simpler to parametrize than the un-scaled version.

\subsubsection{Fitting the Scaled Variable}\label{subsubsec:Figure_66ting}

\begin{figure}[htbp]
  \centering
  \includegraphics[width=0.5\textwidth]{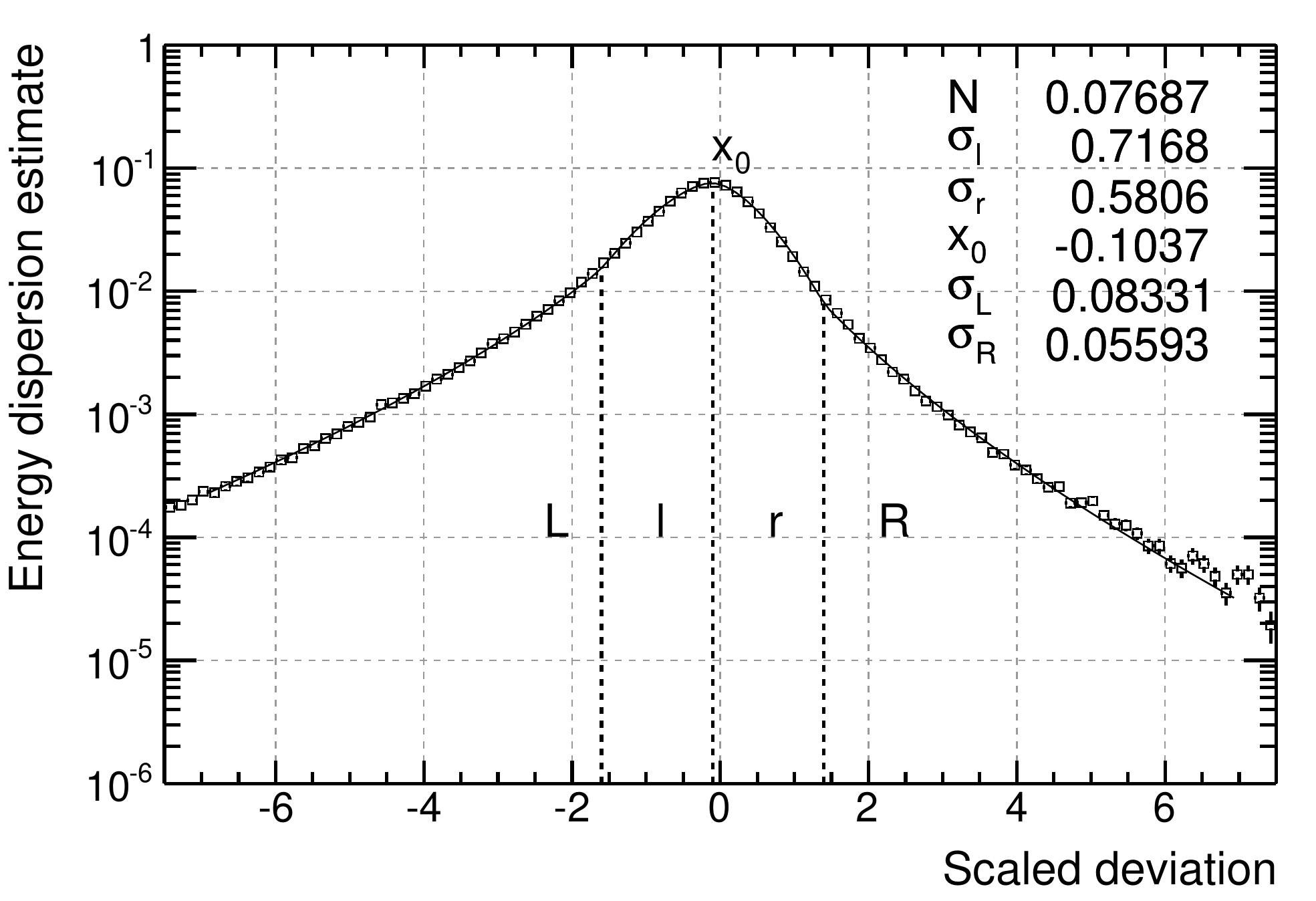}
  \caption{Histogram of the scaled energy deviation, as defined
    in \Eqref{eq:edisp_scale_dev}, fitted with the
    function $\edisp(x)$ in~\Eqref{eq:EdispFunc}.
    The plot refers to the $(E, \theta)$ bin centered at 7.5~GeV and
    $30^\circ$ for the front section.}
  \label{fig:scalededisp}
\end{figure}

The scaled energy deviation, as illustrated by the example in
\Figref{scalededisp}, has a well defined core accompanied by elongated
tails.
We found we can effectively fit this structure with several piecewise functions
of the form:\label{conv:RandoSigma}\label{conv:RandoGamma}
\begin{equation}\label{eq:RandoFunction}
  \Randof(x, x_0, \sigma, \gamma) = N\exp\left(-{\frac{1}{2}}
  \left\vert{\frac{x-x_0}{\sigma}}\right\vert^{\gamma}\right)
\end{equation}
\label{conv:rando}
Indeed we model the energy dispersion $\edisp(x)$ within each
$(E,\theta)$ bin as four piecewise functions:
\begin{equation}\label{eq:EdispFunc}
  \edisp(x) = \left\{\begin{array}{rl}
  N_{L}\Randof(x, x_0, \sigma_{L}, \gamma_{L}) &
  \quad\text{if } (x - x_0) < -\tilde{x}\\
  N_{l}\Randof(x, x_0, \sigma_{l}, \gamma_{l}) & 
  \quad\text{if } (x - x_0) \in [-\tilde{x}, 0]\\
  N_{r}\Randof(x, x_0, \sigma_{r}, \gamma_{r}) &
  \quad\text{if } (x - x_0) \in [0, \tilde{x}]\\
  N_{R}\Randof(x, x_0, \sigma_{R}, \gamma_{R}) &
  \quad\text{if } (x - x_0) > \tilde{x}\\
  \end{array}\right.
\end{equation}
The values of the split point $\tilde{x}$\label{conv:randoBreak} and of the four exponents $\gamma$
of the energy dispersion parametrization in \Eqref{eq:EdispFunc} are
fixed as specified in \tabref{EdispFixParams}.
Moreover, the relative normalizations are set by requiring continuity at
$x = x_0$ and $\vert x - x_0 \vert = \tilde{x}$ and therefore the fit is
effectively performed with a total of six free parameters, which are stored
in the IRF FITS files: the overall normalization $N_{r} = N_{l}$ (\texttt{NORM}),
the centroid position $x_0$ (\texttt{BIAS}), the two core scales $\sigma_{r}$
(\texttt{RS1}) and $\sigma_{l}$ (\texttt{LS1}) and the two tail scales
$\sigma_{R}$ (\texttt{RS2}) and $\sigma_{L}$ (\texttt{LS2}).

\begin{table}[htbp]
  \centering
  \begin{tabular}{ccccc}
    \hline
    $\tilde{x}$ & $\gamma_{L}$ & $\gamma_{l}$ & $\gamma_{r}$ & $\gamma_{R}$\\
    \hline
    \hline
    1.5 & 0.6 & 1.6 & 1.6 & 0.6\\
    \hline
  \end{tabular}
  \caption{Numerical values of the split point $\tilde{x}$ and of the four
    exponents $\gamma$ in \Eqref{eq:EdispFunc} that are fixed when fitting
    the scaled energy deviations.}
  \label{tab:EdispFixParams}
\end{table}

\subsubsection{Energy Resolution}

\begin{figure}[htb!]
  \centering\includegraphics[width=\onecolfigwidth]{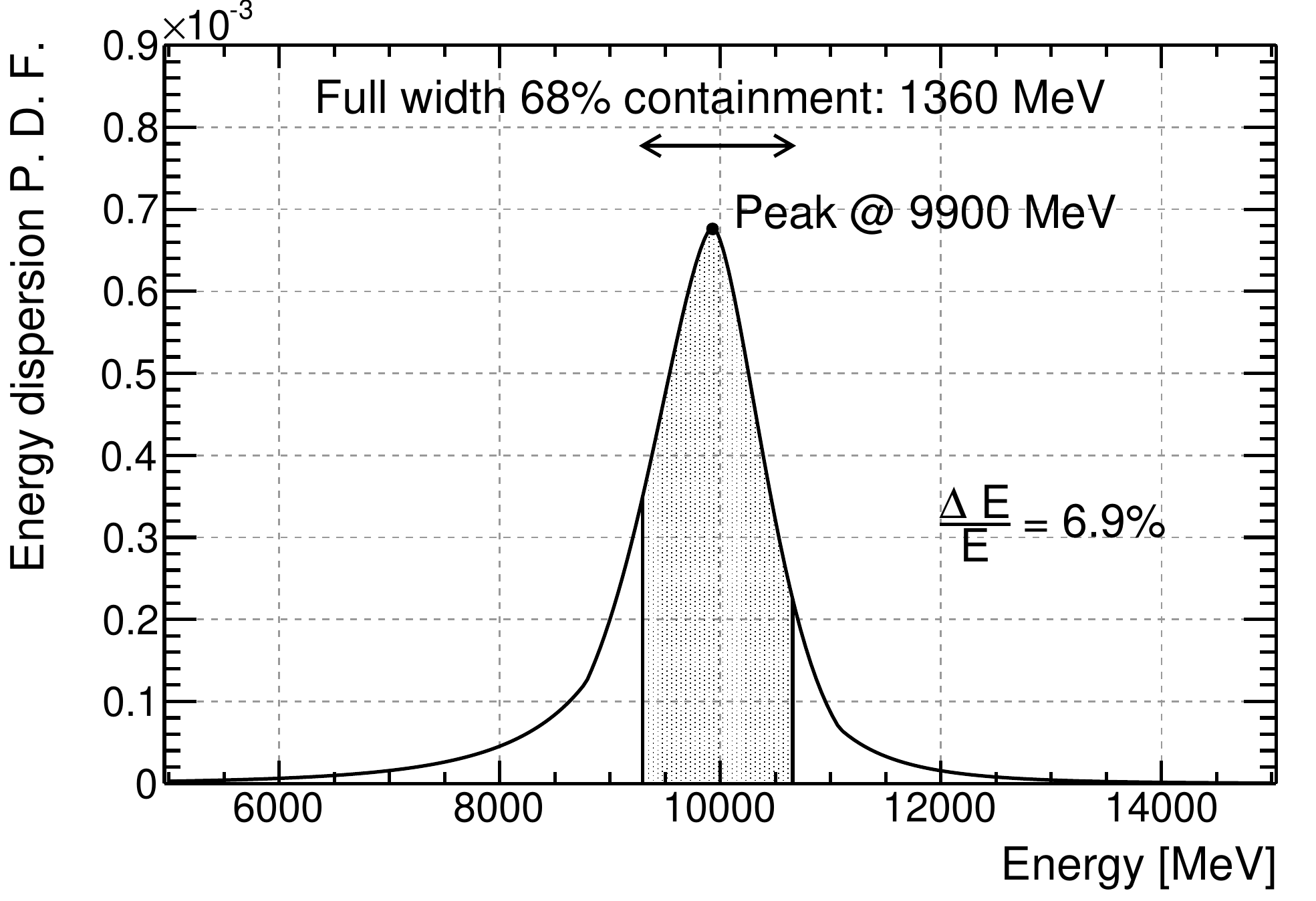}
  \caption{Energy dispersion at 10~GeV for front-converting
    \irf{P7\_SOURCE} events $30^\circ$ off-axis.
    The most probable value of the distribution and the 68\%
    containment window are indicated.}
  \label{fig:energy_resolution}
\end{figure}

The energy resolution is a figure of merit which is customarily used to
summarize in a single number the information contained in the energy dispersion
parametrization.
As illustrated in \figref{energy_resolution} we define the energy resolution as
the half width of the energy window containing $34\% + 34\%$ (i.e., 68\%) of the
energy dispersion on both sides of its most probable value, divided by
the most probable value itself.
We note that this prescription gives slightly larger values of 
energy resolution than using the \emph{smallest} 68\% containment window.

\Figrefs{P7SOURCE_eres2D}{P7SOURCE_eres} shows the energy resolution for \irf{P7SOURCE}
events as a function of energy and incidence angle. As mentioned in
\secref{subsec:LAT_CAL} the energy resolution has a broad minimum between
$\sim1$ and $\sim10$ GeV, degrading at lower energies due to the energy
deposited in the TKR and at higher energies due to the leakage of the
electromagnetic shower out the sides and the back of the CAL.
Conversely the energy resolution tends to improve as the incidence angle
increases. This is especially true at high energy, where a longer path length
in the CAL implies less shower leakage---though front-converting events
more than 55$^{\circ}$ off-axis tend to exit the sides of the CAL and have worse
energy resolution.

\twopanel{htb}{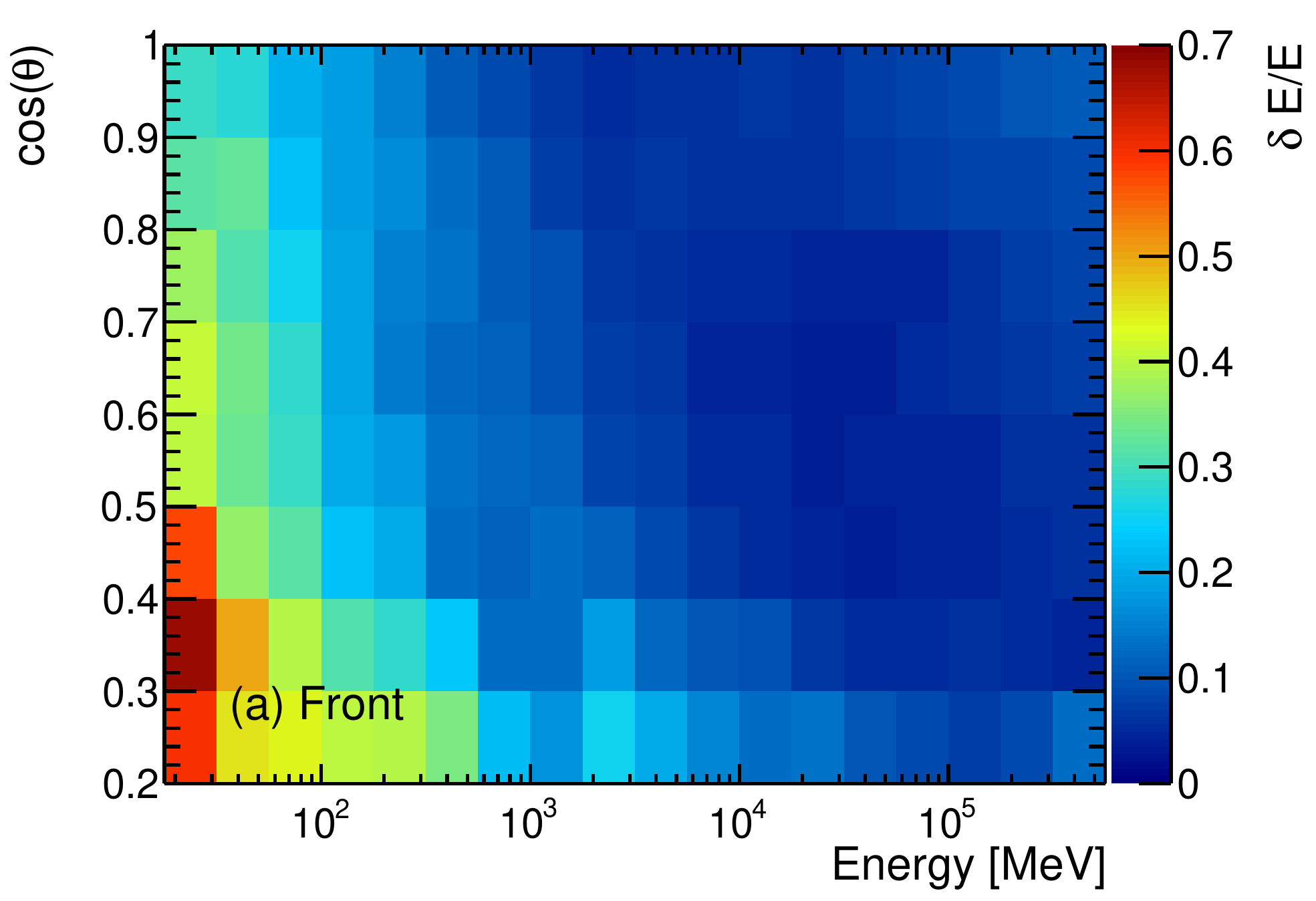}{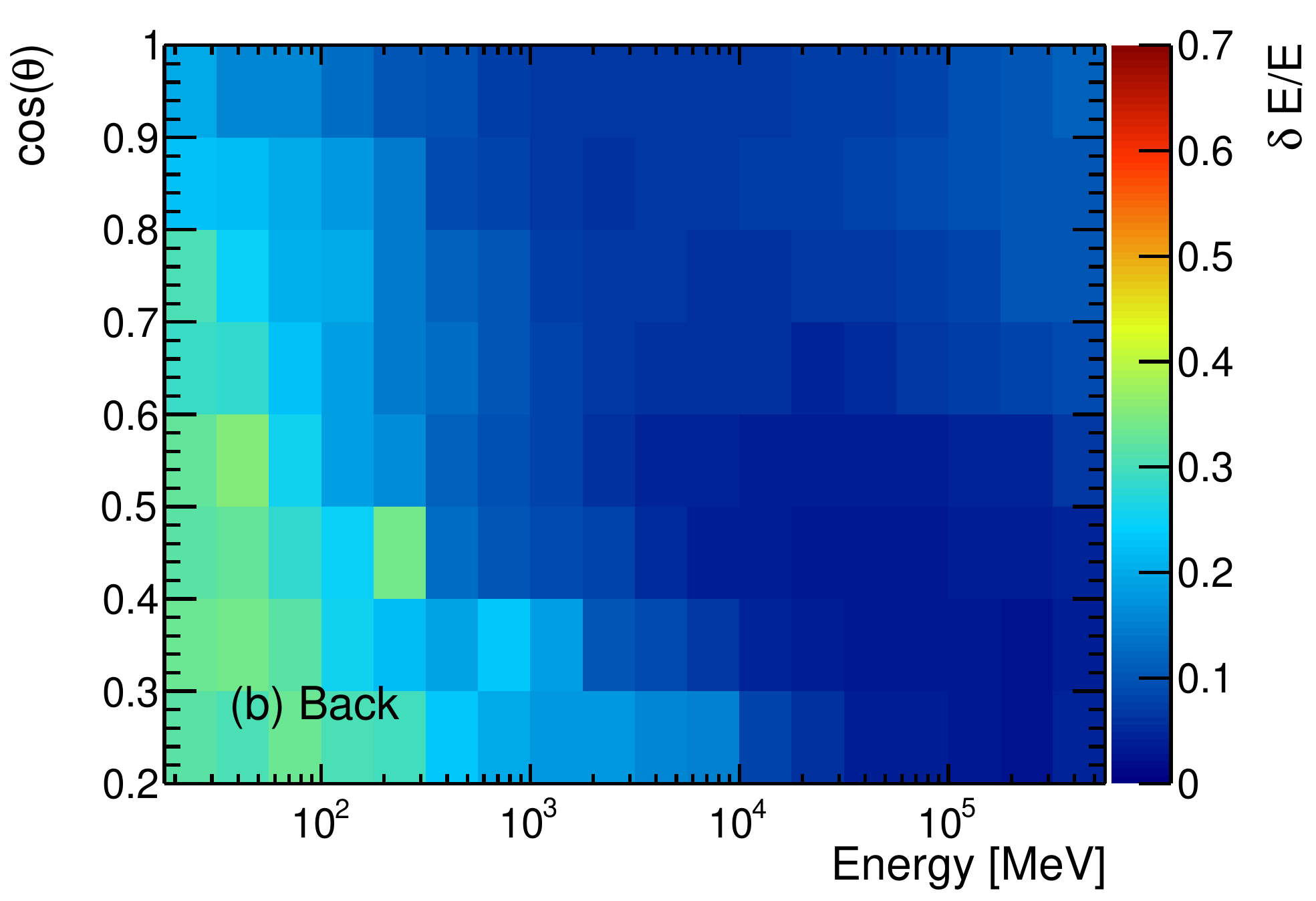}{
  \caption{Energy resolution (68\% containment half-width, as decribed in the text) as
    a function of energy and incidence angle for front (a) and back (b) conversions.}
  \label{fig:P7SOURCE_eres2D}
}

\twopanel{htb!}{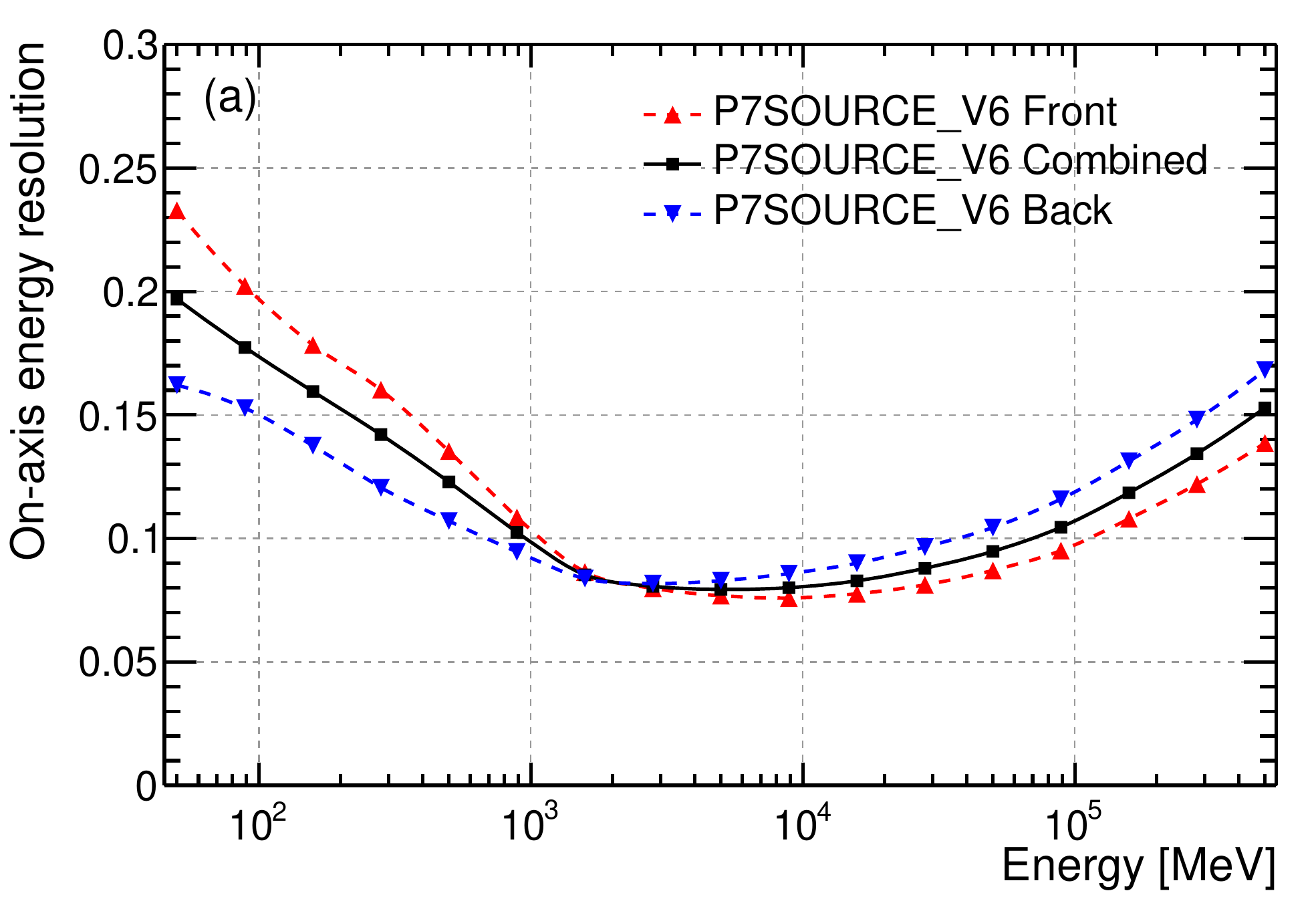}{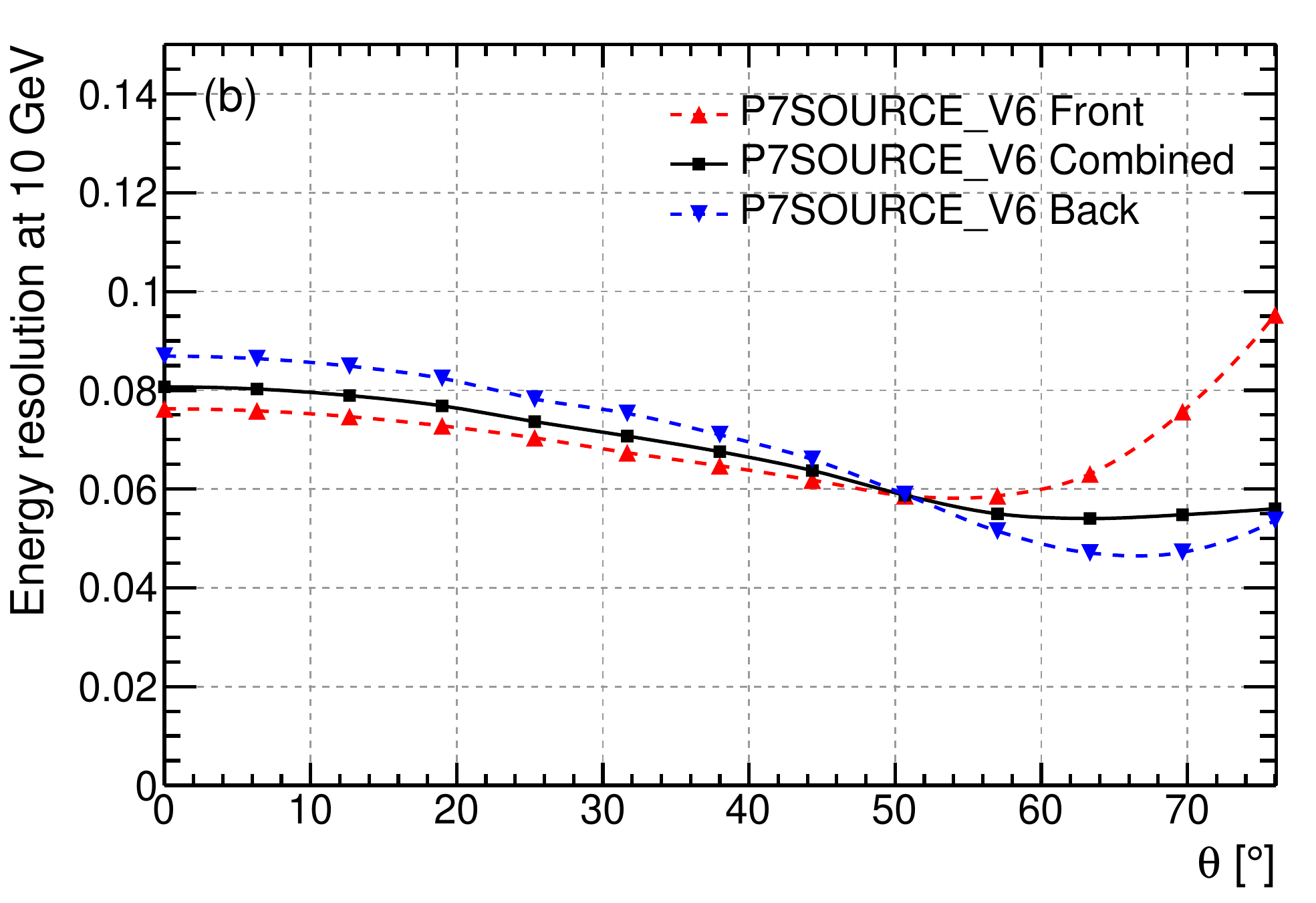}{
  \caption{Energy resolution as a function of energy on-axis (a) and incidence angle
    at 10~GeV (b) for the \irf{P7SOURCE\_V6} event class.}
  \label{fig:P7SOURCE_eres}
}

\subsubsection{Correlation Between Energy Dispersion and PSF}\label{subsec:EDisp_psf_correl}

As stated in \secref{sec:intro}, we factorize the IRFs, effectively assuming
that the energy and direction measurements are uncorrelated.
We use our \allgamma\ MC sample to test this hypothesis.
We locate each event within the cumulative PSF and measured energy
distributions:
\begin{align}
  P_{P} &= \int_{0}^{\alpha} \psf(\alpha;E,\hat{v}) d\alpha \nonumber\\
  P_{D} &= \int_{0}^{E'} \edisp(E';E,\hat{v}) dE'
\end{align}
If the IRFs perfectly described the MC sample, both of these distributions should
be flat. Furthermore, if there are no correlations between the
energy measurement and the direction estimate, the two-dimensional combined distribution should
also be flat. In practice this is nearly the case; the correlation
coefficient between $P_{P}$ and $P_{D}$ is small ($|C_{P,D}|< 0.1$) across
most of the energy and incidence angle range, as shown in~\figref{PlotIRFCorrel}~(a).
However, in many bins of $(E,\theta)$, we do observe a small excess of counts
with $P_{P} \sim 1$ and $P_{D} \sim 0$. Those values correspond to events that are in
the tails of the PSF and have a energy estimate that are significantly lower
than the true \gammaRayHyph\ energy. To quantify the magnitude of this effect, 
we considered the fraction of highly anti-correlated events $f_{a}$ with
$P_{P} > 0.98$ and $P_{D} < 0.02$ as a function of $(E,\theta)$. In the absence
of correlations, this fraction should be $f_{a} \sim 4\times10^{-4}$. In fact,
for highly off-axis \gammaRays\ in some energy ranges this fraction approaches
$f_{a} \sim 0.02$; however, averaged over the \fov, it is in
the 0.001 to 0.0035 range for both front- and back-converting events in at all
energies, as shown in~\figref{PlotIRFCorrel}~(b).

Although we can not reproduce this analysis with flight data, we have studied
the correlations between the energy and direction quality estimators
(\probE, see \secref{subsubsec:energy_analysis} and \probCore, see \secref{subsubsec:psf_analysis}) and found similarly small effects.

\twopanel{htb}{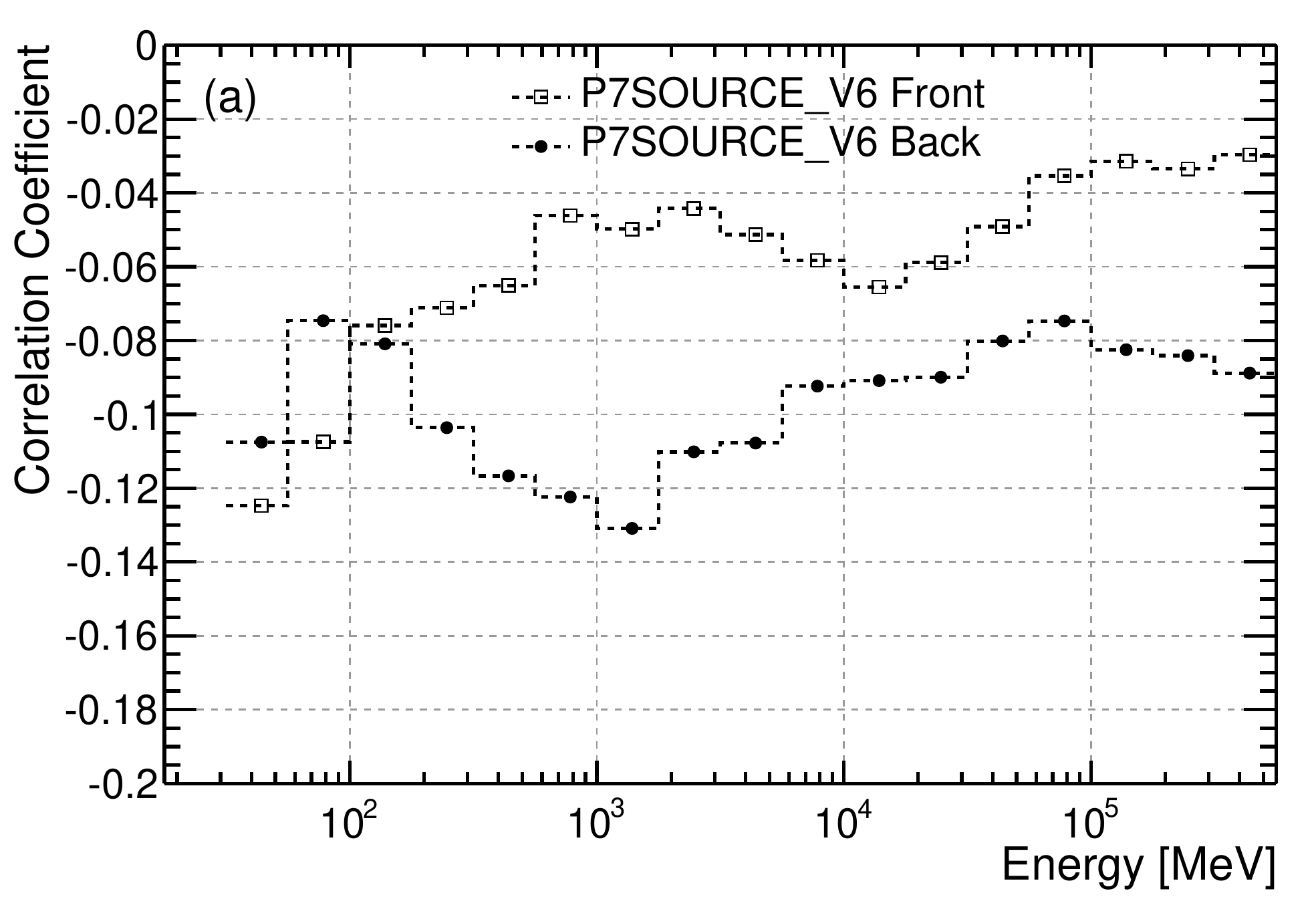}{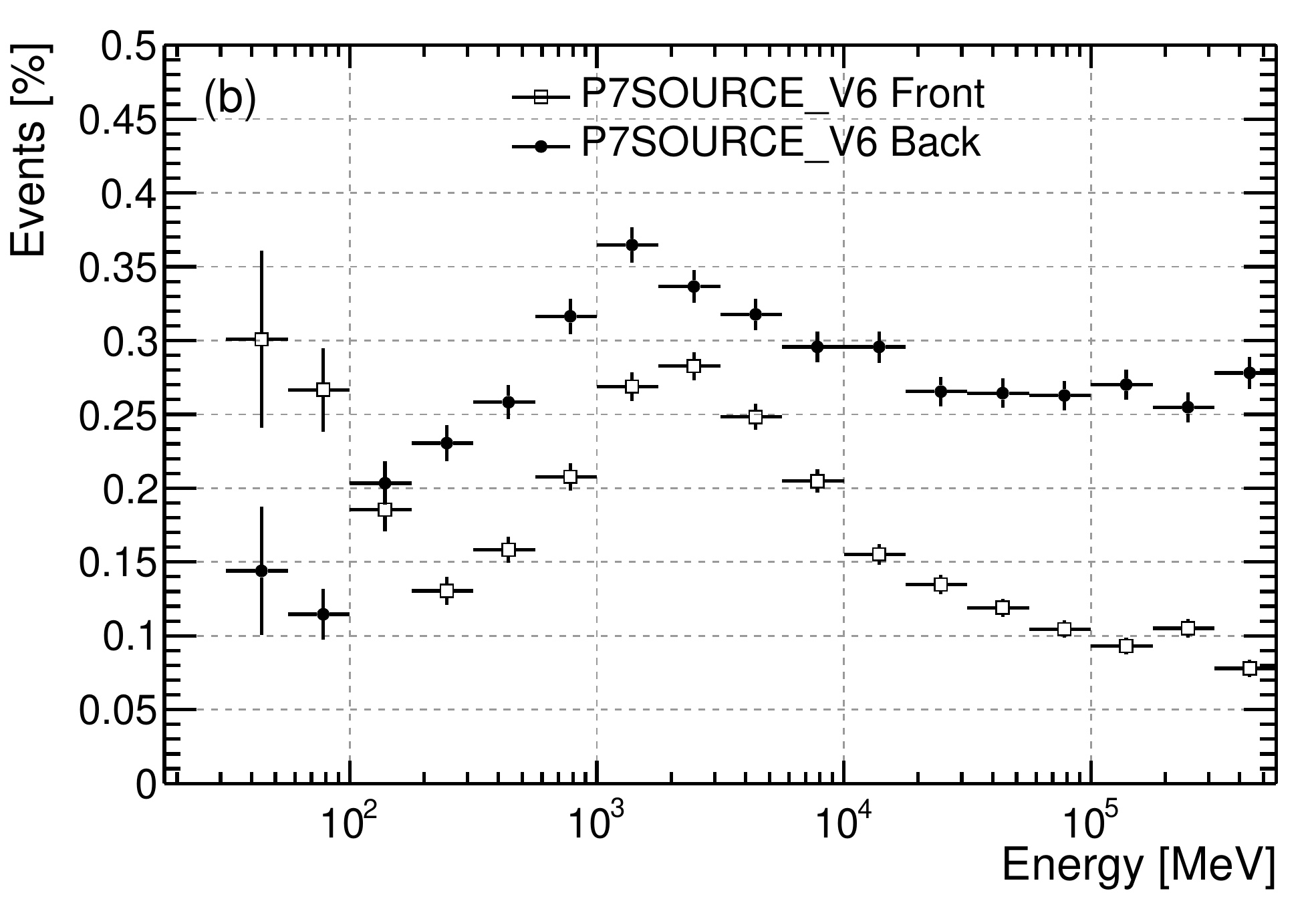}{
  \caption{Correlation between the energy dispersion and the PSF in
    the \allgamma\ (\secref{subsec:simul_allGamma}) sample:
    (a) correlation coefficient $C_{P,D}$ between $P_{P}$ and $P_{D}$ as a
    function of energy and (b) fraction of events with $P_{P} > 0.98$ and
    $P_{D} < 0.02$ (i.e., fraction of events which are in the tail of the PSF
    and also in the low side tail of the measured energy distribution).}
  \label{fig:PlotIRFCorrel}
}

In summary, averaged over several orbital precession periods any biases caused
by the correlation between the energy dispersion and the PSF are negligible
compared to other systematic uncertainties we consider in the paper. However, 
it is certainly a potential contributor to instrument-induced variability
(see \secref{subsec:perf_variability}).

\subsection{Spectral Effects Observed with Simulations}\label{subsec:eDispCorrections}

As noted in \secref{subsubsec:energy_analysis} the maximum likelihood (LH)
energy correction algorithm described in \secref{subsubsec:cal_recon} (and used
only in \psix) is by construction a binned energy estimator.
We have observed that it introduces
spectral artifacts corresponding to the bins used in creating the
likelihood parametrization. In addition, due to the fact that the
correction is not reliable above a few hundred GeV, the method is specifically
designed not to return values above 300~GeV. As a consequence, it tends to
concentrate events into a relatively narrow feature just below this energy.

Both aspects are illustrated in \figref{spectral_effects_mc}, where count
spectra from standard \allgamma\ simulations
(\secref{subsec:simul_allGamma}) are shown for both
\psix\ and \pseven.
While the overall shapes of the spectra reflect the LAT acceptance for
the corresponding event classes (\irf{P6\_V3\_DIFFUSE} and
\irf{P7\_SOURCE\_V6}), a clear sawtooth structure is visible in
the \psix\ count spectrum (with a typical width comparable with the
LAT energy resolution and peak-to-trough amplitude of the order of 
$\sim5\%$), along with a prominent feature at 300~GeV, above which
energies from the LH estimator cannot be selected

\twopanel{htb}{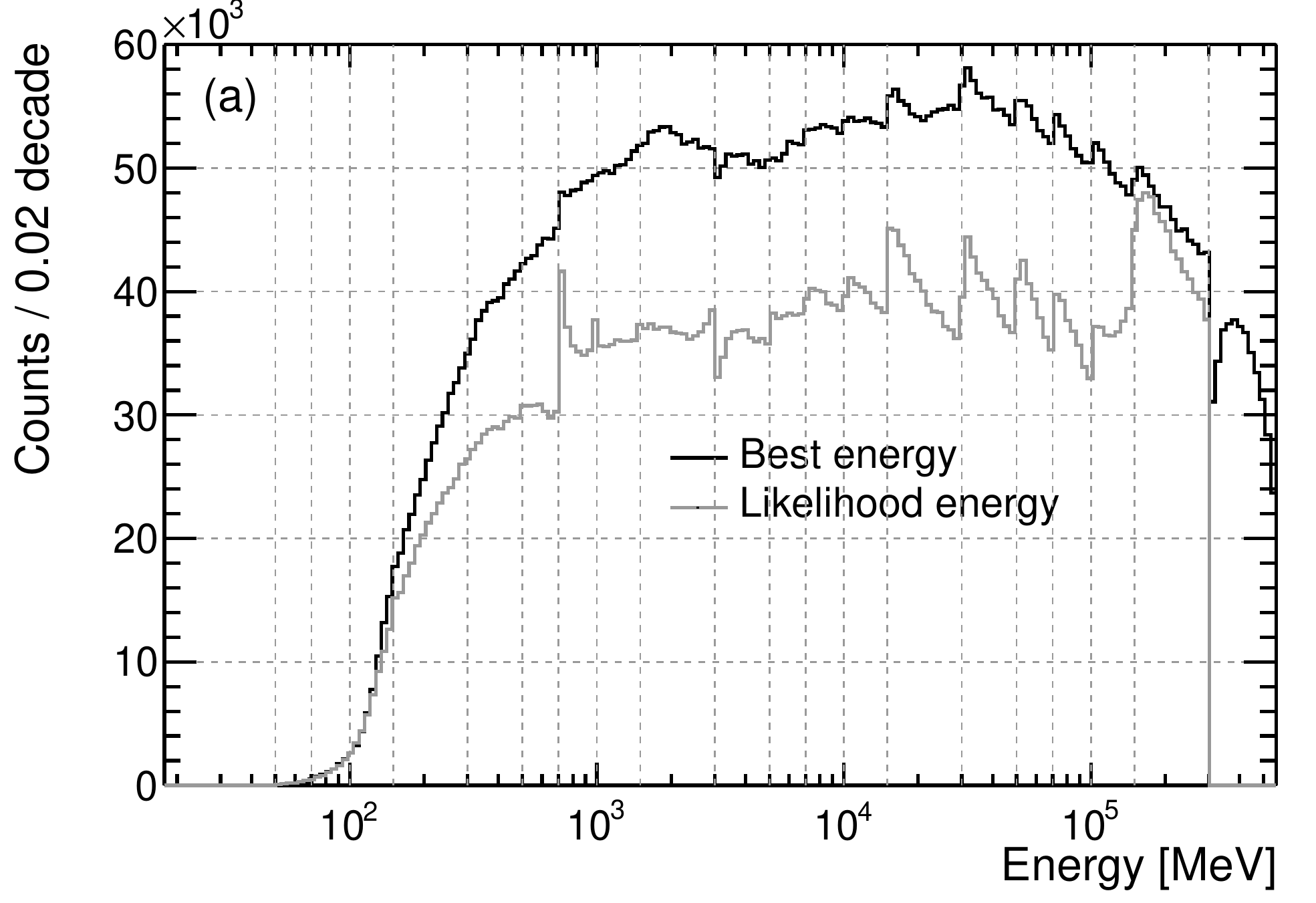}{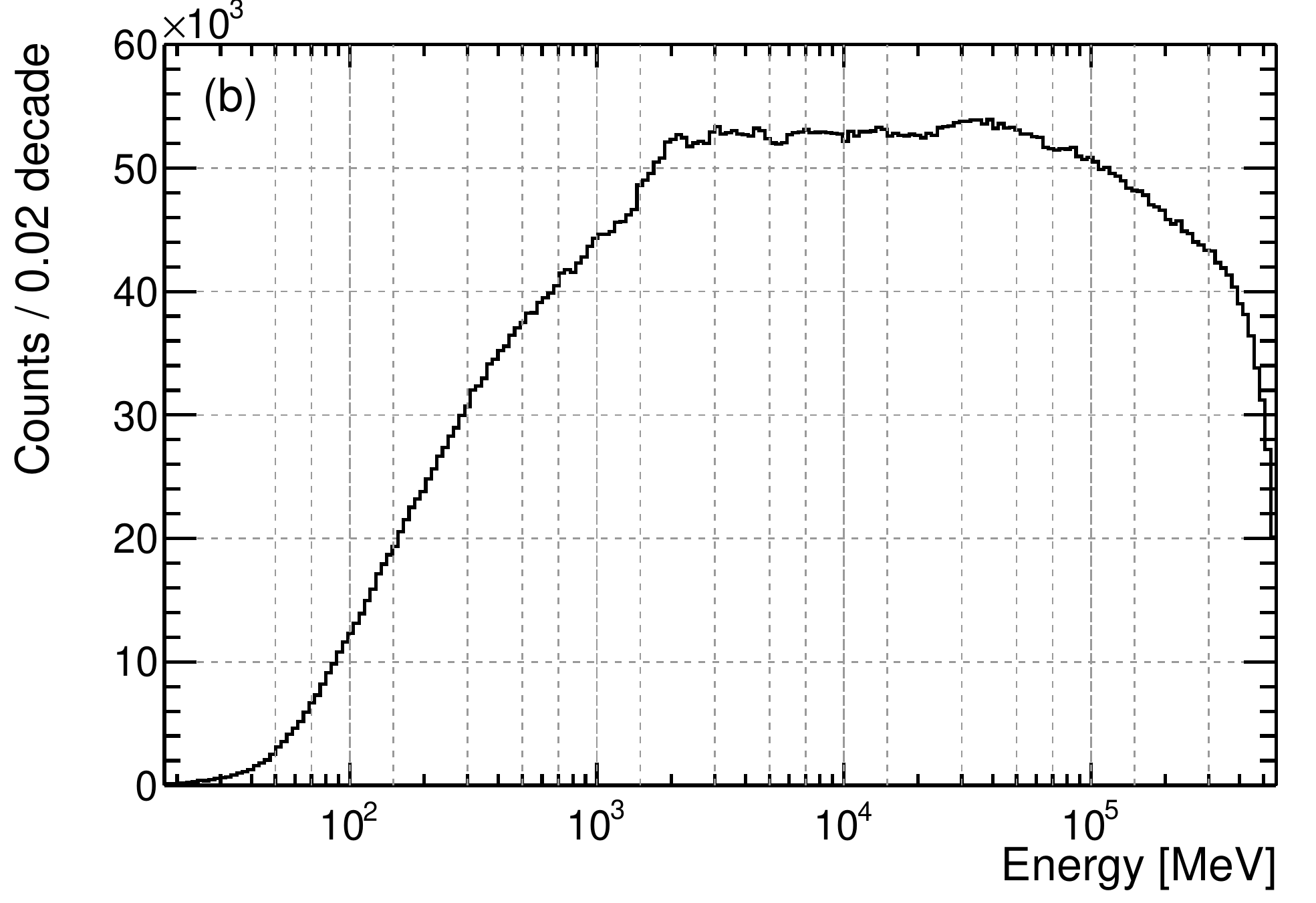}{
  \caption{Finely binned (50 bins per energy decade) count spectra for events
    passing the \irf{P6\_V3\_DIFFUSE} (a) and the \irf{P7\_SOURCE\_V6} (b)
    event selections; (a) also shows the LH energy estimates for the
    subset of events where they are available (note that the LH energy estimate is not always available, nor
    is it always selected when it is available). The final energy measurement
    in \psix\ reflects the artificial sawtooth structures of the underlying LH
    energy estimator, which was removed in \pseven. For both panels the 
    vertical hashing corresponds to the bin boundaries of the LH energy estimator.
  }
  \label{fig:spectral_effects_mc}
}

The binning of the count spectra in \figref{spectral_effects_mc} (50 bins per
energy decade) is deliberately much finer than the instrumental resolution---and
therefore much finer than the binning one would use for a real spectral analysis
of \fermi-LAT data. As a matter of fact, the features of the LH energy
estimator have little or no effect in most practical situations.
One noticeable exception is the search for spectral lines, such as might occur
from the annihilation or decay of a massive particle into a two-body final
state including \gammaRays\ \citep[see][for a more detailed discussion]{REF:2010.LineSearch}.

As already mentioned, in \pseven\ we chose to remove the LH
algorithm from consideration in the energy assignment---see \figref{spectral_effects_mc}~(b).

\subsection{Uncertainties in the Energy Resolution and Scale}%
\label{subsec:EDisp_errors}

In this section we will briefly review the systematic uncertainties on the
absolute energy scale and the energy dispersion. The energy measurement
is a complex process involving several different steps
(see~\secref{subsubsec:cal_recon} and~\secref{subsubsec:energy_analysis}).
Verification of the process is correspondingly complex, involving tests with
sea-level CR muons, tests at accelerators, and analysis of flight data.

\subsubsection{The Calibration Unit Beam Test Campaign}%
\label{subsec:EDisp_beamTest}

Since a direct calibration of the LAT with a particle beam was impractical for
schedule and cost reasons, the LAT Collaboration assembled for this purpose a
dedicated Calibration Unit (CU)\acronymlabel{CU} composed of two complete
flight spare towers, one additional CAL module and five ACD tiles.
An intensive beam test campaign was performed on the CU, between 2006~July and
November, with the primary goal of validating the MC
simulation used to define the event reconstruction and the background
rejection.
The CU was exposed to beams of bremsstrahlung \gammaRays, protons, electrons,
positrons and pions---with energies ranging between 50~MeV and 280~GeV
(\Tabref{tbeam_beams})---produced by the CERN Proton Synchrotron
(PS)\acronymlabel{PS} and Super Proton Synchrotron (SPS)\acronymlabel{SPS}.
In addition, the CU was irradiated with 1--1.5~GeV/n $^{12}$C and $^{131}$Xe
beams in the Gesellschaft f\"ur Schwerionenforschung
(GSI)\acronymlabel{GSI} facility with the purpose of studying the instrument
response to heavy ions.
\begin{table}[ht] 
  \begin{center}
    \begin{tabular}{lcrr}
      \hline
      Particle & Line & Energy & Triggers\\
      \hline
      \hline
      $\gamma$ & PS & 0--2.5~GeV & 12~M \\
      $\gamma$ (tagged) & PS & 0.02--1.5~GeV & 4~M \\
      $e^-$ & PS & 1--5~GeV & 6.4~M \\
      $e^+$ & PS & 1~GeV & 2.5~M \\
      $\pi^-$ & PS & 5~GeV & 0.6~M \\
      $p$ & PS & 6--10~GeV & 0.6~M \\
      \hline
      $e^-$ & SPS & 10--280 GeV & 17.8~M \\
      $\pi^-$ & SPS & 20~GeV & 1.6~M \\
      $p$ & SPS & 20--100~GeV & 0.8~M \\
      \hline
      $^{12}$C & GSI & 1--1.5 GeV/n & 3~M \\
      $^{131}$Xe & GSI & 1--1.5 GeV/n & 1.5~M \\
      \hline
    \end{tabular}
    \caption{Summary of the 2006 Calibration Unit beam test campaign,
      adapted from \citet{REF:2007.PreliminaryCUResults}.}
    \label{tab:tbeam_beams}
  \end{center}
\end{table}

A complete review of the results of the beam test campaign is beyond the scope
of this paper. The reader is referred to~\citet{REF:2007.PreliminaryCUResults}
for a description of the experimental setup and a review of the main results and
to~\citet{REF:2010.CREPRD} for additional related material.

From the standpoint of the energy measurement one of the primary goals of the
beam test campaign was to validate the leakage correction algorithms and,
ultimately, the ability of our MC simulation to reliably predict the
energy resolution at high energy. The tests confirmed that we understand
the overall shower development and the energy resolution to better than 10\% up
the maximum available electron energy, i.e., 280~GeV~\citep{REF:2010.CREPRD}.
As we will see in the next section, this implies that the effect of
systematic uncertainties on the energy resolution itself, when propagated
to the high-level spectral analysis, is essentially irrelevant in the vast
majority of practical cases.

The most significant discrepancy was observed in the raw energy deposited in
the calorimeter, as measured values were on average $\sim9$\% higher when
compared with simulations (with smaller fluctuations around this
value, slightly dependent on the particle energy and incidence angle).
The origin of this discrepancy is unknown, and possibly could be attributed to
residual environmental effects not properly accounted for, although an
imperfect calibration of the CU cannot be excluded.
In \secref{subsec:EDisp_absoluteScale} we will see that we have indications
from flight data that we do understand the absolute energy
scale to a precision better than~9\%.

\subsubsection{Crystal Calibrations with Cosmic-Ray Data}%
\label{subsec:EDisp_xtalCalib}

The calibration of the CAL crystals is the starting point of the
energy reconstruction chain and underlies all the subsequent steps
\citep[a detailed description of LAT on-orbit calibrations is given
in][]{REF:2009.OnOrbitCalib}.
As explained in \secref{subsec:LAT_CAL}, the large dynamic range of the LAT CAL
($2$~MeV--$70$~GeV per crystal) is achieved by means of four independent chains
of electronics per crystal, with different amplifications
(or \emph{ranges}), as summarized in \Tabref{cal_gains}.
In the most common readout mode (the so called \emph{single-range}) the highest
gain range that is not saturated is selected as the best estimate and
converted into a digital signal that is eventually used in the event
reconstruction.

\begin{table}[ht]
  \begin{center}
    \begin{tabular}{lllll}
      \hline
      Range & Diode & Gain & Energy range & MeV/Bin \\
      \hline
      \hline
      \lexeight & Small & High & 2~MeV--100~MeV & 0.033 \\
      \lexone & Small & Low & 2~MeV--1~GeV & 0.30 \\
      \hexeight & Large & High & 30~MeV--7~GeV & 2.3 \\
      \hexone & Large & Low & 30~MeV--70~GeV & 20 \\
      \hline
    \end{tabular}
    \caption{Readout ranges and energy conversion factors for the CAL 
      crystals, adapted from \citet{REF:2009.OnOrbitCalib}.}
    \label{tab:cal_gains}
  \end{center}
\end{table}

\begin{figure}[htbp]
  \centering
  \includegraphics[width=\onecolfigwidth]{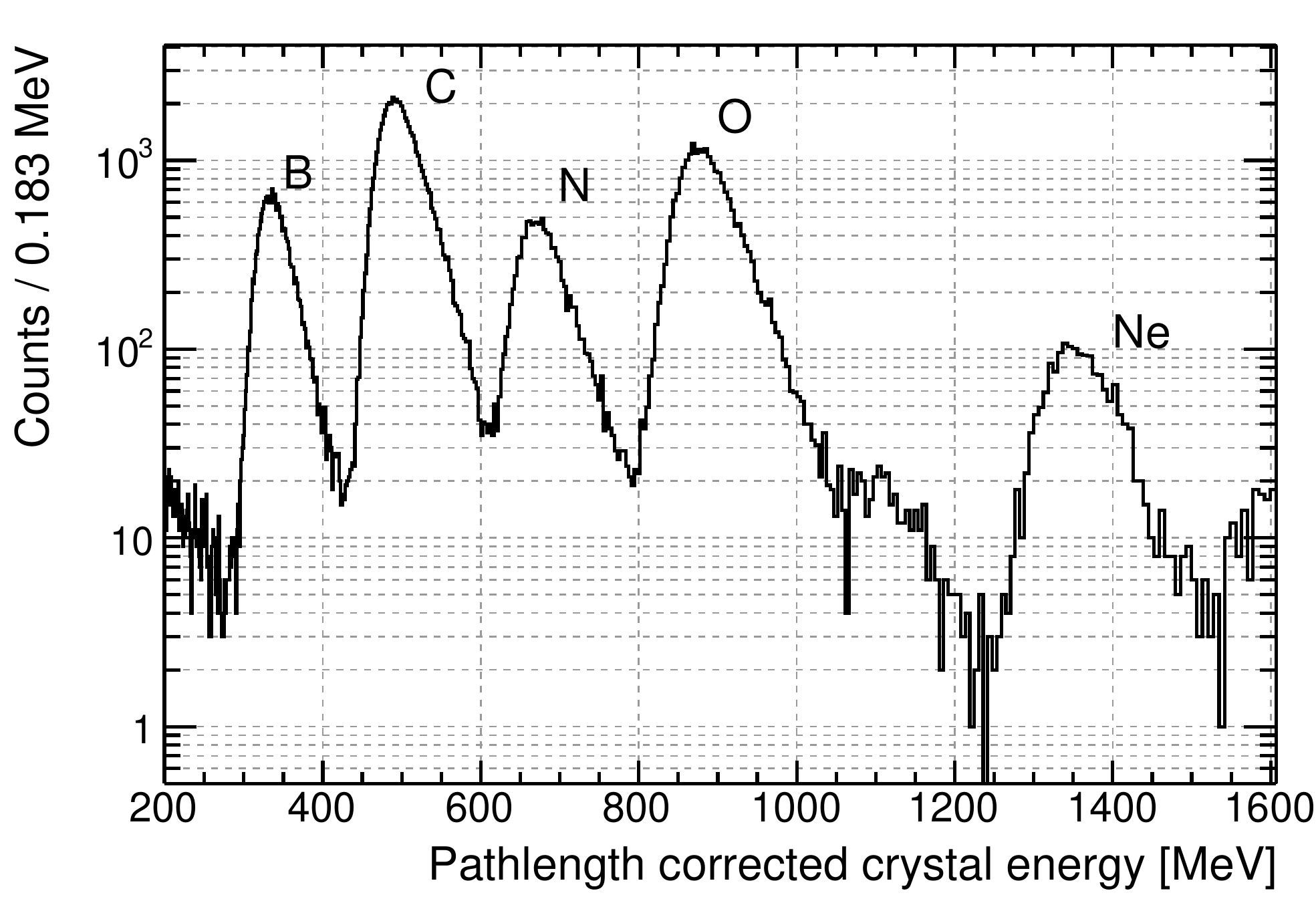}
  \caption{Distribution of the path length-corrected crystal energy deposition 
    for a sample of events used to inter-calibrate the low-energy and
    high-energy CAL energy ranges. The peaks corresponding to the
    most abundant Galactic CR species are clearly visible, though the 
    composition differs somewhat from the Galactic CRs because of secondary
    production in the ACD and TKR.} 
  \label{fig:Figure_72}
\end{figure}

What is relevant for the discussion here is the on-orbit calibration of the
crystal response, as determined by the crystal light yield and the linearity
of the electronics. The lower energy scales are calibrated using primary
protons that do not undergo nuclear interactions in the LAT, which are selected
through a dedicated event analysis.
For each crystal the most probable value of the energy deposition, corrected
for the path length, is compared with the MC prediction and the conversion
factor between digital signal and MeV is computed. This procedure was first
tested and validated with sea-level CR muons.
The high-energy ranges are calibrated in the energy range of overlap between
\lexone\ and \hexeight\ using events collected by a special on-board filter
for selecting heavy ions (see \parenfigref{Figure_72}), read out in full
\emph{four-range} mode (i.e., with all four energy ranges read out at each log
end).
The non-linearity of the electronics is characterized across the entire energy
range using a dedicated internal charge injection system and is corrected for in
the energy measurement.

Though we originally intended to use heavy primary nuclei for an independent
calibration of the high-energy ranges, the uncertainty in the scintillation
efficiency of heavy nuclei in the CsI(Tl) crystals relative to electromagnetic
showers makes the heavy ion peaks unsuitable for an independent cross-check of
the absolute scale.
The magnitude of the effect was measured with  600--1600~MeV/nucleon C and Ni
beams~\citep{REF:2006.CalBeamTest}, but we are uncertain how to scale the effect
to the typical on-orbit energies ($\sim5$~GeV/nucleon or greater) with our
desired accuracy of a few percent. 

We have used protons and Galactic CRs from Be to Fe at
incidence angles ranging from on-axis to $60^\circ$ off axis to demonstrate that the crystal
scintillation signal is proportional to path length for each species---i.e.
that the crystal response is piecewise linear over factors of two in signal.
Because the peaks from Be to O are closely spaced and easily resolved, this
method demonstrates that the CAL energy scale is linear over at least the
range $\sim 180$~MeV to $\sim 1500$~MeV per crystal, so any residual error in
absolute energy scale applies equally over that entire range.
Unfortunately it is difficult to bridge the gap between protons at $60^\circ$
($\sim 22$~MeV) and He on axis ($\sim 45$~MeV), and between He at $60^\circ$
($\sim 90$~MeV) and Be on axis ($\sim 180$~MeV), so we cannot demonstrate
linearity in this region with this technique. We are exploring alternative CR
event selections and geometries to cover these energy regions.
Nonetheless, the relative variations of the peak positions (for both non
interacting protons and heavier nuclei) can be measured with high accuracy,
which effectively allows us to monitor the time drift of the absolute energy
scale and prompt the update of new calibration constants when necessary.
\begin{figure}[htbp]
  \centering
  \includegraphics[width=\onecolfigwidth]{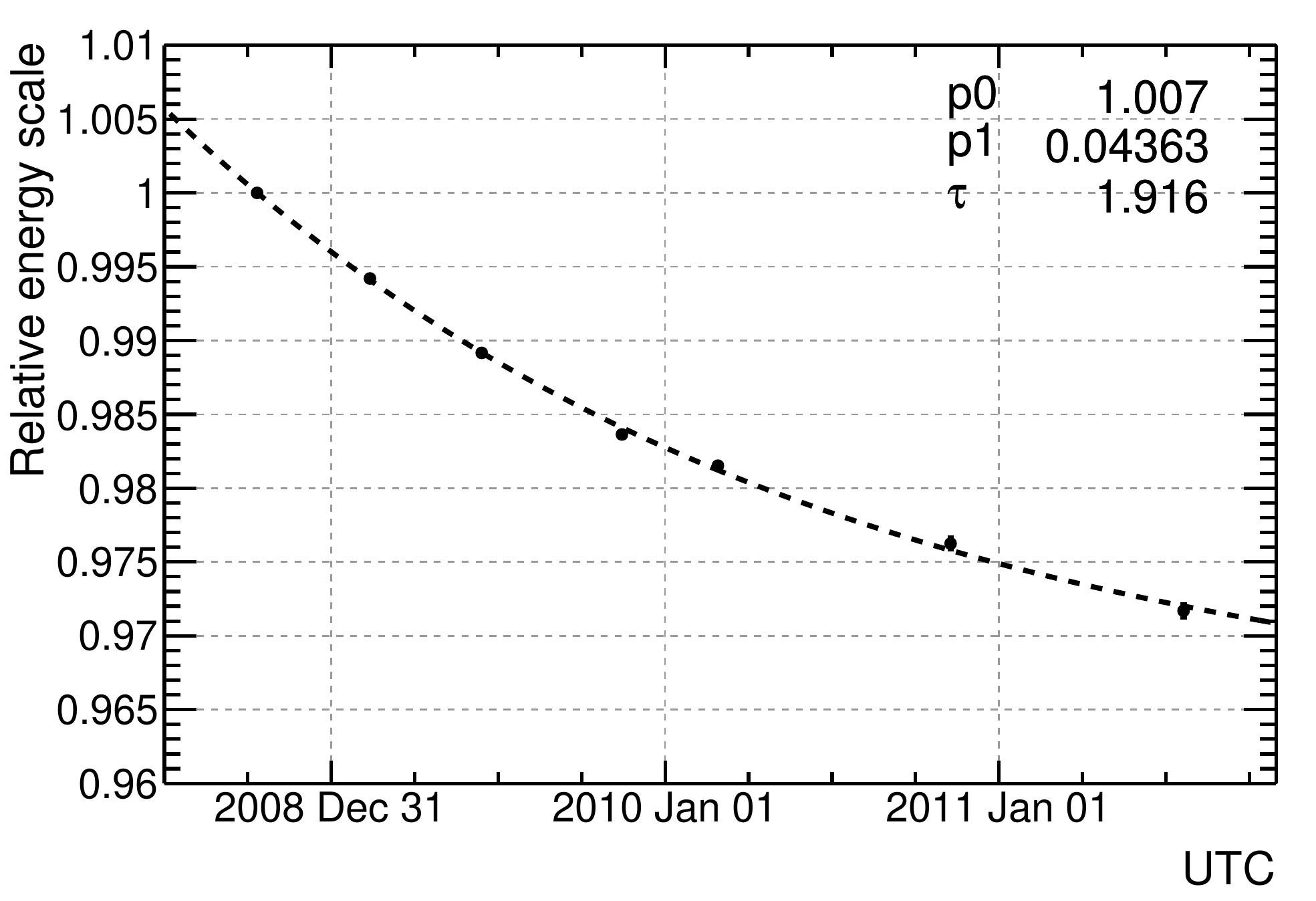}
  \caption{Relative variation of the absolute energy scale, as measured from
  the position of the proton peak position, throughout the first three years
  of the mission.} 
  \label{fig:Figure_73}
\end{figure}
The results in \figref{Figure_73} are in good agreement with the
pre-launch estimates of the crystal light yield attenuation due to radiation
damage, which results predominantly from trapped charged particles in the SAA. A fit with the function
\begin{equation}\label{conv:lightScale}
s(t) = (p_0 - p_1) + p_1 e^{-\frac{(t - t_0)}{\tau}}
\end{equation}
yields a time constant $\tau$ of the order of $\sim 2$~years, and predicts
an overall shift ($\sim p_0 - p_1$) of the energy scale (which can be corrected
for) on the order of 4.5\% after 10 years.

\subsubsection{Absolute Measurement of the Energy Scale Using the Earth's Geomagnetic Cutoff}\label{subsec:EDisp_absoluteScale}

As mentioned at the beginning of this section, there are essentially no
astronomical sources with spectral features that are sharp enough and whose
absolute energies are known to such a level of accuracy that they can be
effectively exploited for an on-orbit validation of the absolute energy scale.

One exception, perhaps unique in the energy range of the LAT, is the narrow 67.5~MeV
pion decay line predicted to originate from interactions of primary
CRs with the surface of the Moon~\citep{REF:MoonSpectrum}.
Unfortunately this feature is located at the lower end of the LAT energy
range. While it is not inconceivable that the LAT possibly could provide the
first evidence for the existence of such a line, the limited energy and angular
resolution at these energies, together with the brightness of the limb of the
Moon in continuum \gammaRays, make the lunar pion line impractical as an
absolute energy calibrator.

\begin{figure}[htbp]
  \centering
  \includegraphics[width=\onecolfigwidth]{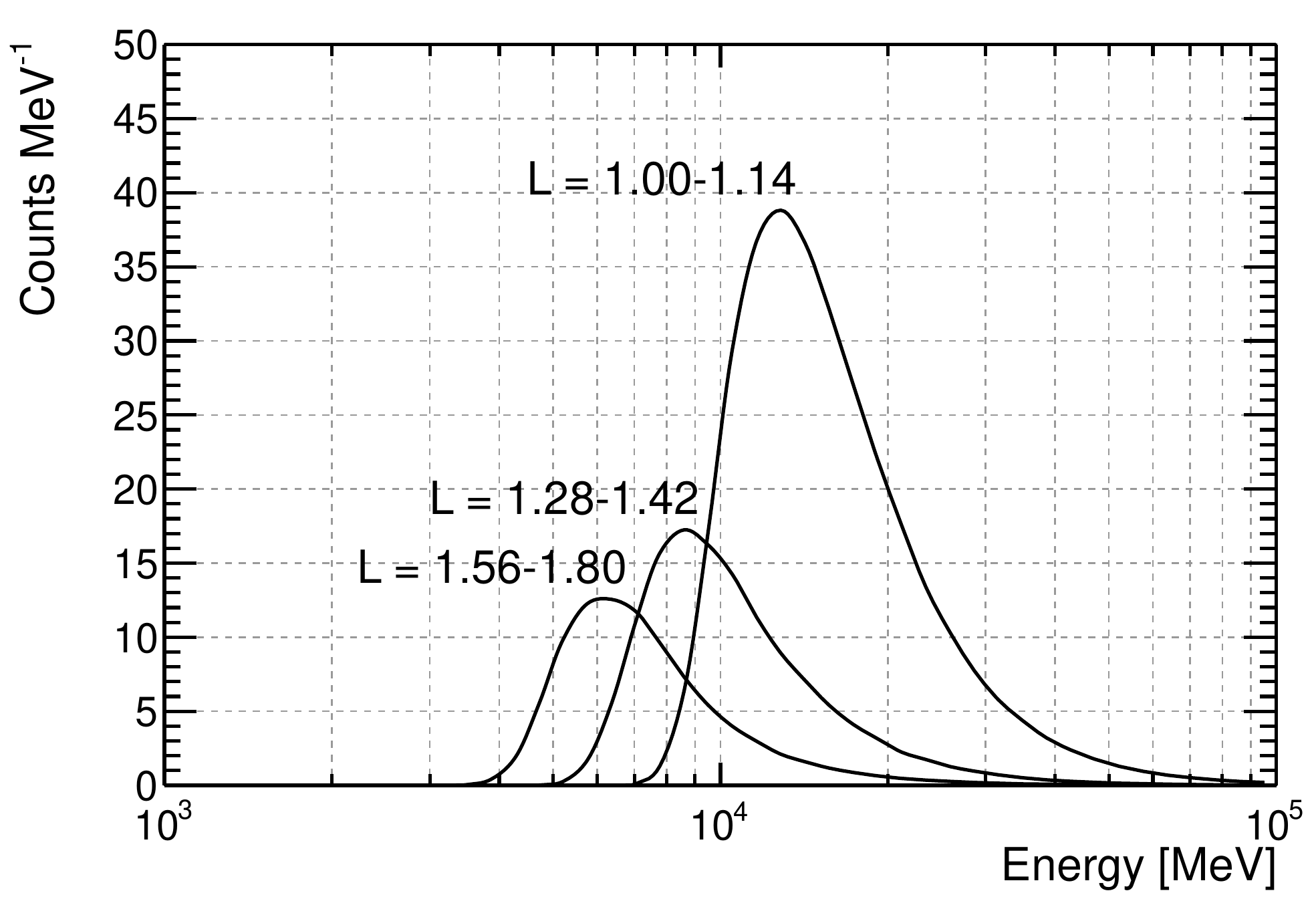}
    \caption{Examples of simulated CR electron count spectra in bins of
    \mcilwainl\ for the \Fermi\ orbit. They are averaged over the \fov\
    and folded with the energy resolution. Effectively these are the
    templates that the electron count spectra from real data are compared with
    in order to measure the absolute energy scale.
    The peaked shape is the result of the power-law spectrum of primary
    electrons being convolved with the screening effect of the Earth's magnetic
    field, which, in each bin of \mcilwainl, is effective below a certain
    cutoff energy.
    The relative normalizations reflect the fact that \Fermi\ spends the most 
    live time at low \mcilwainl.}
  \label{fig:Figure_74}
\end{figure}

A practical alternative is the geomagnetic rigidity cutoff in the CR
electron spectrum.  The LAT is effectively shielded from CRs by the
Earth's magnetic field at a low rigidity. The convolution of this
shielding with the primary power-law spectrum results in a peaked spectral
feature, whose shape and absolute energy can be predicted with good
precision by taking advantage of accurate models of the geomagnetic
field that are available~\citep{REF:IGRF11} and particle tracing computer
programs~\citep{REF:SmartShea}. See~\secref{subsec:bkg_MonteCarlo} for additional details.

\Figref{Figure_74} shows an example of such predictions
in different parts of the \fermi\ orbit. The count spectra,
averaged over the LAT \fov\ and folded with the LAT 
energy resolution, are shown in linear, rather than logarithmic, scale in order
to emphasize the peaked spectral shape. Comparison of the predicted and
measured peak positions is an in-flight measurement of the systematic bias in the energy scale. 
Since the magnitude of the Earth's magnetic field
varies across the \fermi\ orbit, this approach has the potential to
provide a series of calibration points, specifically between $\sim 6$~GeV and
$\sim 13$~GeV. The fact that this is the energy range for which the energy
resolution of the LAT is best is beneficial for the measurement itself.
Moreover, since both electrons and \gammaRays\ generate electromagnetic showers
in the detector, they are effectively indistinguishable from the calorimeter
standpoint---so that energy measurements for one species directly apply to the
other.

We have used this approach for an in-flight calibration of the absolute
energy scale using one year of data. The details of the analysis are beyond the
scope of this paper and are discussed in~\citet{REF:2011.AbsoluteEnergyScale}.
The main conclusion is that the measured cutoff energies exceed the predictions
by $2.6\% \pm 0.5\% \text{ (stat)} \pm 2.5\% \text{ (sys)}$ in the range
$6$--$13$ GeV.

\subsubsection{Summary of the Uncertainties in the Energy Resolution and Scale}\label{subsubsec:eDisp_eScale_syst}

Here we summarize the results related to the systematic uncertainties of
energy measurements with the LAT; in the next section we will discuss how these
impact the high-level spectral analysis.

The results from the CU beam test campaign indicated that the energy
resolution, as predicted by our MC simulations, is accurate to within 10\% up
to the maximum accessible beam energy (i.e., 280~GeV). As we will see in the
next section, this implies that this particular systematic effect is negligible
in most practical situations.

Though we were not able to identify the reason for the $\sim 9\%$ discrepancy
in the energy deposited in the CU calorimeter measured at the beam tests
(\secref{subsec:EDisp_beamTest}), we do have compelling indications from the
measurement of the CR electron geomagnetic cutoff
(\secref{subsec:EDisp_absoluteScale}) that in fact the systematic error on
the absolute energy scale is smaller than that. We stress that this measurement
involves the \emph{real} LAT, with the \emph{real} flight calibrations, in its
on-orbit environment.  Although the measurement derived from the CR electron cutoff 
applies only to a small portion of the LAT energy range, the $\sim 9\%$ effect measured at the beam
test affected essentially the entire range of energy and incidence angle, with
a weak dependence on both (see also the remarks
in~\secref{subsec:EDisp_xtalCalib} about the additional evidence for the
fractional error on the energy scale being constant over a wide energy range).
This evidence is further supported by the excellent internal consistency of the
analysis measuring the electron and positron spectra separately
~\citep{REF:2011.PositronFraction}.  However, it is more difficult to constrain the
energy scale at the low and high ends of the LAT energy range, where the energy resolution degrades 
by a factor of $\sim 2$.  Based on the full body of information currently available we conclude that
that the energy scale for the LAT is correct to $+20/ -50\%$ of the energy resolution of the 
LAT at a given energy.   This corresponds to an uncertainty of $+2 / -5\%$ on energy scale over 
the range 1--100~GeV, and increases to $+4 / -10\%$ below 100~MeV and above 300~GeV.

Finally, the measured energies of the Galactic CR peaks are being monitored to
gauge the time stability of the absolute energy scale, which we can control at
the 1\% level by applying calibration constants on a channel-by-channel basis.

\subsection{Propagating Systematic Uncertainties of the Energy Resolution and Scale to High Level Science Analyses}\label{subsec:EDisp_highLevelAnalysis}

Uncertainties in the energy resolution and scale introduce systematics in
spectral analyses, the magnitude of which depends on energy and the spectral
shape of the sources under study.  Since these systematic effects are a
consequence of event redistribution between energy bins, effects are also
strongly coupled to the energy dependence of \aeff.
As the implications of the systematic uncertainties are so dependent on the
analysis, the main purpose of this section is to illustrate the basic ideas
and the tone of the discussion is deliberately general.

\begin{figure}[htb!]
  \centering
  \includegraphics[width=\onecolfigwidth]{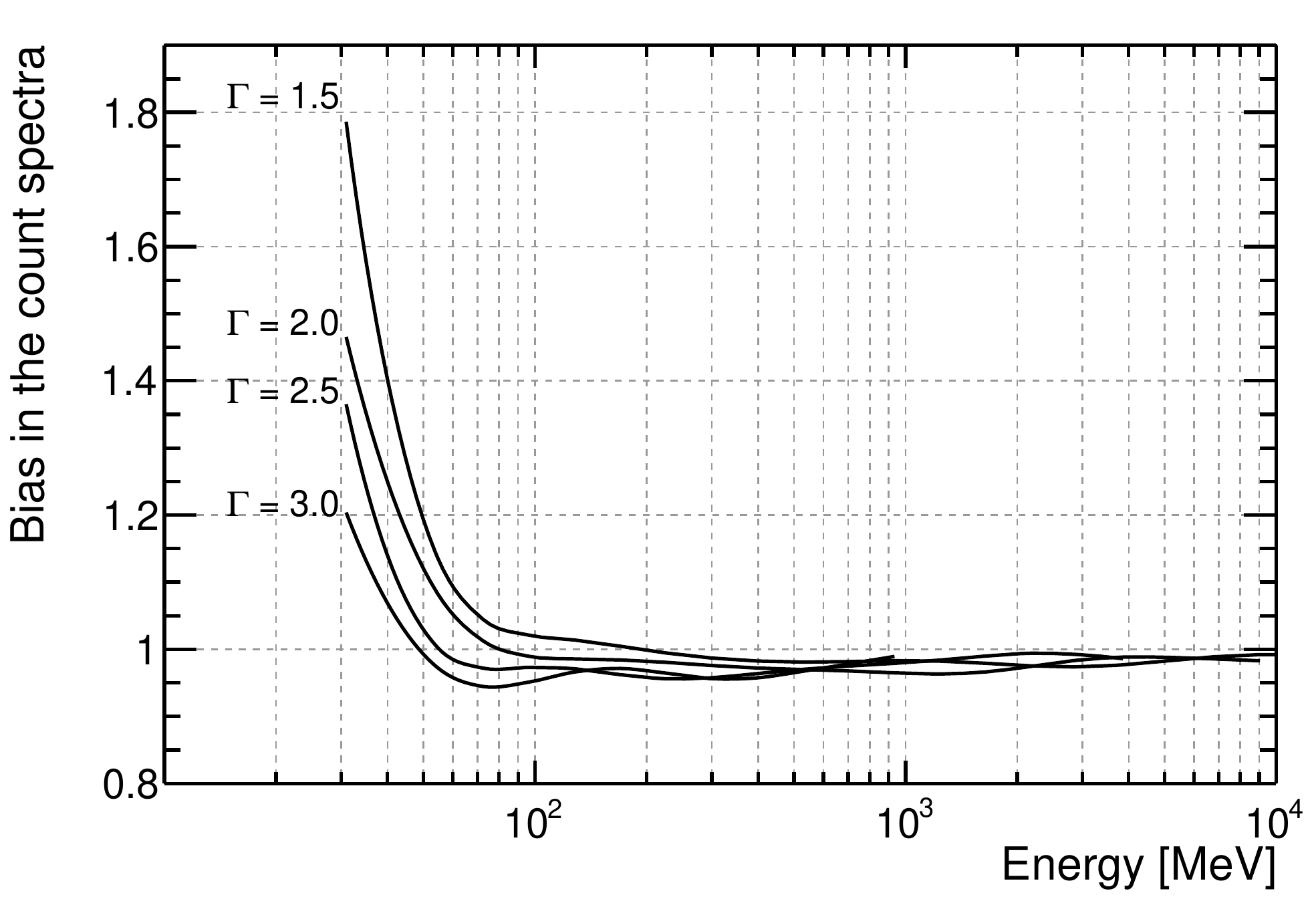}
  \caption{Bias factor for count spectra induced by not taking into account
    the finite energy resolution of the LAT. The plots refer to simulated
    point sources (with no background) with power-law spectra, for several
    different values of the spectral index.}
  \label{fig:Figure_75}
\end{figure}

The effect of the finite energy resolution is generally negligible for spectral
analysis above $\sim 100$~MeV, which as already mentioned is one of the main
reasons why it is not taken into account in the standard likelihood analysis.
Even above $\sim 100$~GeV, where the energy resolution is significantly
degraded by the leakage of the shower from the CAL, the effect is
negligible compared with other sources of systematics---at least for
sources with steep spectra, like primary CR electrons~\citep{REF:2010.CREPRD}.
MC studies with simulated point sources, though, show that the finite
resolution can induce a bias in the count spectra as high as 20--30\% 
at 50~MeV for power-law spectra with $\Gamma\sim 1.5$. The effect is strongly
dependent on the spectral index and is less severe for softer sources, as shown in
\figref{Figure_75}.

A bias in a count spectrum, in general, does not trivially translate into an
effect of the same order of magnitude in the parameter
values derived from a spectral analysis. When fitting a source spectrum with a single
power law, for instance, the combination of the long lever arm of the
high-energy data points and the inability of the spectral model to
accommodate any curvature in the count spectrum, results in the energy
dispersion having very little effect on the fit parameters (even though it can
produce very significant residuals).
\Figref{spectral_fit_erec_sys}~(a) shows that in this setup the bias in the
measured spectral index, when fitting down to 30~MeV, is smaller than 0.03 for
any reasonable input spectrum.
\twopanel{htb!}{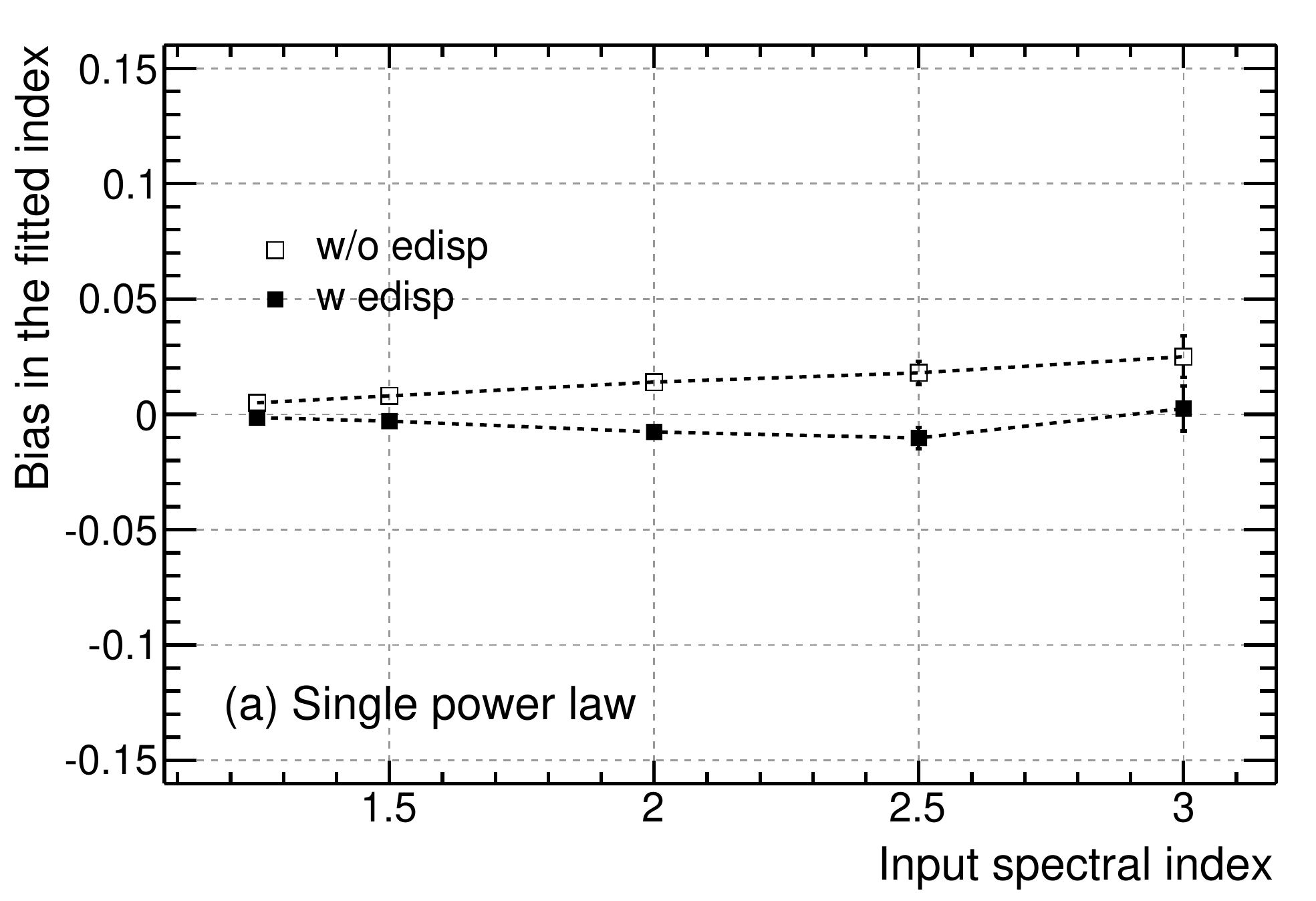}{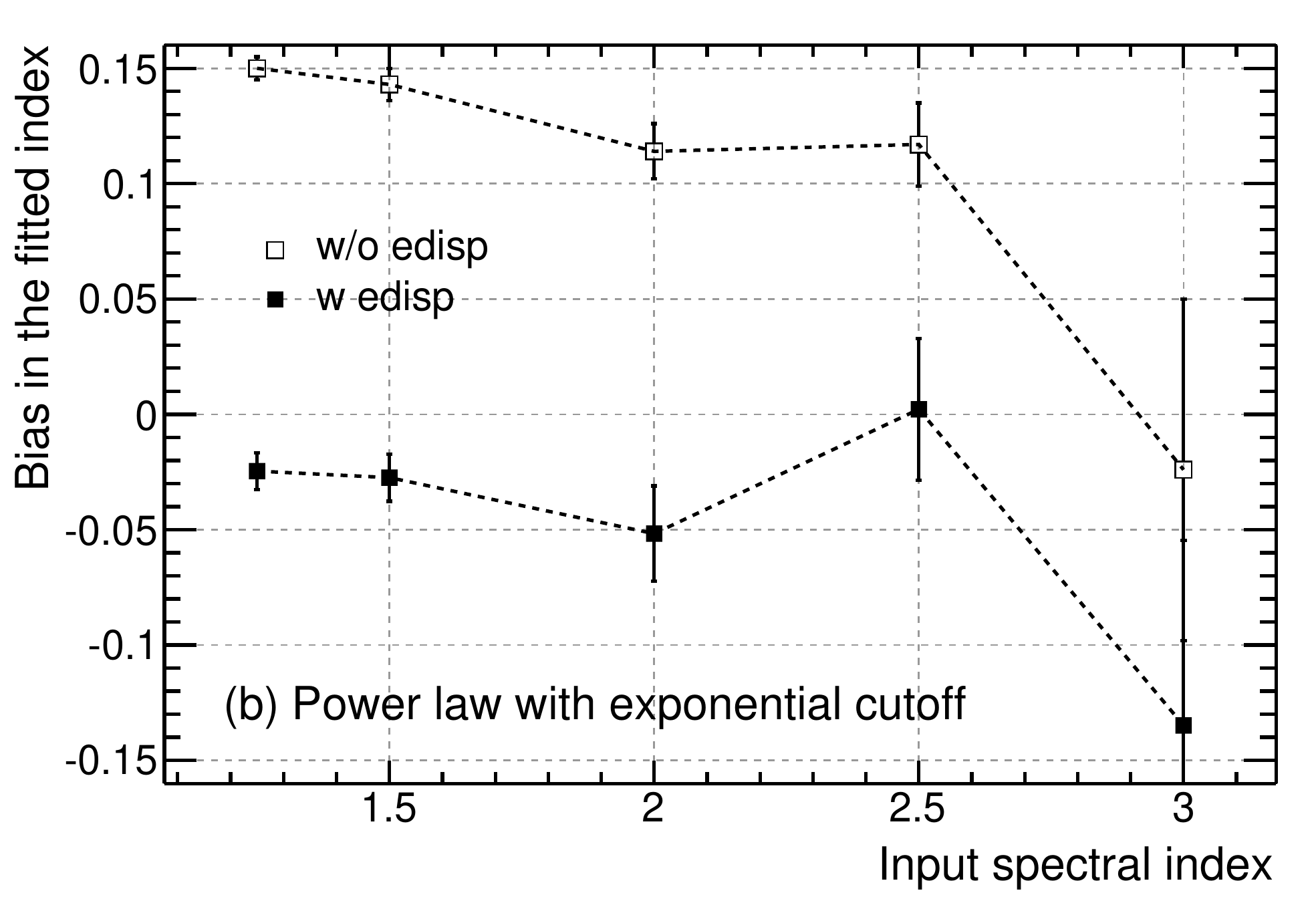}{
  \caption{Bias in the spectral index returned by a binned likelihood
    analysis with a minimum energy of 30~MeV (this setup is taken as a worst
    case scenario; increasing the minimum energy for the fit to 100 MeV
    decreases the bias by a factor of~2 or more).
    The plots refer to simulated point sources (with no background) with two
    different spectral shapes---a single power law (a) and power-law with an
    exponential cutoff at 1~GeV (b)--- for several different values of the
    input index.
    Enabling the energy dispersion handling in the \stools\ significantly
    reduces the bias in the fit parameters.}
  \label{fig:spectral_fit_erec_sys}
}

The situation can be quite different for other spectral shapes---e.g., a
power-law spectrum with exponential cutoff, as illustrated
in \figref{spectral_fit_erec_sys}~(b). Particularly, the bias introduced by the
energy dispersion reaches in this case $\sim 0.15$ for hard spectral indices.
The problem can be studied, for a specific source model, by means of dedicated
MC simulations, and this is strongly recommended, especially for spectral
analysis below 100~MeV. Generally speaking, enabling the energy dispersion
handling in the standard binned likelihood analysis significantly reduces the
bias in the fit parameters.

To quantify the magnitude of the bias introduced by ignoring energy
dispersion in the likelihood fit we have re-analyzed both the
B2~1520+31 and the PG~1553+113 \rois\ with the energy dispersion 
treatment enabled using the \stools\ (version v9r28p0).   We list the changes
with respect to the fit results with the energy dispersion ignored in
\tabref{bracketErrors_eDisp}.

\begin{table}[hp!]
  \begin{center}
    \begin{tabular}{lcc}
      \hline
      Parameter & B2 1520+31 & PG 1553+113 \\
 & (soft) & (hard) \\
\hline
\hline
$\delta N_{0}/N_{0}$ & $+3.9\%$ & $+2.2\%$ \\
$\delta\Gamma$ $\delta\alpha$ & $-0.04$ & $+0.01$ \\
$\delta\beta$ & $+0.02$ & {\hfill-\hfill} \\
$\delta F_{25}/F_{25}$ & $+1.4\%$ & $+3.3\%$ \\
$\delta S_{25}/S_{25}$ & $+2.0\%$ & $+1.1\%$ \\

      \hline
    \end{tabular}
    \caption{Systematic variations arising from ignoring the effects
      of energy dispersion when performing likelihood fitting.  For
      the spectral index ($\Gamma$ or $\alpha$) and spectral
      curvature ($\beta$) we give the absolute variation with respect to the
      value obtained when ignoring the energy dispersion (e.g., $\delta \Gamma$). For the flux prefactor and the
      integral fluxes we give the relative variation with respect to the
      value obtained when ignoring the energy dispersion (e.g., $\delta N_{0} / N_{0}$).}    
    \label{tab:bracketErrors_eDisp}
  \end{center}
\end{table}

MC simulations also show that the effects of systematic uncertainties in
the energy resolution, at the level we understand the energy dispersion of the
detector, are essentially negligible over the entire energy range when studying
\gammaRayHyph\ sources that do not have sharp spectral features---which accounts for
the vast majority of cases of practical interest.
One noticeable exception is the search for \gammaRayHyph\ lines, which
requires a dedicated analysis when evaluating limits.
But even in that case the systematic uncertainty
on the energy resolution is not a major source of concern: if the actual energy
resolution was $\sim 10\%$ broader than that predicted by the MC
the fitted signal counts would be $\sim 10\%$ lower, i.e., not enough to 
dramatically decrease the sensitivity to a spectral line.

The effect of systematic uncertainties on the absolute energy scale is also
strongly dependent on the energy range of interest. 
In the ideal case of an energy-independent effective area, it can be shown
that, for a power-law spectrum with index~$\Gamma$, a relative bias
$b$\label{conv:ebias} in the
absolute scale translates into a rigid shift of the spectrum itself by the
amount $\Delta F/F = (\Gamma - 1)b$ (or $\Delta S_{25}/S_{25} = (\Gamma - 2)b$ if 
we consider the integral energy flux). In short, the measurement
of the spectral index is not affected---while the flux obviously is.  

As a concrete example, we estimate the effect of systematic uncertainties of the 
absolute energy scale on the integral fluxes $F_{25}$ and $S_{25}$ of 
PG~1553+113 and B2~1520+31 by integrating the fitted
source spectra and shifting the limits of integration by the 
uncertainties stated in \secref{subsubsec:eDisp_eScale_syst}. Specifically, 
we consider the integration ranges 96~MeV--95~GeV and 104~MeV--102~GeV:  
the resulting uncertainty estimates are reported in \tabref{eScale_bracket}.  

\begin{table}[htb]
  \begin{center}
  \begin{tabular}{lcc}
\hline
Parameter & B2 1520+31 & PG 1553+113 \\
          & (soft) & (hard) \\
\hline
\hline
$\delta F_{25}/F_{25}$ & $\syserrors{+13.4\%}{-4.6\%}$ & $\syserrors{+3.4\%}{-2.5\%}$ \\
$\delta S_{25}/S_{25}$ & $\syserrors{+3.8\%}{-1.4\%}$ & $\syserrors{+1.7\%}{-0.6\%}$ \\
\hline
\end{tabular}
\caption{Systematic variations of integral fluxes arising from uncertainties in the absolute
         energy scale. We give the relative variation with respect to the
         nominal value (e.g., $\delta F_{25} / F_{25}$).}
\label{tab:eScale_bracket}
  \end{center}
\end{table}

\begin{figure}[htb!]
  \centering
  \includegraphics[width=\onecolfigwidth]{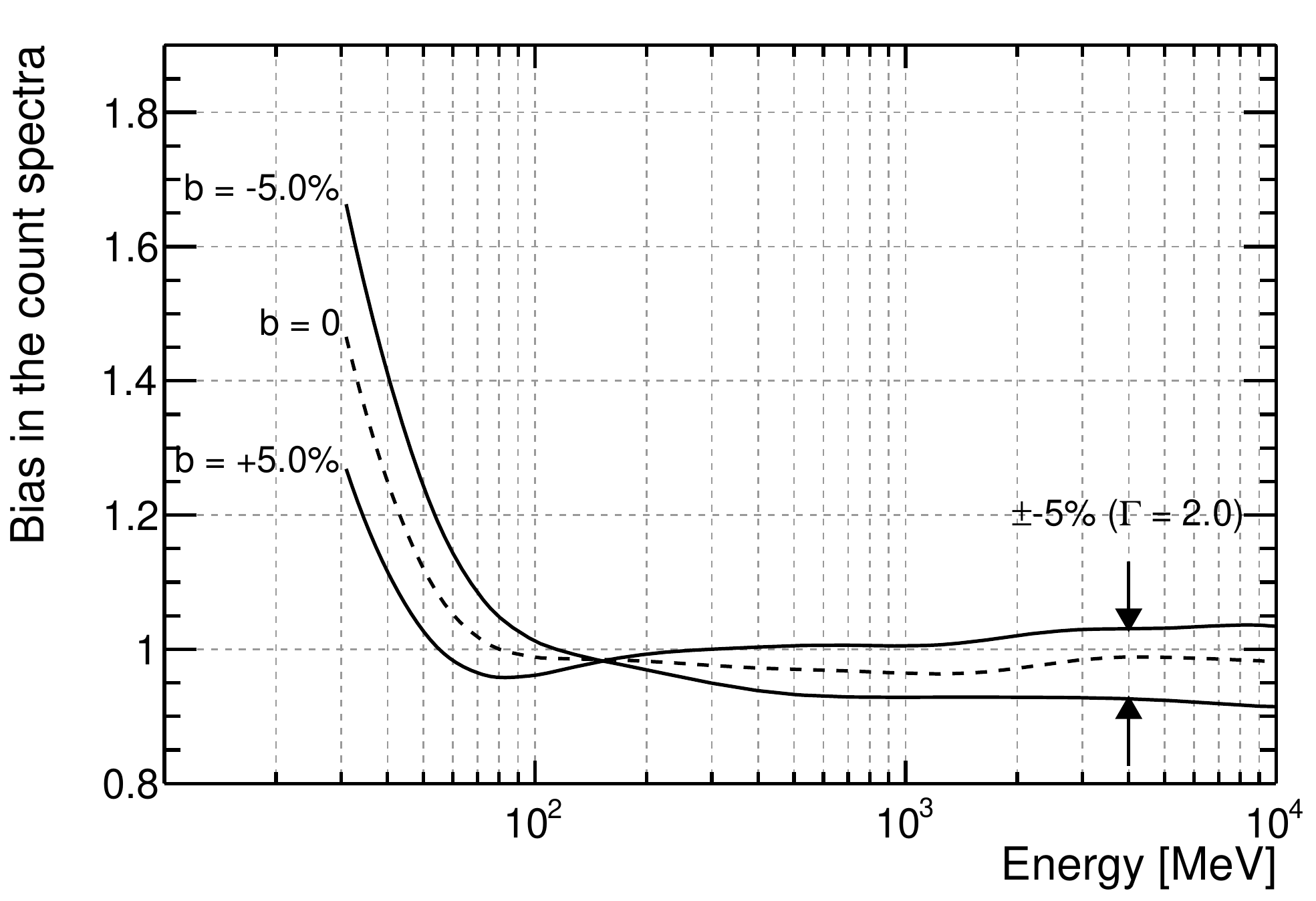}
  \caption{Bias factor on the count spectra induced by systematic uncertainties
    on the absolute energy scale for a simulated point source (with no
    background) with power-law spectrum with $\Gamma = 2$.
    The parameter $b$ indicates the
    relative bias in the absolute energy scale; the case $b = 0$ corresponds
    to the line at $\Gamma = 2$ in \figref{Figure_75}.}
  \label{fig:Figure_77}
\end{figure}

Along the same lines, measured cutoff energies for sources with curved spectra
reflect the bias in the energy scale directly: $\Delta E_c/E_c = b$.
These statements are useful rules of thumbs to estimate the order of magnitude
of the effect above $\approx 1$~GeV, where the effective area is 
not strongly dependent on energy (\parenfigref{Figure_77}). In fact,
since we have good indications that we understand the energy scale of the
detector at the $\sim5$~\% level (see \secref{subsec:EDisp_beamTest} and
\secref{subsec:EDisp_absoluteScale}) at these energies, the effect of systematic
uncertainties in the scale is generally smaller than other sources of
systematics in this energy range, and does not affect the
measurement of the spectral index in the power-law case.

Below a few hundred MeV the systematic uncertainties in the absolute energy 
scale should be carefully considered in any analysis due to the
unfortunate combination of a steeply increasing effective area and
a worsening of the energy resolution at low energies.
As for the energy dispersion the effect is more pronounced for steep sources.
For the typical case of a spectral index of $\Gamma\approx 2$, a $\pm 5\%$
uncertainty in the energy scale is negligible down to $\sim 100$~MeV but at
50~MeV can easily translate into an additional 10--20\% bias in the count
spectrum. Again, the order of magnitude of the effect can be easily assessed
on a case-by-case basis, depending on the spectral shape of the source under
study and energy range of interest.

Spectral unfolding (deconvolution) is an approach for taking into
account the energy dispersion in the spectral fitting. Unfolding is in effect the
opposite of the \emph{forward folding} approach implemented in many popular
astronomical data analysis packages (and used in the standard likelihood
point source analysis implemented in the \stools), where the 
IRFs are folded with the spectral model and the model parameters
are varied until the best match with the measured count spectrum is
found---e.g., this is how XSPEC~\citep{REF:xspec} handles the energy dispersion.

Energy unfolding has been used for several analyses of LAT data, mainly to
cross-check the results of the standard likelihood
analysis~\citep{REF:BlazarSEDs} and especially for the lowest or
highest energies: below $\sim 100$~MeV~\citep{CrabPaper} or above
$\sim 100$~GeV~\citep{REF:2010.CREPRD,REF:2009.Electrons20GeV1TeV}.
The Bayesian approach detailed in~\citet{REF:BayesUnfolding} is that typically
being used by the LAT Collaboration.

\subsection{Event Analysis-Induced Spectral Features}\label{subsec:EDisp_spectralFeatures}

Though we discuss potential event analysis-induced spectral features in the
context of energy reconstruction, we emphasize that the effective area is also
germane to this topic. We conclude that the event analysis and IRFs do not
introduce significant artificial spectral features because bright
\gammaRayHyph\ sources are devoid of spectral features---particularly
the Earth limb (\secref{subsubsec:LAT_earthLimb}), which is the single brightest
source in the LAT energy range.

\begin{figure}[htb!]
  \centering\includegraphics[width=\onecolfigwidth]{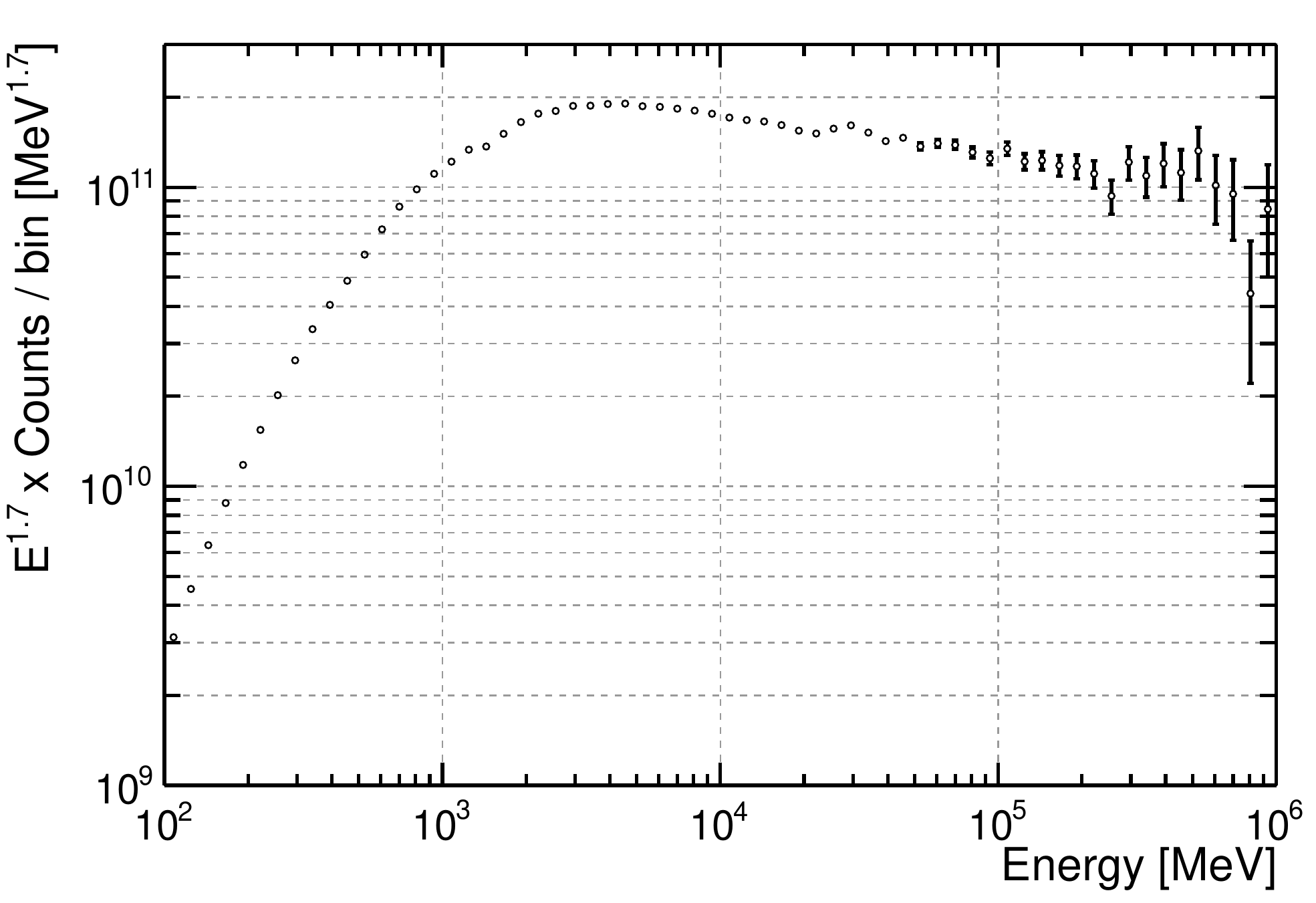}
  \caption{Count spectrum of the Earth limb based on the first two years of
    data, including \irf{P7CLEAN\_V6} events within three times the
    acceptance-averaged PSF 68\% containment from the nominal zenith angle of
    the limb. Since the limb itself is typically observed at large incidence
    angle in the instrument frame, events impinging on the LAT more than
    $65^\circ$ off-axis have been removed in order not to overweight the edge of
    the \fov\ relative to typical analyses of celestial sources.
    The turnover below $\sim 1$~GeV is mostly due to the decrease of the LAT
    acceptance, rather than to an intrinsic roll-off of the source spectrum.}

  \label{fig:limb_count_spec}
\end{figure}

\Figref{limb_count_spec} shows a count spectrum of \irf{P7CLEAN\_V6}
\gammaRays\ from the Earth limb based on the first two years of data.
The spectrum is made with $\sim 16$ bins per energy decade---corresponding
to a bin width slightly smaller than the typical LAT energy resolution---and
its smoothness is a good qualitative indicator of the smallness of any possible
instrument-induced spectral feature.

In order to quantify this we compared the counts $n_i$ in the $i$-th energy
bin with the value $f_i$ returned by log-parabola fit to the four nearest bins
(i.e., two bins on each side). We can construct two obvious metrics, namely the
\emph{normalized} residuals:
\begin{equation}\label{conv:residNorm}
r^n_i  = \frac{(n_i - f_i)}{\sqrt{f_i}}
\end{equation}
and the \emph{fractional} residuals
\begin{equation}\label{conv:residFrac}
r^f_i = \frac{(n_i - f_i)}{f_i}.
\end{equation}

\begin{figure}[htb!]
  \centering\includegraphics[width=\onecolfigwidth]{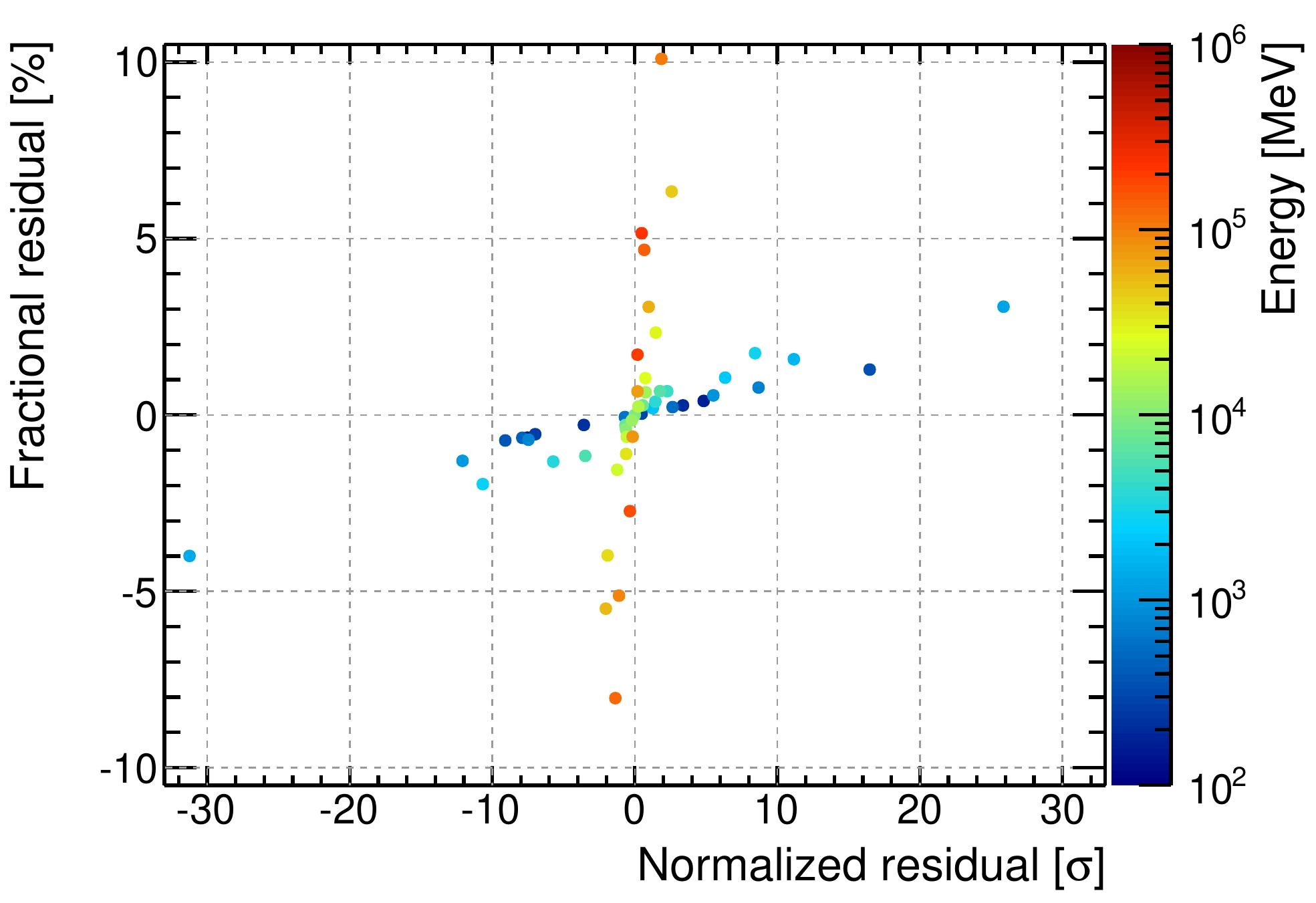}
  \caption{Single-point residuals of the count spectrum in
    \figref{limb_count_spec} with respect to a log-parabola fit of the four
    nearest points.}
  \label{fig:limb_count_spec_res}
\end{figure}

\Figref{limb_count_spec_res} shows a scatter plot of the two metrics, each
point corresponding to an energy bin in the count spectrum
in~\figref{limb_count_spec}.
There are two striking features: (i) a nearly horizontal branch corresponding
to the energy bins below $\sim 1$~GeV, where the statistics are large and the
curvature of the count spectrum makes the quadratic fits less accurate and
(ii) a nearly vertical branch, corresponding to the high-energy points, whose
relatively larger fractional deviations (up to $\sim 10\%$) are just due to
statistical fluctuations.
We note for completeness that there are no points farther than $5\sigma$ from
the fit value above 2~GeV, which implies that we do not expect to be able
to see artificial spectral features above this energy in any practical
case. At lower energies the high counting statistics of the Earth limb emission
allows placing an upper limit on possible narrow (i.e., less than about twice
the energy resolution) spurious features at the level of a few percent.
This method for setting limits on spurious line features is decreasingly
sensitive to broader features, which are more likely to arise from correlated
errors in~\aeff\ (see \secref{subsubsec:aeff_point_to_point}).

\subsection{Comparison of \psix\ and \pseven}\label{subsec:EDisp_comparison_p6}

Though the underlying energy reconstruction algorithms are exactly the same,
\pseven\ differs from \psix\ in (i) the new un-biasing stage described in 
\secref{subsubsec:cal_recon} and (ii) the fact that the LH estimator
is no longer used in the event-level analysis for the final energy assignment
(see \secref{subsubsec:energy_analysis}).

\begin{figure}[htbp]
  \centering\includegraphics[width=\onecolfigwidth]{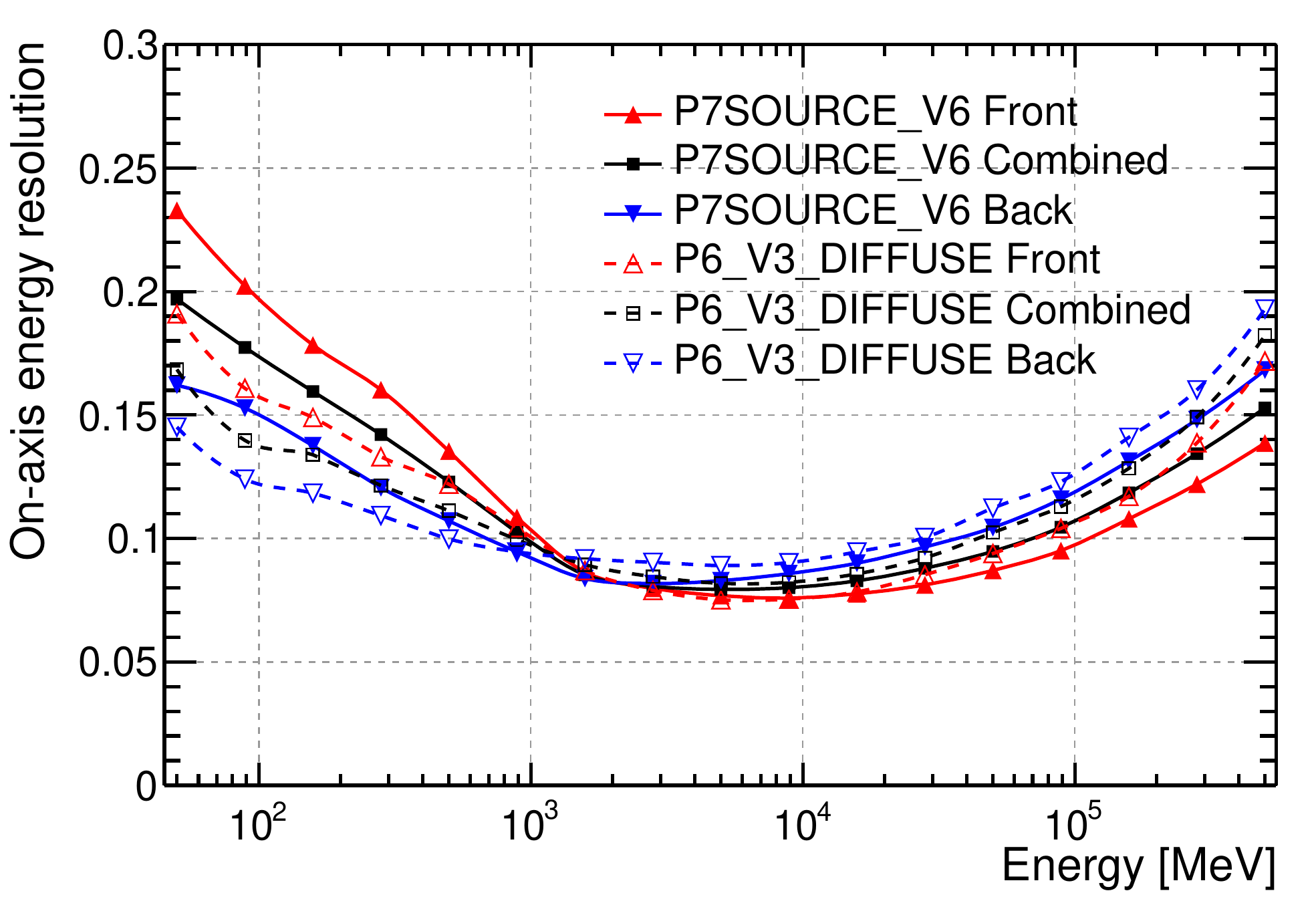}
  \caption{Comparison of the on-axis energy resolutions of the
    \irf{P7SOURCE\_V6} and \irf{P6\_V3\_DIFFUSE} IRFs.}
  \label{fig:compare_edisp_p6}
\end{figure}

\Figref{compare_edisp_p6} shows that the energy resolution for the
\irf{P7SOURCE\_V6} event class is quite comparable to that of
\irf{P6\_V3\_DIFFUSE} over most of the LAT energy range. The most
noticeable difference, namely a slight worsening below 1~GeV (and especially
below 100~MeV), is a small trade-off for the much
greater \pseven\ low-energy acceptance. On the other hand \pseven\ has uniformly
better energy resolution than \psix\ in the energy range above 10 GeV.

\clearpage

\section{PERFORMANCE FOR HIGH LEVEL SCIENCE ANALYSIS}\label{sec:perf}

We report here the high-level science analysis performance of the LAT for the
\pseven\ event selections.

\subsection{Point Source Sensitivity}\label{subsec:perf_sensitivity}

A detailed description of the point-source sensitivity of the LAT is given
in~\citet{REF:2010.1FGL} and \citet{REF:2011.2FGL}. In particular, following
the procedure described in \citet{REF:2010.1FGL}, a semi-analytical estimate 
of the LAT sensitivity for point sources can be calculated for the \pseven\
event analysis.
Here we calculate the LAT sensitivity to a point source for the
\irf{P7SOURCE\_V6} IRFs---as described in \citet{REF:2010.1FGL} for
\irf{P6\_V3}---under the following assumptions:
\begin{fermiitemize}
\item power-law spectrum with $\Gamma = 2$;
\item diffuse Galactic emission as described in the 2-year Galactic diffuse
  model, publicly distributed as \filename{gal\_2yearp7v6\_v0.fits};
\item diffuse isotropic background as described in the 2-year template,
  publicly distributed as \filename{iso\_p7v6source.txt};
\item no confusion with nearby sources;
\item exposure calculated for the first 3 years of nominal science operations.
\end{fermiitemize}

A map of the flux limit is shown in \figref{perf_sensmap3y}.  Away from the
Galactic plane the sensitivity is rather uniform. In reality, the presence
of bright point sources will affect the flux limit and a dedicated analysis is
recommended to evaluate the limit for any particular circumstance. 

\twopanel{htbp}{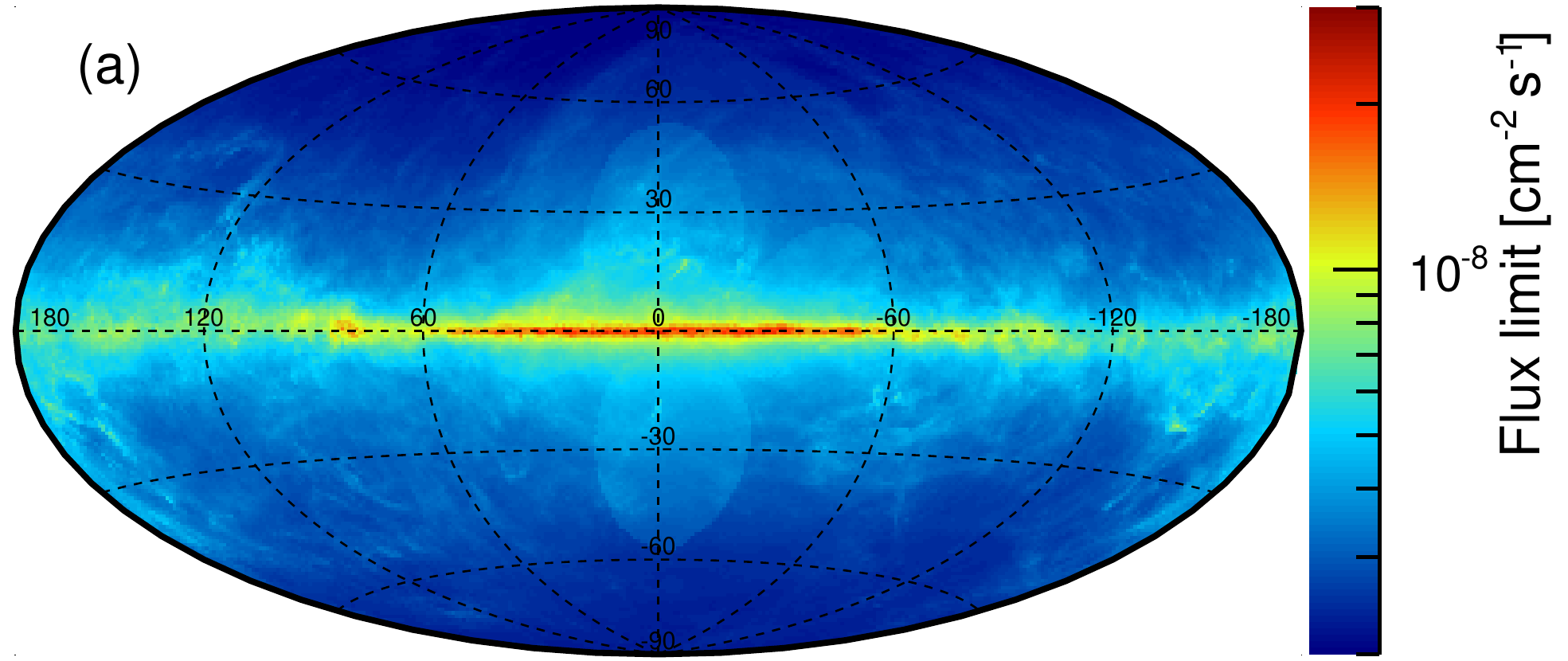}{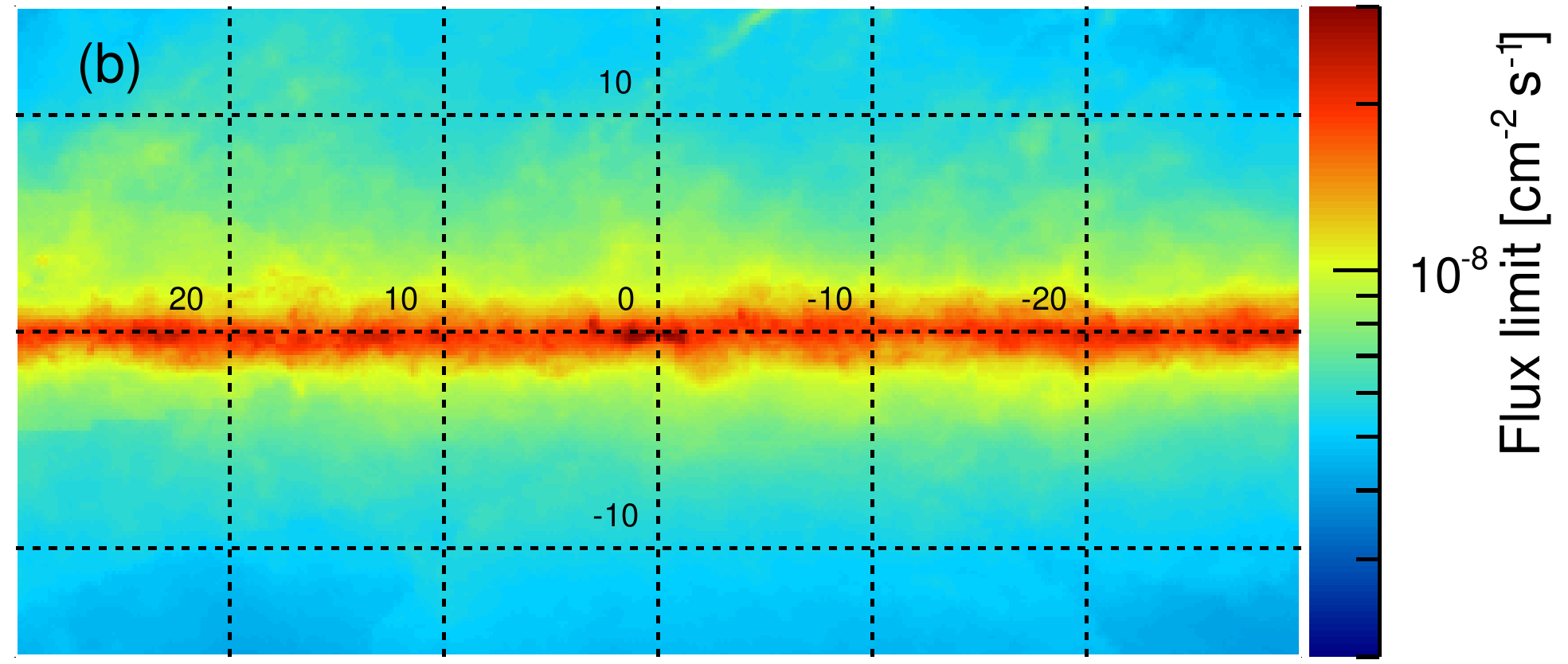}{
  \caption{Flux above 100~MeV required for 5$\sigma$ sensitivity for the
    \irf{P7SOURCE\_V6} event class for a point source with power-law spectrum
    with index $\Gamma = 2$. The calculation assumes a 3-year exposure.
    The entire sky (a) and a zoom on the Galactic center (b) are shown.}
 \label{fig:perf_sensmap3y}
}

\begin{figure}[!ht]
 \centering
  \includegraphics[width=\onecolfigwidth]{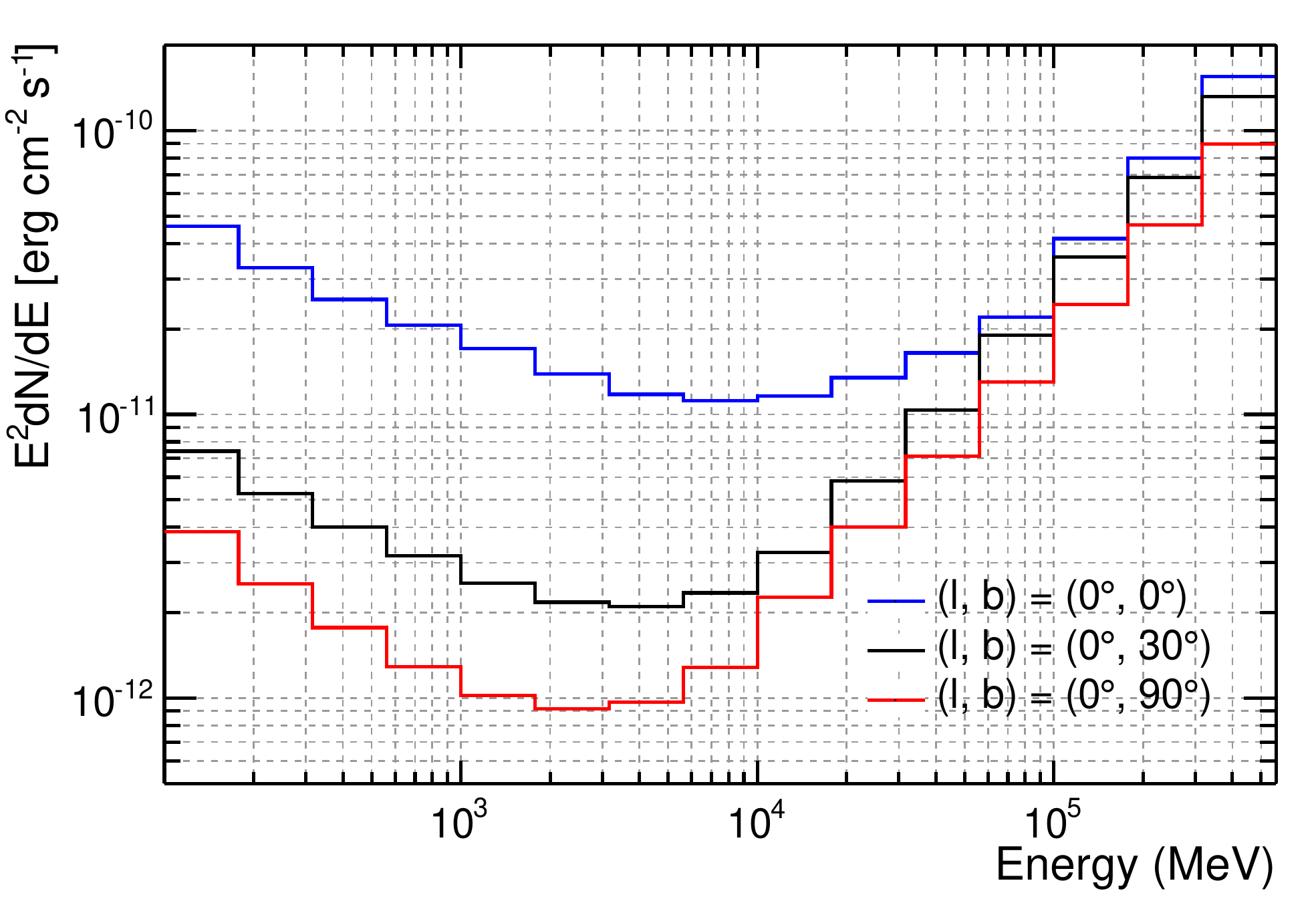}
  \caption{Differential sensitivity for class \irf{P7SOURCE\_V6} for a point
    source; calculation for 3-year exposure, 4 bins per energy decade. Requirements
    are 5$\sigma$ sensitivity and at least 10 counts per bin. The sensitivity is
    calculated at three locations in the sky: at the Galactic pole, at an
    intermediate latitude and on the Galactic plane.}
 \label{fig:perf_diffsens3y}
\end{figure}

In \figref{perf_diffsens3y} we show the corresponding differential sensitivity curves.
The curves show the flux limits for narrow energy ranges and illustrate the
sensitivity for spectral measurements as a function of energy. We require that
in each energy band the source be bright enough to have $5\sigma$ detection
and cause at least 10 \gammaRays\ to be collected.

Finally, we stress that along the Galactic plane the uncertainties of the Galactic diffuse
emission can affect the detection significance of a source, and structured
residuals in the Galactic diffuse model can be mischaracterized as point sources. Further
discussion can be found in \fermi-LAT point source catalogs and 
papers about the Galactic diffuse emission \citep{REF:2010.1FGL,REF:2011.2FGL,REF:Diffuse2:2012}.

\subsection{Point Source Localization}\label{subsec:perf_localization}

The performance of the LAT for the localization of a point source at the detection
threshold also can be evaluated with a semi-analytical approach. The estimated 2-year
performance \citep[see][]{REF:2011.2FGL} indicates that the 95\% localization radius for 
an isolated point source detected at $5\sigma$ significance at high Galactic latitude 
ranges from $\sim 0.1^\circ$ for a hard spectral index ($\Gamma = 1.5$) to
$\sim 0.3^\circ$ for a soft spectral index ($\Gamma = 3.0$).

\begin{figure}[!ht]
  \centering
  \includegraphics[width=\onecolfigwidth]{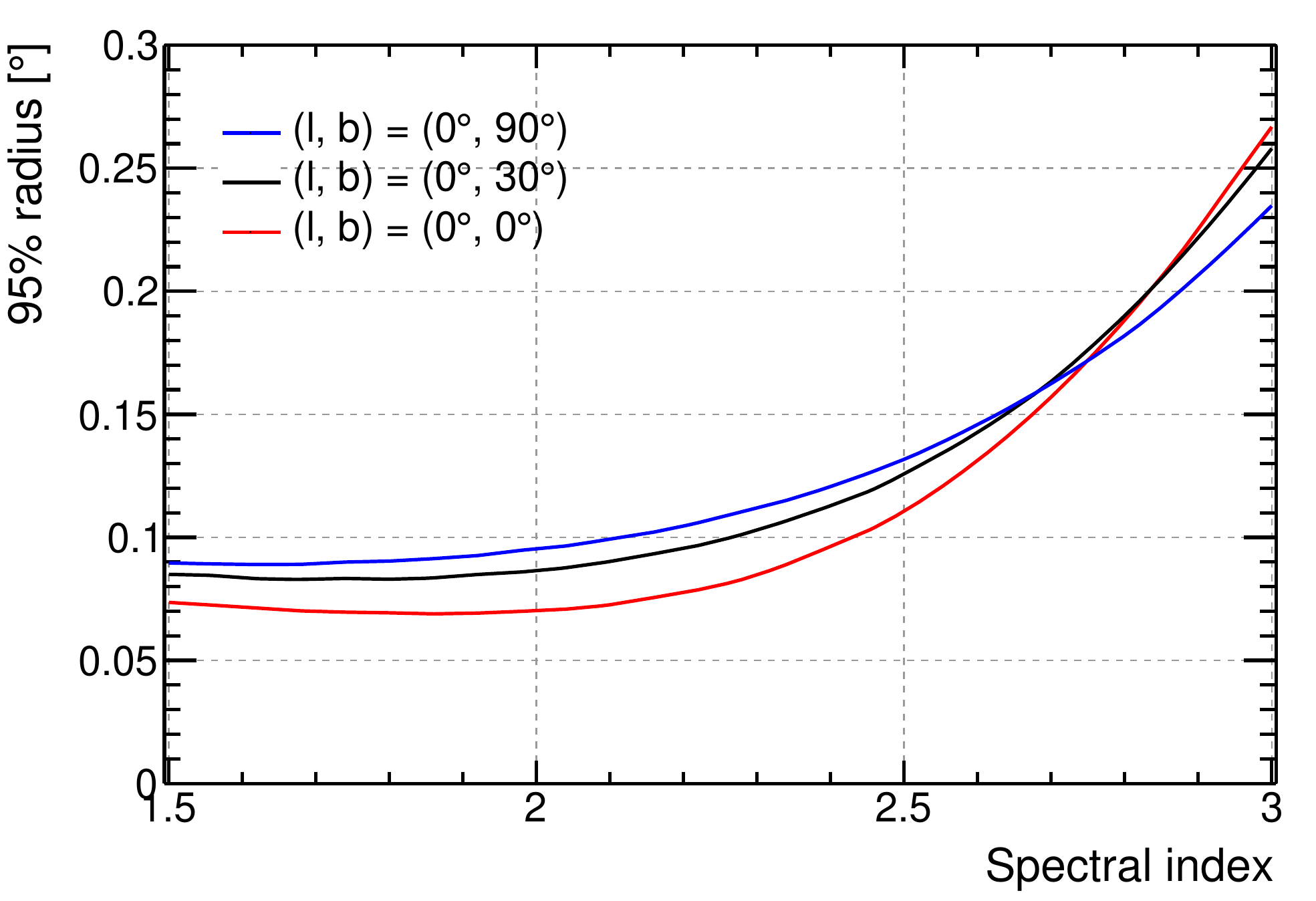}
  \caption{95\% localization radius for class \irf{P6SOURCE\_V6} 
    for a point source at the detection threshold with power-law spectrum, as a function of the spectral
    index (3-year exposure for different locations in the sky).  For sources in the Galactic plane 
    detected at a given significance, the localization performance for those sources which are softer than
    the Galactic diffuse background (which consists of spectral components $\Gamma \in [2.1,2.7]$)
    is significantly degraded with respect to harder sources.}
 \label{fig:perf_psloc}
\end{figure}

Scaling the source location determinations to more intense background
levels is not straightforward.   First of all, localization regions are usually elliptical in shape, as
described in \citet{REF:2011.2FGL}. Secondarily, as the astrophysical
background increases, e.g., in the Galactic plane, for a given value of the
spectral index a higher flux is needed to reach the detection threshold.  As a 
consequence the size of the localization region, which is more sensitive to the number 
of high-energy \gammaRays\ than the point source sensitivity, will be
smaller within the Galactic plane. This apparently counter-intuitive
result is shown in \figref{perf_psloc}, where the 95\% localization radius for
a power-law source is shown as a function of the spectral index for three
different locations.  The improvement of the localization radius with
time (\figref{perf_psloc}) is of the order of 10\% or less going from a 3-year to a 5-year exposure.   

\citet{REF:2011.2FGL} compared the positions and error regions of \gammaRayHyph\ sources to the 
positions of associated multiwavelength counterparts and empirically corrected the localization 
uncertainties by multiplying the error region by a 1.1 scale factor and adding $0.005^\circ$ in 
quadrature to the 95\% error ellipse axes.  Although the uncertainties of the Galactic diffuse
emission are significantly larger along the Galactic plane, we did not observe any systematic increase in the
localization offsets with respect to associated multiwavelength counterparts in the Galactic plane.

\subsection{Flux and Spectral Measurements}\label{subsec:perf_fluxAndIndex}

The precision to which we can measure the flux and spectral index of any
source depends primarily on the counting statistics, which in turn depends on
the flux of the source. However, since the width of the PSF decreases with energy,
high-energy \gammaRays\ contribute more to source detection, so given similar energy
fluxes, we can measure harder sources more precisely. Finally, the
flux of the source relative to nearby diffuse backgrounds and other nearby sources will
also limit the precision of the parameter measurements in a likelihood analysis.
\Figref{perf_syst2FGL_v _EFlux} shows the dependence of the statistical 
uncertainties of spectral index $\Gamma$ and the integral energy flux
between 100~MeV and 100~GeV ($S_{25}$) for the sources in the 2FGL
catalog \citep{REF:2011.2FGL}.  In both cases the minimum statistical uncertainty decreases 
as roughly $S_{25}^{-0.6}$.

\twopanel{ht!}{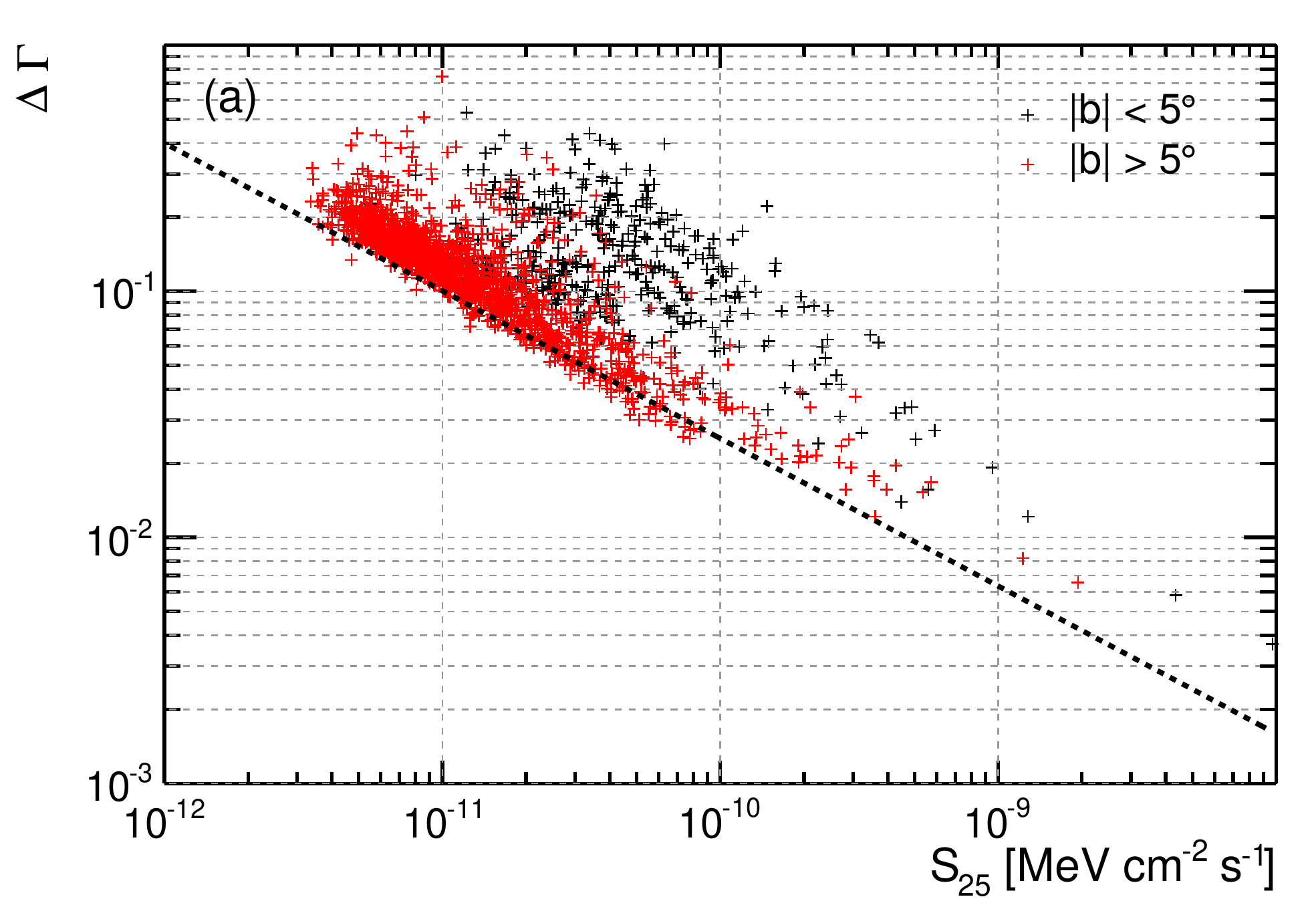}{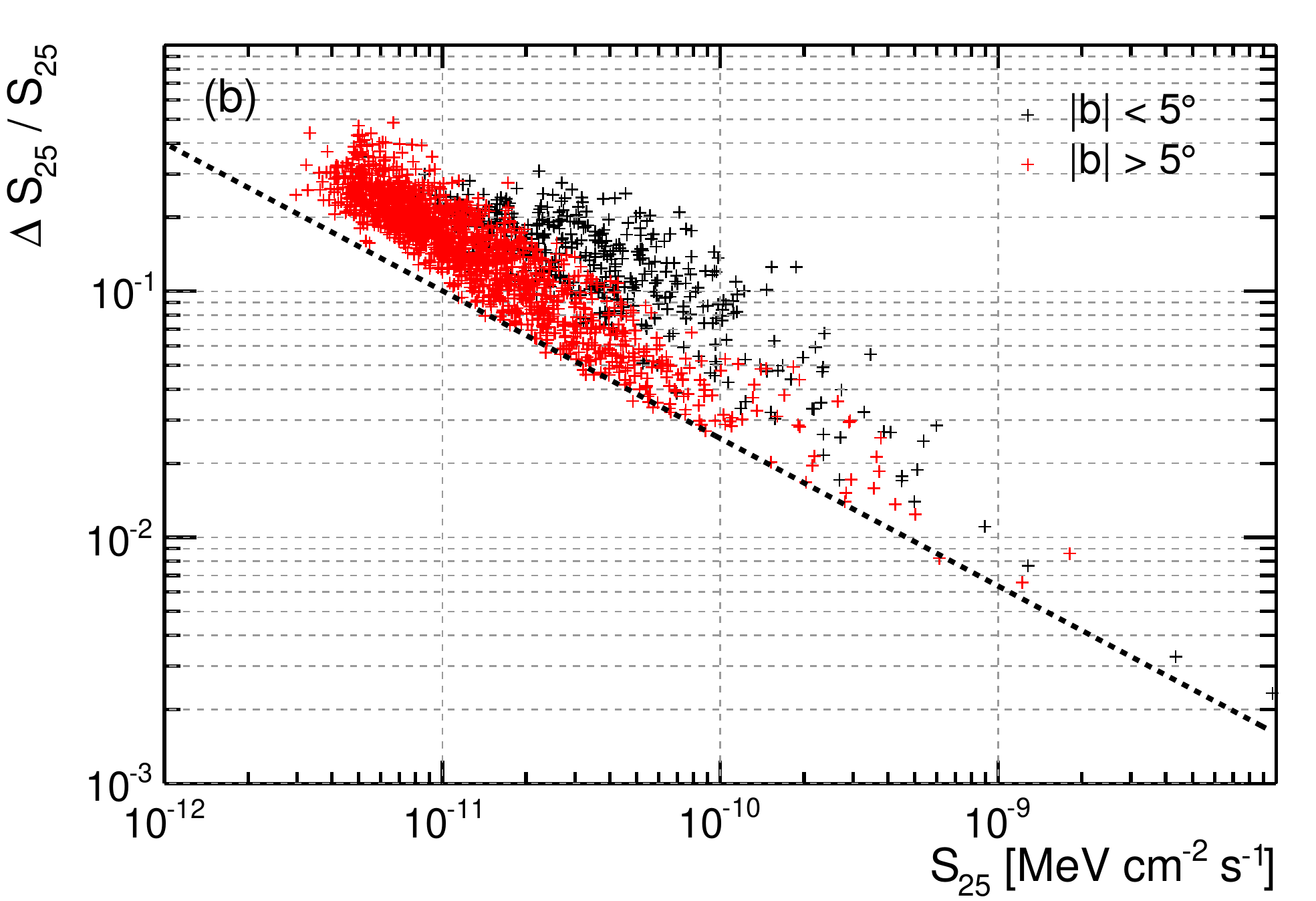}{
  \caption{Statistical uncertainties on spectral index $\Gamma$ (a) and
    integral energy flux above 100~MeV $S_{25}$ (b) as a function of $S_{25}$ 
    for all sources in the 2FGL catalog \citep{REF:2011.2FGL}.  
    The dashed lines show a $S_{25}^{-0.6}$ dependence for comparison.}
  \label{fig:perf_syst2FGL_v _EFlux}
}

For comparison, we combine the systematic uncertainty estimates based on the
bracketing tables for the effective area (\secref{subsec:Aeff_highLevel}), 
the PSF (\secref{subsec:PSF_highLevel}), and the effect of ignoring the energy
dispersion (\secref{subsec:EDisp_highLevelAnalysis}) for the two sources we
studied in detail (B2~1520+31 and PG~1553+113).  Since we have not
seen any evidence of correlated biases in the different terms of the
IRFs (see, for example \secref{subsec:EDisp_psf_correl}) we simply combine 
the uncertainties in quadrature, as summarize in \tabref{fluxAndIndexErrors}).

\begin{table}[htb]
  \centering
  \begin{tabular}{llllll}
  \hline 
  Quantity & \aeff\ & PSF & \multicolumn{2}{c}{Energy} & Total \\
    & & & Dispersion & Scale \\
    \hline
    \hline
  $\delta \Gamma_{\rm{PG 1553+113}}$ & 0.05 & 0.02 & 0.01 & - & 0.05 \\
  $\delta \Gamma_{\rm{B2 1520+31}}$ & 0.09 & 0.07 & 0.04 & - & 0.12 \\
  $\frac{\delta S_{25}}{S_{25}}_{\rm{PG 1553+113}}$ & $10\%$ & $5\%$ & $1\%$ & $2\%$ & $11\%$ \\
  $\frac{\delta S_{25}}{S_{25}}_{\rm{B2 1520+31}}$ & $8\%$ & $6\%$ & $2\%$ & $4\%$ & $11\%$ \\
  \hline
  \end{tabular}
  \caption{Rough estimates of the magnitude of the effects of various sources
    of systematic errors on the integral energy flux and spectral index 
    of B2~1520+31 and PG~1553+113.}
  \label{tab:fluxAndIndexErrors}
\end{table}

\Figref{perf_syst2FGL} shows the distributions of the statistical uncertainties
on $\Gamma$ and $S_{25}$ from the 2FGL catalog, along with our estimates of the
typical systematic uncertainties for hard and soft sources based on our
analysis of B2~1520+31 and PG~1553+113. For the majority of sources in
the 2FGL catalog the measurement precision is still limited by the statistical
uncertainties, though for the brightest sources the systematic uncertainties
dominate. For both $\Gamma$ and $S_{25}$ the transition from statistical
limitation to systematic limitation occurs for
$S_{25} \sim 10^{-11}$~erg~cm$^{-2}$~s$^{-1}$.

\twopanel{ht!}{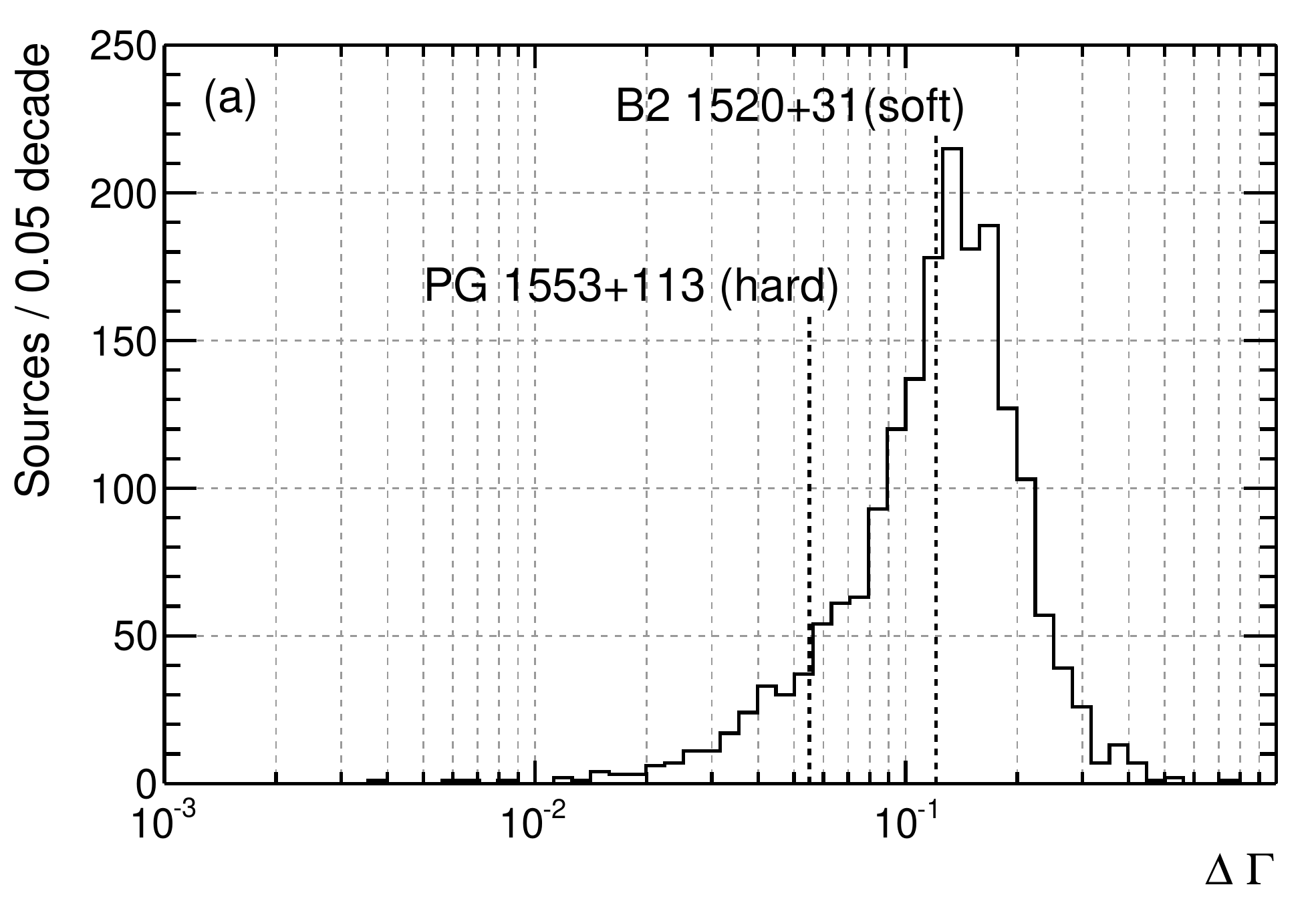}{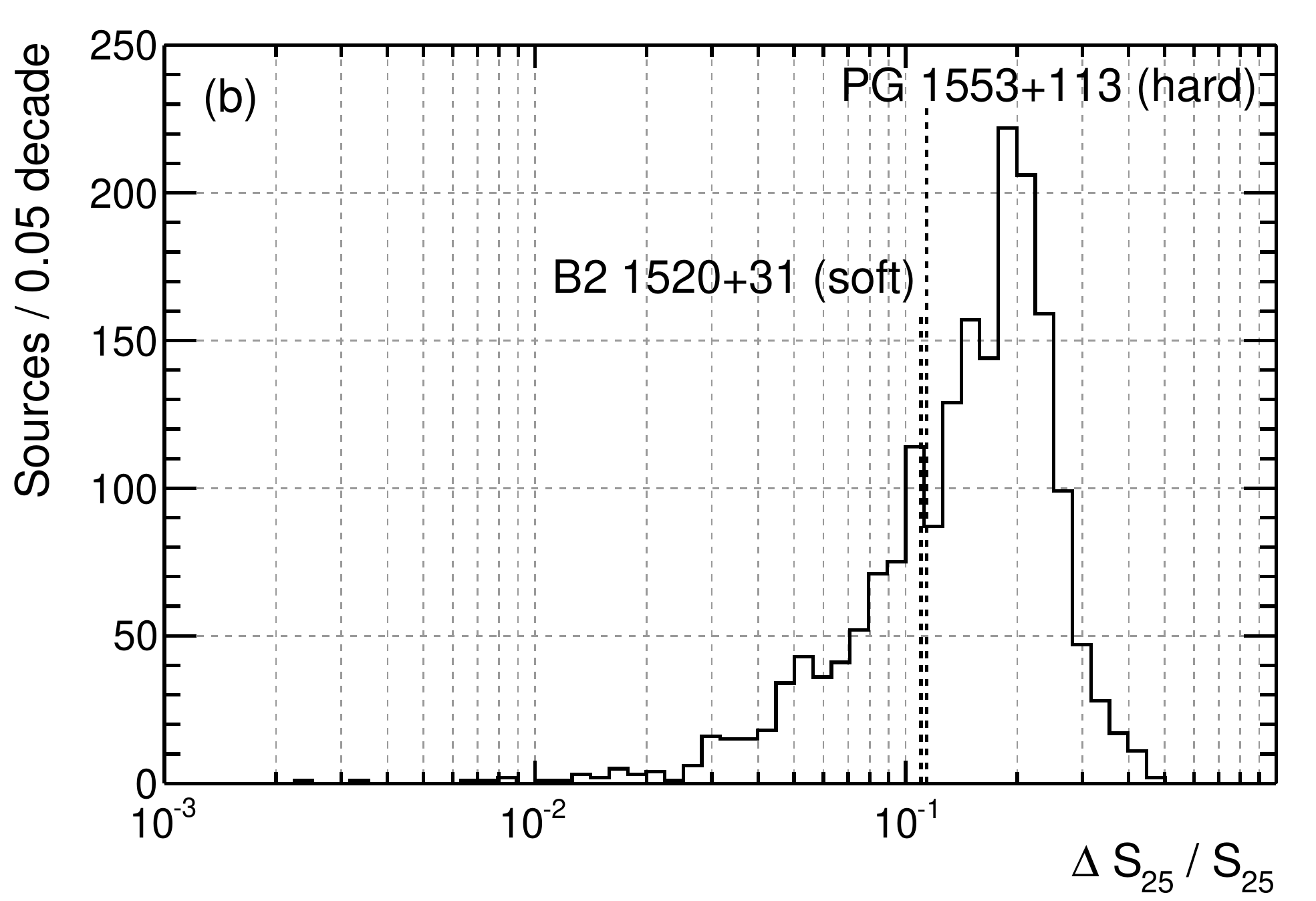}{
  \caption{Statistical uncertainties on $\Gamma$ (a) and $S_{25}$ (b) for all
    2FGL sources. For comparison the estimated systematic uncertainties for
    B2~1520+31 and PG~1553+113 are also shown.}
  \label{fig:perf_syst2FGL}
}

We emphasize that the results in this section were derived using 
bracketing IRFs designed to maximize the variation of the fit results 
for the particular sources under study.  As such, they represent the systematic
uncertainties of measurements on any single source.  Since measurements of different
sources share the same IRFs and associated uncertainties, any relative comparison 
between measurements is significantly more precise.  

Finally, along the Galactic plane the uncertainties of the Galactic diffuse
emission can affect measurements of source parameters and spectral indices. As with the 
question of how these uncertainties affect the LAT point source detection sensitivity, detailed 
discussion is beyond the scope of this paper, but can be found in \citet{REF:2010.1FGL}, 
\citet{REF:2011.2FGL}, and \citet{REF:Diffuse2:2012}.

\subsection{Variability}\label{subsec:perf_variability}

The LAT's ability to measure the variability of any given source
depends on the characteristics of both the source and the
astrophysical backgrounds. 

We can use the detection significances of sources in the 2FGL catalog to derive
a rough estimate of the time-scales at which we can probe the variability of
those sources. For detecting variability, the worst-case scenario is a steady
source, so we ask how long would be required on average to detect any source at a
particular threshold, assuming steady emission. Specifically, we can construct a
metric $\tau_{2\sigma}$, the time required to achieve $2\sigma$ detection of a
source, assuming steady emission at the average 2FGL level, which can be
expressed in terms of the significance of the source in the 2FGL catalog
($\sigma_{\rm 2FGL}$) and the amount of data used to construct the catalog
($\sim 730$~days):
\begin{equation}
  \tau_{2\sigma} \sim 4 \sigma_{\rm 2FGL}^{-2} 730{\rm\ days}.
\end{equation}  
\Figref{perf_DaysTo2Sigma} shows this estimator for all 2FGL sources, we emphasize that 
we are sensitive to more significant flux changes on timescales shorter than $\tau_{2\sigma}$.

\twopanel{ht!}{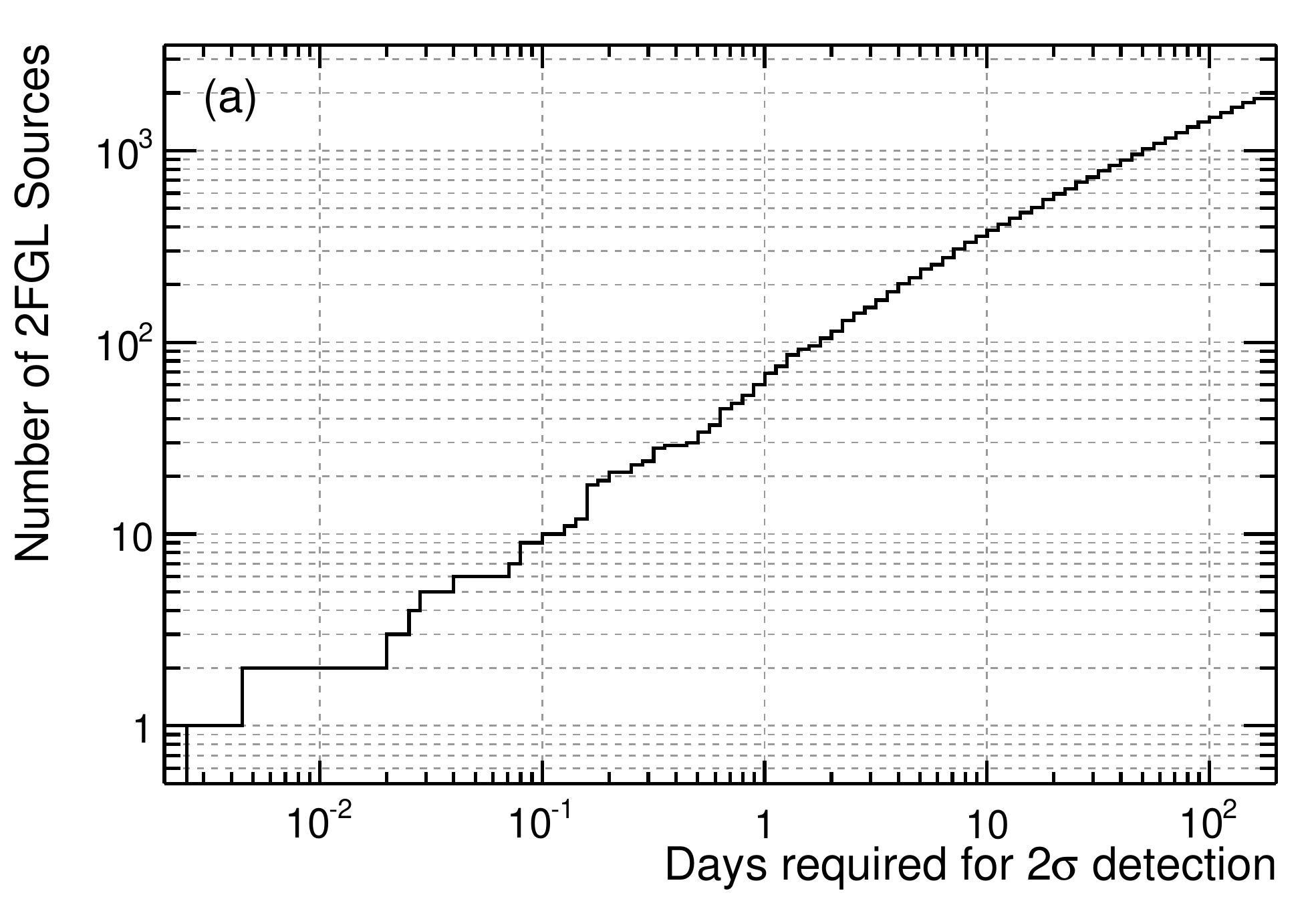}{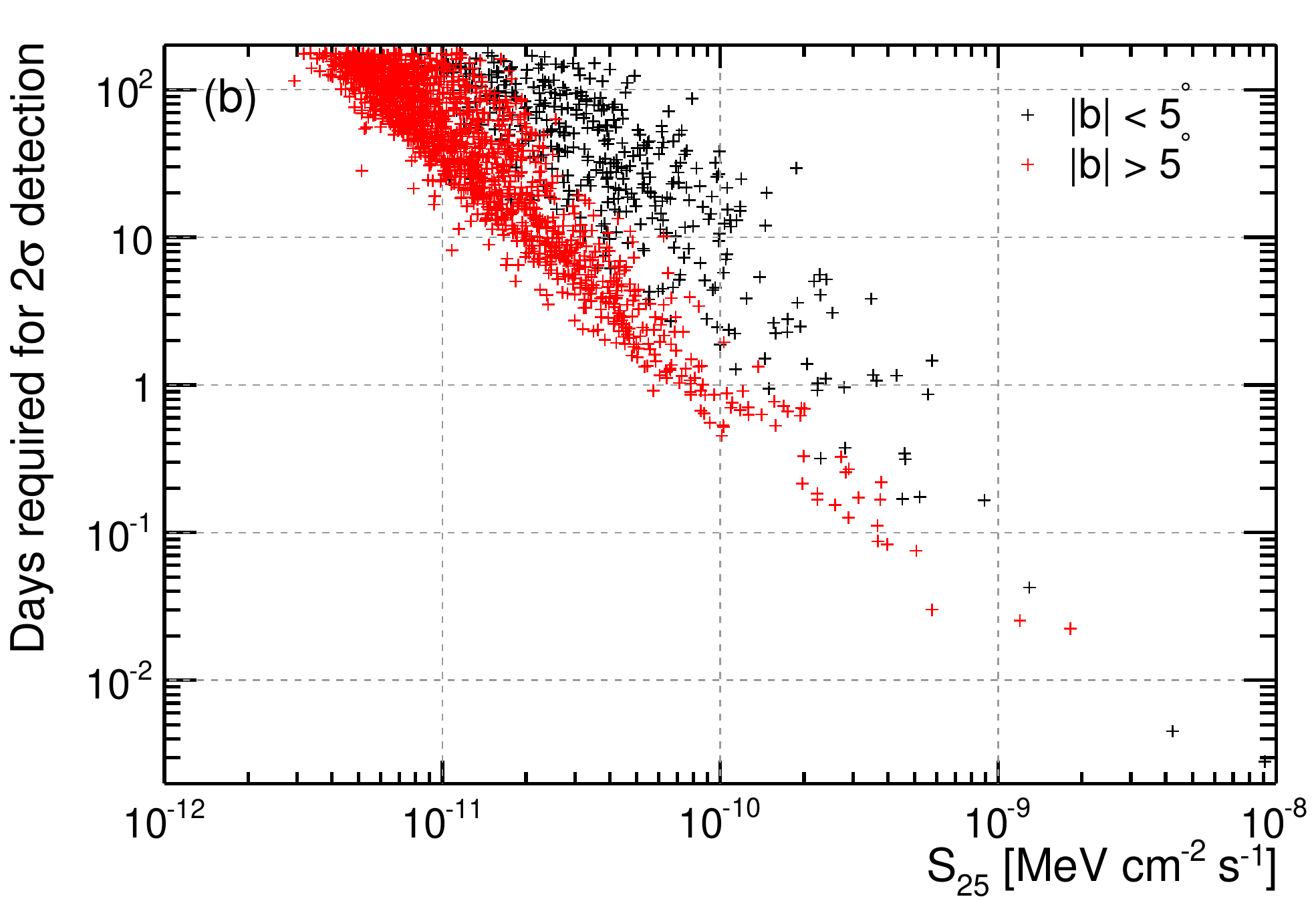}{
  \caption{LAT sensitivity to variability of all 2FGL catalog sources: panel
    (a) shows the cumulative distribution of the estimator $\tau_{2\sigma}$, an
    estimate of the time needed to detect a steady source with $2\sigma$
    significance; panel (b) shows $\tau_{2\sigma}$ as a function of the
    integral energy flux between 100~MeV and 100~GeV ($S_{25}$) for both low- and high- Galactic latitude
    sources.}
  \label{fig:perf_DaysTo2Sigma}
}

In \secref{subsubsec:Aeff_variability} and \secref{subsec:psf_variability} we
examined how instrumental uncertainties in the \aeff\ and PSF can
change over time, and could potentially induce artificial variability in the
measurement of point source fluxes. 

More specifically, in the case of the \aeff-related variability we
showed that the phase-gated counts excess was extremely stable, with the
Fourier transform of $\delta n = n - \tilde{n}$ showing only a single small
feature consistent with the 53.4-day orbital precession period. However,
since that analysis did not treat the PSF, nor allow for the effect of changing 
statistics from differing exposures on the Fourier transform of $\delta n$, we
could not use it to predict the level of instrumentally induced variability we
might measure in a likelihood-based analysis using the \stools.

Similarly, our analysis of PSF-related variability estimated how much the
containment as defined by the $\theta$-averaged $R_{68}$ and $R_{95}$ might vary over
a precession period, but did not attempt to quantify the effect on a likelihood
based analysis using the \stools.  In order to quantify these effects we have studied the Vela and Geminga
\citep[PSR J0633+1746]{REF:Geminga:2010} pulsars in 12-hour time bins with the \stools\ (version 09-26-02)
unbinned likelihood analysis and the following analysis parameters:
\begin{fermienumerate}
\item we started with the \gammaRayHyph\ and time interval selection criteria
  as for the Vela calibration sample (see \secref{subsec:LAT_methods});
  for the Geminga sample we used the same $15^{\circ}$ radius for the \roi;
\item tightened the zenith angle requirement to $\theta_{z} < 95^\circ$ and used all
  \gammaRays\ with energies $E > 70$~MeV;
\item included all 2FGL sources within $20^{\circ}$ in our likelihood model, with
  same spectral parametrizations as were used in the 2FGL catalog;  
\item included models of the Galactic diffuse emission
  (\filename{ring\_2year\_P76\_v0.fits}) and the isotropic diffuse 
  emission (\filename{isotrop\_2year\_P76\_source\_v1.txt});
\item held fixed all of the parameters of the model except for the
  normalization of the flux of the pulsar under test and isotropic diffuse emission.
\end{fermienumerate}

We performed this study with both the \irf{P7SOURCE\_V6} and
\irf{P7SOURCE\_V6MC} IRFs, and included the $\phi$~dependence of the \aeff\ 
in the exposure calculations. In all cases we measured the
pulsars to be almost consistent with having constant fluxes during the
first 700~days of the mission.  Specifically, for Geminga
$P(\chi^{2}) = 0.04$, while for Vela $P(\chi^{2}) = 6.3 \times 10^{-5}$, but adding
$2\%$ error in quadrature brought the $P(\chi^{2})$ up to 0.17.  
Furthermore, no peak was visible in the Fourier spectra of the fitted
fluxes between 12~hours and 700~days.
On the other hand, the fitted normalization for the isotropic diffuse component
was not consistent with constant, and in both cases the Fourier analysis showed
a significant peak at the $53.4$~day orbital precession period
(see \parenfigref{Vela_FittedFlux} for results of the Fourier analysis of the Vela
flux and associated isotropic normalization).
The variability in the normalization of isotropic background is caused by the
CR background leakage not having the same $\theta$~dependence as the
\gammaRayHyph\ effective area, so that exposure calculations used to 
predict the expected counts from the isotropic background template suffer
slightly different biases during different phases of the orbital precession
period. As we have shown here, this can be handled by leaving the
normalization of the isotropic template free, provided that the source
dominates the nearby isotropic background.

\twopanel{htb!}{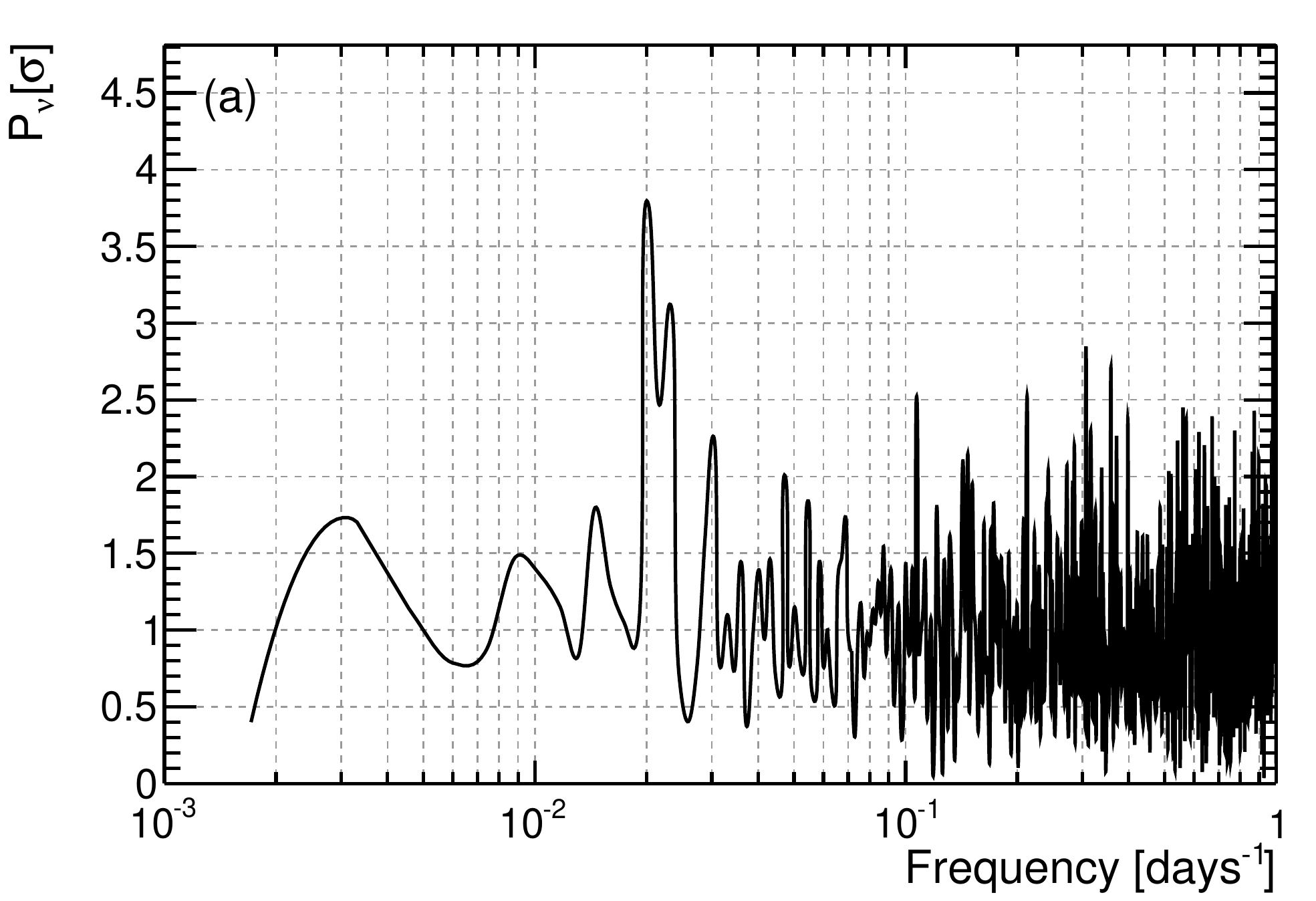}{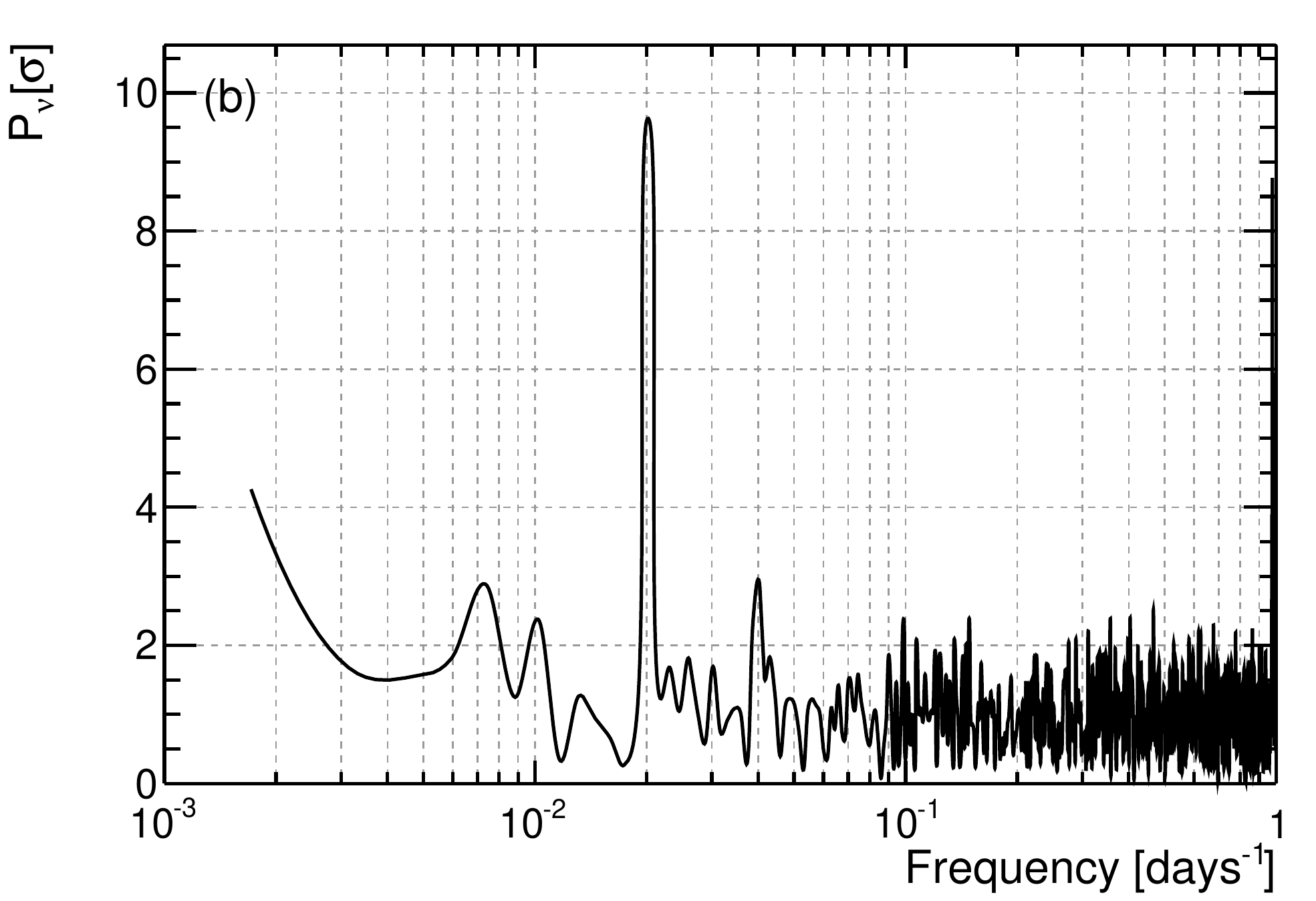}{
  \caption{Fourier transforms of the normalized residuals of the Vela integral
    counts flux (a) and the isotropic normalization factor (b).  Note
    that the figures are normalized and the vertical scale is expressed
    in units of the statistical uncertainty.}
  \label{fig:Vela_FittedFlux}
}

\NEWTEXT{By comparing results obtained on the Vela pulsar with the
  \irf{P7SOURCE\_V6MC} IRFs (which include the $\theta$ dependence of the PSF) 
to results obtained with the \irf{P7SOURCE\_V6} IRFs (which do not),
we estimate that ignoring the $\theta$ dependence of the PSF 
causes a $\sim 4\%$ RMS variation of the flux of Vela when fitted in 12-hour time intervals.  The effect decreases on longer time scales.
Since, as stated in \secref{subsec:PSF_MonteCarlo}, the PSF depends much more strongly on 
$\theta$ than on $\phi$, we neglect the $\phi$ dependence of the PSF as a potential source of instrument-induced 
variability.}

\NEWTEXT{In summary, when the angular dependence of the IRFs is
  properly accounted for, and time variations of the CR-background
leakage are absorbed into the normalization of the isotropic component in the likelihood fit, we find the level of instrument-induced variability 
to be small ($< 5\%$) for all time scales between 12 hours and 2 years.  Since the LAT boresight follows very similar paths
across the sky every two orbits, resulting in similar observing profiles, we believe these systematic uncertainties are also applicable at the 
3-hour time scale.  On the other hand, for observations that do not consist of a complete orbit, and which are therefore more
susceptible to biases in the IRFs at specific incidence angles, the ($\sim 10\%$) uncertainties in the effective area quoted
in \secref{subsubsec:overall_aeff_errors} are more applicable.}

\NEWTEXT{Finally, we note that the slow change in the light yield of the CsI logs of the CAL, and the corresponding change in the energy scale described 
in \secref{subsec:EDisp_xtalCalib} can induce slow shifts in observed fluxes by shifting spectra as a function
of time.  The effect depends on the spectrum of the source under study, as well as on the energy band used, and differs for
counts fluxes and energy fluxes, as summarized in \Tabref{eScale_bracket}.}

\clearpage

\section{SUMMARY}

Since the beginning of science operations in 2008 August, the \fermi-LAT has opened a 
new window on the \gammaRayHyph\ sky, providing the science community with an unrivaled 
set of \gammaRayHyph\ data.   The LAT data set covers the entire sky
over the energy range from $\sim 20$~MeV to greater than $300$~GeV every 3 hours.

During these first years of the mission, the LAT Collaboration has studied the on-orbit performance 
of the LAT, the character of the flight data from the LAT, and the optimization 
of event selections and IRFs for science.  This has led to significant improvements in the event analysis (in particular
the \pseven\ version of the event analysis described in \secref{subsec:event_CT_analysis})
and in the science analysis, and to important clarifications of the systematic uncertainties at each
level of the analysis.  In addition, the procedures we designed and developed
to perform these investigations can be reiterated with minimum modifications
whenever a new event reconstruction and classification is adopted.

We have shown that the LAT has performed extremely well: the data are
of uniformly high quality, we have lost less than 0.5\% of potential
observing time (i.e, when \fermi\ is outside the SAA) to unplanned
outages and instrumental issues.  Furthermore, the LAT response has been 
extremely stable: we are able to use a single set of IRFs for the entire mission to 
date, with any variations in performance contributing negligibly to
systematic uncertainties (\secref{sec:LAT}).  

We have also confirmed that the LAT data reduction and analysis tools
have performed very well.   The combination of a configurable hardware trigger, an
on-board event filter, and ground based event reconstruction and analysis have
allowed us to reduce the CR contamination from 5 - 10~kHz passing
through the instrument to $\sim 1$~Hz in the \irf{P7SOURCE} event
sample (and $\sim 0.1$~Hz in the \irf{P7CLEAN} event sample, \secref{sec:event}) 
while maintaining a peak acceptance of over 
$2~{\rm m}^2\ {\rm sr}$ in the \irf{P7SOURCE} sample
(and $> 1.75~{\rm m}^2\ {\rm sr}$ in the
\irf{P7CLEAN} sample) (\secref{sec:Aeff}).   

We have validated the quality of our MC simulations and found only one
significant discrepancy between the MC simulations and the flight
data.  Specifically, limitations in the pre-launch calibration algorithm of the CAL light
asymmetry produced calibration constants that did not match the MC
predictions of the spatial resolution performance of the CAL.  However, we have quantified the effects of this
discrepancy on the \aeff\  (\secref{subsec:Aeff_flightAEff}) and PSF (\secref{subsec:PSF_onorbit}) and are
currently assessing the improvement of the PSF for data that were
reprocessed with improved calibration constants.

\pseven\ data have been available for public analysis since 2011 August and provide 
substantial improvements over \psix, primarily due to greatly increased \aeff\ below $\sim 300$~MeV 
(\secref{subsec:Aeff_comparison_p6}) and improved modeling of the IRFs.  \NEWTEXT{Coupled with improved understanding 
of the effects of energy dispersion (\S~7) this increase in effective area 
is opening a new window for analysis of LAT data for energies below $100$~MeV.} \Tabref{IRFcomponents} lists the 
features of the components included in each of the IRF sets recommended for analysis with both \psix\ and \pseven. 

(\secref{sec:EDisp})

\begin{table}[htb]
  \centering
  \begin{tabular}{lll}
    \hline
    IRF Set & $\aeff$ & PSF \\
    \hline
    \hline
    \irf{P6\_V3\_TRANSIENT} & $f_{live},\phi$ & Legacy MC \\
    \irf{P6\_V3\_DIFFUSE} & $f_{live},\phi$ & Legacy MC \\
    \irf{P6\_V3\_DATACLEAN} & $f_{live},\phi$ & Legacy MC \\
    \hline
    \irf{P6\_V11\_DIFFUSE} & $f_{live},\phi$,in-flight & in-flight \\
    \hline
    \irf{P7TRANSIENT\_V6} & $f_{live},\phi$ & in-flight \\
    \irf{P7SOURCE\_V6} & $f_{live},\phi$ & in-flight \\
    \irf{P7CLEAN\_V6} & $f_{live},\phi$ & in-flight \\
    \irf{P7ULTRACLEAN\_V6} & $f_{live},\phi$ & in-flight \\
    \hline
    \irf{P7SOURCE\_V6MC} & $f_{live},\phi$ & MC \\
    \irf{P7CLEAN\_V6MC} & $f_{live},\phi$ & MC \\
    \hline
 \end{tabular}
  \caption{IRF sets recommended for data analysis, and the features
    included in the various components.   For \aeff, $f_{live}$ and 
    $\phi$ refer to the live time fraction and $\phi$~dependence,
    respectively (\secref{subsubsec:Aeff_Livetime} and
    \secref{subsubsec:aeff_phi_dep}), 
    and ``in-flight'' refers to flight-based corrections
    (\secref{subsec:Aeff_flightAEff}).  
    For the PSF ``Legacy MC'' refers to an early version of the parametrization fit to 
    \allgamma\ samples (\secref{subsubsec:legacy_psf}), ``in-flight'' refers to the
    flight-based PSF (\secref{subsec:PSF_onorbit}), and ``MC'' refers to the more recent
    parametrization fit to \allgamma\ samples (\secref{subsubsec:psf_prescaling}).
    Finally, we have used the same parametrization of the energy dispersion for all of the
    IRF sets (\secref{subsubsec:Figure_66ting}).}
 \label{tab:IRFcomponents}
\end{table}

In addition to the caveats\footnote{\webpage{http://fermi.gsfc.nasa.gov/ssc/data/analysis/LAT\_caveats.html}} 
and the documentation\footnote{\webpage{http://fermi.gsfc.nasa.gov/ssc/data/analysis/documentation/}}
accompanying the data and the science analysis software,
we provide with this paper a reference document for all currently known
systematic issues.   \Tabref{systErrors} provides numerical estimates
of the residual uncertainties and refers to the sections of this paper where we detail procedures to estimate the 
systematics uncertainties for many analyses.   

\begin{table}[htb]
  \centering
  \begin{tabular}{lllll}
    \hline
    Quantity & \aeff\ & PSF & \multicolumn{2}{c}{Energy} \\
    & & & Dispersion & Scale \\
    \hline
    \hline
    $F_{25}$ & $\sim 8\%$ (\secref{subsec:Aeff_highLevel}) 
    & $\sim 8\%$ (\secref{subsec:PSF_highLevel}) 
    & $\sim 3\%$ (\secref{subsec:EDisp_highLevelAnalysis})
    & $+13\% -5\%$ (\secref{subsec:EDisp_highLevelAnalysis}) \\
    $S_{25}$ & $\sim 10\%$ (\secref{subsec:Aeff_highLevel}) 
    & $\sim 6\%$ (\secref{subsec:PSF_highLevel}) 
    & $\sim 2\%$ (\secref{subsec:EDisp_highLevelAnalysis})
    & $+4\% -2\%$(\secref{subsec:EDisp_highLevelAnalysis}) \\
    $\Gamma $ & $\sim 0.09$ (\secref{subsec:Aeff_highLevel}) 
    & $\sim 0.07$ (\secref{subsec:PSF_highLevel}) 
    & $\sim 0.04$ (\secref{subsec:EDisp_highLevelAnalysis}) 
    & - \\
    Variability & $\sim 3\%$ (\secref{subsec:Aeff_errors}) 
    & $\sim 3\%$ (\secref{subsec:PSF_highLevel}) 
    & - & -\\
    Localization & - 
    & $\sim 0.005^{\circ}$ (\secref{subsec:perf_localization})\tablenotemark{a}
    & - & -\\    
    \hline
  \end{tabular}
  \normalsize
  \tablenotetext{a}{See \citet{REF:2011.2FGL} for a discussion of the
    systematic uncertainties on source localization.} 
  \caption{Rough estimates of the magnitude of the effects of various sources
    of systematic errors for commonly-measured \gammaRayHyph\ source
    properties. We also provide references to the relevant sections with more
    details.}
  \label{tab:systErrors}
\end{table}

Finally, in \secref{sec:perf}, we provide details on the science performance 
we obtain with the \pseven\ event analysis and IRFs.  In particular, we provide 
estimates of the source detection sensitivity threshold (\secref{subsec:perf_sensitivity}), the source localization 
performance (\secref{subsec:perf_localization}), the expected precision and accuracy of measurements of fluxes and 
spectral indices (\secref{subsec:perf_fluxAndIndex}), and the precision and accuracy of variability
 measurements (\secref{subsec:perf_variability}).

As stated in \secref{sec:intro}, the LAT team will continue to make both major
improvements and minor refinements to many aspects of the event reconstruction,
analysis and to the associated IRFs. 
We will continue to keep the \gammaRayHyph\ astronomy community informed
of the state-of-the-art of our understanding of the LAT and issues relating to
analyzing LAT data.

\acknowledgements
The \textit{Fermi} LAT Collaboration acknowledges generous ongoing support
from a number of agencies and institutes that have supported both the
development and the operation of the LAT as well as scientific data analysis.
These include the National Aeronautics and Space Administration and the
Department of Energy in the United States, the Commissariat \`a l'Energie
Atomique and the Centre National de la Recherche Scientifique / Institut
National de Physique Nucl\'eaire et de Physique des Particules in France,
the Agenzia Spaziale Italiana and the Istituto Nazionale di Fisica Nucleare in
Italy, the Ministry of Education, Culture, Sports, Science and Technology
(MEXT), High Energy Accelerator Research Organization (KEK) and Japan Aerospace
Exploration Agency (JAXA) in Japan, and the K.~A.~Wallenberg Foundation,
the Swedish Research Council and the Swedish National Space Board in Sweden.

Additional support for science analysis during the operations phase is
gratefully acknowledged from the Istituto Nazionale di Astrofisica in Italy and
the Centre National d'\'Etudes Spatiales in France.

The Parkes radio telescope is part of the Australia Telescope, which is funded by the 
Commonwealth Government for operation as a National Facility managed by CSIRO. 
We thank our colleagues for their assistance with the radio timing observations.

\clearpage

\appendix

\section{SOLAR FLARES AND BAD TIME INTERVALS}\label{app:bti}

\begin{figure}[!hb]
  \centering
  \includegraphics[width=\onecolfigwidth]{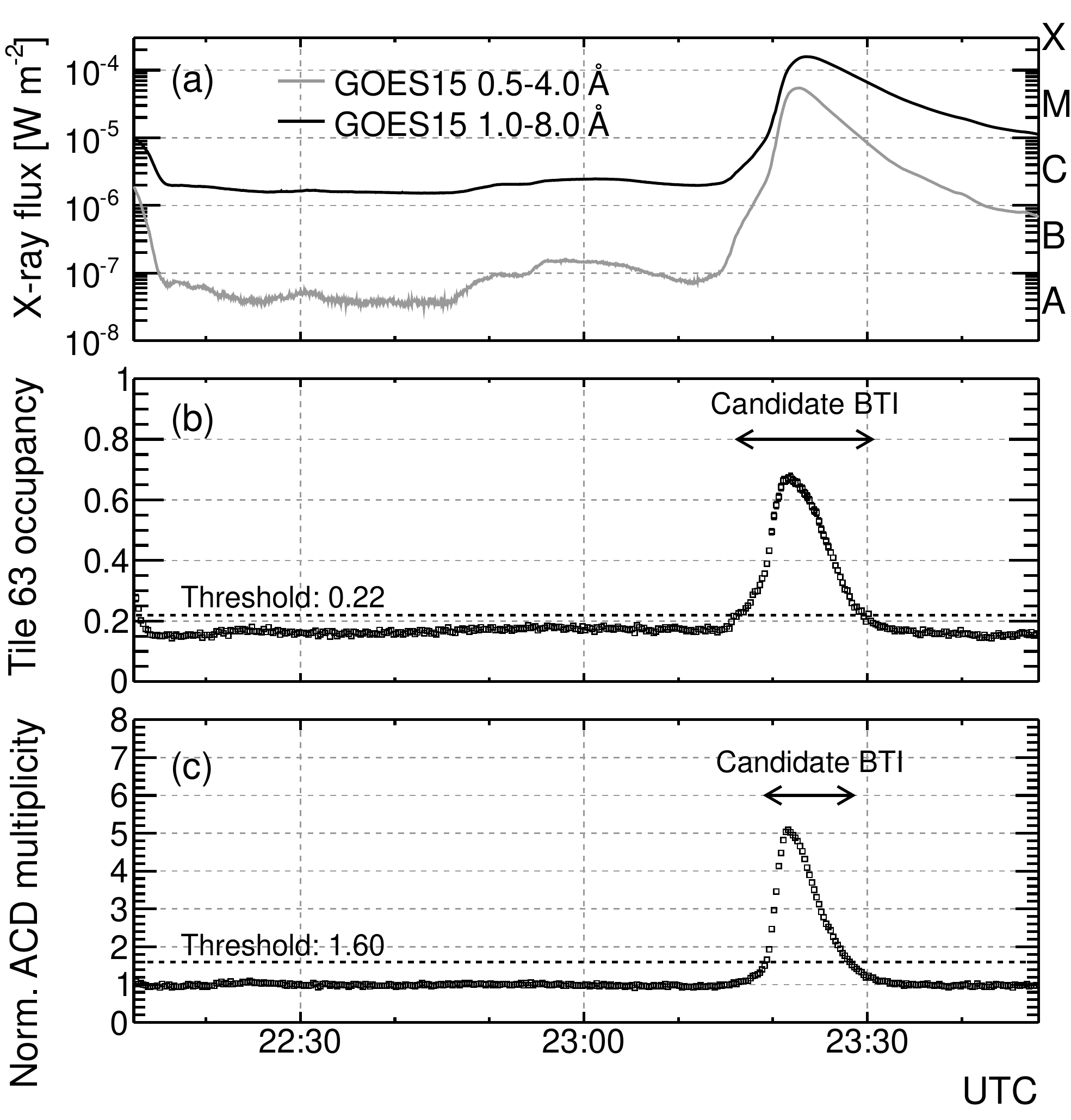}
    \caption{(a) X-ray flux for the X1.5 solar flare on 2011 March 9 measured
      by the GOES-15 satellite in the energy bands 0.5--4.0~\AA\
      (3.1--24.8~keV) and 1.0--8.0~\AA\ (1.5--12.4~keV). The latter is
      customarily used to classify solar flares based on their X-ray radiance,
      as indicated by the letters A, B, C, M, X on the right.
      The effect on two of the basic ACD quantities (see text for more details)
      is clearly visible in panels (b) and (c).}
  \label{fig:bti_seed}
\end{figure}

With the beginning of the solar activity connected with Cycle 24, in early
2009, it became clear that Solar Flares (SFs) can cause significant X-ray pile-up in the ACD,
clearly visible in many low-level ACD quantities (see \parenfigref{bti_seed}).
The on-board ACD veto electronics (see~\tabref{LAT_timeConstants}) are fast
enough that this is not causing any change in the efficiency of the trigger
and/or that of the \obfgam\ (i.e., there is essentially no loss of \gammaRays\
on-board).
However the effects of the pile-up are potentially much more severe at the
level of the event selection on the ground---where the slow signals are
used---as the additional activity in the ACD can cause \gammaRays\ to be 
misclassified as background.

The basic phenomenology is somewhat similar to the ghost-induced loss of
efficiency introduced in~\secref{subsec:LAT_DAQ} and discussed in detail
throughout the paper, the main difference being that during the most
intense SFs the effect can be large enough to make the LAT essentially
\emph{blind} to \gammaRays.
From the point of view of the data analysis the most relevant implication is
that there are time intervals in which the standard IRFs
\emph{do not accurately represent the detector} and therefore the results
of the likelihood analysis are potentially unreliable.
While the LAT collaboration is considering possible modifications to the event
reconstruction and selection aimed at mitigating the problem, these Bad Time
Intervals (BTIs)\acronymlabel{BTI} are being systematically identified and
flagged.

\begin{figure}[htbp]
  \centering
  \includegraphics[width=\onecolfigwidth]{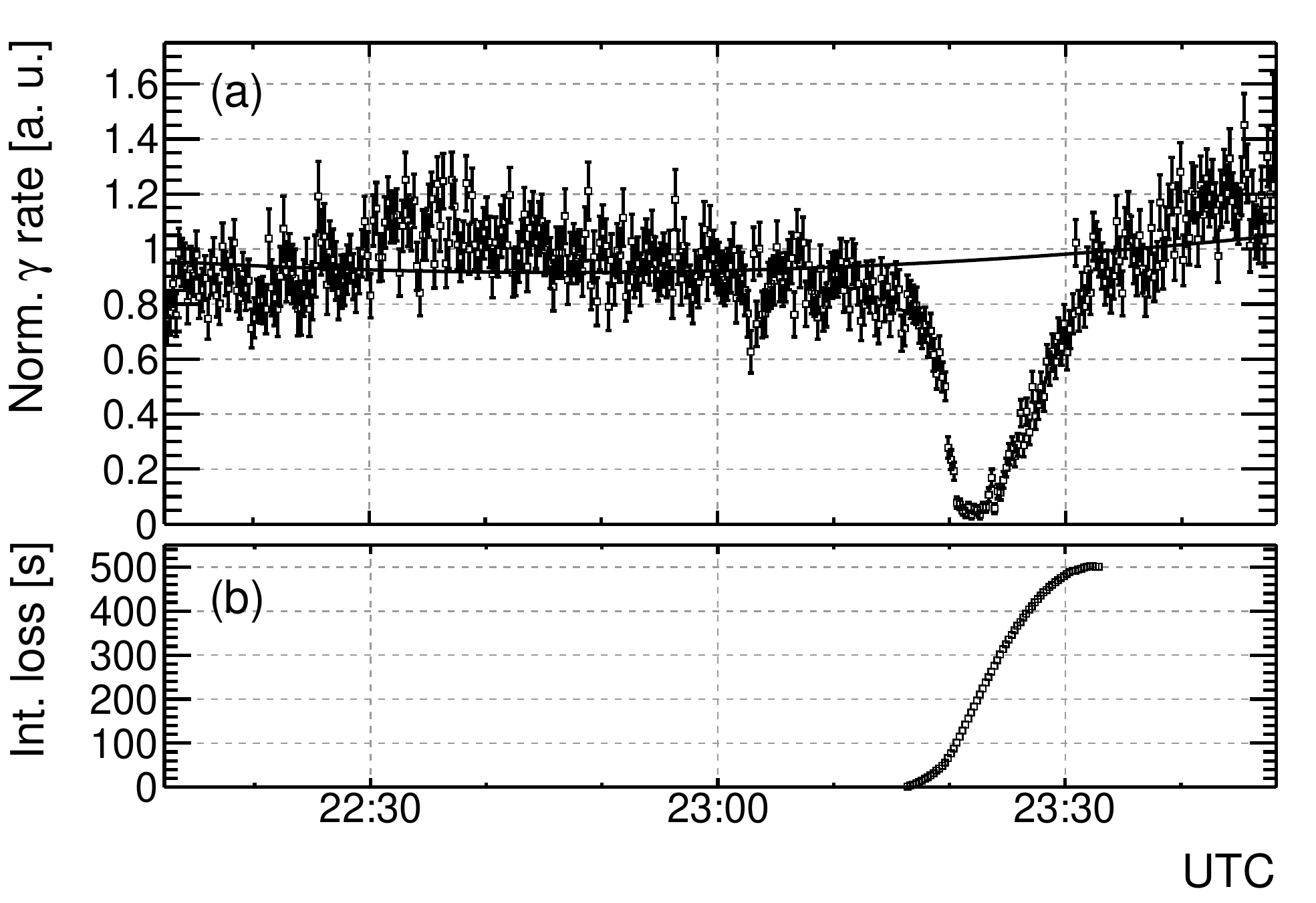}
    \caption{(a) Normalized rate of \irf{P7TRANSIENT} events for the same time
      interval shown in~\figref{bti_seed} and (b) integrated time loss over
      the candidate BTI in~\figref{bti_seed} (b). Each data
      point corresponds to a 15~s time bin. The solid line is a second order
      polynomial fit to the normalized rate outside the candidate BTI and it is
      used to calculate the integrated time loss, as defined
      in~\Eqref{eq:int_time_loss}.}
  \label{fig:bti_trans_rate}
\end{figure}

Operationally a BTI is characterized by a suppression of the rate of events in
the standard \gammaRayHyph\ classes. The \gammaRayHyph\ rates intrinsically
feature large orbital variations depending on both the geomagnetic environment
(through the background leakage and the ghost effect) and the rocking angle of
the observatory (through the change of the arc length of the Earth limb in the
\fov). These variations can be parametrized to an accuracy of 20--30\%
and in fact are accounted for (at a similar level of accuracy) in event rates
normalized to predicted values that are routinely 
accumulated in 15~s time bins for data monitoring purposes (see e.g., \figref{bti_trans_rate}).
Still, the residual variations (especially during rocking maneuvers or
non-standard pointings) make the normalized rates not directly suitable for
identifying the BTIs. We use a two-step procedure instead, in which:
\begin{fermiitemize}
\item[(i)] within each orbit we search for evidence of significant
  X-ray pile-up activity in the ACD in order to define a list of candidate BTIs
  (see~\figref{bti_seed});
\item[(ii)] we search for a temporally coincident decrease in the transient
  \gammaRayHyph\ rate within each of those intervals
  (see~\figref{bti_trans_rate}).
\end{fermiitemize}

As shown in \figref{bti_seed}, X-ray pile-up induced by bright flares is
typically visible both in the normalized ACD hit multiplicity and in
the single tile hit occupancy---particularly in tile 63, which is the largest
tile on the $+x$ side of the observatory (i.e., the side facing the Sun).
Though the correlation with the X-ray radiance measured by GOES is far from
perfect (with a zero-suppression threshold of $\sim 100$~keV the ACD is
sensitive to somewhat higher energies than GOES), both quantities are good
proxies for the increase in the X-ray flux.
\figref{bti_trans_rate} (a) shows the corresponding large suppression of the 
normalized \irf{P7TRANSIENT} rate for the same solar flare.
As shown in~\figref{bti_trans_rate} (b), by fitting this normalized
rate outside the candidate BTI it is also possible to define an
\emph{integrated time loss} $T_{\rm loss}$ corresponding to a particular
candidate BTI
\begin{equation}\label{eq:int_time_loss}
  T_{\rm loss} = \sum_{\rm BTI} (f_i - r_i)\Delta t
\end{equation}
(where $f_i$ and $r_i$ are the fitted and measured normalized rates in the
$i$-th time bin, respectively, and $\Delta t$ is the bin width, i.e., 15~s).
Roughly speaking, if $T_{\rm loss}$ is larger than a few minutes we will mark
the relevant time interval as bad.

All LAT \gammaRayHyph\ data automatically have the \texttt{DATA\_QUAL}
field in the spacecraft pointing and live time history file
(spacecraft file)\footnote{
\webpage{http://fermi.gsfc.nasa.gov/ssc/data/analysis/documentation/Cicerone/Cicerone\_Data/LAT\_Data\_Columns.html\#SpacecraftFile}
} set to 1 by the data processing
system and are immediately exported to the FSSC where they become publicly
available.  The potential for BTIs is reviewed only after the fact.
This is because such a high percentage of the LAT data is
good and making them publicly available as quickly as possible is a
priority.

\begin{figure}[htbp]
  \centering
  \includegraphics[width=\onecolfigwidth]{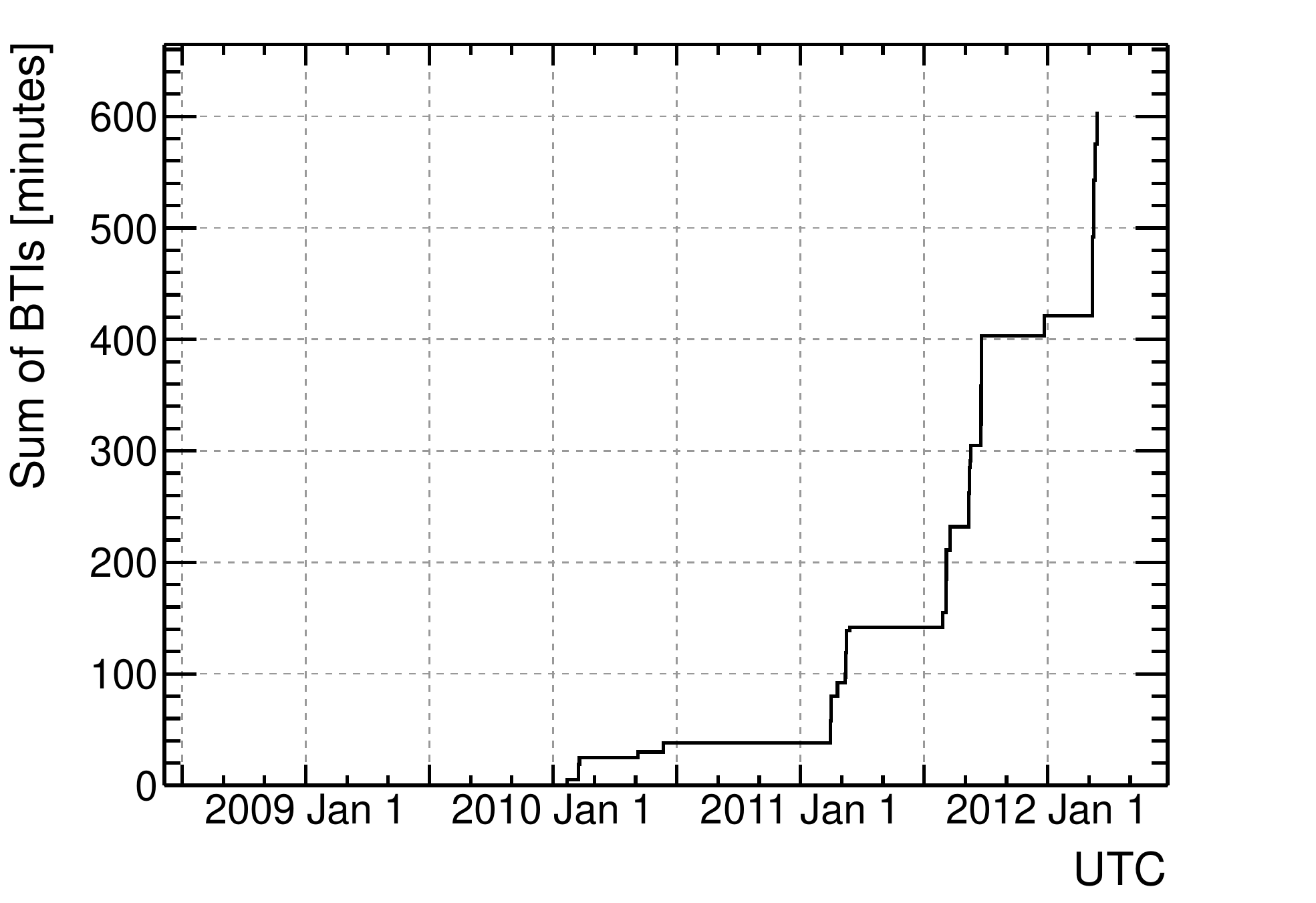}
    \caption{Time history of the sum of BTIs induced by SFs. The sum of
      time intervals marked as \emph{BAD} as of March 2012 is $\sim 600$~minutes
      (i.e., $\sim 10$ hours).}
  \label{fig:bti_list}
\end{figure}

For each flare the entire information available is reviewed manually and if
there is a significant loss of efficiency the corresponding time period is
marked as \emph{BAD} (specifically, the \texttt{DATA\_QUAL} field is set to
$-1$) and a new spacecraft file is generated with the appropriate data quality
flag in the corresponding 30 second time bins. This new file is then exported
to the FSSC and supersedes the original file.
As of 2012 March the sum of time intervals marked as \emph{BAD} is $\sim 10$
hours (\parenfigref{bti_list}).

\clearpage

\section{LIST OF ACRONYMS AND ABBREVIATIONS}\label{app:acronym}

\acronym{2FGL}{\fermi\ LAT Second Source Catalog}
\acronym{ACD}{LAT Anti-Coincidence Detector subsystem}
\acronym{AGN}{Active Galactic Nucleus}
\acronym{BTI}{Bad Time Interval}
\acronym{BSPR}{Blind Search Pattern Recognition}
\acronym{CAL}{LAT imaging Calorimeter subsystem}
\acronym{CNO}{Carbon, Nitrogen, Oxygen [CR species]}
\acronym{CPF}{Charged Particle in the Field of view analysis}
\acronym{CR}{Cosmic Ray}
\acronym{CSPR}{Calorimeter-Seeded Pattern Recognition}
\acronym{CT}{Classification Tree}
\acronym{CU}{LAT Calibration Unit}
\acronym{EGRET}{Energetic Gamma-Ray Experiment Telescope}
\acronym[\fermi]{Fermi}{\emph{Fermi Gamma-ray Space Telescope}}
\acronym[\fov]{FOV}{Field of view}
\acronym{FSSC}{\Fermi\ Science Support Center}
\acronym{GRB}{Gamma-Ray Burst}
\acronym{GSI}{Gesellschaft f\"ur SchwerIonenforschung}
\acronym{IRF}{Instrument Response Function}
\acronym{ISOC}{Instrument Science Operations Center}
\acronym{LAT}{\Fermi\ Large Area Telescope}
\acronym{LH}{Maximum LikeliHood [energy estimation algorithm]}
\acronym{MC}{Monte Carlo}
\acronym{MET}{Mission Elapsed Time}
\acronym{MIP}{Minimum Ionizing Particle}
\acronym{PC}{Parametric correction [energy estimation algorithm]}
\acronym[p.e.]{pe}{photo-electron}
\acronym{PMT}{PhotoMultiplier Tube}
\acronym{PS}{CERN Proton Synchrotron}
\acronym{PSF}{Point-Spread Function}
\acronym[ROI]{\roi}{Region of Interest}
\acronym{SAA}{South Atlantic Anomaly}
\acronym{SF}{Solar Flare}
\acronym{SP}{Shower Profile [energy estimation algorithm]}
\acronym{SSD}{Silicon Strip Detector}
\acronym{SPS}{CERN Super Proton Synchrotron}
\acronym{TKR}{LAT Tracker/converter subsystem}

\clearpage

\section{NOTATION}\label{app:conventions}

\convention{\aeff}{Effective area}{aeff}
\convention{$B$, $B_{\rm a}$, $B_{\rm idx}$}{Bracketing functions for \aeff}{aeff_bracket}
\convention{$B$, $B_{68}$, $B_{r}$}{Bracketing functions for the PSF}{psf_bracket}
\convention{$b$}{Bias in the energy scale}{ebias}
\convention{$b$}{Distribution of CR background events mischaracterized as \gammaRays}{bkgDist}
\notation{$(l,b)$}{Galactic coordinates}{}
\convention{$C_{68}$}{68\% containment radius of the PSF}{psf_68_cont}
\convention{$C_{i}$}{Actual containment at a given nominal containment radius}{psf_actual_cont}
\notation{$c_{i}$}{Generic fitting constants}{}
\convention{$D$}{Energy dispersion parametrization}{edisp}
\convention{$d_{i}$}{Normalized deviations}{norm_dev}
\convention{$E$}{Energy}{edisp}
\convention{$F_{25}$}{Integral counts flux between 100~MeV and 100~GeV}{f25}
\convention{$F_l$}{Live time fraction}{livetime_frac}
\convention{$\core{f}$}{Fraction of counts in the ``core'' King function for the PSF}{king_f_core}
\convention{$\Kingf$}{King function [used to model the PSF]}{king_func}
\convention{$k$}{\aeff\ bracketing smoothing constant}{aeff_bracket_smooth_const}
\convention{$L$}{\mcilwainl}{mcilwainl}
\convention{$L_{ext}$}{Log likelihood for fit as an extended source}{L_ext}
\convention{$L_{pt}$}{Log likelihood for fit as an point source}{L_pt}
\convention{$M$}{Predicted counts distribution of \gammaRays.}{exCountsPred}
\convention{$N_{0}$}{Flux prefactor}{flux_prefactor}
\convention{$N_{\rm gen}$}{Number of events generated [when making IRFs]}{ngen}
\convention{$n$}{Number of observed counts}{nobs}
\convention{$\tilde{n}$}{Expected counts}{npred}
\convention{$P$}{Point-spread function parametrization}{psf}
\convention{$\bar{P}$}{PSF averaged over the observing profile}{psf_bar}
\convention{\probAll}{Combined estimator that the event is a \gammaRay}{probAll}
\convention{\probCAL}{Estimator that the event is a \gammaRay, based on the CAL topology analysis}{probCAL}
\convention{\probCore}{Estimator of the quality of direction measurement}{probCore}
\convention{\probCPF}{Estimator that the event is a \gammaRay, based on the CPF analysis}{probCPF}
\convention{\probE}{Estimate of the quality of energy measurement}{probE}
\convention{\probTKR}{Estimator that the event is a \gammaRay, based on the TKR topology analysis}{probTKR}
\convention{$\hat{p}$}{Incident direction in celestial reference frame}{skyDir}
\convention{$R$}{Rando function, used to model the energy dispersion}{rando}
\convention{$R$}{Ratio of efficiency for flight data to MC simulated data}{ratioEffic}
\convention{$R_{68}$}{68\% containment radius of the PSF (also $R_{95}$)}{R68}
\convention{$r$}{Ratio of size of signal region to background region}{ratioSize}
\convention{$r$}{Ratio of 95\% to 68\% containment radii of the PSF}{psfRatio}
\convention{$r_{i}^{n}$}{Normalized residuals}{residNorm}
\convention{$r_{i}^{f}$}{Fractional residuals}{residFrac}
\convention{$S$}{Source distribution of \gammaRays.}{sourceDistrib}
\convention{$S_{25}$}{Integral energy flux between 100~MeV and 100~GeV}{S25}
\convention{\edispscaling}{Energy dispersion scaling function}{edispscaling}
\convention{\psfscaling}{PSF scaling function}{psf}
\convention{$s$}{Light yield scaling function}{lightScale}
\convention{$TS_{ext}$}{Test statistic for source extension}{ts_ext}
\convention{$t_{obs}$}{Observing time}{tobs}
\convention{$\hat{v}$}{Incident direction in the LAT frame}{LAT_dir}
\convention{$x$}{Scaled angular deviation}{scaledPSFvar}
\convention{$x$}{Scaled energy redistribution}{scaledEDispvar}
\convention{$\tilde{x}$}{Rando function break point}{randoBreak}
\convention{$\alpha$}{Spectral Index in ``log-parabola'' model}{logParAlpha}
\convention{$\beta$}{Spectral curvature parameter in ``log-parabola'' model}{logParBeta}
\convention{$\beta$}{PSF energy scaling index}{psfEScale}
\convention{$\Gamma$}{Spectral index}{powerLawGamma}
\convention{$\gamma$}{King function ``tail'' parameter}{KingGamma}
\convention{$\gamma$}{Rando function ``tail'' parameter}{RandoGamma}
\convention{$\epsilon$}{Relative uncertainty in \aeff}{epsilonBracket}
\convention{$\eta_{data}, \eta_{mc}$}{Efficiencies for flight data and MC simulated data}{aeffEffic}
\convention{$(\theta,\phi)$}{Polar angle and azimuth of the incident direction in the LAT frame}{thetaPhi}
\convention{$(\theta_{z},\phi_{z})$}{Zenith angle and Earth azimuth angle}{earthCoords}
\convention{$\hat{\theta},\hat{\phi}$}{Unit vector along local $\theta$,$\phi$ directions}{localThetaPhi}
\convention{$\xi$}{Folded $\phi$}{foldedPhi}
\convention{$\sigma$}{King function width parameter}{KingSigma}
\convention{$\sigma$}{Rando function width parameter}{RandoGamma}
\convention{$\tau_{n}$}{Correlation metric}{tauCorrel}

\clearpage

\bibliography{Pass7Validation}

\end{document}